\documentclass[12pt, twoside,openright,a4paper]{book}
\usepackage[a4paper,left=3cm,right=3cm,top=2cm,bottom=2cm,includehead,includefoot]{geometry}
\usepackage[latin1]{inputenc}
\usepackage{amssymb, amsmath}
\usepackage[curve]{xypic}
\usepackage{color}

\providecommand{\abs}[1]{\lvert#1\rvert}

\providecommand{\norm}[1]{\lVert#1\rVert}

\providecommand{\Spin}{\textnormal{Spin}}
\providecommand{\Cl}{\textnormal{Cl}}
\providecommand{\CCl}{\mathbb{C}\textnormal{l}}

\providecommand{\Tr}{\textnormal{Tr}}
\providecommand{\Ob}{\textnormal{Ob}}
\providecommand{\id}{\textnormal{id}}

\providecommand{\Ker}{\textnormal{Ker}}
\providecommand{\IIm}{\textnormal{Im}}

\providecommand{\Coker}{\textnormal{Coker}}

\providecommand{\Kom}{\textnormal{Kom}}

\providecommand{\Cyl}{\textnormal{Cyl}}
\providecommand{\Hom}{\textnormal{Hom}}
\providecommand{\Ker}{\textnormal{Ker}}

\providecommand{\Vect}{\textnormal{Vect}}
\providecommand{\rk}{\textnormal{rk}}
\providecommand{\Int}{\textnormal{Int}}
\providecommand{\ch}{\textnormal{ch}}
\providecommand{\PD}{\textnormal{PD}}
\providecommand{\cpt}{\textnormal{cpt}}
\providecommand{\ev}{\textnormal{ev}}
\providecommand{\odd}{\textnormal{odd}}

\providecommand{\Top}{\textnormal{Top}}
\providecommand{\TopCW}{\textnormal{TopCW}}
\providecommand{\TopFCW}{\textnormal{TopFCW}}
\providecommand{\AbGrp}{\textnormal{AbGrp}}
\providecommand{\ExSAbGrp}{\textnormal{ExSAbGrp}}
\providecommand{\con}{\textnormal{con}}
\providecommand{\chr}{\textnormal{char}}
\providecommand{\vol}{\textnormal{vol}}
\providecommand{\Maps}{\textnormal{Maps}}
\providecommand{\SO}{\textnormal{SO}}
\providecommand{\OO}{\textnormal{O}}
\providecommand{\PP}{\textnormal{P}}
\providecommand{\Ab}{\textnormal{Ab}}
\providecommand{\Spin}{\textnormal{Spin}}
\providecommand{\Pin}{\textnormal{Pin}}
\providecommand{\Hol}{\textnormal{Hol}}
\providecommand{\Tor}{\textnormal{Tor}}

\providecommand{\pfaff}{\textnormal{pfaff}}
\providecommand{\Iso}{\textnormal{Iso}}
\providecommand{\End}{\textnormal{End}}
\providecommand{\Aut}{\textnormal{Aut}}
\providecommand{\Ad}{\textnormal{Ad}}

\begin{document}

\begin{titlepage}
\titlepage
\centerline{ \bf \LARGE Topics on Topology and Superstring Theory}
\vskip 1.5truecm

\begin{center}
{\Large Fabio Ferrari Ruffino}
\vskip 1cm
\centerline{\Large Ph.D.\ Thesis}
\vskip 1cm
{\em
International School for Advanced Studies (SISSA/ISAS) \\ 
Via Beirut 2, I-34151, Trieste, Italy\\
Mathematical Physics sector}
\vskip 1.5cm

\vspace{10pt}
\begin{tabular}{ccc}
\textsc{Supervisors} & \phantom{XXXXX} & \textsc{Referee}\\
Prof.\ Loriano Bonora & & Prof.\ Daniel S.\ Freed\\
Prof.\ Ugo Bruzzo
\end{tabular}

\vskip 2.5cm

\large \bf Abstract
\end{center}

\normalsize In this thesis we discuss some topics about topology and superstring backgrounds with D-branes. We start with a mathematical review about generalized homology and cohomology theories and the Atiyah-Hirzebruch spectral sequence, in order to provide an explicit link between such a spectral sequence and the Gysin map. Then we review the basic facts about line bundles and gerbes with connection. In the second part of the thesis we apply the previous material to study the geometry of type II superstring backgrounds. We first present the cohomological discussion about D-brane charges in analogy with classical electromagnetism, then we use the geometry of gerbes to discuss the nature of the A-field and the B-field as follows from the Freed-Witten anomaly, finally we discuss the K-theoretical approaches to classify D-brane charges. In the last part we discuss some topics about spinors and pinors with particular attention to non-orientable manifolds.

\vskip2cm

\vskip1.5\baselineskip

\vfill
 \hrule width 5.cm
\vskip 2.mm
{\small 
\noindent }
\begin{flushleft}
ferrariruffino@gmail.com
\end{flushleft}
\end{titlepage}


\newtheorem{Theorem}{Theorem}[section]
\newtheorem{Lemma}[Theorem]{Lemma}
\newtheorem{Corollary}[Theorem]{Corollary}
\newtheorem{Rmk}[Theorem]{Remark}
\newtheorem{Def}{Definition}[section]
\newtheorem{ThmDef}[Theorem]{Theorem - Defintion}

\newpage \thispagestyle{empty} $ $ \newpage

\thispagestyle{empty}

$ $ \vskip 2 cm
\centerline{\bf \Large Acknowledgments}
\vskip 2 cm
I would like to thank Loriano Bonora for having introduced me to string theory, and I thank both him and Raffaele Savelli for all the time spent working together. I also thank Ugo Bruzzo for his help with scientific discussions and in organizational matters. I am also really grateful to Jarah Evslin for all the topics he taught me about string theory and for having always replied with patience and precision to all my questions. Moreover, a special thank to all my friends in Trieste for these four years in which I never felt alone even for a moment.

\newpage \thispagestyle{empty} $ $ \newpage

\chapter*{Introduction}

In this thesis we discuss some topics about topology and superstring backgrounds with D-branes. The idea is not to solve one single problem, but to deal with the topological and geometrical nature of some ``protagonists'' of superstring backgrounds, mainly for type II theories, solving the problems that appear while describing them. Actually, it happens that some topological statements can be generalized in a more abstract context, so that we are led to deal with purely mathematical topics, independently of their physical meaning. That's why the first two parts are dedicated to algebraic topology, beyond the physical usefulness of the subject, while in the third part we specialize to the string theory context. The forth part deals with pinors and spinors.

Although it may not be evident reading this work, we used Freed-Witten anomaly as a ``guide'' \cite{FW, DFM}. In fact, one of the main physical topics is the use of K-theory to study D-brane charges \cite{MM, Moore, Witten}, which is strongly motivated by Freed-Witten anomaly \cite{MMS, Evslin}, since, as we will see, one advantage of K-theory with respect to ordinary cohomology is that we can use the Atiyah-Hirzebruch spectral sequence to cut anomalous world-volumes. There are two main approaches in the literature, i.e.\ the Gysin map and the Atiyah-Hirzebruch spectral sequence, and we solved the problem of linking these two classifications of the same objects. Since Atiyah-Hirzebruch spectral sequence works for any cohomology theory (not only K-theory), we extended the result to the axiomatic setting. That's why in the first part we start with a description of homology and cohomology theories in general \cite{ES, Whitehead, Hatcher}. Moreover, the Freed-Witten anomaly is a world-sheet anomaly which can be seen from the superstring action, and to state and study it we are forced to consider the geometrical nature of the terms of this action, in particular of the A-field and the B-field. So the Freed-Witten anomaly is at the origin of the other physical topic, the geometrical classification of A-field and B-field configurations which are anomaly-free, using the geometry of gerbes. That's also the reason why in the second part of the thesis we give a description of line bundles and gerbes with the language of Deligne cohomology \cite{Brylinski}. Finally we discuss some topics about pinors and spinors, considering in particular the problem of linking pinors on a non-orientable manifold and spinors on its orientable double-covering.

\paragraph{}The work is organized as follows:
\begin{itemize}
	\item in part I we deal with the foundations of algebraic topology, discussing generalized homology and cohomology theories and presenting Borel-Moore homology, which we will need to deal with D-brane charges; moreover, we give a complete presentation of the topic of spectral sequences, in order to describe the Atiyah-Hirzebruch one, since it will be useful for physical applications; as we already said, the setting of this part is more general than the one needed in string theory;
	\item in part II we review the theory of line bundles and gerbes with connection, using the language of Deligne cohomology;
	\item in part III we apply part of the topological preliminaries to study superstring backgrounds with D-branes; the main topics are the cohomological description of D-brane charges in analogy with classical electromagnetism, the K-theoretical classification of D-brane charges, and the description and classification of A-field and B-field configurations that are anomaly-free;
	\item in part IV we discuss pinors and spinors, with particular attention to non-orientable manifolds.
\end{itemize}
The main results of the thesis are:
\begin{itemize}
	\item an explicit link between the Gysin map and the Atiyah-Hirzebruch spectral sequence for any cohomology theory, which, for the particular case of K-theory, shows how to link the two main K-theoretical classifications of D-brane charges;
	\item the use of the relative Deligne cohomology to classify the allowed configurations of the A-field and the B-field, showing in particular the different nature of the gauge theories on a D-brane;
	\item showing the appropriate version of Borel-Moore homology in order to describe D-brane charges from a cohomological point of view, considering then the analogous discussion within the K-theoretical viewpoint;
	\item the explicit link between pinors on a non-orientable manifold and spinors on its orientable double-cover which are invariant by sheet-exchange.
\end{itemize}
The presentation of all the topics involved is just partial and should be completed in different directions. One relevant example is the fact that, describing D-brane charges via K-theory and cohomology, we always consider the case of vanishing H-flux, both as a form (i.e.\ as a gerbe curvature) and as an integral cohomology class (i.e.\ as the first Chern class of a gerbe). We consider the full generality only describing the nature of the A-field and the B-field. When the H-flux is turned on, K-theory and de-Rahm cohomology must be replaced by their twisted version \cite{Kapustin, Witten}. We will consider in future works these more general situations.

\paragraph{P.S.} The results of the thesis are contained in the following articles, that I quote from the bibliography:
\begin{itemize}
	\item \cite{BFS} L.\ Bonora, F.\ Ferrari Ruffino and R.\ Savelli, \emph{Classifying A-field and B-field configurations in the presence of D-branes}, JHEP 12 (2008) 78, arXiv:0810.4291;
	\item \cite{BFS2} L.\ Bonora, F.\ Ferrari Ruffino and R.\ Savelli, \emph{Revisiting pinors, spinors and orientability}, Bollettino U.M.I.\ (9) 5 (2012), expanded version on arXiv:0907.4334;
	\item \cite{FR} F.\ Ferrari Ruffino, \emph{Gysin map and Atiyah-Hirzebruch spectral sequence}, Bollettino U.M.I.\ (9) 4 (2011), expanded version on arXiv:0904.4103;
	\item \cite{FR2} F.\ Ferrari Ruffino, \emph{Topics on the geometry of D-brane charges and Ramond-Ramond fields}, JHEP 11 (2009) 012, arXiv:0909.0689;
	\item \cite{FS} F.\ Ferrari Ruffino and R.\ Savelli, \emph{Comparing two different K-theoretical classifications of D-branes}, Journal of Geometry and Physics 61 (2011) pp.\ 191-212, arXiv:hep-th/0805.1009.
\end{itemize}
Such articles are almost entirely reproduced in this thesis. I will explicitly cite them when they appear in the following.

\tableofcontents

\part{Homology and cohomology theories}

\chapter{Foundations}

\section{Preliminaries}

\subsection{Singular homology and cohomology}

We assume the reader is familiar with the basic properties of singular homology and cohomology, which can be found in detail in \cite{Hatcher}. We briefly recall the definition. We denote by $\Delta^{n} = \{x \in \mathbb{R}^{n+1}: x_{1} + \cdots + x_{n+1} = 1, \, x_{i} \geq 0 \,\forall i\}$ the $n$-dimensional simplex with the euclidean topology. For $0 \leq k \leq n$, we denote by $(\Delta^{n})^{k}$ the $k$-th face of $\Delta^{n}$ obtained ``removing'' the $k$-vertex, i.e.\ $(\Delta^{n})^{k} = \Delta^{n} \cap \{x: x_{k+1} = 0\}$. Given a topological space $X$, we consider its set of \emph{$n$-chains} defined as the free abelian group generated by the continuous maps from $\Delta^{n}$ to $X$:
	\[C_{n}(X, \mathbb{Z}) := \bigoplus_{\{\sigma^{n}: \Delta^{n} \rightarrow X\}} \mathbb{Z}
\]
and we define a boundary operator $\partial_{n}: C_{n}(X, \mathbb{Z}) \rightarrow C_{n-1}(X, \mathbb{Z})$ as:
	\[\partial_{n}(\sigma^{n}) := \sum_{k=0}^{n} (-1)^{k} (\sigma^{n} \circ i_{k}^{n-1})
\]
where $i_{k}^{n-1}: \Delta^{n-1} \rightarrow (\Delta^{n})^{k}$ is the standard linear immersion. One can prove that $\partial_{n-1} \circ \partial_{n} = 0$, so that one can define the \emph{singular homology groups} of $X$ as:
	\[H_{n}(X, \mathbb{Z}) := \Ker \,\partial_{n} \,/\, \IIm \,\partial_{n+1}.
\]
Moreover we define the set of \emph{$n$-cochains} of $X$ as:
	\[C^{n}(X, \mathbb{Z)} := \Hom(C_{n}(X, \mathbb{Z}), \mathbb{Z}) \simeq \prod_{\{\sigma^{n}: \Delta^{n} \rightarrow X\}} \mathbb{Z}
\]
and the coboundary operator $\delta^{n}: C^{n}(X, \mathbb{Z}) \rightarrow C^{n+1}(X, \mathbb{Z})$ as:
	\[(\delta^{n}\varphi)(x) := \varphi(\partial_{n}x).
\]
The \emph{singular cohomology groups} of $X$ are therefore:
	\[H^{n}(X, \mathbb{Z}) := \Ker \,\delta^{n} \,/\, \IIm \,\delta^{n-1}.
\]

Given a pair of topological spaces $(X,A)$ with $A \subset X$ we define the \emph{relative $n$-chains} of $(X,A)$ as:
	\[C_{n}(X, A) := C_{n}(X) \,/\, C_{n}(A)
\]
and, since $\partial_{n}(C_{n}(A)) \subset C_{n-1}(A)$, we can project the boundary operator to $C_{\bullet}(X,A)$ and define the \emph{relative singular homology groups} as $H_{n}(X,A) = \Ker \, \partial_{n} \,/\, \IIm \,\partial_{n+1}$. Analogously one can define the \emph{relative cohomology groups}.

\paragraph{}The main properties of the singular homology and cohomology are homotopy invariance, the exact sequence of the couple and the excision property: we will deal with these topics discussing generalized homology and cohomology theories. For a proof in the case of singular theory the reader can see \cite{Hatcher}.

We call $\{*\}$ a fixed space with one point, unique up to isomorphism. Given a space $X$ we consider the unique map $f_{0}: X \rightarrow \{*\}$. We then define the \emph{reduced homology groups} as $\tilde{H}_{n}(X) := \Ker ((f_{0})_{*})_{n}: H_{n}(X) \rightarrow H_{n}\{*\}$. Since $H_{0}\{*\} = \mathbb{Z}$ and $H_{n}\{*\} = 0$ for $n \neq 0$, it follows that $\tilde{H}_{n}(X)$ is different from $H_{n}(X)$ only for $n = 0$, and, in this case, a map $\{*\} \rightarrow X$ (inducing $H_{0}\{*\} \rightarrow H_{0}(X)$) determines a non-canonical splitting $H_{0}(X) \simeq \tilde{H}_{0}(X) \oplus \mathbb{Z}$. Similarly, we define \emph{reduced cohomology groups} as $\tilde{H}^{n}(X) := \Coker (f_{0}^{*})_{n}: H^{n}\{*\} \rightarrow H^{n}(X)$, and we have an analogous non-canonical splitting $H^{0}(X) \simeq \tilde{H}^{0}(X) \oplus \mathbb{Z}$. Both the splittings become canonical in the category of \emph{pathwise connected} spaces, since in this case all the maps $\{*\} \rightarrow X$ are homotopic so that they induce the same map in homology and cohomology.\footnote{As we will see, if we consider the category of topological spaces \emph{with marked point} we can define reduced homology and cohomology in such a way that the splitting is always canonical, simply considering the map $\{*\} \rightarrow X$ sending $*$ in the marked point instead of $f_{0}$.}

For $S^{n}$ the $n$-dimensional sphere with the euclidean topology, we recall that $\tilde{H}_{n}(S^{n}) \simeq \tilde{H}^{n}(S^{n}) \simeq \mathbb{Z}$ for every $n \geq 0$ and $\tilde{H}_{n}(S^{k}) = \tilde{H}^{n}(S^{k}) = 0$ for $k \neq n$. Thus, given a map $f: S^{n} \rightarrow S^{n}$ it induces a map in homology $(f_{*})_{n}: \mathbb{Z} \rightarrow \mathbb{Z}$ (choosing the same isomorphism $\tilde{H}_{n}(S^{n}) \simeq \mathbb{Z}$ for both domain and codomain, i.e.\ the same orientation, otherwise it should be defined only up to a sign).
\begin{Def} The \emph{degree} of a continuous map $f: S^{n} \rightarrow S^{n}$ is the integer number $(f_{*})_{n}(1)$.
\end{Def}

\paragraph{}For a fixed abelian group $G$ we can define \emph{singular homology with coefficients in $G$} considering $C_{n}(X, G) := C_{n}(X) \otimes_{\mathbb{Z}} G$ and defining the boundary in the same way on the generators. For cohomology we define $C^{n}(X, G) := \Hom(C_{n}(X), G)$ which is equivalent to define $C^{n}(X, G) := \Hom(C_{n}(X, G), G)$. For the relative case the same construction applies.

\subsection{Borel-Moore homology and cohomology with compact support}\label{BMHomology}

In the ordinary singular homology any chain, thus any cycle, must have compact support. However, there is a suitable notion of homology, called Borel-Moore homology \cite{BM}, which takes into account also non-compact cycles, and, as we now show, it naturally appears on manifolds if we start from cohomology and we want to define the Poincar\'e dual of non compactly-supported classes. It is usually treated in the literature in the sheaf-theoretic or simplicial version, thus we give a description analogous to the one of the singular homology. This section is contained in \cite[Chap.\ 3]{FR2}, with few variations.

\subsubsection{Definition of Borel-Moore homology}

Given an $n$-chain $\sum_{\sigma^{n}} k_{\sigma^{n}}$, we define its \emph{support} as the union of the images of the $n$-simplices with non-zero coefficient, i.e.\ $\bigcup_{\sigma^{n} \,\vert\, k_{\sigma^{n}} \neq 0} \sigma^{n}(\Delta^{n})$. The fact that only finitely many coefficients are non-zero implies that the support of any chain is \emph{compact}. In particular, the support of a cycle is compact without boundary.\footnote{In general the support is not a manifold, it can have singularities. Actually, it can happen that there are homology classes in a smooth manifold which have no representatives made by smooth submanifolds \cite{BHK}.} Thus, for example in $\mathbb{R}^{2}$, the circle $S^{1}$ is the support of some homology cycles (for example, the one obtained triangulating $S^{1}$ with two half-circles), but an infinite line, e.g.\ one of the two coordinate-axes, is not.

There is a different version of homology, called \emph{Borel-Moore homology}, which takes into account also non-compact cycles. To define it, one might think that the right solution is to define chains using the direct product instead of the direct sum (the difference between direct sum and direct product is briefly recalled in the appendix \ref{DirectSumProd}), but in this way we would have no control on the geometry of their support: for example, any subset $A \subset X$, also very irregular, should be the support of a $0$-chain, e.g.\ the one defined giving the coefficient $1$ to the points of $A$ and $0$ to the points of $X \setminus A$ (actually, any subset should be the support of an $n$-chain for any $n$, since one can always consider $n$-simplices whose image is one point). Moreover, in this case we could not define the boundary operator: in fact, let us suppose in $\mathbb{R}^{2}$ to give coefficient $1$ to the $1$-simplices made by the rays of the disc $D^{2}$ (or to infinitely many of them, not necessarily all) oriented from the origin to the boundary, and $0$ to all the others. In this case, applying the boundary operator the origin should have infinite coefficient, thus the boundary is not well-defined. We thus need some conditions. We give the following definitions:
\begin{Def} $ $
\begin{itemize}
	\item A \emph{generalized $n$-chain} on a topological space $X$ is an element of the direct product:
	\[C'_{n}(X, \mathbb{Z}) := \prod_{\{\sigma^{n}: \Delta^{n} \rightarrow X\}} \mathbb{Z}.
\]
	\item The \emph{support} of a generalized $n$-chain $\prod_{\sigma^{n}} k_{\sigma^{n}}$ is $\bigcup_{\sigma^{n} \,\vert\, k_{\sigma^{n}} \neq 0} \sigma^{n}(\Delta^{n})$.
	\item A generalized $n$-chain $\prod_{\sigma^{n}} k_{\sigma^{n}}$ is called \emph{locally finite} if for every $x \in X$ there exists a neighborhood $U \subset X$ of $x$ such that there exist only finitely many simplices $\sigma^{n}$ with non-zero coefficient whose image has non-empty intersection with $U$.
\end{itemize}
\end{Def}
On locally finite chains we can correctly define the boundary operator. In fact, let us consider such a chain $\prod_{\sigma^{n}} k_{\sigma^{n}}$ and let us fix an $(n-1)$-simplex $\tilde{\sigma}^{n-1}$ which lie in the boundary of some $\sigma^{n}$ with non-zero coefficient: we show that it must lie in the boundary of only finitely many of them. In fact, for every $p$ in the image of $\tilde{\sigma}^{n-1}$, we choose a neighborhood realizing the definition of local finiteness, and, being the image compact, we extract a finite subcover. We have thus found a neighborhood of the image of $\tilde{\sigma}^{n-1}$ which intersects only finitely many simplices $\sigma^{n}$ with non-zero coefficient: since any simplex intersects its own boundary, only finitely many $\sigma^{n}$-s can have $\tilde{\sigma}^{n-1}$ as boundary, so that we have no obstructions in extending the boundary operator also to infinite sums of this kind. We can now define Borel-Moore singular homology.
\begin{Def} $ $
\begin{itemize}
	\item A \emph{Borel-Moore $n$-chain} is a generalized $n$-chain which is \emph{locally finite} and has \emph{closed support}.
	\item Calling $\partial_{n}^{BM}$ the boundary operator extended to locally finite generalized $n$-chains and restricted to Borel-Moore ones, we define the \emph{Borel-Moore singular homology groups} as:
		\[H_{n}^{BM}(X, \mathbb{Z}) := \Ker \,\partial_{n}^{BM} \,/\, \IIm \,\partial_{n+1}^{BM}.
\]
\end{itemize}
\end{Def}
Let us consider $\mathbb{R}^{2}$ and a Borel-Moore cycle whose support is a line, e.g.\ the $x$-axes with a suitable triangulation. Of course it is not a cycle in ordinary homology, but if we add a point at infinity, i.e.\ we compactify $\mathbb{R}^{2}$ to $S^{2}$, the line becomes a circle in $S^{2}$, thus a cycle in the ordinary homology. This is a general fact, actually one can prove that, for $X^{+}$ the one-point compactification of a space $X$, there is a canonical isomorphism $H_{n}^{BM}(X, \mathbb{Z}) \simeq H_{n}(X^{+}, \{\infty\}; \mathbb{Z})$. Under suitable hypotheses of regularity (i.e.\ that $\{\infty\}$ is closed and a deformation retract of one of its neighborhoods), $H_{n}(X^{+}, \{\infty\}; \mathbb{Z}) \simeq \tilde{H}_{n}(X^{+}, \mathbb{Z})$. Thanks to this isomorphism we can compute more easily the Borel-Moore homology groups.

\paragraph{}We now see some examples, comparing Borel-Moore homology with the ordinary one. For $\mathbb{R}^{n}$:
	\[H_{n}^{BM}(\mathbb{R}^{n}, \mathbb{Z}) = \mathbb{Z} \qquad H_{k}^{BM}(\mathbb{R}^{n}, \mathbb{Z}) = 0 \; \forall k \neq n.
\]
This immediately follows from that fact that $(\mathbb{R}^{n})^{+} \simeq S^{n}$ so that $H_{k}^{BM}(\mathbb{R}^{n}, \mathbb{Z}) \simeq \tilde{H}_{k}(S^{n}, \mathbb{Z})$. We know that for ordinary homology the only non-zero group is $H_{0}(\mathbb{R}^{n},$ $\mathbb{Z}) = \mathbb{Z}$. The non-trivial cycle in $H_{n}^{BM}(\mathbb{R}^{n}, \mathbb{Z})$ is the whole $\mathbb{R}^{n}$ itself: if we consider an infinite triangulation of it and we give coefficient $1$ to each simplex of the triangulation we describe it as a Borel-Moore cycle, and one can show that it is not a boundary. For ordinary homology it is not a cycle since it is non-compact. Moreover, the origin (or any other point) is a non-trivial cycle in ordinary homology, that's why $H_{0}(\mathbb{R}^{n}, \mathbb{Z}) = \mathbb{Z}$. This cycle becomes trivial in the Borel-Moore homology: in fact, a half-line from the origin to infinity is a $1$-chain whose boundary is exactly the origin,\footnote{One may wonder why the origin becomes trivial in the Borel-Moore homology while, even in the one-point compactification, it remains a non-trivial cycle. The point is that to realize the isomorphism $H_{0}^{BM}(\mathbb{R}^{n}, \mathbb{Z}) \simeq \tilde{H}_{0}((\mathbb{R}^{n})^{+}, \mathbb{Z})$ a cycle in the Borel-Moore homology of $\mathbb{R}^{n}$ becomes a cycle in $(\mathbb{R}^{n})^{+}$ adding the point at infinity, as for the $x$-axes we considered in the previous example. Thus, to the origin of $\mathbb{R}^{n}$ we must also add the infinity point: we thus obtain a couple of points in $S^{n}$, which is the boundary of the segment linking them, and such a segment is exactly the completion of the half-line trivializing the origin in $\mathbb{R}^{n}$.} that's why $H_{0}^{BM}(\mathbb{R}^{n}, \mathbb{Z}) = 0$.

As another example we compute Borel-Moore homology of $\mathbb{R}^{n} \setminus \{0\}$. For this we use another isomorphism, since the one-point compactification is not a good space: if $\overline{X}$ is any compactification of $X$, under suitable hypotheses there is a canonical isomorphism $H_{n}^{BM}(X, \mathbb{Z}) \simeq H_{n}(\overline{X}, \overline{X} \setminus X; \mathbb{Z})$. We thus consider $X = \mathbb{R}^{n} \setminus \{0\}$ and $\overline{X} = S^{n}$ and we call $\overline{X} \setminus X = \{N, S\}$ thinking to north and south poles. We thus have to compute $H_{k}(S^{n}, S^{n} \setminus \{N, S\}; \mathbb{Z})$. We consider the long exact sequence:
	\[\xymatrix{
	\cdots \ar[r] & H_{k}(\{N, S\}) \ar[r] & H_{k}(S^{n}) \ar[r] & H_{k}(S^{n}, \{N, S\}) \ar[r] & H_{k-1}(\{N, S\}) \ar[r] & \cdots
}\]
We suppose $n \geq 2$. Then, for $k \geq 2$ the sequence becomes:
	\[\xymatrix{
	\cdots \ar[r] & 0 \ar[r] & H_{k}(S^{n}) \ar[r] & H_{k}(S^{n}, \{N, S\}) \ar[r] & 0 \ar[r] & \cdots
}\]
so that $H_{k}^{BM}(\mathbb{R}^{n} \setminus \{0\}, \mathbb{Z}) \simeq H_{k}(S^{n})$, i.e.\ $\mathbb{Z}$ for $k = n$ and $0$ for $2 \leq k \leq n-1$. This is different from ordinary homology in which, being $\mathbb{R}^{n} \setminus \{0\}$ homotopic to $S^{n-1}$, we have $H_{k}(\mathbb{R}^{n} \setminus \{0\}, \mathbb{Z}) = \mathbb{Z}$ for $k = n-1$ and $0$ otherwise (we are still in the case $k \geq 2$). The reason of the difference for $k = n$ is still that the whole $\mathbb{R}^{n} \setminus \{0\}$ is a cycle only in Borel-Moore homology, and it turns out that it is non-trivial. For $k = n-1$, a non-trivial cycle for ordinary homology is the sphere $S^{n-1}$ embedded in $\mathbb{R}^{n} \setminus \{0\}$, but it becomes trivial in Borel-Moore homology since it is the boundary of the chain made by the disk without the origin $D^{n} \setminus \{0\}$, which is closed in $\mathbb{R}^{n} \setminus \{0\}$ but it is not compact, thus it is a chain only in Borel-Moore homology.

We remain with the cases $k = 1$ and $k = 0$. For $k = 1$ the sequence becomes:
	\[\xymatrix{
	\cdots \ar[r] & 0 \ar[r] & 0 \ar[r] & H_{1}(S^{n}, \{N, S\}) \ar[r]^{\qquad \alpha} & \mathbb{Z} \oplus \mathbb{Z} \ar[r]^{\quad \beta} & \mathbb{Z} \ar[r] & \cdots
}\]
where the map $\beta$ is given by $\beta(n, m) = n-m$. Thus, $H_{1}^{BM}(\mathbb{R}^{n} \setminus \{0\}, \mathbb{Z}) \simeq \IIm\,\alpha = \Ker\,\beta \simeq \mathbb{Z}$. In ordinary homology $H_{1}(\mathbb{R}^{n} \setminus \{0\}, \mathbb{Z}) = 0$: the non-trivial Borel-Moore cycle is an open half-line from the origin to infinity. Finally, for $k = 0$ the sequence is:
	\[\xymatrix{
	\cdots \ar[r] & \mathbb{Z} \oplus \mathbb{Z} \ar[r]^{\quad \beta} & \mathbb{Z} \ar[r]^{\gamma \qquad\quad} & H_{0}(S^{n}, \{N, S\}) \ar[r] & 0 \ar[r] & \cdots
}\]
so that $H_{0}^{BM}(\mathbb{R}^{n} \setminus \{0\}, \mathbb{Z}) = \IIm\,\gamma$, but $\Ker\,\gamma = \IIm\,\beta = \mathbb{Z}$ so that $\gamma = 0$ thus $H_{0}^{BM}(\mathbb{R}^{n} \setminus \{0\}, \mathbb{Z}) = 0$. In ordinary homology $H_{0}(\mathbb{R}^{n} \setminus \{0\}, \mathbb{Z}) = \mathbb{Z}$: the difference is due to the fact that a point, which is non-trivial in ordinary homology, becomes the boundary of the Borel-Moore cycle made by half a line from it to infinity or from it to the origin.

\paragraph{}We now make an important remark. As one can see from the previous examples, \emph{Borel-Moore homology is not invariant under homotopy}, thus it is not an homology theory in the sense of Eilenberg and Steenrod \cite{ES}. It is invariant under homeomorphism, as one can see from the definition, thus it is a well-defined invariant of a topological space, but not up to homotopy. That's why it is less studied in the mathematical literature; however, as we will see soon, it naturally arises from Poincar\'e duality on manifolds.

\subsubsection{Borel-Moore cohomology}

We can define Borel-Moore cohomology with the same procedure of the ordinary one, i.e.\ considering $\Hom(C_{n}^{BM}(X, \mathbb{Z}), \mathbb{Z})$ and defining the coboundary operator. The cohomology we obtain, under suitable hypotheses that we now state, is the well-known \emph{cohomology with compact support}, i.e.\ the cohomology obtained restricting the boundary operator to cochains $\varphi$ such that there exists a compact subset $K_{\varphi} \subset X$ such that $\varphi$ is zero an all chains with image in $X \setminus K_{\varphi}$. We call the associated cohomology groups $H^{n}_{\cpt}(X, \mathbb{Z})$. The hypotheses we need are that $X$ is Hausdorff and that there exists a countable family of compact sets $\{K_{n}\}_{n \in \mathbb{N}}$ such that $K_{n} \subset \Int(K_{n+1})$ and $\bigcup_{n \in \mathbb{N}} K_{n} = X$. They are always satisfied if $X$ is a manifold.

To prove that compactly-supported cohomology coincides with Borel-Moore cohomology, let us consider a Borel-Moore chain $\prod_{\sigma^{n}} k_{\sigma^{n}}$ and a cochain $\varphi$ with compact support $K_{\varphi}$. Then, for every point of $K_{\varphi}$ we choose a neighborhood realizing the definition of local finiteness and, by compactness, we extract a finite subcover of $K_{\varphi}$: in this way we find a neighborhood of $K_{\varphi}$ intersecting finitely many simplices $\sigma^{n}$ with non-zero coefficient, thus $\varphi\bigl(\prod_{\sigma^{n}} k_{\sigma^{n}}\bigr)$ is well-defined. Viceversa, let us suppose that a cochain $\varphi$ is well-defined on every Borel-Moore chain and has not compact support. Let us consider a countable family of compact sets $\{K_{n}\}_{n \in \mathbb{N}}$ such that $K_{n} \subset \Int(K_{n+1})$ and $\bigcup_{n \in \mathbb{N}} K_{n} = X$. Then, we fix an $n$-simplex $\sigma^{n}_{1}$ such that $\varphi(\sigma^{n}_{1}) \neq 0$: up to change its sign, we can suppose that $\varphi(\sigma^{n}_{1}) > 0$. There exists $n_{1}$ such that $\IIm \sigma^{n}_{1} \subset K_{n_{1}}$. Then, since $\varphi$ has not compact support, we can find another simplex $\sigma^{n}_{2}$ whose image is contained in $X \setminus K_{n_{1}}$ such that $\varphi(\sigma^{n}_{2}) > 0$. Keeping on in this way, we find infinitely many disjoint simplices $\{\sigma^{n}_{k}\}_{k \in \mathbb{N}}$ such that $\varphi(\sigma^{n}_{k}) > 0$ and $\IIm \varphi(\sigma^{n}_{k}) \subset (K_{n_{k}} \setminus K_{n_{k-1}})$. Being them disjoint $\prod_{k} \sigma^{n}_{k}$ is locally finite; we now prove that it is closed. Let us fix $x$ in the complement of the support: there exists $k$ such that $x \in \Int(K_{n_{k}}) \setminus K_{n_{k-2}}$, and the latter is open. In $\Int(K_{n_{k}}) \setminus K_{n_{k-2}}$ there are two simplices, so that their image is closed (since it is compact and $X$ is Hausdorff), so there exists a neighborhood of $x$ contained in the complement. Hence the complement is open so that the support is closed. Therefore $\prod_{k} \sigma^{n}_{k}$ is a Borel-Moore cycle, but $\varphi$ has infinite value on it. That's why $\varphi$ must have compact support.

\paragraph{}For a generic manifold, Poincar\'e duality links ordinary homology (whose chains have compact support) with cohomology with compact support, and Borel-Moore homology with ordinary cohomology: we can say that Poincar\'e duality respects the support. Thus, the Poincar\'e dual of a generic cohomology class is naturally a Borel-Moore homology class. That's why Borel-Moore homology naturally appears on manifolds.

\subsubsection{Modified versions of Borel-Moore homology and cohomology}

We can introduce a suitable variation of Borel-Moore homology and compactly supported cohomology, which can be useful to describe D-brane charges. Let us consider a triple $(X, Y, r)$ where $X$ is a manifold, $Y \subset X$ a submanifold and $r: X \rightarrow Y$ a retraction (i.e.\ a surjective continuous map such that $r(y) = y \,\forall y \in Y$). We want to define a homology whose cycles are ``compact along $Y$ via $r$''. We thus give the following definition:
\begin{Def} $ $
\begin{itemize}
	\item A \emph{$(X, Y, r)$-Borel-Moore $n$-chain} is a generalized $n$-chain on $X$ which is \emph{locally finite}, has \emph{closed support} and is such that the image of its support via $r$ has \emph{compact closure} in $Y$.
	\item Calling $\partial_{n}^{BM (Y, r)}$ the boundary operator extended to locally finite generalized $n$-chains and restricted to $(X, Y, r)$-Borel-Moore ones, we define the \emph{$(X, Y, r)$-Borel-Moore singular homology groups} as:
		\[H_{n}^{BM}(X, Y, r, \mathbb{Z}) := \Ker \,\partial_{n}^{BM (Y, r)} \,/\, \IIm \,\partial_{n+1}^{BM (Y, r)}.
\]
\end{itemize}
\end{Def}
One particular case, which will be the interesting one for D-branes, is the one in which there exists a manifold $Z$ such that $X = Z \times Y$ and $r(z, y) = y$, i.e.\ $r$ is the natural projection. In this case, since we consider cycles which are compact on $Y$, they can go at infinity only along $Z$, that's why we have a canonical isomorphism:
	\[H_{n}^{BM}(Z \times Y, Y, \pi_{Y}, \mathbb{Z}) \simeq H_{n}(Z^{+} \times Y, \{\infty\} \times Y; \mathbb{Z})
\]
or, for a generic compactification $\overline{Z}$ of $Z$, we have $H_{n}^{BM}(Z \times Y, Y, \pi_{Y}, \mathbb{Z}) \simeq H_{n}(\overline{Z} \times Y, (\overline{Z} \times Y) \setminus (Z \times Y); \mathbb{Z})$.

\paragraph{}Let us consider the example of $\mathbb{R}^{n} = \mathbb{R}^{m} \times \mathbb{R}^{n-m}$. In this case we have $H_{k}^{BM}(\mathbb{R}^{n},$ $\mathbb{R}^{n-m},$ $\pi_{n-m}, \mathbb{Z}) \simeq H_{n}(S^{m} \times \mathbb{R}^{n-m}, \{N\} \times \mathbb{R}^{n-m}; \mathbb{Z}) \simeq \tilde{H}_{k}((S^{m} \times \mathbb{R}^{n-m})/(\{N\} \times \mathbb{R}^{n-m}); \mathbb{Z})$, but since the latter space retracts on $S^{m}$ we obtain $\mathbb{Z}$ for $k = m$ and $0$ otherwise. For ordinary homology we would have $\mathbb{Z}$ for $k = 0$ and $0$ otherwise, while for the Borel-Moore homology we would have $\mathbb{Z}$ for $k = n$ and $0$ otherwise. The reason is that, for $k = n$, the whole $\mathbb{R}^{n}$ is a non-trivial Borel-Moore cycle, but it is not a cycle in the modified version (for $m < n$) since it is non-compact also in the last $(n-m)$-directions. For $k = m$, one non-trivial cycle in the modified Borel-Moore homology is $\mathbb{R}^{m} \times \{0\}$, which is not a cycle in ordinary homology since it is non-compact, and which is trivial in standard Borel-Moore homology since it is the boundary of $\{(v, w) \in \mathbb{R}^{m} \times \mathbb{R}^{n-m}: v_{i} \geq 0 \, \forall i = 1, \ldots,m\}$; the latter is not a chain in modified Borel-Moore homology since it non-compact also in the last $(n-m)$ directions, thus it does not make the cycle $\mathbb{R}^{m} \times \{0\}$ trivial in this case. For $k = 0$, the origin, which is a non trivial cycle in ordinary homology, becomes trivial also in the modified Borel-Moore homology: it is enough to take a half-line going to infinity along the first $k$ directions, e.g.\ on the first $k$ coordinate half-axes.

\paragraph{Remark:} We must ask that the projection has compact closure since in general is not closed. For example in $\mathbb{R} \times \mathbb{R}$ the graph of the function $y = \tan(x)$ for $x \in (-\frac{\pi}{2}, \frac{\pi}{2})$ has open projection on the first factor. However, since the closure of a set contains its boundary points, that fact of having compact closure is the right translation of the idea of not to go to infinity along $Y$.

\paragraph{}The cohomological version of this modified theory is defined analogously and it coincides with the cohomology which compact support along $Y$ via $r$. The proof is the same considered for the general case. In particular, Poincar\'e duality gives an isomorphism between modified Borel-Moore homology and cohomology.

\subsubsection{Borel-Moore homology and currents}

Since Borel-Moore homology is isomorphic to ordinary cohomology via Poincar\'e duality, it is also isomorphic to the cohomology of currents, and the same for the modified version. We analyze this isomorphism in more detail. We recall \cite{GH} that, for $X$ an $n$-dimensional manifold, there are two isomorphisms:
	\[\begin{array}{ll}
	\varphi_{1}: &H^{k}_{dR}(X) \overset{\simeq}\longrightarrow H^{k}_{crn}(X)\\
	&[\omega] \longrightarrow [T_{\omega}] \\ \\
	\varphi_{2}: &H_{n-k}(X, \mathbb{R}) \overset{\simeq}\longrightarrow H^{k}_{crn,cpt}(X)\\
	&[\Gamma] \longrightarrow [\delta(\Gamma)]
\end{array}\]
where $T_{\omega}(\varphi) := \int_{X}(\omega \wedge \varphi)$ and $\delta(\Gamma)(\varphi) := \int_{\Gamma}\varphi$ for $\varphi$ compactly-supported $(n-k)$-form. It is easy to verify that $\varphi_{1}^{-1} \circ \varphi_{2}: H_{n-k}(X, \mathbb{R}) \rightarrow H^{k}_{dR,cpt}(X)$ is exactly the Poincar\'e duality. The previous isomorphisms means that currents encodes both homology and cohomology: for example, a $\delta$-current supported over a cycle can be identified both with its support, which is a homology cycle, or with an approximating sequence of bump forms picked over such a supports, which are all cohomologous and determine the Poincar\'e dual of the support.

Of course a $\delta$-current can be picked also over a non-compact cycle, since, being the test form $\varphi$ compactly-supported by definition, the integral is well-defined. That's why currents are more naturally associated to Borel-Moore cycle, i.e.\ we can extend $\varphi_{2}$ to:
	\[\begin{array}{ll}
	\varphi_{2}^{BM}: &H_{n-k}^{BM}(X, \mathbb{R}) \overset{\simeq}\longrightarrow H^{k}_{crn}(X)\\
	&[\Gamma] \longrightarrow [\delta(\Gamma)]
\end{array}\]
and the fact that this is an isomorphism means that every current is cohomologous to a $\delta$-current over a Borel-Moore cycle. The isomorphism $\varphi_{2}^{BM}$ can be defined without problems for the modified versions, assuming in both the l.h.s.\ and the r.h.s.\ the suitable compactness hypotheses.

\subsection{CW-complexes}

We denote by $D^{n} = \{x \in \mathbb{R}^{n+1}: \norm{x} \leq 1\}$ the $n$-dimensional unit disc with the euclidean topology.

\begin{Def} A \emph{CW-complex} is a topological space $X$ obtained inductively from a sequence of sets $\{A_{0}, A_{1}, \ldots\}$ in the following way:
\begin{itemize}
	\item we declare $X^{0} = A_{0}$ with the discrete topology;
	\item given inductively $X^{n-1}$, for $\alpha \in A_{n}$ and $D^{n}_{\alpha} = D^{n} \times \{\alpha\}$ we give a map $\varphi_{\alpha}: \partial D^{n}_{\alpha} \rightarrow X^{n-1}$ and we declare:
	\[X^{n} = \frac{X^{n-1} \sqcup_{\alpha \in A_{n}} D^{n}_{\alpha}}{x \sim \varphi_{\alpha}(x) \, \forall x \in \partial D^{n}_{\alpha} \, \forall \alpha \in A_{n}} \; ;
\]
	\item either $A_{i} = \emptyset$ for $i \geq n$, so that we declare $X = X^{n}$, or we consider $X = \bigcup_{n} X^{n}$ and declare $A \subset X$ open if and only if $A \cap X^{n}$ is open in $X^{n}$ for every $n$.
\end{itemize}
\end{Def}
Given a CW complex, we can define its \emph{cellular homology}. In particular we consider:
	\[C_{n}(X) = \bigoplus_{\alpha \in A_{n}}\mathbb{Z} \qquad \partial_{n}(1 \times \{\alpha\}) = \sum_{\beta \in A_{n-1}} \,\langle D^{n}_{\alpha}, D^{n-1}_{\beta}\rangle\, (1 \times \{\beta\})
\]
for $\langle D^{n}_{\alpha}, D^{n-1}_{\beta}\rangle$ defined in the following way:
\begin{itemize}
	\item we consider $S^{n-1}_{\alpha} = \partial D^{n}_{\alpha}$ and the map $\varphi_{\alpha}: S^{n-1}_{\alpha} \rightarrow X^{n-1}$;
	\item we consider the quotient $X^{n-1} / (X^{n-1} \setminus \Int  D^{n-1}_{\beta})$, which is homeomorphic to a sphere $S'^{\,n-1}_{\beta}$;
	\item we thus consider the composite map $\psi_{\alpha\beta}: S^{n-1}_{\alpha} \rightarrow S'^{\,n-1}_{\beta}$ and define $\langle D^{n}_{\alpha}, D^{n-1}_{\beta}\rangle$ as the degree of $\psi_{\alpha\beta}$.
\end{itemize}
Thus we get the \emph{$n$-th cellular homology group} as $H_{n}(X) = \Ker \, \partial_{n} / \IIm \, \partial_{n-1}$. Dualizing $C_{n}(X)$ and defining the coboundary as in the singular case we can define the \emph{cellular cohomology}. For $A \subset X$ a sub-CW-complex, i.e.\ a subspace defined from subsets $A'_{i} \subset A_{i}$ using the same attaching maps of $X$, we can define \emph{relative cellular (co)homology groups} as in the singular case. Moreover, for an abelian group $G$ we can define \emph{cellular (co)homology with coefficients in $G$} as in the singular case.

\subsection{Simplicial complexes}

We denote by $\Delta^{n} = \{x \in \mathbb{R}^{n+1}: x_{1} + \cdots + x_{n+1} = 1, \, x_{i} \geq 0 \,\forall i\}$ the $n$-dimensional simplex with the euclidean topology. For $0 \leq k \leq n$, we denote by $(\Delta^{n})^{k}$ the $k$-th face of $\Delta^{n}$ obtained ``removing'' the $k$-vertex, i.e.\ $(\Delta^{n})^{k} = \Delta^{n} \cap \{x: x_{k+1} = 0\}$.

\begin{Def} A \emph{simplicial complex} is a topological space $X$ which can be expressed as a union $X = \bigcup_{n \in \mathbb{N}} \bigcup_{i \in A_{n}} X^{n}_{i}$, such that:
\begin{itemize}
	\item $X^{n}_{i}$ is the image via an embedding $\varphi^{n}_{i}: \Delta^{n} \rightarrow X$ of the interior $\Int(\Delta^{n})$; we call $\Delta^{n}_{i} = \varphi^{n}_{i}(\Delta^{n})$;
	\item for every $m$-dimensional face $(\Delta')^{m}_{i} \subset \Delta^{n}_{i}$ there exists $j \in A_{m}$ such that $\varphi^{n}_{i}({\Delta'}^{m}_{i}) = \Delta^{m}_{j}$;
	\item $\Delta^{n}_{i} \cap \Delta^{m}_{j}$ is empty or a face of both; for $j = i$, if $\Delta^{n}_{i} \cap \Delta^{m}_{i} \neq \emptyset$ then it is a \emph{proper} face of both.
\end{itemize}
\end{Def}

We define its \emph{simplicial chains} by:
	\[C_{n}(X) = \bigoplus_{A_{n}}\mathbb{Z}.
\]
For fixed $\Delta^{n}_{i}$ we denote $(\Delta_{i}^{n})^{k} = \varphi^{n}_{i}((\Delta^{n})^{k})$ as an $(n-1)$-simplex, and we denote by $f_{k}(i)$ the corresponding element of $A_{n-1}$. We thus define a \emph{boundary operator} $\partial_{n}: C_{n}(X) \rightarrow C_{n-1}(X)$ by:
	\[\partial_{n}i = \sum_{k=1}^{n+1} (-1)^{k} f_{k}(i) \,.
\]
We thus define the \emph{simplicial $n$-homology group} by $H_{n}(X) = \Ker \,\partial_{n} / \IIm \,\partial_{n+1}$. A simplicial complex is a CW-complex since $\Delta^{n}$ is homeomorphic to $D^{n}$: one can prove that cellular homology and simplicial homology coincide on simplicial complexes, since the two boundaries coincide.

With the same procedure used in the singular and cellular case, we can define \emph{simplicial cohomology groups}, \emph{relative simplicial (co)homology groups} and \emph{simplicial (co)homology groups with coefficients in $G$} for a fixed abelian group $G$.

\paragraph{}For finite simplicial complexes we can define cohomology groups in a more direct way. If we apply the boundary operator to a simplex $\Delta^{n}_{i}$ we obtain the alternated sum of the $(n-1)$-simplicies making the boundary of $\Delta^{n}_{i}$: we can analogously define a \emph{coboundary operator} in $C_{n}(X)$ that to a simplex $\Delta^{n}_{i}$ assigns the alternated sum of the $(n+1)$-simplicies containing $\Delta^{n}_{i}$ in their boundary. For $i \in A_{n}$ we define $A_{n+1}^{(i)}$ as the subset of $A_{n+1}$ made by simplicies containing $i$ in their boundary, and, for $j \in A_{n+1}^{(i)}$, we defined $\rho(j)$ as the vertex to be removed to obtain $i$, i.e.\ $f_{\rho(j)}(j) = i$. Then we define:
\begin{equation}\label{CoboundaryChains}
	\delta_{n}i = \sum_{j \in A_{n+1}^{(i)}} (-1)^{\rho(j)} j.
\end{equation}
Let us show that this coboundary operator is not different from the usual one. For finite simplicial complexes we have a canonical isomorphism:
\begin{equation}\label{IsoFiniteSC}
\begin{array}{rl}
	\eta: & C_{n}(X) \overset{\simeq}\longrightarrow C^{n}(X)\\
	& i \longrightarrow \varphi_{i}
\end{array}
\end{equation}
where $\varphi_{i}(j) = \delta_{i,j}$. This isomorphism is due to the fact that $\Hom(\mathbb{Z}^{k_{n}}, \mathbb{Z}) \simeq \mathbb{Z}^{k_{n}}$ for $k_{n} = \abs{A_{n}}$. We now prove that $\delta^{n}(\varphi_{i}) = \varphi_{\delta^{n}i}$ where in the l.h.s.\ we use the usual coboundary operator while in the r.h.s.\ we use \eqref{CoboundaryChains}. In fact, for $i \in A_{n}$:
	\[\delta^{n}(\varphi_{i})(j) = \varphi_{i}(\partial_{n}j) = \varphi_{i}\biggl( \sum_{k=1}^{n+2} (-1)^{k}f_{k}(j) \biggr) = \left\{ \begin{array}{ll} 0 & \textnormal{for } j \notin A_{n+1}(i) \\ (-1)^{\rho(j)} & \textnormal{for } j \in A_{n+1}(i) \end{array} \right.
\]
thus $\delta^{n}(\varphi_{i}) = \sum_{j \in A_{n+1}^{(i)}} (-1)^{\rho(j)} \varphi_{j}$ which corresponds to \eqref{CoboundaryChains} via \eqref{IsoFiniteSC}.

\paragraph{}For infinite simplicial complexes we do not have the isomorphism \eqref{IsoFiniteSC} but we have the analogous version $\eta': \prod_{A_{n}}\mathbb{Z} \overset{\simeq}\longrightarrow C^{n}(X)$. We can use \eqref{CoboundaryChains} also on $\prod_{A_{n}}\mathbb{Z}$ and with the same proof we see that it coincides with the usual coboundary.\footnote{In this case the single simplices and the morphisms $\varphi_{i}$ are not generators since the product is infinite, but $\delta_{n}$ is by definition extended linearly also to infinite sums, so it is enough to consider them.} One can ask what happens if we use the boundary on the direct product and the coboundary on the direct sum: we obtain Borel-Moore simplicial homology and compactly-supported simplicial cohomology. In fact, in the simplicial context, the conditions of local finiteness and closure are automatic, so that the Borel-Moore $n$-chains are simply the elements of $\prod_{A_{n}}\mathbb{Z}$.

\subsection{Categories of topological spaces}

We define the following categories:
\begin{itemize}
	\item $\Top$ (also denoted by $\Top_{1}$) is the category whose objects are topological spaces and whose morphisms are continuous maps;
	\item $\Top^{+}$ (also denoted by $\Top_{1}^{+}$) is the category whose objects are topological spaces with a marked point $(X, x_{0})$ and whose morphisms $f: (X, x_{0}) \rightarrow (Y, y_{0})$ are continuous maps $f: X \rightarrow Y$ such that $f(x_{0}) = y_{0}$;
	\item $\Top_{n}$ is the category of \emph{$n$-uples of topological spaces}, i.e.\ the category whose objects are $n$-uples of topological spaces $(X, A_{1}, \ldots, A_{n-1})$ such that $A_{n-1} \subset \cdots \subset A_{1} \subset X$, and whose morphisms $f: (X, A_{1}, \ldots, A_{n-1}) \rightarrow (Y, B_{1}, \ldots, B_{n-1})$ are continuous maps $f: X \rightarrow Y$ such that $f(A_{i}) \subset B_{i}$;
	\item $\Top_{n}^{+}$ is the category of $n$-uples of topological spaces with base-point, i.e.\ the category whose objects are $(n+1)$-uples of topological spaces $(X, A_{1}, \ldots, A_{n-1},$ $x_{0})$ such that $x_{0} \in A_{n-1} \subset \cdots \subset A_{1} \subset X$, and whose morphisms $f: (X, A_{1}, \ldots, A_{n-1}, x_{0}) \rightarrow (Y, B_{1}, \ldots, B_{n-1}, y_{0})$ are continuous maps $f: X \rightarrow Y$ such that $f(A_{i}) \subset B_{i}$ and $f(x_{0}) = y_{0}$.
\end{itemize}
There are natural fully faithful embeddings of categories:
\begin{equation}\label{TopEmbeddings}
	\Top_{1} \hookrightarrow \Top_{1}^{+} \hookrightarrow \Top_{2} \hookrightarrow \Top_{2}^{+} \hookrightarrow \cdots
\end{equation}
where the embeddings $\Top_{n}^{+} \hookrightarrow \Top_{n+1}$ are obtained simply considering the marked point as a subspace, while the embeddings $\Top_{n} \hookrightarrow \Top_{n}^{+}$ can be defined sending $(X, A_{1}, \ldots, A_{n-1})$ to $(X \sqcup \{\infty\}, A_{1} \sqcup \{\infty\}, \ldots, A_{n-1} \sqcup \{\infty\}, \infty)$ and asking that the image of every morphism sends $\infty$ to $\infty$.\footnote{We remark that we consider always $X \sqcup \{\infty\}$ and not the one-point compactification $X^{+}$, even if $X$ is non-compact (otherwise they coincide). That's because we get $H_{n}(X) \simeq H_{n}(X \sqcup \{\infty\}, \infty)$ while this is not true for $X^{+}$. For example, $H_{1}(\mathbb{R}) \simeq H_{1}(\mathbb{R} \sqcup \{\infty\}, \{\infty\}) = 0$ while $H_{1}(S^{1}, \{N\}) = \tilde{H}_{1}(S^{1}) = \mathbb{Z}$.} In this way, we define by composition embeddings $\Top_{n} \hookrightarrow \Top_{n+1}$: the latter could also be defined via $(X, A_{1}, \ldots, A_{n-1}) \rightarrow (X, A_{1}, \ldots, A_{n-1}, \emptyset)$, and these two families of embeddings are equivalent for what follows; however, we always think to the one derived from \eqref{TopEmbeddings}.

We consider the following functors for $n \geq 2$:
\begin{equation}\label{FunctPiN}
	\Pi_{n}: \Top_{n} \longrightarrow \Top_{n-1}
\end{equation}
given by $\Pi_{n}(X,A_{1}, \ldots, A_{n-1}) = (A_{1}, \ldots, A_{n-1})$ and $\Pi_{n}(f: (X,A_{1}, \ldots, A_{n-1}) \rightarrow (Y, B_{1}, \ldots,$ $B_{n-1})) = (f\vert_{A_{1}}: (A_{1}, \ldots, A_{n-1}) \rightarrow (B_{1}, \ldots, B_{n-1}))$.

\begin{Def}\label{Lattice} The \emph{lattice} of a pair $(X,A) \in \Ob(\Top_{2})$ is the following diagram:
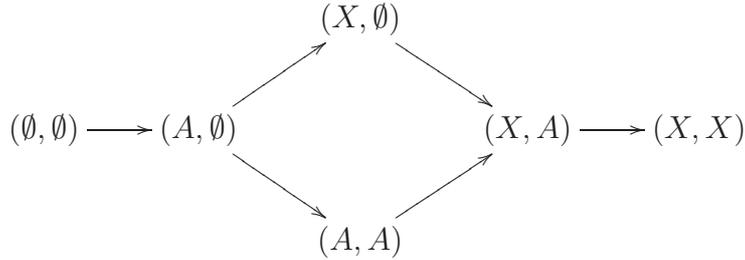
\begin{displaymath}
\xymatrix{
& & (X, \emptyset) \ar[dr] & \\
(\emptyset, \emptyset) \ar[r] & (A, \emptyset) \ar[ru] \ar[rd] & & (X, A) \ar[r] & (X, X)\\
& & (A,A) \ar[ru] & &
}
\end{displaymath}
where all the maps are the natural inclusions.
\end{Def}

\paragraph{}We now consider the categories $\TopCW_{i}$ which are defined as $\Top_{i}$ asking that the $n$-uple $(X, A_{1}, \ldots, A_{n-1})$ is homotopically equivalent to a $CW$-$n$-uple and that $X$ is compactly generated. We put no constraints on the maps, since every map is homotopic to a cellular map. If we ask that the spaces involved are homotopically equivalent to a \emph{finite} CW-complex, we obtain the categories $\TopFCW_{i}$. We thus have a diagram with fully faithful embeddings:
	\[\xymatrix{
	\Top \ar@{^(->}[r] & \Top^{+} \ar@{^(->}[r] & \Top_{2} \ar@{^(->}[r] & \Top_{2}^{+} \ar@{^(->}[r] & \cdots\\
	\TopCW \ar@{^(->}[r] \ar@{^(->}[u] & \TopCW^{+} \ar@{^(->}[r] \ar@{^(->}[u] & \TopCW_{2} \ar@{^(->}[r] \ar@{^(->}[u] & \TopCW_{2}^{+} \ar@{^(->}[r] \ar@{^(->}[u] & \cdots\\
	\TopFCW \ar@{^(->}[r] \ar@{^(->}[u] & \TopFCW^{+} \ar@{^(->}[r] \ar@{^(->}[u] & \TopFCW_{2} \ar@{^(->}[r] \ar@{^(->}[u] & \TopFCW_{2}^{+} \ar@{^(->}[r] \ar@{^(->}[u] & \cdots
	}
\]

\subsection{Basic operations on topological spaces}

\begin{Def} Given a topological space $X$ we define (endowing each real interval with the euclidean topology):
\begin{itemize}
	\item the \emph{cone} of $X$ as $CX = X \times [0,1] \,/\, X \times \{1\}$;
	\item the \emph{unreduced suspension} of $X$ as $\hat{S}X = (X \times [0,1] \,/\, X \times \{1\}) \,/\, X \times \{0\}$;
	\item the \emph{cylinder} of $X$ as $\Cyl(X) = X \times [0,1]$.
\end{itemize}
For a space with marked point $(X, x_{0})$ we define the \emph{reduced suspension} as $SX = \hat{S}X \,/\, \{x_{0}\} \times [0,1]$.
\end{Def}

Given a couple of spaces with marked point $(X, x_{0})$ and $(Y, y_{0})$, we define:
	\[X \vee Y = X \times \{y_{0}\} \cup Y \times \{x_{0}\}
\]
and of course $X \vee Y \subset X \times Y$. We then define:
	\[X \wedge Y = X \times Y \,/\, X \vee Y.
\]
It is easy to prove that $\wedge$ is associative and commutative up to homeomorphism. If we consider the sphere $S^{n} = \partial D^{n}$ with marked point $\{N\} = (0, \ldots, 0, 1)$ (thus, we do not embed $S^{n}$ from $\Top$ to $\Top^{+}$ but we mark one of its points), it is easy to verify the homeomorphisms:
	\[S^{1} \wedge \cdots \wedge S^{1} \simeq S^{n} \qquad S^{n} \wedge X \simeq S^{n}X
\]
(where $S^{n}X = S\cdots SX$ iterated $n$ times).

\paragraph{}We have thus defined the following functors (in all of the cases the maps are trivially extended on $X \times [0,1]$ and projected to the quotient):
\begin{itemize}
	\item $C: \Top_{n} \rightarrow \Top_{n}^{+}$ considering the vertex as the marked point;
	\item $\hat{S}: \Top_{n} \rightarrow \Top_{n}$;
	\item $\Cyl: \Top_{n} \rightarrow \Top_{n}$;
	\item $\vee: \Top_{n}^{+} \times \Top_{n}^{+} \rightarrow \Top_{n}^{+}$ considering as marked point the cross product of the two marked points;
	\item $\wedge: \Top_{n}^{+} \times \Top_{n}^{+} \rightarrow \Top_{n}^{+}$ considering as marked point $X \vee Y / X \vee Y$;
	\item $S^{n}: \Top_{n}^{+} \rightarrow \Top_{n}^{+}$.
\end{itemize}

\section{Eilenberg-Steenrod axioms}

We review the axioms of homology and cohomology, following mainly \cite{ES}, \cite{Hatcher} and \cite{Bredon}. Homology and cohomology theories are defined for appropriate subcategories of $\Top_{2}$, as the following definition states:
\begin{Def}\label{AdmCat} A subcategory $\mathcal{A} \subset \Top_{2}$ is called \emph{admissible for homology and cohomology theories} if the following conditions are satisfied:
\begin{enumerate}
	\item if $(X, A) \in \Ob(\mathcal{A})$, then all the pairs and maps of the lattice\footnote{See definition \ref{Lattice}.} of $(X,A)$ belong to $\mathcal{A}$;
	\item if $(X, A) \in \Ob(\mathcal{A})$ and $I = [0,1]$ with the euclidean topology, then $(X \times I, A \times I) \in \Ob(\mathcal{A})$ and the maps $g_{0}, g_{1}: (X, A) \rightarrow (X \times I, A \times I)$ given by $g_{0}(x) = (x, 0)$ and $g_{1}(x) = (x, 1)$ belongs to the morphisms of $\mathcal{A}$;
	\item there is in $\mathcal{A}$ at least one space $\{*\}$ made by a single point, and for any such space $P \in \Ob(\mathcal{A})$ and any $X \in \Ob(\mathcal{A})$ all the maps $f: P \rightarrow X$ are morphisms in $\mathcal{A}$.
\end{enumerate}
\end{Def}

\begin{Def} $ $
\begin{itemize}
	\item Let $\mathcal{A}$ be an admissible category and $f,g: (X, A) \rightarrow (Y,B)$ two morphisms. Then $f$ and $g$ are called \emph{homotopic} ($f \simeq g$) if there exists a morphism $F: (X \times I, A \times I) \rightarrow (Y , B)$ in $\mathcal{A}$ such that\footnote{The map $F(x,0)$ is surely in $\mathcal{A}$ since it is $F \circ g_{0}$ (see axiom 2 of admissible category). The same for $F(x,1)$.} $F(x, 0) = f(x)$ and $F(x, 1) = g(x)$.
	\item Two couples $(X, A)$ and $(Y, B)$ are called \emph{homotopically equivalent} if there exist two morphisms $f: (X,A) \rightarrow (Y, B)$ and $g: (Y, B) \rightarrow (X, A)$ in $\mathcal{A}$ such that $g \circ f \simeq 1_{(X,A)}$ and $f \circ g \simeq 1_{(Y,B)}$.
	\item A map $f: (X, A) \rightarrow (Y, B)$ \emph{induces a homotopy equivalence} if there exists $g: (Y, B) \rightarrow (X,A)$ such that $f$ and $g$ verify the definition of homotopy equivalence.
\end{itemize}
\end{Def}

\paragraph{}We now state the axioms for homology. We call $\AbGrp$ the category of abelian groups and $\ExSAbGrp$ the category of exact sequences of abelian groups.

\begin{Def} A \emph{homology theory} on an admissible category $\mathcal{A}$ is a sequence of functors $h_{n}: \mathcal{A} \rightarrow \AbGrp$ and morphisms of functors $\beta_{n}: h_{n} \rightarrow h_{n-1} \circ \Pi_{2}$ satisfying the following axioms:
\begin{enumerate}
	\item \emph{(Homotopy axiom)} if $f,g: (X, A) \rightarrow (Y,B)$ are homotopic in $\mathcal{A}$, then $h_{n}(f) = h_{n}(g)$;
	\item \emph{(Excision axiom)} if $(X, A) \in \Ob(\mathcal{A})$ and $U \subset A$ is open and such that $\overline{U} \subset \Int(A)$, and if the inclusion $i: (X \setminus U, A \setminus U) \rightarrow (X,A)$ is a morphism in $\mathcal{A}$, then $i_{*}: h_{*}(X, A) \rightarrow h_{*}(X \setminus U, A \setminus U)$ is a (canonical) isomorphism;
	\item \emph{(Exactness axiom)} the sequences $h_{n}$ and $\beta_{n}$ induce a functor:
	\[h_{*}: \mathcal{A} \longrightarrow \ExSAbGrp
\]
assigning to each pair $(X,A)$ the exact sequence:
	\[\xymatrix{
	\cdots \ar[r] & h_{n}(A) \ar[r]^{(i_{*})_{n}} & h_{n}(X) \ar[r]^{(\pi_{*})_{n}} & h_{n}(X,A) \ar[r]^{\beta_{n}} & h_{n-1}(A) \ar[r] & \cdots
	}
\]
where $(i_{*})_{n}$ and $(\pi_{*})_{n}$ are the image via $h_{n}$ of the inclusions\footnote{Such inclusions are morphisms in $\mathcal{A}$ by the axiom 1 of admissible categories.} $i: (A, \emptyset) \rightarrow (X, \emptyset)$ and $\pi: (X, \emptyset) \rightarrow (X,A)$.
\end{enumerate}
\end{Def}

Reversing the arrows of the exact sequence, we have the corresponding axioms for cohomology:
\begin{Def} A \emph{cohomology theory} on an admissible category $\mathcal{A}$ is a sequence of \emph{contravariant} functors $h^{n}: \mathcal{A} \rightarrow \AbGrp$ and morphisms of functors $\beta^{n}: h^{n} \circ \Pi_{2} \rightarrow h^{n+1}$ satisfying the following axioms:
\begin{enumerate}
	\item \emph{(Homotopy axiom)} if $f,g: (X, A) \rightarrow (Y,B)$ are homotopic in $\mathcal{A}$, then $h^{n}(f) = h^{n}(g)$;
	\item \emph{(Excision axiom)} if $(X, A) \in \Ob(\mathcal{A})$ and $U \subset A$ is open and such that $\overline{U} \subset \Int(A)$, and if the inclusion $i: (X \setminus U, A \setminus U) \rightarrow (X,A)$ is a morphism in $\mathcal{A}$, then $i^{*}: h^{*}(X, A) \rightarrow h^{*}(X \setminus U, A \setminus U)$ is a (canonical) isomorphism;
	\item \emph{(Exactness axiom)} the sequences $h^{n}$ and $\beta^{n}$ induce a functor:
	\[h^{*}: \mathcal{A} \longrightarrow \ExSAbGrp
\]
assigning to each pair $(X,A)$ the exact sequence:
	\[\xymatrix{
	\cdots \ar[r] & h^{n}(X, A) \ar[r]^{(\pi^{*})^{n}} & h^{n}(X) \ar[r]^{(i^{*})^{n}} & h^{n}(A) \ar[r]^{\beta^{n}} & h^{n+1}(X, A) \ar[r] & \cdots
	}
\]
where $(i^{*})^{n}$ and $(\pi^{*})^{n}$ are the image via $h^{n}$ of the inclusions $i: (A, \emptyset) \rightarrow (X, \emptyset)$ and $\pi: (X, \emptyset) \rightarrow (X,A)$.
\end{enumerate}
\end{Def}

\paragraph{}For both homology and cohomology one usually gives the following definition:
\begin{Def} The group $h_{0}\{*\}$ or $h^{0}\{*\}$ is called the \emph{coefficient group} of the (co)homology theory.
\end{Def}

\subsection{Reduced homology and cohomology}

Given a homology theory $h_{*}$ on $\mathcal{A}$, we consider a space $\{*\}$ of one point\footnote{Such a space exists in $\mathcal{A}$ by axiom 3 in definition \ref{AdmCat}.} and, for $X \in \Ob \mathcal{A}$, we consider the unique map $p: X \rightarrow \{*\}$: if such a map belongs to $\mathcal{A}$, the space $X$ is called \emph{collapsable}\footnote{This definition does not depend on the chosen one-point space $\{*\}$: in fact, if $\{*'\}$ is another one-point space, the unique map $\{*\} \rightarrow \{*'\}$ belongs to $\mathcal{A}$ by axiom 3 of definition \ref{AdmCat}, thus the composition $p': X \rightarrow \{*'\}$ is also in $\mathcal{A}$.}. We call $\mathcal{A}^{c}$ the full subcategory of collapsable spaces. For $X \in \Ob \mathcal{A}^{c}$, we define the \emph{reduced homology groups} of $X$ as:
	\[\tilde{h}_{n}(X) := \Ker\bigl( h_{n}(X) \overset{(p_{*})_{n}}\longrightarrow h_{n}\{*\} \bigr).
\]
For a couple $(X, A) \in \mathcal{A}$ with $X, A \in \mathcal{A}^{c}$ there is an exact sequence, called \emph{reduced homology exact sequence}:
	\[\xymatrix{
	\cdots \ar[r] & \tilde{h}_{n}(A) \ar[r]^{(i_{*})_{n}} & \tilde{h}_{n}(X) \ar[r]^{(\pi_{*})_{n}} & h_{n}(X,A) \ar[r]^{\beta_{n}} & \tilde{h}_{n-1}(A) \ar[r] & \cdots.
	}
\]
In fact, if we restrict the map $(i_{*})_{n}$ to $\tilde{h}_{n}(A)$ its image lies in $\tilde{h}_{n}(X)$, since $p^{(X)} \circ i = p^{(A)}$ so that $i_{*}(\Ker\,p^{(A)}_{*}) \subset \Ker\,p^{(X)}_{*}$. It remains to prove that the image of $\beta_{n}$ lies in $\tilde{h}_{n-1}(A)$, i.e.\ that the kernel of $(i_{*})_{n-1}$ is contained in $\tilde{h}_{n-1}(A)$. This is true since, being $p^{(X)} \circ i = p^{(A)}$, if $i_{*}(\alpha) = 0$ then $p^{(A)}_{*}(\alpha) = 0$.

A map $i: \{*\} \rightarrow X$ induces a non-canonical splitting:
\begin{equation}\label{SplitRedHom}
	h_{n}(X) \simeq \tilde{h}_{n}(X) \oplus h_{n}\{*\}
\end{equation}
since, by definition of kernel, we have an exact sequence $0 \longrightarrow \tilde{h}_{n}(X) \longrightarrow h_{n}(X) \overset{(p_{*})_{n}}\longrightarrow h_{n}\{*\}$, moreover, being $p_{*} \circ i_{*} = \id_{h_{n}\{*\}}$, we see that $(p_{*})_{n}$ is surjective and the sequence splits.

\paragraph{Remark:} $\tilde{h}_{*}$ satisfies the three axioms of homology\footnote{This is not true if we include the so-called \emph{dimension axiom}.} if we consider $\mathcal{A}^{c}$ as made by couples $(X, \emptyset)$ with $X$ collapsable, but $\mathcal{A}^{c}$ is not an admissible category, since the pair $(X,X)$, which belongs to the lattice of $X$, is not in $\mathcal{A}^{c}$. If we extend $\mathcal{A}^{c}$ with these pairs we obtain an admissible category $\overline{\mathcal{A}^{c}}$ on which $\tilde{h}_{*}$ is a homology theory, but this is not meaningful since we have just trivial pairs. We cannot extend $\tilde{h}$ to arbitrary pairs, since in the exact sequence we have $h_{n}(X,A)$ not reduced, but for $A = \emptyset$ we have an inconsistency. 

\paragraph{}We define reduced cohomology similarly: 
	\[\tilde{h}^{n}(X) := \Coker\bigl( h^{n}\{*\} \overset{(p^{*})^{n}}\longrightarrow h^{n}(X) \bigr).
\]
For a couple $(X, A) \in \mathcal{A}$ with $X, A \in \mathcal{A}^{c}$ there is an exact sequence, called \emph{reduced cohomology exact sequence}:
	\[\xymatrix{
	\cdots \ar[r] & h^{n}(X, A) \ar[r]^{(\pi^{*})^{n}} & \tilde{h}^{n}(X) \ar[r]^{(i^{*})^{n}} & \tilde{h}^{n}(A) \ar[r]^{\beta^{n}\quad} & h^{n+1}(X, A) \ar[r] & \cdots
	}
\]
The proof is similar the one for homology. Also in this case we have a non-canonical splitting:
\begin{equation}\label{SplitRedCohom}
	h^{n}(X) \simeq \tilde{h}^{n}(X) \oplus h^{n}\{*\}
\end{equation}
which can be proven in the same way.

\paragraph{}We can define in a different way reduced homology and cohomology for spaces with a marked point, in such a way that under the embedding $\Top \hookrightarrow \Top^{+}$ they coincide in $\Top$ with non-reduced homology and cohomology, while, if the marked point belongs to the space, they coincide up to isomorphism with the previous definition. We consider the full subcategory $\mathcal{A}^{c+}$ of $\Top^{+}$ whose objects are couples $(X, x_{0})$ with $X \in \Ob(\mathcal{A}^{c})$. Then, for $(X, x_{0}) \in \mathcal{A}^{c+}$ we consider the map $i_{x_{0}}: \{*\} \rightarrow (X, x_{0})$ such that $i_{x_{0}}\{*\} = x_{0}$. We then define the \emph{reduced cohomology groups} of $X$ with respect to $x_{0}$ as:
	\[\begin{array}{l}
	\tilde{h}_{n}(X)_{x_{0}} := \Coker\bigl( h_{n}(X) \overset{(i_{x_{0}*})_{n}}\longrightarrow h^{n}\{*\} \bigr) \\
	\tilde{h}^{n}(X)_{x_{0}} := \Ker\bigl( h^{n}(X) \overset{(i_{x_{0}}^{*})^{n}}\longrightarrow h^{n}\{*\} \bigr).
\end{array}\]
With these definitions the splittings \eqref{SplitRedHom} and \eqref{SplitRedCohom} become canonical. Actually, these groups depend only on the pathwise connected component of $x_{0}$, not on the point itself: in fact, if $x_{0}$ and $x_{1}$ are connectible by an arc $\varphi: [0,1] \rightarrow X$, then the maps $i_{x_{0}}$ and $i_{x_{1}}$ are homotopic, thus they define the same maps in homology and cohomology by the homotopy axiom. Hence, this version of reduced homology and cohomology is well-defined for pathwise connected spaces with no need of marked points.

\paragraph{Remark:} $\mathcal{A}^{c+}$ is not an admissible category, since it contains the pairs $(X, \{x_{0}\})$ and not the pairs $(X, \emptyset)$, but if we consider the subcategory $\mathcal{A}^{\con}$ of $\mathcal{A} \cap \Top$ made by connected spaces and we complete it to $\overline{\mathcal{A}^{\con}}$ as before, then $\tilde{h}_{*}$ or $\tilde{h}^{*}$ is a (co)homology theory on $\overline{\mathcal{A}^{\con}}$, but this is not meaningful since we have just trivial pairs.

\subsection{First properties}

We now state some basic consequences of the axioms. The following lemma is an immediate consequence of the homotopy axiom:

\begin{Lemma}\label{HomEquivIso} If $(X,A)$ and $(Y, B)$ are homotopically equivalent, then $h_{*}(X,A) \simeq h_{*}(Y,B)$ and $h^{*}(X,A) \simeq h^{*}(Y,B)$ for any homology and cohomology theory. If $f: (X, A) \rightarrow (Y,B)$ induces a homotopy equivalence, then $f_{*}$ and $f^{*}$ are isomorphisms. $\square$
\end{Lemma}

Actually, we can refine the previous lemma. In fact, let us consider the couple $(D^{n}, \partial D^{n})$ and the couple $(D^{n}, D^{n} \setminus \{0\})$. The immersion $i: D^{n} \rightarrow D^{n}$ induces of course a homotopy equivalence of $D^{n}$ with itself, and the same holds for the restriction $i\vert_{\partial D^{n}}: \partial D^{n} \rightarrow D^{n} \setminus \{0\}$ and the retraction $r: D^{n} \setminus \{0\} \rightarrow \partial D^{n}$: however, the two couples are not homotopically equivalent, since $r$ cannot be continuously extended to $D^{n}$, and any map sending $D^{n} \setminus \{0\}$ to $\partial D_{n}$ cannot be extended at $0$. However, also in this case we have equivalence in homology and cohomology:

\begin{Lemma}\label{HomEquivIso2} If there exists a morphism $f: (X, A) \rightarrow (Y, B)$ in $\mathcal{A}$ such that $f: X \rightarrow Y$ and $f\vert_{A}: A \rightarrow B$ induce homotopy equivalences, then $h_{*}(X,A) \simeq h_{*}(Y,B)$ and $h^{*}(X,A) \simeq h^{*}(Y,B)$ via $f_{*}$ and $f^{*}$ for any homology and cohomology theory.
\end{Lemma}
\textbf{Proof:} By the exactness axiom $f$ induces a morphism of exact sequences:
	\[\xymatrix{
	\cdots \ar[r] & h_{n}(A) \ar[d]^{f^{*}}_{\simeq} \ar[r] & h_{n}(X) \ar[d]^{f^{*}}_{\simeq} \ar[r] & h_{n}(X,A) \ar[d]^{f^{*}}\ar[r] & h_{n-1}(A) \ar[d]^{f^{*}}_{\simeq} \ar[r] & h_{n-1}(X) \ar[r] \ar[d]^{f^{*}}_{\simeq} & \cdots\\
	\cdots \ar[r] & h_{n}(B) \ar[r] & h_{n}(Y) \ar[r] & h_{n}(Y,B) \ar[r] & h_{n-1}(B) \ar[r] & h_{n-1}(Y) \ar[r] & \cdots
}\]
where the isomorphisms in the diagram follow from lemma \eqref{HomEquivIso}. By the five lemma \cite{Hatcher} also the central map is an isomorphism. $\square$

\paragraph{}We now state a simple lemma which will be useful sometimes in the future.
\begin{Lemma}\label{TildePt} Let $(X, x_{0}) \in \mathcal{A}^{+}$. Then $\tilde{h}_{*}(X)_{x_{0}} \simeq h_{*}(X, x_{0})$ and $\tilde{h}^{*}(X)_{x_{0}} \simeq h^{*}(X, x_{0})$ canonically, so that also $\tilde{h}_{*}(X) \simeq h_{*}(X, x_{0})$ and $\tilde{h}^{*}(X) \simeq h^{*}(X, x_{0})$.\footnote{The latter are non-canonical since the r.h.s.\ depends on $x_{0}$; actually the l.h.s.\ and the r.h.s.\ are functors with different categories as domain, so canonicity has no meaning.}
\end{Lemma}
\textbf{Proof:} It follows immediately from the exact sequence:
	\[\cdots \longrightarrow \tilde{h}_{n}\{x_{0}\} = 0 \longrightarrow \tilde{h}_{n}(X)_{x_{0}} \longrightarrow h_{n}(X, x_{0}) \longrightarrow \tilde{h}_{n-1}\{x_{0}\} = 0\longrightarrow \cdots
\]
and the same for cohomology. $\square$

\paragraph{}One natural question is the relation between the (co)homology of a couple $(X,A)$ and the one of the quotient $X/A$. We stated the result, the proof can be found in \cite{ES}.
\begin{Def} A pair $(X,A)$ is called \emph{good pair} if $A$ is a non-empty \emph{closed} subspace and it is a deformation retract of some neighborhood in $X$.
\end{Def}
\begin{Lemma}\label{GoodPair} If $(X,A)$ is a good pair, then the projection $\pi: (X,A) \rightarrow (X/A, A/A)$ induces an isomorphism $h_{*}(X,A) \simeq h_{*}(X/A, A/A) = \tilde{h}_{*}(X/A)$. The same for cohomology. $\square$
\end{Lemma}

\paragraph{} We refer to \cite{ES} for \emph{Majer-Vietories sequence}, which is an important tool to concretely compute (co)homology groups.

\begin{Theorem}\label{AdditivityTh} Let $X = X_{1} \sqcup \cdots \sqcup X_{n}$ where each $X_{i}$ is both open and closed, and let $A_{i} \subset X_{i}$ such that $(X_{i}, A_{i})$ is a good pair. Then:
	\[h_{n}(X, A) \simeq \bigoplus_{i = 1, \ldots, n} h_{n}(X_{i}, A_{i})
\]
where the isomorphism is induced by the inclusions. The same holds for cohomology.
\end{Theorem}

\paragraph{}The degree of a map has been defined using singular homology. Actually, we can define it using any (co)homology theory $h$ with coefficient group $\mathbb{Z}$, and the result does not depend on $h$. This is a consequence of the following more general theorem, whose proof can be found in \cite{Bredon}:
\begin{Theorem}\label{DegreeAnyH} Let $f: S^{n} \rightarrow S^{n}$ be a map of degree $k$ and let $h_{*}$ be any homology theory. Then $(f_{*})_{n}: h_{n}(S^{n}) \rightarrow h_{n}(S^{n})$ is $(f_{*})_{n}(\alpha) = k \cdot \alpha$. The same holds for cohomology theories. $\square$
\end{Theorem}

\subsection{Borel-Moore homology and cohomology with compact support}

One can generalize to any (co)homology theory the notions of Borel-Moore homology and cohomology with compact support. We denote by $X^{+}$ the one-point compatification of $X$. For $X$ non compact, it is different from $X \sqcup \{\infty\}$ and it must not be confused with it; instead, for $X$ compact they coincide. We define \emph{cohomology with compact support} as:
	\[h_{n}^{BM}(X) := \tilde{h}_{n}(X^{+}) \qquad h^{n}_{\cpt}(X) := \tilde{h}^{n}(X^{+}).
\]
For $X$ compact we have $h^{n}_{\cpt}(X) = h^{n}(X)$. In fact, $X^{+} = X \sqcup \{\infty\}$ with $\{\infty\}$ isolated point, thus, by theorem \ref{AdditivityTh}, we have that $h^{n}(X^{+}) = h^{n}(X) \oplus h^{n}\{\infty\}$: if we consider the map $i_{\infty}: \{*\} \rightarrow X$ sending $*$ to $\infty$, we clearly have that $\tilde{h}^{n}(X^{+}) = \Ker((i_{\infty}^{*})^{n}) = h^{n}(X)$. For $X$ non compact, instead, they are in general different, as we have seen for singular (co)homology.

\subsection{Multiplicative cohomology theories}

We now introduce the notion of \emph{product} in cohomology, which naturally appears as cup product for singular cohomology, as wedge product for de-Rham cohomology or as tensor product for K-theory.

\begin{Def} A cohomology theory $h^{*}$ on an admissible category $\mathcal{A}$ is called \emph{multiplicative} if there exists an \emph{exterior product}, i.e.\ a natural map:
\begin{equation}\label{ExtProduct}
	\times: h^{i}(X, A) \times h^{j}(Y, B) \longrightarrow h^{i+j}(X \times Y, X \times B \cup A \times Y)
\end{equation}
satisfying the following axioms:
\begin{itemize}
	\item it is bilinear with respect to the sum in $h^{*}$;
	\item it is associative and, for $(X,A) = (Y,B)$, \emph{graded}-commutative;
	\item it admits a unit $1 \in h^{0}\{*\}$;
	\item it is compatible with the Bockstein homomorphisms, i.e.\ the following diagram commutes:
	\[\xymatrix{
	h^{i}(A) \times h^{j}(Y, B) \ar[r]^{\times} \ar[dd]^{\beta^{i} \times 1} & h^{i+j}(A \times Y, A \times B) \ar[d]^{\textnormal{exc}} \\
	& h^{i+j}(X \times B \cup A \times Y, X \times B) \ar[d]^{\beta^{i+j}} \\
	h^{i+1}(X,A) \times h^{j}(Y, B) \ar[r]^{\times} & h^{i+j+1}(X \times Y, X \times B \cup A \times Y).
}\]
\end{itemize}
In this case, we define the \emph{interior product}:
	\[\cdot\,: h^{i}(X, A) \times h^{j}(X, A) \longrightarrow h^{i+j}(X, A)
\]
as $\alpha \cdot \beta := \Delta^{*}(\alpha \times \beta)$ for $\Delta: X \rightarrow X \times X$ the diagonal map.
\end{Def}

\paragraph{Remarks:}
\begin{itemize}
	\item The interior product makes $h^{*}(X,A)$ a ring with unit.
	\item Naturality of the exterior product means that it is a morphisms of functors $\mathcal{A} \times \mathcal{A} \longrightarrow \AbGrp$; thus, given a morphism in $\mathcal{A} \times \mathcal{A}$, i.e.\ a couple of maps $f: (X, A) \rightarrow (Y, B)$ and $g: (X',A') \rightarrow (Y',B')$, and two classes $\alpha \in h^{i}(X,A)$ and $\beta \in h^{j}(Y,B)$ it satisfies:
	\begin{equation}\label{ProductNatural}
		f^{*}_{i}(\alpha) \times g^{*}_{j}(\beta) = (f \times g)^{*}_{i+j}(\alpha \times \beta).
	\end{equation}
	\item Let $(X, x_{0}), (Y, y_{0}) \in \mathcal{A}^{+}$ be spaces with marked point \emph{which are also good pairs} and such that $(X \times Y, X \vee Y)$ is a good pair. Then the exterior product induces a map:
\begin{equation}\label{ProductWedge}
	\tilde{h}^{i}(X)_{x_{0}} \times \tilde{h}^{j}(Y)_{y_{0}} \longrightarrow \tilde{h}^{i+j}(X \wedge Y).
\end{equation}
	In fact, by \eqref{ExtProduct} we have $h^{i}(X, x_{0}) \times h^{i}(Y, y_{0}) \longrightarrow h^{i+j}(X \times Y, X \vee Y)$ which is exactly \eqref{ProductWedge} by lemmas \ref{TildePt} and \ref{GoodPair}.
\end{itemize}

\begin{Lemma}\label{h0Ring} If $h^{*}$ is a multiplicative cohomology theory the coefficient group $h^{0}\{*\}$ is a commutative ring with unit.
\end{Lemma}
\textbf{Proof:} By definition we have a product $h^{0}\{*\} \times h^{0}\{*\} \rightarrow h^{0}\{*\}$ which is associative. Moreover, skew-commutativity in this case coincides with commutativity, and $1$ is a unit also for this product. $\square$

\paragraph{}Given a path-wise connected space $X$, we consider any map $p: \{*\} \rightarrow X$: by the path-wise connectedness of $X$ two such maps are homotopic, thus the pull-back $p^{*}: h^{*}(X) \rightarrow h^{*}\{*\}$ is well defined.
\begin{Def}\label{DefRank} For $X$ a path-connected space we call \emph{rank} of a cohomology class $\alpha \in h^{n}(X)$ the class $\rk(\alpha) := (p^{*})^{n}(\alpha) \in h^{n}\{*\}$ for any map $p: \{*\} \rightarrow X$.
\end{Def}
Let us consider the unique map $P: X \rightarrow \{*\}$.
\begin{Def}\label{TrivialClass} We call a cohomology class $\alpha \in h^{n}(X)$ \emph{trivial} if there exists $\beta \in h^{n}\{*\}$ such that $\alpha = (P^{*})^{n}(\beta)$. We denote by $1$ the class $(P^{*})^{0}(1)$.
\end{Def}
\begin{Lemma} For $X$ a path-wise connected space, a trivial cohomology class $\alpha \in h^{n}(X)$ is the pull-back of its rank.
\end{Lemma}
\textbf{Proof:} Let $\alpha \in h^{n}(X)$ be trivial. Then $\alpha = (P^{*})^{n}(\beta)$ so that $\rk(\alpha) = (p^{*})^{n}(P^{*})^{n}(\beta)$ $= (P \circ p)^{*\,n}(\beta) = \beta$, thus $\alpha = (P^{*})^{n}(\rk(\alpha))$. $\square$

\section{Thom isomorphism and Gysin map}

\subsection{Fiber bundles and module structure}

Let $\pi: E \rightarrow B$ be a fiber bundle with fiber $F$ and let $h^{*}$ be a multiplicative cohomology theory. Then $h^{*}(E)$ has a natural structure of $h^{*}(B)$-module given by:
\begin{equation}\label{ModuleStr}
\begin{split}
	\cdot \;:\; &h^{i}(B) \times h^{j}(E) \longrightarrow h^{i+j}(E)\\
	&a \cdot \alpha \,:=\, (\pi^{*}a) \cdot \alpha.
\end{split}
\end{equation}
In general this is not an algebra structure since, because of skew-commutativity, one has $((\pi^{*}a) \alpha)\beta = \pm \alpha ((\pi^{*}a) \beta)$.

We have an analogous module structure for relative fiber bundles, i.e.\ for pairs $(E, E')$ with $E'$ a sub-bundle of $E$ with fiber $F' \subset F$. In fact, we have a natural diagonal map $\Delta: (E, E') \rightarrow (E \times E, E \times E')$ given by $\Delta(e) = (e, e)$, so that we can define the following module structure:
\begin{equation}\label{Module1}
\begin{split}
	h^{i}(B) \times h^{j}(E, E') \overset{\pi^{*} \times 1}\longrightarrow h^{i}(E) \times h^{j}(E, E') \overset{\times}\longrightarrow h^{i+j}(E \times E, \,&E \times E')\\
	& \overset{\Delta^{*}}\longrightarrow h^{i+j}(E, E').
\end{split}
\end{equation}
Similarly, we can consider the map $\Delta_{\pi}: (E, E') \rightarrow (B \times E, B \times E')$ given by $\Delta_{\pi}(e) = (\pi(e), e)$ and define the module structure:
\begin{equation}\label{Module2}
\begin{split}
	h^{i}(B) \times h^{j}(E, E') \overset{\times}\longrightarrow h^{i+j}(B \times E, B \times E') \overset{\Delta_{\pi}^{*}}\longrightarrow h^{i+j}(E, E').
\end{split}
\end{equation}
To see that these two definitions are equivalent, we consider the following diagram:
\[\xymatrix{
	h^{i}(E) \times h^{j}(E, E') \ar[r]^{(2)} & h^{i+j}(E \times E, E \times E') \ar[dr]^{(5)} & \\
	& & h^{i+j}(E, E') \\
	h^{i}(B) \times h^{j}(E, E') \ar[r]^{(3)} \ar[uu]^{(1)} & h^{i+j}(B \times E, B \times E') \ar[ur]^{(6)} \ar[uu]^{(4)} &
}\]
in which the structure \eqref{Module1} is given by (1)-(2)-(5) and the structure \eqref{Module2} by (3)-(6). The commutativity of the square, i.e.\ (1)-(2) = (3)-(4), follows from the naturality of the product, while the commutativity of the triangle, i.e.\ (6) = (4)-(5), follows from the fact that (4) = $(\pi \times 1)^{*}$, (5) = $\Delta^{*}$, (6) = $\Delta_{\pi}^{*}$, and $\Delta_{\pi} = (\pi \times 1) \circ \Delta$.

\begin{Lemma}\label{Unitarity} The module structure \eqref{Module1} or \eqref{Module2} is unitary, i.e.\ $1 \cdot \alpha = \alpha$ for $1$ defined by \ref{TrivialClass}. More generally, for a trivial class $t = P^{*}(\eta)$, with $\eta \in h^{*}\{*\}$, one has $t \cdot \alpha = \eta \cdot \alpha$.
\end{Lemma}
\textbf{Proof:} We prove for \eqref{Module2}. The thesis follows from the commutativity of the following diagram:
\[\xymatrix{
	h^{i}(B) \times h^{j}(E, E') \ar[r]^{\times} & h^{i+j}(B \times E, B \times E') \ar[r]^{\Delta_{\pi}^{*}} & h^{i+j}(E, E') \\
	h^{i}\{*\} \times h^{j}(E, E') \ar[r]^{\times} \ar[u]^{(P^{*})^{i} \times 1^{j}} & h^{i+j}(\{*\} \times E, \{*\} \times E') \ar[ur]^{\simeq} \ar[u]^{((P \times 1)^{*})^{^{i+j}}} &
}\]
where the commutativity of the square follows directly from the naturality of the product while the commutativity of triangle follows from the fact that $(P \times 1) \circ \Delta_{\pi}$ is exactly the natural map $(E, E') \rightarrow (\{*\} \times E, \{*\} \times E')$ inducing the isomorphism $\simeq$. $\square$

\paragraph{}Let us consider a real \emph{vector} bundle $\pi: E \rightarrow B$ with fiber $\mathbb{R}^{n}$. In this case $\pi^{*}$ is an isomorphism, since $E$ retracts on $B$, thus the module structure \eqref{ModuleStr} is just the product in $h^{*}(B)$ up to isomorphism. Let us instead consider the zero section $B_{0} \simeq B$ and its complement $E_{0} = E \setminus B_{0}$: then \eqref{Module1} or \eqref{Module2} gives a non-trivial module structure on $h^{*}(E, E_{0})$. Defining the cohomology with compact support $h^{*}_{\cpt}(X) := \tilde{h}^{*}(X^{+})$ for $X^{+}$ the one-point compactification of $X$, we have:
	\[h^{*}(E, E_{0}) \simeq h^{*}_{\cpt}(E).
\]
In fact, let us put a metric on $E$ and consider the fiber bundles $D_{E}$ and $S_{E}$ obtained taking respectively the unit disc and the unit sphere in each fiber. Then we have:
\begin{equation}\label{IsoEE0EInfty}
\begin{split}
	h^{*}(E, E_{0}) \overset{(1)}\simeq h^{*}(D_{E}, (D_{E})_{0}) \overset{(2)}\simeq h^{*}(D_{E}, \partial D_{E}) \overset{(3)}\simeq \tilde{h}^{*}\,&(D_{E} / \partial D_{E})\\
	& \overset{(4)}\simeq \tilde{h}^{*}(E^{+}) = h^{*}_{\cpt}(E)
\end{split}
\end{equation}
where (1) follows by excision on the open set $U = E \setminus D_{E}$, (3) from the fact that $(D_{E}, \partial D_{E})$ is a good pair and (4) from the homeomorphism sending $\Int(D_{E})$ to $E$ and $\partial D_{E}$ to $\infty$.

We can also describe a natural module structure:
	\[h^{i}_{\cpt}(B) \times h^{j}_{\cpt}(E) \longrightarrow h^{i+j}_{\cpt}(E)
\]
which, for $B$ compact, coincides with the previous under the isomorphism \eqref{IsoEE0EInfty}. In fact, we consider:
\begin{equation}\label{ModuleCpt}
	h^{i}(B^{+}, \{\infty\}) \times h^{j}(E^{+}, \{\infty\}) \overset{\times}\longrightarrow h^{i+j}(B^{+} \times E^{+}, B^{+} \vee E^{+}) \overset{(\Delta_{\pi}^{+})^{*}}\longrightarrow h^{i+j}(E^{+}, \{\infty\})
\end{equation}
for $\Delta_{\pi}^{+}: (E^{+}, \{\infty\}) \rightarrow (B^{+} \times E^{+}, B^{+} \vee E^{+})$ defined by $\Delta_{\pi}^{+}(e) = (\pi(e), e)$ and $\Delta_{\pi}^{+}(\infty) = \{\infty\} \times \{\infty\}$. For $B$ compact, the module structure \eqref{ModuleCpt} becomes:
\begin{equation}\label{ModuleCptCpt}
	h^{i}(B) \times h^{j}(E^{+}, \{\infty\}) \overset{\times}\longrightarrow h^{i+j}(B \times E^{+}, B \times \{\infty\}) \overset{(\Delta_{\pi}^{+})^{*}}\longrightarrow h^{i+j}(E^{+}, \{\infty\}).
\end{equation}
We now see that \eqref{ModuleCptCpt} coincides with \eqref{Module2} under the isomorphism \eqref{IsoEE0EInfty}. In fact, we consider the following diagram (the arrows with $-1$ are inversions of natural isomorphisms):

{ \tiny \[\xymatrix{
	h^{i}(B) \times h^{j}(E, E_{0}) \ar[d] \ar[r]^{\times} & h^{i+j}(B \times E, B \times E_{0}) \ar[d] \ar[r]^{\Delta_{\pi}^{*}} & h^{i+j}(E, E_{0}) \ar[d] \\
	h^{i}(B) \times h^{j}(D_{E}, (D_{E})_{0}) \ar[d] \ar[r]^{\times} & h^{i+j}(B \times D_{E}, B \times (D_{E})_{0}) \ar[d] \ar[r]^{\Delta_{\pi}^{*}} & h^{i+j}(D_{E}, (D_{E})_{0}) \ar[d] \\
	h^{i}(B) \times h^{j}(D_{E}, \partial D_{E}) \ar[d]^{-1} \ar[r]^{\times} & h^{i+j}(B \times D_{E}, B \times \partial D_{E}) \ar[d]^{-1} \ar[r]^{\Delta_{\pi}^{*}} & h^{i+j}(D_{E}, \partial D_{E}) \ar[d]^{-1} \\
	h^{i}(B) \times h^{j}(D_{E}/\partial D_{E}, \partial D_{E}/\partial D_{E}) \ar[r]^{\times} & h^{i+j}(B \times (D_{E}/\partial D_{E}), B \times (\partial D_{E}/\partial D_{E})) \ar[d] \ar[r]^{\Delta_{\pi}^{*}} & h^{i+j}(D_{E}, \partial D_{E}) \ar[d] \\
	& h^{i+j}(B \times (D_{E}/\partial D_{E}), B \times (\partial D_{E}/\partial D_{E})) \ar[r]^{(\Delta_{\pi}^{+})^{*}} & h^{i+j}(D_{E} / \partial D_{E}, \partial D_{E} / \partial D_{E})
}\] } \\
where the first line is \eqref{Module2} and the sequence made by the last element of each column is \eqref{ModuleCptCpt}.

\paragraph{}We remark for completeness that there is a homeomorphism $B^{+} \wedge E^{+} \simeq (B \times E)^{+}$: in fact, $B^{+} \wedge E^{+} = (B \sqcup \{\infty\}) \times (E \sqcup \{\infty\}) \,/\, (\{\infty\} \times E) \cup (B \times \{\infty\})$ and, at the quotient, $B \times E$ remains unchanged while the denominator $B^{+} \vee E^{+}$ becomes a point which is the $\{\infty\}$ of $(B \times E)^{+}$. Thus the homeomorphism is $\varphi(b, e) = (b,e)$ and $\varphi(\infty, e) = \varphi(b, \infty) = \infty$. We then consider the map $\Delta_{\pi}^{+}: E^{+} \longrightarrow (B \times E)^{+}$ given by $\Delta_{\pi}^{+}(e) = (\pi(e), e)$ and $\Delta_{\pi}^{+}(\infty) = \infty$. Thus, under the hypotheses that $(B^{+}, \{\infty\})$ and $(B^{+} \times E^{+}, B^{+} \vee E^{+})$ are good pairs\footnote{These hypotheses are surely satisfied when $B$ is compact, since $\{\infty\}$ is a neighborhood of itself}, \eqref{ModuleCpt} can also be written as:
\begin{equation}\label{ModuleCptGoodPairs}
	\tilde{h}^{i}(B^{+}) \times \tilde{h}^{j}(E^{+}) \overset{(1)}\longrightarrow \tilde{h}^{i+j}(B^{+} \wedge E^{+}) \simeq \tilde{h}^{i+j}((B\times E)^{+}) \overset{(\Delta_{\pi}^{+})^{*}}\longrightarrow \tilde{h}^{i+j}(E^{+})
\end{equation}
where (1) is given by formula \eqref{ProductWedge}.

\subsection{Orientability and Thom isomorphism}

We now define \emph{orientable} vector bundles with respect to a fixed \emph{multiplicative} cohomology theory. By hypothesis, there exists a unit $1 \in h^{0}\{*\} = \tilde{h}^{0}(S^{0})$. Since $S^{n}$ is homeomorphic to the $n$-th suspension of $S^{0}$, such a homeomorphism defines (via the suspension isomorphism) an element $\gamma^{n} \in \tilde{h}^{n}(S^{n})$ such that $\gamma^{n} = S^{n}(1)$ (clearly $\gamma^{n}$ is not the unit class since the latter does not belong to $\tilde{h}^{n}(S^{n})$). Moreover, given a vector bundle $E \rightarrow B$ with fiber $\mathbb{R}^{k}$, we have the canonical isomorphism \eqref{IsoEE0EInfty} which, in each fiber $F_{x} = \pi^{-1}(x)$, restricts to:
\begin{equation}\label{IsoFF0FInfty}
h^{k}(F_{x}, (F_{x})_{0}) \simeq h^{k}(D^{k}_{x}, \partial D^{k}_{x}) \simeq h^{k}(D^{k}_{x} / \partial D^{k}_{x}, \partial D^{k}_{x} / \partial D^{k}_{x}) \simeq h^{k}(S^{k}, N)
\end{equation}
where the last isomorphism is non-canonical since it depends on the local chart ($N$ is the north pole of the sphere). However, since the homotopy type of a map from $S^{k}$ to $S^{k}$ is uniquely determined by its degree \cite{Hatcher} and a homeomorphism must have degree $\pm 1$, it follows that the last isomorphism of \eqref{IsoFF0FInfty} is canonical up to an overall sign, i.e.\ up to a multiplication by $-1$ in $h^{k}(S^{k}, N)$.
\begin{Def} Let $\pi: E \rightarrow B$ be a vector bundle of rank $k$ and $h^{*}$ a multiplicative cohomology theory in an admissible category $\mathcal{A}$ containing $\pi$. The bundle $E$ is called \emph{$h$-orientable} if there exists a class $u \in h^{k}(E, E_{0})$ such that for each fiber $F_{x} = \pi^{-1}(x)$ it satisfies $u\vert_{F_{x}} \simeq \pm\gamma^{k}$ under the isomorphism \eqref{IsoFF0FInfty}. The class $u$ is called \emph{orientation}.
\end{Def}

We now discuss some properties of $h$-orientations. The following lemma is very intuitive and can be probably deduced by a continuity argument; however, since we have not discussed topological properties of the cohomology groups, we give a proof not involving such problems. For a rank-$k$ vector bundle $\pi: E \rightarrow B$, let $(U_{\alpha}, \varphi_{\alpha})$ be a contractible local chart for $E$, with $\varphi_{\alpha}: \pi^{-1}(U_{\alpha}) \rightarrow U_{\alpha} \times \mathbb{R}^{k}$. Let us consider the compactification $\varphi_{\alpha}^{+}: \pi^{-1}(U_{\alpha})^{+} \rightarrow (U_{\alpha} \times \mathbb{R}^{k})^{+}$, restricting, for $x \in U_{\alpha}$, to $(\varphi_{\alpha})_{x}^{+}: E_{x}^{+} \rightarrow S^{k}$. Then we can consider the map:
\begin{equation}\label{VarphiHat}
	\hat{\varphi}_{\alpha,x} := ((\varphi_{\alpha})_{x}^{+\,-1})^{*\,k}: \tilde{h}^{k}(E_{x}^{+}) \longrightarrow \tilde{h}^{k}(S^{k}).
\end{equation}
\begin{Lemma}\label{Continuity} Let $u$ be an $h$-orientation of a rank-$n$ vector bundle $\pi: E \rightarrow B$, let $(U_{\alpha}, \varphi_{\alpha})$ be a contractible local chart for $E$ and let $\hat{\varphi}_{\alpha,x}$ be defined by \eqref{VarphiHat}. Then $\hat{\varphi}_{\alpha,x}(u\vert_{E_{x}^{+}})$ is constant in $x$ with value $\gamma^{k}$ or $-\gamma^{k}$.
\end{Lemma}
\textbf{Proof:} Let us consider the map $(\varphi_{\alpha}^{+\,-1})^{*\,k}: \tilde{h}^{k}(\pi^{-1}(U_{\alpha})^{+}) \longrightarrow \tilde{h}^{k}((U_{\alpha} \times \mathbb{R}^{k})^{+})$ and let call $\xi := (\varphi_{\alpha}^{+\,-1})^{*\,k}(u\vert_{\pi^{-1}(U_{\alpha})^{+}})$. Since $(U_{\alpha} \times \mathbb{R}^{k})^{+} \simeq U_{\alpha} \times S^{k} \,/\, U_{\alpha} \times \{N\}$ canonically, we can consider the projection $\pi_{\alpha}: U_{\alpha} \times S^{k} \rightarrow U_{\alpha} \times S^{k} \,/\, U_{\alpha} \times \{N\}$. Then $\hat{\varphi}_{\alpha,x}(u\vert_{E_{x}^{+}}) = \xi\vert_{(\{x\} \times \mathbb{R}^{k})^{+}} \simeq \pi_{\alpha}^{*}(\xi)\vert_{\{x\} \times S^{k}}$. But, since $U_{\alpha}$ is contractible, the projection $\pi: U_{\alpha} \times S^{k} \rightarrow S^{k}$ induces an isomorphism in cohomology, so that $\pi_{\alpha}^{*}(\xi) = \pi^{*}(\eta)$ for $\eta \in h^{k}(S^{k})$, so that $\pi_{\alpha}^{*}(\xi)\vert_{\{x\} \times S^{k}} = \pi^{*}(\eta)\vert_{\{x\} \times S^{k}} \simeq \eta$, i.e.\ it is constant in $x$. By definition of orientation, its value must be $\pm\gamma^{k}$. $\square$

\begin{Theorem}\label{OrientedCharts} If a vector bundle $\pi: E \rightarrow B$ of rank $k$ is $h$-orientable, then given trivializing contractible charts $\{U_{\alpha}\}_{\alpha \in I}$ it is always possible to choose trivializations $\varphi_{\alpha}: \pi^{-1}(U_{\alpha}) \rightarrow U_{\alpha} \times \mathbb{R}^{k}$ such that $(\varphi_{\alpha}^{+})_{x}^{*\,k}(\gamma^{k}) = u\vert_{E_{x}^{+}}$. In particular, for $x \in U_{\alpha\beta}$ the homeomorphism $(\varphi_{\beta}\varphi_{\alpha}^{-1})^{+}_{x}: (\mathbb{R}^{k})^{+} \simeq S^{k} \longrightarrow (\mathbb{R}^{k})^{+} \simeq S^{k}$ satisfies $((\varphi_{\beta}\varphi_{\alpha}^{-1})^{+}_{x})^{*}(\gamma^{k}) = \gamma^{k}$.
\end{Theorem}
\textbf{Proof:} Chosen any local trivialization $\varphi_{\alpha}: \pi^{-1}(U_{\alpha}) \rightarrow U_{\alpha} \times \mathbb{R}^{k}$, it verifies $(\varphi_{\alpha}^{+})_{x}^{*\,k}(\gamma^{k}) = \pm u\vert_{E_{x}^{+}}$ by lemma \ref{Continuity}. If the minus sign holds, it is enough to compose $\varphi_{\alpha}$ to the pointwise reflection by an axes in $\mathbb{R}^{k}$, so that the compactified map has degree $-1$. $\square$

\begin{Def} An atlas satisfying the conditions of theorem \ref{OrientedCharts} is called \emph{$h$-oriented atlas}.
\end{Def}

\paragraph{Remark:} the classical definition of orientability, i.e.\ the existence of an atlas with transition functions of pointwise positive determinant, coincides with $H$-orientability for $H$ the singular cohomology with $\mathbb{Z}$-coefficients, as stated in \cite{Dold}. Similarly, an oriented atlas is an $H$-oriented atlas.

\begin{Lemma}\label{TwoOrientations} Let $\pi: E \rightarrow B$ be a rank-$k$ vector bundle which is orientable both for $H^{*}$ and for a multiplicative cohomology theory $h^{*}$, and let $u$ be an orientation with respect to $h^{*}$. Then an $H$-oriented atlas is $h$-oriented with respect to $u$ or $-u$.
\end{Lemma}
\textbf{Proof:} by lemma \ref{Continuity} the value of $u$ is constant in $x$ for each chart, and it is $\pm \gamma^{k}$. Moreover, the compactified transition functions of an $H$-oriented atlas must have degree $1$, thus they send $\gamma^{k}$ in $\gamma^{k}$ for every cohomology theory. Hence, the value of $u$ must be $\gamma^{k}$ or $-\gamma^{k}$ for each chart. The thesis immediately follows. $\square$

\paragraph{}We now state the Thom isomorphism following \cite{Dold}.
\begin{Theorem}\label{Base} Let $(E, E') \rightarrow B$ be a relative fiber bundle with fiber $(F, F')$. Suppose that there exists $a_{1}, \ldots, a_{r} \in h^{*}(E, E')$ such that, for every $x \in B$, their restrictions to $F_{x} = \pi^{-1}(x)$ form a base of $h^{*}(F_{x}, F'_{x})$ as a $h^{*}\{*\}$-module under the module structure \eqref{Module2}. Then $a_{1}, \ldots, a_{r}$ form a base of $h^{*}(E, E')$ as a $h^{*}(B)$-module. $\square$
\end{Theorem}
For the proof see \cite{Dold} page 7.
\begin{Theorem}[Thom isomorphism] Let $\pi: E \rightarrow B$ be a $h$-orientable vector bundle of rank $k$, and let $u \in h^{k}(E, E_{0})$ be an orientation. Then, the map induced by the module structure \eqref{Module2}:
	\[\begin{split}
	T: \;&h^{*}(B) \rightarrow h^{*}(E, E_{0})\\
	&T(\alpha) := \alpha \cdot u
\end{split}\]
is an isomorphism of abelian groups.
\end{Theorem}
\textbf{Proof:} The map $T: h^{*}\{*\} \longrightarrow \tilde{h}^{*}(S^{n})_{N}$ given by $T(\alpha) = \alpha \cdot \gamma^{n}$ is an isomorphism since, up to the suspension isomorphism, it coincides with $T': h^{*}\{*\} \longrightarrow h^{*}\{*\}$ given by $T'(\alpha) = 1\cdot \alpha = \alpha$. Thus, $\gamma^{n}$ is a base of $h^{*}(S^{n}, N)$ as a $h^{*}\{*\}$-module. By definition of h-orientability and theorem \ref{Base}, it follows that $u$ is a base of $h^{*}(E,E_{0})$ as a $h^{*}(B)$-module, i.e.\ $T$ is an isomorphism. $\square$

\subsection{Gysin map}

Let $X$ be a compact smooth $n$-manifold and $Y \subset X$ a compact embedded $r$-dimensional submanifold such that the normal bundle $N(Y) = (TX\,\vert_{Y}) / \,TY$ is $h$-orientable. Then, since $Y$ is compact, there exists a tubular neighborhood $U$ of $Y$ in $X$, i.e.\ there exists an homeomorphism $\varphi_{U}: U \rightarrow N(Y)$.

If $i: Y \rightarrow X$ is the embedding, from this data we can naturally define an homomorphism, called \emph{Gysin map}:
	\[i_{!}: h^{*}(Y) \rightarrow \tilde{h}^{*}(X).
\]
In fact:
\begin{itemize}
	\item we first apply the Thom isomorphism $T: h^{*}(Y) \rightarrow h^{*}_{\cpt}(N(Y)) = \tilde{h}^{*}(N(Y)^{+})$;
	\item then we naturally extend $\varphi_{U}$ to $\varphi_{U}^{+}: U^{+} \rightarrow N(Y)^{+}$ and apply $(\varphi_{U}^{+})^{*}: h^{*}_{\cpt}(N(Y)) \rightarrow h^{*}_{\cpt}(U)$;
	\item there is a natural map $\psi: X \rightarrow U^{+}$ given by:
	\[\psi(x) = \left\{\begin{array}{ll}
	x & \text{if } x \in U \\
	\infty & \text{if } x \in X \setminus U
	\end{array}\right.
\]
hence we apply $\psi^{*}: \tilde{h}^{*}(U^{+}) \longrightarrow \tilde{h}^{*}(X)$.
\end{itemize}
Summarizing:
\begin{equation}\label{GysinMap}
	i_{!}\,(\alpha) = \psi^{*} \circ \bigl(\varphi_{U}^{+}\bigr)^{*} \circ T \, (\alpha).
\end{equation}

\paragraph{Remark:} One could try to use the immersion $i: U^{+} \rightarrow X^{+}$ and the retraction $r: X^{+} \rightarrow U^{+}$ to have a splitting $h(X) = h(U) \oplus h(X, U) = h(Y) \oplus K(X,U)$. But this is false, since the immersion $i: U^{+} \rightarrow X^{+}$ is not continuous: \emph{since $X$ is compact}, $\{\infty\} \subset X^{+}$ is open, but $i^{-1}(\{\infty\}) = \{\infty\}$, and $\{\infty\}$ is not open in $U^{+}$ since $U$ is non-compact.

\paragraph{}One can extend Gysin map to more general maps than embeddings, in particular it can be defined for proper maps. For details the reader can see \cite{Karoubi2}.

\section{Finite CW-complexes}

\subsection{Whitehead axioms}

In this version of the axioms \cite{Whitehead} we consider the category $\TopFCW_{2}$. Since it is an admissible category, we will be able to compare this version with the previous.

\begin{Def} A \emph{homology theory} on $\TopFCW_{2}$ is a sequence of functors $h_{n}: \TopFCW_{2} \rightarrow \AbGrp$ and morphisms of functors $s_{n}: h_{n} \rightarrow h_{n+1} \circ S$ satisfying the following axioms:
\begin{enumerate}
	\item \emph{(Homotopy axiom)} if $f,g: (X, A) \rightarrow (Y,B)$ are homotopic, then $h_{n}(f) = h_{n}(g)$;
	\item \emph{(Suspension axiom)} the morphisms of functors $s_{n}$ are isomorphisms;
	\item \emph{(Exactness axiom)} the sequence of functors $h_{n}$ induces a sequence of functors:
	\[h'_{n}: \mathcal{A} \longrightarrow \ExSAbGrp
\]
assigning to each pair $(X,A)$ the exact sequence:
\begin{displaymath}
\xymatrix{
h_{n}(A) \ar[r]^{(i_{*})_{n}} & h_{n}(X) \ar[r]^{(\pi_{*})_{n}} & h_{n}(X/A)
}
\end{displaymath}
where $(i_{*})_{n}$ and $(\pi_{*})_{n}$ are the image via $h_{n}$ of the inclusions $i: (A, \emptyset) \rightarrow (X, \emptyset)$ and $\pi: (X, \emptyset) \rightarrow (X,A)$.
\end{enumerate}
\end{Def}

Reversing the arrows of the exact sequence, we have the corresponding axioms for cohomology:
\begin{Def} A \emph{cohomology theory} on $\TopFCW_{2}$ is a sequence of \emph{contravariant} functors $h^{n}: \TopFCW_{2} \rightarrow \AbGrp$ and morphisms of functors $s^{n}: h^{n+1} \circ S \rightarrow h^{n}$ satisfying the following axioms:
\begin{enumerate}
	\item \emph{(Homotopy axiom)} if $f,g: (X, A) \rightarrow (Y,B)$ are homotopic, then $h^{n}(f) = h^{n}(g)$;
	\item \emph{(Suspension axiom)} the morphisms of functors $s_{n}$ are isomorphisms;
	\item \emph{(Exactness axiom)} the sequence of functors $h^{n}$ induces a sequence of functors:
	\[(h')^{n}: \mathcal{A} \longrightarrow \ExSAbGrp
\]
assigning to each pair $(X,A)$ the exact sequence:
\begin{displaymath}
\xymatrix{
h^{n}(X/A) \ar[r]^{(\pi^{*})^{n}} & h^{n}(X) \ar[r]^{(i^{*})^{n}} & h^{n}(A)
}
\end{displaymath}
where $(i^{*})^{n}$ and $(\pi^{*})^{n}$ are the image via $h_{n}$ of the inclusions $i: (A, \emptyset) \rightarrow (X, \emptyset)$ and $\pi: (X, \emptyset) \rightarrow (X,A)$.
\end{enumerate}
\end{Def}
One can prove \cite{Whitehead} that these axioms are equivalent to the Eilenberg and Steenrod ones for the category of finite CW-pairs. We will use these axioms to define K-theory as a cohomology theory.

\subsection{S-Duality}

We now study how homology theories are related to cohomology ones and viceversa. It turns out that on the category of pairs having the homotopy type of finite CW-pairs there is a duality between homology and cohomology theories, such that, for compact manifolds orientable with respect to the theory considered, the Poincar\'e duality holds. One natural way to express this duality is to use the theory of spectra, which we do not review here. Otherwise one can use the Alexander duality: a finite CW-complex can be embedded in a sphere $S^{n}$, and, given a cohomology theory $h^{*}$, one can define $h_{p}(X,A) = h^{n-p}(S^{n}\setminus A, S^{n}\setminus X)$ or viceversa. The problem of this construction is that it is not intrinsic, since it requires the embedding in $S^{n}$ for some $n$ which is in general difficult to imagine. We prefer instead to recall the geometric construction of the homology theory dual to a given cohomology one which is described in \cite{Jakob}. The idea of the construction is the following: if we consider singular homology, one could ask if, for any class $A \in H_{n}(X, \mathbb{Z})$, there exists a smooth compact orientable $n$-manifold $M$ and a map $f: M \rightarrow X$ such that $A = f_{*}[M]$, where $[M]$ is the fundamental class of $M$. This is actually not true, but Steenrod proved that there always exists a triple $(M, \alpha, f)$ where $M$ has dimension $n+q$, $\alpha \in H^{q}(M, \mathbb{Z})$ and $A = f_{*}(\alpha \cap [M])$, or equivalently $A = f_{*}(\PD_{M}\alpha)$. This construction can be generalized.

We work on the category $HCW_{f}$ of spaces having the same homotopy type of a finite CW-complex, and we suppose to have fixed a cohomology theory $h^{*}$. For a couple of spaces $(X,A)$ in $HCW_{f}$ we define the group of \emph{$n$-pre-cycles} as the free abelian group $h_{PC,n}(X,A)$ generated by triples $(M, \alpha, f)$ where:
\begin{itemize}
	\item $M$ is a smooth \emph{compact connected $h^{*}$-orientable} manifold of dimension $n+q$ in general with boundary;
	\item $\alpha \in h^{q}(M)$;
	\item $f: M \rightarrow X$ is a continuous map such that $f(\partial M) \subset A$.
\end{itemize}
We define the group of cycles $h_{C,n}(X,A)$ as the quotient of $h_{PC,n}(X,A)$ by the subgroup generated by:
\begin{itemize}
	\item elements of the form $(M, \alpha + \beta, f) - (M, \alpha, f) - (M, \beta, f)$, so that we impose additivity with respect to the cohomology class in the middle;
	\item elements of the form $(M, \varphi_{!}\alpha, f) - (N, \alpha, f \circ \varphi)$ where $\pi: N \rightarrow N$ is a smooth map and $\varphi_{!}$ is the associated Gysin map.
\end{itemize}
Thus a generic $n$-cycle is an equivalence class $[(M, \alpha, f)]$. We define the subgroup of \emph{$n$-boundaries} $h_{B,n}(X,A)$ as the subgroup of $h_{C,n}(X,A)$ generated by the elements $[(M, \alpha, f)]$ such that there exists a precycle $(W, \beta, g) \in h_{PC,n}(X,X)$ such that $M = \partial W$, $\alpha = \beta\vert_{M}$ and $f = g\vert_{M}$. We then define the groups:
	\[h_{n}(X, A) := h_{C,n}(X,A) \,/\, h_{B,n}(X,A).
\]

\paragraph{Remark:} we cannot define chains as for singular homology, since the cohomology class $\alpha$ in the triple can be non-trivial only if $M$ is in general a non-trivial manifold, not necessarily a simplex which is contractible. Thus we define cycles and boundaries but we do not build a graded complex of chains whose homology is isomorphic to the one we are defining.

\paragraph{}In this picture we can naturally define cap product if $h^{*}$ is multiplicative. In fact, for $\beta \in h^{*}(X)$ and $[(M, \alpha, f)] \in h_{*}(X)$ we define:
	\[\beta \cap [(M, \alpha, f)] := [(M, \alpha \cdot f^{*}\beta, f)].
\]

\subsection{Extension}\label{Extension}

As shown in \cite{AH2}, given a cohomology theory on the category of finite CW-pairs we can extend it and associate a group to any pair of spaces, but we do not obtain in general a cohomology theory. In particular, for any pair of topological spaces $(X,A)$ we define $\mathfrak{h}^{*}(X,A)$ as the group whose generic element is a functor\footnote{The domain of this functor is the category $\mathcal{M}_{(X,A)}$ whose objects are maps $f: (Y, B) \rightarrow (X,A)$ with $(Y,B)$ a finite CW-pair, and whose morphisms from $f: (Y, B) \rightarrow (X,A)$ to $g: (Z, C) \rightarrow (X,A)$ are maps $h: (Y,B) \rightarrow (Z,C)$ such that $g \circ h = f$; the codomain of the functor is the category of abelian groups.} $\xi$ which, given a finite CW-pair $(Y,B)$ and a map $f: (Y, B) \rightarrow (X,A)$, assigns a class $f^{!}(\xi) \in h^{*}(X, A)$, satisfying the following hypotheses:
\begin{itemize}
	\item $f^{!}(\xi)$ depends only on the homotopy class of $f$;
	\item given a couple of maps $(Z, C) \overset{g}\longrightarrow (Y, B) \overset{f}\longrightarrow (X, A)$ with $(Z,C)$ and $(Y,B)$ finite CW-pairs, it satisfies $(f \circ g)^{!}(\xi) = g^{!}f^{!}(\xi)$.
\end{itemize}
With this definition the group $\mathfrak{h}^{*}(X,A)$ is homotopy-invariant for any couple of spaces, but actually it is not a generalized cohomology theory since the exactness of the cohomology sequence may fail. Actually, there is not a good way to extend the theory to any couple of spaces. However, at least we obtain the result that for couples having the homotopy type of finite CW-pairs we obtain a cohomology theory, which coincides with the previous for finite CW-pairs.

If $(X,A)$ is a CW-pair, not necessarily finite, there is a canonical isomorphism:
\begin{equation}
	\mathfrak{h}^{*}(X, A) \simeq \varprojlim_{\alpha} h^{*}(X_{\alpha}, A_{\alpha})
\end{equation}
where $\{X_{\alpha}\}$ is the set of \emph{finite} sub-complexes of $X$ and $A_{\alpha} = X_{\alpha} \cap A$. This groups is in general different from $h^{*}(X,A)$ defined in the ordinary way. If $(X,A)$ is a finite CW-pair, then there is a canonical isomorphism $\mathfrak{h}^{*}(X, A) \simeq h^{*}(X,A)$ since $X$ is a maximum in the family $\{X_{\alpha}\}$.

If we consider singular cohomology restricted to finite CW-pairs and we extend it in this way we do not obtain again singular cohomology, actually we get a surjective map $H^{*}(X,A) \rightarrow \mathcal{H}^{*}(X,A)$ which is not injective in general.


\chapter{Spectral sequences}

\section{General setting}

We consider an abelian group $K$ provided with a \emph{filtration}, i.e.\ with a sequence of nested subgroups $\{F^{p}K\}_{p \in \mathbb{Z}}$ such that:
\begin{itemize}
\item $\cdots \supset F^{p-1}K \supset F^{p}K \supset F^{p+1}K \supset \cdots$;
\item $\underset{p \in \mathbb{Z}}\bigcup F^{p}K = K$.
\end{itemize}
We can define $F^{-\infty}K = K$ and $F^{\infty}K = 0$. It is not necessary that the intersection of the groups $F^{p}K$ is $0$. It often happens that $F^{p}K = K$ for $p \leq 0$, and in this case the filtration can be written in the form $K = F^{0}K \supset F^{1}K \supset \cdots$.

Given a group with a filtration, we can construct the groups:
	\[E^{p}_{0}K = F^{p}K / F^{p+1}K
\]
whose direct sum $\underset{p\in \mathbb{Z}}\bigoplus E^{p}_{0}K$ is called the \emph{associated graded group} of the filtration $\{F^{p}K\}_{p \in \mathbb{Z}}$. We start considering spectral sequences without grading, thus, in this context, the languages of homology and cohomology are completely equivalent. We use the cohomological one.

\paragraph{} Let $d: K \rightarrow K$ be a coboundary, i.e.\ a morphism such that $d^{2} = 0$. In this case we can define a cohomology $H(K) = \Ker \, d \,/\, \IIm \, d$. Let us also suppose that $d$ \emph{preserves the filtration}, i.e.\ $d(F^{p}K) \subset F^{p}K$: in this case we have a cohomology $H(F^{p}K)$ for every $p$. We also put:
	\[ZK = \Ker \, d \qquad BK = \IIm \, d \qquad Z^{p}K = \Ker \bigl( d\vert_{F^{p}K} \bigr) \qquad B^{p}K = \IIm \bigl( d\vert_{F^{p}K} \bigr).
\]

\paragraph{}The inclusion $i_{p}: F^{p}K \hookrightarrow K$ induces a morphism in cohomology:
	\[i_{p}^{\#}: H(F^{p}K) \rightarrow H(K)
\]
whose image is given by equivalence classes of cocycles in $F^{p}K$ up to coboundaries coming from elements of all $K$. In particular, for $a \in Z^{p}K$:
\begin{itemize}
\item $[\,a\,] = \{a + d(x) \,\vert\, x \in F^{p}K\} \in H(F^{p}K)$;
\item $i_{p}^{\#}[\,a\,] = \{a + d(x) \,\vert\, x \in K\} \in H(K)$.
\end{itemize}
That's why in general $i_{p}^{\#}$ is not injective: a cocycle $a \in F^{p}K$ can be equal to $d(x)$ for $x \in K \setminus F^{p}K$, and, in this case, $[\,a\,] \neq 0$ in $H(F^{p}K)$ but $i_{p}^{\#}[\,a\,] = [\,0\,]$.

\paragraph{}We present the situation in a diagram, in which between the first two columns we declare when $a$ becomes zero at the quotient (it must be a coboundary) and between the second and the third column we declare the condition that $a$ must satisfy to be a cocycle ($da = 0$):

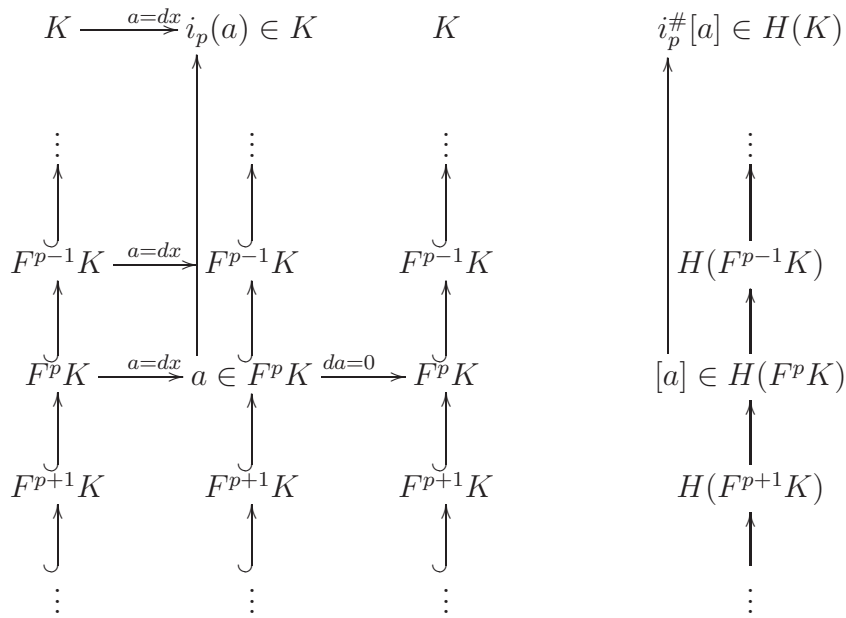
\begin{displaymath}
\xymatrix{
K \ar[r]^{a = dx} & i_{p}(a) \in K & K & & i_{p}^{\#}[a] \in H(K)\\
\vdots & \vdots & \vdots & & \vdots\\
F^{p-1}K \ar@{^(->}[u] \ar[r]^{a = dx} & F^{p-1}K \ar@{^(->}[u] & F^{p-1}K \ar@{^(->}[u] & & H(F^{p-1}K) \ar[u]\\
F^{p}K \ar@{^(->}[u] \ar[r]^{a = dx} & a \in \ar@<4ex>[uuu] F^{p}K \ar@{^(->}[u] \ar[r]^{da = 0} & F^{p}K \ar@{^(->}[u] & & [a] \in \ar@<6ex>[uuu] H(F^{p}K) \ar[u]\\
F^{p+1}K \ar@{^(->}[u] & F^{p+1}K \ar@{^(->}[u] & F^{p+1}K \ar@{^(->}[u] & & H(F^{p+1}K) \ar[u]\\
\vdots \ar@{^(->}[u] & \vdots \ar@{^(->}[u] & \vdots \ar@{^(->}[u] & & \vdots \ar[u]
}
\end{displaymath}
The cocycle $[a]$ is mapped in $H(K)$, hence quotiented out by coboundaries from all $K$. The image of $i_{p}^{\#}$ is thus:
\begin{equation}\label{ImIp}
	\IIm \bigl( \, i_{p}^{\#} \, \bigr) = \langle\, i_{p}Z^{p}K, BK \,\rangle \,/\, BK.
\end{equation}

\paragraph{} We define $F^{p}H(K) = \IIm(i_{p}^{\#})$. In this way, we obtain a filtration of $H(K)$ given by $\cdots \supset F^{p-1}H(K) \supset F^{p}H(K) \supset F^{p+1}H(K) \supset \cdots$, whose associated graded group is the direct sum of:
	\[E_{0}^{p}H(K) = F^{p}H(K) \,/\, F^{p+1}H(K).
\]
The spectral sequence is a sequence of groups which approximates, in a sense still to establish, the graded group $\underset{p\in \mathbb{Z}}\bigoplus E^{p}_{0}H(K)$.

\paragraph{Remark:} the spectral sequence can be built using two very similar viewpoints, which differ by a certain isomorphism in the groups involved. We develop both of them.

\paragraph{Notation:} We previously defined the immersions $i_{p}: F^{p}K \hookrightarrow K$. We also define the immersions:
	\[i_{p,p-r}: F^{p}K \hookrightarrow F^{p-r}K.
\]
When the index $p$ is clear from the context, we denote $i_{p,p-1}$ with $i$, and $i_{p,p-r}$ with $i^{r}$.

\section{Finite filtrations}

\subsection{Preliminaries}

We consider the case of a finite filtration, i.e.\ such that $F^{p}K = K$ for $p \leq 0$ and $F^{p}K = 0$ for $p \geq l$, for $l$ a fixed integer. The filtration is then:
	\[K = F^{0}K \supset \cdots \supset F^{l}K = 0
\]
with the corresponding filtration in cohomology:
	\[H(K) = F^{0}H(K) \supset \cdots \supset F^{l}H(K) = 0.
\]
\begin{Def} Given a \emph{finite} filtration of an abelian group with coboundary $(K, d)$, a \emph{spectral sequence} is a $p$-graded sequence of groups and coboundaries $\{E^{p}_{r}K, d^{p}_{r}\}_{r \in \mathbb{N},\,p \in \mathbb{Z}}$, with $d^{p}_{r}: E^{p}_{r}K \rightarrow E^{p+r}_{r}K$, such that, for $E_{r}K := \bigoplus_{p}E^{p}_{r}K$ and $d_{r} := \bigoplus_{p}d^{p}_{r}$, it satisfies the following conditions:
\begin{itemize}
\item $E^{p}_{0}K$ is the associated graded group of the filtration of $K$;
\item $E^{p}_{r+1}K \simeq \Ker \, d^{p}_{r} \,/\, \IIm \, d^{p-r}_{r}$ canonically, so that $E_{r+1}K \simeq H(E_{r}K, \, d_{r})$ canonically for every $r$;
\item the sequence stabilizes, i.e.\ for $r \geq r_{0}$ one has $d^{p}_{r} = 0$ and $E^{p}_{r}K = E^{p}_{r_{0}}K$; we call $E^{p}_{\infty}K$ the limit of the sequence;
\item $E^{p}_{\infty}K \simeq E^{p}_{0}H(K)$ canonically.
\end{itemize}
\end{Def}
In this case, it is clear in what sense the spectral sequence $\bigl\{ \underset{p\in \mathbb{Z}}\bigoplus E^{p}_{r}K \bigr\}_{r \in \mathbb{N}}$ approximates the associated graded group $\underset{p\in \mathbb{Z}}\bigoplus E^{p}_{0}H(K)$: the sequence stabilizes becoming equal to such a group.

\paragraph{}In the following we will use this simple lemma:

\begin{Lemma}\label{AlgLemma} Let $G$ be an abelian group and $A_{1}, A_{2}, B \leq G$, with $A_{2} \leq A_{1}$. Then:
	\[\frac{\langle A_{1}, B \rangle}{\langle A_{2}, B \rangle} \simeq \frac{A_{1}}{\langle A_{2}, A_{1} \cap B \rangle}.
\]
$\square$
\end{Lemma}

\subsection{First viewpoint}

\subsubsection{Image at level $p$}

As we have seen, $i^{p}_{\#}[\,a\,] = \{a + d(x) \,\vert\, x \in K\} \in H(K)$, hence in the diagram we think of $i^{p}_{\#}[\,a\,]$ as living at the level $-\infty$, i.e.\ for finite filtrations, at level $0$. We can also think of $\IIm(i^{p}_{\#})$ at level $p$. In particular, it is isomorphic to the group of cocycles in $F^{p}K$ up to coboundaries $d(x)$ such that $x \in K$ but $d(x) \in F^{p}K$:
	\[i^{p}_{\#}[\,a\,] \overset{\simeq}\longrightarrow \{a + d(x) \,\vert\, x \in K, d(x) \in F^{p}K\}.
\]
With respect to \eqref{ImIp}, this viewpoint is:
\begin{equation}\label{ImIp2}
	F^{p}H(K) \simeq Z^{p}K / \, (BK \cap Z^{p}K)
\end{equation}
and the isomorphism follows from lemma \ref{AlgLemma} for $A_{1} = Z^{p}K$, $A_{2} = \{0\}$, $B = BK$. In the diagram we can represent \eqref{ImIp2} as follows:
\begin{displaymath}
\xymatrix{
K \ar[r]^{a = dx} & i^{p}(a) \ar@<-4ex>[ddd] \in K & K & & H(K)\\
\vdots & \vdots & \vdots & & \vdots\\
F^{p-1}K \ar@{^(->}[u] \ar[r]^{a = dx} & F^{p-1}K \ar@{^(->}[u] & F^{p-1}K \ar@{^(->}[u] & & H(F^{p-1}K) \ar[u]\\
F^{p}K \ar@{^(->}[u] \ar[r]^{a = dx} & a \in F^{p}K \ar@{^(->}[u] \ar[r]^{da = 0} & F^{p}K \ar@{^(->}[u] & & [\,a\,] \in H(F^{p}K); \quad i^{p}_{\#}[\,a\,] \in \frac{Z^{p}K}{BK \cap Z^{p}K} \ar[u]\\
F^{p+1}K \ar@{^(->}[u] & F^{p+1}K \ar@{^(->}[u] & F^{p+1}K \ar@{^(->}[u] & & H(F^{p+1}K) \ar[u]\\
\vdots \ar@{^(->}[u] & \vdots \ar@{^(->}[u] & \vdots \ar@{^(->}[u] & & \vdots \ar[u]
}
\end{displaymath}

The isomorphism
\begin{equation}\label{IsoFpHK}
	\varphi: \, \langle \,i^{p}Z^{p}K, BK \,\rangle \,/\, BK \, \overset{\simeq}\longrightarrow \, Z^{p}K / \, (BK \cap Z^{p}K)
\end{equation}
is given by $\varphi[\,a\,] = [\,a\,] \cap F^{p}K$, with inverse $\varphi^{-1}[\,b\,] = [\,b\,] + BK$.

\subsubsection{First attempt with $F^{p}H(K)$}

One can ask why we do not search an approximation of the complete groups $F^{p}H(K)$, which should be of course more useful since, in particular, $F^{0}H(K) = H(K)$. The problem is that, in this way, although we can construct a sequence of groups with coboundaries using the same technique, it is not true in general that the group at a certain step is canonically isomorphic to the cohomology of the previous, as we now show.

\paragraph{} We construct a natural approximation of $F^{p}H(K)$, thought at the level $p$, i.e.\ $F^{p}H(K) = Z^{p}K / (BK \cap Z^{p}K)$. In particular, since the filtration has length $l$, for $0 \leq p \leq l-1$ we have $F^{p+l}K = 0$ and $F^{p-l+1}K = K$. Hence:
\begin{equation}\label{LastApprox}
\begin{split}
	&Z^{p}K = \{a \in F^{p}K \,\vert\, d(a) \in F^{p+l}K\}\\
	&BK \cap Z^{p}K = \{a \in F^{p}K \,\vert\, a = d(x),\, x \in F^{p-l+1}K\}.
\end{split}
\end{equation}

\paragraph{Remark:}\label{ShiftCob1} the fact of considering $p-l+1$ instead of $p-l$ is very important, and the meaning of the $+1$ will become clear later.

\paragraph{}In the diagram, as before, between the first two columns we declare when $a$ becomes zero at the quotient and between the second and the third we declare the condition that $a$ must satisfy to be a cocycle:

\begin{displaymath}
\xymatrix{
&\vdots & \vdots & \vdots\\
(p-l+1) & K \ar@{^(->}[u] \ar[r]^{a = dx} & \quad \ar@<-4ex>[ddd] K \quad \ar@{^(->}[u] & K \ar@{^(->}[u]\\
& \vdots \ar@{^(->}[u] & \vdots \ar@{^(->}[u] & \vdots \ar@{^(->}[u]\\
(p-1) & F^{p-1}K \ar@{^(->}[u] \ar[r]^{a = dx} & F^{p-1}K \ar@{^(->}[u] & F^{p-1}K \ar@{^(->}[u]\\
(p) & F^{p}K \ar@{^(->}[u] \ar[r]^{a = dx} & a \in F^{p}K \ar@{^(->}[u] \ar[r]^{da} & \ar@<-4ex>[dd] \; F^{p}K \; \ar@{^(->}[u]\\
& \vdots \ar@{^(->}[u] & \vdots \ar@{^(->}[u] & \vdots \ar@{^(->}[u]\\
(p+l) & 0 \ar@{^(->}[u] & 0 \ar@{^(->}[u] & da \in \{0\} \qquad \ar@{^(->}[u]\\
& \vdots \ar@{^(->}[u] & \vdots \ar@{^(->}[u] & \vdots \ar@{^(->}[u]
}
\end{displaymath}
For this reason we use the following notations:
\begin{itemize}
\item we denote $Z^{p}K$ also by $Z^{p}_{l}K$, meaning that the boundary of an element must live $l$ steps under $p$;
\item we denote $BK \cap Z^{p}K$ also by $B^{p}_{l}K$, meaning that we consider boundaries of elements living $l-1$ steps over $p$.
\end{itemize}
In this way:
	\[F^{p}H(K) = Z^{p}_{l}K \,/\, B^{p}_{l}K.
\]
Now we can give the following generalizations for $r \geq 1$:
\begin{equation}
\begin{split}
	&Z^{p}_{r}(K) = \{a \in F^{p}K \,\vert\, d(a) \in F^{p+r}K\}\\
	&B^{p}_{r}(K) = \{a \in F^{p}K \,\vert\, a = d(x),\, x \in F^{p-r+1}K\}\\
	&F^{p}_{r}H(K) = Z^{p}_{r}K \,/\, B^{p}_{r}K.
\end{split}
\end{equation}
We can also consider $Z^{p}_{0}(K)$, and we trivially find that $Z^{p}_{0}(K) = F^{p}K$. Thus, we also put $B^{p}_{0}(K) = 0$, so that $F^{p}_{0}H(K) = F^{p}K$.

\paragraph{Remark:}\label{ShiftCob2} For coboundaries we still consider $p-r+1$ instead of $p-r$, as anticipated in the remark at page \pageref{ShiftCob1}.

\paragraph{}In this way we obtain two filtrations and a sequence:
	\[\begin{split}
&B^{p}K = B^{p}_{1}(K) \subset \cdots \subset B^{p}_{l}(K) = BK \cap Z^{p}K\\
&\qquad\qquad\qquad \subset Z^{p}K = Z^{p}_{l}K \subset \cdots \subset Z^{p}_{0}K = F^{p}K\\
&F^{p}_{1}H(K), \cdots, F^{p}_{l}H(K) = F^{p}H(K).
\end{split}\]
In particular, for $r$ increasing in $F^{p}_{r}H(K)$, we require that the coboundary of a generalized cocycle to live in smaller and smaller groups $F^{p+r}K$, and we allow that the coboundaries come from bigger and bigger groups $F^{p-r+1}K$. At the end, for $r = l$, we require that the coboundary of a cocycle is zero and we allow that the coboundaries come from all of $K$, hence we obtain $F^{p}H(K)$. In the diagram \ref{fig:Diagram} one can see the first step $F^{p}_{1}H(K)$ and the general step. In particular, passing from $r$ to $r+1$, the reader has to imagine the arrow of the cocycles (between the second and the third column) increasing by one step down, and the arrow of the coboundaries (between the first two columns) coming from one step more up.

\begin{figure}
\emph{First step:}
\begin{displaymath}
\xymatrix{
&\vdots & \vdots & \vdots\\
(p-1) & F^{p-1}K \ar@{^(->}[u] & F^{p-1}K \ar@{^(->}[u] & F^{p-1}K \ar@{^(->}[u]\\
(p) & F^{p}K \ar@{^(->}[u] \ar[r]^{a = dx} & a \in F^{p}K \ar@{^(->}[u] \ar[r]^{da} & \ar@<-4ex>[d] \; F^{p}K \; \ar@{^(->}[u]\\
(p+1) & F^{p+1}K \ar@{^(->}[u] & F^{p+1}K \ar@{^(->}[u] & da \in F^{p+1}K \ar@{^(->}[u]\\
& \vdots \ar@{^(->}[u] & \vdots \ar@{^(->}[u] & \vdots \ar@{^(->}[u]
}
\end{displaymath}
\emph{General step:}
\begin{displaymath}
\xymatrix{
&\vdots & \vdots & \vdots\\
(p-r+1) & F^{p-r+1}K \ar@{^(->}[u] \ar[r]^{a = dx} & \ar@<-4ex>[ddd] F^{p-r+1}K \ar@{^(->}[u] & F^{p-r+1}K \ar@{^(->}[u]\\
& \vdots \ar@{^(->}[u] & \vdots \ar@{^(->}[u] & \vdots \ar@{^(->}[u]\\
(p-1) & F^{p-1}K \ar@{^(->}[u] \ar[r]^{a = dx} & F^{p-1}K \ar@{^(->}[u] & F^{p-1}K \ar@{^(->}[u]\\
(p) & F^{p}K \ar@{^(->}[u] \ar[r]^{a = dx} & a \in F^{p}K \ar@{^(->}[u] \ar[r]^{da} & \ar@<-4ex>[dd] \; F^{p}K \; \ar@{^(->}[u]\\
& \vdots \ar@{^(->}[u] & \vdots \ar@{^(->}[u] & \vdots \ar@{^(->}[u]\\
(p+r) & F^{p+r}K \ar@{^(->}[u] & F^{p+r}K \ar@{^(->}[u] & da \in F^{p+r}K \ar@{^(->}[u]\\
& \vdots \ar@{^(->}[u] & \vdots \ar@{^(->}[u] & \vdots \ar@{^(->}[u]
}
\end{displaymath}
\caption{Diagram}\label{fig:Diagram}
\end{figure}

\paragraph{} We put:
	\[F_{r}H(K) = \bigoplus_{p \in \mathbb{Z}} F^{p}_{r}H(K).
\]
Now we construct a boundary $d^{p}_{r}: F^{p}_{r}H(K) \rightarrow F^{p+r}_{r}H(K)$:
\begin{itemize}
\item by construction, the boundary $d$ induces well-defined maps $\tilde{d}^{p}_{r}: F^{p}_{r}H(K) \rightarrow F^{p+r}K$, as one can see in the diagram \ref{fig:Diagram};
\item since $d^{2} = 0$, in particular one has $\tilde{d}^{p}_{r}: F^{p}_{r}H(K) \rightarrow Z^{p+r}K$;
\item since $Z^{p+r}K \subset Z^{p+r}_{s}K$ for every $s$, for $s = r$ we can think of $\tilde{d}^{p}_{r}: F^{p}_{r}H(K) \rightarrow Z^{p+r}_{r}K$;
\item composing with the projection, we obtain $d^{p}_{r}: F^{p}_{r}H(K) \rightarrow F^{p+r}_{r}H(K)$.
\end{itemize}
In this way, we obtain a coboundary $d_{r}: F_{r}H(K) \rightarrow F_{r}H(K)$. Here we see the importance of the $+1$ shift in the coboundary index, as discussed in the remarks at pages \pageref{ShiftCob1} and \pageref{ShiftCob2}: \emph{without that shift, the imagine of $d^{p}_{r}$ would be $0$}, since the boundary $d^{p}_{r}$ of an element by definition lives at the level $p+r$ and comes from the level $p$, so, quotienting out by coboundaries from $r$ step above $p+r$, we would trivially obtain $0$. With the shift, we quotient out by $r-1$ steps above $p+r$, hence we obtain in general a non-trivial cohomology class.

\paragraph{}Now we would like to show that $F_{r+1}H(K)$ is isomorphic to $H(F_{r}H(K),d_{r})$, but in this setting it is not true in general. Let us see the exact meaning of this isomorphism. Since $F^{p}_{r}H(K) = Z^{p}_{r}K / B^{p}_{r}K$ and $F^{p}_{r+1}H(K) = Z^{p}_{r+1}K / B^{p}_{r+1}K$, and since $B^{p}_{r}K \subset Z^{p}_{r+1}K, B^{p}_{r+1}K \subset Z^{p}_{r}K$, the fact that $F_{r+1}H(K) \simeq H(F_{r}H(K),d_{r})$ should be naturally implied by the following conditions:
\begin{itemize}
\item $\Ker \, d^{p}_{r} \;\; \overset{?}= Z^{p}_{r+1}K / B^{p}_{r}K \subset Z^{p}_{r}K / B^{p}_{r}K$;
\item $\IIm \, d^{p-r}_{r} \overset{?}= B^{p}_{r+1}K / B^{p}_{r}K \subset Z^{p}_{r}K / B^{p}_{r}K$.
\end{itemize}
The second equality is trivially true, but the first one is false: let us consider $[a] \in F^{p}_{r}H(K)$ with $a \in Z^{p}_{r+1}K$. Then by construction one has $\tilde{d}^{p}_{r}(a) \in Z^{p+r}_{r}K$, but the fact that $a \in Z^{p}_{r+1}K$ implies $\tilde{d}^{p}_{r}(a) \in Z^{p+r+1}_{r-1}K \subset Z^{p+r}_{r}K$. However, there is no reason why its equivalence class in $Z^{p+r}_{r}K / B^{p+r}_{r}K$ is zero. \emph{For such a class to be zero, we must quotient out by elements of $Z^{p+r+1}_{r-1}K$, i.e.\ by cocycles living one step under the image of $\tilde{d}^{p}_{r}$}. Thus, instead of considering $Z^{p+r}_{r}K / B^{p+r}_{r}K$, we must consider $Z^{p+r}_{r}K / \, \langle B^{p+r}_{r}K, i(Z^{p+r+1}_{r-1}K) \rangle$. Hence, for generic level $p$, we must consider the groups $Z^{p}_{r}K / \, \langle B^{p}_{r}K, i(Z^{p+1}_{r-1}K) \rangle$, which, for $r = l$, becomes $Z^{p}K / \, \langle BK \cap Z^{p}K, i(Z^{p+1}K) \rangle$. But this is exactly $F^{p}H(K) / F^{p+1}H(K)$, since, using lemma \ref{AlgLemma}:
	\[\frac{F^{p}H(K)}{F^{p+1}H(K)} = \frac{\langle Z^{p}K, BK \rangle / BK}{\langle Z^{p+1}K, BK \rangle / BK} = \frac{\langle Z^{p}K, BK \rangle}{\langle Z^{p+1}K, BK \rangle} = \frac{Z^{p}K}{\langle Z^{p+1}K, BK \cap Z^{p}K \rangle}.
\]

\subsubsection{First correction}

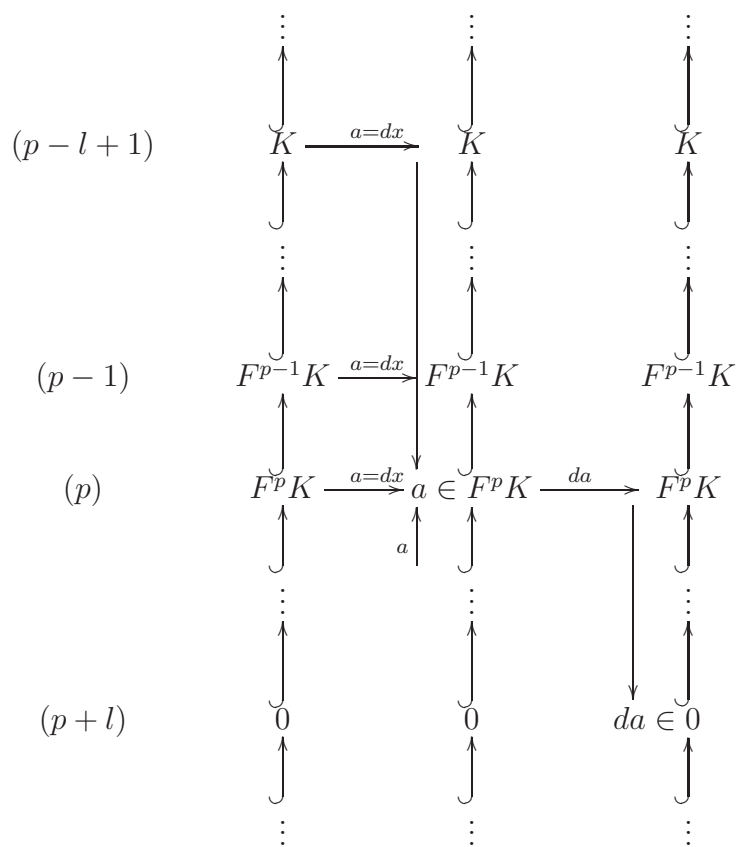
\begin{figure}
\begin{displaymath}
\xymatrix{
&\vdots & \vdots & \vdots\\
(p-l+1) & K \ar@{^(->}[u] \ar[r]^{a = dx} & \quad \ar@<-4ex>[ddd] K \quad \ar@{^(->}[u] & K \ar@{^(->}[u]\\
& \vdots \ar@{^(->}[u] & \vdots \ar@{^(->}[u] & \vdots \ar@{^(->}[u]\\
(p-1) & F^{p-1}K \ar@{^(->}[u] \ar[r]^{a = dx} & F^{p-1}K \ar@{^(->}[u] & F^{p-1}K \ar@{^(->}[u]\\
(p) & F^{p}K \ar@{^(->}[u] \ar[r]^{a = dx} & a \in F^{p}K \ar@{^(->}[u] \ar[r]^{da} & \ar@<-4ex>[dd] \; F^{p}K \; \ar@{^(->}[u]\\
& \vdots \ar@{^(->}[u] & \vdots \ar@<4ex>[u]^{a} \ar@{^(->}[u] & \vdots \ar@{^(->}[u]\\
(p+l) & 0 \ar@{^(->}[u] & 0 \ar@{^(->}[u] & da \in 0 \qquad \ar@{^(->}[u]\\
& \vdots \ar@{^(->}[u] & \vdots \ar@{^(->}[u] & \vdots \ar@{^(->}[u]
}
\end{displaymath}
\caption{Correction}\label{fig:DiagramCorr}
\end{figure}

Thus, we search an approximation not of $F^{p}H(K)$, but of $E_{0}^{p}H(K) = F^{p}H(K)$ $/ F^{p+1}H(K)$.
For $F^{p}H(K)$ thought at the level $p$, i.e.\ $F^{p}H(K) = Z^{p}K / (BK \cap Z^{p}K)$, we have that:
\begin{equation}\label{E0pHK}
E_{0}^{p}H(K) = Z^{p}K / \, \langle BK \cap Z^{p}K, i(Z^{p+1}K) \rangle.
\end{equation}
As one can see in the diagram \ref{fig:DiagramCorr}, there is a new vertical arrow for $Z^{p+1}K$.
We could give the following definitions:
\begin{equation}
\begin{split}
	&\tilde{B}^{p}_{r}K = \langle B^{p}_{r}K, i(Z^{p+1}_{r-1}K) \rangle \subset Z^{p}K\\
	&E^{p}_{r}K = Z^{p}_{r}K / \tilde{B}^{p}_{r}K
\end{split}
\end{equation}
which, for $r = l$, become:
	\[\begin{split}
	&\tilde{B}^{p}_{l}K = \langle BK \cap Z^{p}K, i(Z^{p+1}K) \rangle\\
	&E^{p}_{0}H(K) = Z^{p}_{l}K /\, \tilde{B}^{p}_{l}K.
\end{split}\]
The problem is that, in this way, we do not have a filtration $\tilde{B}^{p}_{1}K \subset \cdots \subset \tilde{B}^{p}_{l}K$, since the groups $Z^{p+1}_{r-1}K$ are \emph{decreasing} in $r$. Thus we introduce a correction on cocycles and coboundaries which does not affect the quotient.

\subsubsection{The right sequence}

There is also another way to think of $E_{0}^{p}H(K)$ (cfr.\ equation \eqref{E0pHK}):
\begin{equation}\label{E0pHK2}
E_{0}^{p}H(K) \simeq \langle Z^{p}K, i(F^{p+1}K) \rangle \, / \, \langle BK \cap Z^{p}K, i(F^{p+1}K) \rangle.
\end{equation}
This equality follows from lemma \ref{AlgLemma} for $A_{1} = Z^{p}K$, $A_{2} = BK \cap Z^{p}K$, $B = i(F^{p+1}K)$. In other words, we consider all the elements of $F^{p+1}K$, not only the cocycles. We use this point of view during the construction of the spectral sequence. We thus give the definitions for $r \geq 1$:
\begin{equation}
\begin{split}
	&\overline{Z}^{p}_{r}K = \langle Z^{p}_{r}K, i(F^{p+1}K) \rangle\\
	&\overline{B}^{p}_{r}K = \langle B^{p}_{r}K, i(F^{p+1}K) \rangle\\
	&E^{p}_{r}K = \overline{Z}^{p}_{r}K \,/\, \overline{B}^{p}_{r}K
\end{split}
\end{equation}
which, for $r = l$, become:
	\[\begin{split}
	&\overline{Z}^{p}_{l}(K) = \langle Z^{p}K, i(F^{p+1}K) \rangle\\
	&\overline{B}^{p}_{l}(K) = \langle BK \cap Z^{p}K, i(F^{p+1}K) \rangle\\
	&E^{p}_{0}H(K) = \overline{Z}^{p}_{l}K \,/\, \overline{B}^{p}_{l}K.
\end{split}\]
In this way we obtain two filtrations and a sequence:
	\[\begin{split}
&\langle B^{p}K, F^{p+1}K \rangle = \overline{B}^{p}_{1}(K) \subset \cdots \subset \overline{B}^{p}_{l}(K) = \langle BK \cap Z^{p}K, i(F^{p+1}K) \rangle\\
&\qquad\qquad\qquad\qquad\qquad \subset \langle Z^{p}K, i(F^{p+1}K) \rangle = \overline{Z}^{p}_{l}K \subset \cdots \subset \overline{Z}^{p}_{0}K = F^{p}K\\
&E^{p}_{0}H(K), \cdots, E^{p}_{l}H(K) = E^{p}_{0}H(K).
\end{split}\]
On the diagram \ref{fig:Diagram2}, one can see the first step $E^{p}_{1}K$, which is exactly $H(F^{p}K/F^{p+1}K)$, and the general step. Since the first step is $H(F^{p}K/F^{p+1}K)$, we can define as before $B^{p}_{0}K = \{0\}$ and $Z^{p}_{0}K = F^{p}K$: in this way, $\overline{Z}^{p}_{0}K = F^{p}K$ and $\overline{B}^{p}_{0}K = F^{p+1}K$, so that $E^{p}_{0}K = F^{p}K / F^{p+1}K$, as previously defined.
\begin{figure}
\emph{First step:}
\begin{displaymath}
\xymatrix{
&\vdots & \vdots & \vdots\\
(p-1) & F^{p-1}K \ar@{^(->}[u] & F^{p-1}K \ar@{^(->}[u] & F^{p-1}K \ar@{^(->}[u]\\
(p) & F^{p}K \ar@{^(->}[u] \ar[r]^{a = dx} & a \in F^{p}K \ar@{^(->}[u] \ar[r]^{da} & \ar@<-4ex>[d] \; F^{p}K \; \ar@{^(->}[u]\\
(p+1) & F^{p+1}K \ar@{^(->}[u] & F^{p+1}K \ar@<4ex>[u]^{a} \ar@{^(->}[u] & da \in F^{p+1}K \ar@{^(->}[u]\\
& \vdots \ar@{^(->}[u] & \vdots \ar@{^(->}[u] & \vdots \ar@{^(->}[u]
}
\end{displaymath}
\emph{General step:}
\begin{displaymath}
\xymatrix{
&\vdots & \vdots & \vdots\\
(p-r+1) & F^{p-r+1}K \ar@{^(->}[u] \ar[r]^{a = dx} & \ar@<-4ex>[ddd] F^{p-r+1}K \ar@{^(->}[u] & F^{p-r+1}K \ar@{^(->}[u]\\
& \vdots \ar@{^(->}[u] & \vdots \ar@{^(->}[u] & \vdots \ar@{^(->}[u]\\
(p-1) & F^{p-1}K \ar@{^(->}[u] \ar[r]^{a = dx} & F^{p-1}K \ar@{^(->}[u] & F^{p-1}K \ar@{^(->}[u]\\
(p) & F^{p}K \ar@{^(->}[u] \ar[r]^{a = dx} & a \in F^{p}K \ar@{^(->}[u] \ar[r]^{da} & \ar@<-4ex>[dd] \; F^{p}K \; \ar@{^(->}[u]\\
& \vdots \ar@{^(->}[u] & \vdots \ar@<4ex>[u]^{a} \ar@{^(->}[u] & \vdots \ar@{^(->}[u]\\
(p+r) & F^{p+r}K \ar@{^(->}[u] & F^{p+r}K \ar@{^(->}[u] & da \in F^{p+r}K \ar@{^(->}[u]\\
& \vdots \ar@{^(->}[u] & \vdots \ar@{^(->}[u] & \vdots \ar@{^(->}[u]
}
\end{displaymath}
\caption{Diagram}\label{fig:Diagram2}
\end{figure}

\paragraph{} We put:
	\[E_{r}K = \bigoplus_{p \in \mathbb{Z}} E^{p}_{r}K.
\]
Now we build the map $d^{p}_{r}: E^{p}_{r}K \rightarrow E^{p+r}_{r}K$. Let $[a] \in E^{p}_{r}K$. Then:
\begin{itemize}
\item $a = z + x$, with $z \in Z^{p}_{r}K$ and $x \in F^{p+1}K$;
\item $d(z) \in Z^{p+r}K$ and $d(x) \in B^{p+1}K$;
\item hence $d(a) \in F^{p+1}K$, in particular $d(a) \in \langle i^{r-1}(Z^{p+r}_{r}K), B^{p+1}K \rangle$;
\item hence we have also $d(a) \in \langle\, i^{r-1}(Z^{p+r}_{r}K), i^{r}(F^{p+r+1}), B^{p+1}K \,\rangle$;
\item we consider:
	\[[d(a)] \in \frac{\langle\, i^{r-1}(Z^{p+r}_{r}K), i^{r}(F^{p+r+1}), B^{p+1}K \,\rangle}{\langle\, B^{p+1}K, i^{r}(F^{p+r+1}) \,\rangle};
\]
\item there is an isomorphism, obtained from lemma \ref{AlgLemma} for $A_{1} = \langle\, Z^{p+r}_{r}, i(F^{p+r+1}) \,\rangle$, $A_{2} = i(F^{p+r+1})$, $B = B^{p+1}K$:
	\[\begin{split}
	\varphi: \, &\frac{\langle\, i^{r-1}(Z^{p+r}_{r}K), i^{r}(F^{p+r+1}), B^{p+1}K \,\rangle}{\langle\, B^{p+1}K, i^{r}(F^{p+r+1}) \,\rangle}\\
	& \phantom{XXXXXXXXX} \overset{\simeq}\longrightarrow \; \frac{\langle\, Z^{p+r}_{r}K, i^{r}(F^{p+r+1}) \,\rangle}{\langle\, (B^{p+1}K \cap Z^{p+r}_{r}K), i^{r}(F^{p+r+1}) \,\rangle}
\end{split}\]
given by $\varphi[\,x\,] = [\,x\,] \cap F^{p+r}K$, and the second member is exactly $E^{p+r}_{r}K$;
\item hence, we put $d^{p}_{r}[a] = \varphi([d(a)]) \in E^{p+r}_{r}K$.
\end{itemize} 

In this setting, we can see that $E_{r+1}K = H(E_{r}K, d_{r})$, i.e.\ $E^{p}_{r+1}K = \Ker \, d^{p}_{r} \,/\, \IIm \, d^{p-r}_{r}$. In fact, since $E^{p}_{r}K = \overline{Z}^{p}_{r}K /\, \overline{B}^{p}_{r}K$ and $E^{p}_{r+1}K = \overline{Z}^{p}_{r+1}K /\, \overline{B}^{p}_{r+1}K$, the fact that $E_{r+1}K = H(E_{r}K,d_{r})$ naturally follows from:
\begin{itemize}
\item $\Ker \, d^{p}_{r} \;\; = \overline{Z}^{p}_{r+1}K /\, \overline{B}^{p}_{r}K \subset \overline{Z}^{p}_{r}K /\, \overline{B}^{p}_{r}K$;
\item $\IIm \, d^{p-r}_{r} = \overline{B}^{p}_{r+1}K /\, \overline{B}^{p}_{r}K \subset \overline{Z}^{p}_{r}K /\, \overline{B}^{p}_{r}K$.
\end{itemize}
Let us prove this (the reader should look at the diagram while following the proof):
\begin{itemize}
\item \emph{Kernel}:
	\begin{itemize}
	\item[$\supset)$] Let $[a] \in \overline{Z}^{p}_{r+1}K /\, \overline{B}^{p}_{r}K$. Hence:
	\begin{itemize}
	\item[$\bullet$] $a = z + x$ for $x \in Z^{p}_{r+1}K$ and $x \in F^{p+1}K$;
	\item[$\bullet$] $d(a) = d(z) + d(x)$, with $d(z) \in F^{p+r+1}K$ and $d(x) \in B^{p+1}K$;
	\item[$\bullet$] hence $[d(a)] = [0]$.
	\end{itemize}
	\item[$\subset)$] Let $[d(a)] = [0]$, i.e.\ $d(a) \in \langle B^{p+1}K, i^{r}(F^{p+r+1}K) \rangle$.
	\begin{itemize}
	\item[$\bullet$] $a = z + x$ for $x \in Z^{p}_{r}K$ and $x \in F^{p+1}K$;
	\item[$\bullet$] $d(a) = d(z) + d(x)$, with $d(z) \in F^{p+r}K$ and $d(x) \in B^{p+1}K$;
	\item[$\bullet$] hence, by the hypothesis on $d(a)$, one has $d(z) \in \langle B^{p+1}K, i^{r}(F^{p+r+1}K) \rangle$, i.e.\ $d(z) = d(f^{p+1}) + g^{p+r+1}$, with $f^{p+1} \in F^{p+1}K$ and $g^{p+r+1} \in F^{p+r+1}K$;
	\item[$\bullet$] thus $z = (z - f^{p+1}) + f^{p+1}$, with $z - f^{p+1} \in Z^{p}_{r+1}K$ and $f^{p+1} \in F^{p+1}K$;
	\item[$\bullet$] hence $a \in \langle Z^{p}_{r+1}K, i(F^{p+1}K) \rangle = \overline{Z}^{p}_{r+1}K$.
	\end{itemize}
	\end{itemize}
\item \emph{Image}: follows directly from the definitions.
\end{itemize}

\subsection{Second viewpoint}

In this viewpoint we do not use isomorphism \eqref{IsoFpHK}, hence we still think of $F^{p}H(K)$ at level $-\infty$, which coincides, for finite filtrations, with level $0$.

\subsubsection{First attempt with $F^{p}H(K)$}

As before, we first consider a natural approximation of $F^{p}H(K) = \langle i^{p}Z^{p}K,$ $BK \rangle / BK$. The following digram represents what we are going to explain:

\begin{displaymath}
\xymatrix{
&\vdots & \vdots & \vdots\\
(p-l+1) & K \ar@{^(->}[u] \ar[r]^{a = dx} & a \in K \quad \ar@{^(->}[u] & K \ar@{^(->}[u]\\
& \vdots \ar@{^(->}[u] & \vdots \ar@{^(->}[u] & \vdots \ar@{^(->}[u]\\
(p-1) & F^{p-1}K \ar@{^(->}[u] \ar[r]^{a = dx} & F^{p-1}K \ar@{^(->}[u] & F^{p-1}K \ar@{^(->}[u]\\
(p) & F^{p}K \ar@{^(->}[u] \ar[r]^{a = dx} & a \in F^{p}K \ar@{^(->}[u] \ar@<4ex>[uuu] \ar[r]^{da} & \ar@<-4ex>[dd] \; F^{p}K \; \ar@{^(->}[u]\\
& \vdots \ar@{^(->}[u] & \vdots \ar@{^(->}[u] & \vdots \ar@{^(->}[u]\\
(p+l) & 0 \ar@{^(->}[u] & 0 \ar@{^(->}[u] & da \in 0 \qquad \ar@{^(->}[u]\\
& \vdots \ar@{^(->}[u] & \vdots \ar@{^(->}[u] & \vdots \ar@{^(->}[u]
}
\end{displaymath}

In particular, since the filtration has length $l$, for $0 \leq p \leq l-1$ one has $F^{p+l}K = 0$ and $F^{p-l+1}K$ $= K$. Hence:
\begin{equation}\label{LastApprox2}
\begin{split}
	&Z^{p}K = \{a \in F^{p}K \,\vert\, d(a) \in F^{p+l}K\}\\
	&BK = B^{p-l+1}K.
\end{split}
\end{equation}
We shift by $+1$ the coboundary index for the same reason as in the first viewpoint. For this reason we use the following notations:
\begin{itemize}
\item we denote $Z^{p}K$ also by $Z^{p}_{l}K$, meaning that the boundary of an element must live $l$ steps under $p$;
\item we denote $BK$ also by $\mathfrak{B}^{p}_{l}K$, meaning that we consider boundaries living $l-1$ steps over $p$.
\end{itemize}
In this way:
	\[F^{p}H(K) = \langle i^{l-1}Z^{p}_{l}K, \mathfrak{B}^{p}_{l}K \rangle / \mathfrak{B}^{p}_{l}K.
\]
Now we can give the following generalizations for $r \geq 1$:
\begin{equation}\label{ApproxFH2}
\begin{split}
	&Z^{p}_{r}K = \{a \in F^{p}K \,\vert\, d(a) \in F^{p+r}K\}\\
	&\mathfrak{B}^{p}_{r}K = B^{p-r+1}K\\
	&Z\mathfrak{B}^{p}_{r}K = \langle i^{r-1}Z^{p}_{r}K, \mathfrak{B}^{p}_{r}K \rangle \subset F^{p-r+1}K\\
	&F^{p}_{r}H(K) = Z\mathfrak{B}^{p}_{r} / \mathfrak{B}^{p}_{r}K.
\end{split}
\end{equation}
As before, we can extend the definition of $Z^{p}_{r}K$ for $r = 0$. In this way we obtain two filtrations (for $Z^{p}_{\bullet}$ and $\mathfrak{B}^{p}_{\bullet}$, not for $Z\mathfrak{B}^{p}_{\bullet}$ !) and a sequence:
	\[\begin{split}
&B^{p}K = \mathfrak{B}^{p}_{1}(K) \subset \cdots \subset \mathfrak{B}^{p}_{l}(K) = BK\\
&Z^{p}K = Z^{p}_{l}K \subset \cdots \subset Z^{p}_{0}K = F^{p}K\\
&F^{p}_{1}H(K), \cdots, F^{p}_{l}H(K) = F^{p}H(K).
\end{split}\]
In particular, for $r$ increasing in $F^{p}_{r}H(K)$, we require that the boundary of a generalized cocycle lives in smaller and smaller groups $F^{p+r}K$, and we quotient by coboundaries living in bigger and bigger groups $F^{p-r+1}K$. At the end, for $r = l$, we require that the boundary of a cocycle is zero and we quotient out by coboundaries from all of $K$, hence we obtain $F^{p}H(K)$.

We remark that this viewpoint is less natural than the previous, since, although the cocycles $Z^{p}_{r}K$ decrease in $r$, they are embedded in $Z\mathfrak{B}^{p}_{r}K$ which are not decreasing in $r$ any more. However, the coboundaries added in $Z\mathfrak{B}^{p}_{r}K$ are quotiented out in $F^{p}_{r}H(K)$, so that, \emph{at the quotient}, we still obtain the same approximation we got before by lemma \ref{AlgLemma}. Instead, the construction of the boundary will be more natural in this viewpoint.

\paragraph{Remark:} we point out that $F^{p}_{r}H(K)$ must be thought at the level $p-r+1$ of the filtration.

\paragraph{}On the diagram \ref{fig:Diagram3}, one can see the first step $F^{p}_{1}H(K)$ and the general step.

\begin{figure}
\emph{First step:}
\begin{displaymath}
\xymatrix{
&\vdots & \vdots & \vdots\\
(p-1) & F^{p-1}K \ar@{^(->}[u] & F^{p-1}K \ar@{^(->}[u] & F^{p-1}K \ar@{^(->}[u]\\
(p) & F^{p}K \ar@{^(->}[u] \ar[r]^{a = dx} & a \in F^{p}K \ar@{^(->}[u] \ar[r]^{da} & \ar@<-4ex>[d] \; F^{p}K \; \ar@{^(->}[u]\\
(p+1) & F^{p+1}K \ar@{^(->}[u] & F^{p+1}K \ar@{^(->}[u] & da \in F^{p+1}K \ar@{^(->}[u]\\
& \vdots \ar@{^(->}[u] & \vdots \ar@{^(->}[u] & \vdots \ar@{^(->}[u]
}
\end{displaymath}
\emph{General step:}
\begin{displaymath}
\xymatrix{
&\vdots & \vdots & \vdots\\
(p-r+1) & F^{p-r+1}K \ar@{^(->}[u] \ar[r]^{a = dx} & F^{p-r+1}K \ar@{^(->}[u] & F^{p-r+1}K \ar@{^(->}[u]\\
& \vdots \ar@{^(->}[u] & \vdots \ar@{^(->}[u] & \vdots \ar@{^(->}[u]\\
(p-1) & F^{p-1}K \ar@{^(->}[u] \ar[r]^{a = dx} & F^{p-1}K \ar@{^(->}[u] & F^{p-1}K \ar@{^(->}[u]\\
(p) & F^{p}K \ar@{^(->}[u] \ar[r]^{a = dx} & \ar@<4ex>[uuu] a \in F^{p}K \ar@{^(->}[u] \ar[r]^{da} & \ar@<-4ex>[dd] \; F^{p}K \; \ar@{^(->}[u]\\
& \vdots \ar@{^(->}[u] & \vdots \ar@{^(->}[u] & \vdots \ar@{^(->}[u]\\
(p+r) & F^{p+r}K \ar@{^(->}[u] & F^{p+r}K \ar@{^(->}[u] & da \in F^{p+r}K \ar@{^(->}[u]\\
& \vdots \ar@{^(->}[u] & \vdots \ar@{^(->}[u] & \vdots \ar@{^(->}[u]
}
\end{displaymath}
\caption{Diagram}\label{fig:Diagram3}
\end{figure}

\paragraph{} We put:
	\[F_{r}H(K) = \bigoplus_{p \in \mathbb{Z}} F^{p}_{r}H(K).
\]
By construction, the boundary $d$ induces well-defined maps $\tilde{d}^{p}_{r}: F^{p}_{r}H(K) \rightarrow F^{p+r}K$, as one can see in the diagram. Since $d^{2} = 0$, in particular one has $\tilde{d}^{p}_{r}: F^{p}_{r}H(K) \rightarrow Z^{p+r}_{r}K$, hence, composing with $i^{r-1}$ and considering the class up to coboundaries in $B^{p+1}K$, we obtain $d^{p}_{r}: F^{p}_{r}H(K) \rightarrow F^{p+r}_{r}H(K)$. In this way, we obtain a cohomology $d_{r}: F_{r}H(K) \rightarrow F_{r}H(K)$. As in the first viewpoint, the $+1$ shift in the coboundary index allows us to obtain a non-trivial cohomology class.

\paragraph{}We would like now to show that $F_{r+1}H(K)$ is isomorphic to $H(F_{r}H(K),d_{r})$, but in this setting it is not true for the same reason discussed in the first viewpoint.

\subsubsection{The right sequence}

Thus, we search an approximation not of $F^{p}H(K)$, but of $E_{0}^{p}H(K) = F^{p}H(K)$ $/ F^{p+1}H(K)$. We think of $E_{0}^{p}H(K)$ as:
\begin{equation}\label{E0pHK4}
E_{0}^{p}H(K) \simeq \langle i^{p}Z^{p}K, BK, i^{p+1}F^{p+1}K \rangle \, / \, \langle BK, i^{p+1}F^{p+1}K \rangle.
\end{equation}
This reduces to \eqref{E0pHK2} using lemma \ref{AlgLemma} for $A_{1} = \langle i^{p}Z^{p}K, BK \rangle$, $A_{2} = BK$, $B = i^{p+1}F^{p+1}K$.

We thus use the definitions:
\begin{equation}
\begin{split}
	&\overline{\mathfrak{B}}^{p}_{r}K = \langle \mathfrak{B}^{p}_{r}K, i^{r}F^{p+1}K \rangle = \langle B^{p-r+1}K, i^{r}F^{p+1}K \rangle\\
	&\overline{Z\mathfrak{B}}^{p}_{r}K = \langle Z\mathfrak{B}^{p}_{r}K, i^{r}F^{p+1}K \rangle = \langle i^{r-1}Z^{p}_{r}K, B^{p-r+1}K, i^{r}F^{p+1}K \rangle\\
	&E^{p}_{r}K = \overline{Z\mathfrak{B}}^{p}_{r}K / \, \overline{\mathfrak{B}}^{p}_{r}K
\end{split}
\end{equation}
which for $r = l$ become:
	\[\begin{split}
	&\overline{\mathfrak{B}}^{p}_{l}K = \langle BK, i^{p+1}F^{p+1}K \rangle\\
	&\overline{Z\mathfrak{B}}^{p}_{l}K = \langle i^{p}Z^{p}K, BK, i^{p+1}F^{p+1}K \rangle\\
	&E^{p}_{l}K = E^{p}_{0}H(K).
\end{split}\]
In diagram \ref{fig:Diagram4} one can see the first step $E^{p}_{1}K$, which is exactly $H(F^{p}K/F^{p+1}K)$, and the general step. As before, we can extend the definition of $Z^{p}_{r}K$ for $r = 0$, obtaining $Z^{p}_{0}K = F^{p}K$. Then, we define $\mathfrak{B}^{p}_{0}K = \{0\}$ and $Z\mathfrak{B}^{p}_{0}K = F^{p}K$, hence $\overline{Z\mathfrak{B}}^{p}_{0}K = F^{p}K$ and $\overline{\mathfrak{B}}^{p}_{0}K = F^{p+1}K$, so that $E^{p}_{0}K = F^{p}K / F^{p+1}K$.
\begin{figure}
\emph{First step:}
\begin{displaymath}
\xymatrix{
&\vdots & \vdots & \vdots\\
(p-1) & F^{p-1}K \ar@{^(->}[u] & F^{p-1}K \ar@{^(->}[u] & F^{p-1}K \ar@{^(->}[u]\\
(p) & F^{p}K \ar@{^(->}[u] \ar[r]^{a = dx} & a \in F^{p}K \ar@{^(->}[u] \ar[r]^{da} & \ar@<-4ex>[d] \; F^{p}K \; \ar@{^(->}[u]\\
(p+1) & F^{p+1}K \ar@{^(->}[u] & F^{p+1}K \ar@<4ex>[u]^{a} \ar@{^(->}[u] & da \in F^{p+1}K \ar@{^(->}[u]\\
& \vdots \ar@{^(->}[u] & \vdots \ar@{^(->}[u] & \vdots \ar@{^(->}[u]
}
\end{displaymath}
\emph{General step:}
\begin{displaymath}
\xymatrix{
&\vdots & \vdots & \vdots\\
(p-r+1) & F^{p-r+1}K \ar@{^(->}[u] \ar[r]^{a = dx} & F^{p-r+1}K \ar@{^(->}[u] & F^{p-r+1}K \ar@{^(->}[u]\\
& \vdots \ar@{^(->}[u] & \vdots \ar@{^(->}[u] & \vdots \ar@{^(->}[u]\\
(p-1) & F^{p-1}K \ar@{^(->}[u] \ar[r]^{a = dx} & F^{p-1}K \ar@{^(->}[u] & F^{p-1}K \ar@{^(->}[u]\\
(p) & F^{p}K \ar@{^(->}[u] \ar[r]^{a = dx} & a \in F^{p}K \ar@{^(->}[u] \ar@<4ex>[uuu] \ar[r]^{da} & \ar@<-4ex>[dd] \; F^{p}K \; \ar@{^(->}[u]\\
& \vdots \ar@{^(->}[u] & \vdots \ar@<4ex>[u]^{a} \ar@{^(->}[u] & \vdots \ar@{^(->}[u]\\
(p+r) & F^{p+r}K \ar@{^(->}[u] & F^{p+r}K \ar@{^(->}[u] & da \in F^{p+r}K \ar@{^(->}[u]\\
& \vdots \ar@{^(->}[u] & \vdots \ar@{^(->}[u] & \vdots \ar@{^(->}[u]
}
\end{displaymath}
\caption{Diagram}\label{fig:Diagram4}
\end{figure}

\paragraph{} We put:
	\[E_{r}K = \bigoplus_{p \in \mathbb{Z}} E^{p}_{r}K.
\]
We have a natural boundary $d^{p}_{r}: E^{p}_{r}K \rightarrow E^{p+r}_{r}K$. In fact, let $[a] \in E^{p}_{r}K$. Then:
\begin{itemize}
\item $a = z + b + x$, with $z \in Z^{p}_{r}K$, $b \in B^{p-r+1}K$ and $x \in F^{p+1}K$, and $[a] = [z]$ in $E^{p}_{r}K$;
\item $d(z) \in Z^{p+r}K \subset Z^{p+r}_{r}K \subset \overline{Z\mathfrak{B}}^{p+r}_{r}K$, \; $d(b) = 0$ and $d(x) \in B^{p+1}K \subset \overline{\mathfrak{B}}^{p+r}_{r}K$;
\item hence $[d(a)] = [i^{r-1}(d(z))] \in E^{p+r}_{r}K$, so that we can define $d^{p}_{r}[a] = [d(a)]$.
\end{itemize} 
It is well defined, since if $[z_{1}] = [z_{2}]$, then $z_{1} - z_{2} = x + dy$ with $x \in F^{p+1}K$, hence $dz_{1} - dz_{2} = dx$ and $[dx] = 0$ in $E^{p+r}_{r}K$.

\paragraph{}In this setting, we can see that $E_{r+1}K = H(E_{r}K, d_{r}$). In fact, we prove that:
\begin{itemize}
\item $\Ker(d^{p}_{r}) = \langle i^{r-1}Z^{p}_{r+1}K, \overline{\mathfrak{B}}^{p}_{r}K \rangle \,/\, \overline{\mathfrak{B}}^{p}_{r}K \subset \overline{Z\mathfrak{B}}^{p}_{r}K \,/\, \overline{\mathfrak{B}}^{p}_{r}K$;
\item $\IIm(d^{p-r}_{r}) = \langle \overline{\mathfrak{B}}^{p}_{r}K, i^{r-1}(B^{p-r}K \cap F^{p}K) \rangle \,/\, \overline{\mathfrak{B}}^{p}_{r}K \subset \overline{Z\mathfrak{B}}^{p}_{r}K \,/\, \overline{\mathfrak{B}}^{p}_{r}K$.
\end{itemize}
Let us prove this (the reader should look at the diagram while following the proof):
\begin{itemize}
\item \emph{Kernel}:
	\begin{itemize}
	\item[$\supset$)] If $a = z + f + b$ corresponding to the decomposition in the statement, then $d(a) = d(z) + d(f)$, with $d(z) \in F^{p+r+1}K$ and $d(f) \in B^{p+1}K$, hence $[d(a)] = 0$ in $E^{p+r}_{r}K$.
	\item[$\subset$)] For $a = z + f + b$ as in the construction of $d^{p}_{r}$, if $[d(a)] = 0$ then $[d(z)] = 0$, i.e.\ $d(z) = d(y) + w$ with $y \in F^{p+1}K$ and $w \in F^{p+r+1}K$. But then $z = (z - y) + y$, with $z - y \in Z^{p}_{r+1}K$ and $y \in F^{p+1}K$. Hence $a \in \langle i^{r-1}Z^{p}_{r+1}K, \overline{\mathfrak{B}}^{p}_{r}K \rangle$.
	\end{itemize}
\item \emph{Image}:
	\begin{itemize}
	\item[$\subset$)] Since $\overline{Z\mathfrak{B}}^{p-r}_{r}K = \langle Z^{p-r}_{r}K, \mathfrak{B}^{p-2r+1}K, F^{p-r+1}K \rangle$, then, for $a = z + b + f$, we obtain $d^{p-r}_{r}[a] = [d(z)] + [d(f)]$, with $d(z) \in B^{p-r}K \cap F^{p}K$ and $d(f) \in B^{p-r+1}K$.
	\item[$\supset$)] We have $\langle \overline{\mathfrak{B}}^{p}_{r}K, i^{r-1}(\mathfrak{B}^{p-r}K \cap F^{p}K) \rangle = \langle B^{p-r+1}K, i^{r}F^{p+1}K, i^{r-1}(B^{p-r}K \cap F^{p}K) \rangle$. Let $a = d(f) + g + d(h)$: then $[a] = [d(h)]$ with respect to $\overline{\mathfrak{B}}^{p}_{r}K$, hence $[a] = d^{p-r}_{r}[h]$.
	\end{itemize}
\end{itemize}

Hence:
	\[\frac{\Ker(d^{p}_{r})}{\IIm(d^{p-r}_{r})} = \frac{\langle i^{r-1}Z^{p}_{r+1}K, \overline{\mathfrak{B}}^{p}_{r}K \rangle} {\langle \overline{\mathfrak{B}}^{p}_{r}K, i^{r-1}(B^{p-r}K \cap F^{p}K) \rangle}.
\]

Let $A_{1}$ and $A_{2}$ be the numerator and the denominator of the previous formula, and let $B = \overline{\mathfrak{B}}^{p}_{r+1}K$. Then:
\begin{itemize}
\item $\langle A_{1}, B \rangle = \langle i^{r-1}Z^{p}_{r+1}K, \overline{\mathfrak{B}}^{p}_{r+1}K \rangle = \overline{Z\mathfrak{B}}^{p}_{r+1}K$;
\item since $B \subset A_{2}$, it follows that:
	\begin{itemize}
	\item[$\bullet$] $\langle A_{2}, B \rangle = B = \overline{\mathfrak{B}}^{p}_{r+1}K$;
	\item[$\bullet$] $A_{1} \cap B \subset A_{2}$.
	\end{itemize}
\end{itemize}
Then, applying lemma $\ref{AlgLemma}$, we obtain that $\Ker(d^{p}_{r}) \, / \, \IIm(d^{p-r}_{r}) = E^{p}_{r+1}K$.

\section{Grading and double complexes}

\subsection{Grading and regular filtrations}

We now suppose that $K$ is a \emph{graded} group, i.e.:
	\[K = \bigoplus_{n\in \mathbb{Z}} K^{n}, \qquad d^{n}: K^{n} \rightarrow K^{n+1}
\]
such that $d = \bigoplus_{n \in \mathbb{Z}} d^{n}$ (equivalently, $K^{\bullet}$ is a complex). Also the filtration is graded, i.e.:
	\[F^{p}K = \bigoplus_{n \in \mathbb{Z}} F^{p}K^{n}
\]
such that $K^{n} = F^{0}K^{n} \supset F^{1}K^{n} \supset \cdots \supset F^{l}K^{n} = 0$. For the associated graded group, we shift the indices in the following way, which will be motivated studying double complexes:
	\[E^{p,\,q}_{0}K = F^{p}K^{p+q} / F^{p+1}K^{p+q}.
\]
\paragraph{Remark:} in the following, we will denote by $n$ the gradation index, and by $q$ the shifted gradation index, i.e.\ $n = p+q$.

\paragraph{} We have cohomology groups $H^{n}(K^{\bullet}) = \Ker \, d^{n} / \IIm \, d^{n-1}$. We still suppose that $d$ preserves the filtration, i.e.\ $d^{n}(F^{p}K^{n}) \subset F^{p}K^{n+1}$: in this case we get cohomology groups $H^{n}(F^{p}K^{\bullet})$ for every $p$. We also put:
	\[ZK^{n} = \Ker \, d^{n} \quad BK^{n} = \IIm \, d^{n-1} \quad Z^{p}K^{n} = \Ker \bigl( d\vert_{F^{p}K^{n}} \bigr) \quad B^{p}K^{n} = \IIm \bigl( d\vert_{F^{p}K^{n-1}} \bigr).
\]

\paragraph{}The inclusions $i_{p,\,n}: F^{p}K^{n} \hookrightarrow K^{n}$ induces morphisms in cohomology:
	\[i_{p,\,n}^{\#}: H^{n}(F^{p,\,\bullet}K) \rightarrow H^{n}(K^{\bullet})
\]
whose image is given by equivalence classes of cocycles in $F^{p}K^{n}$ up to coboundaries coming from elements of all of $K$.

\paragraph{}We present the situation in a diagram: it is the same as above, but with the gradation included.

\begin{displaymath}
\xymatrix{
K^{n-1} \ar[r]^{a = d^{n-1}x} & i_{p}(a) \in K^{n} & K^{n+1} & & i_{p}^{\#}[a] \in H^{n}(K^{\bullet})\\
\vdots & \vdots & \vdots & & \vdots\\
F^{p-1}K^{n-1} \ar@{^(->}[u] \ar[r]^{a = d^{n-1}x} & F^{p-1}K^{n} \ar@{^(->}[u] & F^{p-1}K^{n+1} \ar@{^(->}[u] & & H^{n}(F^{p-1,\,\bullet}K) \ar[u]\\
F^{p}K^{n-1} \ar@{^(->}[u] \ar[r]^{a = d^{n-1}x} & a \in \ar@<4.5ex>[uuu] F^{p}K^{n} \ar@{^(->}[u] \ar[r]^{d^{n}a = 0} & F^{p}K^{n+1} \ar@{^(->}[u] & & [a] \in \ar@<7.5ex>[uuu] H^{n}(F^{p,\,\bullet}K) \ar[u]\\
F^{p+1}K^{n-1} \ar@{^(->}[u] & F^{p+1}K^{n} \ar@{^(->}[u] & F^{p+1}K^{n+1} \ar@{^(->}[u] & & H^{n}(F^{p+1,\,\bullet}K) \ar[u]\\
\vdots \ar@{^(->}[u] & \vdots \ar@{^(->}[u] & \vdots \ar@{^(->}[u] & & \vdots \ar[u]
}
\end{displaymath}
The image of $i_{p,\,n}^{\#}$ is thus:
\begin{equation}\label{ImIpGraded}
\begin{split}
	\IIm (i_{p,\,n}^{\#}) = \langle\, i_{p,\,n}Z^{p}K^{n}, &BK^{n} \,\rangle / BK^{n}\\
	&\simeq Z^{p}K^{n} \,/\, BK^{n} \cap Z^{p}K^{n}.
\end{split}
\end{equation}

\paragraph{} We define $F^{p}H^{n}(K^{\bullet}) = \IIm(i_{p,\,n}^{\#})$. In this way, we obtain a filtration of $H^{n}(K^{\bullet})$ given by $\cdots \supset F^{p-1}H^{n}(K^{\bullet}) \supset F^{p}H^{n}(K^{\bullet}) \supset F^{p+1}H^{n}(K^{\bullet}) \supset \cdots$, whose associated graded group is the direct sum of:
	\[E_{0}^{p,\,q}H(K) = F^{p}H^{p+q}(K) / F^{p+1}H^{p+q}(K).
\]

\paragraph{Notation:} We have previously defined the immersions $i_{p,\,n}: F^{p}K^{n} \hookrightarrow K^{n}$. We also define the immersions:
	\[i_{p,p-r,n}: F^{p}K^{n} \hookrightarrow F^{p-r}K^{n}.
\]
When the index $p$ is clear from the context, we denote $i_{p,p-1,n}$ by $i_{n}$, and $i_{p,p-r,n}$ by $i_{n}^{r}$.

\subsubsection{First viewpoint}

We now give the same definitions as above for $r \geq 1$, taking into account the grading:
\begin{equation}
\begin{split}
	&Z^{p,\,q}_{r}K = \{a \in F^{p}K^{p+q} \,\vert\, d(a) \in F^{p+r}K^{p+q+1}\}\\
	&B^{p,\,q}_{r}K = \{a \in F^{p}K^{p+q} \,\vert\, a = d(x), x \in F^{p-r+1}K^{p+q-1}\}\\
	&F^{p,\,q}_{r}H(K^{\bullet}) = Z^{p,\,q}_{r}K / B^{p,\,q}_{r}K\\
	& \\
	&\overline{Z}^{p,\,q}_{r}K = \langle Z^{p,\,q}_{r}K, i(F^{p+1}K^{p+q}) \rangle\\
	&\overline{B}^{p,\,q}_{r}K = \langle B^{p,\,q}_{r}K, i(F^{p+1}K^{p+q}) \rangle\\
	&E^{p,\,q}_{r}K = \overline{Z}^{p,\,q}_{r}K /\, \overline{B}^{p,\,q}_{r}K
\end{split}
\end{equation}
which, for $r = l$, become:
	\[\begin{split}
	&\overline{Z}^{p,\,q}_{l}(K) = \langle Z^{p,\,q}K, i(F^{p+1}K^{p+q}) \rangle\\
	&\overline{B}^{p,\,q}_{l}(K) = \langle BK^{p+q} \cap Z^{p,\,q}K, i(F^{p+1}K^{p+q}) \rangle\\
	&E^{p,\,q}_{0}H(K^{\bullet}) = \overline{Z}^{p,\,q}_{l}K /\, \overline{B}^{p,\,q}_{l}K
\end{split}\]
thus the sequence stabilizes to $E^{p,\,q}_{\infty}K = E_{0}^{p,\,q}H(K) = F^{p}H^{p+q}(K) / F^{p+1}H^{p+q}(K)$. In this way we obtain two filtrations and a sequence:
	\[\begin{split}
&\langle B^{p,\,q}K, F^{p+1}K^{p+q} \rangle = \overline{B}^{p,\,q}_{1}(K) \subset \cdots \subset \overline{B}^{p,\,q}_{l}(K) = \langle BK^{p+q} \cap Z^{p,\,q}K, i(F^{p+1}K^{p+q}) \rangle\\
&\qquad\qquad\qquad\qquad\qquad \subset \langle Z^{p,\,q}K, i(F^{p+1}K^{p+q}) \rangle = \overline{Z}^{p,\,q}_{l}K \subset \cdots \subset \overline{Z}^{p,\,q}_{0}K = F^{p}K^{p+q}\\
&E^{p,\,q}_{0}H(K^{\bullet}), \cdots, E^{p,\,q}_{l}H(K^{\bullet}) = E^{p,\,q}_{0}H(K^{\bullet}).
\end{split}\]
In the diagram \ref{fig:Diagram2Graded} one can see the first step $E^{p,\,q}_{1}K$, which is exactly $H^{p+q}(F^{p}K^{\bullet}/$ $F^{p+1}K^{\bullet})$, and the general step. Since the first step is $H^{p+q}(F^{p}K^{\bullet}/F^{p+1}K^{\bullet})$, we can define $B^{p,\,q}_{0}K = \{0\}$, and, as previously remarked, $Z^{p,\,q}_{0}K = F^{p}K^{p+q}$: in this way, $\overline{Z}^{p,\,q}_{0}K = F^{p}K^{p+q}$ and $\overline{B}^{p,\,q}_{0}K = F^{p+1}K^{p+q}$, so that $E^{p,\,q}_{0}K = F^{p}K^{p+q} / F^{p+1}K^{p+q}$, as previously defined.
\begin{figure}
\emph{First step:}
\begin{displaymath}
\xymatrix{
&\vdots & \vdots & \vdots\\
(p-1) & F^{p-1}K^{p+q-1} \ar@{^(->}[u] & F^{p-1}K^{p+q} \ar@{^(->}[u] & F^{p-1}K^{p+q+1} \ar@{^(->}[u]\\
(p) & F^{p}K^{p+q-1} \ar@{^(->}[u] \ar[r]^{a = dx} & a \in F^{p}K^{p+q} \ar@{^(->}[u] \ar[r]^{da} & \ar@<-4ex>[d] \; F^{p}K^{p+q+1} \; \ar@{^(->}[u]\\
(p+1) & F^{p+1}K^{p+q-1} \ar@{^(->}[u] & F^{p+1}K^{p+q} \ar@<5.5ex>[u]^{a} \ar@{^(->}[u] & da \in F^{p+1}K^{p+q+1} \ar@{^(->}[u]\\
& \vdots \ar@{^(->}[u] & \vdots \ar@{^(->}[u] & \vdots \ar@{^(->}[u]
}
\end{displaymath}
\emph{General step:}
\begin{displaymath}
\xymatrix{
&\vdots & \vdots & \vdots\\
(p-r+1) & F^{p-r+1}K^{p+q-1} \ar@{^(->}[u] \ar[r]^{a = dx} & \ar@<-5.5ex>[ddd] F^{p-r+1}K^{p+q} \ar@{^(->}[u] & F^{p-r+1}K^{p+q+1} \ar@{^(->}[u]\\
& \vdots \ar@{^(->}[u] & \vdots \ar@{^(->}[u] & \vdots \ar@{^(->}[u]\\
(p-1) & F^{p-1}K^{p+q-1} \ar@{^(->}[u] \ar[r]^{a = dx} & F^{p-1}K^{p+q} \ar@{^(->}[u] & F^{p-1}K^{p+q+1} \ar@{^(->}[u]\\
(p) & F^{p}K^{p+q-1} \ar@{^(->}[u] \ar[r]^{a = dx} & a \in F^{p}K^{p+q}\ar@{^(->}[u] \ar[r]^{da} & \ar@<-6.5ex>[dd] \; F^{p}K^{p+q+1} \; \ar@{^(->}[u]\\
& \vdots \ar@{^(->}[u] & \vdots \ar@<5.5ex>[u]^{a} \ar@{^(->}[u] & \vdots \ar@{^(->}[u]\\
(p+r) & F^{p+r}K^{p+q-1} \ar@{^(->}[u] & F^{p+r}K^{p+q} \ar@{^(->}[u] & da \in F^{p+r}K^{p+q+1} \ar@{^(->}[u]\\
& \vdots \ar@{^(->}[u] & \vdots \ar@{^(->}[u] & \vdots \ar@{^(->}[u]
}
\end{displaymath}
\caption{Diagram}\label{fig:Diagram2Graded}
\end{figure}

\paragraph{} Let us now consider the boundary, in particular its behavior with respect to the grading. We know that $d^{p}_{r}: E^{p}_{r}K \rightarrow E^{p+r}_{r}K$, i.e.\ sends $p$ to $p'=p+r$; moreover, a boundary sends $n$ in $n'=n+1$. Being $n=p+q$, one has $q' = n'-p' = (p+q+1)-(p+r) = q-r+1$. Hence:
	\[d^{p,\,q}_{r}: \; E^{p,\,q}_{r}K \longrightarrow E^{p+r,\,q-r+1}_{r}K.
\]
We can show in a picture the behavior of $d^{p,\,q}_{r}$:

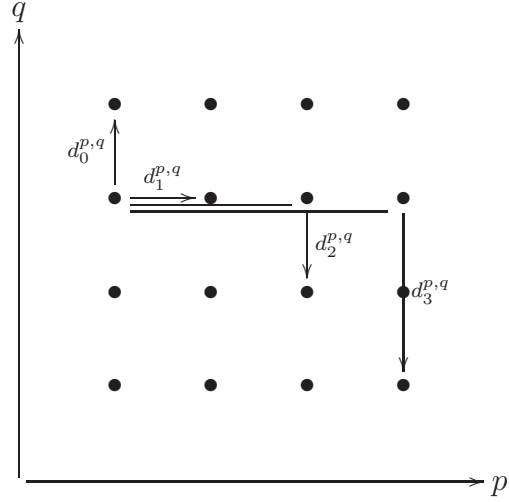
\begin{displaymath}
\xymatrix{
q & & & &\\
& \bullet & \bullet & \bullet & \bullet &\\
& \bullet \ar[u]^{d_{0}^{p,q}} \ar[r]^{d_{1}^{p,q}} \ar@<-0.5ex>@{-}[rr] \ar@<-1ex>@{-}[rrr] & \bullet & \bullet \ar[d]^{d_{2}^{p,q}} & \bullet \ar[dd]^{d_{3}^{p,q}} &\\
& \bullet & \bullet & \bullet & \bullet &\\
& \bullet & \bullet & \bullet & \bullet &\\
\ar[uuuuu]\ar[rrrrr] & & & & & p
}
\end{displaymath}

\paragraph{}As before, $E_{r+1}K = H(E_{r}K, d_{r})$, i.e.\ $E^{p,\,q}_{r+1}K = \Ker \, d^{p,\,q}_{r} \,/\, \IIm \, d^{p-r,\,q+r-1}_{r}$. Since $E^{p,\,q}_{r}K = \overline{Z}^{p,\,q}_{r}K /$ $\overline{B}^{p,\,q}_{r}K$ and $E^{p,\,q}_{r+1}K = \overline{Z}^{p,\,q}_{r+1}K /\, \overline{B}^{p,\,q}_{r+1}K$, the fact that $E_{r+1}K = H(E_{r}K,d_{r})$ naturally follows from:
\begin{itemize}
\item $\Ker \, d^{p,q}_{r} \;\; = \overline{Z}^{p,\,q}_{r+1}K /\, \overline{B}^{p,\,q}_{r}K \subset \overline{Z}^{p,\,q}_{r}K /\, \overline{B}^{p,\,q}_{r}K$;
\item $\IIm \, d^{p-r}_{r} = \overline{B}^{p,\,q}_{r+1}K /\, \overline{B}^{p,\,q}_{r}K \subset \overline{Z}^{p,\,q}_{r}K /\, \overline{B}^{p,\,q}_{r}K$.
\end{itemize}
This can been proven as for the ungraded case.

\subsubsection{Second viewpoint}

We define for $r \geq 1$:
\begin{equation}
\begin{split}
	&Z^{p,\,q}_{r}K = \{a \in F^{p}K^{p+q} \,\vert\, d(a) \in F^{p+r}K^{p+q+1}\}\\
	&\mathfrak{B}^{p,\,q}_{r}K = B^{p-r+1}K^{p+q}\\
	&Z\mathfrak{B}^{p,\,q}_{r}K = \langle i^{r-1}Z^{p,\,q}_{r}K, \mathfrak{B}^{p,\,q}_{r}K \rangle \subset F^{p-r+1}K^{p+q}\\
	&F^{p,\,q}_{r}H(K^{\bullet}) = Z\mathfrak{B}^{p,\,q}_{r}K / \mathfrak{B}^{p,\,q}_{r}K\\
	&\\
	&\overline{\mathfrak{B}}^{p,\,q}_{r}K = \langle \mathfrak{B}^{p,\,q}_{r}K, i^{r}F^{p+1}K^{p+q} \rangle = \langle B^{p-r+1}K^{p+q}, i^{r}F^{p+1}K^{p+q} \rangle\\
	&\overline{Z\mathfrak{B}}^{p,\,q}_{r}K = \langle Z\mathfrak{B}^{p,\,q}_{r}K, i^{r}F^{p+1}K^{p+q} \rangle = \langle i^{r-1}Z^{p,\,q}_{r}K, B^{p-r+1}K^{p+q}, i^{r}F^{p+1}K^{p+q} \rangle\\
	&E^{p,\,q}_{r}K = \overline{Z\mathfrak{B}}^{p,\,q}_{r}K / \, \overline{\mathfrak{B}}^{p,\,q}_{r}K
\end{split}
\end{equation}
which, for $r = l$, become:
	\[\begin{split}
	&\overline{\mathfrak{B}}^{p,\,q}_{l}K = \langle BK^{p+q}, i^{p+1}F^{p+1}K^{p+q} \rangle\\
	&\overline{Z\mathfrak{B}}^{p,\,q}_{l}K = \langle i^{p}Z^{p,\,q}K, BK^{p+q}, i^{p+1}F^{p+1}K^{p+q} \rangle\\
	&E^{p,\,q}_{l}K = E^{p,\,q}_{0}H(K).
\end{split}\]
thus the sequence stabilizes to $E^{p,\,q}_{\infty}K = E_{0}^{p,\,q}H(K) = F^{p}H^{p+q}(K) / F^{p+1}H^{p+q}(K)$. For the boundary, the same considerations of the first viewpoint apply.

\subsubsection{Regular filtrations}

Up to now we considered only finite filtrations. When grading is introduced, there is an important kind of filtrations, intermediate between finite and infinite filtrations.

\begin{Def} Let $(K^{\bullet}, d)$ be a complex and $\{F^{p}K^{\bullet}\}_{p \in \mathbb{N}}$ be a filtration. The filtration is called \emph{regular} if, for any $n$ fixed, the filtration $\{F^{p}K^{n}\}_{p \in \mathbb{N}}$ is finite.
\end{Def}

For regular filtrations all the theory developed works in the same way, with the exception that, only for a fixed $n = p + q$, we have that eventually $E^{p,\,q}_{r}K$ becomes stationary and equal to $E^{p,\,q}_{\infty}K$.

\subsection{Double complexes}

\subsubsection{Basic definitions}

\begin{Def} Given an additive category $\mathcal{A}$, a \emph{double complex} is a set of objects $\{K^{p,\,q}\}_{p,q \in \mathbb{Z}}$ of $\mathcal{A}$ with two morphisms $\delta_{1}^{p,\,q}: K^{p,\,q} \rightarrow K^{p+1,\,q}$ and $\delta_{2}^{p,\,q}: K^{p,\,q} \rightarrow K^{p,\,q+1}$ such that:
	\[\delta_{1}^{2} = 0 \qquad \delta_{2}^{2} = 0 \qquad \delta_{1}\delta_{2} + \delta_{2}\delta_{1} = 0.
\]
\end{Def}
\begin{displaymath}
\xymatrix{
q & & & \\
& \bullet & \bullet & \bullet &\\
& \bullet & \bullet & \bullet &\\
& \bullet & \bullet \ar[r]^{\delta_{1}^{p,q}} \ar[u]^{\delta_{2}^{p,q}} & \bullet &\\
\ar[uuuu]\ar[rrrr] & & & & p
}
\end{displaymath}

\begin{Def} Let $\{K^{p,\,q}\}_{p,\,q \in \mathbb{N}}$ be a double complex. The \emph{associated total complex} is the complex $(T^{n}, d^{n})_{n \in \mathbb{Z}}$ such that:
	\[T^{n} = \bigoplus_{p+q=n}K^{p,\,q} \qquad d^{n} = \bigoplus_{p+q=n} (\delta_{1}^{p,\,q} + \delta_{2}^{p,\,q}).
\]
\end{Def}

The conditions $\delta_{1}\delta_{2} + \delta_{2}\delta_{1} = 0$ in the definition of double complex ensures that $d^{2} = 0$. We now construct a filtration of $(T^{n}, d^{n})$ and the corresponding spectral sequence.

\paragraph{}We put:
	\[F^{p}T^{n} = \bigoplus_{\substack{i \geq p\\i+q=n}} K^{i,\,q}
\]
i.e.\ the $p$-th group of the filtration at degree $n$ is made by the direct sum of the groups $K^{i,\,q}$, for $i+q = n$, with $i \geq p$. We can also define:
	\[F^{p}T = \bigoplus_{\substack{i \geq p\\q \in \mathbb{Z}}} K^{i,\,q} \qquad F^{p}T^{n} = F^{p}T \cap T^{n}.
\]
Since the role of the two indexes $p, \,q$ is symmetric, we can also consider the filtration:
	\['F^{p}T^{n} = \bigoplus_{\substack{i \geq p\\p'+i=n}} K^{p',\,i}.
\]

\begin{Def} Let $(K^{\bullet,\,\bullet}, \delta_{1}, \delta_{2})$ and $(L^{\bullet,\,\bullet}, \delta'_{1}, \delta'_{2})$ be two double complexes in an additive category $\mathcal{A}$. A \emph{morphism} of double complexes is a set of morphisms $f^{p,\,q}: K^{p,\,q} \rightarrow L^{p,\,q}$ such that:
	\[f^{p+1,\,q} \circ \delta_{1}^{p,\,q} = {\delta'}_{1}^{p,\,q} \circ f^{p,\,q} \qquad f^{p,\,q+1} \circ \delta_{2}^{p,\,q} = {\delta'}_{2}^{p,\,q} \circ f^{p,\,q}.
\]
\end{Def}
In this way we define the \emph{category of double complexes} in $\mathcal{A}$ denoted by $\Kom^{\bullet \bullet}(\mathcal{A})$. In particular, we can define the subcategory $\Kom^{++}(\mathcal{A})$ made by double complexes $K^{\bullet,\,\bullet}$ such that $K^{p,\,q} = 0$ for $i < 0$ or $j < 0$.

\begin{Def} Let $(K^{\bullet,\,\bullet}, \delta_{1}, \delta_{2})$ and $(L^{\bullet,\,\bullet}, \delta'_{1}, \delta'_{2})$ be two double complexes in an additive category $\mathcal{A}$ and $f^{\bullet,\,\bullet}, g^{\bullet,\,\bullet}: K^{\bullet,\,\bullet} \rightarrow L^{\bullet,\,\bullet}$ two morphisms. An \emph{homotopy} between $f$ and $g$ is a set of morphisms:
	\[h^{p,\,q}_{1}: K^{p,\,q} \rightarrow L^{p-1,\,q} \qquad h^{p,\,q}_{2}: K^{p,\,q} \rightarrow L^{p,\,q-1}
\]
such that:
	\[g-f = (\delta_{1} + \delta_{2})(h_{1} + h_{2}) + (h_{1} + h_{2})(\delta_{1} + \delta_{2}).
\]
\end{Def}
We can thus define the category $K^{\bullet \bullet}(\mathcal{A})$ obtained from $\Kom^{\bullet \bullet}(\mathcal{A})$ quotienting out by morphisms homotopic to zero. Similarly we define $K^{++}(\mathcal{A})$.

\subsubsection{Spectral sequences of double complexes}

Let us consider $K^{\bullet,\,\bullet} \in \Ob(\Kom^{++}(\mathcal{A}))$, whose associated total complex we denote by $(T^{\bullet}, d^{\bullet})$. With the filtration $\{F^{p}T^{n}\}_{p,\,n \in \mathbb{Z}}$, which is \emph{regular} since $K^{p,\,q} = 0$ for negative $p$ or $q$, we build the corresponding spectral sequence. In particular, we have:
	\[E^{p,\,q}_{0} = F^{p}T^{p+q} \,/\, F^{p+1}T^{p+q} = K^{p,\,q}
\]
and this justifies the shift in the indices for the associated graded group. The first boundary is $d^{p,\,q}_{0}: E^{p,\,q}_{0} \rightarrow E^{p,\,q+1}_{0}$, i.e.\ $d^{p,\,q}_{0}: K^{p,\,q} \rightarrow K^{p,\,q+1}$. We now prove that $d^{p,\,q}_{0} = \delta_{2}^{p,\,q}$:
\begin{itemize}
	\item let $[a] \in E^{p,\,q}_{0} = F^{p}T^{p+q} / F^{p+1}T^{p+q} = K^{p,\,q}$;
	\item $a \in F^{p}T^{p+q} = \bigoplus_{i \geq p, \,i+j=p+q} K^{i,\,j}$, i.e.\ $a = \oplus\, a^{i,j}$;
	\item $[a] = [a^{p,q}]$, since the other components $a^{i,j}$ are quotiented out (in fact, $E^{p,\,q}_{0} = K^{p,\,q}$);
	\item $[da] = [d(a^{p,q})] = [\delta_{1}^{p,\,q}a^{p,q} + \delta_{2}^{p,\,q}a^{p,q}] \in E^{p,\,q+1}_{0} = F^{p}T^{p+q+1} / F^{p+1}T^{p+q+1}$;
	\item $\delta_{1}^{p,\,q}a^{p,q} \in F^{p+1}T^{p+q+1}$, thus it is quotiented out;
	\item hence $[da] = [\delta_{2}^{p,\,q}a^{p,q}]$.
\end{itemize}
Thus:
	\[E^{p,\,q}_{1} \simeq H^{q}(K^{p,\,\bullet}, \delta_{2}).
\]
In fact, we have seen that, in general, $E^{p,\,q}_{1} \simeq H^{p+q}(F^{p}T^{\bullet}/F^{p+1}T^{\bullet})$, which, in this case, becomes exactly the previous formula.

\paragraph{}We now discuss the second boundary $d_{1}^{p,\,q}: E^{p,\,q}_{1} \rightarrow E^{p+1,\,q}_{1}$: we prove that it is exactly induced by $\delta_{1}^{p,\,q}$ acting on $E^{p,\,q}_{1}$. In fact:
\begin{itemize}
	\item let $[a] \in E^{p,\,q}_{1} = H^{q}(K^{p,\,\bullet}, \delta_{2})$, so that $a \in \Ker(\delta_{2}^{p,\,q}) \subset K^{p,\,q}$;
	\item then $[da] = [\delta_{1}^{p,\,q}a + \delta_{2}^{p,\,q}a]$; but $\delta_{2}^{p,\,q}a \in K^{p,\,q+1}$ is zero in $H^{q+1}(K^{p,\,\bullet}, \delta_{2}) = E^{p,\,q+1}_{1}$;
	\item hence $[da] = [\delta_{1}^{p,\,q}a]$.
\end{itemize}
Thus:
	\[E^{p,\,q}_{2} = H^{p}(E^{\bullet,\,q}_{1}, \delta_{1}).
\]

\paragraph{}This spectral sequence converges by construction to the cohomology of the total complex with respect to the given filtration $F^{p}T^{n}$. Let us see how the various steps of the spectral sequence are done. Since an element of $T^{n}$ is made by sum of elements with total index $n$, then $F^{p}H^{n}(T^{\bullet})$ is given by the cocycles whose direct summands have first index bigger or equal to $p$, up to any coboundary. In the following diagram, the points linked by a double continuous line form a cocycle in $F^{p}T^{n}$, while the ones linked by a double dotted line form an element of $F^{p+q-1}K$, whose boundary is an element of $B^{p+q}K$. The group $F^{p}H^{n}(T^{\bullet})$ is made by such cocycles up to such coboundaries. If we were considering $H^{n}(T^{\bullet})$, between the points linked by dotted line only the lowest one would have remained, since we are considering the $p$-step, i.e.\ the first index must be at least $p$. When we build the spectral sequence, in $E^{p}_{r}H(K)$ we consider generalized cocycles whose boundary is non-zero only for horizontal index bigger or equal to $p+r$, and we allow coboundaries of elements which are non-zero for horizontal index at lest $p-r+1$.
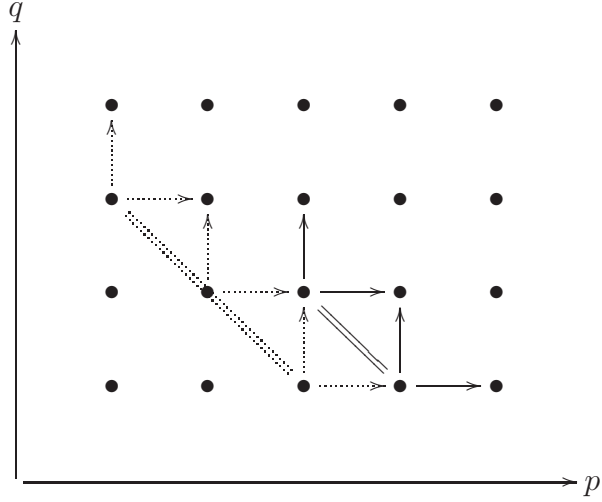
\begin{displaymath}
\xymatrix{
q & & & & &\\
& \bullet & \bullet & \bullet & \bullet & \bullet &\\
& \bullet \ar@{.>}[r]\ar@{.>}[u] \ar@{:}[ddrr] & \bullet & \bullet & \bullet & \bullet &\\
& \bullet & \bullet \ar@{.>}[r]\ar@{.>}[u] & \bullet \ar@{=}[dr] \ar[r]\ar[u] & \bullet & \bullet &\\
& \bullet & \bullet & \bullet \ar@{.>}[r]\ar@{.>}[u] & \bullet \ar[r]\ar[u] & \bullet &\\
\ar[uuuuu]\ar[rrrrrr] & & & & & & p
}
\end{displaymath}
The limit of the sequence is $E_{\infty}^{p,\,q}K = E_{0}^{p,\,q}H(K^{\bullet}) = F^{p}H^{p+q}(K^{\bullet}) / F^{p+1}H^{p+q}(K^{\bullet})$.

\paragraph{}If we consider the filtration $'F^{p}T^{n}$, we obtain a spectral sequence $\{'E^{p,\,q}_{r}\}_{r \in \mathbb{N}}$ such that:
	\[\begin{array}{lll}
	'E^{p,\,q}_{0} = K^{p,\,q} & & 'd^{p,\,q}_{0} = \delta_{1}^{p,\,q}\\
	'E^{p,\,q}_{1} = H^{p}(K^{\bullet,\,q}, \delta_{1}) & & 'd^{p,\,q}_{1} \simeq \delta_{2}^{p,\,q}\\
	'E^{p,\,q}_{2} = H^{q}(E^{p,\,\bullet}_{1}, \delta_{2})
\end{array}\]
and $'E_{\infty}^{p,\,q}K = 'F^{p}H^{p+q}(K^{\bullet}) / 'F^{p+1}H^{p+q}(K^{\bullet})$.

\section{Generalization}

\subsection{Cohomology of the quotients}

We can interpret cocycles and coboundaries in another way (v.\ \cite{CE} chap. XV). We still consider \emph{regular} filtrations (in particular, they can be finite), i.e.\ $F^{p}K^{n} = K^{n}$ for $n \leq 0$ and for any fixed $n$ there exists $l \in \mathbb{N}$ such that $F^{p}K^{n} = 0$ for $p \geq l$. Using the convention $F^{-\infty}K^{n} = K^{n}$ and $F^{+\infty}K^{n} = 0$, we define, for $-\infty \leq p \leq t \leq +\infty$:
	\[\begin{split}
	&H^{n}(p, t) = H^{n}(F^{p}K^{\bullet} / F^{t}K^{\bullet})\\
	&H^{n}(p) = H^{n}(p, +\infty) = H^{n}(F^{p}K^{\bullet}).
\end{split}\]
For $p \leq t \leq u$, $a, b \geq 0$, $p+a \leq t+b$, we define two maps:
\begin{equation}\label{PsiDelta}
\begin{split}
	&\Psi^{n}: H^{n}(p+a,t+b) \rightarrow H^{n}(p,t)\\
	&\Delta^{n}: H^{n}(p,t) \rightarrow H^{n+1}(t,u)
\end{split}
\end{equation}
where:
\begin{itemize}
	\item $\Psi^{n}$ is induced in cohomology by the natural map $F^{p+a}K / F^{t+b}K \rightarrow F^{p}K / F^{t}K$, induced by the inclusions of the numerators and the denominators;
	\item $\Delta^{n}$ is the composition of the Bockstein map $\beta: H^{n}(p, t) \rightarrow H^{n+1}(t)$ and the map induced in cohomology by $\pi: F^{t}K \rightarrow F^{t}K / F^{u}K$.
\end{itemize}

Let us consider $H^{p+q}(p, p+r)$: it is given by cocycles in $F^{p}K^{p+q}/F^{p+r}K^{p+q}$, i.e.\ $Z^{p,\,q}_{r}K/$ $F^{p+r}K^{p+q}$, up to $\langle B^{p}K^{p+q}, i^{r}(F^{p+r}K^{p+q}) \rangle$, as shown in the diagram \ref{fig:DiagramQuotient}. Applying lemma \ref{AlgLemma} we get:
\begin{equation}\label{CohQuot}
	H^{p+q}(p, p+r) = Z^{p,\,q}_{r}K \,/\, \langle B^{p}K^{p+q}, i^{r}(F^{p+r}K^{p+q}) \rangle.
\end{equation}
We remark that $F^{p+r}K^{p+q} \subset Z^{p,\,q}_{r}K$ (so we do not need the explicit intersection) since every $x \in F^{p+r}K^{p+q}$ is such that $d(x) \in F^{p+r}K^{p+q}$.
\begin{figure}
\begin{displaymath}
\xymatrix{
& \vdots & \vdots & \vdots\\
(p-1) & F^{p-1}K^{n-1} \ar@{^(->}[u] & F^{p-1}K^{n} \ar@{^(->}[u] & F^{p-1}K^{n+1} \ar@{^(->}[u]\\
(p) & F^{p}K^{n-1} \ar@{^(->}[u] \ar[r]^{a = dx} & a \in F^{p}K^{n} \ar@{^(->}[u] \ar[r]^{da} & \ar@<-5ex>[dd] \; F^{p}K^{n+1} \; \ar@{^(->}[u]\\
& \vdots \ar@{^(->}[u] & \vdots \ar@{^(->}[u] & \vdots \ar@{^(->}[u]\\
(p+r) & F^{p+r}K^{n-1} \ar@{^(->}[u] & \ar@<4.5ex>[uu] F^{p+r}K^{n} \ar@{^(->}[u] & da \in F^{p+r}K^{n+1} \ar@{^(->}[u]\\
& \vdots \ar@{^(->}[u] & \vdots \ar@{^(->}[u] & \vdots \ar@{^(->}[u]
}
\end{displaymath}
\caption{Quotient cohomology}\label{fig:DiagramQuotient}
\end{figure}

\paragraph{}We can now reinterpret spectral sequences only using such groups and maps, and this is more natural in the second viewpoint, since we now prove that for $r \geq 1$:
\begin{equation}\label{Epr}
\IIm \bigl( H^{p+q}(p, p+r) \overset{\Psi^{p+q}}\longrightarrow H^{p+q}(p-r+1, p+1) \bigr) = \overline{Z\mathfrak{B}}^{p,\,q}_{r}K / \overline{\mathfrak{B}}^{p,\,q}_{r}K = E^{p,\,q}_{r}K
\end{equation}
as shown in the diagram \ref{fig:DiagramImQuotient}. 
\begin{figure}
\begin{displaymath}
\xymatrix{
&\vdots & \vdots & \vdots\\
(p-r+1) & F^{p-r+1}K^{n-1} \ar@{^(->}[u] \ar[r]^{a = dx} & F^{p-r+1}K^{n} \ar@{^(->}[u] & F^{p-r+1}K^{n+1} \ar@{^(->}[u]\\
& \vdots \ar@{^(->}[u] & \vdots \ar@{^(->}[u] & \vdots \ar@{^(->}[u]\\
(p) & F^{p}K^{n-1} \ar@{^(->}[u] \ar[r]^{a = dx} & a \in F^{p}K^{n} \ar@{^(->}[u] \ar@<4.5ex>[uu] \ar[r]^{da} & \ar@<-4,5ex>[dd] \; F^{p}K^{n+1} \; \ar@{^(->}[u]\\
& \vdots \ar@{^(->}[u] & \vdots \ar@<4.5ex>[u]^{a} \ar@{^(->}[u] & \vdots \ar@{^(->}[u]\\
(p+r) & F^{p+r}K^{n-1} \ar@{^(->}[u] & F^{p+r}K^{n} \ar@{^(->}[u] & da \in F^{p+r}K^{n+1} \ar@{^(->}[u]\\
& \vdots \ar@{^(->}[u] & \vdots \ar@{^(->}[u] & \vdots \ar@{^(->}[u]
}
\end{displaymath}
\caption{Image}\label{fig:DiagramImQuotient}
\end{figure}
In fact, considering \ref{CohQuot}, the image of $(\Psi^{p+q})^{p,\,p+r}_{p-r+1,\,p+1}$ can be described as:
	\[\begin{split}
	\IIm \bigl( \; Z^{p,\,q}_{r}K \,/\, \langle B^{p}K^{p+q}, &i^{r}(F^{p+r}K^{p+q}) \rangle \overset{\Psi^{p+q}}\longrightarrow\\
	&Z^{p-r+1,\,q+r-1}_{r}K \,/\, \langle B^{p-r+1}K^{p+q}, i^{r}(F^{p+1}K^{p+q}) \rangle \; \bigr)
\end{split}\]
which is:
	\[\langle i^{r-1}Z^{p,\,q}_{r}K, B^{p-r+1}K^{p+q}, i^{r}(F^{p+1}K^{p+q}) \rangle \,/\, \langle B^{p-r+1}K^{p+q}, i^{r}(F^{p+1}K^{p+q}) \rangle
\]
i.e.\ $\overline{Z\mathfrak{B}}^{p,\,q}_{r}K / \overline{\mathfrak{B}}^{p,\,q}_{r}K$.

\paragraph{}We have a commutative diagram:
\begin{equation}\label{BoundaryDiagram}
\xymatrix{
H^{p+q}(p, p+r) \ar[r]^{\Psi_{1}^{p+q}} \ar[d]^{\Delta_{1}^{p+q}} & H^{p+q}(p-r+1, p+1) \ar[d]_{\Delta_{2}^{p+q}}\\
H^{p+q+1}(p+r, p+2r) \ar[r]^{\Psi_{2}^{p+q+1}} & H^{p+q+1}(p+1,p+r+1).
}
\end{equation}
We have that:
\begin{itemize}
	\item $\IIm(\Psi_{1}^{p+q}) = E^{p,\,q}_{r}K$ and $\IIm(\Psi_{2}^{p+q}) = E^{p+r,\,q}_{r}K$;
	\item $d^{p,\,q}_{r} = \Delta_{2}^{p+q} \,\big\vert_{\IIm(\Psi_{1}^{p+q})}\,:\, E^{p,\,q}_{r}K \rightarrow E^{p+r,\,q-r+1}_{r}K$.
\end{itemize}
We have already proven the first part. For the boundary, let us consider $[a] \in E^{p,\,q}_{r}K$ with $a = z+b+x \in \langle i^{r-1}Z^{p,\,q}_{r}K, B^{p-r+1}K^{p+q}, i^{r}(F^{p+1}K^{p+q}) \rangle$. Then we know that $d^{p,\,q}_{r}[a] = [d(z)] \in E^{p+r,\,q-r+1}_{r}K$. Let us compute $\Delta_{2}^{p+q}([a])$: first we compute the Bockstein map to $H^{p+q+1}(p+1)$, which consists of applying the boundary to get $d(z) + d(x)$ and considering the class $[d(z)] \in H^{p+q+1}(p+1)$; then we compose with the map in cohomology induced by $\pi: F^{p+1}K^{p+q+1} \rightarrow F^{p+1}K^{p+q+1}/F^{p+r+1}K^{p+q+1}$, to get $[d(z)] \in H^{p+q+1}(p+1,p+r+1) = Z^{p+1,\,q}_{r} \,/\, \langle B^{p+1}K^{p+q+1}, i^{r}(F^{p+r+1}K^{p+q+1})\rangle$. But, being $d(z)$ a boundary, we have $d(z) \in B^{p}K^{p+q+1} \cap Z^{p+1,\,q}_{r}$, thus $d(z) \in i^{r-1}Z^{p+r,\,q-r+1}_{r}$ so that we can consider $[d(z)] \in \IIm(\Psi_{2}^{p+q}) = E^{p+r,\,q-r+1}_{r}$.

\paragraph{}This approach works for $r \geq 1$, since, for $r = 0$, we get $0$ in \eqref{CohQuot}, so that the l.h.s.\ of \eqref{Epr} is zero and not equal $E^{p,\,q}_{0}$. Thus, we start from $r = 1$. The limit of the sequence can be obtain putting $r = +\infty$ in \eqref{Epr}:
\begin{equation}\label{Limit}
	E^{p,\,q}_{\infty}K = E^{p,\,q}_{0}H(K) = \IIm \bigl( H^{p+q}(p, +\infty) \overset{\Psi^{p+q}}\longrightarrow H^{p+q}(0, p+1) \bigr).
\end{equation}
In fact, since $H^{p+q}(p, +\infty) = H^{p+q}(F^{p}K^{\bullet})$ and $H^{p+q}(0, +\infty) = H^{p+q}(K^{\bullet})$ we have that:
\begin{equation}\label{Filtr}
	F^{p}H^{p+q}(K^{\bullet}) = \IIm \bigl( H^{p+q}(p, +\infty) \overset{\Psi^{p+q}}\longrightarrow H^{p+q}(0, +\infty) \bigr).
\end{equation}
Let us see that the associated graded group of this filtration of $H^{p+q}(K^{\bullet}) = H^{p+q}(0,$ $+\infty)$ is given by \eqref{Limit}. In fact, by \eqref{CohQuot} we have $H^{p+q}(0, +\infty) = Z^{0,\,p+q}_{+\infty} / BK^{p+q} = ZK^{p+q} / BK^{p+q} = H^{p+q}(K^{\bullet})$ and $H^{p+q}(0, p+1) = Z^{0,\,p+q}_{p+1} / \langle BK^{p+q}, i^{p+1}(F^{p+1}K^{p+q})\rangle = H^{p+q}(K^{\bullet} / F^{p+1}K^{\bullet})$: then \eqref{Limit} is obtained by \eqref{Filtr} via the immersion $ZK^{p+q} \rightarrow Z^{0,\,p+q}_{p+1}$, so that we get $\langle ZK^{p+q},$ $i^{p+1}(F^{p+1}K^{p+q}) \rangle \,/\, \langle BK^{p+q}, i^{p+1}(F^{p+1}K^{p+q}) \rangle$, which is exactly $F^{p}H^{p+q}(K^{\bullet}) / F^{p+1}H^{p+q}(K^{\bullet})$.

\paragraph{}Using this new language, we never referred to the groups $F^{p}K^{n}$, but only to $H^{n}(p,q)$: thus, we can provide the groups $H^{n}(p,q)$ axiomatically, without referring to the filtered groups $K^{n}$. The main advantage of this axiomatization is the possibility to build a spectral sequence for a generic cohomology theory, not necessarily induced by a coboundary.

\subsubsection{Description of the isomorphisms}

We now explicitly describe, in this language of cohomology of quotients, the canonical isomorphisms involved in the definition of a spectral sequence, i.e.\ $E^{p,q}_{r+1}K \simeq \Ker d^{p,q}_{r} / \IIm d^{p-r,q+r-1}_{r}$. Considering \eqref{BoundaryDiagram}, from the two diagrams:
\begin{equation}\label{BoundaryDiagram2}
\xymatrix{
H^{p+q-1}(p-r, p) \ar[r]^{\Psi_{0}^{p+q}} \ar[d]^{\Delta_{-1}^{p+q}} & H^{p+q-1}(p-2r+1, p-r+1) \ar[d]_{\Delta_{0}^{p+q}}\\
H^{p+q}(p, p+r) \ar[r]^{\Psi_{1}^{p+q}} \ar[d]^{\Delta_{1}^{p+q}} & H^{p+q}(p-r+1, p+1) \ar[d]_{\Delta_{2}^{p+q}}\\
H^{p+q+1}(p+r, p+2r) \ar[r]^{\Psi_{2}^{p+q+1}} & H^{p+q+1}(p+1,p+r+1)\\
H^{p+q}(p, p+r+1) \ar[r]^{\Psi_{3}^{p+q}} & H^{p+q}(p-r, p+1).
}
\end{equation}
we have that:
\begin{itemize}
	\item $\IIm(\Psi_{1}^{p+q}) = E^{p,\,q}_{r}K$;
	\item $d^{p,\,q}_{r} = \Delta_{2}^{p+q} \,\big\vert_{\IIm(\Psi_{1}^{p+q})}\,:\, E^{p,\,q}_{r}K \longrightarrow E^{p+r,\,q-r+1}_{r}K$ and $d^{p-r,\,q+r-1}_{r} = \Delta_{0}^{p+q} \,\big\vert_{\IIm(\Psi_{0}^{p+q})}\,:\, E^{p-r,\,q+r-1}_{r}K \longrightarrow E^{p,\,q}_{r}K$;
	\item $\IIm(\Psi_{3}^{p+q}) = E^{p,\,q}_{r+1}K$.
\end{itemize}
To find the isomorphism $E^{p,\,q}_{r+1}K \simeq \Ker d^{p,\,q}_{r} \,/\, \IIm d^{p-r,\,q+r-1}_{r}$ we thus need a map $\varphi^{p,\,q}_{r}: H^{p+q}(p-r+1, p+1) \rightarrow H^{p+q}(p-r, p+1)$ which is naturally given by the immersion $i: F^{p-r+1}K^{p+q} / F^{p+1}K^{p+q} \rightarrow F^{p-r}K^{p+q} / F^{p+1}K^{p+q}$. We claim that this map induces a surjection:
\begin{equation}\label{VarphiPqr}
	\varphi^{p,\,q}_{r}: \Ker (\Delta_{2}^{p+q} \,\vert_{\IIm(\Psi_{1}^{p+q})}) \longrightarrow \IIm \Psi_{3}^{p+q}.
\end{equation}
In fact:
\begin{itemize}
	\item $\IIm \Psi_{1}^{p+q} = \langle i^{r-1}Z^{p,\,q}_{r}K, i^{r}F^{p+1}K^{p+q}, B^{p-r+1}K^{p+q} \rangle \,/\, \langle i^{r}F^{p+1}K^{p+q}, B^{p-r+1}K^{p+q} \rangle$;
	\item $\Ker (\Delta_{2}^{p+q} \,\vert_{\IIm(\Psi_{1}^{p+q})}) = \langle i^{r-1}Z^{p,q}_{r+1}K, i^{r}F^{p+1}K^{p+q}, B^{p-r+1}K^{p+q} \rangle \,/\, \langle i^{r}F^{p+1}K^{p+q},$ \\ $B^{p-r+1}K^{p+q} \rangle$;
	\item $\IIm \Psi_{3}^{p+q} = \langle i^{r}Z^{p,\,q}_{r+1}K, i^{r+1}F^{p+1}K^{p+q}, B^{p-r}K^{p+q} \rangle \,/\, \langle i^{r+1}F^{p+1}K^{p+q}, B^{p-r}K^{p+q} \rangle$.
\end{itemize}
The map $\varphi^{p,\,q}_{r}$ is induced by the immersion of the numerators, and it is surjective since the only elements in the numerator of $\IIm \Psi_{3}^{p+q}$ which can not to be in the image are the elements of $B^{p-r}K$ which are quotiented out. Moreover, the kernel of $\varphi^{p,\,q}_{r}$ is made by classes of elements in $i^{r-1}Z^{p,\,q}_{r+1}K$ which, after the immersion, belongs to $B^{p-r}K$, i.e.:
\begin{equation}
\begin{split}
	\Ker \varphi^{p,\,q}_{r} &= \langle i^{r-1}(Z^{p,\,q}_{r+1}K) \cap B^{p-r}K^{p+q}, i^{r}F^{p+1}K^{p+q}, B^{p-r+1}K^{p+q} \rangle \,/ \\
	&\phantom{XXXXXXXXXXXXXXXXXXX} \langle i^{r}F^{p+1}K^{p+q}, B^{p-r+1}K^{p+q} \rangle\\
	&= \langle i^{r-1}(F^{p}K^{p+q}) \cap B^{p-r}K^{p+q}, i^{r}F^{p+1}K^{p+q}, B^{p-r+1}K^{p+q} \rangle \,/ \\
	&\phantom{XXXXXXXXXXXXXXXXXXX} \langle i^{r}F^{p+1}K^{p+q}, B^{p-r+1}K^{p+q} \rangle
\end{split}
\end{equation}
but the latter is exactly $\IIm (\Delta_{0}^{p+q} \,\vert_{\IIm(\Psi_{0}^{p+q})})$, thus $\varphi^{p,\,q}_{r}$ induces the isomorphism $E^{p,\,q}_{r+1}K$ $\simeq \Ker d^{p,\,q}_{r} \,/\, \IIm d^{p-r,\,q+r-1}_{r}$.

\paragraph{}Let us consider $E^{p,q}_{1} = H^{p+q}(p, p+1)$. Some elements will lie in $\Ker \, d^{p,\,q}_{1}$, and they are mapped to $H^{p+q}(p-1, p+1)$ by $\varphi^{p,\,q}_{1}$, which is induced by $i_{1}: F^{p}K^{p+q} \,/\, F^{p+1}K^{p+q}$ $\rightarrow F^{p-1}K^{p+q} \,/\, F^{p+1}K^{p+q}$. We iterate the procedure: some elements will lie in $\Ker \, d^{p,\,q}_{2}$ and are mapped to $H^{p+q}(p-2, p+1)$ by $\varphi^{p,\,q}_{2}$, which is induced by $i_{2}: F^{p-1}K^{p+q} \,/\, F^{p+1}K^{p+q} \rightarrow F^{p-2}K^{p+q} \,/\, F^{p+1}K^{p+q}$. Thus, in the original group $E^{p,q}_{1} = H^{p+q}(p, p+1)$ we can consider the elements that survives to both these steps and maps them directly to $H^{p+q}(p-2, p+1)$ via $i_{1,2}: F^{p}K^{p+q} \,/\, F^{p+1}K^{p+q} \rightarrow F^{p-2}K^{p+q} \,/\, F^{p+1}K^{p+q}$. This procedure stops after $l$ steps where $l$ is the length of the filtration. In particular, we obtain a subset $A^{p,\,q} \subset E^{p,\,q}_{1}$ of \emph{surviving elements}, and a map:
\begin{equation}\label{MapAInfty}
	\varphi^{p,\,q}: A^{p,\,q} \subset E^{p,\,q}_{1} \longrightarrow E^{p,\,q}_{\infty}
\end{equation}
assigning to each surviving element its class in the last step. The map is simply induced by $i_{1\cdots l}: F^{p}K^{p+q} / F^{p+1}K^{p+q} \rightarrow K^{p+q} / F^{p+1}K^{p+q}$. We now prove that the surviving elements are classes in $H^{p+q}(F^{p}K^{\bullet}/F^{p+1}K^{\bullet})$ represented by elements which are in $F^{p+1}K^{p+q}$ or by elements whose boundary is $0$ (not in $F^{p+1}K^{p+q+1}$!), or more generally that the elements surviving for $r$ steps (thus from $1$ to $r+1$) are represented by elements of $F^{p+1}K^{p+q}$ or by elements whose boundary is in $F^{p+r+1}K^{p+q+1}$. In fact, the first boundary is given by $\Delta^{p,\,q}: H^{p+q}(p, p+1) \rightarrow H^{p+q}(p+1, p+2)$ so that the elements in its kernel are classes in $H^{p+q}(p,p+1)$ represented by elements of $F^{p+1}K^{p+q}$ or by elements whose boundary lives in $F^{p+2}K^{p+q}$. Then the isomorphism \eqref{VarphiPqr} sends such elements to $H^{p+q}(p-1,p+1)$ by immersion and the second boundary is given by $\Delta^{p,\,q}: H^{p+q}(p-1,p+1) \rightarrow H(p-1,p+3)$ restricted to the image, thus the elements in its kernel must have boundary in $F^{p+3}K^{p+q}$, and so on. Thus we have that:
	\[A^{p,\,q} = \IIm(H^{p+q}(p, +\infty) \rightarrow H^{p+q}(p, p+1))
\]
and we have a commutative diagram:
\begin{equation}\label{DiagrE1EInfty}
\xymatrix{
	H^{p+q}(p, +\infty) \ar[rr]^{\Psi} \ar[dr]_{\pi^{*}} & & H^{p+q}(0, p+1) \\
	& H^{p+q}(p, p+1) \ar[ur]_{i^{*}} &
}
\end{equation}
with $A^{p,\,q} = \IIm \, \pi^{*}$ and $i^{*}\vert_{\IIm \, \pi^{*}} = \varphi^{p,\,q}$.

\subsection{Axiomatization}\label{Axiomatization}

Let us consider the following assignments, for $p, t, u \in \mathbb{Z} \cup \{-\infty, +\infty\}$:
\begin{itemize}
	\item for $-\infty \leq p \leq t \leq \infty$ and $n \in \mathbb{Z}$ an abelian group $h^{n}(p,t)$, such that $h^{n}(p, t) = h^{n}(0, t)$ for $p \leq 0$ and there exists $l \in \mathbb{N}$ such that $h^{n}(p, t) = h^{n}(p, +\infty)$ for $t \geq l$;
	\item for $p \leq t \leq u$, $a, b \geq 0$, two maps:
	\[\begin{split}
	&\Psi^{n}: h^{n}(p+a,t+b) \rightarrow h^{n}(p,t)\\
	&\Delta^{n}: h^{n}(p,t) \rightarrow h^{n+1}(t,u)
\end{split}\]
\end{itemize}
satisfying the following axioms:
\begin{enumerate}
	\item $\Psi^{n}: h^{n}(p, t) \rightarrow h^{n}(p,t)$ is the identity;
	\item the following diagram, when defined, commutes:
\begin{displaymath}
\xymatrix{
h^{n}(p'', t'') \ar[rr]^{\Psi^{n}} \ar[dr]^{\Psi^{n}} & & h^{n}(p,t)\\
& h^{n}(p',t'); \ar[ur]^{\Psi^{n}}
}
\end{displaymath}
	\item the following diagram, when defined, commutes:
\begin{displaymath}
\xymatrix{
h^{n}(p', t') \ar[r]^{\Delta^{n}} \ar[d]_{\Psi^{n}} & h^{n+1}(t',u') \ar[d]^{\Psi^{n+1}}\\
h^{n}(p,t) \ar[r]^{\Delta^{n}} & h^{n+1}(t,u);
}
\end{displaymath}
	\item the following triangle, when defined, is exact:
\begin{displaymath}
\xymatrix{
h^{\bullet}(t, u) \ar[rr]^{\Psi^{\bullet}} & & h^{\bullet}(p,u) \ar[dl]^{\Psi^{\bullet}}\\
& h^{\bullet}(p,t) \ar[ul]^{\Delta^{\bullet(+1)}}.
}
\end{displaymath}
\end{enumerate}
Then we define for $r \geq 1$:
\begin{equation}\label{DefSS}
\begin{split}
	&E^{p,\,q}_{r} = \IIm\bigl( h^{p+q}(p, p+r) \overset{\Psi^{p+q}} \longrightarrow h^{p+q}(p-r+1, p+1) \bigr)\\ \\
	&d^{p,\,q}_{r} \,=\, (\Delta^{p+q})^{p-r+1,p+1,p+r+1} \,\big\vert_{\IIm((\Psi^{p+q})^{p,p+r}_{p-r+1,p+1})}\,:\, E^{p,\,q}_{r}K \longrightarrow E^{p+r,\,q-r+1}_{r}K \\ \\
	&F^{p}h^{p+q} = \IIm \bigl( h^{p+q}(p, +\infty) \overset{\Psi}\longrightarrow h^{p+q}(0, +\infty) \bigr). \\ \text{ }
\end{split}
\end{equation}
In this way:
\begin{itemize}
	\item the groups $F^{p}h^{n}$ are a filtration of $h^{n}(0, +\infty)$;
	\item for $E_{r} = \bigoplus_{p,\,q}E^{p,\,q}_{r}$ and $d_{r} = \bigoplus_{p,\,q}d^{p,\,q}_{r}$ one has $E_{r+1} = h(E_{r}, d_{r})$;
	\item for every $n = p+q$ fixed, the sequence $\{E^{p,\,q}_{r}\}_{r \in \mathbb{N}}$ stabilizes to $F^{p}h^{p+q} / F^{p+1}h^{p+q}$.
\end{itemize}
The canonical isomorphisms $E^{p,q}_{r+1}K \simeq \Ker d^{p,q}_{r} / \IIm d^{p-r,q+r-1}_{r}$ are induced by the $\Psi$-map $\varphi^{p,q}_{r}: h^{p+q}(p-r+1, p+1) \rightarrow h^{p+q}(p-r, p+1)$, which induces a surjection in the diagram \eqref{BoundaryDiagram2}:
	\[\varphi^{p,q}_{r}: \Ker (\Delta_{2}^{p+q} \,\vert_{\IIm(\Psi_{1}^{p+q})}) \rightarrow \IIm \Psi_{3}^{p+q}.
\]
Moreover, we have the commutative diagram \eqref{DiagrE1EInfty}, with $\varphi^{p,q}$ sending the surviving elements, i.e.\ $\IIm \, \pi^{*}$, to $E^{p,q}_{\infty}$, i.e.\ to $\IIm \Psi$.

\paragraph{}As we have seen, the case of a graded abelian group $K^{\bullet}$ with a finite filtration $\{F^{p}K^{\bullet}\}_{p\in \mathbb{Z}}$ and a filtration-preserving boundary operator $d^{\bullet}$, can be considered as a particular case of this axiomatization putting:
\begin{itemize}
	\item $h^{n}(p, t) = H(F^{p}K^{n}/F^{t}K^{n})$;
	\item $\Psi$ and $\Delta$ as in \eqref{PsiDelta}
\end{itemize}
and the axioms are satisfied. In particular, axiom 4 follows from the short exact sequence $0 \rightarrow F^{t}K/F^{u}K \rightarrow F^{p}K/F^{u}K \rightarrow F^{p}K/F^{t}K \rightarrow 0$.

\subsection{Generic cohomology theory}

Given a topological space $X$ with a finite filtration:
	\[\emptyset = X^{-1} \subset X^{0} \subset \cdots \subset X^{m} = X
\]
we can consider a cohomology theory $h^{\bullet}$, not necessarily induced by a coboundary operator. In this case we define:
\begin{itemize}
	\item $h^{n}(p,t) = h^{n}(X^{t-1}, X^{p-1})$;
	\item $\Psi^{n}: h^{n}(p+a, t+b) \rightarrow h^{n}(p,t)$ is induced (by the axioms of cohomology) by the map of couples $i: (X^{t-1}, X^{p-1}) \rightarrow (X^{t+b-1}, X^{p+a-1})$;
	\item $\Delta^{n}: h^{n}(p,t) \rightarrow h^{n}(t,u)$ is the composition of the map $\pi^{*}: h^{n}(X^{t-1}, X^{p-1}) \rightarrow h^{n}(X^{t-1})$ induced by $\pi: (X^{t-1}, \emptyset) \rightarrow (X^{t-1}, X^{p-1})$, and the Bockstein map $\beta: h^{n}(X^{t-1}) \rightarrow h^{n+1}(X^{u-1},X^{t-1})$.
\end{itemize}
In this case one can verify that the previous axioms are satisfied, so that we obtain a spectral sequence called Atiyah-Hirzebruch spectral sequence, which we now study in detail.

\paragraph{Remark:} the shift by $-1$ in the definition of $h^{n}(p,t)$ is necessary to have $h^{n}(0, +\infty)$ $= h^{n}(X)$. It would not be necessary if we declared $X^{0} = \emptyset$ instead of $X^{-1} = \emptyset$, but it is better to use a notation coherent with the case of a finite simplicial complex filtered by its skeletons (in that case $X^{0}$ denotes the $0$-skeleton, not the empty set).


\chapter{Atiyah-Hirzebruch spectral sequence}

\section{Description of the spectral sequence}

The Atiyah-Hirzebruch spectral sequence relates the cellular cohomology of a finite CW-complex (or any space homotopic to it) to a generic cohomology theory. In particular, let $h^{\bullet}$ be a cohomology theory defined on an admissible category $\mathcal{A}$ containing $\TopFCW_{2}$. We start from finite simplicial complexes and then we will extend the sequence to all $\TopFCW$. For a finite simplicial complex $X$ we consider the natural filtration:
	\[\emptyset = X^{-1} \subset X^{0} \subset \cdots \subset X^{m} = X
\]
where $X^{i}$ is the $i$-th skeleton of $X$. Under these hypotheses, we can build a spectral sequence $E^{p,\,q}_{r}(X)$ such that:
\begin{itemize}
	\item $E^{p,\,q}_{2}(X) \simeq H^{p}(X, h^{q}\{*\})$ canonically, where $H^{\bullet}$ denotes the simplicial cohomology and $\{p\}$ the space made by one point;
	\item the sequence converges to $h^{\bullet}(X)$.
\end{itemize}
Such a sequence is called \emph{Atiyah-Hirzebruch spectral sequence}.

\paragraph{}As anticipated in the previous section, we can define the following groups and maps:
\begin{itemize}
	\item $h^{n}(p,t) = h^{n}(X^{t-1}, X^{p-1})$;
	\item $\Psi^{n}: h^{n}(p+a, t+b) \rightarrow h^{n}(p,t)$ is induced (by the axioms of cohomology) by the map of couples $i: (X^{t-1}, X^{p-1}) \rightarrow (X^{t+b-1}, X^{p+a-1})$;
	\item $\Delta^{n}: h^{n}(p,t) \rightarrow h^{n}(t,u)$ is the composition of the map $\pi^{*}: h^{n}(X^{t-1}, X^{p-1}) \rightarrow h^{n}(X^{t-1})$ induced by $\pi: (X^{t-1}, \emptyset) \rightarrow (X^{t-1}, X^{p-1})$, and the Bockstein map $\beta: h^{n}(X^{t-1}) \rightarrow h^{n+1}(X^{u-1},X^{t-1})$.
\end{itemize}
With these definitions all the axioms of section \ref{Axiomatization} are satisfied, so that we can consider the corresponding spectral sequence $E^{p,\,q}_{r}(X)$. We now analyze the first two and the last steps of such a sequence, expanding what is summarized in \cite{FR}.

\subsection{The first step}

From equation \eqref{DefSS} with $r = 1$:
	\[E^{p,\,q}_{1}(X) = \IIm\bigl( h^{p+q}(p, p+1) \overset{\Psi^{p+q}} \longrightarrow h^{p+q}(p, p+1) \bigr) = h^{p+q}(p, p+1)
\]
where the last equality is due to axiom 1. We now have:
	\[\begin{split}
	h^{p+q}(p, p+1) &= h^{p+q}(X^{p}, X^{p-1}) \simeq \tilde{h}^{p+q}(X^{p} / X^{p-1}) \simeq \bigoplus_{A_{p}} \tilde{h}^{p+q}(S^{p})\\
	&\simeq \bigoplus_{A_{p}} \tilde{h}^{q}(S^{0}) \simeq \bigoplus_{A_{p}} h^{q}\{p\} = C^{p}(X, h^{q}\{*\})
\end{split}\]
where $C^{p}(X, h^{q}\{*\})$ is the group of simplicial cochains with coefficients in $h^{q}\{*\}$.

\paragraph{}From equation \eqref{DefSS} we get:
	\[d^{p,\,q}_{1} \,=\, (\Delta^{p+q}) \,\big\vert_{\IIm(\Psi^{p+q})}\,:\, E^{p,\,q}_{1}K \longrightarrow E^{p+1,\,q}_{1}K
\]
i.e.:
\begin{equation}\label{d1Cell}
	d^{p,\,q}_{1} \,=\, \Delta^{p+q} \,:\, h^{p+q}(p, p+1) \longrightarrow h^{p+q+1}(p+1, p+2)
\end{equation}
which becomes:
\begin{equation}\label{d1pq}
	d^{p,\,q}_{1} \,:\, C^{p}(X, H^{q}\{*\}) \longrightarrow C^{p+1}(X, H^{q}\{*\}).
\end{equation}
It follows from \eqref{d1Cell} that this is exactly the coboundary of cellular cohomology, thus for simplicial complexes it coincides with the simplicial coboundary.

\paragraph{}We can write down functorially the canonical isomorphism $(E^{\bullet,\,q}_{1}(X), d^{\bullet,\,q}_{1}) \simeq (C^{\bullet}(X, h^{q}\{*\}),$ $\delta^{\bullet})$. In fact:
\begin{itemize}
	\item we consider the index set of $p$-simplices $A_{p}$, and we consider it as a topological space with the discrete topology; thus, we have $h^{q}(A_{p}) = \bigoplus_{i \in A_{p}}\mathbb{Z} = C^{p}(X, h^{q}\{*\})$;
	\item we consider the $p$-fold suspension of $A_{p}^{+}$, i.e.\ $S^{p}A_{p}^{+} = S^{p} \times (A_{p} \sqcup \{\infty\}) \,/\, (S^{p} \times \{\infty\} \cup \{N\} \times A_{p})$: we have the homeomorphisms $S^{p} \simeq D^{p} / \partial D^{p} \simeq \Delta^{p} / \dot{\Delta}^{p}$ sending $N \in S^{p}$ to $\dot{\Delta}^{p} / \dot{\Delta}^{p} \in \Delta^{p} / \dot{\Delta}^{p}$ for $\dot{\Delta}^{p}$ the $(p-1)$-skeleton of $\Delta^{p}$, thus we have a canonical homeomorphism $S^{p}A_{p}^{+} \simeq \Delta^{p} \times A^{p} \,/\, \dot{\Delta}^{p} \times A_{p}$;
	\item we have a canonical homeomorphism $\varphi: \Delta^{p} \times A^{p} \,/\, \dot{\Delta}^{p} \times A_{p} \rightarrow \Delta^{p}X / \Delta^{p-1}X$ obtained applying $\varphi^{p}_{i}$ to each $\Delta^{p} \times \{i\}$;
	\item thus we have canonical isomorphisms $h^{q}(A_{p}) \simeq \tilde{h}^{q}(A_{p}^{+}) \simeq \tilde{h}^{p+q}(S^{p}A_{p}^{+}) \simeq \tilde{h}^{p+q}(\Delta^{p} \times A^{p} \,/\, \dot{\Delta}^{p} \times A_{p}) \simeq \tilde{h}^{p+q}(\Delta^{p}X / \Delta^{p-1}X)$.
\end{itemize}
In a diagram:
\begin{equation}\label{IsoCpE1p}
	\xymatrix{
	A_{p} \ar[r]^{S^{p}} \ar[d]^{h^{q}} & S^{p}A^{+} \ar[r]^{\simeq} & \frac{\Delta^{p} \times A^{p}}{\dot{\Delta}^{p} \times A_{p}} \ar[r]^{\sqcup \varphi^{i}_{n}} & \frac{\Delta^{p}X}{\Delta^{p-1}X} \ar[d]^{\tilde{h}^{p+q}}\\
	C^{p}(X,h^{q}\{*\}) \ar[rrr]^{\simeq} & & & \tilde{h}^{p+q}(\Delta^{p}X / \Delta^{p-1}X).
}
\end{equation}

\subsection{The second step}

From equation \eqref{DefSS} with $r = 2$:
\begin{equation}\label{Epq2}
\begin{split}
	E^{p,\,q}_{2}(X) &= \IIm\bigl( h^{p+q}(p, p+2) \overset{\Psi^{p+q}} \longrightarrow h^{p+q}(p-1, p+1) \bigr)\\
	&= \IIm\bigl( h^{p+q}(X^{p+1}, X^{p-1}) \overset{\Psi^{p+q}} \longrightarrow h^{p+q}(X^{p}, X^{p-2}) \bigr)
\end{split}
\end{equation}
and for what we have seen about the first coboundary we have a canonical isomorphism:
	\[E^{p,\,q}_{2}(X) \simeq H^{p}(X, h^{q}\{*\}).
\]

\subsubsection{Cocycles and coboundaries}

We now consider the maps:
	\[\begin{split}
	&j: X^{p}/X^{p-1} \longrightarrow X^{p+1}/X^{p-1}\\
	&\pi: X^{p}/X^{p-2} \longrightarrow X^{p}/X^{p-1} = \frac{X^{p}/X^{p-2}}{X^{p-1}/X^{p-2}}\\
	&i: X^{p}/X^{p-2} \longrightarrow X^{p+1}/X^{p-1}.
\end{split}\]
These maps induce a commutative diagram:
\begin{equation}\label{DiagramE1E2}
\xymatrix{
E^{p, q}_{1} = \tilde{h}^{p+q}(X^{p}/X^{p-1}) \ar[dr]^{\pi^{*}} & \\
\tilde{h}^{p+q}(X^{p+1}/X^{p-1}) \ar[u]^{j^{*}} \ar[r]^{i^{*}} & \tilde{h}^{p+q}(X^{p}/X^{p-2})
}
\end{equation}
where $i^{*}, j^{*}, \pi^{*}$ are maps of the $\Psi$-type. We have that $E^{p, q}_{2} = \IIm \, i^{*}$ by \eqref{Epq2}.
\\ \\
We now prove that:
\begin{equation}\label{d1pqCocCob}
	\Ker \, d^{p, q}_{1} = \IIm \, j^{*} \qquad \IIm \, d^{p-1, q}_{1} = \Ker \, \pi^{*}
\end{equation}
The first statement follows from \eqref{d1pq} using the exact sequence:
	\[\xymatrix{
	\cdots \ar[r] & \tilde{h}^{p+q}(X^{p+1}/X^{p-1}) \ar[r]^{j^{*}} & \tilde{h}^{p+q}(X^{p}/X^{p-1}) \ar[r]^{d^{p, q}_{1}} & \tilde{h}^{p+q+1}(X^{p+1}/X^{p}) \ar[r] & \cdots
}\]
and the second by the exact sequence:
	\[\xymatrix{
	\cdots \ar[r] & \tilde{h}^{p+q-1}(X^{p-1}/X^{p-2}) \ar[r]^{d^{p-1, q}_{1}} & \tilde{h}^{p+q}(X^{p}/X^{p-1}) \ar[r]^{\pi^{*}} & \tilde{h}^{p+q}(X^{p}/X^{p-2}) \ar[r] & \cdots.
}\]
From \eqref{d1pqCocCob} we get:
\begin{itemize}
	\item cocycles in $C^{p}(X, h^{q}\{*\})$ correspond to classes in $\IIm \, j^{*}$, i.e.\ to classes in $\tilde{h}^{p+q}(X^{p}/X^{p-1})$ that are restriction of classes in $\tilde{h}^{p+q}(X^{p+1}/X^{p-1})$;
	\item coboundaries in $C^{p}(X, h^{q}\{*\})$ corresponds to classes in $\Ker \, \pi^{*}$, i.e.\ to classes in $\tilde{h}^{p+q}(X^{p}/X^{p-1})$ that are $0$ when lifted to $\tilde{h}^{p+q}(X^{p}/X^{p-2})$;
	\item $\IIm \, \pi^{*}$ corresponds to cochains (not only cocycles) up to coboundaries and its subset $\IIm \, i^{*}$ corresponds to cohomology classes;
	\item given $\alpha \in \IIm \, i^{*}$, we can lift it to elements $\tilde{h}^{p+q}(X^{p} / X^{p-1})$ which are the representative cocycles of the class.
\end{itemize}

\subsection{The last step}

\emph{Notation:} we denote $i^{p}: X^{p} \rightarrow X$ and $\pi^{p} : X \rightarrow X/X^{p}$ for any $p$.

\paragraph{}We recall equation \eqref{Limit}:
	\[E^{p,q}_{\infty} = \IIm \bigl( h^{p+q}(p, +\infty) \overset{\Psi^{p+q}}\longrightarrow h^{p+q}(0, p+1) \bigr)
\]
which, for Atiyah-Hirzebruch spectral sequence, becomes:
\begin{equation}\label{EpinftyA}
	E^{p, q}_{\infty} = \IIm \bigl( h^{p+q}(X, X^{p-1}) \overset{\Psi}\longrightarrow h^{p+q}(X^{p}) \bigr)
\end{equation}
where $\Psi$ is obtained by the pull-back of $i: X^{p} \rightarrow X / X^{p-1}$. Since $i = \pi_{p-1} \circ i_{p}$, the following diagram commutes:
\begin{equation}\label{EpinftyB}
\xymatrix{
\tilde{h}^{p+q}(X/X^{p-1}) \ar[dr]^{\pi^{*}_{p-1}} \ar[rr]^{\Psi} & & \tilde{h}^{p+q}(X^{p})\\
& \tilde{h}^{p+q}(X) \ar[ur]^{i^{*}_{p}}.
}
\end{equation}

\paragraph{Remark:} in the previous triangle we cannot say that $i_{p}^{*} \circ \pi_{p-1}^{*} = 0$ by exactness, since by exactness $i_{p}^{*} \circ \pi_{p}^{*} = 0$  at the same level $p$, as follows by $X^{p} \rightarrow X \rightarrow X/X^{p}$.

\paragraph{}By exactness of $h^{p+q}(X, X^{p-1}) \overset{\pi_{p-1}^{*}}\longrightarrow h^{p+q}(X) \overset{i_{p-1}^{*}}\longrightarrow h^{p+q}(X^{p-1})$, we deduce that:
	\[\IIm \, \pi_{p-1}^{*} = \Ker \, i_{p-1}^{*}.
\]
Since trivially $\Ker \, i_{p}^{*} \subset \Ker \,i_{p-1}^{*}$, we obtain that $\Ker \, i_{p}^{*} \subset \IIm \, \pi_{p-1}^{*}$. Moreover:
	\[\IIm \, \Psi = \IIm \, \bigl( i_{p}^{*} \circ \pi_{p-1}^{*} \bigr) = \IIm \, \Bigl( i_{p}^{*} \; \big\vert_{\IIm \, \pi_{p-1}^{*}} \Bigr) \simeq \frac{\, \IIm \, \pi_{p-1}^{*} \,} {\Ker \,i_{p}^{*}} = \frac{\, \Ker \, i_{p-1}^{*} \,} {\Ker \,i_{p}^{*}}
\]
hence, finally:
\begin{equation}\label{EpqInfty}
	E^{p, q}_{\infty} = \frac{\, \Ker \bigl( h^{p+q}(X) \longrightarrow h^{p+q}(X^{p-1}) \bigr) \,} {\Ker \bigl( h^{p+q}(X) \longrightarrow h^{p+q}(X^{p}) \bigr)}
\end{equation}
i.e.\ $E^{p, q}_{\infty}$ is made by $(p+q)$-classes on $X$ which are $0$ on $X^{p-1}$, up to classes which are $0$ on $X^{p}$. In fact, the direct sum over $p$ of \eqref{EpqInfty} is the associated graded group of the filtration $F^{p}h^{p+q} = \Ker (h^{p+q}(X) \longrightarrow h^{p+q}(X^{p-1}))$.

\subsection{From the first to the last step}

We now see how to link the first and the last step of the sequence. In the diagram \eqref{DiagrE1EInfty}, we know that an element $\alpha \in E^{p,q}_{1}$ survives until the last step if and only if $\alpha \in \IIm \, \pi^{*}$ and its class in $E^{p,q}_{\infty}$ is $\varphi^{p,q}(\alpha)$. We thus define, for $\alpha \in A^{p,q} = \IIm \, \pi^{*} \subset E^{p,q}_{1}$:
	\[\{\alpha\}^{(1)}_{E^{p,q}_{\infty}} := \varphi^{p,q}(\alpha)
\]
where the upper $1$ means that we are starting from the first step.

\paragraph{}For AHSS this becomes:
	\[E^{p, \,q}_{1} = h^{p+q}(X^{p}, X^{p-1}) \qquad E^{p, \,q}_{\infty} = \IIm \bigl( h^{p+q}(X, X^{p-1}) \overset{\Psi}\longrightarrow h^{p+q}(X^{p}) \bigr)
\]
and:
	\[\iota: h^{p+q}(X, X^{p-1}) \rightarrow h^{p+q}(X^{p}, X^{p-1}).
\]
In this case, $\iota = i^{*}$ for $i: X^{p} / X^{p-1} \rightarrow X / X^{p-1}$. Thus, the classes in $E^{p, \,q}_{1}$ surviving until the last step are the ones which are restrictions of a class defined on all $X / X^{p-1}$. Moreover, $\Psi = j^{*}$ for $j: X^{p} \rightarrow X / X^{p-1}$, and $j = i \circ \pi^{p}$ for $\pi^{p}: X^{p} \rightarrow X^{p} / X^{p-1}$. Hence $\Psi = (\pi^{p})^{*} \circ \iota$, so that, for $\alpha \in \IIm \, \iota \subset E^{p, \,0}_{1}$:
\begin{equation}\label{FromOneToInfty}
	\{\alpha\}^{(1)}_{E^{p, \,q}_{\infty}} = (\pi^{p})^{*}(\alpha).
\end{equation}

\paragraph{}Since in the following we will need to start from an element $\beta \in E^{p, \,0}_{2}$ which survives until the last step, we also define in the same way:
	\[\{\beta\}^{(2)}_{E^{p, \,q}_{\infty}}
\]
as the class in $E^{p, \,q}_{\infty}$ corresponding to $\beta$.

\paragraph{Remark:} one can easily find a definition of the spectral sequence which is independent of the simplicial structure and homotopy-invariant, as explained in \cite{AH}.

\section{Gysin map and Atiyah-Hirzebruch spectral sequence}\label{GysinAHSS}

The following section contains \cite[Chap.\ 3]{FR}, and almost coincides with chap.\ 4 of the expanded version on arXiv. We call $X$ a compact smooth $n$-dimensional manifold and $Y$ a compact embedded $p$-dimensional submanifold. We choose a finite triangulation of $X$ which restricts to a triangulation of $Y$ \cite{Munkres}. We use the following notation:
\begin{itemize}
	\item we denote the triangulation of $X$ by $\Delta = \{\Delta^{m}_{i}\}$, where $m$ is the dimension of the simplex and $i$ enumerates the $m$-simplices;
	\item we denote by $X_{\Delta}^{p}$ the $p$-skeleton of $X$ with respect to $\Delta$.
\end{itemize}
The same notation is used for other triangulations or simplicial decompositions of $X$ and $Y$. In the following theorem we need the definition of ``dual cell decomposition'' with respect to a triangulation: we refer to \cite{GH} pp.\ 53-54.
\begin{Theorem}\label{Triangulation} Let $X$ be an $n$-dimensional compact manifold and $Y \subset X$ a $p$-dimensional embedded compact submanifold. Let:
\begin{itemize}
	\item $\Delta = \{\Delta^{m}_{i}\}$ be a triangulation of $X$ which restricts to a triangulation $\Delta' = \{\Delta^{m}_{i'}\}$ of $Y$;
	\item $D = \{D^{n-m}_{i}\}$ be the dual decomposition of $X$ with respect to $\Delta$;
	\item $\tilde{D} \subset D$ be subset of $D$ made by the duals of the simplices in $\Delta'$.
\end{itemize}
Then, calling $\abs{\tilde{D}}$ the support of $\tilde{D}$:
\begin{itemize}
	\item the interior of $\abs{\tilde{D}}$ is a tubular neighborhood of $Y$ in $X$;
	\item the interior of $\abs{\tilde{D}}$ does not intersect $X_{D}^{n-p-1}$, i.e.:
	\[\abs{\tilde{D}} \cap X_{D}^{n-p-1} \subset \partial \abs{\tilde{D}}.
\]
\end{itemize}
\end{Theorem}
\textbf{Proof:} The $n$-simplices of $\tilde{D}$ are the duals of the vertices of $\Delta'$. Let $\tau = \{\tau^{m}_{j}\}$ be the first barycentric subdivision of $\Delta$ \cite{GH,Hatcher}. For each vertex $\Delta^{0}_{i'}$ in $Y$ (thought of as an element of $\Delta$), its dual is:
\begin{equation}\label{DTilde}
	\tilde{D}^{n}_{i'} = \bigcup_{\Delta^{0}_{i'} \in \tau^{n}_{j}} \tau^{n}_{j}.
\end{equation}
Moreover, if $\tau' = \{{\tau'}^{m}_{j'}\}$ is the first barycentric subdivision of $\Delta'$ (of course $\tau' \subset \tau$) and $D' = \{{D'}^{m}_{i'}\}$ is the dual of $\Delta'$ in $Y$, then (reminding that $p$ is the dimension of $Y$):
\begin{equation}\label{DPrimo}
	D'^{\,p}_{\;\,i'} = \bigcup_{\Delta^{0}_{i'} \in {\tau'}^{p}_{j'}} {\tau'}^{p}_{j'}
\end{equation}
and:
	\[\tilde{D}^{n}_{i'} \cap Y = D'^{\,p}_{\;\,i'}.
\]
Moreover, let us consider the $(n-p)$-simplices in $\tilde{D}$ contained in $\partial \tilde{D}^{n}_{i'}$ (for a fixed $i'$ in formula \eqref{DTilde}), i.e.\ $X^{n-p}_{\tilde{D}} \cap \tilde{D}^{n}_{i'}$: they intersect $Y$ transversally in the barycenters of each $p$-simplex of $\Delta'$ containing $\Delta^{0}_{i'}$: we call such barycenters $\{b_{1}, \ldots, b_{k}\}$ and the intersecting $(n-p)$-cells $\{\tilde{D}^{n-p}_{l}\}_{l = 1, \ldots, k}$. Since (for a fixed $i'$) $\tilde{D}^{n}_{i'}$ retracts on $\Delta^{0}_{i'}$, we can consider a local chart $(U_{i'}, \varphi_{i'})$, with $U_{i'} \subset \mathbb{R}^{n}$ neighborhood of $0$, such that:
\begin{itemize}
	\item $\varphi_{i'}^{-1}(U_{i'})$ is a neighborhood of $\tilde{D}^{n}_{i'}$;
	\item $\varphi_{i'}(D'^{\,p}_{\;\,i'}) \subset U_{i'} \cap (\{0\} \times \mathbb{R}^{p})$, for $0 \in \mathbb{R}^{n-p}$ (v.\ eq.\ \eqref{DPrimo});
	\item $\varphi_{i'}(\tilde{D}^{n-p}_{l}) \subset U_{i'} \cap \bigl(\mathbb{R}^{n-p} \times \pi_{p}(\varphi_{i'}(b_{l}))\bigr)$, for $\pi_{p}: \mathbb{R}^{n} \rightarrow \{0\} \times \mathbb{R}^{p}$ the projection.
\end{itemize}
We now consider the natural foliation of $U_{i'}$ given by the intersection with the hyperplanes $\mathbb{R}^{n-p} \times \{x\}$ and its image via $\varphi_{i'}^{-1}$: in this way, we obtain a foliation of $\tilde{D}^{n}_{i'}$ transversal to $Y$. If we do this for any $i'$, by construction the various foliations glue on the intersections, since such intersections are given by the $(n-p)$-cells $\{\tilde{D}^{n-p}_{l}\}_{l = 1, \ldots, k}$, and the interior gives a $C^{0}$-tubular neighborhood of $Y$.

Moreover, a $(n-p-r)$-cell of $\tilde{D}$, for $r > 0$, cannot intersect $Y$ since it is contained in the boundary of a $(n-p)$-cell, and such cells intersect $Y$, which is done by $p$-cells, only in their interior points $b_{j}$. Being the simplicial decomposition finite, it follows that the interior of $\abs{\tilde{D}}$ does not intersect $X_{D}^{n-p-1}$. \\
$\square$

\paragraph{}We now consider quintuples $(X, Y, \Delta, D, \tilde{D})$ satisfying the following condition:
\begin{itemize}
	\item[$(\#)$] $X$ is an $n$-dimensional compact manifold and $Y \subset X$ a $p$-dimensional embedded compact submanifold such that $N(Y)$ is $h$-orientable. Moreover, $\Delta$, $D$ and $\tilde{D}$ are defined as in theorem \ref{Triangulation}.
\end{itemize}

\begin{Lemma}\label{TrivialityXnp1} Let $(X,Y,\Delta,D,\tilde{D})$ be a quintuple satisfying $(\#)$, $U = \Int\abs{\tilde{D}}$ and $\alpha \in h^{*}(Y)$. Then:
\begin{itemize}
	\item there exists a neighborhood $V$ of $X \setminus U$ such that $i_{!}(\alpha)\vert_{V} = 0$;
	\item in particular, $i_{!}(\alpha) \,\vert_{X^{n-p-1}_{D}} = 0$.
\end{itemize}
\end{Lemma}
\textbf{Proof:} By equation \eqref{GysinMap}:
	\[i_{!}(\alpha) = \psi^{*} \beta \qquad \beta = (\varphi_{U}^{+})^{*} \circ \,T (\alpha) \in \tilde{h}^{*}(U^{+}).
\]
Let $V_{\infty} \subset U^{+}$ be a contractible neighborhood of $\infty$, which exists since $U$ is a tubular neighborhood of a smooth manifold, and let $V = \psi^{-1}(V_{\infty})$. Then $\tilde{h}^{*}(V_{\infty}) \simeq \tilde{h}^{*}\{*\} = 0$, thus $\beta\vert_{V_{\infty}} = 0$ so that $(\psi^{*}\beta) \vert_{V} = 0$. By theorem \ref{Triangulation} $X^{n-p-1}_{D}$ does not intersect the tubular neighborhood $\Int \abs{\tilde{D}}$ of $Y$, hence $X^{n-p-1}_{D} \subset V$, so that $(\psi^{*} \beta)\vert_{X^{n-p-1}_{D}} = 0$. $\square$

\subsection{Unit class}

We start by considering the case of the unit class $1 \in h^{0}(Y)$ (see def.\ \ref{TrivialClass}). Before we have assumed $X$ orientable for simplicity. We denote by $H$ the singular cohomology with coefficients in $h^{0}\{*\}$: then the correct hypothesis is that $X$ must by $H$-orientable, since we need the Poincar\'e duality with respect to $H$. Therefore, the orientability of $X$ is necessary only if $\chr\,h^{0}\{*\} > 2$. If the normal bundle $N_{Y}X$ of $Y$ in $X$ is $h$-orientable, as in our hypotheses, then it is also $H$-orientable, thanks to Lemma \ref{TwoOrientations}. Actually, it also follows from the following argument. $Y$ is an $H$-orientable manifold: for $\chr\,h^{0}\{*\} = 2$ any bundle is orientable (thus also the tangent bundle $TY$), otherwise, being $Y$ a simplicial complex, in order to be a cycle in $C_{p}(X, h^{0}\{*\})$ it must be oriented as a simplicial complex, thus also as a manifold. Since also $X$ is $H$-orientable, it follows that both $TX\vert_{Y}$ and $TY$ are $H$-orientable, hence also $N_{Y}X$ is. Moreover, the atlas arising in the proof of theorem \ref{Triangulation} is naturally $H$-oriented, as follows from the construction of the dual cell decomposition.

\begin{Theorem}\label{FirstTheorem} Let $(X,Y,\Delta,D,\tilde{D})$ be a quintuple satisfying $(\#)$ and $\Phi^{n-p}_{D}: C^{n-p}$ $(X, h^{q}(\{*\}))$ $\rightarrow h^{n-p+q}(X_{D}^{n-p}, X_{D}^{n-p-1})$ be the standard canonical isomorphism. Let us define the natural projection and immersion:
	\[\pi^{n-p,\,n-p-1}: X_{D}^{n-p} \longrightarrow X_{D}^{n-p} / X_{D}^{n-p-1} \qquad\quad i^{n-p}: X_{D}^{n-p} \longrightarrow X
\]
and let $\PD_{\Delta}(Y)$ be the representative of $\PD_{X}[Y]$ given by the sum of the cells dual to the $p$-cells of $\Delta$ covering $Y$. Then:
	\[(i^{n-p})^{*}(i_{!}(1)) = (\pi^{n-p,\,n-p-1})^{*} ( \Phi_{D}^{n-p} (\PD_{\Delta}(Y) ) ).
\]
\end{Theorem}
\textbf{Proof:} Let $U$ be the tubular neighborhood of $Y$ in $X$ stated in theorem \ref{Triangulation}. We define the space $(U^{+})^{n-p}_{D}$ obtained considering the interior of the $(n-p)$-cells intersecting $Y$ transversally and compactifying this space to one point. The interiors of such cells forms exactly the intersection between the $(n-p)$-skeleton of $D$ and $U$, i.e.\ $X^{n-p}_{D} \, \vert_{U}$, since the only $(n-p)$-cells intersecting $U$ are the ones intersecting $Y$, and their interior is completely contained in $U$, as stated in theorem \ref{Triangulation}. If we close this space in $X$ we obtain the closed cells intersecting $Y$ transversally, whose boundary lies entirely in $X^{n-p-1}_{D}$. Thus the one-point compatification of the interior is:
	\[(U^{+})^{n-p}_{D} = \frac{\overline{X^{n-p}_{D} \, \vert_{U}}^{X}}{X^{n-p-1}_{D} \, \vert_{\partial U}}
\]
so that there is a natural inclusion $(U^{+})^{n-p}_{D} \subset U^{+}$ sending the denominator to $\infty$ (the numerator is exactly $X_{\tilde{D}}^{n-p}$ of theorem \ref{Triangulation}). We also define:
	\[\psi^{n-p} = \psi \,\big\vert_{X^{n-p}_{D}}: \, X^{n-p}_{D} \longrightarrow (U^{+})^{n-p}_{D}.
\]
The latter is well-defined since the $(n-p)$-simplices outside $U$ and all the $(n-p-1)$-simplices are sent to $\infty$ by $\psi$. Calling $I$ the set of indices of the $(n-p)$-simplices in $D$, calling $S^{k}$ the $k$-dimensional sphere and denoting by $\dot{\cup}$ the one-point union of topological spaces, there are the following canonical homeomorphisms:
	\[\begin{split}
	& \xi^{n-p}_{X}: \pi^{n-p}(X_{D}^{n-p}) \overset{\simeq}\longrightarrow \dot{\bigcup_{i \in I}} \; S^{n-p}_{i} \\
	& \xi^{n-p}_{U^{+}}: \psi^{n-p}(X_{D}^{n-p}) \overset{\simeq}\longrightarrow \dot{\bigcup_{j \in J}} \; S^{n-p}_{j}
\end{split}\]
where $\{S^{n-p}_{j}\}_{j \in J}$, with $J \subset I$, is the set of $(n-p)$-spheres corresponding to the $(n-p)$-simplices with interior contained in $U$, i.e.\ corresponding to $\pi^{n-p}\bigl(\overline{X^{n-p}_{D} \, \big\vert_{U}}\,\bigr)$. The homeomorphism $\xi^{n-p}_{U^{+}}$ is due to the fact that the boundary of the $(n-p)$-cells intersecting $U$ is contained in $\partial U$, hence it is sent to $\infty$ by $\psi^{n-p}$, while all the $(n-p)$-cells outside $U$ are sent to $\infty$: hence, the image of $\psi^{n-p}$ is homeomorphic to $\dot{\bigcup}_{j \in J} \; S^{n-p}_{j}$ sending $\infty$ to the attachment point. We define:
	\[\begin{split}
	\rho: \, \dot{\bigcup_{i \in I}} \; S^{n-p}_{i} \longrightarrow \dot{\bigcup_{j \in J}} \; S^{n-p}_{j}
\end{split}\]
as the natural projection, i.e.\ $\rho$ is the identity of $S^{n-p}_{j}$ for every $j \in J$ and sends all the spheres in $\{S^{n-p}_{i}\,\}_{i \in I \setminus J}$ to the attachment point. We have that:
	\[\xi^{n-p}_{U^{+}} \circ \psi^{n-p} = \rho \circ \xi^{n-p}_{X} \circ \pi^{n-p,\,n-p-1}
\]
hence:
\begin{equation}\label{PsiPiRho}
	(\psi^{n-p})^{*} \circ (\xi^{n-p}_{U^{+}})^{*} = (\pi^{n-p,\,n-p-1})^{*} \circ (\xi^{n-p}_{X})^{*} \circ \rho^{*}.
\end{equation}
We put $N = N(Y)$ and $\tilde{u}_{N} = (\varphi_{U}^{+})^{*}(u_{N})$, where $u_{N}$ is the Thom class of the normal bundle. By lemma \ref{Unitarity} and equation \eqref{GysinMap} we have $i_{!}(1) = \psi^{*} \circ (\varphi_{U}^{+})^{*} (u_{N})$.
Then:
	\[(i^{n-p})^{*}(i_{!}(1)) = (i^{n-p})^{*} \psi^{*} (\tilde{u}_{N}) = (\psi^{n-p})^{*} \bigl( \tilde{u}_{N} \,\big\vert_{(U^{+})^{n-p}_{D}} \bigr)
\]
and
	\[(\xi^{n-p}_{X})^{*} \circ \rho^{*} \circ ((\xi^{n-p}_{U^{+}})^{-1})^{*} \bigl(\, \tilde{u}_{\mathcal{N}} \,\big\vert_{(U^{+})^{n-p}_{D}} \,\bigr) = \Phi_{D}^{n-p}(\PD_{\Delta}Y)
\]
since:
\begin{itemize}
	\item $\PD_{\Delta}(Y)$ is the sum of the $(n-p)$-cells intersecting $U$, oriented as the normal bundle;
	\item hence $((\xi^{n-p}_{X})^{-1})^{*} \circ \Phi_{D}^{n-p}(\PD_{\Delta}(Y))$ gives a $\gamma^{n-p}$ factor to each sphere $S^{n-p}_{j}$ for $j \in J$ and $0$ otherwise, orienting the sphere orthogonally to $Y$;
	\item but this is exactly $\rho^{*} \circ ((\xi^{n-p}_{U^{+}})^{-1})^{*} ( \tilde{u}_{N} \,\vert_{(U^{+})^{n-p}_{D}} )$ since by definition of orientability the restriction of $\tilde{\lambda}_{N}$ must be $\pm\gamma^{n}$ for each fiber of $N^{+}$. We must show that the sign ambiguity is fixed: this follows from the fact that the atlas arising from the tubular neighborhood in theorem \ref{Triangulation} is $H$-oriented, as we pointed out at the beginning of this section. For the spheres outside $U$, that $\rho$ sends to $\infty$, we have that:
	\[\begin{split}
	\rho^{*} \bigl( \tilde{u}_{N} \,\big\vert_{(U^{+})^{n-p}_{D}} \bigr) \Big\vert_{\dot{\bigcup}_{i \in I \setminus J} \; S^{n-p}_{i}}
	&= \rho^{*} \bigl( \tilde{u}_{N} \,\big\vert_{\rho(\dot{\bigcup}_{i \in I \setminus J} \; S^{n-p}_{i})} \bigr)\\
	&= \rho^{*} \bigl( \tilde{u}_{N} \,\big\vert_{\{\infty\}} \bigr) = \rho^{*}(0) = 0.
\end{split}\]
\end{itemize}
Hence, from equation \eqref{PsiPiRho}:
	\[\begin{split}
	i_{!}(Y \times \mathbb{C}) \, \big\vert_{X_{D}^{n-p}} &= (\psi^{n-p})^{*} \bigl( \tilde{u}_{N} \,\big\vert_{(U^{+})^{n-p}_{D}} \bigr)\\
	&= (\pi^{n-p,\,n-p-1})^{*} \circ (\xi^{n-p}_{X})^{*} \circ \rho^{*} \circ ((\xi^{n-p}_{U^{+}})^{-1})^{*}\bigl( \tilde{u}_{N} \,\big\vert_{(U^{+})^{n-p}_{D}} \bigr)\\
	&= (\pi^{n-p,\,n-p-1})^{*} \Phi_{D}^{n-p}(\PD_{\Delta}Y).
\end{split}\]
$\square$

\paragraph{}Let us now consider any trivial class $P^{*}\eta \in h^{q}(Y)$. By lemma \ref{Unitarity} we have that $P^{*}\eta \cdot u_{N} = \eta \cdot u_{N}$, hence theorem \ref{FirstTheorem} becomes:
	\[(i^{n-p})^{*}(i_{!}(P^{*}\eta)) = (\pi^{n-p,\, n-p-1})^{*} ( \Phi_{D}^{n-p}( \PD_{\Delta}(Y \otimes \eta) )).
\]
In fact, the same proof applies considering that $\eta \cdot u_{N}$ provides a factor $\eta \cdot \gamma^{n-p}$ instead of $\gamma^{n-p}$ for each sphere of $N^{+}$, with $\eta \in h^{q}(\{*\}) \simeq \tilde{h}^{q}(S^{q})$.

\paragraph{}The following theorem encodes the link between Gysin map and AHSS.
\begin{Theorem}\label{SecondTheorem} Let $(X,Y,\Delta,D,\tilde{D})$ be a quintuple satisfying $(\#)$ and $\Phi^{n-p}_{D}: C^{n-p}$ $(X, h^{q}(\{*\}))$ $\rightarrow h^{n-p+q}(X_{D}^{n-p}, X_{D}^{n-p-1})$ be the standard canonical isomorphism. Let us suppose that $\PD_{\Delta}Y$ is contained in the kernel of all the boundaries $d^{n-p, \,q}_{r}$ for $r \geq 1$. Then it defines a class:
	\[\{\Phi^{n-p}_{D}(\PD_{\Delta}(Y \otimes \eta))\}_{E^{n-p, \,q}_{\infty}} \in E^{n-p, \,q}_{\infty} \simeq \frac{\, \Ker ( h^{n-p+q}(X) \longrightarrow h^{n-p+q}(X^{n-p-1}) ) \,} {\Ker ( h^{n-p+q}(X) \longrightarrow h^{n-p+q}(X^{n-p}) )}.
\]
The following equality holds:
	\[\{\Phi^{n-p}_{D}(\PD_{\Delta}(Y \otimes \eta))\}_{E^{n-p, \,q}_{\infty}} = [i_{!}(P^{*}\eta)].
\]
\end{Theorem}
\textbf{Proof:} By equations \eqref{EpinftyA} and \eqref{EpinftyB} we have:
\begin{equation}\label{Epinfty}
\xymatrix{
	E^{n-p, \,q}_{\infty} = \IIm \bigl( \tilde{h}^{n-p+q}(X/X_{D}^{n-p-1}) \ar[dr]_{(\pi^{n-p-1})^{*}} \ar[rr]^{\qquad\quad (f^{n-p})^{*}} & & \tilde{h}^{n-p+q}(X_{D}^{n-p}) \bigr)\\
& \tilde{h}^{n-p+q}(X) \ar[ur]_{(i^{n-p})^{*}}
}
\end{equation}
and, given a representative $\alpha \in \IIm \, (\pi_{n-r-1})^{*} = \Ker ( h^{n-p+q}(X) \longrightarrow h^{n-p+q}(X_{D}^{n-p-1}) )$, we have that $\{\alpha\}_{E^{n-p, \,q}_{\infty}} = (i^{n-p})^{*}(\alpha) = \alpha\,\vert_{X_{D}^{n-p}}$. Moreover, from \eqref{DiagrE1EInfty} we have the diagram:
\begin{equation}\label{DiagrE1EInfty2}
\xymatrix{
	E^{n-p, \,q}_{\infty} = \IIm \bigl( \tilde{h}^{n-p+q}(X/X^{n-p-1}_{D}) \ar[rr]^{\qquad\quad (f^{n-p})^{*}} \ar[dr]_{(i^{n-p,\,n-p-1})^{*}} & & \tilde{h}^{n-p+q}(X^{n-p}_{D}) \bigr)\\
	& \tilde{h}^{n-p+q}(X^{n-p}_{D} / X^{n-p-1}_{D}) \ar[ur]_{(\pi^{n-p,\,n-p-1})^{*}}. &
}
\end{equation}
where $i^{n-p,\,n-p-1}: X^{n-p}_{D} / X^{n-p-1}_{D} \rightarrow X/X^{n-p-1}$ is the natural immersion. We have that:
\begin{itemize}
	\item by formula \eqref{FromOneToInfty} the class $\{\Phi_{D}^{n-p} (\PD_{\Delta}(Y \otimes \eta))\}_{E^{n-p, \,q}_{\infty}}$ is given in diagram \eqref{DiagrE1EInfty2} by $(\pi^{n-p,\,n-p-1})^{*}(\Phi_{D}^{n-p}(\PD_{\Delta}(Y \otimes \eta)))$;
	\item by lemma \ref{TrivialityXnp1} we have $i_{!}(1) \in \Ker ( h^{n-p+q}(X) \rightarrow h^{n-p+q}(X_{D}^{n-p-1}) )$, hence the class $[i_{!}(P^{*}\eta)]$ is well-defined in $E^{n-p, \,q}_{\infty}$, and, by exactness, $i_{!}(P^{*}\eta) \in \IIm \, (\pi^{n-p-1})^{*}$;
	\item by theorem \ref{FirstTheorem} we have $(i^{n-p})^{*}( i_{!}(P^{*}\eta)) = (\pi^{n-p,\,n-p-1})^{*}(\Phi_{D}^{n-p}(\PD_{\Delta}(Y \otimes \eta)))$;
	\item hence $\{\Phi_{D}^{n-p} (\PD_{\Delta}(Y \otimes \eta))\}_{E^{n-p, \,q}_{\infty}} = [i_{!}(P^{*}\eta)]$.
\end{itemize}
$\square$

\begin{Corollary}\label{OrientableSurvives} Assuming the same data of the previous theorem, the fact that $Y$ has orientable normal bundle with respect to $h^{*}$ is a sufficient condition for $\PD_{\Delta}(Y)$ to survive until the last step of the spectral sequence. Thus, the Poincar\'e dual of any homology class $[\,Y\,] \in H_{p}(X, h^{q}\{*\})$ having a smooth representative with $h$-orientable normal bundle survives until the last step.
\end{Corollary}
\textbf{Proof:} we put together the diagrams \eqref{Epinfty} and \eqref{DiagrE1EInfty2}:
\begin{equation}\label{TwoDiagrams}
\xymatrix{
\tilde{h}^{n-p}(X/X^{n-p-1}_{D}) \ar[d]_{(i^{n-p,\, n-p-1})^{*}} \ar[rrr]^{\qquad (\pi^{n-p-1})^{*}} \ar[drrr]^{(f^{n-p})^{*}} & & & \tilde{h}^{n-p}(X) \ar[d]^{(i^{n-p})^{*}} \\
\tilde{h}^{n-p}(X^{n-p}_{D} / X^{n-p-1}_{D}) \ar[rrr]^{\qquad (\pi^{n-p,\,n-p-1})^{*}} & & & \tilde{h}^{n-p}(X^{n-p}_{D})
}
\end{equation}
and the diagram commutes being $\pi^{n-p,\,n-p-1} \circ i^{n-p,\, n-p-1} = i^{n-p} \circ \pi^{n-p-1}$. Under the hypotheses stated, we have that $i_{!}(1) \in \IIm (\pi^{n-p-1})^{*}$, so that $i_{!}(1) = (\pi^{n-p-1})^{*}(\alpha)$. Then $(i^{n-p})^{*}(\alpha) \in A^{n-p,\,0}$, so that it survives until the last step giving a class $(i^{n-p})^{*}(\pi^{n-p})^{*}(\alpha)$ in the last step. $\square$

\paragraph{}One could inquire if the condition of having $h$-orientable normal bundle is homology invariant. This is not true: let us consider the example of K-theory, for which a bundle is orientable if and only if it is a spin$^{c}$ bundle. In \cite{BHK} the authors show that in general, for a manifold $X$, there exist homologous submanifolds $Y$ and $Y'$, such that the normal bundle of $Y$ is spin$^{c}$, while the normal bundle of $Y'$ is not. Since the second step of the Atiyah-Hirzebruch spectral sequence coincides with the cohomology of $X$, this means that both $\PD_{\Delta}Y$ and $\PD_{\Delta'}Y'$ (for suitable $\Delta$ and $\Delta'$) survive until the last step, even if the normal bundle of $Y'$ is not orientable. Then, it is natural to inquire if it is true that a cohomology class survives until the last step if and only if it admits smooth representatives with orientable normal bundle, but we do not know the answer.

\subsection{Generic cohomology class}

If we consider a generic class $\alpha$ over $Y$ of rank $\rk(\alpha)$, we can prove that $i_{!}(E)$ and $i_{!}(P^{*}\rk(\alpha))$ have the same restriction to $X^{n-p}_{D}$: in fact, the Thom isomorphism gives $T(\alpha) = \alpha \cdot u_{N}$ and, if we restrict $\alpha \cdot u_{N}$ to a \emph{finite} family of fibers, which are transversal to $Y$, the contribution of $\alpha$ becomes trivial, so it has the same effect of the trivial class $P^{*}\rk(\alpha)$. We now prove this.

\begin{Lemma}\label{LineBundleXnp} Let $(X,Y,\Delta,D,\tilde{D})$ be a quintuple satisfying $(\#)$ and $\alpha \in h^{*}(Y)$ a class of rank $\rk(\alpha)$. Then:
	\[(i^{n-p})^{*}(i_{!}\alpha) = (i^{n-p})^{*}(i_{!}(P^{*}\rk\,\alpha)).
\]
\end{Lemma}
\textbf{Proof:} Since $X^{n-p}_{D}$ intersects the tubular neighborhood in a finite number of cells corresponding under $\varphi_{U}^{+}$ to a finite number of fibers of the normal bundle $N$ attached to one point, it is sufficient to prove that, for any $y \in Y$, $(\alpha \cdot u_{N}) \, \vert_{N_{y}^{+}} = P^{*}\rk(\alpha) \cdot u_{N} \, \vert_{N_{y}^{+}}$. Let us consider the following diagram for $y \in B$:
\[\xymatrix{
	h^{i}(Y) \times h^{n}(N_{y}, N_{y}') \ar[r]^{\times \quad} \ar[d]_{(i^{*})^{i} \times (i^{*})^{n}} & h^{i+n}(Y \times N, Y \times N') \ar[d]^{(i\times i)^{*\,i+n}} \\
	h^{i}\{y\} \times h^{n}(N_{y}, N_{y}') \ar[r]^{\times \qquad} & h^{i+n}(\{y\} \times N_{y}, \{*\} \times N'_{y}).
}\]
The diagram commutes by naturality of the product, thus $(\alpha \cdot u_{N}) \, \vert_{N_{y}^{+}} = \alpha\vert_{\{y\}} \cdot u_{N}\vert_{N_{y}^{+}}$. Thus, we just have to prove that $\alpha\vert_{\{y\}} = (P^{*}\rk(\alpha))\,\vert_{\{y\}}$, i.e. that $i^{*}\alpha = i^{*}P^{*}p^{*}\alpha = (p \circ P \circ i)^{*}\alpha$. This immediately follows from the fact that $p \circ P \circ i = i$.
$\square$

\paragraph{}In the previous theorems we started from the first step of the spectral sequence, therefore we had to choose a simplicial decomposition of $X$. Anyway, if we start from the second step, we loose the dependence on the triangulation \cite{AH}.

\chapter{K-theory}

\section{Basic notions of K-theory}

\subsection{General definitions}

We recall the basic definitions and theorems of K-theory, referring mainly to \cite{Atiyah}, and also to \cite{Karoubi}, \cite{LM} and \cite{Piazza}.

\begin{Def} Let $X$ be a \emph{compact Hausdorff} topological space and $(\Vect(X), \oplus)$ the semigroup of vector bundles (up to isomorphism) on $X$. The \emph{K-theory group} of $X$ is the associated Grothendieck group $K(X)$.
\end{Def}

It is easy to prove that in $K(X)$ one has that $[E] = [F]$ if and only if there exists $G$ such that $E \oplus G \simeq F \oplus G$, and this happens if and only if there exists a trivial bundle $n$ such that $E \oplus n \simeq F \oplus n$. In particular, it is necessary that $\rk\, E = \rk\, F$. Moreover, every K-theory class $[E] - [F]$ can be represented in the form $[E'] - [n]$ trivializing $F$ with an appropriate direct summand.

\begin{Def} Let $(X, x_{0})$ be a compact Hausdorff topological space with a marked point. Let $i: \{x_{0}\} \rightarrow X$ the immersion. We define:
	\[\tilde{K}(X) = \Ker(i^{*}: K(X) \rightarrow K(x_{0})).
\]
\end{Def}
In other words, $\tilde{K}(X)$ is made by the K-theory classes $[E] - [F]$ such that $\rk_{x_{0}}(E) = \rk_{x_{0}}(F)$. If $X$ is connected, the condition becomes simply $\rk\, E = \rk\, F$, so that any class can be represented in the form $[E] - [\rk\, E]$. In this case, $\tilde{K}(X)$ does not depend on the marked point $x_{0}$.

\paragraph{}The relation between $K(X)$ and $\tilde{K}(X)$ is:
	\[K(X) \simeq \tilde{K}(X) \oplus \mathbb{Z}
\]
and the isomorphism is given by $\alpha \rightarrow (\alpha - (\rk\, \alpha)_{x_{0}}) \oplus (\rk\, \alpha)_{x_{0}}$, where $(\rk\, \alpha)_{x_{0}}$ is the bundle which is trivial with rank $\rk\, \alpha$ on the connected component of $x_{0}$, and $0$ on the other components.

\paragraph{}Let $(X, Y)$ be a couple of compact topological spaces (i.e.\ $Y \subset X$). The projection map $\pi: X \rightarrow X/Y$ induces a pull-back $\pi^{*}: \Vect(X/Y) \rightarrow \Vect(X)$. For $E$ bundle on $X/Y$, $\pi^{*}E$ is a bundle on $X$ such that $E\vert_{Y}$ is trivial. However, this map is not injective: given a bundle $E$ on $X$ which is trivial on $Y$, there is always a way to build from $E$ a bundle $E'$ on $X/Y$ such that $\pi^{*}E' = E$, but $E'$ in general is not unique. In order to build such a bundle, we must consider a trivialization $\alpha: E\vert_{Y} \rightarrow Y \times \mathbb{C}^{n}$, and consider the following equivalence relation:
\begin{itemize}
	\item let $e_{y_{1}}, e_{y_{2}} \in E$, with $\pi_{E}(e_{y_{i}}) = y_{i}$, for $y_{1}, y_{2} \in Y$;
	\item let us consider $\alpha(e_{y_{1}}) = (y_{1}, z_{1})$ and $\alpha(e_{y_{2}}) = (y_{2}, z_{2})$;
	\item we declare $e_{y_{1}} \sim_{\alpha} e_{y_{2}}$ if and only if $z_{1} = z_{2}$.
\end{itemize}
Then $E/\sim_{\alpha}$ is a bundle on $X/Y$. One can prove that the isomorphism class of $E/\sim_{\alpha}$ only depends on the \emph{homotopy} class of $\alpha$ \cite{Atiyah}. However, in general, choosing two non-homotopic trivializations one obtains non-isomorphic bundles: that's why $\pi^{*}$ is not injective.

\label{Contractible}There is a remarkable exception: when $Y$ is \emph{contractible}, then $\pi^{*}$ is a \emph{bijection}: in fact, it is injective since two trivializations $\alpha, \beta: E\vert_{Y} \rightarrow Y \times \mathbb{C}^{n}$ are necessarily homotopic (the codomain is contractible, hence they are both homotopic to a constant function), and it is surjective since on a contractible space every bundle is trivial (hence, for every $E$ in $X$, $E\vert_{Y}$ is necessarily trivial).

\begin{Def} Let $(X, Y)$ be a couple of compact topological spaces. We define:
	\[K(X, Y) = \tilde{K}(X/Y)
\]
considering $Y/Y$ as the marked point of $X/Y$.
\end{Def}

If $\pi: X \rightarrow X/Y$ is the projection, let us consider the pull-back $\pi^{*}: \tilde{K}(X/Y) \rightarrow \tilde{K}(X)$: its image is given by classes $[E] - [F] \in \tilde{K}(X)$ such that $[E\vert_{Y}] - [F\vert_{Y}] = 0 \in K(Y)$. In fact:
\begin{itemize}
	\item if $[E'] - [F'] \in \tilde{K}(X/Y)$, then $E = \pi^{*}E'$ and $F = \pi^{*}F'$ are trivial when restricted to $Y$; since $Y/Y$ is the marked point of $X/Y$, they also have the same rank on $Y$, hence $[E\vert_{Y}] - [F\vert_{Y}] = 0 \in K(Y)$;
	\item if $[E] - [n] \in \tilde{K}(X)$ and $[E\vert_{Y}] - [n\vert_{Y}] = 0 \in K(Y)$, then $(E \oplus m)\vert_{Y} \simeq (n\oplus m) \vert_{Y}$. Let $\alpha: (E \oplus m)\vert_{Y} \rightarrow Y \times \mathbb{C}^{n+m}$ be a trivialization: then $E' = (E \oplus m) / \sim_{\alpha}$ is a bundle on $X/Y$, and $\pi^{*}([E'] - [n\oplus m]) = [E] - [n]$.
\end{itemize}

However, since $\pi^{*}$ is in general not injective, the subgroup of $K(X)$ made by such classes is not isomorphic to $K(X, Y)$. There are two remarkable exceptions:
\begin{itemize}
	\item when $Y$ is a retract of $X$, we'll prove that $\pi^{*}$ is injective (theorem \ref{Retract});
	\item when $Y$ is contractible, as we have seen, $\pi^{*}$ is bijective.
\end{itemize}
What we said up to now can be summarized by the following exact sequence:
\begin{equation}\label{ExSeqKTilde}
	K(X, Y) \overset{\pi^{*}}\longrightarrow \tilde{K}(X) \overset{i^{*}}\longrightarrow \tilde{K}(Y)
\end{equation}
($i^{*}$ is the restriction) from which we trivially deduce the exactness of:
	\[K(X, Y) \overset{\pi^{*}}\longrightarrow K(X) \overset{i^{*}}\longrightarrow K(Y)
\]
since the $\mathbb{Z}$-factor of $K(X)$ is sent by $i^{*}$ to the $\mathbb{Z}$-factor of $K(Y)$.

\paragraph{}If $Y$ is a retract of $X$, we have a \emph{split} short exact sequence:
	\[0 \longrightarrow K(X, Y) \overset{\pi^{*}}\longrightarrow \tilde{K}(X) \overset{i^{*}}\longrightarrow \tilde{K}(Y) \longrightarrow 0
\]
and the same for $K$.

\paragraph{}Let $(X, x_{0})$ and $(Y, y_{0})$ be topological spaces with marked points. We define:
\begin{itemize}
	\item $X \vee Y = (X \times \{y_{0}\}) \cup (\{x_{0}\} \times Y)$;
	\item $X \wedge Y = X \times Y \,/\, X \vee Y$.
\end{itemize}
Equivalently, $X \wedge Y = \bigl( X \times Y / X \bigr) / Y$, where the embeddings $i_{1}: X \rightarrow X \times Y$ and $i_{2}: Y \rightarrow \bigl( X \times Y / X \bigr)$ are defined via marked points.

\begin{Def} Let $(X, x_{0})$ be a space with marked point, and $(S^{n}, \{*\})$ be the $n$-sphere. Then we define the \emph{$n$-th reduced suspension}:
	\[S^{n}X = S^{n} \wedge X = S^{1} \wedge \ldots \wedge S^{1} \wedge X.
\]
\end{Def}

\paragraph{}We can also define the \emph{unreduced suspension} of $X$ as the double cone over $X$:
	\[\hat{S}^{1}X = \frac{X \times [-1,1]}{(X \times \{-1\}) \cup (X \times \{1\})}.
\]
Then $S^{1}X = \hat{S}^{1}X \,/\, (\{x_{0}\} \times [-1,1])$, and, since the denominator is contractible, we have $K(S^{1}X) = K(\hat{S}^{1}X)$.

\begin{Def} We define:
\begin{itemize}
	\item $K^{-n}(X) = \tilde{K}(S^{n}(X^{+}))$;
	\item $\tilde{K}^{-n}(X) = \tilde{K}(S^{n}X)$;
	\item $K^{-n}(X, Y) = \tilde{K}^{-n}(X/Y) = \tilde{K}(S^{n}(X/Y))$.
\end{itemize}
\end{Def}

\paragraph{}The following relations hold \cite{OS}:
\begin{equation}\label{KKTilde}
	K^{-2n}(X) = \tilde{K}^{-2n}(X) \oplus \mathbb{Z} \qquad K^{-2n-1}(X) = \tilde{K}^{-2n-1}(X).
\end{equation}
which can be summarized by $K^{-n}(X) = \tilde{K}^{-n}(X) \oplus K^{-n}(\{p\})$. Table \ref{fig:KSpheres} recalls K-theory groups for spheres.
\begin{table*}[h]
	\centering
		\begin{tabular}{|c|c|c|c|}
			\hline & & & \\
			$n$ & $k$ & $\tilde{K}^{-n}(S^{k})$ & $K^{-n}(S^{k})$ \\
			& & & \\ \hline
			\text{even} & \text{even} & $\mathbb{Z}$ & $\mathbb{Z} \oplus \mathbb{Z}$ \\ \hline
			\text{even} & \text{odd} & $0$ & $\mathbb{Z}$ \\ \hline
			\text{odd} & \text{even} & $0$ & $0$ \\ \hline
			\text{odd} & \text{odd} & $\mathbb{Z}$ & $\mathbb{Z}$ \\ \hline
		\end{tabular}
\caption{K-theory groups of spheres}\label{fig:KSpheres}
\end{table*}

\subsection{Products in K-theory}

$K(X)$ has a natural ring structure given by tensor product: $[E] \otimes [F] := [E \otimes F]$. Such a product restricts to $\tilde{K}(X)$. In general, we can define a product:
\begin{equation}\label{SquareProduct}
	K(X) \otimes K(Y) \overset{\boxtimes}\longrightarrow K(X \times Y)
\end{equation}
where, if $\pi_{1}: X \times Y \rightarrow X$ and $\pi_{2}: X \times Y \rightarrow Y$ are the projections, $E \boxtimes F = \pi_{1}^{*}E \otimes \pi_{2}^{*}F$. The fiber of $E \boxtimes F$ at $(x,y)$ is $E_{x} \otimes E_{y}$\footnote{If $X = Y$ and $\Delta: X \rightarrow X \times X$ is the diagonal embedding, then $E \otimes F = \Delta^{*}(E \boxtimes F)$.}. We now prove that, fixing a marked point for $X$ and $Y$, the product restricts to \cite{Piazza}:
\begin{equation}\label{ProductWedgeK}
	\tilde{K}(X) \otimes \tilde{K}(Y) \overset{\boxtimes}\longrightarrow \tilde{K}(X \wedge Y).
\end{equation}
For this, we first state that:\footnote{\eqref{TimesSplitting} is actually true for $\tilde{K}^{-n}(X \times Y)$ for any $n$, with the same proof.}
\begin{equation}\label{TimesSplitting}
	\tilde{K}(X \times Y) \simeq \tilde{K}(X \wedge Y) \oplus \tilde{K}(Y) \oplus \tilde{K}(X).
\end{equation}
In fact:
\begin{itemize}
	\item since $X$ is a retract of $X \times Y$ via the projection, we have that $\tilde{K}(X \times Y) = K(X \times Y, X) \oplus \tilde{K}(X) = \tilde{K}(X \times Y / X) \oplus \tilde{K}(X)$;
	\item since $Y$ is a retract $X \times Y / X$ via the projection, we also have $\tilde{K}(X \times Y / X) = K(X \times Y / X, \, Y) \oplus \tilde{K}(Y) = \tilde{K}(X \wedge Y) \oplus \tilde{K}(Y)$.
\end{itemize}
Combining we obtain \eqref{TimesSplitting}. The explicit isomorphism in \eqref{TimesSplitting} is given, for $\alpha = [E] - [F] \in \tilde{K}(X \times Y)$, by:
	\[\alpha \longrightarrow \bigl(\alpha - \pi_{1}^{*}\, \alpha\vert_{X} - \pi_{2}^{*}\, \alpha\vert_{Y} \bigr) \oplus \, \pi_{2}^{*}\, \alpha\vert_{Y} \oplus\, \pi_{1}^{*}\, \alpha\vert_{X}.
\]

Let $\alpha \in \tilde{K}(X)$ and $\beta \in \tilde{K}(Y)$: then $\alpha \boxtimes \beta \vert_{X} = 0$ and $\alpha \boxtimes \beta \vert_{Y} = 0$. In fact:
	\[\alpha \boxtimes \beta \vert_{X} = \alpha \otimes (\pi_{2}^{*} \,\beta)\vert_{X} = \alpha \otimes i_{1}^{*}\pi_{2}^{*} \,\beta = \alpha \otimes (\pi_{2}i_{1})^{*} \,\beta.
\]
But $\pi_{2}i_{1}: X \rightarrow Y$ is the constant map with value $y_{0}$, and the pull-back of a bundle by a constant map is trivial. Hence $(\pi_{2}i_{1})^{*} \,\beta = 0$. Similarly for $Y$. Hence, by \eqref{TimesSplitting}, we obtain $\alpha \boxtimes \beta \in \tilde{K}(X \wedge Y)$.

\subsection{Bott periodicity}

Let us consider on $S^{2}$ the bundle $\eta = \mathcal{O}_{S^{2}}(1)$. Then, since $\tilde{K}^{-2}(X) = \tilde{K}(S^{2} \wedge X)$, by \eqref{ProductWedgeK} we obtain a map:
	\[\begin{split}
	B: \;&\tilde{K}(X) \rightarrow \tilde{K}^{-2}(X)\\
	& \alpha \rightarrow (\eta - 1) \boxtimes \alpha.
\end{split}\]
The following fundamental theorem holds \cite{Atiyah, LM, Piazza}:
\begin{Theorem}[Bott periodicity]\label{Bott} For any compact space $X$, the map $B$ is an isomorphism.
\end{Theorem}

\paragraph{}Similarly, replacing $X$ by $S^{n}X$ in theorem \ref{Bott}, we have:
	\[\tilde{K}^{-n}(X) \simeq \tilde{K}^{-n-2}(X)
\]
then, by \eqref{KKTilde}:
	\[K^{-n}(X) \simeq K^{-n-2}(X)
\]
and, finally, replacing $X$ with $X/Y$:
	\[K^{-n}(X,Y) \simeq K^{-n-2}(X,Y).
\]
For this reason we extend the definition of $K^{n}(X)$, $\tilde{K}^{n}(X)$ and $K^{n}(X,Y)$ to $n > 0$, declaring that $K^{2n}(X) = K(X)$ and $K^{2n+1}(X) = K^{-1}(X)$, and similarly for the other cases.

\subsection{K-theory as a cohomology theory}

Given a space $X$, we define its \emph{cone} as:
	\[CX = \frac{X \times [0, 1]}{X \times \{1\}}.
\]
Moreover, when we need another cone not intersecting $CX$, we consider:
	\[C'X = \frac{X \times [-1, 0]}{X \times \{-1\}}.
\]
We consider two useful isomorphisms in K-theory:
\begin{itemize}
	\item since a cone is contractible (cfr.\ page \pageref{Contractible}) and since $X/Y = X \cup CY / CY$, by the previous discussion we obtain that $\tilde{K}(X/Y) = \tilde{K}(X \cup CY / CY) \simeq \tilde{K}(X \cup CY)$, hence $K(X, Y) \simeq \tilde{K}(X \cup CY)$;
	\item since $S^{1}Y = X \cup CY / X = C'X \cup CY / C'X$, for the same reason $\tilde{K}(\hat{S}^{1}Y) = \tilde{K}(C'X \cup CY / C'X) \simeq \tilde{K}(C'X \cup CY)$, hence $\tilde{K}^{-1}(Y) \simeq \tilde{K}(C'X \cup CY)$.
\end{itemize}
From the three sequences of topological spaces:
	\[\begin{split}
	&Y \overset{i}\longrightarrow X \overset{\pi}\longrightarrow X/Y\\
	&X \overset{i'}\longrightarrow X \cup CY \overset{\pi'}\longrightarrow X \cup CY / X = \hat{S}^{1}Y\\
	&X \cup CY \overset{i''}\longrightarrow C'X \cup CY \overset{\pi''}\longrightarrow (C'X \cup CY) / (X \cup CY) = \hat{S}^{1}X
\end{split}\]
we obtain three exact sequences in K-theory:
	\[\begin{split}
	&K(X,Y) \overset{\pi^{*}}\longrightarrow \tilde{K}(X) \overset{i^{*}}\longrightarrow \tilde{K}(Y)\\
	&\tilde{K}^{-1}(Y) \overset{\pi'^{*}}\longrightarrow K(X,Y) \overset{i'^{*}}\longrightarrow \tilde{K}(X)\\
	&\tilde{K}^{-1}(X) \overset{\pi''^{*}}\longrightarrow \tilde{K}^{-1}(Y) \overset{i''^{*}}\longrightarrow K(X,Y).
\end{split}\]
One can prove that $i'^{*} \simeq \pi^{*}$ and $i''^{*} \simeq \pi'^{*}$ under the isomorphisms considered in K-theory \cite{Atiyah}, so that we obtain a five-term exact sequence:
	\[K^{-1}(X) \overset{\pi''^{*}}\longrightarrow K^{-1}(Y) \overset{\delta}\longrightarrow K(X,Y) \overset{\pi^{*}}\longrightarrow K(X) \overset{i^{*}}\longrightarrow K(Y).
\]
Since $\tilde{K}(S^{n}X / S^{n}Y) \simeq \tilde{K}(S^{n}(X/Y))$,\footnote{In fact, $\hat{S}^{1}X / \hat{S}^{1}Y = S^{1}(X/Y)$, thus, by induction, we obtain the thesis for any $n$.} replacing $X$ and $Y$ with $S^{n}X$ and $S^{n}Y$ we obtain a five-term exact sequence:
	\[\tilde{K}^{-n-1}(X) \overset{\pi''^{*}}\longrightarrow \tilde{K}^{-n-1}(Y) \overset{\delta}\longrightarrow K^{-n}(X,Y) \overset{\pi^{*}}\longrightarrow \tilde{K}^{-n}(X) \overset{i^{*}}\longrightarrow \tilde{K}^{-n}(Y)
\]
and, gluing such sequences, we obtain a long exact sequence:
	\[\begin{split}
	\cdots \tilde{K}^{-n-1}(Y) &\overset{\delta}\longrightarrow K^{-n}(X,Y) \overset{\pi^{*}}\longrightarrow \tilde{K}^{-n}(X) \overset{i^{*}}\longrightarrow \tilde{K}^{-n}(Y) \longrightarrow \cdots\\
&\cdots \longrightarrow \tilde{K}^{-1}(Y) \overset{\delta}\longrightarrow K(X,Y) \overset{\pi^{*}}\longrightarrow \tilde{K}(X) \overset{i^{*}}\longrightarrow \tilde{K}(Y).
\end{split}\]
Similarly, we obtain the corresponding sequence for K-groups. By the previous discussion, it follows that \emph{K-theory is a cohomology theory} associating to a pair of topological spaces $(X,A)$ the cohomology groups $K^{n}(X,A)$.

\paragraph{}We can now consider the case of $(X,Y)$ for $Y$ retract of $X$:

\begin{Theorem}\label{Retract} Let $Y$ be a retract of $X$. Then, for any $n$, there is a short split exact sequence:
	\[0 \longrightarrow K^{-n}(X,Y) \overset{\pi^{*}}\longrightarrow K^{-n}(X) \overset{i^{*}}\longrightarrow K^{-n}(Y) \longrightarrow 0.
\]
Hence $K^{-n}(X) \simeq K^{-n}(X,Y) \oplus K^{-n}(Y)$.
\end{Theorem}

\paragraph{Proof:} we have already proven the exactness in the middle (eq.\ \eqref{ExSeqKTilde}). The retraction $r: X \rightarrow Y$ gives a map $r^{*}: K^{-n}(Y) \rightarrow K^{-n}(X)$, and since by definition $r \circ i = \id$, one has $i^{*} \circ r^{*} = \id^{*}$, hence $i^{*}$ is surjective. Moreover, $i^{*}$ is surjective also at the level $-n-1$, hence, by exactness, we obtain $\Ker\, \delta = K^{-n-1}(Y) \Rightarrow \IIm\, \delta = 0 \Rightarrow \Ker\, \pi^{*} = 0$. The splitting is induced by $r^{*}$. $\square$

\paragraph{}In particular, the isomorphism $K^{-n}(X) \simeq K^{-n}(X,Y) \oplus K^{-n}(Y)$ is obtained by:
	\[\alpha \rightarrow (\alpha - r^{*}(\alpha\vert_{Y})) \oplus r^{*}(\alpha\vert_{S^{n}Y}).
\]

\subsection{Non-compact case}

Up to now we considered \emph{compact} topological spaces. For a generic space $X$, we use K-theory with compact support, which we now define. We denote by $X^{+}$ the one-point compactification of $X$. In particular, if $X$ is compact, $X^{+} = X \sqcup \{\infty\}$. We also consider $\{\infty\}$ as marked point of $X^{+}$. For $X$ compact we trivially have that $\tilde{K}(X^{+}) = K(X)$: we assume this as the general definition.

\begin{Def} Let $X$ be a generic topological space (also non-compact). We define:
	\[K(X) = \tilde{K}(X^{+}).
\]
\end{Def}

\paragraph{}One can easily prove that $X^{+} \wedge Y^{+} = (X \times Y)^{+}$. Hence, product \eqref{ProductWedgeK} exactly becomes:
\begin{equation}\label{ProductNonCompact}
	K(X) \otimes K(Y) \overset{\boxtimes}\longrightarrow K(X \times Y)
\end{equation}
also for the non-compact case.

\subsection{Thom isomorphism}

Let $X$ be a \emph{compact} topological space and $\pi: E \rightarrow X$ a vector bundle (real or complex): we show that $K(E)$ has a natural structure of $K(X)$-module. It seems natural to use the pull-back $\pi^{*}: K(X) \rightarrow K(E)$, but this is not possible: in fact, the group $K(E)$ is defined as the reduced K-theory group of $E^{+}$, and in general there are no possibilities to extend continuously the projection $\pi$ to $E^{+}$. Hence we use the product \eqref{ProductNonCompact}: considering the embedding $i: E \rightarrow X \times E$ defined by $i(e) = (\pi(e), e)$,\footnote{For such an embedding it is not necessary to have a marked point on $X$.} which trivially extends to $i: E^{+} \rightarrow (X \times E)^{+}$ requiring that $i(\infty) = \infty$, we can define a product:
\begin{equation}\label{ProductModule}
\begin{split}
	K(&X) \otimes K(E) \rightarrow K(E)\\
	&\alpha \otimes \beta \rightarrow i^{*}(\alpha \boxtimes \beta).
\end{split}
\end{equation}
This product defines a structure of $K(X)$-module on $K(E)$.

\begin{Lemma}\label{UnitarityK} $K(E)$ is \emph{unitary} as a $K(X)$-module.
\end{Lemma}
\textbf{Proof:} Let us consider the following maps:
	\[\begin{split}
	&\pi_{1}: X^{+} \times E^{+} \rightarrow X^{+}\\
	&\pi_{2}: X^{+} \times E^{+} \rightarrow E^{+}\\
	&i: E^{+} \rightarrow (X \times E)^{+}\\
	&\tilde{\pi}: X^{+} \times E^{+} \rightarrow X^{+} \wedge E^{+} = (X \times E)^{+}\\
	&\tilde{\pi}_{2}: (X \times E)^{+} \rightarrow E^{+}
\end{split}\]
where $i(e) = (\pi(e), e)$ and the others are defined in the obvious way. Since the map:
	\[r: X^{+} \times E^{+} \rightarrow (X^{+} \times \{\infty\}) \cup (\{\infty\} \times E^{+})
\]
given by $r(x, e) = (x, \infty)$ and $r(\infty, e) = (\infty, e)$ \footnote{The map $r$ is continuous because $X$ is compact, so that its $\infty$-point is disjoint from it.} is a retraction, $\tilde{\pi}^{*}: \tilde{K}((X \times E)^{+}) \rightarrow \tilde{K}(X^{+} \times E^{+})$ is injective \cite{Atiyah}. Then, by the definition of the module structure, for $\alpha \in K(X) = \tilde{K}(X^{+})$ and $\beta \in K(E) = \tilde{K}(E^{+})$ we reformulate \eqref{ProductModule} as:\footnote{With respect to \eqref{ProductModule} we think $\alpha \boxtimes \beta \in \tilde{K}(X^{+} \times E^{+})$ and we write explicitly $(\tilde{\pi}^{*})^{-1}$.}
	\[\alpha \cdot \beta = i^{*} (\tilde{\pi}^{*})^{-1}(\alpha \boxtimes \beta) = i^{*} (\tilde{\pi}^{*})^{-1} (\pi_{1}^{*}\alpha \otimes \pi_{2}^{*}\beta).
\]
For $\alpha = 1$ one has $\alpha\vert_{X} = X \times \mathbb{C}$ and $\alpha\vert_{\{\infty\}} = 0$. Hence:
	\[\begin{split}
	&(1 \boxtimes \beta)\vert_{X \times E^{+}} = \pi_{2}^{*}\beta\,\big\vert_{X \times E^{+}}\\
	&(1 \boxtimes \beta)\vert_{\{\infty\} \times E^{+}} = 0.
\end{split}\]
But:
\begin{itemize}
	\item since $\pi_{2} \vert_{X \times E^{+}} = (\tilde{\pi}_{2} \circ \tilde{\pi}) \vert_{X \times E^{+}}$, one has $\pi_{2}^{*}\beta\vert_{X \times E^{+}} = \tilde{\pi}^{*}\tilde{\pi}_{2}^{*} \beta\vert_{X \times E^{+}}$;
	\item since $\tilde{\pi}_{2} \circ\, \tilde{\pi}\, (\{\infty\} \times E^{+}) = \{\infty\}$ and $\beta \in \tilde{K}(E^{+})$, one has $(\tilde{\pi}^{*}\tilde{\pi}_{2}^{*} \beta) \vert_{\{\infty\} \times E^{+}} = 0$.
\end{itemize}
Hence $1 \boxtimes \beta = \tilde{\pi}^{*}\tilde{\pi}_{2}^{*} \beta$, so that:
	\[1 \cdot \beta = i^{*} (\tilde{\pi}^{*})^{-1}\tilde{\pi}^{*}\tilde{\pi}_{2}^{*} \beta = i^{*} \tilde{\pi}_{2}^{*} \beta = (\tilde{\pi}_{2} \circ i)^{*}\beta = \id^{*}\beta = \beta.
\]
$\square$

\paragraph{}Let us consider a vector space $\mathbb{R}^{2n}$ as a vector bundle on a point $\{x\}$. Then we have:
\begin{itemize}
	\item $K(\{x\}) = \mathbb{Z}$;
	\item $K(\mathbb{R}^{2n}) = \tilde{K}((\mathbb{R}^{2n})^{+}) = \tilde{K}(S^{2n}) = \mathbb{Z}$.
\end{itemize}
Hence $K(\{x\}) \simeq K(\mathbb{R}^{2n})$. The idea of the Thom isomorphism is to extend this isomorphism to a generic bundle $E \rightarrow X$ with fiber $\mathbb{R}^{2n}$. To achieve this, we try to write such an isomorphism in a way that extends to a generic bundle. Actually, this generalization works for $E$ a spin$^{c}$-bundle of even dimension.

Let us consider the spin group $\Spin(2n)$ \cite{LM}. The spin representation acts on $\mathbb{C}^{2^{n}}$, and it splits in the two irreducible representations of positive and negative chirality, acting on the subspaces $S^{+}$ and $S^{-}$ of $\mathbb{C}^{2^{n}}$ of dimension $2^{n-1}$. Also the group $\Spin^{c}(2n)$, defined as $\Spin(2n) \otimes_{\mathbb{Z}_{2}} U(1)$, acts on $\mathbb{C}^{2^{n}}$ via the standard spin$^{c}$ representation, and the same splitting in chirality holds: we call the two corresponding subspaces $S^{+}_{\mathbb{C}}$ and $S^{-}_{\mathbb{C}}$ when we think of them as $\Spin^{c}(2n)$-modules instead of $\Spin(2n)$-modules. For $\CCl(2n)$ the complex Clifford algebra of dimension $2n$, $\mathbb{C}^{2^{n}}$ is also a $\CCl(2n)$-module, and, for $v \in \mathbb{R}^{2n} \subset \CCl(2n)$, we have $v \cdot S_{\mathbb{C}}^{+} = S_{\mathbb{C}}^{-}$. We thus consider the following complex:
	\[0 \longrightarrow \mathbb{R}^{2n} \times S_{\mathbb{C}}^{+} \overset{c}\longrightarrow \mathbb{R}^{2n} \times S_{\mathbb{C}}^{-} \longrightarrow 0
\]
where $c$ is the Clifford multiplication by the first component: $c(v, z) = (v, v \cdot z)$. Such a sequence of trivial bundles on $\mathbb{R}^{2n}$ is exact when restricted to $\mathbb{R}^{2n} \setminus \{0\}$, hence the alternated sum:
	\[\lambda_{\mathbb{R}^{2n}} = \bigl[ \mathbb{R}^{2n} \times S_{\mathbb{C}}^{-} \bigr] - \bigl[ \mathbb{R}^{2n} \times S_{\mathbb{C}}^{+} \bigr]
\]
naturally gives a class in $K(\mathbb{R}^{2n}, \mathbb{R}^{2n} \setminus \{0\})$  \cite{Atiyah}. The sequence is exact in particular in $\mathbb{R}^{2n} \setminus B^{2n}$, where $B^{2n}$ is the open ball of radius $1$ in $\mathbb{R}^{2n}$, hence it defines a class:
	\[\lambda_{\mathbb{R}^{2n}} \in K(\mathbb{R}^{2n}, \mathbb{R}^{2n} \setminus B^{2n}) = \tilde{K}( \overline{B^{2n}} / S^{2n-1} ) = \tilde{K}(S^{2n}).
\]
One can prove that, for $\eta$ the dual of the tautological line bundle on $\mathbb{CP}^{1}$, whose sheaf of sections is usually denoted as $\mathcal{O}_{\mathbb{CP}^{1}}(1)$, if we identify $S^{2}$ with $\mathbb{CP}^{1}$ topologically, we have that:
\begin{equation}\label{LambdaR2n}
	\lambda_{\mathbb{R}^{2n}} = (-1)^{n} \cdot (\eta - 1)^{\boxtimes n}
\end{equation}
i.e.\ $\lambda_{\mathbb{R}^{2n}}$ is a generator of $\tilde{K}(S^{2n}) \simeq \mathbb{Z}$ \cite{Atiyah}.

We now show the generalization to a spin$^{c}$-bundle $\pi: E \rightarrow X$ of dimension $2n$. Let $S^{\pm}_{\mathbb{C}}(E)$ be the bundles of complex chiral spinors associated to $E$: to define them, we consider a spin$^{c}$-lift of the orthogonal frame bundle $\SO(E)$, which we call $\Spin^{c}(E)$, and we define $S_{\mathbb{C}}(E)$ as the vector bundle with fiber $\mathbb{C}^{2^{n}}$ associated to the spin$^{c}$ representation, the latter being induced by the action of the complex Clifford algebra via the inclusion $\Spin^{c}(2n) \subset \CCl(2n) \hookrightarrow \mathbb{C}^{2^{n}}$. This bundle splits into $S_{\mathbb{C}}(E) = S^{+}_{\mathbb{C}}(E) \oplus S^{-}_{\mathbb{C}}(E)$; moreover, $S_{\mathbb{C}}(E)$ is naturally a $\CCl(E)$-module. We can lift $S^{\pm}_{\mathbb{C}}(E)$ to $E$ by $\pi^{*}$. Then we consider the complex:
	\[0 \longrightarrow \pi^{*}S^{+}_{\mathbb{C}}(E) \overset{c}\longrightarrow \pi^{*}S^{-}_{\mathbb{C}}(E) \longrightarrow 0
\]
where $c$ is the Clifford multiplication given by the structure of $\CCl(E)$-module: for $e \in E$ and $s_{e} \in (\pi^{*}S^{+}_{\mathbb{C}}(E))_{e}$, we define $c(s_{e}) = e \cdot s_{e}$. Such a sequence is exact when restricted to $E \setminus B(E)$, where, for any fixed metric on $E$, $B(E)$ is the union of the open balls of radius $1$ on each fiber. Hence we can define the \emph{Thom class}:
\begin{equation}\label{ThomClass}
	\lambda_{E} = [\pi^{*}S^{-}_{\mathbb{C}}(E)] - [\pi^{*}S^{+}_{\mathbb{C}}(E)]
\end{equation}
as a class in $K(\,E ,\, E \setminus B(E)\,) = \tilde{K}( \, \overline{B(E)} \,/\, S(E) \, ) = \tilde{K}(E^{+}) = K(E)$. The following fundamental theorem holds (\cite{LM,Karoubi} and, only for the complex case, \cite{Atiyah, Piazza}):

\begin{Theorem}[Thom isomorphism] Let $X$ be a compact topological space and $\pi: E \rightarrow X$ an \emph{even} dimensional spin$^{c}$-bundle. For
	\[\lambda_{E} = [\pi^{*}S^{-}_{\mathbb{C}}(E)] - [\pi^{*}S^{+}_{\mathbb{C}}(E)] \in K(E)
\]
the map, defined using the module structure \eqref{ProductModule}:
	\[\begin{split}
	T: \; &K(X) \longrightarrow K(E)\\
	& \alpha \rightarrow \alpha \cdot \lambda_{E}
\end{split}\]
is a group isomorphism.
\end{Theorem}

\paragraph{}We can now see that the construction for a generic $2n$-dimensional spin$^{c}$-bundle $E \rightarrow X$ is a generalization of the construction for $\mathbb{R}^{2n}$. In fact, for $x \in X$:
\begin{itemize}
	\item $\bigl( \pi^{*}S^{\pm}_{\mathbb{C}}(E) \bigr) \big\vert_{E_{x}} = E_{x} \times \bigl( S^{\pm}_{\mathbb{C}}(E) \bigr)_{x} \simeq \mathbb{R}^{2n} \times S^{\pm}_{\mathbb{C}}(\mathbb{R}^{2n})$;
	\item the Clifford multiplication restricts on each fiber $E_{x}$ to the Clifford multiplication in $\mathbb{R}^{2n} \times S_{\mathbb{C}}(\mathbb{R}^{2n})$.
\end{itemize}
Hence:
\begin{equation}\label{RestrictionThom}
\lambda_{E} \vert_{E_{x}} \simeq \lambda_{\mathbb{R}^{2n}}.
\end{equation}
In particular, we see that, for $i: E_{x}^{+} \rightarrow E^{+}$, the restriction $i^{*}: K(E) \rightarrow K(E_{x}) \simeq \mathbb{Z}$ is surjective.

\subsection{Gysin map}

Let $X$ be a compact smooth $n$-manifold and $Y \subset X$ a compact embedded $p$-submanifold such that $n - p$ is even and the normal bundle $\mathcal{N}(Y) = (TX\,\vert_{Y}) / \,TY$ is spin$^{c}$. Then, since $Y$ is compact, there exists a tubular neighborhood $U$ of $Y$ in $X$, i.e.\ there exists an homeomorphism $\varphi_{U}: U \rightarrow \mathcal{N}(Y)$.

If $i: Y \rightarrow X$ is the embedding, from this data we can naturally define a group homomorphism, called \emph{Gysin map}:
	\[i_{!}: K(Y) \longrightarrow \tilde{K}(X).
\]
In fact:
\begin{itemize}
	\item we first apply the Thom isomorphism $T: K(Y) \longrightarrow K(\mathcal{N}(Y)) = \tilde{K}(\mathcal{N}(Y)^{+})$;
	\item then we naturally extend $\varphi_{U}$ to $\varphi_{U}^{+}: U^{+} \longrightarrow \mathcal{N}(Y)^{+}$ and apply $(\varphi_{U}^{+})^{*}: K(\mathcal{N}(Y)) \longrightarrow K(U)$;
	\item there is a natural map $\psi: X \rightarrow U^{+}$ defined by:
	\[\psi(x) = \left\{\begin{array}{ll}
	x & \text{if } x \in U \\
	\infty & \text{if } x \in X \setminus U
	\end{array}\right.
\]
hence we apply $\psi^{*}: K(U) \rightarrow \tilde{K}(X)$.
\end{itemize}
Summarizing:
\begin{equation}\label{GysinMapK}
	i_{!}\,(\alpha) = \psi^{*} \circ (\varphi_{U}^{+})^{*} \circ T \, (\alpha).
\end{equation}

\paragraph{Remark:} One could try to use the immersion $i: U^{+} \rightarrow X^{+}$ and the retraction $r: X^{+} \rightarrow U^{+}$ to have a splitting $K(X) = K(U) \oplus K(X, U) = K(Y) \oplus K(X,U)$. This is false, since the immersion $i: U^{+} \rightarrow X^{+}$ is not continuous: \emph{since $X$ is compact}, $\{\infty\} \subset X^{+}$ is open, but $i^{-1}(\{\infty\}) = \{\infty\}$, and $\{\infty\}$ is not open in $U^{+}$ since $U$ is not compact.

\section{Atiyah-Hirzebruch spectral sequence}

We briefly recall the main notions about Atiyah-Hirzebruch spectral sequence in the context of K-theory. We reproduce in this section \cite[Chap.\ 4]{FS}.

\subsection{K-theory and simplicial cohomology}

In the proof of the following lemma we will need the definition of \emph{reduced} and \emph{unreduced suspension} of a topological space $X$. We recall that the unreduced suspension is defined as $\hat{S}^{1}X = (X \times [-1,1]) / (X \times \{-1\}, X \times \{1\})$, i.e.\ as the double cone built on $X$. Instead, fixing a marked point $x_{0} \in X$, the reduced suspension is defined as $S^{1}X = \hat{S}^{1}X / (\{x_{0}\} \times [-1,1])$. The group $K^{1}(X)$ is defined as $K(S^{1}X)$, but, since $S^{1}X$ is obtained from $\hat{S}^{1}X$ quotienting out by a contractible subspace, it follows that $K(S^{1}X) \simeq K(\hat{S}^{1}X)$ \cite{Atiyah}.

\begin{Lemma}\label{OnePointUnion} For $k \in \mathbb{N}$ and $0 \leq i \leq k$, let:
	\[X = \dot{\bigcup_{i = 0, \ldots, k}} \; X_{i}
\]
be the one-point union of $k$ topological spaces. Then:
	\[\tilde{K}^{n}(X) \simeq \bigoplus_{i = 0}^{k} \tilde{K}^{n}(X_{i}).
\]
\end{Lemma}
\textbf{Proof:} For $n = 0$, let us construct the isomorphism $\varphi: \tilde{K}(X) \rightarrow \bigoplus \tilde{K}(X_{i})$: it is simply given by $\varphi(\alpha)_{i} = \alpha\vert_{X_{i}}$, where $\alpha\vert_{X_{i}}$ is the pull-back via the immersion $X_{i} \rightarrow X$. To build $\varphi^{-1}$, let us consider $\{ [E_{i}] - [n_{i}] \} \in \bigoplus \tilde{K}(X_{i})$, where $[n_{i}]$ is the K-theory class represented by the trivial bundle of rank $n_{i}$. Since the sum is finite, by adding and subtracting a trivial bundle we can suppose $n_{i} = n_{j}$ for every $i,j$, so that we consider $\{ [E_{i}] - [n]\}$. Since the intersection of the $X_{i}$ is a point and the bundles $E_{i}$ have the same rank, we can glue them to a bundle $E$ on $X$ (see \cite{Atiyah} pp.\ 20-21): then we declare $\varphi^{-1} ( \, \{ [E_{i}] - [n] \} \, ) = ([E] - [n])$.

For $n = 1$, we first note that $\tilde{K}(\hat{S}^{1}(X_{1} \,\dot{\cup}\, X_{2})) = \tilde{K} (\hat{S}^{1}X_{1} \,\dot{\cup}\, \hat{S}^{1}X_{2})$, since quotienting by a contractible space (the linking between vertices of the cones and the joining point) we obtain the same space. Hence $\tilde{K}^{1}(X_{1} \,\dot{\cup}\, X_{2}) \simeq \tilde{K}^{1}(X_{1}) \oplus \tilde{K}^{1}(X_{2})$. Then, by induction, the thesis extends to finite families. Hence we have proven the result for $\tilde{K}^{n}$ with $n = 0$ and $n = 1$: by Bott periodicity \cite{Atiyah} the result holds for any $n$. $\square$

\paragraph{Remark:} we stress the fact that the previous lemma holds only for the one-point union of a \emph{finite} number of spaces.

\paragraph{}In the following theorem we suppose that the group of simplicial cochains $C^{p}(X, \mathbb{Z})$ of a finite simplicial complex coincides with the group of chains $C_{p}(X, \mathbb{Z})$: that's because, being the dimension finite, we can define the coboundary operator $\delta^{p}$ directly on chains, asking that the coboundary of a simplicial $p$-simplex $\sigma^{p}$ is the alternated sum of the $(p+1)$-simplices whose boundary contains $\sigma^{p}$ (while the boundary operator $\partial^{p}$ gives the alternated sum of the $(p-1)$-simplices contained in the boundary of $\sigma^{p}$). We can use this definition since the group of $p$-cochains as usually defined, i.e.\ $\Hom(C_{p}(X, \mathbb{Z}), \mathbb{Z})$, is canonically isomorphic to $C_{p}(X, \mathbb{Z})$ in the case of finite simplicial complexes, and the usual coboundary operator corresponds to the one we defined above under such an isomorphism.

\begin{Theorem}\label{KTheoryCohomology} Let $X$ be a $n$-dimensional finite simplicial complex, $X^{p}$ be the $p$-skeleton of $X$ for $0 \leq p \leq n$ and $C^{p}(X, \mathbb{Z})$ be the group of simplicial $p$-cochains. Then, for any $p$ such that $0 \leq 2p \leq n$ or $0 \leq 2p+1 \leq n$, there are isomorphisms:
	\[\begin{split}
	&\Phi^{2p}: C^{2p}(X, \mathbb{Z}) \overset{\simeq}\longrightarrow K(X^{2p}, X^{2p-1})\\
	&\Phi^{2p+1}: C^{2p+1}(X, \mathbb{Z}) \overset{\simeq}\longrightarrow K^{1}(X^{2p+1}, X^{2p})
\end{split}\]
which can be summarized by:
	\[\Psi^{p}: C^{p}(X, \mathbb{Z}) \overset{\simeq}\longrightarrow K^{p}(X^{p}, X^{p-1}).
\]
Moreover:
	\[K^{1}(X^{2p}, X^{2p-1}) = K(X^{2p+1}, X^{2p}) = 0.
\]
\end{Theorem}
\paragraph{Proof:} We denote the simplicial structure of $X$ by $\Delta = \{\Delta^{m}_{i}\}$, where $m$ is the dimension of the simplex and $i$ enumerates the $m$-simplices, so that $X^{2p} = \displaystyle\bigcup_{i = 0}^{k} \Delta_{i}^{2p}$. Then the quotient by $X^{2p-1}$ is homeomorphic to $k$ spheres of dimension $2p$ attached to a point:
	\[X^{2p} / X^{2p-1} = \dot{\bigcup_{i}} \; S^{2p}_{i}.
\]
By lemma \ref{OnePointUnion} we obtain $\tilde{K}(X^{2p}/X^{2p-1}) \simeq \underset{i}\bigoplus \tilde{K}(S^{2p})$, and, by Bott periodicity, $\tilde{K}(S^{2p}) = \tilde{K}(S^{0}) = \mathbb{Z}$. Hence:
	\[K(X^{2p}, X^{2p-1}) \simeq \bigoplus_{i} \mathbb{Z} = C^{2p}(X, \mathbb{Z}).
\]
For the odd case, let $X^{2p+1} = \displaystyle\bigcup_{j = 0}^{h} \Delta_{j}^{2p+1}$. We have by lemma \ref{OnePointUnion}:
	\[\begin{split}
	K^{1}(X^{2p+1}, X^{2p}) &= \tilde{K}^{1}\Bigl(\dot{\bigcup_{j}} \; S^{2p+1}_{j} \Bigr) = \bigoplus_{j} \tilde{K}^{1}\bigl(S^{2p+1}_{j}\bigr)\\
	&= \bigoplus_{j} \tilde{K}(S^{2p+2}_{j}) = \bigoplus_{j} \mathbb{Z} = C^{2p+1}(X, \mathbb{Z}).
\end{split}\]
In the same way, $K^{1}(X^{2p}, X^{2p-1}) = \bigoplus_{j} \tilde{K}^{1}(S^{2p}_{j}) = \bigoplus_{j} \tilde{K}(S^{2p+1}_{j}) = 0$, and similarly for $K(X^{2p+1}, X^{2p})$. $\square$

\paragraph{}For $\eta$ the dual of the tautological line bundle on $\mathbb{CP}^{1}$, whose sheaf of sections is usually denoted as $\mathcal{O}_{\mathbb{CP}^{1}}(1)$, if we identify $S^{2}$ with $\mathbb{CP}^{1}$ topologically, the explicit isomorphisms $\Phi^{2p}$ and $\Phi^{2p+1}$ of theorem \ref{KTheoryCohomology} are:
	\[\Phi^{2p} \bigl(\Delta^{2p}_{i}\bigr) = \left\{\begin{matrix}
	(-1)^{p}(\eta - 1)^{\boxtimes p} & \in \tilde{K} \bigl(S^{2p}_{i}\bigr) &\\ \\
	0 & \in \tilde{K} \bigl(S^{2p}_{j}\bigr) & \text{ for $j \neq i$}
	\end{matrix}\right.
\]
and:
	\[\Phi^{2p+1} \bigl(\Delta^{2p+1}_{i}\bigr) = \left\{\begin{matrix}
	(-1)^{p+1}(\eta - 1)^{\boxtimes (p+1)} & \in \tilde{K}^{1} \bigl(S^{2p+1}_{i}\bigr) &\\ \\
	0 & \in \tilde{K}^{1} \bigl(S^{2p+1}_{j}\bigr) & \text{ for $j \neq i$}
	\end{matrix}\right.
\]
where we put the overall factors $(-1)^{p}$ and $(-1)^{p+1}$ for coherence with \eqref{LambdaR2n}.

\paragraph{Remark:} such isomorphisms are canonical, since every simplex is supposed to be oriented and $\eta - 1$ is distinguishable from $1 - \eta$ also up to automorphisms of $X$ (in the first case the trivial bundle has negative coefficient, in the second case the non-trivial one, so that, for example, they have opposite first Chern class).

\subsection{The spectral sequence}

We now recall how to build the spectral sequence. The assigned groups are:
	\[H^{n}(p,p') = K^{n}(X^{p'-1}, X^{p-1}).
\]

\subsubsection{The first step}

The first step, from \eqref{Epr}, is:
	\[E^{p,\,q}_{1} = H^{p+q}(p, p+1) = K^{p+q}(X^{p}, X^{p-1}).
\]
By theorem \ref{KTheoryCohomology} we have the isomorphisms:
	\[\begin{array}{lllllll}
	E^{2p, \,0}_{1} & \simeq & C^{2p}(X, \mathbb{Z}) & & E^{2p, \,1}_{1} & = & 0\\
	E^{2p+1, \,0}_{1} &\simeq & C^{2p+1}(X, \mathbb{Z}) & & E^{2p+1, \,1}_{1} & = & 0.
\end{array}\]
Since, for a point $x_{0}$, $K(\{x_{0}\}) = \mathbb{Z}$ and $K^{1}(\{x_{0}\}) = 0$, we can write these isomorphisms in a compact form:
\begin{equation}\label{IsomCp}
	E^{p, \,q}_{1} \simeq C^{p}(X, K^{q}(x_{0})).
\end{equation}
Anyway, since $E^{p, \,1}_{1} = 0$ for every $p$, and since only the parity of $q$ is meaningful, the only interesting case is $q = 0$. Therefore, from now on we deal only with the groups $E^{p, \,0}_{r}$. For the coboundaries, since $d^{p,\,q}_{r}: E^{p, \,q}_{r} \rightarrow E^{p+r, \,q-r+1}_{r}$, in particular $d^{p,\,0}_{r}: E^{p, \,0}_{r} \rightarrow E^{p+r, \,-r+1}_{r}$, if $r$ is even the coboundary is surely $0$, thus only the odd coboundaries are interesting. Therefore, from now on we deal only with the coboundaries $d^{p, \,0}_{r}$ with $r$ odd.

For $r = 1$, in the diagram \eqref{BoundaryDiagram} one has $\psi_{1} = \psi_{2} = \id$, hence $d^{p,\,0}_{1} = \delta_{2}$, i.e.\ $d^{p,\,0}_{1} = (\delta^{p})^{p,p+1,p+2}$. In particular:
	\[d^{p,\,0}_{1}: K^{p}(X^{p}, X^{p-1}) \longrightarrow K^{p+1}(X^{p+1}, X^{p})
\]
is the composition:
\begin{displaymath}
\xymatrix{
\tilde{K}^{p}(X^{p}/X^{p-1}) \ar[rr]^{d^{p, \,0}_{1}} \ar[dr]_{(\pi^{p,p-1})^{*}} & & \tilde{K}^{p+1}(X^{p+1}/X^{p})\\
& \tilde{K}^{p}(X^{p}). \ar[ur]_{\delta^{p}}
}
\end{displaymath}
for $\pi^{p,p-1}: X^{p} \rightarrow X^{p}/X^{p-1}$ the natural projection and $\delta^{p}$ is the Bockstein map. Another way to describe $d_{1}^{p,\,0}$ can be obtained considering the exact sequence induced by $X^{p}/X^{p-1} \longrightarrow X^{p+1}/X^{p-1} \longrightarrow X^{p+1}/X^{p}$: then $d_{1}^{p, \,0}$ is the corresponding Bockstein map:
\begin{equation}\label{d1p}
	d_{1}^{p, \,0}: \, \tilde{K}^{p}(X^{p}/X^{p-1}) \longrightarrow \tilde{K}^{p+1}(X^{p+1}/X^{p}).
\end{equation}

\subsubsection{The second step}

We have shown that $E^{p, \,0}_{1} \simeq C^{p}(X, \mathbb{Z})$; we also have that $E^{p, \,0}_{2} \simeq H^{p}(X, \mathbb{Z})$ (see \cite{AH}), i.e.\ $d^{p, \,0}_{1}$ is the simplicial coboundary operator under the isomorphism \eqref{IsomCp}. By the first formula of \eqref{Epr} we have $E^{p,\,0}_{2} = \IIm \bigl( H^{p}(p, p+2) \overset{\psi^{p}}\longrightarrow H^{p}(p-1, p+1) \bigr)$, i.e.:
\begin{equation}\label{Ep2}
	E^{p, \,0}_{2} = \IIm \bigl( K^{p}(X^{p+1}, X^{p-1}) \overset{\psi^{p}}\longrightarrow K^{p}(X^{p}, X^{p-2}) \bigr).
\end{equation}
Thus there is a canonical isomorphism:
\begin{equation}\label{IsomXi}
	\Xi^{p}: H^{p}(X, \mathbb{Z}) \longrightarrow \IIm \, \psi^{p} \subset K^{p}(X^{p}, X^{p-2}).
\end{equation}

\paragraph{Cocycles and coboundaries} We now consider the maps:
	\[\begin{split}
	&\tilde{j}^{p}: X^{p}/X^{p-1} \longrightarrow X^{p+1}/X^{p-1}\\
	&\tilde{\pi}^{p}: X^{p}/X^{p-2} \longrightarrow X^{p}/X^{p-1} = \frac{X^{p}/X^{p-2}}{X^{p-1}/X^{p-2}}\\
	&\tilde{f}^{p}: X^{p}/X^{p-2} \longrightarrow X^{p+1}/X^{p-1}
\end{split}\]
These maps induce a commutative diagram:
\begin{equation}\label{DiagramE1E2K}
\xymatrix{
E^{p, \,0}_{1} = \tilde{K}^{p}(X^{p}/X^{p-1}) \ar[dr]^{(\tilde{\pi}^{p})^{*}} & \\
\tilde{K}^{p}(X^{p+1}/X^{p-1}) \ar[u]^{(\tilde{j}^{p})^{*}} \ar[r]^{(\tilde{f}^{p})^{*}} & \tilde{K}^{p}(X^{p}/X^{p-2})
}
\end{equation}
where $(\tilde{f}^{p})^{*}, (\tilde{j}^{p})^{*}, (\tilde{\pi}^{p})^{*}$ are maps of the $\psi$-type. We have that $E^{p, \,0}_{2} = \IIm (\tilde{f}^{p})^{*}$ by \eqref{Ep2}. We now prove that:
\begin{enumerate}
	\item $\Ker \, d^{p, \,0}_{1} = \IIm (\tilde{j}^{p})^{*}$;
	\item $\IIm \, d^{p-1, \, 0}_{1} = \Ker (\tilde{\pi}^{p})^{*}$.
\end{enumerate}
The first statement follows directly from \eqref{d1p} using the exact sequence:
	\[\cdots \longrightarrow \tilde{K}^{p}(X^{p+1}/X^{p-1}) \overset{(\tilde{j}^{p})^{*}}\longrightarrow \tilde{K}^{p}(X^{p}/X^{p-1}) \overset{d^{p, \,0}_{1}}\longrightarrow \tilde{K}^{p+1} (X^{p+1}/X^{p}) \longrightarrow \cdots
\]
and the second by the exact sequence:
	\[\cdots \longrightarrow \tilde{K}^{p-1}(X^{p-1}/X^{p-2}) \overset{d^{p-1, \, 0}_{1}}\longrightarrow \tilde{K}^{p}(X^{p}/X^{p-1}) \overset{(\tilde{\pi}^{p})^{*}}\longrightarrow \tilde{K}^{p}(X^{p}/X^{p-2}) \longrightarrow \cdots.
\]
Since $\IIm (\tilde{f}^{p})^{*} \simeq H^{p}(X, \mathbb{Z})$ and $d^{p,\,0}_{1}$ corresponds to the simplicial coboundary under this isomorphism, it follows that:
\begin{itemize}
	\item cocycles in $C^{p}(X, \mathbb{Z})$ correspond to classes in $\IIm (\tilde{j}^{p})^{*}$, i.e.\ to classes in $\tilde{K}^{p}$ $(X^{p}/X^{p-1})$ that are restriction of classes in $\tilde{K}^{p}(X^{p+1}/X^{p-1})$;
	\item coboundaries in $C^{p}(X, \mathbb{Z})$ corresponds to classes in $\Ker (\tilde{\pi}^{p})^{*}$, i.e.\ to classes in $\tilde{K}^{p}(X^{p}/X^{p-1})$ that are $0$ when lifted to $\tilde{K}^{p}(X^{p}/X^{p-2})$;
	\item $\IIm \, \pi^{*}$ corresponds to cochains (not only cocycles) up to coboundaries and its subset $\IIm (\tilde{f}^{p})^{*}$ corresponds to cohomology classes;
	\item given $\alpha \in \IIm (\tilde{f}^{p})^{*}$, we can lift it to a class in $\tilde{K}^{p}(X^{p} / X^{p-1})$ choosing different trivializations on $X^{p-1}/X^{p-2}$, and the different homotopy classes of such trivializations determine the different representative cocycles of the class.
\end{itemize}

\subsubsection{The last step}

We recall equation \eqref{Limit}:
	\[E^{p,\,q}_{\infty} = \IIm \bigl( H^{p+q}(p, +\infty) \overset{\psi^{p+q}}\longrightarrow H^{p+q}(0, p+1) \bigr)
\]
which, in our case, becomes:
\begin{equation}\label{EpinftyAK}
	E^{p, \,0}_{\infty} = \IIm \bigl( K^{p}(X, X^{p-1}) \overset{\psi^{p}}\longrightarrow K^{p}(X^{p}) \bigr)
\end{equation}
where $\psi$ is obtained by the pull-back of $f^{p}: X^{p} \rightarrow X / X^{p-1}$. Since, for $i^{p}: X^{p} \rightarrow X$ the natural immersion and $\pi^{p} : X \rightarrow X/X^{p}$ the natural projection, $f^{p} = \pi^{p-1} \circ i^{p}$ holds, the following diagram commutes:
\begin{equation}\label{EpinftyBK}
\xymatrix{
\tilde{K}^{p}(X/X^{p-1}) \ar[dr]_{(\pi^{p-1})^{*}} \ar[rr]^{(f^{p})^{*}} & & \tilde{K}^{p}(X^{p})\\
& \tilde{K}^{p}(X) \ar[ur]_{(i^{p})^{*}}.
}
\end{equation}

\paragraph{Remark:} in the previous triangle we cannot say that $(i^{p})^{*} \circ (\pi^{p-1})^{*} = 0$ by exactness, since by exactness $(i^{p})^{*} \circ (\pi^{p})^{*} = 0$  at the same level $p$, as follows from $X^{p} \rightarrow X \rightarrow X/X^{p}$.

\paragraph{}The sequence $K^{p}(X, X^{p-1}) \overset{(\pi^{p-1})^{*}}\longrightarrow K^{p}(X) \overset{(i^{p-1})^{*}}\longrightarrow K^{p}(X^{p-1})$ is exact, i.e.:
	\[\IIm (\pi^{p-1})^{*} = \Ker (i^{p-1})^{*}.
\]
Since trivially $\Ker (i^{p})^{*} \subset \Ker (i^{p-1})^{*}$, we obtain that $\Ker (i^{p})^{*} \subset \IIm (\pi^{p-1})^{*}$. Moreover:
	\[\IIm \, \psi = \IIm ( (i^{p})^{*} \circ (\pi^{p-1})^{*} ) = \IIm  \bigl( (i^{p})^{*} \vert_{\IIm (\pi^{p-1})^{*}} \bigr) \simeq \frac{ \IIm (\pi^{p-1})^{*} } {\Ker (i^{p})^{*}} = \frac{ \Ker  (i^{p-1})^{*} } {\Ker (i^{p})^{*}}
\]
hence, finally:
\begin{equation}
	E^{p, \,0}_{\infty} \simeq \frac{\, \Ker \bigl( K^{p}(X) \longrightarrow K^{p}(X^{p-1}) \bigr) \,} {\Ker \bigl( K^{p}(X) \longrightarrow K^{p}(X^{p}) \bigr)}\\
\end{equation}
i.e.\ $E^{p, \, 0}_{\infty}$ is made, up to canonical isomorphism, by $p$-classes on $X$ which are $0$ on $X^{p-1}$, up to classes which are $0$ on $X^{p}$.

\subsubsection{From the first to the last step}

We now see how to link the first and the last step of the sequence. In general we have:
	\[E^{p,\,q}_{1} = H^{p+q}(p, p+1) \qquad E^{p,\,q}_{\infty} = \IIm \bigl( H^{p+q}(p, +\infty) \overset{\psi_{1}}\longrightarrow H^{p+q}(0, p+1) \bigr).
\]
for $\psi_{1} = (\psi^{p+q})^{p, +\infty}_{0, p+1}$. We also consider the map:
	\[\psi_{2}: H^{p+q}(p, +\infty) \longrightarrow H^{p+q}(p, p+1)
\]
where $\psi_{2} = (\psi^{p+q})^{p, +\infty}_{p, p+1}$. An element $\alpha \in E^{p,\,q}_{1}$ survives until the last step if and only if $\alpha \in \IIm \, \psi_{2}$ and its class in $E^{p,\,q}_{\infty}$ is $\psi_{1} \circ (\psi_{2}^{-1})(\alpha)$, which is well-defined since $\Ker \, \psi_{2} \subset \Ker \, \psi_{1}$. For:
	\[\psi_{3}: H^{p+q}(p, p+1) \longrightarrow H^{p+q}(0, p+1)
\]
i.e.\ $\psi_{3} = (\psi^{p+q})^{p, p+1}_{0, p+1}$, it holds that $\psi_{1} = \psi_{3} \circ \psi_{2}$, so that $\psi_{1} \circ (\psi_{2}^{-1}) = \psi_{3}$. For $\alpha \in \IIm \, \psi_{2} \subset E^{p,\,q}_{1}$, we call $\{\alpha\}_{E^{p,\,q}_{\infty}}$ the class it reaches in $E^{p,\,q}_{\infty}$. Then we have:
	\[\{\alpha\}_{E^{p,\,q}_{\infty}} = \psi_{3}(\alpha).
\]

\paragraph{}For AHSS this becomes:
	\[E^{p, \,0}_{1} = K^{p}(X^{p}, X^{p-1}) \qquad E^{p, \,0}_{\infty} = \IIm \bigl( K^{p}(X, X^{p-1}) \overset{\psi_{1}}\longrightarrow K^{p}(X^{p}) \bigr)
\]
and:
	\[\psi_{2}: K^{p}(X, X^{p-1}) \longrightarrow K^{p}(X^{p}, X^{p-1}).
\]
In this case, $\psi_{2} = (i^{p,p-1})^{*}$ for $i^{p,p-1}: X^{p} / X^{p-1} \rightarrow X / X^{p-1}$. Thus, the classes in $E^{p, \,0}_{1}$ surviving until the last step are the ones which are restrictions of a class defined on all $X / X^{p-1}$. Moreover, $\psi_{1} = (f^{p})^{*}$ for $f^{p}: X^{p} \rightarrow X / X^{p-1}$, and $f^{p} = i^{p,p-1} \circ \pi^{p,p-1}$ for $\pi^{p,p-1}: X^{p} \rightarrow X^{p} / X^{p-1}$. Hence $\psi_{1} = (\pi^{p,p-1})^{*} \circ \psi_{2}$, and, in fact, $\psi_{3} = (\pi^{p,p-1})^{*}$. This implies that, for $\alpha \in \IIm \, \psi_{2} \subset E^{p, \,0}_{1}$:
\begin{equation}\label{FromOneToInftyK}
	\{\alpha\}_{E^{p, \,0}_{\infty}} = (\pi^{p,p-1})^{*}(\alpha).
\end{equation}

\subsection{Rational K-theory and cohomology}

We now consider the Atiyah-Hirzebruch spectral sequence in the rational case \cite{AH}. In particular, we consider the groups:
	\[H^{n}(p,p') = K_{\mathbb{Q}}^{n}(X^{p'-1}, X^{p-1})
\]
where $K_{\mathbb{Q}}^{n}(X, Y) := K^{n}(X, Y) \otimes_{\mathbb{Z}} \mathbb{Q}$. In this case the sequence is made by the groups $Q^{p, \, q}_{r} = E^{p, \, q}_{r} \otimes_{\mathbb{Z}} \,\mathbb{Q}$. In particular:
\begin{equation}\label{QSequence}
\begin{array}{lll}
		Q^{p, \, 0}_{1} \simeq C^{p}(X, \mathbb{Q}) & & Q^{p, \, 1}_{1} = 0\\ \\
		Q^{p, \, 0}_{2} \simeq H^{p}(X, \mathbb{Q}) & & Q^{p, \, 1}_{2} = 0\\ \\
		Q^{p, \, 0}_{\infty} \simeq \displaystyle\frac{\, \Ker \bigl( K_{\mathbb{Q}}^{p}(X) \longrightarrow \displaystyle K_{\mathbb{Q}}^{p}(X^{p-1}) \bigr) \,} {\Ker \bigl( K_{\mathbb{Q}}^{p}(X) \longrightarrow \displaystyle K_{\mathbb{Q}}^{p}(X^{p}) \bigr)} & & Q^{p, \, 1}_{\infty} = 0.
	\end{array}
\end{equation}
Such a sequence collapses at the second step \cite{AH}, hence $Q^{p, \, 0}_{\infty} \simeq Q^{p, \, 0}_{2}$. Since:
\begin{itemize}
	\item $\bigoplus_{p} Q^{p, \, 0}_{\infty}$ is the graded group associated to the chosen filtration of $K(X) \oplus K^{1}(X)$;
	\item in particular, by \eqref{QSequence}, $\bigoplus_{2p} Q^{2p, \, 0}_{\infty}$ is the graded group of $K_{\mathbb{Q}}(X)$ and $\bigoplus_{2p+1} Q^{2p+1, \, 0}_{\infty}$ is the graded group of $K^{1}_{\mathbb{Q}}(X)$;
	\item $Q^{p, \, 0}_{\infty} \simeq H^{p}(X, \mathbb{Q})$, thus it has no torsion;
\end{itemize}
it follows that:
	\[K_{\mathbb{Q}}(X) = \bigoplus_{2p} Q^{2p, \, 0}_{\infty} \qquad K_{\mathbb{Q}}^{1}(X) = \bigoplus_{2p+1} Q^{2p+1, \, 0}_{\infty}
\]
hence:
	\[K_{\mathbb{Q}}(X) \simeq H^{\ev}(X, \mathbb{Q}) \qquad K^{1}_{\mathbb{Q}}(X) \simeq H^{\odd}(X, \mathbb{Q}).
\]
In particular, the isomorphisms of the last equation are given by the Chern character:
	\[\begin{split}
	&\ch: K_{\mathbb{Q}}(X) \longrightarrow H^{\ev}(X, \mathbb{Q})\\
	&\ch: K^{1}_{\mathbb{Q}}(X) \longrightarrow H^{\ev}(S^{1}X, \mathbb{Q}) \simeq H^{\odd}(X, \mathbb{Q})
\end{split}\]
and they are also isomorphism of rings.

\section{Gysin map and AHSS}

We are now ready to describe the explicit link between the Gysin map and the Atiyah-Hirzebruch spectral sequence, specializing to K-theory what discussed in general in section \ref{GysinAHSS}. We start with the case of an embedded sumbanifold of even codimension, corresponding, from a physical point of view, to a D-brane world-volume in type IIB superstring theory, then we reproduce the same result in the case of odd codimension, corresponding to a D-brane world-volume in type IIA superstring theory. We reproduce in this section \cite[Chap.\ 5]{FS}.

\subsection{Even case}

We call $X$ a compact smooth $n$-dimensional manifold and $Y$ a compact embedded $p$-dimensional submanifold. We choose a finite triangulation of $X$ which restricts to a triangulation of $Y$ \cite{Munkres}. We use the following notation:
\begin{itemize}
	\item we denote the triangulation of $X$ by $\Delta = \{\Delta^{m}_{i}\}$, where $m$ is the dimension of the simplex and $i$ enumerates the $m$-simplices;
	\item we denote by $X_{\Delta}^{p}$ the $p$-skeleton of $X$ with respect to $\Delta$.
\end{itemize}
The same notation is used for other triangulations or simplicial decompositions of $X$ and $Y$. In the following theorem we need the definition of ``dual cell decomposition'' with respect to a triangulation: we refer to \cite{GH} pp.\ 53-54.
\begin{Theorem}\label{TriangulationK} Let $X$ be an $n$-dimensional compact manifold and $Y \subset X$ a $p$-dimensional embedded compact submanifold. Let:
\begin{itemize}
	\item $\Delta = \{\Delta^{m}_{i}\}$ be a triangulation of $X$ which restricts to a triangulation $\Delta' = \{\Delta^{m}_{i'}\}$ of $Y$;
	\item $D = \{D^{n-m}_{i}\}$ be the dual decomposition of $X$ with respect to $\Delta$;
	\item $\tilde{D} \subset D$ be subset of $D$ made by the duals of the simplices in $\Delta'$.
\end{itemize}
Then, calling $\abs{\tilde{D}}$ the support of $\tilde{D}$:
\begin{itemize}
	\item the interior of $\abs{\tilde{D}}$ is a tubular neighborhood of $Y$ in $X$;
	\item the interior of $\abs{\tilde{D}}$ does not intersect $X_{D}^{n-p-1}$, i.e.:
	\[\abs{\tilde{D}} \cap X_{D}^{n-p-1} \subset \partial \abs{\tilde{D}}.
\]
\end{itemize}
\end{Theorem}
\textbf{Proof:} The $n$-simplices of $\tilde{D}$ are the duals of the vertices of $\Delta'$. Let $\tau = \{\tau^{m}_{j}\}$ be the first barycentric subdivision of $\Delta$ \cite{GH,Hatcher}. For each vertex $\Delta^{0}_{i'}$ in $Y$ (thought of as an element of $\Delta$), its dual is:
\begin{equation}\label{DTildeK}
	\tilde{D}^{n}_{i'} = \bigcup_{\Delta^{0}_{i'} \in \tau^{n}_{j}} \tau^{n}_{j}.
\end{equation}
Moreover, if $\tau' = \{{\tau'}^{m}_{j'}\}$ is the first barycentric subdivision of $\Delta'$ (of course $\tau' \subset \tau$) and $D' = \{{D'}^{m}_{i'}\}$ is the dual of $\Delta'$ in $Y$, then (reminding that $p$ is the dimension of $Y$):
\begin{equation}\label{DPrimoK}
	D'^{\,p}_{\;\,i'} = \bigcup_{\Delta^{0}_{i'} \in {\tau'}^{p}_{j'}} {\tau'}^{p}_{j'}
\end{equation}
and:
	\[\tilde{D}^{n}_{i'} \cap Y = D'^{\,p}_{\;\,i'}.
\]
Moreover, let us consider the $(n-p)$-simplices in $\tilde{D}$ contained in $\partial \tilde{D}^{n}_{i'}$ (for a fixed $i'$ in formula \eqref{DTildeK}), i.e.\ $X^{n-p}_{\tilde{D}} \cap \tilde{D}^{n}_{i'}$: they intersect $Y$ transversally in the barycenters of each $p$-simplex of $\Delta'$ containing $\Delta^{0}_{i'}$: we call such barycenters $\{b_{1}, \ldots, b_{k}\}$ and the intersecting $(n-p)$-cells $\{\tilde{D}^{n-p}_{l}\}_{l = 1, \ldots, k}$. Since (for a fixed $i'$) $\tilde{D}^{n}_{i'}$ retracts on $\Delta^{0}_{i'}$, we can consider a local chart $(U_{i'}, \varphi_{i'})$, with $U_{i'} \subset \mathbb{R}^{n}$ neighborhood of $0$, such that:
\begin{itemize}
	\item $\varphi_{i'}^{-1}(U_{i'})$ is a neighborhood of $\tilde{D}^{n}_{i'}$;
	\item $\varphi_{i'}(D'^{\,p}_{\;\,i'}) \subset U_{i'} \cap (\{0\} \times \mathbb{R}^{p})$, for $0 \in \mathbb{R}^{n-p}$ (v.\ eq.\ \eqref{DPrimoK});
	\item $\varphi_{i'}(\tilde{D}^{n-p}_{l}) \subset U_{i'} \cap \bigl(\mathbb{R}^{n-p} \times \pi_{p}(\varphi_{i'}(b_{l}))\bigr)$, for $\pi_{p}: \mathbb{R}^{n} \rightarrow \{0\} \times \mathbb{R}^{p}$ the projection.
\end{itemize}
We now consider the natural foliation of $U_{i'}$ given by the intersection with the hyperplanes $\mathbb{R}^{n-p} \times \{x\}$ and its image via $\varphi_{i'}^{-1}$: in this way, we obtain a foliation of $\tilde{D}^{n}_{i'}$ transversal to $Y$. If we do this for any $i'$, by construction the various foliations glue on the intersections, since such intersections are given by the $(n-p)$-cells $\{\tilde{D}^{n-p}_{l}\}_{l = 1, \ldots, k}$, and the interior gives a $C^{0}$-tubular neighborhood of $Y$.

Moreover, a $(n-p-r)$-cell of $\tilde{D}$, for $r > 0$, cannot intersect $Y$ since it is contained in the boundary of a $(n-p)$-cell, and such cells intersect $Y$, which is done by $p$-cells, only in their interior points $b_{j}$. Being the simplicial decomposition finite, it follows that the interior of $\abs{\tilde{D}}$ does not intersect $X_{D}^{n-p-1}$. \\
$\square$

\paragraph{}We now consider quintuples $(X, Y, \Delta, D, \tilde{D})$ satisfying the following condition:
\begin{itemize}
	\item[$(\#)$] $X$ is an $n$-dimensional compact manifold and $Y \subset X$ a $p$-dimensional embedded compact submanifold, such that $n - p$ is even and the normal bundle $\mathcal{N}(Y)$ is spin$^{c}$. Moreover, $\Delta$, $D$ and $\tilde{D}$ are defined as in theorem \ref{TriangulationK}.
\end{itemize}

\begin{Lemma}\label{TrivialityXnp1K} Let $(X,Y,\Delta,D,\tilde{D})$ be a quintuple satisfying $(\#)$, $U = \Int\abs{\tilde{D}}$ and $\alpha \in K(Y)$. Then:
\begin{itemize}
	\item there exists a neighborhood $V$ of $X \setminus U$ such that $i_{!}(\alpha) \vert_{V} = 0$;
	\item in particular, $i_{!}(\alpha) \,\vert_{X^{n-p-1}_{D}} = 0$.
\end{itemize}
\end{Lemma}
\textbf{Proof:} By equation \eqref{GysinMapK}:
	\[i_{!}(\alpha) = \psi^{*} \beta, \qquad \beta = (\varphi_{U}^{+})^{*} \circ T(\alpha) \in \tilde{K}(U^{+}).
\]
Let $\beta = [E] - [n]$, and let $V_{\infty} \subset U^{+}$ be a neighborhood of $\infty$ which trivializes $E$. Then $(\psi^{*}E) \,\big\vert_{\psi^{-1}(V_{\infty})}$ is trivial. Hence, for $V = \psi^{-1}(V_{\infty})$:
	\[(\psi^{*} \beta)\vert_{V} = [(\psi^{*}E)\vert_{V}] - [n] = [n] - [n] = 0.
\]
By theorem \ref{TriangulationK}, $X^{n-p-1}_{D}$ does not intersect the tubular neighborhood $\Int \abs{\tilde{D}}$ of $Y$, hence $X^{n-p-1}_{D} \subset \psi^{-1}(V_{\infty}) = V$, so that $(\psi^{*} \beta)\big\vert_{X^{n-p-1}_{D}} = 0$. $\square$

\subsubsection{Trivial bundle}

We start considering the case of a trivial bundle.

\begin{Theorem}\label{FirstTheoremK} Let $(X,Y,\Delta,D,\tilde{D})$ be a quintuple satisfying $(\#)$ and $\Phi^{n-p}_{D}: C^{n-p}$ $(X, \mathbb{Z}) \rightarrow K(X_{D}^{n-p},$ $X_{D}^{n-p-1})$ be the isomorphism stated in theorem \ref{KTheoryCohomology}. Let:
	\[\pi^{n-p,\,n-p-1}: X_{D}^{n-p} \longrightarrow X_{D}^{n-p} / X_{D}^{n-p-1}
\]
be the projection and $\PD_{\Delta}Y$ be the representative of $\PD_{X}[Y]$ (for $[Y]$ the homology class of $Y$) given by the sum of the cells dual to the $p$-cells of $\Delta$ covering $Y$. Then:
	\[i_{!}(Y \times \mathbb{C}) \vert_{X_{D}^{n-p}} = (\pi^{n-p,\,n-p-1})^{*}( \Phi_{D}^{n-p}(\PD_{\Delta}Y)).
\]
\end{Theorem}
\textbf{Proof:} We define:
	\[(U^{+})^{n-p}_{D} = \frac{\overline{X^{n-p}_{D} \, \vert_{U}}}{X^{n-p-1}_{D} \, \vert_{\partial U}}
\]
so that there is a natural immersion $(U^{+})^{n-p}_{D} \subset U^{+}$ defined sending the denominator to $\infty$ (the numerator is exactly $X^{n-p}_{\tilde{D}}$ of theorem \ref{TriangulationK}). We also define, considering the map $\psi$ of equation \eqref{GysinMapK}:
	\[\psi^{n-p} = \psi\big\vert_{X^{n-p}_{D}}: \, X^{n-p}_{D} \longrightarrow (U^{+})^{n-p}_{D}.
\]
The latter is well-defined since the $(n-p)$-simplices outside $U$ and all the $(n-p-1)$-simplices are sent to $\infty$ by $\psi$. Calling $I$ the set of indices of the $(n-p)$-simplices in $D$, calling $S^{k}$ the $k$-dimensional sphere and denoting by $\dot{\cup}$ the one-point union of topological spaces, there are the following canonical homeomorphisms:
	\[\begin{split}
	& \xi^{n-p}_{X}: \pi^{n-p}(X_{D}^{n-p}) \overset{\simeq}\longrightarrow \dot{\bigcup_{i \in I}} \; S^{n-p}_{i} \\
	& \xi^{n-p}_{U^{+}}: \psi^{n-p}(X_{D}^{n-p}) \overset{\simeq}\longrightarrow \dot{\bigcup_{j \in J}} \; S^{n-p}_{j}
\end{split}\]
where $\{S^{n-p}_{j}\}_{j \in J}$, with $J \subset I$, is the set of $(n-p)$-spheres corresponding to the $(n-p)$-simplices with interior contained in $U$, i.e.\ corresponding to $\pi^{n-p}\bigl(\overline{X^{n-p}_{D} \, \big\vert_{U}}\,\bigr)$. The homeomorphism $\xi^{n-p}_{U^{+}}$ is due to the fact that the boundary of the $(n-p)$-cells intersecting $U$ is contained in $\partial U$, hence it is sent to $\infty$ by $\psi^{n-p}$, while all the $(n-p)$-cells outside $U$ are sent to $\infty$: hence, the image of $\psi^{n-p}$ is homeomorphic to $\dot{\bigcup}_{j \in J} \; S^{n-p}_{j}$ sending $\infty$ to the attachment point. We define:
	\[\begin{split}
	\rho: \, \dot{\bigcup_{i \in I}} \; S^{n-p}_{i} \longrightarrow \dot{\bigcup_{j \in J}} \; S^{n-p}_{j}
\end{split}\]
as the natural projection, i.e.\ $\rho$ is the identity of $S^{n-p}_{j}$ for every $j \in J$ and sends all the spheres in $\{S^{n-p}_{i}\,\}_{i \in I \setminus J}$ to the attachment point. We have that:
	\[\xi^{n-p}_{U^{+}} \circ \psi^{n-p} = \rho \circ \xi^{n-p}_{X} \circ \pi^{n-p,\,n-p-1}
\]
hence:
\begin{equation}\label{PsiPiRhoK}
	(\psi^{n-p})^{*} \circ (\xi^{n-p}_{U^{+}})^{*} = (\pi^{n-p,\,n-p-1})^{*} \circ (\xi^{n-p}_{X})^{*} \circ \rho^{*}.
\end{equation}
We put $\mathcal{N} = \mathcal{N}(Y)$ and $\tilde{\lambda}_{\mathcal{N}} = (\varphi_{U}^{+})^{*} (\lambda_{\mathcal{N}})$, where $\lambda_{\mathcal{N}}$ is the Thom class of the normal bundle defined in equation \eqref{ThomClass}. By lemma \ref{UnitarityK} and equation \eqref{GysinMapK} we have $i_{!}(Y \times \mathbb{C}) = \psi^{*} \circ (\varphi_{U}^{+})^{*} (\lambda_{\mathcal{N}})$. Then:
	\[i_{!}(Y \times \mathbb{C}) \, \big\vert_{X_{D}^{n-p}} = \psi^{*} (\tilde{\lambda}_{\mathcal{N}}) \, \big\vert_{X_{D}^{n-p}} = (\psi^{n-p})^{*} \bigl( \tilde{\lambda}_{\mathcal{N}} \,\big\vert_{(U^{+})^{n-p}_{D}} \bigr)
\]
and
	\[(\xi^{n-p}_{X})^{*} \circ \rho^{*} \circ ((\xi^{n-p}_{U^{+}})^{-1})^{*} \bigl(\, \tilde{\lambda}_{\mathcal{N}} \,\big\vert_{(U^{+})^{n-p}_{D}} \,\bigr) = \Phi_{D}^{n-p}(\PD_{\Delta}Y)
\]
since:
\begin{itemize}
	\item $\PD_{\Delta}Y$ is the sum of the $(n-p)$-cells intersecting $U$;
	\item hence $((\xi^{n-p}_{X})^{-1})^{*} \circ \Phi_{D}^{n-p}(\PD_{\Delta}Y)$ gives a $(-1)^{\frac{n-p}{2}} (\eta - 1)^{\boxtimes \frac{n-p}{2}}$ factor to each sphere $S^{n-p}_{j}$ for $j \in J$ and $0$ otherwise;
	\item but this is exactly $\rho^{*} \circ ((\xi^{n-p}_{U^{+}})^{-1})^{*} \bigl( \tilde{\lambda}_{\mathcal{N}} \,\big\vert_{(U^{+})^{n-p}_{D}} \bigr)$ since by equation \eqref{RestrictionThom} we have, for $y \in Y$:
	\[(\lambda_{\mathcal{N}})\, \big\vert_{\mathcal{N}_{y}^{+}} = \lambda_{\mathbb{R}^{n-p}} \simeq (-1)^{\frac{n-p}{2}}(\eta - 1)^{\boxtimes \frac{n-p}{2}}
\]
and for the spheres outside $U$, that $\rho$ sends to $\infty$, we have that:
	\[\begin{split}
	\rho^{*} \bigl( \tilde{\lambda}_{\mathcal{N}} \,\big\vert_{(U^{+})^{n-p}_{D}} \bigr) \Big\vert_{\dot{\bigcup}_{i \in I \setminus J} \; S^{n-p}_{i}}
	&= \rho^{*} \bigl( \tilde{\lambda}_{\mathcal{N}} \,\big\vert_{\rho(\dot{\bigcup}_{i \in I \setminus J} \; S^{n-p}_{i})} \bigr)\\
	&= \rho^{*} \bigl( \tilde{\lambda}_{\mathcal{N}} \,\big\vert_{\{\infty\}} \bigr) = \rho^{*}(0) = 0.
\end{split}\]
\end{itemize}
Hence, from equation \eqref{PsiPiRhoK}:
	\[\begin{split}
	i_{!}(Y \times \mathbb{C}) \, \big\vert_{X_{D}^{n-p}} &= (\psi^{n-p})^{*} \bigl( \tilde{\lambda}_{\mathcal{N}} \,\big\vert_{(U^{+})^{n-p}_{D}} \bigr)\\
	&= (\pi^{n-p,\,n-p-1})^{*} \circ (\xi^{n-p}_{X})^{*} \circ \rho^{*} \circ ((\xi^{n-p}_{U^{+}})^{-1})^{*}\bigl( \tilde{\lambda}_{\mathcal{N}} \,\big\vert_{(U^{+})^{n-p}_{D}} \bigr)\\
	&= (\pi^{n-p,\,n-p-1})^{*} \Phi_{D}^{n-p}(\PD_{\Delta}Y).
\end{split}\]
$\square$

\paragraph{}The following theorem encodes the link between the Gysin map and the AHSS.

\begin{Theorem}\label{SecondTheoremK} Let $(X,Y,\Delta,D,\tilde{D})$ be a quintuple satisfying $(\#)$ and $\Phi^{n-p}_{D}: C^{n-p}$ $(X, \mathbb{Z}) \rightarrow K(X_{D}^{n-p}, X_{D}^{n-p-1})$ be the isomorphism stated in theorem \ref{KTheoryCohomology}. Let us suppose that $\PD_{\Delta}Y$ is contained in the kernel of all the boundaries $d^{n-p, \,0}_{r}$ for $r \geq 1$. Then it defines a class:
	\[\{\Phi^{n-p}_{D}(\PD_{\Delta}Y)\}_{E^{n-p, \,0}_{\infty}} \in E^{n-p,\,0}_{\infty} \simeq \frac{\, \Ker ( K(X) \longrightarrow K(X^{n-p-1}) ) \,} {\Ker ( K(X) \longrightarrow K(X^{n-p}) )}.
\]
The following equality holds:
	\[\{\Phi^{n-p}_{D}(\PD_{\Delta}Y)\}_{E^{n-p, \,0}_{\infty}} = [i_{!}(Y \times \mathbb{C})].
\]
\end{Theorem}
\textbf{Proof:} By equations \eqref{EpinftyAK} and \eqref{EpinftyBK} we have the following commutative diagram:
\begin{equation}\label{EpinftyK}
\xymatrix{
	E^{n-p, \,0}_{\infty} = \IIm \bigl( \tilde{K}(X/X_{D}^{n-p-1}) \ar[dr]_{(\pi^{n-p-1})^{*}} \ar[rr]^{\qquad\qquad (f^{n-p})^{*}} & & \tilde{K}(X_{D}^{n-p}) \bigr)\\
& \tilde{K}(X) \ar[ur]_{(i^{n-p})^{*}}
}
\end{equation}
and, given a representative $\alpha \in \IIm (\pi^{n-p-1})^{*} = \Ker ( \tilde{K}(X) \rightarrow \tilde{K}(X_{D}^{n-p-1}) )$, we have that $\{\alpha\}_{E^{n-p, \,0}_{\infty}} = (i^{n-p})^{*}(\alpha) = \alpha\vert_{X_{D}^{n-p}}$. Moreover:
\begin{itemize}
	\item the class $\{\Phi^{n-p}_{D}(\PD_{\Delta}Y)\}_{E^{n-p, \,0}_{\infty}}$, by formula \eqref{FromOneToInftyK}, corresponds to the element of $\tilde{K}(X_{D}^{n-p})$ defined by $(\pi^{n-p,\,n-p-1})^{*}(\Phi^{n-p}_{D}(\PD_{\Delta}Y))$, for $\pi^{n-p,\,n-p-1}: X^{n-p}_{D} \rightarrow X^{n-p}_{D} / X^{n-p-1}_{D}$;
	\item by lemma \ref{TrivialityXnp1K} we have $i_{!}(Y \times \mathbb{C}) \in \Ker(K(X) \rightarrow K(X_{D}^{n-p-1}))$, hence $[i_{!}(Y \times \mathbb{C})]$ is well-defined as an element of $E^{n-p, \,0}_{\infty}$ and, by exactness, $i_{!}(Y \times \mathbb{C}) \in \IIm (\pi^{n-p-1})^{*}$;
	\item by theorem \ref{FirstTheoremK} we have $(i^{n-p})^{*}(i_{!}(Y \times \mathbb{C})) = (\pi^{n-p,\,n-p-1})^{*}(\Phi_{D}^{n-p}(\PD(Y)))$;
	\item hence $\{\Phi^{n-p}_{D}(\PD_{\Delta}Y)\}_{E^{n-p, \,0}_{\infty}} = [i_{!}(Y \times \mathbb{C})]$.
\end{itemize}
$\square$

\paragraph{}Let us consider a trivial vector bundle of generic rank $Y \times \mathbb{C}^{r}$. We denote by $[r]$ its K-theory class on $Y$. By lemma \ref{UnitarityK} we have that $[r] \cdot \lambda_{\mathcal{N}} = \lambda_{\mathcal{N}}^{\oplus r}$, hence theorem \ref{FirstTheoremK} becomes:
	\[i_{!}(Y \times \mathbb{C}^{r}) \, \big\vert_{X_{D}^{n-p}} = (\pi^{n-p,\,n-p-1})^{*} \bigl( \Phi_{D}^{n-p}(\PD_{\Delta}(r \cdot Y)) \bigr)
\]
and theorem \ref{SecondTheorem} becomes:
	\[\{\Phi^{n-p}_{D}(\PD_{\Delta}(r \cdot Y))\}_{E^{n-p, \,0}_{\infty}} = [i_{!}(Y \times \mathbb{C}^{r})].
\]

\subsubsection{Generic bundle}

If we consider a generic bundle $E$ over $Y$ of rank $r$, we can prove that $i_{!}(E)$ and $i_{!}(Y \times \mathbb{C}^{r})$ have the same restriction to $X^{n-p}_{D}$: in fact, the Thom isomorphism gives $T(E) = E \cdot \lambda_{\mathcal{N}}$ and, if we restrict $E \cdot \lambda_{\mathcal{N}}$ to a finite family of fibers, which are transversal to $Y$, the contribution of $E$ becomes trivial, so it has the same effect of the trivial bundle $Y \times \mathbb{C}^{r}$. We now give a precise proof of this statement.

\begin{Lemma}\label{LineBundleXnpK} Let $(X,Y,\Delta,D,\tilde{D})$ be a quintuple satisfying $(\#)$ and $\pi: E \rightarrow Y$ a vector bundle of rank $r$. Then:
	\[i_{!}(E) \, \big\vert_{X_{D}^{n-p}} = i_{!}(Y \times \mathbb{C}^{r}) \, \big\vert_{X_{D}^{n-p}}.
\]
\end{Lemma}
\textbf{Proof:} referring to the notations in the proof of lemma \ref{UnitarityK}, we have that:
	\[E \cdot \lambda_{\mathcal{N}} = i^{*} (\tilde{\pi}^{*})^{-1}(E \boxtimes \lambda_{\mathcal{N}}) = i^{*} (\tilde{\pi}^{*})^{-1} ( \pi_{1}^{*}E \otimes \pi_{2}^{*}\lambda_{\mathcal{N}} ).
\]
Since $X^{n-p}_{D}$ intersects the tubular neighborhood in a finite number of cells, corresponding under $\varphi_{U}^{+}$ to a finite number of fibers of $\mathcal{N}$, it is sufficient to prove that, for any $y \in Y$, $(E \cdot \lambda_{\mathcal{N}}) \, \big\vert_{\mathcal{N}_{y}^{+}} = \lambda_{\mathcal{N}}^{\oplus r} \, \big\vert_{\mathcal{N}_{y}^{+}}$. First of all:
\begin{itemize}
	\item $i(\mathcal{N}_{y}^{+}) = (\{y\} \times \mathcal{N}_{y})^{+} \subset (\{y\} \times \mathcal{N})^{+}$;
	\item $E \cdot \lambda_{\mathcal{N}} \, \big\vert_{\mathcal{N}_{y}^{+}} = (i \vert_{\mathcal{N}_{y}^{+}})^{*} \bigl\{ \bigl[(\tilde{\pi}^{*})^{-1} (\pi_{1}^{*}E \otimes \pi_{2}^{*}\lambda_{\mathcal{N}})\bigr] \,\big\vert_{i(\mathcal{N}_{y}^{+})} \bigr\}$.
\end{itemize}
To obtain the bundle $\bigl[(\tilde{\pi}^{*})^{-1} (\pi_{1}^{*}E \otimes \pi_{2}^{*}\lambda_{\mathcal{N}} )\bigr] \,\big\vert_{i(\mathcal{N}_{y}^{+})}$, we can restrict $\tilde{\pi}$ to:
	\[A = \tilde{\pi}^{-1}[i(\mathcal{N}_{y}^{+})] = \tilde{\pi}^{-1}\bigl[(\{y\} \times \mathcal{N}_{y})^{+}\bigr] = \bigl(\{y\} \times \mathcal{N}_{y}^{+}\bigr) \,\cup\, \bigl(Y \times \{\infty\} \bigr) \,\cup\, \bigl(\{\infty\} \times \mathcal{N}^{+}\bigr)
\]
and consider $(\tilde{\pi} \, {\vert_{A}}^{*})^{-1} \bigl[( \pi_{1}^{*}E \otimes \pi_{2}^{*}\lambda_{\mathcal{N}} ) \,\big\vert_{A} \bigr]$. Moreover:
\begin{itemize}
	\item $(\pi_{1}^{*}E \otimes \pi_{2}^{*}\lambda_{\mathcal{N}}) \,\big\vert_{\{y\} \times \mathcal{N}_{y}^{+}} = (\mathbb{C}^{r} \otimes \pi_{2}^{*}\lambda_{\mathcal{N}} ) \,\big\vert_{\{y\} \times \mathcal{N}_{y}^{+}} \simeq \lambda_{\mathcal{N}}^{\oplus r} \,\big\vert_{\mathcal{N}_{y}^{+}}$;
	\item $(\pi_{1}^{*}E \otimes \pi_{2}^{*}\lambda_{\mathcal{N}}) \,\big\vert_{Y \times \{\infty\}} = (\pi_{1}^{*}E \otimes 0) \,\big\vert_{Y \times \{\infty\}} = 0$;
	\item $(\pi_{1}^{*}E \otimes \pi_{2}^{*}\lambda_{\mathcal{N}}) \,\big\vert_{\{\infty\} \times \mathcal{N}^{+}} = (0 \otimes \pi_{2}^{*}\lambda_{\mathcal{N}}) \,\big\vert_{\{\infty\} \times \mathcal{N}^{+}} = 0$.
\end{itemize}
Hence, since the three components of $A$ intersect each other at most at one point, by lemma \ref{OnePointUnion} we obtain:
	\[(\pi_{1}^{*}E \otimes \pi_{2}^{*}\lambda_{\mathcal{N}}) \,\big\vert_{A} = \bigl( \pi_{1}^{*}(Y \times \mathbb{C}^{r}) \otimes \pi_{2}^{*}\lambda_{\mathcal{N}} \bigr) \,\big\vert_{A}.
\]
$\square$

\paragraph{Remark:} In the statement of theorem \ref{SecondTheorem} (and of its generalization to any vector bundle) it was necessary to explicitly introduce a triangulation $\Delta$ on $X$, since the first step of the spectral sequence consists of simplicial cochains, which by definition depend on the simplicial structure chosen. Anyway, the groups $E^{p, \,0}_{r}$ for $r \geq 2$ and the filtration $\Ker( K(X) \rightarrow K(X^{n-p}))$ of $K(X)$ do not depend on the particular simplicial structure chosen \cite{AH}, thus, if we start from the cohomology class $\PD_{X}[Y]$ at the second step of the spectral sequence (which is the D-brane charge density with respect to the cohomological classification) we can drop the dependence on $\Delta$, $D$ and $\tilde{D}$. Therefore the choice of the triangulation has no effect on the physical classification of D-brane charges.

\subsection{Odd case}

We now consider the case of $n - p$ odd (for $n$ the dimension of $X$ and $p$ the dimension of $Y$), corresponding by a physical point of view to type IIA superstring theory. In this case the Gysin map takes value in $K^{1}(X)$, which is isomorphic to $K(\hat{S}^{1}X)$, for $\hat{S}^{1}X$ the unreduced suspension of $X$ defined as:
	\[\hat{S}^{1}X = (X \times [-1,1]) / (X \times \{-1\}, X \times \{1\})
\]
i.e.\ as a double cone built on $X$. We thus consider the natural embedding $i^{1}: Y \rightarrow \hat{S}^{1}X$ and the corresponding Gysin map:
	\[(i^{1})_{!}: K(Y) \rightarrow K(\hat{S}^{1}X) \simeq K^{1}(X).
\]
Let $U$ be a tubular neighborhood of $Y$ in $X$, and let $U^{1} \subset \hat{S}^{1}X$ be the tubular neighborhood of $Y$ in $\hat{S}^{1}X$ defined removing the vertices of the double cone to $\hat{S}^{1}U$. We have that $\overline{\hat{S}^{1}(X^{n-p}_{D}\vert_{U})} \subset \overline{U^{1}}$ and $\hat{S}^{1}(X^{n-p-1}_{D}\vert_{\partial U}) \subset \partial U^{1}$, where $\partial U^{1}$ contains also the vertices of the double cone. In this way we can reformulate the previous results in the odd case, considering $\hat{S}^{1}(X^{n-p}_{D})$ and $\hat{S}^{1}(X^{n-p-1}_{D})$ rather than $X^{n-p}_{D}$ and $X^{n-p-1}_{D}$.

\paragraph{}We consider quintuples $(X, Y, \Delta, D, \tilde{D})$ satisfying the following condition:
\begin{itemize}
	\item[$(\#^{1})$] $X$ is an $n$-dimensional compact manifold and $Y \subset X$ a $p$-dimensional embedded compact submanifold, such that $n - p$ is \emph{odd} and $\mathcal{N}(Y)$ is spin$^{c}$. Moreover, $\Delta$, $D$ and $\tilde{D}$ are defined as in theorem \ref{TriangulationK}.
\end{itemize}

\paragraph{}We now reformulate the same theorems stated for the even case, which can be proved in the same way. We remark that $\mathcal{N}_{\hat{S}^{1}X}Y$ is spin$^{c}$ if and only $\mathcal{N}_{X}Y$ is, since $\mathcal{N}_{\hat{S}^{1}X}Y = \mathcal{N}_{X}Y \oplus 1$ so that, by the axioms of characteristic classes \cite{MS}, $W_{3}$ is the same.

\begin{Lemma}\label{TrivialityXnp1Odd} Let $(X, Y, \Delta, D, \tilde{D})$ be a quintuple satisfying $(\#^{1})$ and $\alpha \in K(Y)$. Then:
\begin{itemize}
	\item there exists a neighborhood $V$ of $\hat{S}^{1}X \setminus U^{1}$ such that $i^{1}_{!}(\alpha) \, \big\vert_{V} = 0$;
	\item in particular, $i^{1}_{!}(\alpha) \, \big\vert_{\hat{S}^{1}(X^{n-p-1}_{D})} = 0$.
\end{itemize}
$\square$
\end{Lemma}

\begin{Theorem}\label{FirstTheoremOdd} Let $(X, Y, \Delta, D, \tilde{D})$ be a quintuple satisfying $(\#^{1})$ and $\Phi^{n-p}_{D}: C^{n-p}$ $(X, \mathbb{Z}) \rightarrow K^{1}(X_{D}^{n-p}, X_{D}^{n-p-1}) \simeq K(\hat{S}^{1}(X_{D}^{n-p}), \hat{S}^{1}(X_{D}^{n-p-1}))$ be the isomorphism stated in theorem \ref{KTheoryCohomology}. Let:
	\[\pi^{n-p,\,n-p-1}: \hat{S}^{1}(X_{D}^{n-p}) \longrightarrow \hat{S}^{1}(X_{D}^{n-p}) / \hat{S}^{1}(X_{D}^{n-p-1})
\]
be the projection and $\PD_{\Delta}Y$ be the representative of $\PD_{X}[Y]$ (for $[Y]$ the homology class of $Y$) given by the sum of the cells dual to the $p$-cells of $\Delta$ covering $Y$. Then:
	\[i^{1}_{!}\,(Y \times \mathbb{C}) \, \big\vert_{\hat{S}^{1}(X_{D}^{n-p})} = (\pi^{n-p,\,n-p-1})^{*}( \Phi_{D}^{n-p}(\PD_{\Delta}Y)).
\]
$\square$
\end{Theorem}

\begin{Theorem}\label{SecondTheoremOdd} Let $(X,Y,\Delta,D,\tilde{D})$ be a quintuple satisfying $(\#^{1})$ and $\Phi^{n-p}_{D}: C^{n-p}$ $(X, \mathbb{Z}) \rightarrow K^{1}(X_{D}^{n-p}, X_{D}^{n-p-1})$ be the isomorphism stated in theorem \ref{KTheoryCohomology}. Let us suppose that $\PD_{\Delta}Y$ is contained in the kernel of all the boundaries $d^{n-p, \,0}_{r}$ for $r \geq 1$. Then it defines a class:
	\[\{\Phi^{n-p}_{D}(\PD_{\Delta}Y)\}_{E^{n-p, \,0}_{\infty}} \in E^{n-p,\,0}_{\infty} \simeq \frac{\, \Ker ( K^{1}(X) \longrightarrow K^{1}(X^{n-p-1}) ) \,} {\Ker ( K^{1}(X) \longrightarrow K^{1}(X^{n-p}) )}.
\]
The following equality holds:
	\[\{\Phi^{n-p}_{D}(\PD_{\Delta}Y)\}_{E^{n-p, \,0}_{\infty}} = [(i^{1})_{!}(Y \times \mathbb{C})].
\]
$\square$
\end{Theorem}

\subsection{The rational case}

\subsubsection{Even case}

We now analyze the case of rational coefficients. We define:
	\[K_{\mathbb{Q}}(X) := K(X) \otimes_{\mathbb{Z}} \mathbb{Q}.
\]
We can thus classify the D-brane charge density at rational level as $i_{!}(E) \otimes_{\mathbb{Z}} \mathbb{Q}$. The Chern character provides an isomorphism $\ch: K_{\mathbb{Q}}(X) \rightarrow H^{\ev}(X, \mathbb{Q})$. Since the square root of $\hat{A}(TX)$ is a polyform starting with 1, it also defines an isomorphism, so that the composition:
	\[\begin{split}
	\widehat{\ch}: \,&K_{\mathbb{Q}}(X) \longrightarrow H^{\ev}(X, \mathbb{Q})\\
	&\widehat{\ch}(\alpha) = \ch(\alpha) \wedge \sqrt{\hat{A}(TX)}
\end{split}\]
remains an isomorphism. Thus, the classifications with rational K-theory and rational cohomology are completely equivalent.

We can also define the rational Atiyah-Hirzebruch spectral sequence $Q^{2k, \,0}_{r}(X) := E^{2k, \,0}_{r}(X)$ $\otimes_{\mathbb{Z}} \mathbb{Q}$. Such a sequence collapses at the second step \cite{AH}, i.e.\ at the level of cohomology: thus $Q^{2k, \,0}_{\infty}(X) \simeq Q^{2k, \,0}_{2}(X)$. An explicit isomorphism is given by the appropriate component of the Chern character:
	\[\ch_{\frac{n-p}{2}}: \frac{\, \Ker \bigl( K_{\mathbb{Q}}(X) \longrightarrow K_{\mathbb{Q}}(X^{n-p-1}) \bigr) \,} {\Ker \bigl( K_{\mathbb{Q}}(X) \longrightarrow K_{\mathbb{Q}}(X^{n-p}) \bigr)} \longrightarrow H^{n-p}(X, \mathbb{Q}).
\]
This map is well-defined since, for a bundle which is trivial on the $(n-p)$-skeleton, the Chern characters of degree less or equal to $\frac{n-p}{2}$ are zero \cite{AH} (in particular $\ch_{\frac{n-p}{2}} = \widehat{\ch}_{\frac{n-p}{2}}$ for a bundle which is trivial on the $(n-p-1)$-skeleton). Moreover, since $Q^{2k,0}_{\infty}$ has no torsion:
	\[K_{\mathbb{Q}}(X) = \bigoplus_{2k}Q^{2k,0}_{\infty}
\]
and an isomorphism can be obtained splitting $\alpha \in K_{\mathbb{Q}}(X)$ as $\alpha = \sum_{2k}\alpha_{2k}$ where $\ch(\alpha_{2k}) = \ch_{k}(\alpha)$.

\subsubsection{Odd case}

In this case, we have the isomorphism $\ch: K^{1}_{\mathbb{Q}}(X) \rightarrow H^{\odd}(X, \mathbb{Q})$. Moreover, $H^{\odd}(X, \mathbb{Q}) \simeq H^{\ev}(\hat{S}^{1}X, \mathbb{Q})$. Hence we have the correspondence among:
\begin{itemize}
	\item $i^{1}_{!}(E) \in K^{1}_{\mathbb{Q}}(X)$;
	\item $\widehat{\ch}(i^{1}_{!}E) \in H^{\ev}(\hat{S}^{1}X, \mathbb{Q}) \simeq H^{\odd}(X, \mathbb{Q})$;
	\item $\oplus_{2k} \, \bigl[(i^{1}_{k})_{!}(Y_{k} \times \mathbb{C}^{q_{k}})\bigr]_{Q^{2k+1,0}_{\infty}}\,$.
\end{itemize}

\section{K-homology and its non-compact versions}

\subsection{K-homology with compact support}

As for any cohomology theory we can define a corresponding homology theory for K-theory, called K-homology. We follow \cite{Jakob} and \cite{RZ}. We work on the category $HCW_{f}$ of spaces having the same homotopy type of a finite CW-complex. K-homology can be geometrically described as follows: for a couple of spaces $(X,A)$ in $HCW_{f}$ we define the group of K-homology \emph{$n$-pre-cycles} as the free abelian group $K_{PC,n}(X,A)$ generated by triples $(M, \alpha, f)$ where:
\begin{itemize}
	\item $M$ is a smooth \emph{compact connected spin$^{c}$} manifold of dimension $n+q$ in general with boundary;
	\item $\alpha \in K^{q}(M)$;
	\item $f: M \rightarrow X$ is a continuous map such that $f(\partial M) \subset A$.
\end{itemize}
We define the group of cycles $K_{C,n}(X,A)$ as the quotient of $K_{PC,n}(X,A)$ by the subgroup generated by:
\begin{itemize}
	\item elements of the form $(M, \alpha + \beta, f) - (M, \alpha, f) - (M, \beta, f)$, so that we impose additivity with respect to the K-theory class in the middle;
	\item elements of the form $(M, \varphi_{!}\alpha, f) - (N, \alpha, f \circ \pi)$ where $f: N \rightarrow M$ is a smooth map and $\varphi_{!}$ is the Gysin map.
\end{itemize}
Thus a generic $n$-cycle is an equivalence class $[(M, \alpha, f)]$. We define the subgroup of \emph{$n$-boundaries} $K_{B,n}(X,A)$ as the subgroup of $K_{C,n}(X,A)$ generated by the elements $[(M, \alpha, f)]$ such that there exists a precycle $(W, \beta, g) \in K_{PC,n}(X,X)$ such that $M = \partial W$, $\alpha = \beta\vert_{M}$ and $f = g\vert_{M}$. We then define the K-homology $n$-group:
	\[K_{n}(X, A) := K_{C,n}(X,A) \,/\, K_{B,n}(X,A).
\]
These are the standard K-homology groups, which have compact support as for singular homology.

\paragraph{Remark:} we cannot define K-homology chains as for singular homology, since the K-theory class $\alpha$ in the triple can be non-trivial only if $M$ is in general a non-trivial manifold, not necessarily a simplex which is contractible. Thus we define cycles and boundaries but we do not build a graded complex of chains whose homology is isomorphic K-homology.

\subsection{Borel-Moore K-homology and variants}

We now define K-homology with generic support, which we call Borel-Moore K-homology by analogy with singular homology. The idea of course is to define it as the usual one without assuming that $M$ is compact in the definition of precycles. However, for generic backgrounds, we need an hypothesis on the manifolds involved to avoid irregular behaviors: we assume that the space-time manifold $S$ has a well-defined ``infinity'' on which we classify Ramond-Ramond charges \cite{MW}, i.e.\ we suppose that $S$ can be embedded as the interior of a manifold with boundary $\overline{S}$, so that infinity is $\partial \overline{S}$. We assume for simplicity the manifold to be smooth, although what we are going to say holds also if there are some singularities. Under this assumption $S$ and $\overline{S}$ are homotopic: in fact, by the collar neighborhood theorem \cite{Hirsh} there is a collar neighborhood of $\partial \overline{S}$ in $\overline{S}$, which is by definition a neighborhood $U$ diffeomorphic to $\partial \overline{S} \times [0,1)$. Now we can retract both $S$ and $\overline{S}$ to the same compact submanifold obtained retracting $U$ to the image of $\partial \overline{S} \times [\frac{1}{2},1)$ under the diffeomorphism, thus $S$ and $\overline{S}$ are homotopic. This will have some important consequences.

We can show that this assumption does not hold for any manifold. In fact, if it holds, we have shown that $S$ is homotopic to $\overline{S}$, thus, being $\overline{S}$ compact, it is homotopic to a finite CW-complex. There are manifold not homotopic to a finite CW-complex: one counterexample is given by a surface with infinite genus. In this case $H_{1}(S, \mathbb{Z}) = \mathbb{Z}^{\aleph_{0}}$, where $\aleph_{0}$ is the cardinality of countable infinite sets as $\mathbb{N}$ or $\mathbb{Z}$, in particular $H_{1}(S, \mathbb{Z})$ is not finitely generated, hence $S$ cannot be homotopic to a finite CW-complex.

If there is an embedding of $S$ in $\overline{S}$ such that $S = \overline{S} \setminus \partial \overline{S}$ it is unique, as we prove in appendix \ref{AppCompactifications}. We call \emph{manifolds with collar at infinity} the manifolds satisfying this assumption (they contain as particular case the compact ones, for which $\overline{S} = S$). We call \emph{couple of manifolds with collar at infinity} a couple $(X,A)$ such that both $X$ and $A$ are manifolds with infinity and the closure of $A$ is $\overline{X}$ is diffeomorphic to $\overline{A}$. We can now define:
	\[K^{BM}_{n}(X) := K_{n}(\overline{X}, \overline{X} \setminus X).
\]
The hypothesis can be relaxed requiring that $X$ is homotopic to a finite CW-complex and admits any compactificaton $\overline{X}$ which is also homotopic to a CW-complex. The result is independent on the compactification chosen. Similarly for a couple:
	\[K^{BM}_{n}(X, A) := K_{n}(\overline{X}, (\overline{X} \setminus X) \cup A).
\]
As for singular homology, if $X = Y \times Z$ we can define homology with compact support along $Y$:
	\[K^{BM}_{n}(Y \times Z; Y) := K_{n}(Y \times \overline{Z}, (Y \times \overline{Z}) \setminus (Y \times Z)).
\]

\section{K-theory and compactness}

\subsection{Definition of K-theory}

Let us consider a finite CW-complex $X$ \cite{Hatcher}; in particular $X$ is compact. The K-theory group of $X$ is the Grothendieck group associated to the semigroup of complex vector bundles on $X$. For $\{*\}$ a space with one point, we consider the unique map $p: X \rightarrow \{*\}$ and we define $\tilde{K}(X) := \Coker(i_{*}: K(\{*\}) \rightarrow K(X))$; if $X$ is connected $\tilde{K}(X)$ is made by the K-theory classes $[E] - [F]$ such that $E$ and $F$ have the same rank. Moreover, given a finite CW-pair $(X,A)$, i.e.\ a pair of finite CW-complexes such that $A$ is a subcomplex of $X$, we define the \emph{relative K-theory group} as $K(X,A) := \tilde{K}(X/A)$. In this way we define a generalized cohomology theory \cite{ES, Whitehead} on the category of finite CW-pairs.

In order to obtain a cohomology theory from these definitions (thus, in order to have the tools we need in string theory as Gysin map and Atiyah-Hirzebruch spectral sequence) it is important that the spaces involved are finite CW-complexes and not generic spaces. In fact, for example, if we consider the exact sequence in cohomology $K(X, A) \rightarrow K(X) \rightarrow K(A)$ we see that all the K-theory classes on $X$ which are trivial when restricted to $A$ are pull-back of a $\tilde{K}$-class on $X/A$: the proof of these \cite{Atiyah} requires that every bundle $E$ is a direct summand of a trivial bundle, which is not true for generic spaces. Moreover, defining the relative K-theory group of a pair $(X,A)$ as $\tilde{K}(X/A)$, we need to assume that $A$ is a closed and sufficiently regular subset of $X$, since, for example, if we consider the pair $(\mathbb{R}^{n}, \mathbb{R}^{n} \setminus \{0\})$ we have that the quotient is the same non-Hausdorff space with two points for every $n$, while the correct definition of K-theory should be homotopy-invariant, thus it should coincide with $K(\mathbb{R}^{n}, \mathbb{R}^{n} \setminus B^{n}) = \tilde{K}(S^{n})$. We thus should consider the extension as defined in \ref{Extension}.

We now consider manifolds with collar at infinity. It is easy to show that under these hypotheses there are canonical isomorphisms $K(S) \simeq \mathcal{K}(S) \simeq K(\overline{S}) \simeq \mathcal{K}(\overline{S})$. In fact, $K(S) \simeq K(\overline{S})$ since $S$ and $\overline{S}$ are paracompact and this is enough to guarantee that the semigroups of vector bundles are isomorphic \cite{Husemoller}, thus also the associated Grothendieck groups are. Moreover $K(\overline{S}) \simeq \mathcal{K}(\overline{S})$ because $\overline{S}$ is a finite CW-complex\footnote{Any compact smooth manifold, with or without boundary, admits a CW-complex structure, as a consequence of Morse theory \cite{Milnor2}.} and $\mathcal{K}(\overline{S}) \simeq \mathcal{K}(S)$ since the $\mathcal{K}$-groups are homotopy-invariant. Thus, we can deal with the usual K-theory group $K(S)$ since it gives a well-behaved cohomology theory for the manifolds we are considering.

\paragraph{Remarks:} $ $
\begin{itemize}
	\item Since $K$-theory is a cohomology (not homology) theory the K-theory groups with any support must be homotopy invariant, while the groups with compact support are not; it is the opposite for homology.
	\item The question when $K(X) \simeq \mathcal{K}(X)$ is not generalizable to any cohomology theory: in fact, if we have a cohomology theory $h^{*}$ defined for finite CW-complexes, then the only way to extend it is to use the generalized definition, there is no meaning for the $h$-groups of other spaces; instead, for K-theory, we can define K-theory groups independently on the fact that they define or not a cohomology theory, thus we can compare them with the extensions of the ones defined only on CW-complexes.
\end{itemize}

\subsection{K-theory with compact support and variants}

Also for K-theory there is the analogue of the different versions of Borel-Moore cohomology considered in \cite{FR2}. In particular, we can consider K-theory with compact support $K_{\cpt}(S)$ made by classes $\alpha \in K(S)$ such that there exists a compact set $K \subset S$ such that $\alpha\vert_{S \setminus K} = 0$, and we have that $K_{\cpt}(S) \simeq \tilde{K}(S^{+})$ if $S^{+}$ is homotopic to a finite CW-complex. Moreover, for $S = Y \times Z$ we can consider $K_{\cpt(Y)}(S)$ made by classes $\alpha \in K(S)$ such that for any slice $Y \times {z}$ with $z \in \mathbb{Z}$ the restriction $\alpha\vert_{W \times X}$ has compact support.

We described K-theory group with compact or partially compact support intrinsically, without referring to the compactification $\overline{S}$. Actually, we have canonical isomorphisms $K_{\cpt}(S) \simeq K(\overline{S}, \overline{S} \setminus \partial \overline{S})$ and $K_{\cpt(Y)}(Y \times Z) \simeq K(\overline{Y} \times Z, (\overline{Y} \times Z) \setminus (Y \times Z))$.

\subsection{K-homology and Gysin map}\label{KHomologyGysinMap}

Let us consider K-homology cycle $[(M, \alpha, i)] \in K_{0}(S)$ such that $M$ is a compact submanifold of $S$ and $i: M \hookrightarrow S$ the embedding. Let us suppose for the moment that also $S$ is compact. We can consider the Gysin map $i_{!}: K(M) \rightarrow K(S)$.\footnote{Actually it is not necessary to consider embeddings, it is enough to suppose that the map $f: M \rightarrow S$ is proper, i.e.\ that the counter-image of a compact subset is compact. However, we will need to deal only with the embeddings of the D-brane world-volumes in space-time.} It turns out that if we consider $i_{!}(\alpha)$, where $\alpha$ is the middle term of the triple, then we obtain exactly the Poincar\'e dual of $[(M, \alpha, i)]$ in $S$. In fact, the idea of K-homology, as explained in \cite{Jakob}, is the following: if we consider singular homology, it is not true that any $n$-class $A$ can be represented by a smooth manifold, in the sense that there exists a smooth orientable $n$-manifold $M$ and a continuous map $f: M \rightarrow S$ such that $A = f_{*}([M])$ for $[M]$ the fundamental class of $M$. However, there always exists a triple $(M, \alpha, f)$ with $M$ an orientable $(n+q)$-manifold and $\alpha$ a $q$-class in $M$ such that $A = f_{*}(\alpha \cap [M])$. If we consider the equivalence relations analogous to the one we recalled defining K-homology, we obtain that the map $[(M, \alpha, f)] \rightarrow f_{*}(\alpha \cap [M])$ is an isomorphism between equivalence classes of such triples and the $n$-homology group of $S$. But for an orientable manifold $M$ by definition $\PD_{M}(\alpha) := \alpha \cap [M]$, thus the class corresponding to $[(M, \alpha, f)]$ is exactly $f_{*}\PD_{M}(\alpha)$. Since Gysin map commutes with Poincar\'e duality, in the sense that $\PD_{S}f_{!}(\alpha) = f_{*}\PD_{M}(\alpha)$, we obtain exactly that $f_{!}(\alpha) = \PD_{S}f_{*}\PD_{M}(\alpha) \simeq \PD_{S}[(M, \alpha, f)]$. The situation is completely analogous for any cohomology theory, so also for K-theory, provided that we give the corresponding definition of Poicar\'e dual and that we replace the orientability of $M$ with the orientability with respect to the cohomology theory considered, which for K-theory is equivalent for $M$ to be spin$^{c}$.

We now consider $S$ not necessarily compact. In this case, as for singular homology, the Poincar\'e dual of a K-homology class is a compactly supported K-theory class, while the Poincar\'e dual of a Borel-Moore K-homology class is an ordinary K-theory class. Similarly, for the variants with partially compact support Poincar\'e duality respects the directions in which we assume compactness.


\part{Line bundles and gerbes}

\chapter{Topological preliminaries}

\section{Triangulations}\label{TriangCovers}

Let us fix a topological manifold $X$ equipped with a good cover $\mathfrak{U} = \{U_{i}\}_{i \in I}$. We now construct, starting from $\mathfrak{U}$, a natural open cover for the space of maps from a curve or a surface to $X$.

\subsection{Loop space}

\begin{Def} Given a topological space $X$, the \emph{loop space} $LX$ is the set of continuous maps:
	\[\gamma: S^{1} \rightarrow X
\]
equipped with the compact-open topology.
\end{Def}
We now describe a natural open cover for the loop space. In particular:
\begin{itemize}
	\item let us fix a triangulation $\tau$ of $S^{1}$, i.e.\ a set of vertices $\sigma^{0}_{1}, \ldots, \sigma^{0}_{l} \in S^{1}$ and of edges $\sigma^{1}_{1}, \ldots, \sigma^{1}_{l} \subset S^{1}$ such that $\partial \sigma^{1}_{i} = \sigma^{0}_{i+1} - \sigma^{0}_{i}$ for $1 \leq i < l$ and $\partial \sigma^{1}_{l} = \sigma^{0}_{1} - \sigma^{0}_{l}$;
	\item we consider the following set of indices:
	\[J = \left\{(\tau, \varphi): \quad \begin{array}{l}
	    \bullet \textnormal{ $\tau = \{\sigma^{0}_{1}, \ldots, \sigma^{0}_{l(\tau)}; \sigma^{1}_{1}, \ldots, \sigma^{1}_{l(\tau)}\}$ is a triangulation of $S^{1}$} \\
	    \bullet \textnormal{ $\varphi: \{1, \ldots, l(\tau)\} \longrightarrow I$ is a function}
	\end{array} \right\}.
\]
\end{itemize}
We obtain a covering $\mathfrak{V} = \{V_{(\tau,\sigma)}\}_{(\tau,\sigma) \in J}$ of $LX$ by:
	\[V_{(\tau, \varphi)} = \{\gamma \in LX: \; \gamma(\sigma^{1}_{i}) \subset U_{\varphi(i)} \}.
\]
One can prove that these sets are open in the compact-open topology and that they cover $LX$.

\begin{Def} Given a topological space $X$, the \emph{space of open curves} $CX$ is the set of continuous maps:
	\[\gamma: [\,0,1] \rightarrow X
\]
equipped with the compact-open topology.
\end{Def}
We now describe an analogous open covering for the space of open curves. In particular:
\begin{itemize}
	\item let us fix a triangulation $\tau$ of $[\,0,1]$, i.e.\ a set of vertices $\sigma^{0}_{1}, \ldots, \sigma^{0}_{l}, \sigma^{0}_{l+1} \in [\,0,1]$ and of edges $\sigma^{1}_{1}, \ldots, \sigma^{1}_{l} \subset [\,0,1]$ such that:
\begin{itemize}
	\item $\partial \sigma^{1}_{i} = \sigma^{0}_{i+1} - \sigma^{0}_{i}$ for $1 \leq i \leq l$;
	\item $\sigma^{0}_{1} = 0$ and $\sigma^{0}_{l+1} = 1$; these are called \emph{boundary vertices};
\end{itemize}
	\item we consider the following set of indices:
	{\footnotesize
	\[J = \left\{(\tau, \varphi): \; \begin{array}{l}
	    \bullet \textnormal{ $\tau = \{\sigma^{0}_{1}, \ldots, \sigma^{0}_{l(\tau)}, \sigma^{0}_{l(\tau) + 1}; \sigma^{1}_{1}, \ldots, \sigma^{1}_{l(\tau)}\}$ is a triangulation of $[\,0,1]$} \\
	    \bullet \textnormal{ $\varphi: \{1, \ldots, l(\tau)\} \longrightarrow I$ is a function}
	\end{array} \right\}.
\] }
\end{itemize}
We obtain a covering $\{V_{(\tau,\sigma)}\}_{(\tau,\sigma) \in J}$ of $CX$ by:
	\[V_{(\tau, \varphi)} = \{\gamma \in CX: \; \gamma(\sigma^{1}_{i}) \subset U_{\varphi(i)} \}.
\]
One can prove that these sets are open in the compact-open topology and that they cover $CX$. Notice that in these triangulations the number of vertices is one more with respect to the number of edges, since now $[\,0,1]$ is not closed, hence $\sigma^{0}_{l+1} \neq \sigma^{0}_{1}$.

\subsection{Two-dimensional case}

While for curves we had only one type of closed curves, i.e.\ $S^{1}$, and one type of open curves, i.e.\ $[\,0,1]$, for surfaces the situation is different, therefore we consider different kind of spaces depending on the surface we start from. We assume that our surfaces are compact.

\begin{Def} Given a topological space $X$ and a closed compact surface $\Sigma$, the \emph{space of maps from $\Sigma$ to $X$}, called $\Sigma X$, is the set of continuous maps:
	\[\Gamma: \Sigma \longrightarrow X
\]
equipped with the compact-open topology.
\end{Def}
We now describe a natural open covering for the space of maps. In particular:
\begin{itemize}
	\item let us fix a triangulation $\tau$ of $\Sigma$, i.e.:
\begin{itemize}
	\item a set of vertices $\sigma^{0}_{1}, \ldots, \sigma^{0}_{l} \in \Sigma$;
	\item a subset $E \subset \{1, \ldots, l\}^{2}$, determining a set of oriented edges $\{\sigma^{1}_{(a,b)} \subset \Sigma\}_{(a,b) \in E}$ such that $\partial \sigma^{1}_{(a,b)} = \sigma^{0}_{b} - \sigma^{0}_{a}$; if $(a,b) \in E$ then $(b,a) \notin E$ and we declare $\sigma^{1}_{(b,a)} := -\sigma^{1}_{(a,b)}$;
	\item a subset $T \subset \{1, \ldots, l\}^{3}$, determining a set of oriented triangles $\{\sigma^{2}_{(a,b,c)} \subset \Sigma\}_{(a,b,c) \in T}$ such that $\partial \sigma^{2}_{(a,b,c)} = \sigma^{1}_{(a,b)} + \sigma^{1}_{(b,c)} + \sigma^{1}_{(c,a)}$; given $a,b,c$ only one permutation of them belongs to $T$ and for a permutation $\rho$ we declare $\sigma^{2}_{\rho(a),\rho(b),\rho(c)} := (-1)^{\rho}\sigma^{2}_{(a,b,c)}$;
\end{itemize}
satisfying the following conditions:
\begin{itemize}
	\item every point $P \in \Sigma$ belongs to at least a triangle, and if it belongs to more than one triangle then it belongs to the boundary of each of them;
	\item every edge $\sigma^{1}_{(a,b)}$ lies in the boundary of exactly two triangles $\sigma^{2}_{(a,b,c)}$ and $\sigma^{2}_{(b, a, d)}$, inducing on it opposite orientations, and $\sigma^{2}_{(a,b,c)} \cap \sigma^{2}_{(b, a, d)} = \sigma^{1}_{(a,b)}$; if a point $p \in \Sigma$ belongs to an edge $\sigma^{1}_{(a,b)}$ and it is not a vertex, than the only two triangles containing it are the ones having $\sigma^{1}_{(a,b)}$ as common boundary; thus, there exists a function $b: E \rightarrow T^{2}$ such that $\sigma^{1}_{(a,b)} \subset \partial \sigma^{2}_{b^{1}(a,b)}$ and $-\sigma^{1}_{(a,b)} \subset \partial \sigma^{2}_{b^{2}(a,b)}$;
	\item for every vertex $\sigma^{0}_{i}$ there exists a finite set of triangles $\{\sigma^{2}_{(i,a_{1},a_{2})}, \ldots,$ $\sigma^{2}_{(i,a_{k_{i}},a_{1})}\}$ having $\sigma^{0}_{i}$ as vertex, such that $\sigma^{2}_{(i,a_{j},a_{j+1})} \cap \sigma^{2}_{(i,a_{j+1},a_{j+2})} = \sigma^{1}_{(i,a_{j+1})}$ (we use the notation $k_{i} + 1 = 1$), these triangles are the only one containing $\sigma^{0}_{i}$ and their union is a neighborhood of it; thus, there exists a function $B: \{1, \ldots, l\} \rightarrow \coprod_{i=1}^{l} T^{k_{i}}$, such that $B(i) \in T^{k_{i}}$ and $B(i) = \{\sigma^{2}_{(i,a_{1},a_{2})}, \ldots, \sigma^{2}_{(i,a_{k_{i}},a_{1})}\}$;
\end{itemize}
	\item we consider the following set of indices:
	\[J = \left\{(\tau, \varphi): \quad \begin{array}{l}
	    \bullet \textnormal{ $\tau = \bigl\{\sigma^{0}_{1}, \ldots, \sigma^{0}_{l(\tau)}, E, T \bigr\}$ is a triangulation of $\Sigma$} \\
	    \bullet \textnormal{ $\varphi: T \longrightarrow I$ is a function}
	\end{array} \right\}		 
\]
and a covering $\{V_{(\tau,\sigma)}\}_{(\tau,\sigma) \in J}$ of $\Sigma X$ is given by:
	\[V_{(\tau, \varphi)} = \{\Gamma \in \Sigma X: \; \Gamma(\sigma^{2}_{(a,b,c)}) \subset U_{\varphi(a,b,c)} \}.
\]
\end{itemize}
One can prove that these sets are open in the compact-open topology and that they cover $\Sigma X$.

\begin{Def} Given a topological space $X$ and a compact surface \emph{with boundary} $\Sigma$, the \emph{space of maps from $\Sigma$ to $X$}, called $\Sigma X$, is the set of continuous maps:
	\[\Gamma: \Sigma \longrightarrow X
\]
equipped with the compact-open topology.
\end{Def}
We now describe an analogous open covering for the space of maps. In particular:
\begin{itemize}
	\item let us fix a triangulation $\tau$ of $\Sigma$, i.e.:
	\begin{itemize}
	\item a set of vertices $\sigma^{0}_{1}, \ldots, \sigma^{0}_{l} \in \Sigma$;
	\item a subset $E \subset \{1, \ldots, l\}^{2}$, determining a set of oriented edges $\{\sigma^{1}_{(a,b)} \subset \Sigma\}_{(a,b) \in E}$ such that $\partial \sigma^{1}_{(a,b)} = \sigma^{0}_{b} - \sigma^{0}_{a}$; if $(a,b) \in E$ then $(b,a) \notin E$ and we declare $\sigma^{1}_{(b,a)} := -\sigma^{1}_{(a,b)}$;
	\item a subset $T \subset \{1, \ldots, l\}^{3}$, determining a set of oriented triangles $\{\sigma^{2}_{(a,b,c)} \subset \Sigma\}_{(a,b,c) \in T}$ such that $\partial \sigma^{2}_{(a,b,c)} = \sigma^{1}_{(a,b)} + \sigma^{1}_{(b,c)} + \sigma^{1}_{(c,a)}$; given $a,b,c$ only one permutation of them belongs to $T$ and for a permutation $\rho$ we declare $\sigma^{2}_{\rho(a),\rho(b),\rho(c)} := (-1)^{\rho}\sigma^{2}_{(a,b,c)}$;
\end{itemize}
satisfying the following conditions:
\begin{itemize}
	\item every point $P \in \Sigma$ belongs to at least a triangle, and if it belongs to more than one triangle then it belongs to the boundary of each of them; if $P \in \partial \Sigma$ and it is not a vertex, then it belongs to just one triangle and lies on the boundary of it;
	\item for every edge $\sigma^{1}_{(a,b)}$  there are two possibilities:
\begin{itemize}
	\item or it contains no points of $\partial \Sigma$ except possibly the vertices; in this case it lies in the boundary of exactly two triangles $\sigma^{2}_{(a,b,c)}$ and $\sigma^{2}_{(b, a, d)}$, inducing on it opposite orientations, and $\sigma^{2}_{(a,b,c)} \cap \sigma^{2}_{(b, a, d)} = \sigma^{1}_{(a,b)}$; if a point $p \in \Sigma$ belongs to such an edge $\sigma^{1}_{(a,b)}$ and it is not a vertex, than the only two triangles containing it are the ones having $\sigma^{1}_{(a,b)}$ as common boundary;
	\item or it is entirely contained in $\partial \Sigma$; in this case it lies in the boundary of just one triangle $\sigma^{2}_{(a,b,c)}$; if a point $p \in \Sigma$ belongs to such an edge $\sigma^{1}_{(a,b)}$ and it is not a vertex, than the only triangle containing it is the one containing $\sigma^{1}_{(a,b)}$ in its boundary;
\end{itemize}
thus, there exists a partition $E = BE \,\dot{\cup}\, IE$ in \emph{boundary edges} and \emph{internal edges}, and two functions:
\begin{itemize}
	\item $b: IE \rightarrow T^{2}$ such that $\sigma^{1}_{(a,b)} \subset \partial \sigma^{2}_{b^{1}(a,b)}$ and $-\sigma^{1}_{(a,b)} \subset \partial \sigma^{2}_{b^{2}(a,b)}$;
	\item $b: BE \rightarrow T$ such that $\sigma^{1}_{(a,b)} \subset \partial \sigma^{2}_{b(a,b)}$;
\end{itemize}
	\item there exists a partition $\{0, \ldots, l\} = BV \,\dot{\cup}\, IV$ in \emph{boundary vertices} and \emph{internal vertices}, such that:
\begin{itemize}
	\item for $i \in IV$, there exists a finite set of triangles $\{\sigma^{2}_{(i,a_{1},a_{2})}, \ldots, \sigma^{2}_{(i,a_{k_{i}},a_{1})}\}$ having $\sigma^{0}_{i}$ as vertex; $\sigma^{2}_{(i,a_{j},a_{j+1})} \cap \sigma^{2}_{(i,a_{j+1},a_{j+2})} = \sigma^{1}_{(i,a_{j+1})}$ with a cyclic order (i.e.\ we use the notation $k_{i} + 1 = 1$); these triangles are the only one containing $\sigma^{0}_{i}$ and their union is a neighborhood of it; thus, there exists a function $B: IV \rightarrow \coprod_{i\in IV} T^{k_{i}}$, such that $B(i) \in T^{k_{i}}$ and $B(i) = \{\sigma^{2}_{(i,a_{1},a_{2})}, \ldots, \sigma^{2}_{(i,a_{k_{i}},a_{1})}\}$.
	\item for $i \in BV$, there exists a finite set of triangles $\{\sigma^{2}_{(i,a_{1},a_{2})}, \ldots,$ \\ $\sigma^{2}_{(i,a_{k_{i}-1},a_{k_{i}})}\}$ (without $\sigma^{2}_{(i,a_{k_{i}},a_{1})}$) having $\sigma^{0}_{i}$ as vertex; $\sigma^{2}_{(i,a_{j},a_{j+1})} \cap$ \\ $\sigma^{2}_{(i,a_{j+1},a_{j+2})} = \sigma^{1}_{(i,a_{j+1})}$ for $1 < i < k_{i}$, these triangles are the only one containing $\sigma^{0}_{i}$ and their union is a neighborhood of it; thus, there exists a function $B: BV \rightarrow \coprod_{i=\in BV} T^{k_{i}-1}$, such that $B(i) \in T^{k_{i}-1}$ and $B(i) = \{\sigma^{2}_{(i,a_{1},a_{2})}, \ldots, \sigma^{2}_{(i,a_{k_{i}-1},a_{k_{i}})}\}$;
\end{itemize}
\end{itemize}

	\item we consider the following set of indices:
	\[J = \left\{(\tau, \varphi): \quad \begin{array}{l}
	    \bullet \textnormal{ $\tau = \bigl\{\sigma^{0}_{1}, \ldots, \sigma^{0}_{l(\tau)}, E, T \bigr\}$ is a triangulation of $\Sigma$} \\
	    \bullet \textnormal{ $\varphi: T \longrightarrow I$ is a function}
	\end{array} \right\}		 
\]
and a covering $\{V_{(\tau,\sigma)}\}_{(\tau,\sigma) \in J}$ of $\Sigma X$ is given by:
	\[V_{(\tau, \varphi)} = \{\Gamma \in \Sigma X: \; \Gamma(\sigma^{2}_{(a,b,c)}) \subset U_{\varphi(a,b,c)} \}.
\]
\end{itemize}
One can prove that these sets are open in the compact-open topology and that they cover $\Sigma X$.

\section{De-Rham theorem}

There is a canonical isomorphism between the $\rm\check{C}$ech cohomology of the constant sheaf $\mathbb{R}$ and the De-Rham cohomology. We give as an example the construction for a 2-form, since the general case is completely analogous. Let us consider a closed 2-form $\omega^{2}$: to find the corresponding $\rm\check{C}$ech class, we use iteratively the Poincar\'e lemma. Given a good cover $\mathfrak{U} = \{U_{\alpha}\}_{\alpha \in I}$ of $X$, we consider the restrictions $\omega^{2}_{\alpha} = \omega^{2}\vert_{U_{\alpha}}$, so that, being $U_{\alpha}$ contractible, we have that $\omega^{2}_{\alpha} = d\omega^{1}_{\alpha}$. Thus, considering the restriction to $U_{\alpha\beta}$, we have $d\omega^{1}_{\alpha} - d\omega^{1}_{\beta} = \omega^{2}_{\alpha\beta} - \omega^{2}_{\alpha\beta} = 0$, thus, being $U_{\alpha\beta}$ contractible, we have $\omega^{1}_{\alpha} - \omega^{1}_{\beta} = d \omega^{0}_{\alpha\beta}$. Moreover, $d(\omega^{0}_{\alpha\beta} + \omega^{0}_{\beta\gamma} + \omega^{0}_{\gamma\alpha}) = \omega^{1}_{\alpha} - \omega^{1}_{\beta} + \omega^{1}_{\beta} - \omega^{1}_{\gamma} + \omega^{1}_{\gamma} - \omega^{1}_{\alpha} = 0$. Now, since we started from a 2-form, we have that $\omega^{0}_{\alpha\beta}$ are functions, so that the last expression becomes $\omega^{0}_{\alpha\beta} + \omega^{0}_{\beta\gamma} + \omega^{0}_{\gamma\alpha} = c_{\alpha\beta\gamma} \in \mathbb{R}$.  One can prove that this correspondence sends cohomologous cocycles in cohomologus forms, thus the isomorphism we are searching is $[\,\omega^{2}\,] \in H^{2}_{dR}(X) \longrightarrow [\{c_{\alpha\beta\gamma}\}] \in \check{H}^{2}(\mathfrak{U}, \mathbb{R})$.

\paragraph{}Viceversa, given $\{c_{\alpha\beta\gamma}\} \in \check{Z}^{2}(\mathfrak{U}, \mathbb{R})$, we consider the sheaves immersion $\mathbb{R} \subset \underline{\mathbb{R}}$, and we know that $\underline{\mathbb{R}}$ is fine, thus $\{c_{\alpha\beta\gamma}\} = \check{\delta}^{1}(\{\omega^{0}_{\alpha\beta}\})$. But $\check{\delta}^{1}(\{d\,\omega^{0}_{\alpha\beta}\}) = \{d \, c_{\alpha\beta\gamma}\} = 0$, thus, since also $\Omega^{1}_{\mathbb{R}}$ is fine, we obtain $d\omega^{0}_{\alpha\beta} = \omega^{1}_{\alpha} - \omega^{1}_{\beta}$. Thus $\check{\delta}^{0}(d\,\omega^{1}_{\alpha}\}) = \{d^{2}\,\omega^{0}_{\alpha\beta}\} = 0$, thus $\{d\,\omega^{1}_{\alpha}\}$ defines a form $\omega \in \check{H}^{0}(\mathfrak{U}, \Omega^{2}_{\mathbb{R}}) = \Gamma(X, \Omega^{2}_{\mathbb{R}}) = \Omega^{2}(X, \mathbb{R})$.

\paragraph{}Given a de-Rham 1-class $[\omega^{1}]$, let us consider its corresponding $\rm\check{C}$ech class $[\{c_{\alpha\beta}\}]$. Let us fix a curve $\gamma: S^{1} \rightarrow X$. There exists $(\tau, \varphi) \in J$ such that $i \in V_{(\tau, \varphi)}$. Thus, since $\omega^{1} \vert_{U_{\alpha}} = d \omega^{0}_{\alpha}$, one has:
	\[\int_{\gamma} \omega^{1} = \sum_{i=1}^{l} \int_{\sigma^{1}_{i}} d\omega^{0}_{\varphi(i)} = \sum_{i=1}^{l} \, \Bigl[ \, \omega^{0}_{\varphi(i)}(\sigma^{0}_{i+1}) - \omega^{0}_{\varphi(i)}(\sigma^{0}_{i}) \, \Bigr].
\]
The last sum can be written as:
\begin{equation}\label{IntDeRhamCech}
	\int_{\gamma} \omega^{1} = \sum_{i=1}^{l} \, \Bigl[ \, \omega^{0}_{\varphi(i-1)}(\sigma^{0}_{i}) - \omega^{0}_{\varphi(i)}(\sigma^{0}_{i}) \, \Bigr] = \, \sum_{i=1}^{l} \, c_{\varphi(i-1)\varphi(i)}(\sigma^{0}_{i})
\end{equation}
so that \emph{the integral of a 1-form corresponds to the sum of the $\rm\check{C}$ech representatives on the vertices of a triangulation of the curve}. In particular, we see that $\int_{\gamma} \omega^{1} \in \mathbb{Z}$ for every $\gamma$ if and only if $c_{\alpha\beta} \in \mathbb{Z}$ for every $\alpha, \beta$, thus integer de-Rham classes corresponds to integer $\rm\check{C}$ech classes.

\paragraph{}For a de-Rham 2-class $[\omega^{2}]$ corresponding to $[\{c_{\alpha\beta\gamma}\}]$, given a closed surface $\Sigma \subset X$, we consider it as an immersion $i: \Sigma \rightarrow X$, thus there exists $(\tau, \varphi) \in J$ such that $i \in V_{(\tau, \varphi)}$. Thus, since $\omega^{2} \vert_{U_{\alpha}} = d \omega^{1}_{\alpha}$, one has:
	\[\int_{\Sigma} \omega^{2} = \sum_{(a,b,c) \in T_{\tau}} \int_{\sigma^{2}_{(a,b,c)}} d\omega^{1}_{\varphi(a,b,c)} = \sum_{(a,b,c) \in T_{\tau}} \int_{\partial\sigma^{2}_{(a,b,c)}} \omega^{1}_{\varphi(a,b,c)} \, .
\]
For every edge $\sigma^{1}_{(a,b)}$ we thus obtain:
	\[\begin{split}
	\int_{\sigma^{1}_{(a,b)}} \Bigl( &\omega^{1}_{\varphi(b^{1}(a,b))} - \omega^{1}_{\varphi(b^{2}(a,b))} \Bigr) = \int_{\sigma^{1}_{(a,b)}} d\omega^{0}_{\varphi(b^{1}(a,b)), \varphi(b^{2}(a,b))}\\
	&= \int_{\partial\sigma^{1}_{(a,b)}} \omega^{0}_{\varphi(b^{1}(a,b)), \varphi(b^{2}(a,b))} = \omega^{0}_{\varphi(b^{1}(a,b)), \varphi(b^{2}(a,b))}(b) - \omega^{0}_{\varphi(b^{1}(a,b)), \varphi(b^{2}(a,b))}(a).
\end{split}\]
Thus we reduce the integral to the contribution of the vertices. For a vertex $a$, we obtain:
	\[\sum_{i=1}^{k_{i}} \omega^{0}_{\varphi(a, a_{i}, a_{i+1}), \varphi(a, a_{i+1}, a_{i+2})} (a)
\]
but:
	\[\begin{split}
	&\omega^{0}_{\varphi(a, a_{i}, a_{i+1}), \varphi(a, a_{i+1}, a_{i+2})} + \omega^{0}_{\varphi(a, a_{i+1}, a_{i+2}), \varphi(a, a_{i+2}, a_{i+3})}\\
	&\phantom{XXXXXXXX} = \omega^{0}_{\varphi(a, a_{i}, a_{i+1}), \varphi(a, a_{i+2}, a_{i+3})} + c_{\varphi(a, a_{i}, a_{i+1}), \varphi(a, a_{i+1}, a_{i+2}), \varphi(a, a_{i+2}, a_{i+3})}
\end{split}\]
thus, after the cycle, we remains just with the $c$-terms. In particular, $\int_{\Sigma} F \in \mathbb{Z}$ for every $\Sigma$ if and only if $c_{\alpha\beta\gamma} \in \mathbb{Z}$ for every $(\alpha, \beta, \gamma)$.

\paragraph{}We also add the following lemma for future reference.
\begin{Lemma}\label{Integral1Forms} A 1-form $\omega \in \Omega^{1}_{\mathbb{C}}(X)$ represents an integral cohomology class if and only there exists a nowhere vanishing function $f$ such that $\omega = \frac{1}{2\pi i}f^{-1}df$.
\end{Lemma}
\textbf{Proof:} If $\omega = \frac{1}{2\pi i}f^{-1}df$, then $\omega_{\alpha} = \frac{1}{2\pi i}d\log f\vert_{U_{\alpha}}$, thus $c_{\alpha\beta}$ is the difference between two logarithms of $f\vert_{U_{\alpha\beta}}$ divided by $\frac{1}{2\pi i}$, which is integral. Viceversa, if $\omega$ is integral, then $\omega_{\alpha} = \frac{1}{2\pi i}d\varphi_{\alpha}$ and $c_{\alpha\beta} = \frac{1}{2\pi i}(\varphi_{\alpha} - \varphi_{\beta}) \in \mathbb{Z}$, thus the local functions $f_{\alpha} = e^{i\varphi_{\alpha}}$ glue to a global one $f$ being $e^{2\pi i c_{\alpha\beta}} = 1$. But $\omega_{\alpha} = d\varphi_{\alpha} = d\log f_{\alpha} = f_{\alpha}^{-1}df_{\alpha}$ and, since $f$ is global, we get $\omega = f^{-1}df$. $\square$

\chapter{Line bundles}

\section{Cohomology and line bundles}

\subsection{Bundles and $\rm\check{C}$ech cohomology}\label{CocyclesCohomology}

For $X$ a paracompact space, let us consider a complex line bundle $L \rightarrow X$ and let us fix a \emph{good cover} $\mathfrak{U} = \{U_{\alpha}\}_{\alpha \in I}$ of $X$. By definition of line bundle, $L$ is isomorphic to a bundle of the form:
\begin{equation}\label{BundleCharts}
	\Bigl( \, \bigsqcup \, (\,U_{\alpha} \times \mathbb{C}\,) \, \Bigr) \Big/ \sim \;, \qquad (x, z)_{\alpha} \sim (x, g_{\alpha\beta}(x) \cdot z)_{\beta}, \textnormal{ for }x \in U_{\alpha\beta}.
\end{equation}
The transition functions $\{g_{\alpha\beta}\} \in \check{C}^{1}(\mathfrak{U}, \underline{\mathbb{C}}^{*})$ satisfy cocycle condition, which exactly means that $\check{\delta}^{1} \{g_{\alpha\beta}\} = 0$, so that they determine a cohomology class $[\{g_{\alpha\beta}\}] \in \check{H}^{1}(\mathfrak{U}, \underline{\mathbb{C}}^{*})$.

If $(LB(X), \otimes)$ is the group of \emph{isomorphism classes} of line bundles over $X$, we obtain in this way a group isomorphism $(LB(X), \otimes) \simeq (\check{H}^{1}(\mathfrak{U}, \underline{\mathbb{C}}^{*}), \,\cdot\,)$. In fact, let us suppose that $L \in LB(X)$ is isomorphic to two bundles $L_{g}$ and $L_{h}$ of the form \eqref{BundleCharts}, with transition functions respectively $\{g_{\alpha\beta}\}$ and $\{h_{\alpha\beta}\}$. Then there exists an isomorphism $\varphi: L_{g} \rightarrow L_{h}$, which must be of the form $\varphi(x,z)_{g,\alpha} = (x, f_{\alpha}(x) \cdot z)_{h,\alpha}$. Thus:
	\[\begin{split}
	&\varphi(x, z)_{g,\alpha} = (x, f_{\alpha}(x) \cdot z)_{h,\alpha} = (x, h_{\alpha\beta}(x) \cdot f_{\alpha}(x) \cdot z)_{h,\beta}\\
	&\varphi(x, g_{\alpha\beta}(x) \cdot z)_{g,\beta} = (x, f_{\beta}(x) \cdot g_{\alpha\beta}(x) \cdot z)_{h,\beta}
\end{split}\]
thus $h_{\alpha\beta} \cdot g_{\alpha\beta}^{-1} = f_{\alpha}^{-1} \cdot f_{\beta}$, so that $[\{h_{\alpha\beta}\}] = [\{g_{\alpha\beta}\}]$. Viceversa, given a class $[\{g_{\alpha\beta}\}]$, formula \eqref{BundleCharts} gives a bundle associated to such a class. In particular, a bundle is trivial if and only if it is represented by the zero-class, since $X \times \mathbb{C}$ is a representative and all the transition functions are $1$.

\paragraph{}If we give a line bundle $L$ with a fixed set of local sections $\{s_{\alpha}: U_{\alpha} \rightarrow L\}$, it is canonically isomorphic to a line bundle of the form \eqref{BundleCharts} by $\varphi(s_{\alpha})_{x} = (x, 1)_{\alpha}$. This isomorphism can be applied to any bundle isomorphic to $L$ using the pull-back of the sections $\{s_{\alpha}\}$ via the isomorphism. In this case we have $g_{\alpha\beta} = s_{\alpha} / s_{\beta}$, since, for $x \in U_{\alpha\beta}$, one has $(s_{\alpha})_{x} = (x, 1)_{\alpha} = (x, g_{\alpha\beta}(x))_{\beta} = g_{\alpha\beta}(x)(s_{\beta})_{x}$ (of course the sections $\{s_{\alpha}\}$ does not make $\{g_{\alpha\beta}\}$ a coboundary since they are not functions, they are sections of a bundle). 

\paragraph{}Summarizing:
\begin{itemize}
	\item when we give a \emph{cohomology class} $\alpha = [\{g_{\alpha\beta}\}] \in \check{H}^{1}(\mathfrak{U}, \underline{\mathbb{C}}^{*})$, we associate to it an \emph{equivalence class up to isomorphism} of line bundles represented by \eqref{BundleCharts};\footnote{This equivalence class is much larger than the class made by the bundles of the form \eqref{BundleCharts} for the various representatives $\{g_{\alpha\beta}\}$ of $\alpha$, since there are all the bundles which are not of the form \eqref{BundleCharts} but only isomorphic to one of them.}
	\item when we give a \emph{cocycle} $\{g_{\alpha\beta}\} \in \check{Z}^{1}(\mathfrak{U}, \underline{\mathbb{C}}^{*})$, we associate to it the \emph{equivalence class} of a line bundle \emph{with a fixed set of local sections} $\{s_{\alpha}: X \rightarrow L\}$ \emph{up to isomorphism with relative pull-back of the sections}. In this case we have dependence on the covering $\mathfrak{U}$, but this is obvious since the local sections themselves determines the covering by their domains. We have a canonical representative for each of these classes given by \eqref{BundleCharts}.
\end{itemize}

\paragraph{}Let us consider $g = \{g_{\alpha\beta}\}, \,h = \{h_{\alpha\beta}\} \in \check{Z}^{1}(\mathfrak{U}, \underline{\mathbb{C}}^{*})$ and the representative bundles $L_{g}$ and $L_{h}$. As we have seen, if $[\,g\,] = [\,h\,]$, an isomorphism $\varphi: L_{g} \rightarrow L_{h}$ is given by $\varphi(x,z)_{g,\alpha} = (x, f_{\alpha}(x) \cdot z)_{h,\alpha}$ with $f_{\alpha}^{-1} \cdot f_{\beta} = h_{\alpha\beta} \cdot g_{\alpha\beta}^{-1}$, i.e.\ by $\{f_{\alpha}\} \in \check{C}^{0}(\mathfrak{U}, \underline{\mathbb{C}}^{*})$ such that $\check{\delta}^{0}\{f_{\alpha}\} = h \cdot g^{-1}$. By an \emph{active} point of view, we can see $L_{h} = L_{g} \otimes L_{\check{\delta}^{0}f}$, where $L_{\check{\delta}^{0}f}$ is the \emph{trivialized} bundle $X \times \mathbb{C}$ with sections $\{f_{\alpha}\}$ fixed, considering the tensor product of the sections. Thus, we have $\Iso(L_{g}, L_{h}) \simeq (\check{\delta}^{0})^{-1}(hg^{-1})$.

Similarly, an \emph{automorphism} of $L_{g}$ is given by $\{f_{\alpha}\} \in \check{Z}^{0}(\mathfrak{U}, \underline{\mathbb{C}}^{*})$, i.e.\ by a function $f: X \rightarrow \mathbb{C}^{*}$: it is of the form $\varphi(x, z)_{\alpha} = (x, f(x) \cdot z)_{\alpha}$. Clearly, $\Iso(L_{g}, L_{h})$ and $\Aut(L_{g})$ are in bijection but \emph{not canonically}, since any isomorphism can be written as a fixed one composed with an automorphism; in fact, any $\{f_{\alpha}\}$ whose coboundary is $h \cdot g^{-1}$ can be written as a fixed one multiplied by a cocycle.

The isomorphism class of \emph{trivial} line bundles correspond to the zero class\footnote{We use multiplicative notation since the group is $\mathbb{C}^{*}$, hence the zero-class is $1$ and not $0$.} $1 \in \check{H}^{1}(\mathfrak{U}, \underline{\mathbb{C}}^{*})$, which is represented by coboundaries. One preferred coboundary is $\{1\} \in \check{B}^{1}(\mathfrak{U}, \underline{\mathbb{C}}^{*})$, which represents the class of a trivial bundle \emph{with a global section}. We define a \emph{trivialization} of a trivial bundle as an isomorphism from it to $X \times \mathbb{C}$.

The coboundary $\{1\}$ determines the bundle $L_{1} = \bigsqcup \, (\,U_{\alpha} \times \mathbb{C}\,) \, / \sim\,$, with $(x, z)_{\alpha} \sim (x, z)_{\beta}$ for $x \in U_{\alpha\beta}$, which is \emph{canonically} isomorphic to $X \times \mathbb{C}$ by $\varphi(x,z)_{\alpha} = (x,z)$, so that, for a coboundary $b = \{g_{\alpha\beta}\} \in \check{B}^{1}(\mathfrak{U}, \underline{\mathbb{C}}^{*})$, we can see a trivialization of $L_{b}$ as an isomorphism $\varphi: L_{b} \rightarrow L_{1}$. Hence, a trivialization $\varphi: L_{b} \rightarrow L_{1}$ corresponds to a cochain $\{g_{\alpha}\} \in \check{C}^{0}(\mathfrak{U}, \underline{\mathbb{C}}^{*})$ such that $\check{\delta}^{0}\{g_{\alpha}\} = b^{-1}$, i.e.\ for $\{g_{\alpha\beta}\} = \check{\delta}^{0}\{g_{\alpha}\}$, a trivialization of $L_{b}$ is given by $\varphi(x, z)_{\alpha} = (x, z \cdot g_{\alpha}(x))$. In particular one has $\varphi(x, g_{\alpha}(x)^{-1})_{\alpha} = (x, 1)_{\alpha}$ and the sections $(x,1)_{\alpha}$ glue to a global one in $X \times \mathbb{C}$, thus $g_{\alpha}^{-1}$ determines in $L_{b}$ the local expression of a global section, that's why it is a trivialization. Summarizing, the group of trivializations of $L_{b}$ is $(\check{\delta}^{0})^{-1}(b^{-1}) \subset \check{C}^{0}(\mathfrak{U}, \underline{\mathbb{C}}^{*})$. In particular, the trivializations of $L_{1}$, i.e.\ its automorphisms, are given by $\check{H}^{0}(\mathfrak{U}, \underline{\mathbb{C}}^{*}) = \check{Z}^{0}(\mathfrak{U}, \underline{\mathbb{C}}^{*})$: for a function $f: X \rightarrow \mathbb{C}^{*}$ the automorphism is $\varphi(x,z)_{\alpha} = (x,z / f(x))$.

Clearly, $\Iso(L_{g}, L_{h})$ is in canonical bijection with the trivializations of $L_{h \cdot g^{-1}}$, and $\Aut(L_{g})$ is canonically isomorphic to the trivializations of $L_{1}$, i.e.\ with $\Aut(L_{1})$. By an \emph{active} point of view, we can see a trivialization of $L_{b}$ as a tensor product by $L_{g^{-1}}$, i.e.\ the trivialized bundle $X \times \mathbb{C}$ with sections $\{g_{\alpha}^{-1}\}$, for $\check{\delta}^{0}g = b$.

\paragraph{}At the end we have the following picture:
\begin{itemize}
	\item isomorphism classes of line bundles are in bijection (actually, it is a group isomorphism) with $\check{H}^{1}(\mathfrak{U}, \underline{\mathbb{C}}^{*})$; the class of trivial bundles corresponds to the zero-class;
	\item an element of $\zeta \in \check{Z}^{1}(\mathfrak{U}, \underline{\mathbb{C}}^{*})$ determines a class of a line bundle with local sections, canonically represented by $L_{\zeta}$;
	\item if $g, h \in \check{Z}^{1}(\mathfrak{U}, \underline{\mathbb{C}}^{*})$ are cohomologous, then $\Iso(L_{g}, L_{h}) \simeq (\check{\delta}^{0})^{-1}(hg^{-1})$;
	\item if $\zeta$ is a coboundary, the trivializations of $L_{\zeta}$, i.e.\ the isomorphisms $\varphi: L_{\zeta} \rightarrow X \times \mathbb{C}$, are in natural bijection with $(\check{\delta}^{0})^{-1}(\zeta^{-1}) \subset \check{C}^{0}(\mathfrak{U}, \underline{\mathbb{C}}^{*})$; in particular, the trivializations of $L_{1}$ are in natural bijection with $\check{Z}^{0}(\mathfrak{U}, \underline{\mathbb{C}}^{*}) = \check{H}^{0}(\mathfrak{U}, \underline{\mathbb{C}}^{*})$, which are the functions $f: X \rightarrow \mathbb{C}^{*}$.
\end{itemize}
This picture will be analogue raising by 1 the degree in cohomology: a gerbe will be an element of $\check{H}^{2}(\mathfrak{U}, \underline{\mathbb{C}}^{*})$, a trivialization of the trivial gerbe $G_{1}$ and element of $\check{H}^{1}(\mathfrak{U}, \underline{\mathbb{C}}^{*})$ and a trivialization of a trivial gerbe $G_{b}$ will be an element of $(\check{\delta}^{1})^{-1}(b^{-1}) \subset \check{C}^{1}(\mathfrak{U}, \underline{\mathbb{C}}^{*})$.

\subsubsection{Hermitian metrics}

If we put an hermitian metric on a bundle, we can locally find a section of unit norm, thus it is isomorphic to a bundle of the form \eqref{BundleCharts} such that $\norm{(x,1)_{\alpha}} = 1$ for every $\alpha$. Thus $1 = \norm{(x,1)_{\alpha}} = \norm{(x,g_{\alpha\beta}(x))_{\beta}} = \abs{g_{\alpha\beta}(x)} \cdot \norm{(x,1)_{\beta}} = \abs{g_{\alpha\beta}(x)}$, so that the transition functions have unit modulus. Since every line bundle has an hermitian metric, every bundle is isomorphic to a bundle determined by a cocycle in $\check{Z}^{1}(\mathfrak{U}, \underline{S}^{1})$. Viceversa, given a cocycle in $\check{Z}^{1}(\mathfrak{U}, \underline{S}^{1})$ we determine an hermitian metric by $\langle\, (x,z)_{\alpha}, (x,w)_{\alpha} \,\rangle := z \cdot \overline{w}$. It is well defined since $g_{\alpha\beta} = e^{2\pi i \cdot \rho_{\alpha\beta}}$, thus $\overline{g_{\alpha\beta}} = g_{\alpha\beta}^{-1}$, so that $\langle\, (x,g_{\alpha\beta} \cdot z)_{\beta}, (x,g_{\alpha\beta} \cdot w)_{\beta} \,\rangle = g_{\alpha\beta} \cdot \overline{g_{\alpha\beta}} \cdot \langle\, (x, z)_{\beta}, (x, w)_{\beta} \,\rangle = z \cdot \overline{w}$.

Actually, if we put two hermitian metrics $\langle \cdot, \cdot \rangle_{1}$ and $\langle \cdot, \cdot \rangle_{2}$ on the same line bundle, there exists an automorphism $\varphi$ such that $\varphi^{*}\langle \cdot, \cdot \rangle_{2} = \langle \cdot, \cdot \rangle_{1}\,$: in fact, if, for fixed non-zero $v_{x},w_{x} \in L_{x}$, we put $f(x) = \langle v_{x}, w_{x} \rangle_{1} \,/\, \langle v_{x}, w_{x} \rangle_{2}$, we have a well defined function $f: X \longrightarrow \mathbb{R}^{+} \subset \mathbb{C}^{*}$ independent by the various $v_{x}$ and $w_{x}$ chosen, since for $v'_{x} = \lambda \cdot  v_{x}$ and $w'_{x} = \mu \cdot w_{x}$ we get a factor $\lambda \cdot \overline{\mu}$ both at the numerator and denominator. If we put $\varphi(x, z) = (x, \sqrt{f(x)} \cdot z)$ we obtain the desired automorphism, since $(\varphi^{*}\langle \cdot, \cdot \rangle_{2})(v_{x}, w_{x}) = \langle \varphi(v_{x}), \varphi(w_{x}) \rangle_{2} = f(x) \langle v_{x}, w_{x} \rangle_{2} = \langle v_{x}, w_{x} \rangle_{1}$. Thus, once we fix an equivalence class of bundles there is only one metric up to equivalence, thus \emph{the isomorphism classes of line bundles are the same as equivalence classes of line bundles with hermitian metric}, hence $\check{H}^{1}(\mathfrak{U}, \underline{S}^{1}) \simeq \check{H}^{1}(\mathfrak{U}, \underline{\mathbb{C}}^{*})$. In fact, we have a \emph{splitting} exact sequence:
	\[0 \longrightarrow \underline{S}^{1} \longrightarrow \underline{\mathbb{C}}^{*} \longrightarrow \underline{\mathbb{R}}^{+} \longrightarrow 0
\]
and $\underline{\mathbb{R}}^{+} \simeq \underline{\mathbb{R}}$ via the logarithm, thus it is acyclic. Since the sequence splits, we have $\check{H}^{n}(\mathfrak{U}, \underline{\mathbb{C}}^{*}) \simeq \check{H}^{n}(\mathfrak{U}, \underline{S}^{1}) \oplus \check{H}^{n}(\mathfrak{U}, \underline{\mathbb{R}}^{+})$ for every $n$ and, for $n \geq 1$, the $\mathbb{R}^{+}$-factor is zero.\footnote{It seems that, at degree 1, the map in cohomology from $S^{1}$ to $\mathbb{C}^{*}$ is not injective because $\check{H}^{0}(\mathfrak{U}, C^{\infty}(\,\cdot\,, \mathbb{R}^{+})) \neq 0$, but the Bockstein map is zero. In fact, its kernel is the image of $\check{H}^{0}(\mathfrak{U}, \underline{\mathbb{C}}^{*}) \rightarrow \check{H}^{0}(\mathfrak{U}, \underline{\mathbb{R}}^{+})$ which is surjective.}

\subsubsection{First Chern class}

We fix a \emph{good cover} $\mathfrak{U}$ of $X$ and consider $[L] = [\{g_{\alpha\beta}\}] \in \check{H}^{2}(\mathfrak{U}, \underline{\mathbb{C}}^{*})$ or $[L] = [\{g_{\alpha\beta}\}] \in \check{H}^{2}(\mathfrak{U}, \underline{S}^{1})$. We can write $g_{\alpha\beta} = e^{2\pi i \cdot \rho_{\alpha\beta}}$, so that $\rho_{\alpha\beta} + \rho_{\beta\gamma} + \rho_{\gamma\alpha} = c_{\alpha\beta\gamma} \in C^{\infty}(U_{\alpha\beta\gamma}, \mathbb{Z}) \simeq \mathbb{Z}$. We thus obtain a class $\mathcal{C} = [\{c_{\alpha\beta\gamma}\}] \in \check{H}^{2}(\mathfrak{U}, \mathbb{Z})$, called \emph{first Chern class} of $L$. This operation is exactly the Bockstein homomorphism of the exact sequence of sheaves:
	\[0 \longrightarrow \mathbb{Z} \longrightarrow \underline{\mathbb{C}} \overset{e^{2\pi i \, \cdot\,}}\longrightarrow \underline{\mathbb{C}}^{*} \longrightarrow 0 \; \,
\]
or of the exact sequence of sheaves:
	\[0 \longrightarrow \mathbb{Z} \longrightarrow \underline{\mathbb{R}} \overset{e^{2\pi i \, \cdot\,}}\longrightarrow \underline{S}^{1} \longrightarrow 0 \; ,
\]
inducing the same result by the inclusion of the second sequence into the first, which is the identity on $\mathbb{Z}$.

\subsubsection{Torsion line bundles}

A \emph{torsion line bundle} is a line bundle $L$ such that $c_{1}(L) \in \check{H}^{2}(\mathfrak{U}, \mathbb{Z})$ is a torsion class. Let us consider the following exact sequences of sheaves:
\begin{equation}\label{ExactSeqZRS}
\xymatrix{
0 \ar[r] & \mathbb{Z} \ar[r] & \underline{\mathbb{R}} \ar[r]^{e^{2\pi i \,\cdot\,}} & \underline{S}^{1} \ar[r] & 0\\
0 \ar[r] & \mathbb{Z} \ar[r] \ar@{=}[u] & \mathbb{R} \ar[r]^{e^{2\pi i \,\cdot\,}} \ar@{^(->}[u] & S^{1} \ar[r] \ar@{^(->}[u] & 0
}
\end{equation}
and the corresponding degree-1 Bockstein homomorphisms $\beta_{1}$ and $\beta_{2}$. For $L \in \check{H}^{1}(\mathfrak{U}, \underline{S}^{1})$, one has $\beta_{1}(L) = c_{1}(L) \in \check{H}^{2}(\mathfrak{U}, \mathbb{Z})$, and $\beta_{1}$ is an isomorphism. Considering the second sequence, $c_{1}(L)$ is torsion if and only if its image in $\check{H}^{2}(\mathfrak{U}, \mathbb{R})$ is zero, thus, by exactness, if and only is if can be lifted to $\tilde{L} \in \check{H}^{1}(\mathfrak{U}, S^{1})$ such that $\beta_{2}(\tilde{L}) = c_{1}(L)$. Moreover, $\tilde{L}$ is unique up to $\alpha \in \check{H}^{1}(\mathfrak{U}, \mathbb{R})$. By commutativity, the possible $\tilde{L}$ are all sent to $L$ by the inclusion $S^{1} \hookrightarrow \underline{S}^{1}$. We will discuss the meaning of this liftings dealing with holonomy.

This means that \emph{a line bundle is torsion if and only if it can be realized by constant transition functions}. Of course, not all the representatives of $L \in \check{H}^{1}(\mathfrak{U}, \underline{S}^{1})$ will be constant, since we are free to add any coboundary, but there exists a constant representative in the class if and only if $L$ is torsion. This is not a surprise: for a trivial bundle, we can realize it by transition functions all equal to $1$, and, if $L^{\otimes n}$ is trivial, we can realize $L$ by transition functions which are $n$-roots on unity, thus, since the set of $n$-roots of unity is discrete, they are constant. The non trivial fact is the opposite: if there exists a representative with constant transition functions, they must be root of unity up to $\alpha \in \check{H}^{1}(\mathfrak{U}, \mathbb{R})$, i.e.\ there is a representative made by root of unity and the bundle is torsion.

\subsection{Connection and field strength}\label{sec:ConnectionFS}

Given a line bundle $L \rightarrow X$, we choose local sections $\{s_{\alpha}\}_{\alpha \in I}$, with $s_{\alpha}: U_{\alpha} \rightarrow L$ and define:
	\[iA_{\alpha}(X) = \frac{\nabla_{X}s_{\alpha}}{s_{\alpha}}
\]
obtaining a well-defined function since it is the ratio of two sections, so that $A_{\alpha} \in \Omega^{1}_{\mathbb{C}}(U_{\alpha})$. In this way:
	\[\nabla_{X}(f \cdot s_{\alpha}) = \partial_{X} f \cdot s_{\alpha} + f \cdot \nabla_{X} s_{\alpha} = (\partial_{X} f + f \cdot i A_{\alpha}(X)) \cdot s_{\alpha}.
\]
If we put an hermitian metric and choose unitary sections and a compatible connection, $A_{\alpha}$ becomes real since:
	\[\begin{split}
	&\partial_{X} \langle s_{\alpha}, s_{\alpha} \rangle = \langle \nabla_{X} s_{\alpha}, s_{\alpha} \rangle + \langle s_{\alpha}, \nabla_{X} s_{\alpha} \rangle\\
	&0 = \langle \nabla_{X} s_{\alpha}, s_{\alpha} \rangle + \overline{\langle \nabla_{X} s_{\alpha}, s_{\alpha} \rangle}\\
	&0 = \Re \; \langle \nabla_{X} s_{\alpha}, s_{\alpha} \rangle = \Re \; \bigl[\, i A_{\alpha}(X) \cdot \langle s_{\alpha}, s_{\alpha} \rangle \,\bigr] = \Re \; i A_{\alpha}(X)
\end{split}\]
thus $A_{\alpha} \in \Omega^{1}_{\mathbb{R}}(U_{\alpha})$.

\paragraph{}We now study the transition functions for the connection. On $U_{\alpha\beta}$ there are two sections $s_{\alpha}$ and $s_{\beta}$ such that $s_{\alpha} = g_{\alpha\beta} \cdot s_{\beta}$, and $[L] \simeq [\{g_{\alpha\beta}\}]$ since $L$ is isomorphic to \eqref{BundleCharts} via $s_{\alpha}(x) \simeq (x, 1)_{\alpha}$. Then:
\begin{equation}\label{TransitionConnection2}
	i A_{\alpha}(X) = \frac{\nabla_{X} s_{\alpha}}{s_{\alpha}} = \frac{\partial_{X} g_{\alpha\beta} \cdot s_{\beta} + g_{\alpha\beta} \cdot \nabla_{X} s_{\beta}}{g_{\alpha\beta} \cdot s_{\beta}} = g_{\alpha\beta}^{-1} \cdot \partial_{X}g_{\alpha\beta} + i A_{\beta}(X)
\end{equation}
so that we obtain the transition function for the local connection:
\begin{equation}\label{TransitionConnection}
	i A_{\alpha} = i A_{\beta} + g_{\alpha\beta}^{-1} \cdot d g_{\alpha\beta}.
\end{equation}
Since $\mathfrak{U}$ is a good cover $U_{\alpha\beta}$ is contractible, thus we can write $g_{\alpha\beta}^{-1} \cdot d g_{\alpha\beta} = d \log g_{\alpha\beta}$, i.e.\ for $g_{\alpha\beta} = e^{2\pi i \cdot \rho_{\alpha\beta}}$, one gets $g_{\alpha\beta}^{-1} \cdot d g_{\alpha\beta} = 2\pi i \cdot d \rho_{\alpha\beta}$, so that $A_{\alpha} = A_{\beta} + 2\pi \cdot d \rho_{\alpha\beta}$. Thus $dA_{\alpha}\vert_{U_{\alpha\beta}} = dA_{\beta}\vert_{U_{\alpha\beta}}$. We define $F_{\alpha} = dA_{\alpha}$, so that the local forms $F_{\alpha}$ glue to a global form $F \in \Omega^{2}_{\mathbb{R}}(X)$ called \emph{curvature} or \emph{field strength}. Clearly $dF = 0$, thus we can consider $[\,F\,] \in H^{2}_{dR}(X)$. In this way we have described a connection on a fixed bundle as $\{A_{\alpha}\} \in \check{C}^{0}(\mathfrak{U}, \Omega^{1}_{\mathbb{C}})$ or, if compatible with an hermitian metric, as $\{A_{\alpha}\} \in \check{C}^{0}(\mathfrak{U}, \Omega^{1}_{\mathbb{R}})$, such that $\check{\delta}^{0}\{dA_{\alpha}\} = 0$.

\subsubsection{First Chern class and field strength: link}

We now prove that:
\begin{equation}\label{ChernF}
	\bigl[ \, \textstyle\frac{1}{2\pi} \cdot F \,\bigr]_{H^{2}_{dR}(X)} \simeq \bigl[\, c_{1}(L) \otimes_{\mathbb{Z}} \mathbb{R} \,\bigr]_{\check{H}^{2}(\mathfrak{U}, \mathbb{R})}
\end{equation}
under the standard canonical isomorphism between de-Rham and $\rm\check{C}$ech cohomology. In fact, let us consider $F$: to find the corresponding $\rm\check{C}$ech class, we consider the restrictions $F_{\alpha} = F\vert_{U_{\alpha}}$ so that, being $U_{\alpha}$ contractible, we have $F_{\alpha} = dA_{\alpha}$, and, by definition, $A_{\alpha}$ is exactly the local expression of a connection with curvature $F$. Thus, $dA_{\alpha} - dA_{\beta} = F_{\alpha\beta} - F_{\alpha\beta} = 0$, thus, being $U_{\alpha\beta}$ contractible, we have $A_{\alpha} - A_{\beta} = d x_{\alpha\beta}$. We know that $dx_{\alpha\beta} = 2\pi \cdot d \rho_{\alpha\beta}$. Now, since we started from a 2-form, we have that $\rho_{\alpha\beta}$ are functions. Moreover, $d(\rho_{\alpha\beta} + \rho_{\beta\gamma} + \rho_{\gamma\alpha}) = A_{\alpha} - A_{\beta} + A_{\beta} - A_{\gamma} + A_{\gamma} - A_{\alpha} = 0$, thus $2\pi(\rho_{\alpha\beta} + \rho_{\beta\gamma} + \rho_{\gamma\alpha}) = 2\pi \, c_{\alpha\beta\gamma} \in \mathbb{R}$ constant. By construction we arrived exactly to the first Chern class multiplied by $2\pi$.

\paragraph{}Viceversa, given $\{2\pi \,c_{\alpha\beta\gamma}\} \in \check{Z}^{2}(\mathfrak{U}, \mathbb{R})$, we consider the sheaves immersion $\mathbb{R} \subset \underline{\mathbb{R}}$, and we know that $\underline{\mathbb{R}}$ is fine, thus $\{c_{\alpha\beta\gamma}\} = \check{\delta}^{1}(\{\rho_{\alpha\beta}\})$. But $\check{\delta}^{1}(\{d\rho_{\alpha\beta}\}) = \{d \, c_{\alpha\beta\gamma}\} = 0$, thus, since also $\Omega^{1}_{\mathbb{R}}$ is fine, we obtain $2\pi \, d\rho_{\alpha\beta} = A_{\alpha} - A_{\beta}$. Thus $\check{\delta}^{0}(\{dA_{\alpha}\}) = \{2\pi\, d^{2}\rho_{\alpha\beta}\} = 0$, thus $\{dA_{\alpha}\}$ defines a global form $F \in \check{H}^{0}(\mathfrak{U}, \Omega^{2}_{\mathbb{R}}) = \Gamma(X, \Omega^{2}_{\mathbb{R}}) = \Omega^{2}_{\mathbb{R}}(X)$. One can prove that this correspondence sends cohomologous cocycles in cohomologous forms. From \eqref{ChernF} and formula \eqref{IntDeRhamCech}, it follows that the fact that $[\, c_{1}(L) \,]$ is an integral class is equivalent to the fact that $\bigl[\frac{1}{2\pi} \cdot F\bigr]$ is, thus in the case of a line bundle we have to start from $\{c_{\alpha\beta\gamma}\} \in \check{Z}^{2}(\mathfrak{U}, \mathbb{Z})$ (we will discuss in the following the meaning of non-integral Chern classes).

\subsubsection{Affine structure on connections}

Let us now consider two different connections $\nabla$ and $\nabla'$ on the same line bundle $L$: then, as it is easy to verify, their difference $\nabla - \nabla'$ is, for a fixed $X$, an endomorphism of $L$, thus it is a 1-form, i.e.\ $(\nabla - \nabla')_{X}s = \omega(X) \cdot s$. In fact, if $\{A_{\alpha}\}$ and $\{A'_{\alpha}\}$ are the local expressions of $\nabla$ and $\nabla'$ with respect to a fixed set of sections $\{s_{\alpha}\}$, we have that $A_{\alpha} - A_{\beta} = A'_{\alpha} - A'_{\beta} = 2\pi\, d\rho_{\alpha\beta}$, thus $A_{\alpha} - A'_{\alpha} = A_{\beta} - A'_{\beta}$, so that $\omega := A_{\alpha} - A'_{\alpha} \in \Omega^{1}_{\mathbb{R}}(X)$, and $(\nabla - \nabla')_{X}s_{\alpha} = \omega(X) \cdot s_{\alpha}$, thus $(\nabla - \nabla')_{X}s = \omega(X) \cdot s$ for every local section $s$. Hence \emph{the set of connections on a fixed bundle $L$ is an affine space whose underlying vector space is $\Omega^{1}_{\mathbb{R}}(X)$}. Moreover, if we fix the curvature, the difference of two connection is a \emph{closed} global 1-form, since $0 = dA_{\alpha} - dA'_{\alpha} = d\omega$, thus \emph{the set of connections on a fixed bundle $L$ with a fixed curvature $F$ is an affine space whose underlying vector space is $\mathcal{Z}^{1}_{\mathbb{R}}(X)$}. However, these are single connections and it is natural to ask when they are equivalent, i.e.\ when one is the pull-back by an automorphism of another.

\subsection{Group of bundles with connections}

The equivalence classes of line bundles with connection form an abelian group $(LB\nabla(X),$ $\otimes)$. In fact, given two bundles with compatible connections $(L, \nabla)$ and $(L',\nabla')$, we can consider the product $(L \otimes L', \nabla \otimes \nabla')$ for:
	\[(\nabla \otimes \nabla')_{X}(s_{\alpha} \otimes s'_{\alpha}) \,:=\, \nabla_{X} s_{\alpha} \otimes s'_{\alpha} + s_{\alpha} \otimes \nabla'_{X} s'_{\alpha}.	
\]
The zero-element of this group is $[\,(X \times \mathbb{C}, \partial_{X})\,]$, as it is easy to verify. If we express the connection with respect to sections $\{s_{\alpha}\}$ and $\{s'_{\alpha}\}$ we have that the expression of $\nabla \otimes \nabla'$ with respect to $\{s_{\alpha} \otimes s'_{\alpha}\}$ is exactly the sum:
\begin{equation}\label{TensorConnections}
\begin{split}
	\frac{(\nabla \otimes \nabla')_{X}(s_{\alpha} \otimes s'_{\alpha})}{s_{\alpha} \otimes s'_{\alpha}} &= \frac{\nabla_{X}s_{\alpha} \otimes s'_{\alpha} + s_{\alpha} \otimes \nabla'_{X}s'_{\alpha}}{s_{\alpha} \otimes s'_{\alpha}}\\
	&= \frac{iA_{\alpha}(X) \cdot s_{\alpha} \otimes s'_{\alpha} + s_{\alpha} \otimes iA'_{\alpha}(X) \cdot s'_{\alpha}}{s_{\alpha} \otimes s'_{\alpha}}\\
	&= iA_{\alpha}(X) + iA'_{\alpha}(X).
\end{split}
\end{equation}
There is a natural forgetful morphism $\varphi: LB\nabla(X) \longrightarrow LB(X)$. There are important subgroups of $LB\nabla(X)$:
\begin{itemize}
	\item the classes of \emph{trivial} bundles with connection are a subgroup $(TrLB\nabla(X), \otimes)$, in fact, $TrLB\nabla(X) = \Ker \, \varphi$; the classes of a fixed class of bundles with any connection are cosets of this subgroup;
	\item the classes of \emph{trivial} bundles with \emph{flat} connection are a subgroup $(TrFLB\nabla(X),$ $\otimes)$ of the previous; the classes of trivial bundles with connection of a fixed curvature are cosets of this group in the previous;
	\item the classes of \emph{torsion} bundles with connection are a subgroup $(TLB\nabla(X),\otimes)$, in fact, this group is $\varphi^{-1}(TLB(X))$;
	\item the classes of \emph{torsion} bundles with \emph{flat} connection are a subgroup $(TFLB\nabla(X),$ $\otimes)$ of the previous; the classes of torsion bundles with connection of a fixed curvature are cosets of this group in the previous.
\end{itemize}
Summarizing, we have the following scheme for subgroups of $LB\nabla(X)$:
\begin{equation}\label{SubgroupsScheme}
\xymatrix{
TrLB\nabla(X) \ar@{^(->}[r] & TLB\nabla(X) \ar@{^(->}[r] & LB\nabla(X)\\
TrFLB\nabla(X) \ar@{^(->}[u] \ar@{^(->}[r] & TFLB\nabla(X). \ar@{^(->}[u]
}
\end{equation}
Since a flat connection can exist only on a torsion bundle, the classes made by bundles with connection of a fixed curvature are cosets of $(TFLB\nabla(X),\otimes)$, thus, denoting by $\mathcal{I}^{2}_{\mathbb{R}}(X)$ the set of \emph{integral} real 2-forms on $X$, there exist an exact sequence:
\begin{equation}\label{ExSeqLBNabla}
0 \longrightarrow TFLB\nabla(X) \longrightarrow LB\nabla(X) \longrightarrow \mathcal{I}^{2}_{\mathbb{R}}(X) \longrightarrow 0.
\end{equation}
Instead, classes made by a fixed class of bundles with connection of a fixed curvature are cosets of $(TrFLB\nabla(X), \otimes)$. Moreover, since on a flat bundle the curvature is cohomologous to 0, so that the connection can be globally defined, \eqref{ExSeqLBNabla} restricts to:
\begin{equation}\label{ExSeqTLBNabla}
0 \longrightarrow TFLB\nabla(X) \longrightarrow TLB\nabla(X) \longrightarrow d\Omega^{1}_{\mathbb{R}}(X) \longrightarrow 0 \; \,
\end{equation}
and also to:
\begin{equation}\label{ExSeqTrLBNabla}
0 \longrightarrow TrFLB\nabla(X) \longrightarrow TrLB\nabla(X) \longrightarrow d\Omega^{1}_{\mathbb{R}}(X) \longrightarrow 0.
\end{equation}

\subsubsection{Local description and equivalence}

Since two connections are equivalent when one is the pull-back of the other by a bundle isomorphism, we now want to see how to read this from the local expression $A_{\alpha} \in \Omega^{1}(U_{\alpha}, \mathbb{R})$. If we have an isomorphism $\varphi: L \rightarrow L'$, let us fix section $\{s'_{\alpha}\}$ and a connection $\nabla'$ on $L'$. Let us now define $s_{\alpha} = \varphi^{-1}(s'_{\alpha})$ and $\nabla = \varphi^{*}\nabla'$. Then we obtain the same local forms:
	\[\frac{\nabla'_{X} s_{\alpha}'}{s_{\alpha}'} = \frac{\varphi(\nabla_{X}\varphi^{-1}(s_{\alpha}'))}{s_{\alpha}'} = \frac{\varphi(\nabla_{X}s_{\alpha})}{\varphi(s_{\alpha})} = \frac{\nabla_{X}s_{\alpha}}{s_{\alpha}}
\]
where the last equality is due to the fact that, if $s_{x} = f(x) \cdot t_{x}$, then $\varphi(s)_{x} = f(x) \cdot \varphi(t)_{x}$, thus the ratio of two sections remains constant under line bundle maps. Thus, any connection equivalent to $\nabla$ on a bundle $L'$ can be realized by the same local forms $A_{\alpha}$ of $\nabla$ with respect to certain local sections of $L$. Hence, the only freedom left is the choice of the local sections $s_{\alpha}$ of $L$ to define $A_{\alpha}$. If we choose another section $t_{\alpha}$, then, if $t_{\alpha} = f_{\alpha} \cdot s_{\alpha}$, by the same computation of \eqref{TransitionConnection2} we obtain:
	\[iA_{\alpha}^{(t)} = iA_{\alpha}^{(s)} + f_{\alpha}^{-1} \cdot df_{\alpha}
\]
so that $A_{\alpha}^{(t)} = A_{\alpha}^{(s)} - i \cdot d\log f_{\alpha}$. Thus, the local forms change by \emph{exact forms}, or, equivalently, by closed forms since $U_{\alpha}$ is contractible. Viceversa, if we consider $\tilde{A}_{\alpha} = A^{(s_{\alpha})}_{\alpha} + d\varphi_{\alpha}$, we consider the section $t_{\alpha} = e^{i\varphi_{\alpha}} \cdot s_{\alpha}$ and we have that $\tilde{A}_{\alpha} = A_{\alpha}^{(t_{\alpha})}$. This means that, if we consider the connection up to isomorphism, then $A_{\alpha} \in \Omega^{1}_{\mathbb{R}}(U_{\alpha}) / d\Omega^{0}_{\mathbb{R}}(U_{\alpha})$, where $\Omega^{0}_{\mathbb{R}}(U_{\alpha}) = C^{\infty}(U_{\alpha}, \mathbb{R})$. Then we have the condition that $\check{\delta}^{0}\{dA_{\alpha}\} = 0$, but, since the intersections $U_{\alpha\beta}$ are contractible, this exactly means that $A_{\alpha} - A_{\beta} \in d\Omega^{0}_{\mathbb{R}}(U_{\alpha\beta})$. This means that we obtain \emph{a section of the sheaf $(\,\Omega^{1}_{\mathbb{R}} \,/\, d\Omega^{0}_{\mathbb{R}})^{\natural}= \,\Omega^{1}_{\mathbb{R}} \,/\, \mathcal{Z}^{1}_{\mathbb{R}}$} in the sense of sheaf quotient (so the sheafification is part of the definition). This result is not so interesting: the exterior differential $d$ induces a sheaf isomorphism $d: \;\Omega^{1}_{\mathbb{R}} \,/\, \mathcal{Z}^{1}_{\mathbb{R}} \longrightarrow \mathcal{Z}^{2}_{\mathbb{R}}$ and this is exactly the first step of the isomorphism between de-Rham and $\rm\check{C}$ech cohomology. We obtain the only possible class of forms $A_{\alpha}$ up to closed ones, such that $dA_{\alpha} = F\vert_{U_{\alpha}}$. Thus, the class we obtain is exactly the curvature, so the local description of the connection itself contains information only about the curvature. This mean that \emph{there is no possibility to completely recover the class of the connection from the local expression of the connection, actually neither the complete information about the topology of the bundle}, since we recover only the curvature, so that \emph{we miss flat connection}. In fact, a connection is flat if and only if there exist local parallel sections $\{s_{\alpha}\}$: with respect to these sections, $iA_{\alpha}(X) = (\nabla_{X}s_{\alpha}) / s_{\alpha} = 0$, thus a flat connection can be realized by $\{0\}$, that's why its equivalence class does not contribute at all. Thus, to recover the complete information about a bundle with connection \emph{we must consider both the transition function and the local representation of the connection}: this what we do using hypercohomology of an appropriate complex.

\subsubsection{Cohomological description}

Let us consider the complex of sheaves $\underline{S}^{1} \overset{i\, d \,\circ\, \log}\longrightarrow \Omega^{1}_{\mathbb{R}}$ and the relative $\rm\check{C}$ech double complex $\check{C}^{\bullet,\bullet}(\mathfrak{U}, \underline{S}^{1} \rightarrow \Omega^{1}_{\mathbb{R}})$ (v.\ appendix \ref{AppHyperC} for more details), with associated total complex $\check{T}^{\bullet}(\,\mathfrak{U}\,)$. We have that:
	\[\check{T}^{1}(\mathfrak{U}) = \check{C}^{1}(\mathfrak{U}, \underline{S}^{1}) \,\oplus\, \check{C}^{0}(\mathfrak{U}, \Omega^{1}_{\mathbb{R}})
\]
thus, given a line bundle with sections and the relative connection, we can consider $(g, -A) = (\{g_{\alpha\beta}\},$ $\{-A_{\alpha}\}) \in \check{T}^{1}(\,\mathfrak{U}\,)$: we claim that it is a cocycle. In fact, we have that:
	\[\check{T}^{2}(\,\mathfrak{U}\,) \,=\, \check{C}^{2}(\mathfrak{U}, \underline{S}^{1}) \,\oplus\, \check{C}^{1}(\mathfrak{U}, \Omega^{1}_{\mathbb{R}})
\]
and $\check{\delta}^{1}(g, -A) = (\check{\delta}^{1}g, i\, d\log g - \check{\delta}^{0}A)$: but $\check{\delta}^{1}g = 0$ since it is a line bundle, and $i\,d\log g = \check{\delta}^{0}A$ by \eqref{TransitionConnection}.

We now claim that $(g, -A) \in \check{T}^{1}(\,\mathfrak{U}\,)$ is a coboundary if and only if it represents the trivial class $[\,(X \times \mathbb{C}, \partial_{X})\,]$. In fact, $\check{T}^{0}(\,\mathfrak{U}\,) \,=\, \check{C}^{0}(\mathfrak{U}, \underline{S}^{1})$ and $\check{\delta}^{0}(\{g_{\alpha}\}) = (\check{\delta}^{0}\{g_{\alpha}\}, \{i\,d\log g_{\alpha}\})$ $= (\{g_{\alpha}^{-1} g_{\beta}\}, \{i\,d\log g_{\alpha}\})$, so that $A_{\alpha} = -i\,d\log g_{\alpha}$ with respect to the local sections $(x, 1)_{\alpha}$ in $L_{g_{\alpha}^{-1} g_{\beta}}$. We now express the connection with respect to the global sections $(x, g_{\alpha})_{\alpha}$ which glue to a global one: in this case, $A'_{\alpha} = A_{\alpha} + i\, d\log g_{\alpha} = 0$, so that we obtain the local representation $(1, 0)$. Viceversa, we can represent $[\,(X \times \mathbb{C}, \partial_{X})\,]$ as $(1, 0)$ and, if we consider a different set of local sections $\{g_{\alpha}\}$, we exactly obtain $(\{g_{\alpha}^{-1} g_{\beta}\}, \{i\,d\log g_{\alpha}\})$.

\paragraph{}We call $\check{Z}_{T}^{n}(\,\mathfrak{U}\,)$ and $\check{B}_{T}^{n}(\,\mathfrak{U}\,)$ cocycles and coboundaries of $\check{T}^{n}(\,\mathfrak{U}\,)$: a cocycle $(g, -A) \in \check{Z}_{T}^{1}(\,\mathfrak{U}\,)$ represent a bundle \emph{with local sections} and the relative local expression, denoted by $L_{g,A}$, while the cohomology class represents the class up to isomorphism and pull-back of the connection (not of the sections, although the pull-back connection is expressed with respect to the pull-back of the sections). We consider automorphisms for $(g, -A) \in \check{Z}_{T}^{1}(\,\mathfrak{U}\,)$ fixed:
\begin{itemize}
	\item automorphisms $\varphi: L_{g} \rightarrow L_{g}$ with pull-back of the connection correspond to $\check{H}^{0}(X, \underline{S}^{1})$, as for simple bundles, acting as $\varphi(x, z)_{\alpha} = (x, f(x) \cdot z)_{\alpha}$ and $A'_{\alpha} = A_{\alpha} + d\log f\vert_{U_{\alpha}}$;
	\item automorphisms $\varphi: L_{g} \rightarrow L_{g}$ fixing the representation $\{A_{\alpha}\}$ of the connection correspond to $\check{H}^{0}(X, S^{1})$, i.e.\ to constant functions, since we must have $d\log f = 0$: this is not surprising, since such an automorphism can be seen by an active point of view as a tensor product by a flat trivialized bundle with sections.
\end{itemize}
Similarly for $g$ and $h$ cohomologous:
\begin{itemize}
	\item isomorphisms $\varphi: L_{g} \rightarrow L_{h}$ with pull-back of connections correspond to $(\check{\delta}^{0})^{-1}(h \cdot g^{-1}) \subset \check{C}^{0}(X, \underline{S}^{1})$, as for simple bundles, acting as $\varphi(x,z)_{g,\alpha} = (x, f_{\alpha}(x) \cdot z)_{h,\alpha}$ and $A'_{\alpha} = A_{\alpha} + i\,d\log f_{\alpha}$; by an \emph{active} point of view we can see $L_{h} = L_{g, A} \otimes L_{f,\,i\, d\log f}$, so that an isomorphism can be seen as the tensor product by a trivialized flat bundle with sections;
	\item isomorphisms $\varphi: L_{g} \rightarrow L_{h}$ sending the representation $\{A_{\alpha}\}$ in the representation $\{A'_{\alpha}\}$ corresponds to $(\check{\delta}^{1})^{-1}(h \cdot g^{-1}) \,\cap\, (i\,d\log)^{-1}(A' - A) \subset \check{C}^{0}(X, \underline{S}^{1})$: this set is actually a coset of $\check{H}^{0}(X, S^{1})$ in $\check{C}^{0}(X, \underline{S}^{1})$, since the transition functions are determined up to constants; by an \emph{active} point of view we can see $L_{h} = L_{g, A} \otimes L_{f, 0}$, so that an isomorphism can be seen as the tensor product by a trivialized flat bundle with parallel sections.
\end{itemize}
We now consider \emph{trivial} line bundles. We have two kinds of trivialization, correspondingly to generic isomorphisms:
\begin{itemize}
	\item a trivialization $\varphi: L_{b} \rightarrow L_{1}$ with pull-back of connection corresponds, as for simple bundles, to a cocycle $\{g_{\alpha}^{-1}\} \in \check{C}^{0}(\mathfrak{U}, \underline{S}^{1})$ such that $\check{\delta}^{0}\{g_{\alpha}^{-1}\} = b^{-1}$, i.e.\ for $\{g_{\alpha\beta}\} = \check{\delta}^{0}\{g_{\alpha}\}$, a trivialization of $L_{b}$ is given by $\varphi(x, z)_{\alpha} = (x, z / g_{\alpha}(x))$ and $A'_{\alpha} = A_{\alpha} - d\log g_{\alpha}$;
	\item a trivialization $\varphi: L_{b} \rightarrow L_{1}$ sending the representation $\{A_{\alpha}\}$ in the representation $\{A'_{\alpha}\}$ corresponds to $(\check{\delta}^{1})^{-1}(b^{-1}) \,\cap\, (d\log)^{-1}(A' - A) \subset \check{C}^{0}(X, \underline{S}^{1})$: this set is actually a coset of $\check{H}^{0}(X, S^{1})$ in $\check{C}^{0}(X, \underline{S}^{1})$.
\end{itemize}
In particular, the trivializations of $L_{1}$, i.e.\ its automorphisms, are given by $\check{H}^{0}(\mathfrak{U}, \underline{S}^{1})$ and the one fixing a representation by $\check{H}^{0}(\mathfrak{U}, S^{1})$.

\subsubsection{Connections on torsion line bundles}

From \eqref{ChernF}, it follows that $L$ is torsion if and only if $[\,F\,] = 0 \in \check{H}^{2}(\mathfrak{U}, \mathbb{R})$, in particular, \emph{a flat line bundle is torsion}, but not vice-versa, since \emph{a bundle is flat when $F = 0$ as a single form, not as an equivalence class in cohomology} (also a trivial bundle can be non-flat). When a bundle is torsion, since $[\,F\,] = 0$ there exists a global 1-form $A$ on $X$ such that $F = dA$. In fact, let us consider \eqref{TransitionConnection}. If we realize the bundle by constant transition functions (thus, we are dealing with a bundle with a fixed set of sections), we have that $dg_{\alpha\beta} = 0$, so that \eqref{TransitionConnection} becomes $A_{\alpha} = A_{\beta}$, and viceversa. This means that \emph{we can define the connection as a global 1-form $A$ if and only if the line bundle is torsion, coherently with the fact that $[\,F\,] = 0$}. However, this does not mean that in any realization of the bundle the connection is globally defined, since, if we add a generic coboundary, we still have a non trivial expression for \eqref{TransitionConnection}.

We can see that flat connections can exist only on torsion line bundles also from the local forms: since $dA_{\alpha} = 0$ for every $\alpha$, we have $A_{\alpha} = da_{\alpha}$; since $da_{\alpha} - da_{\beta} = d\rho_{\alpha\beta}$, we have $d(\rho_{\alpha\beta} - (a_{\alpha} - a_{\beta})) = 0$, thus $\rho_{\alpha\beta} = a_{\alpha} - a_{\beta} + c_{\alpha\beta}$ with $c_{\alpha\beta}$ constant, thus $g_{\alpha\beta} = (e^{a_{\beta}})^{-1} \cdot (e^{a_{\alpha}}) \cdot e^{c_{\alpha\beta}}$, thus $g_{\alpha\beta}$ is constant up to a coboundary.

\paragraph{}Let us consider a torsion non-trivial line bundle with local sections, with a globally defined connection $A$: it is impossible to distinguish it from a globally defined connection on any other topologically different torsion line bundle, since we have however a globally defined form. We have seen that flat connections (up to equivalence) are not classified by the local description, since can be realized locally by $\{0\}$: now we know that flat connections can exists only on torsion line bundles, and they can be globally defined if we realize the bundle by constant transition functions. Viceversa, if we consider a set of \emph{parallel} local sections $\{s_{\alpha}\}$, so that the corresponding representative is $\{0\}$, then the corresponding transition functions are constant. In fact, if $s_{\beta} = g_{\alpha\beta} \cdot s_{\alpha}$ and $\nabla_{X}s_{\alpha} = \nabla_{X}s_{\beta} = 0$, then one has $0 = \nabla_{X}(g_{\alpha\beta} \cdot s_{\alpha}) = \partial_{X}g_{\alpha\beta} \cdot s_{\alpha}$, thus $\partial_{X}g_{\alpha\beta} = 0$. Thus, if we give a cocycle $\{g_{\alpha\beta}\} \in \check{Z}^{1}(\mathfrak{U}, S^{1})$ we determine a flat connection on the corresponding bundle $L_{g}$ represented by $\{0\}$, and all the flat connections can be obtained in this way. For a fixed $c_{1}(L) \in \Tor\,\check{H}^{2}(\mathfrak{U}, \mathbb{Z})$ we thus have a map:
	\[(\,\textnormal{Flat connections on $L$}\,) \;\longrightarrow\; \beta^{-1}(c_{1}(L)) \subset \check{H}^{1}(\mathfrak{U}, S^{1}).
\]
We now prove that the equivalence classes of flat connections on $[\,L\,]$ is given by $\beta^{-1}(c_{1}(L)) \subset \check{H}^{1}(\mathfrak{U}, S^{1})$, so that we have a group isomorphism:
\begin{equation}\label{IsoTFLB}
	TFLB\nabla(X) \,\simeq\, \check{H}^{1}(\mathfrak{U}, S^{1}).
\end{equation}

\paragraph{}In fact, as we have seen, describing locally a push-forward of a connection $\nabla$ on $L$ by any isomorphism (or automorphism) is equivalent to consider $\nabla$ with respect to other sections. Thus, having two realization of $[\,\nabla\,]$ by parallel sections, is equivalent to realize $\nabla$ with respect to two different set of parallel sections $\{s_{\alpha}\}$ and $\{s'_{\alpha}\}$. Then $s'_{\alpha} = f_{\alpha} \cdot s_{\alpha}$ and $0 = \nabla_{X} s'_{\alpha} = \partial_{X} f \cdot s_{\alpha}$, thus $f$ is constant. Hence, the possible realizations of a flat connection with respect to parallel sections correspond to its realization with respect to multiples of a fixed set of sections $\{s_{\alpha}\}$. But, for $s'_{\alpha} = \lambda_{\alpha} \cdot s_{\alpha}$, we obtain:
	\[s'_{\beta} = g'_{\alpha\beta} \cdot s'_{\alpha} \;\Longrightarrow\; \lambda_{\beta} \cdot s_{\beta} = g'_{\alpha\beta} \cdot \lambda_{\alpha} \cdot s_{\alpha} \;\Longrightarrow\; g_{\alpha\beta} = \frac{\lambda_{\alpha}}{\lambda_{\beta}} \cdot g'_{\alpha\beta}
\]
thus we obtain another representative of $[\{g_{\alpha\beta}\}] \in \check{H}^{1}(\mathfrak{U}, S^{1})$ and viceversa.

\paragraph{}Thanks to \eqref{IsoTFLB} we have a cohomological description of $TFLB\nabla(X)$. The quotient $LB\nabla(X) \,/\, TFLB\nabla(X)$ is made by the possible curvatures, since two connections have a flat quotient if and only if they have the same curvature. We will prove that the class in $\check{H}^{1}(\mathfrak{U}, S^{1})$ actually corresponds to the \emph{holonomy} of the flat connection.

\subsubsection{Connections on trivial line bundles}

Let us suppose that a connection $\nabla$ on a trivial bundle is realized by $A \in \Omega^{1}_{\mathbb{R}}(X)$ with respect to a global section $s$: the automorphisms of a trivial bundle correspond to multiplications by a global section. Then, for $s' = f \cdot s$, we obtain $A' = A - i f^{-1} \cdot df$, but, since, $f$ is global, we cannot in general extract the logarithm. But a closed 1-form $\omega$ is integral if and only if it can be expressed in the form $\omega = f^{-1} \cdot df$, thus a global defined connection change under isomorphisms by an integer form. In particular, \emph{the global connections $A$ with $[\,A\,]$ integral are equivalent to $\partial_{X}$ by an automorphism of $X \times \mathbb{C}$}, coherently with the fact that they have trivial holonomy. Hence we have a group isomorphism:
\begin{equation}
TrLB\nabla(X) \;\simeq\; \Omega^{1}(X, \mathbb{R}) \,/\, \mathcal{I}^{\,1}(X, \mathbb{R})
\end{equation}
for $\mathcal{I}^{\,1}(X, \mathbb{R})$ the group of integral 1-forms. For flat connections the representative form is closed, thus we have a group isomorphism:
\begin{equation}\label{TfFLBForms}
TrFLB\nabla(X) \;\simeq\; \mathcal{Z}^{1}(X, \mathbb{R}) \,/\, \mathcal{I}^{\,1}(X, \mathbb{R}) \, .
\end{equation}
In particular, sequence \eqref{ExSeqTrLBNabla} corresponds to:
\begin{equation}\label{ExSeqTrLBNabla2}
0 \longrightarrow \mathcal{Z}^{1}(X, \mathbb{R}) \,/\, \mathcal{I}^{\,1}(X, \mathbb{R}) \longrightarrow \Omega^{1}(X, \mathbb{R}) \,/\, \mathcal{I}^{\,1}(X, \mathbb{R}) \longrightarrow d\Omega^{1}(X, \mathbb{R}) \longrightarrow 0.
\end{equation}

\paragraph{}Let us consider the exact sequence:
	\[0 \longrightarrow \mathbb{Z} \longrightarrow \mathbb{R} \overset{e^{2\pi i \cdot}}\longrightarrow S^{1} \longrightarrow 0.
\]
Flat connections on trivial bundles are realized by the classes $\alpha \in \check{H}^{1}(\mathfrak{U}, S^{1})$ such that $\beta(\alpha) = 0 \in \check{H}^{2}(\mathfrak{U}, \mathbb{Z})$. By exactness, these are the classes such that there exists $[\,A\,] \in \check{H}^{1}(\mathfrak{U}, \mathbb{R})$ whose image in $S^{1}$ is $\alpha$. We prove that these real classes corresponds exactly to the classes of the connections globally defined on the trivial bundle, which are closed forms by flatness. In fact, let us consider $\alpha \in \check{H}^{1}(\mathfrak{U}, S^{1})$, corresponding to parallel sections $\{s_{\alpha}\}$ of the trivial bundle $L$, so that $\alpha = [\,\{s_{\alpha}^{-1} \cdot s_{\beta}\}\,] = [\,\{g_{\alpha\beta}\}\,]$ (clearly $\{s_{\alpha}\}$ is not a trivialization of $\{g_{\alpha\beta}\}$ since they are not functions, they are sections of the bundle). Since the bundle is trivial, there exists a global section $s$ such that $s\vert_{U_{\alpha}} = f_{\alpha} \cdot s_{\alpha}$. In particular, $f_{\alpha} \cdot s_{\alpha} = f_{\beta} \cdot s_{\beta}$, thus $s_{\alpha}^{-1} \cdot s_{\beta} = f_{\alpha} \cdot f_{\beta}^{-1}$, thus $\{f_{\alpha}^{-1}\}$ is a trivialization of $\{g_{\alpha\beta}\}$ as a cocycle for $\underline{S}^{1}$. We realize the connection as a global form $A$ with respect to $s$, i.e.\ $iA(X) = \nabla_{X} s \,/\, s$. Then:
	\[A_{\alpha}(X) := A\vert_{U_{\alpha}}(X) = \frac{\nabla_{X} s\vert_{U_{\alpha}}}{s\vert_{U_{\alpha}}} = \frac{\partial_{X}f_{\alpha} \cdot s_{\alpha}}{f_{\alpha} \cdot s_{\alpha}} = d\log f_{\alpha}(X).
\]
Let us now realize $A$ as a $\rm\check{C}$ech class: we have $A_{\alpha} = d\log f_{\alpha}$, thus $\log f_{\alpha} - \log f_{\beta} = \log g_{\alpha\beta} = \rho_{\alpha\beta}$ constant, and $[A]_{H^{1}_{dR}(X)} \simeq [\{\rho_{\alpha\beta}\}]_{\check{H}^{1}(\mathfrak{U}, \mathbb{R})}$: thus, the image in $S^{1}$ is $\{e^{2\pi i \cdot \rho_{\alpha\beta}}\} = \{g_{\alpha\beta}\}$, i.e.\ exactly $\alpha$.

Viceversa, given a closed connection $A$, with respect to a section $s$, we realize it in $\rm\check{C}$ech cohomology as $[\{\rho_{\alpha\beta}\}]$: in particular, $A_{\alpha} = da_{\alpha}$ and $a_{\alpha} - a_{\beta} = \rho_{\alpha\beta}$. We consider $g_{\alpha\beta} = e^{2\pi i \cdot \rho_{\alpha\beta}} = e^{2\pi i \cdot a_{\alpha}} \cdot (e^{2\pi i \cdot a_{\beta}})^{-1}$. Then we consider the sections $s_{\alpha} := (s\vert_{U_{\alpha}}) \,/\, e^{2\pi i \cdot a_{\alpha}}$, and these section are parallel, since:
	\[\begin{split}
	\nabla_{X} s_{\alpha} &= \nabla_{X} \, (e^{-2\pi i \cdot a_{\alpha}} \cdot s)\\
	&= -2\pi i \cdot \partial_{X} a_{\alpha} \cdot e^{-2\pi i \cdot a_{\alpha}} \cdot s + e^{-2\pi i \cdot a_{\alpha}} \cdot i\, da_{\alpha}(X) \cdot s\\
	&= -2\pi i \cdot \partial_{X} a_{\alpha} \cdot e^{-2\pi i \cdot a_{\alpha}} \cdot s + e^{-2\pi i \cdot a_{\alpha}} \cdot i\, \partial_{X}a_{\alpha} \cdot s\\
	&= 0.
\end{split}\]
Thus, $A$ is the global realization of a connection which is realized by $0$ in the bundle whose transition functions are the image of $A$, as $\rm\check{C}$ech class, in $S^{1}$. Given $[\{g_{\alpha\beta}\}] \in \check{H}^{1}(\mathfrak{U}, S^{1})$, which determines the equivalence class of the flat connection, by exactness its realization as a global form is unique up to integer classes: this coincides with \eqref{TfFLBForms}.

\subsubsection{Large gauge transformations}

Since we have chosen a good cover, we worked just with contractible sets. However, it can happen to consider non-contractible open sets of $X$ that trivialize $L$ however. Let us consider two such sets $W_{a}$ and $W_{b}$ such that $W_{ab} \neq \emptyset$, and let us consider the corresponding transition function $g_{ab}$. In this case, with the same proof we obtain that $A_{a} = A_{b} + g_{ab}^{-1} \cdot d g_{ab}$, but we cannot extract the logarithm any more since $W_{ab}$ is not necessarily contractible. Of course, $g_{ab}^{-1} \cdot d g_{ab}$ still remains closed, since it is \emph{locally} $d \log g_{ab}$. This transition function defines necessarily an \emph{integral} class in $W_{ab}$ thanks to lemma \ref{Integral1Forms}, i.e.\ large gauge transformations are quantized. We can see this also using the notion of holonomy which we introduce in the next paragraph. In fact, if $\{A_{\alpha}\} \in \check{C}^{0}(\mathfrak{U}, \Omega^{1}_{\mathbb{R}})$ is such that the induced $F$ is an integral class, then the corresponding class in $\check{\rm{C}}$ech hypercohomology represents a line bundle, and we have just proven that large gauge transformations are quantized. Viceversa, let us suppose that for two generic open sets $W_{a}$ and $W_{a}$ such that $F \vert_{W_{a}} = d A_{a}$ and $F \vert_{W_{b}} = d A_{b}$, it happens that $A_{a} = A_{b} + \Phi_{ab}$ on $W_{ab}$ with $d\Phi_{ab} = 0$ but $\Phi_{ab}$ not integral. If we consider a 1-cycle $C \subset W_{ab}$ such that $\int_{C}\Phi_{ab} \in \mathbb{R} \setminus \mathbb{Z}$, then the exponential of both $\int_{C} A_{a}$ and $\int_{C} A_{b}$ gives the holonomy of $A$ along $C$ (since we can refine the cover to a good one and on each subset of $W_{a}$ we always get $A_{a}$), but the two expressions differ by the exponential of $\int_{C} \Phi_{ab}$ which is not $1$. Thus the holonomy is not well-defined, so that $A$ is not a connection on a bundle, i.e.\ $F$ is not quantized.

Actually quantization of large gauge transformations is equivalent to quantization of the field strength $F$. One could ask what happens for $F$ not quantized since, as we will see, the relations $A_{\alpha} - A_{\beta} = g_{\alpha\beta}^{-1}dg_{\alpha\beta}$ holds also for trivialization of flat gerbes. The point is that the functions $g_{\alpha\beta}$ does not glue to a unique transition function $g_{ab}$ on a non-contractible intersection $U_{ab}$. In fact, if we write $A_{\alpha} - A_{\beta} = d\log g_{\alpha\beta}$ for a contractible cover refining the given one, then $g_{\alpha\beta}$ glue to $g_{ab}$ if and only if $A_{a} - A_{b}$ is integral on $W_{ab}$ by lemma \ref{Integral1Forms}, i.e.\ if and only if large gauge transformations are quantized.

\section{Holonomy and Wilson loop}

\subsection{Global description}

Let us consider a line bundle with connection $(L, \nabla)$ on $X$. Let us consider a \emph{closed} curve $\gamma: S^{1} \rightarrow X$ and fix a point $x = \gamma(e^{2\pi i \cdot t})$. Parallel transport along $\gamma$ gives a linear map $t_{x}: L_{x} \rightarrow L_{x}$, which can be thought of as a number $\Hol_{\nabla}(\gamma) \in S^{1}$ thanks to the canonical isomorphism $L_{x}^{\checkmark} \otimes L_{x} \simeq \mathbb{C}$ given by $\varphi \otimes v \simeq \varphi(v)$. Thus, parallel transport defines a function $\Hol_{\nabla}: LX \rightarrow S^{1}$ called \emph{holonomy} of $\nabla$.

What can we say about open curves? Given a curve $\gamma: [0,1] \rightarrow X$, let us put $x = \gamma(0)$ and $y = \gamma(1)$: parallel transport defines a linear map $t_{x,y}: L_{x} \rightarrow L_{y}$, which is not a number any more since $L_{x}^{\checkmark} \otimes L_{y}$ is not canonically isomorphic to $\mathbb{C}$. Thus, given a curve $\gamma \in CX$, holonomy is an element of a fiber $CL_{\gamma} = L_{x}^{\checkmark} \otimes L_{y}$: we now see that indeed holonomy defines a section of a line bundle $CL \rightarrow CX$. In fact, let us consider the bundle $L^{\checkmark} \boxtimes L \rightarrow X \times X$, i.e.\ $L^{\checkmark} \boxtimes L = \pi_{1}^{*}L^{\checkmark} \otimes \pi_{2}^{*}L$. We have a natural map $\pi: CX \rightarrow X \times X$ given by $\pi(\gamma) = (\gamma(0), \gamma(1))$, so that we can define $CL = \pi^{*}(L^{\checkmark} \boxtimes L)$. By construction $CL_{\gamma} = (L^{\checkmark} \boxtimes L)_{\pi(\gamma)} = (L^{\checkmark} \boxtimes L)_{(\gamma(0),\gamma(1))} = L^{\checkmark}_{\gamma(0)} \otimes L_{\gamma(1)}$, so we obtain exactly the desired fiber. Thus holonomy defines a section $\Hol_{\nabla}: CX \rightarrow LX$. By construction $c_{1}(CL) = \pi^{*}(\pi_{2}^{*}\,c_{1}(L) - \pi_{1}^{*}\,c_{1}(L))$.

If we consider piecewise smooth curves, we can define an embedding $LX \hookrightarrow CX$ since $\gamma$ is closed for $\gamma(0) = \gamma(1)$: in fact, by construction $CL$ is canonically trivial when restricted to $CX$, since it corresponds to $L^{\checkmark} \otimes L \simeq X \times \mathbb{C}$ canonically, thus we restore case of closed curves as a particular case.

As one can see from the expression of $c_{1}(CL)$, if $L$ is trivial so is $CL$. There is more: a trivialization of $L$ determines \emph{canonically} a trivialization of $CL$. In fact, is $s: X \rightarrow L$ is a global section, then it determines canonically a global section $s^{\checkmark}: X \rightarrow L^{\checkmark}$ given by $s^{\checkmark}(s) = X \times \{1\}$, thus a section $s^{\checkmark} \boxtimes s: X \times X \rightarrow L^{\checkmark} \boxtimes L$, thus, by pull-back, a global section $\pi^{*}(s^{\checkmark} \boxtimes s): CX \rightarrow CL$. What is happening geometrically? It happens that a global section $s: X \rightarrow L$ provides a way to identify the fibers of $L$, which exactly means that it makes $L$ isomorphic to $X \times \mathbb{C}$, thus $L_{x}^{\checkmark} \otimes L_{y}$ becomes isomorphic to $\mathbb{C}$ via $\lambda \cdot s_{x}^{\checkmark} \otimes \mu \cdot s_{y} \simeq \lambda \cdot \mu$, or, equivalently, a linear map $L_{x} \rightarrow L_{y}$ is the number $\lambda$ such that $s_{x} \rightarrow \lambda \cdot s_{y}$. Thus, \emph{for a trivial bundle with a global section holonomy is a well-defined function also over the space of open curves}.

In the same way, \emph{a system of local sections of $L$ with respect to a cover $\mathfrak{U}$ of $X$ determines a system of local sections of $CL$ with respect to the cover $\mathfrak{V}$ defined in section \ref{TriangCovers}}: in fact, let us consider $\gamma \in V_{(\tau, \varphi)}$: then $L_{\gamma(0)}^{\checkmark} \otimes L_{\gamma(1)}$ is isomorphic to $\mathbb{C}$ via $s_{\varphi(1)}$ and $s_{\varphi(l(\tau))}$, so that we have a local trivialization $V_{(\tau, \varphi)} \times \mathbb{C}$ corresponding to the local section $V_{(\tau, \varphi)} \times \{1\}$. Thus, we can describe transition functions of $CL$ for $\mathfrak{V}$ in terms of the ones of $L$ for $\mathfrak{U}\,$. In particular, the local expression of parallel transport along $\gamma$ with respect to the local sections fixed is given by $\{\rho_{(\tau,\varphi)}\}$ such that $t_{\gamma(0), \gamma(1)}(x, z)_{\varphi(1)} = (x, \rho_{(\tau,\varphi)} \cdot z)_{\varphi(l)}$. Then, if $\gamma \in V_{(\tau, \varphi)} \cap V_{(\tau', \varphi')}$, we have, with respect to the second chart, $t_{\gamma(0), \gamma(1)}(x, z)_{\varphi'(1)} = (x, \rho_{(\tau',\varphi')} \cdot z)_{\varphi(l')}$. Then, since $(x, z)_{\varphi(1)} = (x, g_{\varphi(1), \varphi'(1)} \cdot z)_{\varphi'(1)}$, one has:
	\[\begin{split}
	&t_{\gamma(0), \gamma(1)}(x, z)_{\varphi(1)} = (x, \rho_{(\tau,\varphi)} \cdot z)_{\varphi(l)} = (x, g_{\varphi(l), \varphi'(l')} \cdot \rho_{(\tau,\varphi)} \cdot z)_{\varphi'(l')}\\
	&t_{\gamma(0), \gamma(1)}(x, g_{\varphi(1), \varphi'(1)} \cdot z)_{\varphi'(1)} = (x, \rho_{(\tau',\varphi')} \cdot g_{\varphi(1), \varphi'(1)} \cdot z)_{\varphi'(l)}
\end{split}\]
so that $g_{\varphi(l), \varphi'(l')} \cdot \rho_{(\tau,\varphi)} = \rho_{(\tau',\varphi')} \cdot g_{\varphi(1), \varphi'(1)}$, thus, $\rho_{(\tau,\varphi)} = \rho_{(\tau',\varphi')} \cdot (g_{\varphi(l), \varphi'(l')}^{-1} \cdot g_{\varphi(1), \varphi'(1)})$. Hence the transition functions of $CL$ are exactly $g_{(\tau, \varphi), (\tau', \varphi')}(\gamma) := g_{\varphi(l), \varphi'(l')}^{-1}\gamma(1) \cdot g_{\varphi(1), \varphi'(1)}\gamma(0)$.

\paragraph{}We can give a generalization of this construction: let us consider a line bundle $L \rightarrow X$ and a subset $Y \subset X$: we can consider the space $C_{Y}X$ of open curves in $X$ with boundary in $Y$, i.e.\ such that $\gamma(0), \gamma(1) \in Y$. In this case, we have $\pi: C_{Y}X \rightarrow Y \times Y$ and holonomy is a section of the bundle $C_{Y}L = \pi^{*}({L\vert_{Y}}^{\checkmark} \boxtimes L\vert_{Y})$. Thus, to have a function we only need the triviality of $L\vert_{Y}$ and one of its global sections, it is not necessary that the whole $L$ is trivial.

\subsection{Local description}

\subsubsection{Overview of local description}

We want to define Wilson loop, which is the integral of the connection over a curve \emph{as an $\mathbb{R}/\mathbb{Z}$-valued function}, and it will turn out that its exponential is exactly the holonomy:
	\[\Hol_{A}(\gamma) \,=\, \exp\biggl( \, 2\pi i \cdot \int_{\gamma} A \, \biggr).
\]
As we have seen in the global description, for generic bundles with connection $(\{g_{\alpha\beta}\}, \{-A_{\alpha}\}) \in \check{H}^{1}(\mathfrak{U},\, \underline{S}^{1} \rightarrow \Omega^{1}_{\mathbb{R}})$, we can define Wilson loop as an $\mathbb{R}/\mathbb{Z}$-valued function only on \emph{closed curves}. In fact, let us consider $\gamma: S^{1} \rightarrow X$ and, once we have fixed a good cover $\mathfrak{U} = \{U_{\alpha}\}_{\alpha \in I}$ of $X$, let us suppose for the moment that $\gamma(S^{1}) \subset U_{\alpha}$ for a fixed $\alpha \in I$. In this case we define:
	\[\int_{\gamma} A \; := \; \int_{\gamma} A_{\alpha}.
\]
If it happens that $\gamma(S^{1}) \subset U_{\alpha\beta}$, we have two definitions of Wilson loop, but they coincide:
\begin{equation}\label{WLOneChart}
	\int_{\gamma} A_{\alpha} = \int_{\gamma} A_{\beta} + 2\pi\int_{\gamma} d\rho_{\alpha\beta} = \int_{\gamma} A_{\beta} + 2\pi\int_{\partial\gamma} \rho_{\alpha\beta} = \int_{\gamma} A_{\beta}.
\end{equation}
In the last expression it was crucial that $\gamma$ was a closed curve.\footnote{In this case we have an $\mathbb{R}$-valued function, but this is due to the assumption that $\gamma$ is contained in only one chart. We will see in the following that without this hypotheses we have the $\mathbb{Z}$-uncertainty.}

\paragraph{}For open curves, we can give a good definition of Wilson loop only for \emph{trivial} line bundles. Actually, it seems from \eqref{WLOneChart} that it is sufficient that the the bundle is torsion: in fact, as we have seen, we can realize it with constant transition functions and, in this case, $d\rho_{\alpha\beta} = 0$ and $A$ is globally defined. Actually, we will see that this is not a good definition of Wilson loop if the bundle is non-trivial: in fact, if we assume this definition for torsion non-trivial bundles, the exponential is not the holonomy.

For a trivial bundle $L_{g} = \{g_{\alpha\beta}\} \in \check{B}^{1}(\mathfrak{U}, \underline{S}^{1})$, we fix a trivialization $\{g_{\alpha}\} \in \check{C}^{0}(\mathfrak{U}, \underline{S}^{1})$ so that $g_{\alpha\beta} = g_{\alpha}^{-1} \cdot g_{\beta}$. In particular, for $g_{\alpha\beta} = e^{2\pi i \cdot \rho_{\alpha\beta}}$ and $g_{\alpha} = e^{2\pi i \cdot \rho_{\alpha}}$, we have that $\rho_{\alpha\beta} = \rho_{\alpha} - \rho_{\beta} + 2\pi i \cdot n$, with $n \in \mathbb{Z}$. Thus \eqref{TransitionConnection} becomes:
	\[\begin{split}
	&A_{\alpha} = A_{\beta} + 2\pi \,d(\rho_{\alpha} - \rho_{\beta})\\
	&A_{\alpha} - 2\pi\, d\rho_{\alpha} = A_{\beta} - 2\pi\, d\rho_{\beta}
\end{split}\]
so that we obtain a global form (or gauge-invariant form) $\tilde{A} \in \Omega^{1}_{\mathbb{R}}(X)$ such that $\tilde{A}\vert_{U_{\alpha}} = A_{\alpha} - 2\pi\, d\rho_{\alpha}$. This is equivalent to having fixed a global section $s$. We thus define, for a curve $\gamma: [\,0,1] \rightarrow X$ with respect to the global section $s$:
	\[\int_{\gamma} A \,:=\, \int_{\gamma} \tilde{A}
\]
which, for $\gamma([\,0,1]) \subset U_{\alpha}$, becomes:
\begin{equation}\label{WLTrivial}
\begin{split}
	&\int_{\gamma} A = \int_{\gamma} \bigl( A_{\alpha} - 2\pi\, d\rho_{\alpha} \bigr) = \int_{\gamma} A_{\alpha} - 2\pi\, \rho_{\alpha}(\gamma(1)) + 2\pi\, \rho_{\alpha}(\gamma(0))\\
	&\Hol_{A}(\gamma) \,=\, \exp\biggl( 2\pi i \cdot \int_{\gamma} A_{\alpha} \biggr) \cdot g_{\alpha}(\gamma(1))^{-1} \cdot g_{\alpha}(\gamma(0)).
\end{split}
\end{equation}
If we started with the trivial bundle with a global section, so that $g_{\alpha\beta} = 1$, then $\rho_{\alpha\beta} = 0$, thus $\tilde{A} = A$ and we recover the expression with the globally-defined connection. Actually, $\tilde{A}$ is exactly the expression of $A$ after a local coordinate change which makes the transition functions equal to $1$.

For torsion line bundles the definition of the global form works however, since $g_{\alpha\beta} = g_{\alpha}^{-1} \cdot g_{\beta} \cdot c_{\alpha\beta}$ with $c_{\alpha\beta}$ constant, so that $d\rho_{\alpha\beta} = d\rho_{\alpha} - d\rho_{\beta}$ anyway. But, as we said, we have problems in the definition of the holonomy.

\paragraph{}What happens for generic bundles with sections? It seems natural to find a generalization of the definition given for trivial bundles. Since we do not have the trivialization $\{g_{\alpha}\} \in \check{C}^{0}(\mathfrak{U}, \underline{S}^{1})$, i.e.\ we do not have a global section, for $\gamma: [\,0,1] \rightarrow X$ with $\gamma([\,0,1]) \subset U_{\alpha}$ we can only keep the fixed sections and give the same definition as for closed curves:
	\[\int_{\gamma} A \; := \; \int_{\gamma} A_{\alpha}.
\]
Of course it is not a well-defined function, but, as we have seen, it is a section of a line bundle. In fact, let us suppose that $\gamma([\,0,1]) \subset U_{\alpha\beta}$. Then we have:
	\[\begin{split}
	&\int_{\gamma} A_{\alpha} = \int_{\gamma} A_{\beta} + \int_{\gamma} d\rho_{\alpha\beta} = \int_{\gamma} A_{\beta} + \rho_{\alpha\beta}(\gamma(1)) - \rho_{\alpha\beta}(\gamma(0))\\
	&\exp\biggl( 2\pi i \cdot \int_{\gamma} A_{\alpha} \biggr) \,=\, \exp\biggl( 2\pi i \cdot \int_{\gamma} A_{\beta} \biggr) \cdot g_{\alpha\beta}(\gamma(1)) \cdot g_{\alpha\beta}(\gamma(0))^{-1}.
\end{split}\]
To interpret this expression, let us consider the space $CU_{\alpha}$ of curves contained in $U_{\alpha}$ (for $\alpha$ fixed), i.e.\ the space of maps $\gamma: [\,0,1] \rightarrow U_{\alpha}$ with the compact-open topology. If the fixed good covering of $X$ is $\mathfrak{U} = \{U_{\beta}\}_{\beta \in I}$, we can define a covering $\mathcal{V} = \{V_{\rho}\}_{\rho \in I}$ of $CU_{\alpha}$ given by $V_{\rho} = \{\rho \in CU_{\alpha}: \, \gamma([\,0,1]) \subset U_{\alpha\rho}\}$, with the notation $U_{\alpha\alpha} = U_{\alpha}$. Of course, $V_{\alpha}$ is the whole $CU_{\alpha}$. Thus we have an expression of the holonomy for every $V_{\rho}$:
	\[\Hol^{\rho}_{A}(\gamma) = \exp\biggl( 2\pi i \cdot \int_{\gamma} A_{\rho} \biggr)
\]
linked by the expression:
	\[\Hol^{\rho}_{A}(\gamma) \,=\, \Hol^{\eta}_{A}(\gamma) \cdot h_{\rho\eta}(\gamma) \,, \qquad h_{\rho\eta}(\gamma) = g_{\rho\eta}(\gamma(1)) \cdot g_{\rho\eta}(\gamma(0))^{-1}.
\]
Moreover, we have:
	\[\begin{split}
	h_{\rho\eta}(\gamma) &\cdot h_{\eta\chi}(\gamma) \cdot h_{\chi\rho}(\gamma)\\
	& = g_{\rho\eta}(\gamma(1)) \cdot g_{\rho\eta}(\gamma(0))^{-1} \cdot g_{\eta\chi}(\gamma(1)) \cdot g_{\eta\chi}(\gamma(0))^{-1} \cdot g_{\chi\rho}(\gamma(1)) \cdot g_{\chi\rho}(\gamma(0))^{-1}\\
	& = \bigl[ g_{\rho\eta}(\gamma(1)) \cdot g_{\eta\chi}(\gamma(1)) \cdot g_{\chi\rho}(\gamma(1)) \bigr] \cdot \bigl[ g_{\rho\eta}(\gamma(0)) \cdot g_{\eta\chi}(\gamma(0)) \cdot g_{\chi\rho}(\gamma(0)) \bigr]^{-1}\\
	& = 1
\end{split}\]
thus we can think $\Hol_{A}(\gamma)$ as \emph{a section of a line bundle $CL_{\alpha}$ over $CU_{\alpha}$} with transition functions $\{h_{\rho\eta}\} \in \check{H}^{1}(CU_{\alpha}, \underline{S}^{1})$. In this way we have a clear meaning for the holonomy: the fact that it is not well-defined as a function does not mean that it is not well-defined at all, it is just a section of a line bundle, hence it is naturally different with respect to different charts. Under the hypothesis that $\gamma$ is contained in $U_{\alpha}$ we have that $V_{\alpha}$ is a global chart for $CL_{\alpha}$, hence the bundle is trivial: we will see that for generic curves we obtain a non-trivial bundle, but the picture is the same.

\paragraph{}Let us now suppose that $L$ is trivial with sections, so that we think to is as $X \times \mathbb{C}$ with a trivialization $\{g_{\alpha}\} \in \check{C}^{0}(\mathfrak{U}, \underline{S}^{1})$. Then we have:
	\[\begin{split}
	h_{\rho\eta}(\gamma) &= g_{\rho\eta}(\gamma(1)) \cdot g_{\rho\eta}(\gamma(0))^{-1}\\
	& = \bigl[ g_{\rho}(\gamma(1))^{-1} \cdot g_{\eta}(\gamma(1)) \bigr] \cdot \bigl[ g_{\rho}(\gamma(0)) \cdot g_{\eta}(\gamma(0))^{-1} \bigr]\\
	&= \bigl[ g_{\rho}(\gamma(0))^{-1} \cdot g_{\rho}(\gamma(1)) \bigr]^{-1} \cdot \bigl[ g_{\eta}(\gamma(0))^{-1} \cdot g_{\eta}(\gamma(1)) \bigr]\\
	&= h_{\rho}(\gamma)^{-1} \cdot h_{\eta}(\gamma)
\end{split}\]
for $h_{\rho}(\gamma) = g_{\rho}(\gamma(0))^{-1} \cdot g_{\rho}(\gamma(1))$. Thus, a trivialization of $L$ determines \emph{canonically} a trivialization of $CL_{\alpha}$: in this way, the section $\Hol_{A}$ becomes a function. Let us see which function: when we have a section $\{s_{\alpha}\} \in \check{C}^{0}(\mathfrak{U}, \underline{S}^{1})$ of a trivial bundle $\{g_{\alpha\beta}\} \in \check{B}^{1}(\mathfrak{U}, \underline{S}^{1})$, so that $\check{\delta}^{0}(\{s_{\alpha}\}) = \{g_{\alpha\beta}\}$, and a trivialization $\{g_{\alpha}\} \in \check{C}^{0}(\mathfrak{U}, \underline{S}^{1})$, so that $\check{\delta}^{0}(\{g_{\alpha}\}) = \{g_{\alpha\beta}\}$, we obtain a function $f: X \rightarrow S^{1}$ by $f_{\alpha} = s_{\alpha} \cdot g_{\alpha}^{-1}$, since:
	\[\begin{split}
	&s_{\alpha}^{-1} \cdot s_{\beta} = g_{\alpha\beta} = g_{\alpha}^{-1} \cdot g_{\beta} \; \Longrightarrow \; s_{\alpha} \cdot g_{\alpha}^{-1} = s_{\beta} \cdot g_{\beta}^{-1}
\end{split}\]
or, equivalently, $\check{\delta}^{0}(\{s_{\alpha} \cdot g_{\alpha}^{-1}\}) = \{g_{\alpha\beta} \cdot g_{\alpha\beta}^{-1}\} = 1$ (this is the usual statement that the ratio of two trivializations is a function). Thus, for $CL_{\alpha}$, we obtain the function:
	\[\Hol_{A}(\gamma) = \Hol^{\rho}_{A}(\gamma) \cdot h_{\rho}(\gamma)^{-1} = \exp\biggl( \int_{\gamma} 2\pi i \cdot A_{\rho} \biggr) \cdot g_{\rho}(\gamma(1))^{-1} \cdot g_{\rho}(\gamma(0))
\]
which is exactly \eqref{WLTrivial}.

\paragraph{}Thus we recovered the picture stated in the global description: the holonomy of a line bundle $L \rightarrow X$ is a well-defined function over the loop space, while for open curves, it is a section of a line bundle $CL \rightarrow CX$; if $L$ is trivial then $CL$ is trivial too, and a trivialization of $L$ determines canonically\footnote{In general, when we have a trivial bundle $\{g_{\alpha\beta}\} \in \check{B}^{1}(X, C^{0}(S^{1}))$ with a section $\{s_{\alpha}\} \in \check{C}^{0}(X, C^{0}(\mathbb{C}))$, so that $s_{\beta} = g_{\alpha\beta}s_{\alpha}$, given a trivialization $\{g_{\alpha}\} \in \check{C}^{0}(X, C^{0}(S^{1}))$ we obtain a function, since $s_{\beta} = g_{\alpha}g_{\beta}^{-1}s_{\alpha}$, hence $g_{\beta}s_{\beta} = g_{\alpha}s_{\alpha}$. However, this function depends on the trivialization: if we have $\rho \in \check{H}^{0}(X, C^{0}(S^{1}))$, then also $\{g_{\alpha} \cdot \rho\vert_{U_{\alpha}}\}$ is a trivialization, and the function we obtain is $\rho\vert_{U_{\alpha}} \cdot g_{\alpha}s_{\alpha}$, corresponding to the fact that $\rho$ is a non-zero section of $X \times \mathbb{C}$, hence an automorphism of $X \times \mathbb{C}$ giving a different trivialization of the original bundle. Thus, to have a function is not sufficient to have a section of a trivial bundle, we must also have a canonical trivialization.} a trivialization of $CL$, so that for any trivial bundle we have a canonically defined holonomy also for open curves. We now give the same picture without assuming that the curve (closed or open) lies in only one chart: we can describe the integral of $A$ in two ways using local charts, or globally using the principal bundle formulation.

\subsubsection{Local description: first way}

\paragraph{Closed curves} Closed curves are elements of the \emph{loop space} of $X$. We start from the good cover $\{U_{i}\}_{i \in I}$ of $X$ and we consider the open covering $\{V_{(\tau,\sigma}\}_{(\tau,\sigma) \in J}$ of $LX$ defined in section \ref{TriangCovers}. Then, for a fixed $\gamma \in LX$, there exists $(\tau, \varphi) \in J$ such that $\gamma \in V_{(\tau, \varphi)}$. We define:
\begin{equation}\label{WilsonLoop}
\int_{\gamma} A \;:=\; \sum_{i=1}^{l(\tau)} \;\; \biggl[ \, \biggl( \int_{\gamma(\sigma^{1}_{i})}A_{\varphi(i)} \biggr) + \textstyle\frac{1}{2\pi i}\displaystyle \log g_{\varphi(i),\varphi(i+1)}\gamma(\sigma^{0}_{i+1}) \, \biggr]
\end{equation}
using the notation $l+1 = 1$. The logarithm can be taken since we have chosen a good covering, so the intersections are contractible. Of course, it is  defined up to an integer, so the quantity that can be well-defined as a number is $\exp\bigl(2\pi i \int_{\gamma} A\bigr)$.

\paragraph{}We now prove that $\int_{\gamma} A$ is well-defined in $\mathbb{R} / \mathbb{Z}$: it must be invariant under both the choice of the open set $V_{(\tau, \varphi)}$ to which $\gamma$ belongs and the choice of the open cover $\{U_{i}\}_{i \in I}$ of $X$. We prove it in steps:
\begin{itemize}
	\item We consider $V_{(\tau, \varphi')}$ such that $\varphi'$ differs from $\varphi$ just on $i \in \{1, \ldots, l\}$ fixed, so that $\gamma(\sigma^{1}_{i}) \subset U_{\varphi(i)\varphi'(i)}$. We suppose for simplicity $1 < i <l$. Then, the $i$-th summand becomes (we omit the $2\pi i$ factor):
	\[\begin{split}
	\int_{\gamma(\sigma^{1}_{i})} &A_{\varphi'(i)} + \log g_{\varphi'(i),\varphi(i+1)}\gamma(\sigma^{0}_{i+1})\\
	& = \int_{\gamma(\sigma^{1}_{i})} \Bigl( A_{\varphi(i)} + d\log g_{\varphi(i),\varphi'(i)} \Bigr) + \log g_{\varphi'(i),\varphi(i+1)}\gamma(\sigma^{0}_{i+1})\\
	& = \int_{\gamma(\sigma^{1}_{i})} A_{\varphi(i)} + \textcolor{red}{ \log g_{\varphi(i),\varphi'(i)}\gamma(\sigma^{0}_{i+1}) } - \textcolor{blue}{ \log g_{\varphi(i),\varphi'(i)}\gamma(\sigma^{0}_{i}) } \\
	& \phantom{XXXXXXXXXXXXXXXXXXX} + \textcolor{red}{ \log g_{\varphi'(i),\varphi(i+1)}\gamma(\sigma^{0}_{i+1}) }
\end{split}\]
while the $(i-1)$-th summand becomes:
	\[\int_{\gamma(\sigma^{1}_{i-1})}A_{\varphi(i-1)} + \textcolor{blue}{ \log g_{\varphi(i-1),\varphi'(i)} \gamma(\sigma^{0}_{i}) }.
\]
Now we the red terms give:
	\[\begin{split}
	\textcolor{red}{ \log g_{\varphi(i),\varphi'(i)} } & \textcolor{red}{ \gamma(\sigma^{0}_{i+1}) } + \textcolor{red}{ \log g_{\varphi'(i),\varphi(i+1)}\gamma(\sigma^{0}_{i+1}) }\\
	&= \log \bigl(g_{\varphi(i),\varphi'(i)}\gamma(\sigma^{0}_{i+1}) \cdot g_{\varphi'(i),\varphi(i+1)} \gamma(\sigma^{0}_{i+1}) \bigr)\\
	&= \log g_{\varphi(i),\varphi(i+1)} \gamma(\sigma^{0}_{i+1})
\end{split}\]
while the blue ones give:
	\[\begin{split}
	\textcolor{blue}{ \log g_{\varphi(i-1),\varphi'(i)} } & \textcolor{blue}{ \gamma(\sigma^{0}_{i}) } - \textcolor{blue}{ \log g_{\varphi(i),\varphi'(i)} \gamma(\sigma^{0}_{i}) }\\
	&= \log \bigl(g_{\varphi(i-1),\varphi'(i)} \gamma(\sigma^{0}_{i}) \cdot g_{\varphi'(i),\varphi(i)} \gamma(\sigma^{0}_{i+1}) \bigr)\\
	&= \log g_{\varphi(i-1),\varphi(i)} \gamma(\sigma^{0}_{i+1})
\end{split}\]
so that the sum of is equal for $\varphi$ and $\varphi'$.
\item In general, let us consider $\gamma \in V_{(\tau, \varphi)} \cap V_{(\tau', \varphi')}$. The first triangulation is $\tau = \{\sigma^{0}_{1}, \ldots, \sigma^{0}_{l};$ $\sigma^{1}_{1}, \ldots, \sigma^{1}_{l}\}$. We can group the vertices of $\tau'$ as divided by the vertices of $\tau$:
	\[\tau' = \{\; \eta^{0}_{1,1}, \ldots, \eta^{0}_{k_{1},1}, \; \ldots, \; \eta^{0}_{1,l}, \ldots, \eta^{0}_{k_{l},l}; \quad \eta^{1}_{1,1}, \ldots, \eta^{1}_{k_{1},1}, \; \ldots, \; \eta^{1}_{1,l}, \ldots, \eta^{1}_{k_{l},l} \;\}
\]
with the counterclockwise ordering $\sigma^{0}_{i} \leq \eta^{0}_{1,i} < \cdots < \eta^{0}_{k_{i},i} < \sigma^{0}_{i+1}$, still using the notation $l+1 = 1$. (Of course it can happen that $k_{i} = 0$ for some $i$.) We can now consider the triangulation given by the union of the vertices of $\tau$ and $\tau'$, with the corresponding edges:
	\[\begin{split}
	\tau \cup \tau' = \{\; &\sigma^{0}_{1}, \eta^{0}_{1,1}, \ldots, \eta^{0}_{k_{1},1}, \quad \sigma^{0}_{2}, \eta^{0}_{1,2}, \ldots, \eta^{0}_{k_{2},2}, \quad \ldots, \quad \sigma^{0}_{l}, \eta^{0}_{1,l}, \ldots, \eta^{0}_{k_{l},l};\\
	&\sigma^{1}_{1}, \eta^{1}_{1,1}, \ldots, \eta^{1}_{k_{1},1}, \quad \sigma^{1}_{2}, \eta^{1}_{1,2}, \ldots, \eta^{1}_{k_{2},2}, \quad \ldots, \quad \sigma^{1}_{l}, \eta^{1}_{1,l}, \ldots, \eta^{1}_{k_{l},l} \;\}.
\end{split}\]
We call for simplicity $\sigma^{0}_{i} = \eta^{0}_{0,i}$ and $\sigma^{1}_{i} = \eta^{1}_{0,i}$, and, if $\eta^{0}_{0,i} = \eta^{0}_{1,i}$, we simply rescale the indexes from $1$ to $k_{i}$. We can now redefine $\varphi$ on $\tau \cup \tau'$ declaring $\varphi(i,j) := \varphi(j)$. Supposing $g_{\alpha\alpha} = 1$, thus $\log g_{\alpha\alpha} = 0$, we obtain the same value of $\int_{\gamma} A$ considering $(\tau \cup \tau', \varphi)$ instead of $(\tau, \varphi)$.

We can apply the same procedure starting from $(\tau', \varphi')$, obtaining $(\tau \cup \tau', \varphi')$. Now we have that the two values of $\int_{\gamma} A$ corresponding to $V_{(\tau, \varphi)}$ and $V_{(\tau', \varphi')}$ are equal to the values corresponding to $V_{(\tau \cup \tau', \varphi)}$ and $V_{(\tau \cup \tau', \varphi')}$, i.e.\ we use the same triangulation. Now, applying the previous step chart by chart we obtain that the two values are the same.

\item If we choose a different covering of $X$, considering a common refinement of the two covering we can actually see that the result is the same.
\end{itemize}

\paragraph{Open curves} For $\gamma$ open, we first define Wilson loop for \emph{trivial} line bundles with a fixed trivialization. For a fixed $\gamma \in CX$, there exists $(\tau, \varphi) \in J$ such that $\gamma \in V_{(\tau, \varphi)}$. We define:
\begin{equation}\label{WilsonLine}
\begin{split}
	\int_{\gamma} A \;:=\; \sum_{i=1}^{l(\tau)} \; \biggl[ \, \int_{\gamma(\sigma^{1}_{i})}A_{\varphi(i)} \,+ & \,\textstyle \frac{1}{2\pi i} \displaystyle\, \log g_{\varphi(i)}\gamma(\sigma^{0}_{i+1}) \,- \textstyle \frac{1}{2\pi i} \displaystyle\, \log g_{\varphi(i)}\gamma(\sigma^{0}_{i}) \, \biggr] \\
	&= \sum_{i=1}^{l(\tau)} \; \biggl[ \, \int_{\gamma(\sigma^{1}_{i})}A_{\varphi(i)} + \,\textstyle \frac{1}{2\pi i} \displaystyle\, \Big\vert_{\gamma(\partial \sigma^{1}_{i})} \log g_{\varphi(i)} \, \biggr].
\end{split}
\end{equation}
Notice that $g_{\varphi(i)}$ has one pedix, so it is not a transition function but an element of the trivialization. As before, the logarithm can be taken since we have chosen a good cover and it is  defined up to $2\pi i \,\mathbb{Z}$, so that $\exp\bigl(\int_{\gamma} A\bigr)$ is defined as a number.

\paragraph{}For $1 \leq i < l$, we have that $\log g_{\varphi(i)} \gamma(\sigma^{0}_{i+1}) - \log g_{\varphi(i+1)} \gamma(\sigma^{0}_{i+1}) = \log g_{\varphi(i),\varphi(i+1)} \gamma(\sigma^{0}_{i+1})$, so that we recover exactly the expression of the closed case for these summand, but with a crucial difference at the extrema $\sigma^{0}_{1}$ and $\sigma^{0}_{l+1}$: when the curve is open, we cannot write $\log g_{\varphi(l),\varphi(1)} \gamma(\sigma^{0}_{i+1})$, since the curve does not close. Thanks to the trivialization we have the two pieces $\log g_{\varphi(l)} \gamma(\sigma^{0}_{l+1})$ and $-\log g_{\varphi(1)} \gamma(\sigma^{0}_{1})$.

\paragraph{}We now prove that $\int_{\gamma} A$ is well-defined in $\mathbb{C} / \mathbb{Z}$: it must be invariant under both the choice of the open set $V_{(\tau, \varphi)}$ to which $\gamma$ belongs and the choice of the open cover $\{U_{i}\}_{i \in I}$ of $X$. We prove it in steps:
\begin{itemize}
	\item We consider $V_{(\tau, \varphi')}$ such that $\varphi'$ differs from $\varphi$ just on $i \in \{1, \ldots, l\}$ fixed, so that $\gamma(\sigma^{1}_{i}) \subset U_{\varphi(i)\varphi'(i)}$. Then:
	\[\begin{split}
	\int_{\gamma(\sigma^{1}_{i})}A_{\varphi'(i)} &= \int_{\gamma(\sigma^{1}_{i})}\Bigl(A_{\varphi(i)} + d\log g_{\varphi(i)\varphi'(i)} \Bigr) = \int_{\gamma(\sigma^{1}_{i})}A_{\varphi(i)} + \Big\vert_{\gamma(\partial \sigma^{1}_{i})} \log g_{\varphi(i)\varphi'(i)}\\
	&= \int_{\gamma(\sigma^{1}_{i})}A_{\varphi(i)} + \Big\vert_{\gamma(\partial \sigma^{1}_{i})} \log g_{\varphi(i)} - \Big\vert_{\gamma(\partial \sigma^{1}_{i})} \log g_{\varphi'(i)}
\end{split}\]
thus:
	\[\int_{\gamma(\sigma^{1}_{i})}A_{\varphi'(i)} + \Big\vert_{\gamma(\partial \sigma^{1}_{i})} \log g_{\varphi'(i)} = \int_{\gamma(\sigma^{1}_{i})}A_{\varphi(i)} + \Big\vert_{\gamma(\partial \sigma^{1}_{i})} \log g_{\varphi(i)}
\]
so the $i$-th summand does not change.
\item In general, let us consider $\gamma \in V_{(\tau, \varphi)} \cap V_{(\tau', \varphi')}$. The first triangulation is $\tau = \{\sigma^{0}_{1}, \ldots, \sigma^{0}_{l}, \sigma^{0}_{l+1};$ $\sigma^{1}_{1}, \ldots, \sigma^{1}_{l}\}$. We can group the vertices of $\tau'$ as divided by the vertices of $\tau$:
{ \small
	\[\tau' = \{\; \eta^{0}_{1,1}, \ldots, \eta^{0}_{k_{1},1}, \; \ldots, \; \eta^{0}_{1,l}, \ldots, \eta^{0}_{k_{l},l}, \; \eta^{0}_{1,l+1}; \quad \eta^{1}_{1,1}, \ldots, \eta^{1}_{k_{1},1}, \; \ldots, \; \eta^{1}_{1,l}, \ldots, \eta^{1}_{k_{l},l} \;\}
\] }
with $\sigma^{0}_{i} \leq \eta^{0}_{1,i} < \cdots < \eta^{0}_{k_{i},i} < \sigma^{0}_{i+1}$. (Of course it can happen that $k_{i} = 0$ for some $i$.) We can now consider the triangulation given by the union of the vertices of $\tau$ and $\tau'$, with the corresponding edges:
{ \small
	\[\begin{split}
	\tau \cup \tau' = \{\; &\sigma^{0}_{1}, \eta^{0}_{1,1}, \ldots, \eta^{0}_{k_{1},1}, \quad \sigma^{0}_{2}, \eta^{0}_{1,2}, \ldots, \eta^{0}_{k_{2},2}, \quad \ldots, \quad \sigma^{0}_{l}, \eta^{0}_{1,l}, \ldots, \eta^{0}_{k_{l},l}, \quad \sigma^{0}_{l+1};\\
	&\sigma^{1}_{1}, \eta^{1}_{1,1}, \ldots, \eta^{1}_{k_{1},1}, \quad \sigma^{1}_{2}, \eta^{1}_{1,2}, \ldots, \eta^{1}_{k_{2},2}, \quad \ldots, \quad \sigma^{1}_{l}, \eta^{1}_{1,l}, \ldots, \eta^{1}_{k_{l},l} \qquad\qquad\}.
\end{split}\] }
We call for simplicity $\sigma^{0}_{i} = \eta^{0}_{0,i}$ and $\sigma^{1}_{i} = \eta^{1}_{0,i}$, and, if $\eta^{0}_{0,i} = \eta^{0}_{1,i}$, we simply rescale the indexes from $1$ to $k_{i}$. We can now redefine $\varphi$ on $\tau \cup \tau'$ declaring $\varphi(i,j) := \varphi(j)$. Since the boundary terms $\vert_{\gamma(\partial \sigma^{1}_{i})} \log g_{\varphi(i)}$ simplifies for each group $\eta^{1}_{\,\cdot\,,i}$, we obtain the same value of $\int_{\gamma} A$ considering $(\tau \cup \tau', \varphi)$ instead of $(\tau, \varphi)$.

We can apply the same procedure starting from $(\tau', \varphi')$, obtaining $(\tau \cup \tau', \varphi')$. Now we have that the two values of $\int_{\gamma} A$ corresponding to $V_{(\tau, \varphi)}$ and $V_{(\tau', \varphi')}$ are equal to the values corresponding to $V_{(\tau \cup \tau', \varphi)}$ and $V_{(\tau \cup \tau', \varphi')}$, i.e.\ we use the same triangulation. Now, applying the previous step chart by chart we obtain that the two values are the same.
\end{itemize}

\paragraph{}For non-trivial bundles, the only possibility is to use the same definition as for closed curves, but without the term $\log g_{\varphi(l),\varphi(1)}(\sigma^{0}_{i+1})$, since this term requires the closure of the curve. Thus we obtain:
\begin{equation}\label{WilsonLineBundle}
\int_{\gamma} A \;:=\; \Biggl(\sum_{i=1}^{l(\tau)-1} \; \int_{\gamma(\sigma^{1}_{i})}A_{\varphi(i)} + \log g_{\varphi(i),\varphi(i+1)}\gamma(\sigma^{0}_{i+1}) \Biggr) + \int_{\gamma(\sigma^{1}_{l})}A_{\varphi(l)}.
\end{equation}
In this case the integral is not well-defined as a function, but we now prove that it is \emph{a section of a line bundle}. The steps of the proof of well-definiteness for close curves works also in this case, except for the extrema of $\gamma$. In particular:
\begin{itemize}
	\item it is still true that, given two charts $V_{(\tau, \varphi)}$ and $V_{(\tau', \varphi')}$, we can consider $V_{(\tau \cup \tau', \varphi)}$ and $V_{(\tau \cup \tau', \varphi')}$ obtaining the same values; in fact, the boundary vertices are the same for $\tau, \, \tau'$ and $\tau \cup \tau'$, and they have no role in unifying $\tau$ and $\tau'$ in the integral (see previous step);
	\item the sum of blue and red terms works in the same way for all the intermediate summands $1 < i < l(\tau)$.
\end{itemize}
Thus the only difference are the terms $\textcolor{red}{ \log g_{\varphi(l),\varphi'(l)}\gamma(1) } - \textcolor{blue}{ \log g_{\varphi(1),\varphi'(1)}\gamma(0) }$, in particular:
	\[\biggl(\int_{\gamma} A\biggr)_{(\tau', \varphi')} = \biggl(\int_{\gamma} A\biggr)_{(\tau, \varphi)} + \textcolor{red}{ \log g_{\varphi(l),\varphi'(l)}\gamma(1) } - \textcolor{blue}{ \log g_{\varphi(1),\varphi'(1)} \gamma(0) }
\]
where $l$ is the length of $\tau \cup \tau'$, $g_{\varphi(l),\varphi'(l)}$ is the transition function from the last chart of $\tau$ to the last chart of $\tau'$ and $g_{\varphi(1),\varphi'(1)}$ is the transition function from the first chart of $\tau$ to the first chart of $\tau'$. In particular:
	\[\exp\biggl(\int_{\gamma} A\biggr)_{(\tau', \varphi')} = \exp\biggl(\int_{\gamma} A\biggr)_{(\tau, \varphi)} \cdot \textcolor{red}{ g_{\varphi(l),\varphi'(l)}\gamma(1) } \cdot \textcolor{blue}{ g_{\varphi(1),\varphi'(1)}\gamma(0)^{-1} }.
\]
Thus, we can interpret this expression thinking to have a transition function from $V_{(\tau, \varphi)}$ to $V_{(\tau', \varphi')}$ given by:
	\[\tilde{g}_{(\tau, \varphi), (\tau', \varphi')}(\gamma) = g_{\varphi(l),\varphi'(l)}\gamma(1) \cdot g_{\varphi(1),\varphi'(1)}\gamma(0)^{-1}.
\]
If we consider three charts $V_{(\tau, \varphi)}$, $V_{(\tau', \varphi')}$ and $V_{(\tau'', \varphi'')}$, these transition functions satisfy a cocycle condition:
	\[\begin{split}
	\bigl( g_{\varphi(l),\varphi'(l)}&\gamma(1) \cdot g_{\varphi(1),\varphi'(1)}\gamma(0)^{-1} \bigr) \cdot \bigl( g_{\varphi'(l),\varphi''(l)}\gamma(1) \cdot g_{\varphi'(1),\varphi''(1)}\gamma(0)^{-1} \bigr)\\
	& \phantom{XXXXXXXXXXXXX} \cdot \bigl( g_{\varphi''(l),\varphi(l)}\gamma(1) \cdot g_{\varphi''(1),\varphi(1)}\gamma(0)^{-1} \bigr)\\
	& = \bigl( g_{\varphi(l),\varphi'(l)}\gamma(1) \cdot g_{\varphi'(l),\varphi''(l)}\gamma(1) \cdot g_{\varphi''(l),\varphi(l)}\gamma(1) \bigr)\\
	& \phantom{XXXXXXXXXXXXX} \cdot \bigl( g_{\varphi''(1),\varphi(1)}\gamma(0) \cdot g_{\varphi'(1),\varphi''(1)}\gamma(0) \cdot g_{\varphi(1),\varphi'(1)}\gamma(0) \bigr)^{-1}\\
	& = 1.
\end{split}\]
Thus we obtain:
\begin{Theorem} For $A$ connection on a bundle $L \rightarrow X$, $\exp \int_{\gamma} A$ is a section of a line bundle $CL \rightarrow CX$, whose transition function from $V_{(\tau, \varphi)}$ to $V_{(\tau', \varphi')}$ is given by $\tilde{g}_{(\tau, \varphi), (\tau', \varphi')}(\gamma) = g_{\varphi(l),\varphi'(l)}\gamma(1) \cdot g_{\varphi(1),\varphi'(1)}\gamma(0)^{-1}$.
\end{Theorem}

\paragraph{}From the transition functions we see that \emph{if the bundle $L$ is trivial, then $CL$ is trivial too}, since, for $g_{\alpha\beta} = g_{\alpha} \cdot g_{\beta}^{-1}$ on $X$, we have:
	\[\begin{split}
	\tilde{g}_{(\tau, \varphi), (\tau', \varphi')}(\gamma) &= g_{\varphi(l)}\gamma(1) \cdot g_{\varphi'(l)}\gamma(1)^{-1} \cdot g_{\varphi'(1)}\gamma(0) \cdot g_{\varphi(1)}\gamma(0)^{-1}\\
	&= \bigl(g_{\varphi(l)}\gamma(1) \cdot g_{\varphi(1)}\gamma(0)^{-1}\bigr) \cdot \bigl( g_{\varphi'(l)}\gamma(1) \cdot g_{\varphi'(1)}\gamma(0)^{-1} \bigr)^{-1}
\end{split}\]
so that we have a trivialization:
\begin{equation}\label{Trivialization}
\tilde{g}_{(\tau, \varphi)}(\gamma) = g_{\varphi(l)}\gamma(1) \cdot g_{\varphi(1)}\gamma(0)^{-1}.
\end{equation}
Moreover, as we see from the previous expression, a trivialization of $L$ determines \emph{canonically} a trivialization of $CL$. Thus, when $L$ is given with a trivialization, i.e.\ when $A$ is a connection on $X \times \mathbb{C}$, then $CL$ is canonically trivialized, i.e.\ $\exp \int_{\gamma} A$ is a section of $CX \times \mathbb{C}$, so it is a function.

\paragraph{}We now see that, under the trivialization \eqref{Trivialization}, the function we obtain from $\exp \int_{\gamma} A$ is exactly \eqref{WilsonLine}. In fact, from the section $\{s_{\alpha}\}$ and the trivialization $\{g_{\alpha}\}$ we obtain the function $\{g_{\alpha}s_{\alpha}\}$, in this case we obtain from \eqref{WilsonLineBundle} and \eqref{Trivialization}:
	\[\begin{split}
	\exp &\biggl(\int_{\gamma} A\biggr)_{(\tau, \varphi)} \cdot \tilde{g}_{(\tau, \varphi)}(\gamma)\\
	& = \prod_{i=1}^{l(\tau)-1}\; \exp \Biggl[ \Biggl(\int_{\gamma(\sigma^{1}_{i})}A_{\varphi(i)} \Biggr) \cdot g_{\varphi(i),\varphi(i+1)}\gamma(\sigma^{0}_{i+1}) \Biggr] \cdot \exp\Biggl( \int_{\gamma(\sigma^{1}_{l})}A_{\varphi(l)} \Biggr)\\
	&\phantom{XXXXXXXXXXXXXXXXXXXXXXX} \cdot g_{\varphi(l)}\gamma(1) \cdot g_{\varphi(1)}\gamma(0)^{-1}\\
	& = \prod_{i=1}^{l(\tau)-1} \; \exp \Biggl[ \Biggl( \int_{\gamma(\sigma^{1}_{i})}A_{\varphi(i)} \Biggr) \cdot g_{\varphi(i)}\gamma(\sigma^{0}_{i+1}) \cdot g_{\varphi(i+1)}\gamma(\sigma^{0}_{i+1})^{-1} \Biggr] \cdot \exp\Biggl( \int_{\gamma(\sigma^{1}_{l})}A_{\varphi(l)} \Biggr)\\
	&\phantom{XXXXXXXXXXXXXXXXXXXXXXX} \cdot g_{\varphi(l)}\gamma(\sigma^{0}_{l+1}) \cdot g_{\varphi(1)}\gamma(\sigma^{0}_{1})^{-1}\\
	&= \prod_{i=1}^{l(\tau)} \; \exp \Biggl[ \Biggl( \int_{\gamma(\sigma^{1}_{i})}A_{\varphi(i)} \Biggr) \cdot g_{\varphi(i)}\gamma(\sigma^{0}_{i+1}) \cdot g_{\varphi(i)}\gamma(\sigma^{0}_{i})^{-1} \Biggr]
\end{split}\]
and the last expression is exactly the exponential of \eqref{WilsonLine}.

\subsubsection{Local description: second way}

\paragraph{Closed curves} We notice that any (complex) line bundle, restricted to a curve $\gamma$, is trivial since $H^{2}(\gamma, \mathbb{Z}) = 0$ for dimensional reasons. Thus we can choose a globally defined form $A_{0}: T\gamma \rightarrow \mathbb{R}$ and define:
	\[\int_{\gamma} A \,:=\, \int_{\gamma} A_{0}.
\]
If we choose a different trivialization of the bundle, we obtain a globally defined connection $A_{1}$ such that $A_{1} = A_{0} + g_{01}^{-1}\cdot dg_{01}$, but now, since $\gamma$ is not necessarily contractible, we cannot say that $g_{01}^{-1}\cdot dg_{01} = d\log g_{01}$: we can only consider it as a closed, but not necessarily exact, 1-form (it is a large gauge transformation). Hence:
	\[\int_{\gamma} A_{1} = \int_{\gamma} A_{0} + \int_{\gamma} g_{01}^{-1}\cdot dg_{01}.
\]
However, as we have seen, large gauge transformations are quantized, thus holonomy is well-defined also in this way. One can give the analogue definition for open curves and prove that this definition gives the same result of the previous one.

\subsection{Torsion line bundles}

We said in the first section that, for torsion non-trivial bundles, although the connection can be globally defined, holonomy is not a well-defined function for open curves. Let us consider the expression for closed curves \eqref{WilsonLoop}, in the case of a torsion line bundle realized with constant transition functions: in this case $A$ is globally defined, so that the first summand becomes exactly $\int_{\gamma} A$, but the second summand is however non-trivial, thus the integral of $A$ is just a piece of the holonomy. The same applies for open curves in \eqref{WilsonLineBundle}, so that we obtain a line bundle over loop space, which is torsion since the transition functions are constant. Thus, for $L$ torsion line bundle, also $CL$ is torsion, but not trivial if $L$ is not. This solves the problem stated in the first section.

\paragraph{}Let us consider the exact sequence:
	\[0 \longrightarrow \mathbb{Z} \longrightarrow \mathbb{R} \overset{e^{2\pi i \,\cdot}}\longrightarrow S^{1} \longrightarrow 0.
\]
A torsion line bundle can be realized by $[\,\{g_{\alpha\beta}\}\,] \in \check{H}^{1}(\mathfrak{U}, S^{1})$, such that $\beta([\,\{g_{\alpha\beta}\}\,]) = c_{1}(L) \in \check{H}^{2}(\mathfrak{U}, \mathbb{Z})$, and these transition functions gives a non trivial summand in \eqref{WilsonLoop} and \eqref{WilsonLineBundle}. If the bundle is trivial, by exactness we find a class $[\,\{\rho_{\alpha\beta}\}\,] \in \check{H}^{1}(\mathfrak{U}, \mathbb{R})$ whose image in $S^{1}$ is $[\,\{g_{\alpha\beta}\}\,]$, i.e.\ such that $g_{\alpha\beta} = e^{2\pi i\cdot \rho_{\alpha\beta}}$. If the corresponding class of $[\,\{\rho_{\alpha\beta}\}\,]$ in the de-Rham cohomology is $[\,A_{\rho}\,]$, we see from formula \eqref{IntDeRhamCech} that the second summand of \eqref{WilsonLoop} is exactly $\int_{\gamma} A_{\rho}$. Thus, if a \emph{trivial} bundle is realized with constant transition functions, we have a global connection $A$ but the holonomy is given by:
	\[\Hol_{A}(\gamma) = \exp \biggl[ 2\pi i \int_{\gamma} \bigl( A + A_{\rho} \bigr) \biggr].
\]
Thus, we cannot find the connection using the exact sequence, since such a sequence describes the bundle topologically, but a connection is not determined by the topology. The global real class we find for the trivial bundle is just the correction due to the fact that we have not chosen the transition functions all equal to 1. If we do this, we find a connection equal to $A + \tilde{A}_{\rho}$ where $\tilde{A}_{\rho}$ is a certain representative of $[\,A_{\rho}\,]$.

\paragraph{}If we have a \emph{trivial flat} bundle, we realize it with transition functions equal to $1$, so that we have a globally defined connection $A$. Now we have that:
\begin{itemize}
	\item since the bundle is flat, $dA = 0$, thus $A$ determines a de-Rham class $[\,A\,] \in H^{1}_{dR}(X)$;
	\item since the bundle is trivial, the holonomy is exactly $\Hol_{A}(\gamma) = \int_{\gamma} A$.
\end{itemize}
Thus, $\Hol_{A}$ can be seen exactly as the cohomology class $[\,A\,]$ as function over $H_{1}(X, \mathbb{R})$, i.e.\ as element of $H^{1}(X, \mathbb{R})$. Thus, for trivial flat bundle the holonomy is a real singular cohomology class corresponding to the de-Rham class of the globally-defined closed connection.

For non-trivial \emph{torsion flat} bundles, the globally-defined connection determines a de-Rham class $[\,A\,]$ since $dA = 0$, but, as we have seen, the holonomy does not coincide with the integral of $A$ and it is not in general a real cohomology class.

\section{Summary about connections and holonomy}

Summarizing, we have that classes of line bundles with connections form a group $(LB\nabla(X),$ $\otimes)$ with the following important subgroups:
\[\xymatrix{
TrLB\nabla(X) \ar@{^(->}[r] & TLB\nabla(X) \ar@{^(->}[r] & LB\nabla(X) \, .\\
TrFLB\nabla(X) \ar@{^(->}[u] \ar@{^(->}[r] & TFLB\nabla(X) \ar@{^(->}[u]
}
\]
We can describe cohomologically these scheme as:
\[\xymatrix{
\Omega^{1}_{\mathbb{R}}(X) \,/\, \mathcal{I}^{1}_{\mathbb{R}}(X) \ar@{^(->}[r] & TLB\nabla(X) \ar@{^(->}[r] & \check{H}^{1}\bigl(\, \mathfrak{U},\, \underline{S}^{1} \overset{d \,\circ\, \log}\longrightarrow \Omega^{1}_{\mathbb{R}} \,\bigr) \, .\\
\mathcal{Z}^{1}_{\mathbb{R}}(X) \,/\, \mathcal{I}^{1}_{\mathbb{R}}(X) \ar@{^(->}[u] \ar@{^(->}[r] & \check{H}^{1}(\mathfrak{U}, S^{1}) \ar@{^(->}[u]
}
\]

Thus the sequence \eqref{ExSeqLBNabla} is isomorphic to:
\begin{equation}\label{ExSeqLBNabla2}
0 \longrightarrow \check{H}^{1}(\mathfrak{U}, S^{1}) \longrightarrow \check{H}^{1}\bigl(\, \mathfrak{U},\, \underline{S}^{1} \rightarrow \Omega^{1}_{\mathbb{R}} \,\bigr) \longrightarrow \mathcal{I}^{2}(X, \mathbb{R}) \longrightarrow 0.
\end{equation}

\paragraph{}We have seen that connections on a fixed line bundle are an affine space whose underlying vector space is $\Omega^{1}_{\mathbb{R}}(X)$ since two connections differ by a 1-form with values in the endomorphisms of the line bundle, the latter being the multiplication by a function. Thus $\nabla'_{X}s = \nabla_{X}s + \omega(X)s$ for $\omega \in \Omega^{1}_{\mathbb{R}}(X)$. We can ask what is the quotient of this affine space by equivalence of connection with respect to automorphisms of the bundle: the quotient is an affine space with underlying vector space $\Omega^{1}_{\mathbb{R}}(X) \,/\, \mathcal{I}^{1}_{\mathbb{R}}(X)$ for $\mathcal{I}^{1}_{\mathbb{R}}(X)$ the space of integral 1-forms. In fact, if $\nabla'$ is the pull-back of $\nabla$ by an automorphism $\varphi(s) \rightarrow f\cdot s$ for $f$ non-zero function, we get $\nabla'_{X}s = \varphi^{-1}\nabla_{X}\varphi(s) = \frac{1}{f}(\partial_{X}f \cdot s + f\nabla_{X}s) = \nabla_{X}s + \frac{1}{f}df(X)s$ and we have seen that integral 1-forms are the one which can be expressed as $f^{-1}df$.

\section{Real Chern classes}

We now consider a more general object than a line bunlde, such that the first Chern class is not necessarily integral, as discussed in \cite[Chap.\ 6]{BFS}.

\subsection{Definition}

Let us consider the definition of Chern class of a trivial bundle: we have a bundle $[\,\{g_{\alpha\beta}\}\,] \in \check{H}^{1}(\mathfrak{U}, \underline{S}^{1})$, so that $g_{\alpha\beta} \cdot g_{\beta\gamma} \cdot g_{\gamma\alpha} = 1$; if $g_{\alpha\beta} = e^{2\pi i \cdot \rho_{\alpha\beta}}$, one has $\rho_{\alpha\beta} + \rho_{\beta\gamma} + \rho_{\gamma\alpha} = \rho_{\alpha\beta\gamma} \in \mathbb{Z}$, so that we obtain a class $[\,\{\rho_{\alpha\beta\gamma}\}\,] \in \check{H}^{2}(\mathfrak{U}, \mathbb{Z})$ which is the first Chern class.

Let us call $\Gamma_{n}$ the subgroup of $S^{1}$ given by the $n$-th root of unity. If we call $\frac{1}{n}\mathbb{Z}$ the subgroup of $\mathbb{R}$ made by the fractions $\frac{k}{n}$ for $k \in \mathbb{Z}$, we have that $\Gamma_{n} = e^{2\pi i \cdot \frac{1}{n}\mathbb{Z}}$. Let us suppose we have a cochain $\{g_{\alpha\beta}\} \in \check{C}^{1}(\mathfrak{U}, \underline{S}^{1})$ such that $g_{\alpha\beta} \cdot g_{\beta\gamma} \cdot g_{\gamma\alpha} = g_{\alpha\beta\gamma} \in \Gamma_{n}$. Then, for $g_{\alpha\beta} = e^{2\pi i \cdot \rho_{\alpha\beta}}$, we have that $\rho_{\alpha\beta} + \rho_{\beta\gamma} + \rho_{\gamma\alpha} = \rho_{\alpha\beta\gamma} \in \frac{1}{n}\mathbb{Z}$, so that we obtain a rational class $c_{1} = [\,\{\rho_{\alpha\beta\gamma}\}\,] \in \check{H}^{2}(\mathfrak{U}, \mathbb{Q})$ such that $n \cdot c_{1}$ is an integral class. We thus have a Chern class for certain cochains (up to coboundaries), which is integral for cocycles, i.e.\ for line bundles. Can we give a geometric interpretation of these classes?

\paragraph{}A 2-cochain can be thought of as a trivialization of a gerbe, so that a line bundle is a trivialization (or section) of a gerbe represented by transition functions equal to $1$. Since we are not dealing with gerbes yet, we need to lower by 1 the degree in cohomology. In particular, as line bundles, which are classes in $\check{H}^{1}(\mathfrak{U}, \underline{S}^{1})$, are trivializations of gerbes, similarly a section of a line bundle, represented by transition functions equal to 1, is a class in $\check{H}^{0}(\mathfrak{U}, \underline{S}^{1})$, i.e.\ a function $f: X \rightarrow S^{1}$. A cochain $\{f_{\alpha}\} \in \check{C}^{0}(\mathfrak{U}, \underline{S}^{1})$ is a section of a trivial bundle represented by transition functions $f_{\alpha}^{-1} \cdot f_{\beta}$.

Given a function $f: X \rightarrow S^{1}$, we can naturally define a Chern class $c_{1}(f) \in H^{1}(\mathfrak{U}, \mathbb{Z})$, which is the image under the Bockstein map of $f = [\,\{f_{\alpha}\}\,] \in \check{H}^{0}(\mathfrak{U}, \underline{S}^{1})$. We compute it as for bundles: since $f_{\alpha}^{-1} \cdot f_{\beta} = 1$, for $f_{\alpha} = e^{2\pi i \cdot \rho_{\alpha}}$ we have $\rho_{\beta} - \rho_{\alpha} = \rho_{\alpha\beta} \in \mathbb{Z}$, so that we have a class $c_{1}(f) = [\,\{\rho_{\alpha\beta}\}\,] \in \check{H}^{1}(\mathfrak{U}, \mathbb{Z})$. The geometric interpretation is very simple: $c_{1}(f)$ is the pull-back under $f$ of the generator of $H^{1}(S^{1}, \mathbb{Z}) \simeq \mathbb{Z}$. As for bundles, let us suppose we have a cochain $[\,\{f_{\alpha}\}\,] \in \check{C}^{0}(\mathfrak{U}, \underline{S}^{1})$ such that $f_{\alpha}^{-1} \cdot f_{\beta} = f_{\alpha\beta} \in \Gamma_{n}$. Then $\rho_{\beta} - \rho_{\alpha} = \rho_{\alpha\beta} \in \frac{1}{n} \mathbb{Z}$, so that we obtain a class $c_{1} = [\,\{\rho_{\alpha\beta}\}\,] \in \check{H}^{1}(\mathfrak{U}, \mathbb{Q})$ such that $n \cdot c_{1}$ is an integral class.

\paragraph{}By the exact sequences point of view, the Chern class is the image of the Bockstein map of the sequence:
	\[0 \longrightarrow \mathbb{Z} \longrightarrow \underline{\mathbb{R}} \overset{e^{2\pi i \, \cdot}}\longrightarrow \underline{S}^{1} \longrightarrow 0.
\]
In the fractionary case, since the coboundary lies in $\Gamma_{n}$, the cochain is a cocycle in $S^{1} / \,\Gamma_{n}$. Thus, we consider the sequence:
	\[0 \longrightarrow \textstyle \frac{1}{n} \displaystyle \mathbb{Z} \longrightarrow \underline{\mathbb{R}} \overset{\pi_{\Gamma_{n}} \,\circ\, e^{2\pi i \, \cdot}}\longrightarrow \underline{S}^{1} / \,\Gamma_{n} \longrightarrow 0
\]
and the image of the Bockstein map is exactly the fractionary Chern class. Actually, it is not necessary to obtain a rational class. It is sufficient that $\rho_{\alpha\beta}$ is constant for every $\alpha, \beta$ to obtain a real Chern class. The corresponding sequence, which contains all the previous by inclusion, is:
	\[0 \longrightarrow \textstyle \mathbb{R} \longrightarrow \underline{\mathbb{R}} \overset{\pi_{S^{1}} \,\circ\, e^{2\pi i \, \cdot}}\longrightarrow \underline{S}^{1} / \,S^{1} \longrightarrow 0.
\]
In other words, if the cochain is a cocycle up to constant functions, we obtain a real Chern class. If these constant functions belongs to $\Gamma_{n}$, we obtain a rational Chern class in $\frac{1}{n}\mathbb{Z}$. We now want to give a geometric interpretation of these classes.

\subsection{Geometric interpretation}

If we think of the cochain as a trivialization of a bundle, we have that different trivializations have different Chern classes, depending on the realization of the trivial bundle as $\rm\check{C}$ech coboundary. This seems quite unnatural, since the particular realization should not play any role. Actually, if we fix a flat connection, we can distinguish some trivializations which are parallel with respect to such a connection.

Let us consider a trivial line bundle with a global section and a flat connection $\nabla$, which we think of as $X \times \mathbb{C}$ with a globally defined by form $A$, expressing $\nabla$ with respect to the global section $1$. We know the following facts:
\begin{itemize}
	\item if choose parallel sections $\{s_{\alpha}\}$, we obtain a trivialization with a real Chern class $c_{1}(s) \in \check{H}^{1}(X, \mathbb{R})$, and the local expression of the connection becomes $\{0\}$;
	\item the globally defined connection $A$, expressed with respect to $1$, is closed by flatness, thus it determines a de-Rham cohomology class $[\,A\,] \in H^{1}_{dR}(X)$.
\end{itemize}
We now prove that these two classes coincide under the standard isomorphism. This is the geometric interpretation of real Chern classes: \emph{the real Chern class of a trivialization of $X \times \mathbb{C}$ corresponds to the cohomology class of a globally-defined flat connection, expressed with respect to $1$, for which the trivialization is parallel}. If the trivial bundle is also \emph{geometrically trivial}, by definition the holonomy is $1$: in fact, we can find a global parallel section, so that there exists a \emph{cocycle} $\{s_{\alpha}\} \in \check{Z}^{1}(X, \underline{S}^{1})$ trivializing the bundle, and the Chern class of a cocycle is integral. Thus, \emph{the image in $\mathbb{R}/\mathbb{Z}$ of the real Chern class of a trivialization of a bundle is the obstruction for the bundle to be also geometrically trivial, when endowed with the connection for which the trivialization is parallel}, which is exactly the holonomy.

We will find exactly the same situation for degree 2: the real Chern class of a 2-cochain is the class of a connection on a trivial gerbe for which this trivialization is parallel.

\paragraph{}We now prove the statement. Given $\{f_{\alpha}\} \in \check{C}^{0}(\mathfrak{U}, \underline{S}^{1})$ such that $\check{\delta}^{0}\{f_{\alpha}\} \in \check{C}^{0}(\mathfrak{U}, S^{1})$, we consider the connection $\nabla$ on $X \times \mathbb{C}$ which is represented by $0$ with respect to $\{f_{\alpha}\}$. If we represent $\nabla$ with respect to $X \times \{1\}$ we obtain $A_{\alpha} = d\log f_{\alpha}$, and $A_{\beta} - A_{\alpha} = d\log (f_{\alpha}^{-1} \cdot f_{\beta}) = 0$. We thus realize the 1-form $A$ as a $\rm\check{C}$ech cocycle: we have that $A_{\alpha} = d\log f_{\alpha}$ and $\log f_{\beta} - \log f_{\alpha} = \log g_{\alpha\beta} = \rho_{\alpha\beta}$ which is constant, so that $[\,A\,]_{H^{1}_{dR}(X)} \simeq [\,\{\rho_{\alpha\beta}\}\,]_{\check{H}^{1}(\mathfrak{U}, \mathbb{R})}$. By definition, the first Chern class of $\{f_{\alpha}\}$ is computed by taking logarithms so that $\log f_{\beta} - \log f_{\alpha} = \rho_{\alpha\beta}$ constant and $c_{1}(\{f_{\alpha}\}) = [\,\{\rho_{\alpha\beta}\}\,]$. Thus $[\,A\,]_{H^{1}_{dR}(X)} \simeq c_{1}(\{f_{\alpha}\})_{\check{H}^{1}(\mathfrak{U}, \mathbb{R})}$.

\chapter{Gerbes}

\section{Cohomology and gerbes}

Gerbes are the concept equivalent to line bundles with hermitian metric, but increasing by $1$ the degree in cohomology. We define an \emph{equivalence class of gerbes} as a cohomology class $[\,\{g_{\alpha\beta\gamma}\}\,] \in \check{H}^{2}(\mathfrak{U}, \underline{S}^{1})$. We define a \emph{gerbe with sections} as a cocycle $\{g_{\alpha\beta\gamma}\} \in \check{Z}^{2}(\mathfrak{U}, \underline{S}^{1})$, so that its equivalence class is the corresponding cohomology class. Actually, there is not a good notion of sections, but we use this term by analogy with vector bundles. By the Bockstein homomorphism of the sequence:
	\[0 \longrightarrow \mathbb{Z} \longrightarrow \underline{\mathbb{R}} \overset{e^{2\pi i \, \cdot\,}}\longrightarrow \underline{S}^{1} \longrightarrow 0
\]
we describe an equivalence class of gerbes as an element of $\check{H}^{3}(\mathfrak{U}, \mathbb{Z})$, which we call the \emph{first Chern class of the gerbe}. It can be directly computed as for line bundles: we have $g_{\alpha\beta\gamma} = e^{2\pi i \cdot \rho_{\alpha\beta\gamma}}$ so that $\check{\delta}^{2}\{\rho_{\alpha\beta\gamma}\} = \{\rho_{\alpha\beta\gamma\delta}\} \in \check{Z}^{3}(\mathfrak{U}, \mathbb{Z})$, and we have $c_{1}(\mathcal{G}) = [\,\{\rho_{\alpha\beta\gamma\delta}\}\,] \in \check{H}^{3}(\mathfrak{U}, \mathbb{Z})$.

\paragraph{}We now discuss connections, starting with the case of connections up to flat ones. For a line bundle with sections, it turns out that the connections $\{A_{\alpha}\} \in \check{C}^{0}(\mathfrak{U}, \; \Omega^{1}_{\mathbb{R}})$ are the first step in the passage from the curvature $F \in H^{2}_{dR}(X)$ to its $\rm\check{C}$ech realization $[\,\{\rho_{\alpha\beta\gamma}\}\,] \in \check{H}^{2}(\mathfrak{U}, \mathbb{R})$, which is exactly $c_{1}(L) \otimes_{\mathbb{Z}} \mathbb{R}$. We adopt the same point of view for gerbes:
\begin{itemize}
	\item the \emph{curvature} of a connection is a form $G \in \Omega^{3}_{\mathbb{R}}$ such that $[\,G\,]_{H^{3}_{dR}(X)} \simeq c_{1}(L) \otimes_{\mathbb{Z}} \mathbb{R}$;
	\item a \emph{connection} with curvature $G$ is a couple $\{ \{F_{\alpha}\}, \{A_{\alpha\beta}\} \} \in \check{C}^{0}(\mathfrak{U}, \; \Omega^{2}_{\mathbb{R}}) \oplus \check{C}^{1}(\mathfrak{U},$ $\Omega^{1}_{\mathbb{R}})$ such that $dF_{\alpha} = G\vert_{U_{\alpha}}$ and $dA_{\alpha\beta} = F_{\beta} - F_{\alpha}$. By the hypothesis on $G$, it turns out that $A_{\alpha\beta} + A_{\beta\gamma} + A_{\gamma\alpha} = d\rho_{\alpha\beta\gamma}$;
	\item an \emph{equivalence class of connections} is an element of the hypercohomology group (v.\ appendices \ref{AppHyperC} and \ref{AppGerbes} for more details):
	\[\check{H}^{2}\bigl(\, \mathfrak{U}, \underline{S}^{1} \overset{d \circ \log}\longrightarrow \Omega^{1}_{\mathbb{R}} \overset{d}\longrightarrow \Omega^{2}_{\mathbb{R}} \,\bigr).
\]
\end{itemize}
Thus a gerbe with connection, as a cocycle (equivalent to a bundle with fixed sections, thus with fixed local representation of the connection), is represented by $(\{g_{\alpha\beta\gamma}\}, \{-A_{\alpha\beta}\}, \{-F_{\alpha}\})$ $\in \check{C}^{2}(\mathfrak{U}, \underline{S}^{1}) \,\oplus\, \check{C}^{1}(\mathfrak{U}, \Omega^{1}_{\mathbb{R}}) \,\oplus\, \check{C}^{0}(\mathfrak{U}, \Omega^{2}_{\mathbb{R}})$. Its boundary is given by:
	\[0 = \check{\delta}(g, -A, F) = (\, \check{\delta}g, d\log g - \check{\delta}A, F_{\beta} - F_{\alpha} - dA_{\alpha\beta} \,)
\]
so that we recover $F_{\beta} - F_{\alpha} = dA_{\alpha\beta}$ and $d\log g = \check{\delta}A$. In particular, $dF_{\beta} - dF_{\alpha} = 0$, so that we recover the curvature $G = dF$.

By definition, a trivial gerbe with trivial connection is represented by coboundaries:
\begin{equation}\label{TrivialGerbeConnection}
\check{\delta}^{0}(\,\{g_{\alpha\beta}\}, \{A_{\alpha}\}\,) = (\,\{g_{\alpha\beta}g_{\beta\gamma}g_{\gamma\alpha}\}, \{d\log g_{\alpha\beta} + A_{\beta} - A_{\alpha}\}, \{-dA_{\alpha}\}\,).
\end{equation}
Of course the curvature $G$ is zero, so that the 2-form of the connection is locally exact. We are instead completely free to choose the 1-forms $A_{\alpha}$, thus when they have the form \eqref{TrivialGerbeConnection} they are immaterial for the gerbe geometry. Since \eqref{TrivialGerbeConnection} is a coboundary, it is equivalent to the realization $(1, 0, 0)$.

\subsection{Torsion gerbes}

A gerbe is called \emph{torsion gerbe} if its first Chern class is torsion. As for bundles, using \eqref{ExactSeqZRS} one can prove that a gerbe is torsion if and only if it can be realized by constant transition functions. Let us see what happens for connections. We realize the gerbe as $(\{g_{\alpha\beta\gamma}\}, \{-A_{\alpha\beta}\}, \{F_{\alpha}\})$ with $g_{\alpha\beta\gamma}$ torsion. In this case, its coboundary is $0 = (\check{\delta}g, - \check{\delta}A, F_{\beta} - F_{\alpha} - dA_{\alpha\beta})$: thus $\check{\delta}A = 0$ and, since $\Omega^{1}_{\mathbb{R}}$ is acyclic, one has $A_{\alpha\beta} = A_{\beta} - A_{\alpha}$. We now have $F_{\beta} - F_{\alpha} = dA_{\beta} - dA_{\alpha}$, so that $F_{\alpha} - dA_{\alpha}$ is globally defined: we call it $\tilde{F}$ and we try to realize the connection with it. In fact, let us consider the coboundary $\check{\delta}(1, \{A_{\alpha}\}) = (1, A_{\beta} - A_{\alpha}, -dA_{\alpha})$: then we add it to the original representative of the gerbe obtaining $(\{g_{\alpha\beta\gamma}\}, \{-A_{\alpha\beta}\}, \{F_{\alpha}\}) + (1, A_{\beta} - A_{\alpha}, -dA_{\alpha}) = (\{g_{\alpha\beta\gamma}\}, 0, \tilde{F})$. Thus, a torsion gerbe can always be realized in this form.

\section{Real Chern classes and curvatures}

Let us now consider a trivialization of a gerbe $\{g_{\alpha\beta}\} \in \check{C}^{1}(X, \underline{S}^{1})$ such that $\check{\delta}^{1}\{g_{\alpha\beta}\} \in \check{B}^{2}(X, \Gamma_{n})$. We can consider a connection $\{A_{\alpha}\}$ such that $A_{\beta} - A_{\alpha} = d\rho_{\alpha\beta}$, as for an ordinary bundle. We have $dA_{\alpha} = dA_{\beta}$ so that $F = dA_{\alpha}$ is a global closed form whose class $[\,F\,]$ is exactly the fractionary Chern class of $[\,\{g_{\alpha\beta}\}\,] \in \check{\delta}^{-1}(\Gamma_{n}) \,/\, \check{B}^{1}$. We define such a trivialization with connection as an element of the hypercohomology group:
	\[\check{H}^{1}(\mathfrak{U}, \underline{S}^{1} / \Gamma_{n} \overset{d \circ \log}\longrightarrow \Omega^{1}_{\mathbb{R}}).
\]
What happens for the holonomy of such connections? They are not well-defined as a function on closed curves, but they are a section of a line bundle with sections which, on curves which are boundary of open surfaces, is canonically trivial and coincides with the one determined by the flat gerbe realized by $(1, 0, F)$ but with respect to the sections $\check{\delta}g$. In fact, the expression of the holonomy of $A$ on $\partial\Sigma$ coincides with the holonomy of $(\check{\delta}g, A_{\beta} - A_{\alpha}, dA_{\alpha})$ on $\Sigma$, but $\check{\delta}(g, 0) = (\check{\delta}g, d\log g_{\alpha\beta}, 0)$ and the sum is $(1, 0, dA_{\alpha})$, thus the gerbe is $(1, 0, F)$ but it is realized on open surfaces with respect to $\check{\delta}g$.

\section{Wilson loop}

\subsection{Overview}

We want to define Wilson loop, which is the integral of the connection over a surface \emph{as an $\mathbb{R}/\mathbb{Z}$-valued function}, and it will turn out that its exponential is exactly the holonomy:
	\[\Hol_{F}(\Sigma) \,=\, \exp\biggl( \, 2\pi i \cdot \int_{\Sigma} F \, \biggr).
\]
For generic gerbes we can define Wilson loop as an $\mathbb{R}/\mathbb{Z}$-valued function only on \emph{closed surfaces}. In fact, once we have fixed a good cover $\mathfrak{U} = \{U_{\alpha}\}_{\alpha \in I}$, let us suppose for the moment that $\Sigma \subset U_{\alpha}$ for a fixed $\alpha \in I$. In this case we define:
	\[\int_{\Sigma} F \; := \; \int_{\Sigma} F_{\alpha}.
\]
If it happens that $\Sigma \subset U_{\alpha\beta}$, we have two definitions of Wilson loop, but they coincide:
\begin{equation}\label{WLOneChartGerbes}
	\int_{\Sigma} F_{\alpha} = \int_{\Sigma} F_{\beta} + 2\pi\int_{\Sigma} dA_{\alpha\beta} = \int_{\Sigma} F_{\beta} + 2\pi\int_{\partial\Sigma} A_{\alpha\beta} = \int_{\Sigma} F_{\beta}.
\end{equation}
In the last expression it was crucial that $\Sigma$ was a closed surface\footnote{In this case we have an $\mathbb{R}$-valued function, but this is due to the assumption that $\gamma$ is contained in only one chart. We will see in the following that, without this hypothesis, we have the $\mathbb{Z}$-uncertainty.}.

\paragraph{}For open surfaces, we can give a good definition of Wilson loop only for \emph{trivial} gerbes. For a trivial gerbe $\mathcal{G} = [\,\{g_{\alpha\beta}\}\,] \in \check{H}^{1}(\mathfrak{U}, \underline{S}^{1})$, we have that $g_{\alpha\beta\gamma} = g_{\alpha\beta} \cdot g_{\beta\gamma} \cdot g_{\gamma\alpha}$ for a certain trivialization $\{g_{\alpha\beta}\} \in \check{C}^{1}(\mathfrak{U}, \underline{S}^{1})$. In particular, for $g_{\alpha\beta\gamma} = e^{2\pi i \cdot \rho_{\alpha\beta\gamma}}$ and $g_{\alpha\beta} = e^{2\pi i \cdot \rho_{\alpha\beta}}$, we have that $\rho_{\alpha\beta\gamma} = \rho_{\alpha\beta} + \rho_{\beta\gamma} + \rho_{\gamma\alpha} + 2\pi i \cdot n$, with $n \in \mathbb{Z}$. Thus:
	\[\begin{split}
	&F_{\beta} - F_{\alpha} = dA_{\alpha\beta}\\
	&A_{\alpha\beta} + A_{\beta\gamma} + A_{\gamma\alpha} = d\rho_{\alpha\beta} + d\rho_{\beta\gamma} + d\rho_{\gamma\alpha}\\
	&\check{\delta}^{1}\{A_{\alpha\beta} - d\rho_{\alpha\beta}\} = 0\\
	&A_{\alpha\beta} - d\rho_{\alpha\beta} = A_{\alpha} - A_{\beta}
\end{split}\]
so that we obtain a global form (or gauge-invariant form) $\tilde{F} \in \Omega^{2}(X, \mathbb{R})$ such that $\tilde{F}\vert_{U_{\alpha}} = F_{\alpha} - dA_{\alpha}$ (this works also for torsion gerbes, since $g_{\alpha\beta\gamma} = g_{\alpha\beta} \cdot g_{\beta\gamma} \cdot g_{\gamma\alpha} \cdot c_{\alpha\beta\gamma}$ for $c_{\alpha\beta\gamma}$ constant, thus $d\rho_{\alpha\beta\gamma} = d \rho_{\alpha\beta} + d\rho_{\beta\gamma} + d\rho_{\gamma\alpha}$ anyway). We thus define, for a surface $\Sigma$:
	\[\int_{\Sigma} F \,:=\, \int_{\Sigma} \tilde{F}
\]
which, for $\Sigma \subset U_{\alpha}$, becomes:
\begin{equation}\label{WLTrivialGerbe}
\begin{split}
	&\int_{\Sigma} F = \int_{\Sigma} \bigl( F_{\alpha} - dA_{\alpha} \bigr) = \int_{\Sigma} F_{\alpha} - \int_{\partial \Sigma} A_{\alpha}\\
	&\Hol_{A}(\gamma) \,=\, \exp\biggl( \int_{\Sigma} F_{\alpha} - \int_{\partial \Sigma} A_{\alpha} \biggr).
\end{split}
\end{equation}
If we consider the trivial gerbe with global section $g_{\alpha\beta\gamma} = 1$, then $\rho_{\alpha\beta\gamma} = 0$, so that $\tilde{F} = F$ and we recover the expression with the globally-defined connection.

\paragraph{}For generic gerbes, the situation is analogue to the case of bundles. We give the same definition as for closed curves:
	\[\int_{\Sigma} F \; := \; \int_{\Sigma} F_{\alpha}.
\]
Of course it is not a well-defined function, but let us suppose that $\Sigma \subset U_{\alpha\beta}$. Then we have:
	\[\begin{split}
	&\int_{\Sigma} F_{\alpha} = \int_{\Sigma} F_{\beta} + \int_{\Sigma} dA_{\alpha\beta} = \int_{\Sigma} F_{\beta} + \int_{\partial \Sigma} A_{\alpha\beta}\\
	&\exp\biggl( \int_{\Sigma} F_{\alpha} \biggr) \,=\, \exp\biggl( \int_{\Sigma} F_{\beta} \biggr) \cdot \exp\biggl( \int_{\partial \Sigma} A_{\alpha\beta} \biggr).
\end{split}\]
To interpret this expression, let us consider the space $\Sigma U_{\alpha}$ of surface diffeomorphic to $\Sigma$ contained in $U_{\alpha}$ (for $\alpha$ fixed), i.e.\ the space of maps $\Gamma: \Sigma \rightarrow U_{\alpha}$ with the compact-open topology. If the fixed good covering of $X$ is $\mathfrak{U} = \{U_{\beta}\}_{\beta \in I}$, we can define a covering $\mathcal{V} = \{V_{\rho}\}_{\rho \in I}$ of $\Sigma U_{\alpha}$ given by $V_{\rho} = \{\rho \in \Sigma U_{\alpha}: \, \Gamma(\Sigma) \subset U_{\alpha\rho}\}$, with the notation $U_{\alpha\alpha} = U_{\alpha}$. Of course, $V_{\alpha}$ is the whole $\Sigma U_{\alpha}$. Thus we have an expression of the holonomy for every $V_{\rho}$:
	\[\Hol^{\rho}_{F}(\Gamma) = \exp\biggl( \int_{\Gamma} F_{\rho} \biggr)
\]
linked by the expression:
	\[\Hol^{\rho}_{F}(\Gamma) \,=\, \Hol^{\eta}_{F}(\Gamma) \cdot h_{\rho\eta}(\Gamma) \,, \qquad h_{\rho\eta}(\Gamma) = \exp\biggl( \int_{\partial \Gamma} A_{\rho\eta} \biggr).
\]
Moreover, we have:
	\[\begin{split}
	h_{\rho\eta}(\Gamma) &\cdot h_{\eta\chi}(\Gamma) \cdot h_{\chi\rho}(\Gamma)\\
	& = \exp\biggl( \int_{\partial \Gamma} A_{\rho\eta} + A_{\eta\chi} + A_{\chi\rho} \biggr) = \exp\biggl( \int_{\partial \Gamma} d\rho_{\rho\eta\chi} \biggr)\\
	& = 1
\end{split}\]
thus we can think $\Hol_{A}(\Gamma)$ as \emph{a section of a line bundle $\Sigma L_{\alpha}$ over $\Sigma U_{\alpha}$} with transition functions $\{h_{\rho\eta}\} \in \check{H}^{1}(\Sigma U_{\alpha}, \underline{S}^{1})$. In this way we have a clear meaning for the holonomy: the fact that it is not well-defined as a function does not mean that it is not well-defined at all, it is just a section of a line bundle, hence it is naturally different with respect to different charts. Under the hypothesis that $\Gamma$ is contained in $U_{\alpha}$ we have that $V_{\alpha}$ is a global chart for $\Sigma L_{\alpha}$, hence the bundle is trivial: we will see that for generic curves we obtain a non-trivial bundle, but the picture is the same.

\paragraph{}Let us now suppose that $\mathcal{G}$ is trivial, with a trivialization $\{g_{\alpha\beta}\} \in \check{C}^{1}(\mathfrak{U}, \underline{S}^{1})$. Then we have:
	\[\begin{split}
	h_{\rho\eta}(\Gamma) &= \exp\biggl( \int_{\partial \Gamma} A_{\rho\eta} \biggr)\\
	& = \exp\biggl( \int_{\partial \Gamma} \bigl( d\rho_{\rho\eta} + A_{\rho} - A_{\eta} \bigr) \biggr)\\
	&= \exp\biggl( \int_{\partial \Gamma} A_{\rho} \biggr) \cdot \exp\biggl( \int_{\partial \Gamma} A_{\eta} \biggr)^{-1}\\
	&= h_{\rho}(\gamma)^{-1} \cdot h_{\eta}(\gamma)
\end{split}\]
for $h_{\rho}(\gamma) = \exp(\int_{\partial \Gamma} A_{\rho})$. Thus, a trivialization of $L$ determines \emph{canonically} a trivialization of $\Sigma L_{\alpha}$: in this way, the section $\Hol_{A}$ becomes a function, in particular:
	\[\Hol_{A}(\Gamma) = \Hol^{\rho}_{A}(\Gamma) \cdot h_{\rho}(\Gamma)^{-1} = \exp\biggl( \int_{\Gamma} F_{\rho} + \int_{\partial \Gamma} A_{\rho} \biggr)
\]
which is exactly \eqref{WLTrivialGerbe}.

\paragraph{}Thus we obtain the following picture: the holonomy of a gerbe with connection is a well-defined function over the space of maps from a closed surface, while for open surfaces, it is a section of a line bundle $\Sigma L \rightarrow \Sigma X$; if the gerbe is trivial then $\Sigma L$ is trivial too, and a trivialization of the gerbes determines canonically a trivialization of $\Sigma L$, so that for any trivial bundle we have a canonically defined holonomy also for open surfaces. We now give the same picture without assuming that the curve (closed or open) lies in only one chart: we can describe the integral of $F$ in two ways using local charts.

\subsection{Local description}

\subsubsection{Closed surfaces}

We can describe the integral of $F$ in two ways.

\paragraph{First way:} we choose an open covering for the space of closed surfaces, starting from the covering $\{U_{i}\}_{i \in I}$ of $X$. 

\paragraph{}We now define the integral of the connection. For a fixed $\Gamma \in \Sigma X$, there exists $(\tau, \varphi) \in J$ such that $\Gamma \in V_{(\tau, \varphi)}$. The function $\varphi: T \rightarrow I$ induces two functions:
\begin{itemize}
	\item $\varphi^{E}: E \rightarrow I^{2}$, given by $\varphi^{E}(a,b) = \bigl( \varphi(b^{1}(a,b)), \varphi(b^{2}(a,b) \bigr)$;
	\item $\varphi^{V}: \{1, \ldots, l\} \rightarrow \coprod_{i=1}^{l} (I^{3})^{k_{i}-2}$, such that $\varphi^{V}(i) \in (I^{3})^{k_{i}-2}$ and $\bigl(\varphi^{V}(i)\bigr)^{j} = \bigl( \varphi(B^{1}(i)),$ $\varphi(B^{j}(i)), \varphi(B^{j+1}(i)) \bigr)$.
\end{itemize}
We define:
\begin{equation}\label{WilsonLoopGerbe}
\begin{split}
\int_{\Gamma} F \;:=\; \sum_{(a,b,c) \in T_{\tau}} \; \int_{\Gamma(\sigma^{2}_{(a,b,c)})} F_{\varphi(a,b,c)} \; + \; \sum_{(a,b) \in E_{\tau}} \;& \int_{\Gamma(\sigma^{1}_{(a,b)})} A_{\varphi^{E}(a,b)}\\
& + \; \sum_{i=1}^{l} \sum_{j=1}^{k_{i}} \; \log g_{(\varphi^{V}(i))^{j}}\Gamma(\sigma^{0}_{i}).
\end{split}
\end{equation}
The last term seems more artificial: we briefly discuss it. The logarithm can be taken since we have chosen a good covering, so the intersections are contractible. Of course, it is  defined up to $2\pi i \,\mathbb{Z}$, so the quantity that can be well-defined as a number is $\exp\bigl(\int_{\Gamma} F\bigr)$. The sum is taken in the following way: we consider the star of triangles having $\sigma^{0}_{i}$ as common vertex (each of them associated to a chart via $\varphi$) and, since we are considering 0-simplices, that corresponds to 2-cochains, we consider the possible triads with first triangle fixed $\bigl(\varphi^{V}(i)\bigr)^{j} = \bigl( \varphi(B^{1}(i)),$ $\varphi(B^{j}(i)), \varphi(B^{j+1}(i)) \bigr)$ and sum over them. The fact we fixed $B^{1}(i)$ as first triangle has no effect, since we could consider any other possibility $\bigl(\varphi^{V}_{\alpha}(i)\bigr)^{j} = \bigl( \varphi(B^{\alpha}(i)),$ $\varphi(B^{j}(i)), \varphi(B^{j+1}(i)) \bigr)$. In fact, by cocycle condition with indices $(1, i, i+1, \alpha)$ we have that $g_{1,i+1,\alpha} \cdot g_{1, i, i+1} = g_{i,i+1,\alpha} \cdot g_{1, i, \alpha}$, thus $g_{\alpha,i,i+1} = g_{1, i, \alpha}^{-1} \cdot g_{1, i, i+1} \cdot g_{1,i+1,\alpha}$, but in the cyclic sum the extern terms simplify, hence the sum involving $g_{\alpha,i,i+1}$ is equal to the sum involving $g_{1,i,i+1}$. Finally, we summed over $j = 1, \ldots, k_{i}$, but for $j = 1$ and $j = k_{i}$ we obtain trivial terms, hence the real sum is for $j = 2, \ldots, k_{i}-1$.

\paragraph{}We now prove that $\int_{\Gamma} F$ is well-defined in $\mathbb{C} / 2\pi i\,\mathbb{Z}$: it must be invariant under both the choice of the open set $V_{(\tau, \varphi)}$ to which $\Gamma$ belongs and the choice of the open cover $\{U_{i}\}_{i \in I}$ of $X$. We prove it in steps:
\begin{itemize}
	\item We consider $V_{(\tau, \varphi')}$ such that $\varphi'$ differs from $\varphi$ just on $(a,b,c) \in T$ fixed, so that $\gamma(\sigma^{2}_{(a,b,c)}) \subset U_{\varphi(a,b,c)\varphi'(a,b,c)}$. Then, the summands involving $U_{\varphi(a,b,c)}$ becomes:
\begin{equation}\label{VarphiPrime}
\begin{split}
	&\textcolor{blue}{ \int_{\Gamma(\sigma^{2}_{(a,b,c)})} F_{\varphi'(a,b,c)} } + \textcolor{red}{ \int_{\Gamma(\sigma^{1}_{(a,b)})} A_{\varphi'^{E}(a,b)} } + \textcolor{red}{ \int_{\Gamma(\sigma^{1}_{(b,c)})} A_{\varphi'^{E}(b,c)} } + \textcolor{red}{ \int_{\Gamma(\sigma^{1}_{(c,a)})} A_{\varphi'^{E}(c,a)} }\\
	&\phantom{XXX} + \; \textcolor{green}{ \sum_{j=1}^{k_{a}} \; \log g_{(\varphi'^{V}(a))^{j}}\Gamma(\sigma^{0}_{a}) } \; + \; \textcolor{green}{ \sum_{j=1}^{k_{b}} \; \log g_{(\varphi'^{V}(b))^{j}}\Gamma(\sigma^{0}_{b}) } \; \\
	&\phantom{XXXXXXXXXXXXXXXX} + \; \textcolor{green}{ \sum_{j=1}^{k_{c}} \; \log g_{(\varphi'^{V}(c))^{j}}\Gamma(\sigma^{0}_{c}) }.
	\end{split}
\end{equation}
Then we have:
	\[\begin{split}
	\textcolor{blue}{ \int_{\Gamma(\sigma^{2}_{(a,b,c)})} } \textcolor{blue}{F_{\varphi'(a,b,c)} } &= \int_{\Gamma(\sigma^{2}_{(a,b,c)})}  \Bigl( F_{\varphi(a,b,c)} + dA_{\varphi(a,b,c)\varphi'(a,b,c)} \Bigr)\\
	&= \int_{\Gamma(\sigma^{2}_{(a,b,c)})} F_{\varphi(a,b,c)} + \textcolor{red}{ \int_{\Gamma(\sigma^{1}_{(a,b)}) + \Gamma(\sigma^{1}_{(b,c)}) + \Gamma(\sigma^{1}_{(c,a)})} A_{\varphi(a,b,c)\varphi'(a,b,c)} }.
\end{split}\]
The sum of the red terms gives, e.g.\ for the $(a,b)$-edge:
\begin{equation}\label{ABEdge}
\begin{split}
	\textcolor{red}{ \int_{\Gamma(\sigma^{1}_{(a,b)})} } & \textcolor{red}{ A_{\varphi(a,b,c)\varphi'(a,b,c)} } + \textcolor{red}{ \int_{\Gamma(\sigma^{1}_{(a,b)})} A_{\varphi'^{E}(a,b)} }\\
	&= \int_{\Gamma(\sigma^{1}_{(a,b)})} \Bigl( A_{\varphi(a,b,c)\varphi'(a,b,c)} + A_{( \varphi'(b^{1}(a,b)), \varphi'(b^{2}(a,b))} \Bigl)\\
	&= \int_{\Gamma(\sigma^{1}_{(a,b)})} \Bigl( A_{\varphi(a,b,c)\varphi'(a,b,c)} + A_{( \varphi'(a,b,c), \varphi(b^{2}(a,b))} \Bigr)\\
	&= \int_{\Gamma(\sigma^{1}_{(a,b)})} \Bigl( A_{\varphi(a,b,c), \varphi(b^{2}(a,b))} + d\log g_{\varphi(a,b,c) \varphi'(a,b,c) \varphi(b^{2}(a,b))} \Bigr)\\
	&= \int_{\Gamma(\sigma^{1}_{(a,b)})} \Bigl( A_{\varphi^{E}(a,b)} \Bigr) + \textcolor{green}{ \log g_{\varphi(a,b,c) \varphi'(a,b,c) \varphi(b^{2}(a,b))}\Gamma(\sigma^{0}_{b}) }\\
	&\phantom{XXXXXXXXXXXX} - \textcolor{green}{ \log g_{\varphi(a,b,c) \varphi'(a,b,c) \varphi(b^{2}(a,b))}\Gamma(\sigma^{0}_{a}) }.
\end{split}
\end{equation}
The same for $(b,c)$ and $(c,a)$ edges. We now consider the green terms. We consider the vertex $a$, then the same computation applies for $b$ and $c$. In the green sum of \eqref{VarphiPrime} for $j = 1, \ldots, k_{a}$, the only two terms in which $\varphi'$ is different from $\varphi$ are the terms $(g_{1,j-1,j} \cdot g_{1,j,j+1}) \Gamma(\sigma^{0}_{a})$ for $B^{j}(a) = \sigma^{2}_{(a,b,c)}$. By cocycle condition applied to $(1, j-1, j, j+1)$ we have $g_{1,j-1,j} \cdot g_{1,j,j+1} = g_{j-1,j,j+1} \cdot g_{1,j-1,j+1}$, thus the only term involved is $g_{j-1,j,j+1} \Gamma(\sigma^{0}_{a})$. We write it $g_{j-1,j',j+1}$ to mean that we apply $\varphi'$ at the $j$-th position. But with the green terms of \eqref{ABEdge} (and of the analogous formula for $(a,c)$-edge) we have $g_{j,j',j-1}$ and $g_{j,j',j+1}$, so that we obtain $(g_{j-1,j',j+1} \cdot g_{j,j',j-1} \cdot g_{j,j',j+1})\Gamma(\sigma^{0}_{a}) = g_{j-1,j,j+1}\Gamma(\sigma^{0}_{a})$ and the final sum is equal for $\varphi$ and $\varphi'$.
\item In general, let us consider $\gamma \in V_{(\tau, \varphi)} \cap V_{(\tau', \varphi')}$. We proceed as in the one-dimensional case building a triangulation $\tau \cup \tau'$ refining $\tau$ and $\tau'$ (of course it is technically more complicated than for $S^{1}$, but the result is the same), such that the values of $\int_{\Sigma} F$ for $(\tau \cup \tau', \varphi)$ and $(\tau \cup \tau', \varphi')$ are the same. Then we apply the previous step chart by chart.
\item One can prove that we obtain the same result if we choose a different covering of $X$.
\end{itemize}

\paragraph{Second way:} we notice that any gerbe, restricted to a surface $\Sigma$, is trivial since $H^{3}(\Sigma, \mathbb{Z}) = 0$. Thus, we can choose a globally defined form $F_{0}: T\Sigma \rightarrow \mathbb{R}$ and define:
	\[\int_{\Sigma} F \,:=\, \int_{\Sigma} F_{0}.
\]
If we choose a different trivialization of the gerbe, we obtain a globally defined connection $F_{1}$ such that $F_{1} = F_{0} + g_{01}^{-1}\cdot dg_{01}$, but now, since $\Sigma$ is not necessarily contractible, we cannot say that $g_{01}^{-1}\cdot dg_{01} = d\log g_{01}$: we can only consider it as a closed, but not necessarily exact, 2-form (it is a large gauge transformation). Hence:
	\[\int_{\Sigma} F_{1} = \int_{\Sigma} F_{0} + \int_{\Sigma} g_{01}^{-1}\cdot dg_{01}.
\]
However, the last summand belongs to $2\pi i \, \mathbb{Z}$: this is a general fact, since $g_{01}$ a transition function of a gerbe. We can see it in the following way: we choose a triangulation $\tau = \{\sigma^{0}_{1}, \ldots, \sigma^{0}_{l(\tau)}, E, T \}$ of $\Sigma$, so that, since the triangles are contractible, one has $\int_{\Sigma} g_{01}^{-1}\cdot dg_{01} = \sum_{(a,b,c) \in T} \int_{\sigma^{2}_{(a,b,c)}} g_{01}^{-1}\cdot dg_{01} = \sum_{(a,b,c) \in T} \int_{\sigma^{2}_{(a,b,c)}} d\log g_{01} = \sum_{(a,b,c) \in T} \bigl( \log g_{01}(\sigma_{i+1}) - \log g_{01}(\sigma_{i}) \bigr)$, putting for simplicity $l+1 = 1$. Since $\gamma$ is closed, the sum simplifies to $0$ up to the integer multiples of $2\pi i$ depending on the logarithms chosen at every piece. Thus, Wilson loop is well-defined also in this way.

\subsubsection{Surfaces with boundary}

For $\Sigma$ open, the previous definitions does not apply. Exactly as for line bundles, we now see that the integral of $F$ can be well-defined for \emph{trivial} gerbes, while, in general, it defines a \emph{line bundle} over the space of maps, which is canonically trivial for trivial gerbes.

\paragraph{}Let us consider a trivial gerbe $\{g_{\alpha\beta\gamma}\} \in \check{B}^{2}(H, C^{0}(S^{1}))$, and let $g_{\alpha\beta\gamma} = g_{\alpha\beta} \cdot g_{\beta\gamma} \cdot g_{\gamma\alpha}$. We have:
	\[\begin{split}
	&F_{\alpha} - F_{\beta} = dA_{\alpha\beta}\\
	&A_{\alpha\beta} +  A_{\beta\gamma} + A_{\gamma\alpha} = d \log g_{\alpha\beta} + d \log g_{\beta\gamma} + d \log g_{\gamma\alpha}\\
	&\bigl( A_{\alpha\beta} - d \log g_{\alpha\beta} \bigr) + \bigl( A_{\beta\gamma} - d \log g_{\beta\gamma} \bigr) + \bigl( A_{\gamma\alpha} - d \log g_{\gamma\alpha} \bigr) = 0\\
	&\delta \, \bigl\{ A_{\alpha\beta} - d \log g_{\alpha\beta} \bigr\} = 0
\end{split}\]
and, since the sheaf of 1-forms is fine, hence acyclic, we obtain:
	\[A_{\alpha\beta} - d \log g_{\alpha\beta} = A_{\alpha} - A_{\beta}.
\]

\paragraph{}We now define the integral of the connection. For a fixed $\Gamma \in \Sigma X$, there exists $(\tau, \varphi) \in J$ such that $\gamma \in V_{(\tau, \varphi)}$. We define:
\begin{equation}\label{WilsonLineGerbe}
	\int_{\Gamma} F \;:=\; \sum_{(a,b,c) \in T_{\tau}} \; \biggl( \int_{\Gamma(\sigma^{2}_{(a,b,c)})} F_{\varphi(a,b,c)} + \int_{\Gamma(\partial \sigma^{2}_{(a,b,c)})} A_{\varphi(a,b,c)} \biggr).
\end{equation}
As before, the logarithm can be taken since we have chosen a good cover and it is defined up to $2\pi i \,\mathbb{Z}$, so that $\exp\bigl(\int_{\Gamma} F\bigr)$ is defined as a number. The contribution of $A$ to the internal edges cancel in pairs, so only the integral of $A$ on boundary terms remains. That's why this expression is usually denoted by:
	\[\int_{\Gamma} F + \oint_{\partial \Gamma} A.
\]

\paragraph{}For an internal edge $(a,b)$, let us suppose that $b(a,b) = ((a,b,c), (b,a,d))$. Then, from these two triangles we have the contribution:
	\[\begin{split}
	\int_{\sigma^{1}_{(a,b)}} \Bigl( &A_{\varphi(a,b,c)} - A_{\varphi(b,a,d)} \Bigr) = \int_{\sigma^{1}_{(a,b)}} \Bigl( A_{\varphi(a,b,c),\varphi(b,a,d)} - d \log g_{\varphi(a,b,c), \varphi(b,a,d)} \Bigr)\\
	&= \int_{\sigma^{1}_{(a,b)}} A_{\varphi^{E}(a,b)} + \log g_{\varphi(a,b,c), \varphi(b,a,d)} \Gamma(\sigma^{0}_{a}) - \log g_{\varphi(a,b,c), \varphi(b,a,d)} \Gamma(\sigma^{0}_{b}).
\end{split}\]
Thus the contribution of the internal edges is the same of \eqref{WilsonLoopGerbe}. For the vertices, let us consider $a$: form the previous expression we had the contribution $\log g_{\varphi(a,b,c), \varphi(b,a,d)} \Gamma(\sigma^{0}_{a})$. Doing the same computation for all triangles of the star containing $a$, $\{\sigma^{2}_{(a,a_{1},a_{2})}, \ldots, \sigma^{2}_{(a,a_{k_{a}},a_{1})}\}$, we obtain, exponentiating:
	\[\begin{split}
	\bigl( g_{\varphi(a,a_{1},a_{2}), \varphi(a,a_{2},a_{3})} \cdot g_{\varphi(a,a_{2},a_{3}), \varphi(a,a_{3},a_{4})} \cdots &g_{\varphi(a,a_{k_{a}-1},a_{k_{a}}), \varphi(a,a_{k_{a}},a_{1})}\\
	&\cdot g_{\varphi(a,a_{k_{a}},a_{1}), \varphi(a,a_{1},a_{2})} \bigr) \Gamma(\sigma^{0}_{a}).
\end{split}\]
Each triangle $\varphi(a,a_{i},a_{i+1})$ is present twice since two of its edges contain the vertex. We now fix as base triangle $B^{1}(a) = (a,a_{1},a_{2})$, and we rewrite the previous expression as:
	\[\begin{array}{ccc}
	\textcolor{red}{ g_{\varphi(a,a_{1},a_{2}), \varphi(a,a_{1},a_{2})} } & g_{\varphi(a,a_{1},a_{2}), \varphi(a,a_{2},a_{3})} & \textcolor{blue}{ g_{\varphi(a,a_{2},a_{3}), \varphi(a,a_{1},a_{2})} }\\
	\textcolor{blue}{ g_{\varphi(a,a_{1},a_{2}), \varphi(a,a_{2},a_{3})} } & g_{\varphi(a,a_{2},a_{3}), \varphi(a,a_{3},a_{4})} & \textcolor{green}{ g_{\varphi(a,a_{3},a_{4}), \varphi(a,a_{1},a_{2})} } \\
	\textcolor{green}{ g_{\varphi(a,a_{1},a_{2}), \varphi(a,a_{3},a_{4})} } & \cdots & \cdots\\
	\vdots & \vdots & \vdots\\
	\cdots & \cdots & \textcolor{cyan}{ g_{\varphi(a,a_{k_{a}-1},a_{k_{a}}), \varphi(a,a_{1},a_{2})} } \\
	\textcolor{cyan}{ g_{\varphi(a,a_{1},a_{2}), \varphi(a,a_{k_{a}-1},a_{k_{a}})} } & g_{\varphi(a,a_{k_{a}-1},a_{k_{a}}), \varphi(a,a_{k_{a}},a_{1})} & \textcolor{magenta}{ g_{\varphi(a,a_{k_{a}},a_{1}), \varphi(a,a_{1},a_{2})} }\\
	\textcolor{magenta}{ g_{\varphi(a,a_{1},a_{2}), \varphi(a,a_{k_{a}},a_{1})} } & g_{\varphi(a,a_{k_{a}},a_{1}), \varphi(a,a_{1},a_{2})} & \textcolor{red}{ g_{\varphi(a,a_{1},a_{2}), \varphi(a,a_{1},a_{2})} }
\end{array}\]
so that we obtain:
	\[\begin{split}
	\bigl( &g_{\varphi(a,a_{1},a_{2}), \varphi(a,a_{1},a_{2}), \varphi(a,a_{2},a_{3})} \cdot g_{\varphi(a,a_{1},a_{2}), \varphi(a,a_{2},a_{3}), \varphi(a,a_{3},a_{4})}\\
	&\phantom{XXXXX} \cdots g_{\varphi(a,a_{1},a_{2}), \varphi(a,a_{k_{a}-1},a_{k_{a}}), \varphi(a,a_{k_{a}},a_{1})} \cdot g_{\varphi(a,a_{1},a_{2}), \varphi(a,a_{k_{a}},a_{1}), \varphi(a,a_{1},a_{2})} \bigr) \Gamma(\sigma^{0}_{a})
\end{split}\]
and, if we put $\bigl(\varphi^{V}(a)\bigr)^{j} = (\varphi(a,a_{1},a_{2}), \varphi(a,a_{j},a_{j+1}), \varphi(a,a_{j+1},a_{j+2}))$ we exactly obtain:
	\[\prod_{j=1}^{k_{i}} \; g_{(\varphi^{V}(i))^{j}}\Gamma(\sigma^{0}_{a}).
\]
Thus, for internal edges and vertices, we recover the same expression as for closed surfaces.

\paragraph{}We now prove that $\int_{\Gamma} F$ is well-defined in $\mathbb{C} / 2\pi i\,\mathbb{Z}$: it must be invariant under both the choice of the open set $V_{(\tau, \varphi)}$ to which $\Gamma$ belongs and the choice of the open cover $\{U_{i}\}_{i \in I}$ of $X$. We prove it in steps:
\begin{itemize}
	\item We consider $V_{(\tau, \varphi')}$ such that $\varphi'$ differs from $\varphi$ just on $(a,b,c) \in T_{\tau}$ fixed. Then:
	\[\begin{split}
	\int_{\Gamma(\sigma^{2}_{(a,b,c)})}F_{\varphi'(a,b,c)} &= \int_{\Gamma(\sigma^{2}_{(a,b,c)})} \Bigl(F_{\varphi(a,b,c)} + d A_{\varphi(a,b,c)\varphi'(a,b,c)} \Bigr)\\
	&= \int_{\Gamma(\sigma^{2}_{(a,b,c)})} \Bigl(F_{\varphi(a,b,c)} + d A_{\varphi(a,b,c)} - d A_{\varphi'(a,b,c)} \Bigr)\\
	&= \int_{\Gamma(\sigma^{2}_{(a,b,c)})} F_{\varphi(a,b,c)} + \int_{\Gamma(\partial \sigma^{2}_{(a,b,c)})} A_{\varphi(a,b,c)} - \int_{\Gamma(\partial \sigma^{2}_{(a,b,c)})} A_{\varphi'(a,b,c)}
\end{split}\]
thus:
	\[\int_{\Gamma(\sigma^{2}_{(a,b,c)})}F_{\varphi'(a,b,c)} + \int_{\Gamma(\partial \sigma^{2}_{(a,b,c)})} A_{\varphi'(a,b,c)} = \int_{\Gamma(\sigma^{2}_{(a,b,c)})} F_{\varphi(a,b,c)} + \int_{\Gamma(\partial \sigma^{2}_{(a,b,c)})} A_{\varphi(a,b,c)}
\]
so the $(a,b,c)$-summand does not change.

\item In general, let us consider $\gamma \in V_{(\tau, \varphi)} \cap V_{(\tau', \varphi')}$. We proceed as in the one-dimensional case building a triangulation $\tau \cup \tau'$ refining $\tau$ and $\tau'$ (of course it is technically more complicated than for $S^{1}$, but the result is the same), such that the values of $\int_{\Sigma} F$ for $(\tau \cup \tau', \varphi)$ and $(\tau \cup \tau', \varphi')$ are the same. Then we apply the previous step chart by chart.

\end{itemize}

\paragraph{}What happens for non-trivial gerbes? In this case, the only possibility is to use the same definition as for closed surfaces, but the boundary terms will forbid the well-definiteness of the integral as a function. We now prove that we obtain a line bundle over the space of maps. The steps of the proof of well-definiteness for close curves works also in this case, except at the last step for vertices on $\partial\Sigma$ (there are no problems for boundary edges, since they appear as boundary of triangles anyway). For every boundary vertex, there is a term $g_{1,j-1,j}$ which has not the following $g_{1,j,j+1}$, and from the green terms of \eqref{ABEdge} we get $g_{j,j',j-1}$. Thus, we have:
\begin{itemize}
	\item $g_{1,j-1,j}$ with $\varphi$;
	\item $g_{1,j-1,j'} \cdot g_{j-1,j,j'}$ with $\varphi'$ (after we reduced to $\varphi$ for triangles and edges).
\end{itemize}
Thus the difference is:
	\[g_{1,j-1,j'} \cdot g_{j-1,j,j'} \cdot g_{1,j-1,j}^{-1} = g_{1,j,j'}.
\]
Moreover, the same contribution of $\varphi'$ comes from the vertex for which there is a term $g_{1,j,j+1}$ without the following, and from the green terms of \eqref{ABEdge} we get $g_{j,j',j+1}^{-1}$. Thus:
\begin{itemize}
	\item $g_{1,j,j+1}$ with $\varphi$;
	\item $g_{1,j',j+1} \cdot g_{j,j',j+1}^{-1}$ with $\varphi'$ (after we reduced to $\varphi$ for triangles and edges).
\end{itemize}
Thus the difference is:
	\[g_{1,j',j+1} \cdot g_{j,j',j+1}^{-1} \cdot g_{1,j,j+1}^{-1} = g_{1,j,j'}^{-1}.
\]
These terms seems to simplify, but it is false since the index $1$ refers to the fixed triangle for start around different vertices, so it is not the same. But they satisfy a cocycle condition, since:
	\[g_{1,j,j'} \cdot g_{1,j',j''} \cdot g_{1,j'',j} = g_{j,j',j''}
\]
thus there is no index $1$ any more, so that the two contribution becomes $g_{j,j',j''} \cdot  g_{j,j',j''}^{-1} = 1$.

\paragraph{}Thus:
	\[\begin{split}
	\exp\biggl(\int_{\Gamma} F\biggr)_{(\tau, \varphi')} \; = \; \exp & \biggl(\int_{\Gamma} F\biggr)_{(\tau, \varphi)} \\
	&\cdot \prod_{i \in BV} g_{\varphi(B^{\alpha}(i)), \varphi(B^{1}(i)), \varphi'(B^{1}(i))} \cdot g_{\varphi(B^{\alpha}(i)), \varphi(B^{k_{i}}(i)), \varphi'(B^{k_{i}}(i))}^{-1}.
\end{split}\]
Thus, we can interpret this expression thinking to have a transition function from $V_{(\tau, \varphi)}$ to $V_{(\tau', \varphi')}$ given by:
	\[\tilde{g}_{(\tau, \varphi), (\tau, \varphi')}(\Gamma) = \prod_{i \in BV} g_{\varphi(B^{\alpha}(i)), \varphi(B^{1}(i)), \varphi'(B^{1}(i))} \cdot g_{\varphi(B^{\alpha}(i)), \varphi(B^{k_{i}}(i)), \varphi'(B^{k_{i}}(i))}^{-1} \Gamma(\sigma^{0}_{i}).
\]
One can prove that independence from the triangulation.

\paragraph{}If we consider three charts $V_{(\tau, \varphi)}$, $V_{(\tau', \varphi')}$ and $V_{(\tau'', \varphi'')}$, these transition functions satisfy a cocycle condition. Thus we obtain:
\begin{Theorem} For $F$ connection on a gerbe $\mathcal{G} \in H^{2}(X, C^{0}(S^{1}))$, $\exp \int_{\Gamma} F$ is a section of a line bundle $\Sigma \mathcal{G} \rightarrow \Sigma X$, whose transition function from $V_{(\tau, \varphi)}$ to $V_{(\tau', \varphi')}$ is given by $\tilde{g}_{(\tau, \varphi), (\tau', \varphi')}(\Gamma)$.
\end{Theorem}

\paragraph{}From the transition functions we see another important fact: \emph{if the gerbe $\mathcal{G}$ is trivial, then $\Sigma \mathcal{G}$ is trivial too}, since, for $g_{\alpha\beta\gamma} = g_{\alpha\beta} \cdot g_{\beta\gamma} \cdot g_{\gamma\alpha}$ on $X$, we have:
	\[\begin{split}
	\tilde{g}_{(\tau, \varphi), (\tau', \varphi')}(\gamma) &= g_{\varphi(l)}\gamma(1) \cdot g_{\varphi'(l)}\gamma(1)^{-1} \cdot g_{\varphi'(1)}\gamma(0) \cdot g_{\varphi(1)}\gamma(0)^{-1}\\
	&= \bigl(g_{\varphi(l)}\gamma(1) \cdot g_{\varphi(1)}\gamma(0)^{-1}\bigr) \cdot \bigl( g_{\varphi'(l)}\gamma(1) \cdot g_{\varphi'(1)}\gamma(0)^{-1} \bigr)^{-1}
\end{split}\]
so that we have a trivialization:
\begin{equation}\label{TrivializationGerbe}
\tilde{g}_{(\tau, \varphi)}(\gamma) = g_{\varphi(l)}\gamma(1) \cdot g_{\varphi(1)}\gamma(0)^{-1}.
\end{equation}
Moreover, as we see from the previous expression, a trivialization of $L$ determines \emph{canonically} a trivialization of $CL$. Thus, when $L$ is given with a trivialization, i.e.\ when $A$ is a connection on $X \times \mathbb{C}$, then $CL$ is canonically trivialized, i.e.\ $\exp \int_{\gamma} A$ is a section of $CX \times \mathbb{C}$, so it is a function.

\paragraph{}We now see that, under the trivialization \eqref{TrivializationGerbe}, the function we obtain from $\exp \int_{\gamma} A$ is exactly \eqref{WilsonLoopGerbe}. In fact, from the section $\{s_{\alpha}\}$ and the trivialization $\{g_{\alpha}\}$ we obtain the function $\{g_{\alpha}s_{\alpha}\}$, in this case we obtain from \eqref{WilsonLoopGerbe} and \eqref{TrivializationGerbe}:
	\[\begin{split}
	\exp &\biggl(\int_{\gamma} A\biggr)_{(\tau, \varphi)} \cdot \tilde{g}_{(\tau, \varphi)}(\gamma)\\
	& = \prod_{i=1}^{l(\tau)-1}\; \exp \Biggl[ \Biggl(\int_{\gamma(\sigma^{1}_{i})}A_{\varphi(i)} \Biggr) \cdot g_{\varphi(i),\varphi(i+1)}\gamma(\sigma^{0}_{i+1}) \Biggr] \cdot \exp\Biggl( \int_{\gamma(\sigma^{1}_{l})}A_{\varphi(l)} \Biggr)\\
	&\phantom{XXXXXXXXXXXXXXXXXXXXXXX} \cdot g_{\varphi(l)}\gamma(1) \cdot g_{\varphi(1)}\gamma(0)^{-1}\\
	& = \prod_{i=1}^{l(\tau)-1} \; \exp \Biggl[ \Biggl( \int_{\gamma(\sigma^{1}_{i})}A_{\varphi(i)} \Biggr) \cdot g_{\varphi(i)}\gamma(\sigma^{0}_{i+1}) \cdot g_{\varphi(i+1)}\gamma(\sigma^{0}_{i+1})^{-1} \Biggr] \cdot \exp\Biggl( \int_{\gamma(\sigma^{1}_{l})}A_{\varphi(l)} \Biggr)\\
	&\phantom{XXXXXXXXXXXXXXXXXXXXXXX} \cdot g_{\varphi(l)}\gamma(\sigma^{0}_{l+1}) \cdot g_{\varphi(1)}\gamma(\sigma^{0}_{1})^{-1}\\
	&= \prod_{i=1}^{l(\tau)} \; \exp \Biggl[ \Biggl( \int_{\gamma(\sigma^{1}_{i})}A_{\varphi(i)} \Biggr) \cdot g_{\varphi(i)}\gamma(\sigma^{0}_{i+1}) \cdot g_{\varphi(i)}\gamma(\sigma^{0}_{i})^{-1} \Biggr]
\end{split}\]
and the last expression is exactly the exponential of \eqref{WilsonLineGerbe}.


\part{Type II superstring backgrounds}

\chapter{D-brane charge and Ramond-Ramond fields}

\section{D-brane charge}

We now want to discuss the D-brane charge from the homological point of view. We reproduce here \cite[Chap.\ 2,4,5]{FR2}. Since this is a generalizations of electromagnetism theory with higher-dimensional sources, we start with a brief review of classical electromagnetism theory in four dimensions. For details the reader can see \cite{Naber}.

\subsection{Preliminaries of electromagnetism}

Let us consider an empty minkowskian space-time $\mathbb{R}^{1,3}$. Then Maxwell equations are:
\begin{equation}\label{MaxwellEmpty}
	dF = 0 \qquad d*F = 0
\end{equation}
whose solutions represent electric and magnetic fields without sources. In particular, in a fixed reference frame:
\begin{equation}\label{FMatrix}
	F = \begin{bmatrix} 0 & E^{1} & E^{2} & E^{3} \\ -E^{1} & 0 & B^{3} & -B^{2} \\ -E^{2} & -B^{3} & 0 & B^{1} \\ -E^{3} & B^{2} & -B^{1} & 0
	\end{bmatrix}
\end{equation}
and equations \eqref{MaxwellEmpty} assume their classical form $\nabla \times \underline{E} + \frac{\partial \underline{B}}{\partial t} = 0$ and $\nabla \cdot \underline{B} = 0$ for $dF = 0$, and $\nabla \times \underline{B} - \frac{\partial \underline{E}}{\partial t} = 0$ and $\nabla \cdot \underline{E} = 0$ for $d*F = 0$. Since $\mathbb{R}^{1,3}$ is contractible so that the cohomology is zero, both $F$ and $*F$ are exact: $F = dA$, where $A$ is the scalar potential, i.e.\ $A = (V, \underline{A})$ with $\underline{E} = -\frac{\partial \underline{A}}{\partial t} - \nabla V$ and $\underline{B} = \nabla \times \underline{A}$. Similarly we can find a potential $A'$ such that $*F = dA'$, satisfying the same equations replacing $\underline{B}$ by $\underline{E}$ and $\underline{E}$ by $-\underline{B}$: electric and magnetic fields are interchangeable by Hodge-duality, in fact the matrix representation of $*F$ can be obtained from \eqref{FMatrix} again replacing $\underline{B}$ by $\underline{E}$ and $\underline{E}$ by $-\underline{B}$ (the minus is due to the fact that $**F = -F$ in the minkowskian signature). Thus, up to exchange $F$ and $*F$, electric and magnetic fields without sources are are equivalent.

We now consider an electric charge $q$, moving without accelerating, as a source for the electric field. In this case, Maxwell equations becomes:
\begin{equation}\label{MaxwellCharge}
	dF = 0 \qquad d*F = q \cdot \delta(w)
\end{equation}
where $w$ is the world-line of the particle. This means that we interpret $*F$ not as a form any more but as a current, which is singular in $w$, while in $\mathbb{R}^{1,3} \setminus w$ it is regular and, by equations \eqref{MaxwellCharge}, closed. Instead, $F$ is a closed current on all $\mathbb{R}^{1,3}$, thus it is also exact. Since $H^{2}_{dR}(\mathbb{R}^{1,3} \setminus w) \simeq \mathbb{R}$ being $\mathbb{R}^{1,3} \setminus w$ homotopic to $S^{2}$, the form $*F$ is in general not exact, actually, as it follows from equations \eqref{MaxwellCharge}, if we consider a linking surface $S^{2} \subset \mathbb{R}^{1,3}$ of $w$ we have that $\int_{S^{2}} *F = q$, thus $[*F]_{dR} \simeq q$ under the isomorphism $H^{2}_{dR}(\mathbb{R}^{1,3} \setminus w) \simeq \mathbb{R}$. Instead $F$, being exact on the whole $\mathbb{R}^{1,3}$, is exact also when restricted to $\mathbb{R}^{1,3} \setminus w$, so that it is topologically trivial. That's the well-known fact that the electric charge, represented by $F$, is not topological, while the magnetic charge, which is the electric one for $*F$, is encoded in the topology of space-time. Here we see the difference between electric and magnetic charges. In particular, considering a charged particle moving in this background, its actions minimally couples to a potential $A$ of $F$ if we consider the field as electric, in which case $A$ can be globally defined, or to a potential $A'$ of $*F$ (in $\mathbb{R}^{1,3} \setminus w$) if we consider the field as magnetic, in which case $A$ is only local and we must consider gauge transformations (or viceversa if we exchange $F$ and $*F$ up to a sign). In particular, only from magnetic fields we can find Dirac quantization condition, i.e.\ $q \in \mathbb{Z}$ up to a normalization constant, not from electric ones.

The solutions of \eqref{MaxwellCharge} can all be obtained from a particular one adding the solutions of \eqref{MaxwellEmpty}. One particular solution of \eqref{MaxwellCharge}, in a reference frame in which the charge is fixed in the origin so that $w = \mathbb{R} \times \{0\}$, is:
	\[F = dA, \quad A = \textstyle-\frac{q}{r}dt
\]
where $r$ is the distance of a point from the origin in $\mathbb{R}^{3}$ (thus $A$ is constant in time). Note that the potential $A$ is a $L^{1}_{loc}$-form an all $\mathbb{R}^{3}$, thus $F$ is an exact current in $\mathbb{R}^{3}$. In this way, calling $\vol_{S^{2}} := x_{1}dx_{2} \wedge dx_{3} - x_{2}dx_{1} \wedge dx_{3} + x_{3}dx_{1} \wedge dx_{2}$ the 2-form restricting on $S^{2} \subset \mathbb{R}^{3}$ to the volume form, we get:
	\[\begin{array}{ll}
	& F = dA = \frac{q}{r^{2}} dr \wedge dt = \frac{q}{r^{3}} rdr \wedge dt\\
	& *F = \frac{q}{r^{3}}\vol_{S^{2}}\\
	\textnormal{For $r \neq 0$: } & d*F = -\frac{3q}{r^{4}}dr\wedge \vol_{S^{2}} + \frac{q}{r^{3}}3\vol_{\mathbb{R}^{3}} = \frac{3q}{r^{4}}(r\vol_{\mathbb{R}^{3}} - dr \wedge \vol_{S^{2}}) = 0.
\end{array}\]
Instead, as a current in the whole $\mathbb{R}^{3}$, $d*F = q \cdot \delta(0)$, since:
	\[\begin{split}
	 \langle d*F, \varphi \rangle &= -\langle *F, d\varphi \rangle = -q \int_{\mathbb{R}^{3}} \frac{1}{r^{3}} \vol_{S^{2}} \wedge d\varphi \\
	 & = -q \int_{\mathbb{R}^{3}} \frac{1}{r^{3}} r \frac{d\varphi}{dr}\vol_{\mathbb{R}^{3}} = -q\int_{0}^{+\infty} \frac{d\varphi}{dr}dr = q \cdot \varphi(0)
\end{split}\]
up to a normalization constant. This solution is static.

\paragraph{}We now make some topological remarks. We consider the following cohomology groups for a manifold $X$:
\begin{itemize}
	\item $H^{n}_{dR}(X)$ is the de-Rham $n$-cohomology group, i.e.\ the group of closed $n$-forms up to the exact ones;
	\item $H^{n}_{crn}(X)$ is the $n$-cohomology group of \emph{currents} on $X$; 
	\item $H^{n}(X, \mathbb{R})$ is the singular $n$-cohomology group with real coefficients.
\end{itemize}
These three groups are canonically isomorphic. In particular, the natural map $H^{n}_{dR}(X) \rightarrow H^{n}_{crn}(X)$, obtained thinking of a form as a current, is a canonical isomorphism. To realize an isomorphism between $H^{n}(X, \mathbb{R})$ and $H^{n}_{dR}(X)$ we can use iteratively the Poincar\'e lemma, as explained in \cite{BFS}. For all of these three groups we can consider the compactly-supported version, which we call respectively $H^{n}_{dR,cpt}(X)$, $H^{n}_{crn,cpt}(X)$ and $H^{n}_{cpt}(X, \mathbb{R})$. They are still isomorphic via the restrictions of the previous isomorphisms.

We can define the singular cohomology groups with integral coefficients $H^{n}(X, \mathbb{Z})$, and there is a natural map (not injective in general) $H^{n}(X, \mathbb{Z}) \rightarrow H^{n}(X, \mathbb{R})$ whose image is made by quantized real cohomology classes: the latter correspond in the de-Rahm cohomology to the forms which give an integral value when integrated over a cycle. Poincar\'e duality provides \emph{on a manifold} a canonical isomorphism $\PD: H_{n}(X, \mathbb{Z}) \overset{\simeq}\longrightarrow H^{\dim(X) - n}_{cpt}(X, \mathbb{Z})$ with the analogous version for real coefficients.

Coming back to the electric source in $\mathbb{R}^{1,3}$, if we restrict the second equation of \eqref{MaxwellCharge} to a fixed instant of time, we get $[d(*F)\vert_{\{t\} \times \mathbb{R}^{3}}]_{cpt} = q \cdot \delta(\{p\})$, for $p = w \cap (\{t\} \times \mathbb{R}^{3})$. Since the point $p$ is compact (contrary to $w$), it defines an homology class $[p] \in H_{0}(\mathbb{R}^{3}, \mathbb{Z})$, thus we can define a compactly-supported cohomology class $\PD_{\mathbb{R}^{3}}([p])$. Under the isomorphism $H^{n}_{cpt}(X, \mathbb{R}) \simeq H^{n}_{crn,cpt}(X)$ one has $\PD_{\mathbb{R}^{3}}([p]) \simeq [\delta(p)]$, hence we obtain from Maxwell equations:
\begin{equation}\label{MaxwellCohomology}
	[d(*F)\vert_{\{t\} \times \mathbb{R}^{3}}]_{cpt} = q \cdot \PD_{\{t\} \times \mathbb{R}^{3}}([p]).
\end{equation}
This identity seems meaningless because we are identifying the class of an exact form with a cohomology class which is in general non-trivial. Actually, we are dealing with \emph{compactly supported} cohomology classes, which can be trivial when considered as generic cohomology classes. Thus, the identity is meaningful and implies that the support of $(*F)\vert_{\{t\} \times \mathbb{R}^{3}}$ is not compact. In this way, we can see the electric (or magnetic) source as a homology cycle conserved in time whose coefficient is the charge; its Poincar\'e dual measures the non-closure of the associated magnetic field strength as a current.\footnote{We can consider an inertial reference frame in which the charge is fixed in the origin, so that we consider the electric field it creates in $\mathbb{R}^{1,3} \setminus \{(t, 0, 0, 0)\}$. In this such a frame we can choose the solution of \eqref{MaxwellCharge} given by $\underline{E} = \frac{q}{r^{2}}\underline{u}_{0}$ and $\underline{B} = 0$. As it follows from \eqref{FMatrix}, $F$ is then of the form $F = dt \wedge F'$, thus $*F$ is time-independent and its restriction to any space-slice is the same. In this way we can simply write $d*F = q\cdot \delta(0)$ and $[d*F]_{cpt} = \PD_{\mathbb{R}^{3}}[\{0\}]$, but this picture is not Lorentz invariant.} This viewpoint seems redundant for a point-charge, but for an extended object as a D-brane, which can be topologically non-trivial, it is much more natural.

The cohomological expression \eqref{MaxwellCohomology} is not Lorentz-invariant, since we must fix an instant of time. If we were able to treat $w$ as a homology cycle, we could get from Maxwell equations a Lorentz-invariant expression:
	\[[d*F] = q \cdot \PD_{\mathbb{R}^{1,3}}(w)
\]
without fixing a particular reference frame. We will develop the suitable homology theory to do this.

\paragraph{}We remark that since the de-Rahm cohomology and the cohomology of currents are isomorphic, we can also think of $(*F)_{0}$ as a compactly-supported form whose support is contained in a small neighborhood of the origin. Similarly, the whole $*F$ is a form whose support is contained in a small neighborhood of $t_{0}$. In this case, when we compute the charge as $q = \int_{S^{2}} *F$, we must take $S^{2}$ outside the neighborhood. Using currents or forms is not important, since their cohomology are canonically isomorphic; what really counts is that we consider compactly-supported classes, which can be non-trivial also in $\mathbb{R}^{3}$. However, it is more natural to use currents since Maxwell equations are naturally formulated with a $\delta$-function.

\subsection{Charge of a D-brane}

We consider type II superstring theory in a ten-dimensional space-time of the form $S = \mathbb{R}^{1,3} \times X$ for $X$ in general compact but not necessarily, such that the background metric in $\mathbb{R}^{1,3}$ is the standard minkowski metric $\eta^{\mu\nu}$ and \emph{the $H$-flux is zero}. A Dp-brane $Y_{p}$ has a $(p+1)$-dimensional world-volume $WY_{p} \subset S$, which represents a classical trajectory in space-time. To define the charge of the D-brane, as for a particle we think that it is moving without accelerating in the non-compact directions $\mathbb{R}^{1,3}$ (so the projection on $X$ is fixed), so that the violated Bianchi identity becomes:
\begin{equation}\label{MaxwellDBrane}
	dG_{8-p} = q \cdot \delta(WY_{p}) \qquad dG_{p+2} = 0
\end{equation}
where $q$ is the charge, or equivalently, the number of D-branes in the stack. To compute the charge from the background data, we consider a linking manifold\footnote{A linking manifold is the boundary of a manifold intersecting $WY_{p}$ transversally in isolated points of its interior.} $L$ of $WY_{p}$ in $S$ with linking number $l$, so that we have:
	\[q = \frac{1}{l} \int_{L} G_{8-p}.
\]
We can always choose a linking sphere (so that $l = 1$): in fact, we choose near a point $p \in WY_{p}$ a reference frame such that $WY_{p}$ corresponds to the first $p+1$ coordinates, then we take a small sphere in the transverse coordinates. From Dirac quantization condition (v.\ section \ref{WessZumino}, \cite{Freed2}) we know that the charge is quantized, thus $G_{8-p}$ must be an integral form. In particular, since by \eqref{MaxwellDBrane} we see that $G_{8-p}$ is not closed, we should say that $G_{8-p}$ restricted to the complement of $WY_{p}$ represents an integral cohomology class. Actually the quantization of Ramond-Ramond field seems inconsistent with the duality relation $G = *G$ in the democratic formulation of supergravity. This is the so called $*$-problem, whose solution can be found in \cite{BM3}.

We now suppose that the brane is a particle in $\mathbb{R}^{1,3}$. In a fixed reference frame we call $M$ the space manifold $M = \mathbb{R}^{3} \times X$. We fix at an instant of time $t$ the D-brane volume $Y_{p,t} \subset \{t\} \times M$. We call $M_{t} := \{t\} \times M$. Then, the violated Bianchi identity becomes $d_{M_{t}}(G_{8-p}\vert_{M_{t}}) = q \cdot \delta(Y_{p,t})$ so that, \emph{if $Y_{p,t}$ is compact} (which is always the case when the brane is a particle in $\mathbb{R}^{1,3}$ if $X$ is compact), we obtain:
\begin{equation}\label{BianchiId}
	[\, d_{M_{t}}(G_{8-p}\vert_{M_{t}}) \,]_{\cpt} = \PD_{M_{t}}(q \cdot Y_{p,t}).
\end{equation}
As pointed out before, it is important that the space manifold $M$ is non-compact \cite{MW, FS}, so that the Poincar\'e dual of the brane volume is a \emph{compactly supported} cohomology class, which can be trivial as a generic cohomology class. Thus, the identity \eqref{BianchiId} implies that, for homologically non-trivial branes, the support of $G_{8-p}\vert_{\{t\} \times M}$ is not compact. In particular, $\PD_{M}(q \cdot Y_{p})$ must live in the kernel of the natural map $\iota: H^{9-p}_{\cpt}(M) \rightarrow H^{9-p}(M)$. We could also write the first equation as $dG_{8-p} = \PD_{S}(q \cdot WY_{p})$, but, since $WY_{p}$ is in general non-compact\footnote{If the brane is stable it exists for all the time, from $-\infty$ to $+\infty$, thus the world-volume is non compact.} so that it does not define an homology cycle, we postpone this discussion.

We can compute the charge $q$ at any fixed instant: if we consider a linking surface $L_{t}$ of $Y_{p,t}$ in $M_{t}$ with linking number $l$, we have $q = \frac{1}{l} \int_{L_{t}} (G_{8-p}\vert_{M_{t}})$. The charge $q$ is conserved in time, actually all the homology class of the D-brane is conserved. In fact, let us consider two volumes $Y_{p,t_{1}}$ and $Y_{p,t_{2}}$. Then we can consider the piece of the world-volume linking them, which is $(WY_{p})\vert_{[t_{1}, t_{2}] \times M}$. If we consider the canonical identification $M_{t_{1}} \simeq M_{t_{2}} \simeq M$, we can consider both $Y_{t_{1}}$ and $Y_{t_{2}}$ as cycles in $M$. If we consider the projection $\pi: [t_{1}, t_{2}] \times M \rightarrow M$, then $\pi((WY_{p})\vert_{[t_{1}, t_{2}] \times M})$ is a singular chain in $M$ which makes $Y_{t_{1}}$ and $Y_{t_{2}}$ homologous. Thus they have the same Poincar\'e dual and they define the same charge.

\paragraph{}As for classical electromagnetism, the solutions of \eqref{MaxwellDBrane} can be obtained from a fixed one adding the solution to the equations in the empty space:
\begin{equation}\label{MaxwellDEmpty}
	dG_{8-p} = 0 \qquad dG_{p+2} = 0.
\end{equation}
We study a particular static solution, which we aspect to be similar to the one of classical electromagnetism. Let us consider a brane that is a particle in $\mathbb{R}^{1,3}$ and a reference frame in which it is fixed in the origin. Thus we have a cycle $Y_{p} \subset \{0\} \times X$. We consider the case in which there is a foliation of $(\mathbb{R}^{3} \times X) \setminus Y_{p}$ made by manifolds of points at a fixed distance from $Y_{p}$, as in classical electromagnetism where the origin foliates $\mathbb{R}^{3} \setminus \{0\}$ in spheres: for example, if we imagine a torus embedded in $\mathbb{R}^{3}$ in the standard way and we consider a vertical circle as a cycle, it foliates the torus in couples of circles parallel to it at a fixed distance (with the exception of the opposite one, in which case the two circles of the couple collapse to the same one). We now consider for a point $x \in (\mathbb{R}^{3} \times X) \setminus Y_{p}$ the manifold $Z_{x}$ of the foliation containing $x$, which has dimension 8 (independently on $p$), since to cover a neighborhood of it we need the coordinates on $Z_{x}$ and only one parameter more, the distance from $Y_{p}$. Then, in $T_{x}(Z_{x})$, we consider the subspace $V_{x}$ parallel to the D-brane, i.e.\ if $d(x, Y_{p}) = d(x,y)$, we consider the submanifold of points in $Z_{x}$ distant $r$ from $y$ and we consider its orthogonal. We call $z_{1}, \ldots, z_{p}$ an orthonormal system of generators of $V_{x}$. Then we define:
	\[G_{p+2} = dA, \quad A = -\frac{q}{9-p-2} \cdot \frac{1}{r^{9-p-2}} \, dz_{1} \wedge \ldots \wedge dz_{p} \wedge dt
\]
so that:
	\[\begin{array}{rcl}
	G_{p+2} & = & dA \,=\, \frac{q}{r^{9-p-1}} \, dr \wedge dz_{1} \wedge \ldots \wedge dz_{p} \wedge dt\\
	& = & \frac{q}{r^{9-p}} \, r dr \wedge dz_{1} \wedge \ldots \wedge dz_{p} \wedge dt \\
	G_{8-p} & = & \frac{q}{r^{9-p}} \vol_{Z}
\end{array}\]
where $Z$ is a submanifold of points at fixed distance with respect to $Y_{p}$. In this way, as before, $dG_{8-p} = \delta(Y_{p})$. In this solution $G_{p+2}$ is exact, while $G_{8-p}$ is non-trivial only on the cycle given by a linking sphere of $Y_{p}$ with weight $q$: all such linking spheres are homologous, since, if they have both the same radius, they are linked by a piece of the suitable leaf $Z$ of the foliation. We remand to the next paragraph for a more complete discussion about this.

\paragraph{}For what concerns the solutions of \eqref{MaxwellDEmpty}, from a Matrix representation analogous to \eqref{FMatrix} we get usual Maxwell equations. In particular, we split the Ramond-Ramond fields in the following way:
\begin{equation}\label{RRSplit}
	G_{p} = G^{s}_{p} + dt \wedge G^{t}_{p-1}
\end{equation}
so that, calling $*_{9}$ the Hodge-$*$ in $M$, which is euclidean:
\begin{equation}\label{RRHodge}
	*G_{p} = -*_{9}G^{t}_{p-1} + dt \wedge (-1)^{p}*_{9}G^{s}_{p}.
\end{equation}
In fact, all the terms of $G^{s}_{p}$ are of the form $f \cdot dx_{i_{1}} \wedge \ldots \wedge dx_{i_{p}}$, and their Hodge-$*$ is $\varepsilon^{i_{1}, \ldots, i_{p}, 0, j_{1}, \ldots, j_{9-p}} f \cdot dt \wedge dx_{j_{1}} \wedge \ldots \wedge dx_{j_{9-p}} = (-1)^{p}\varepsilon^{i_{1}, \ldots, i_{p}, j_{1}, \ldots, j_{9-p}} f \cdot dt \wedge dx_{j_{1}} \wedge \ldots \wedge dx_{j_{9-p}} = (-1)^{p}dt \wedge *_{9} f \cdot dx_{i_{1}} \wedge \ldots \wedge dx_{i_{p}}$. Similarly, all the terms of $dt \wedge G^{t}_{p-1}$ are of the form $f \cdot dt \wedge dx_{i_{1}} \wedge \ldots \wedge dx_{i_{p-1}}$ and their Hodge-$*$ is $-\varepsilon^{0,i_{1}, \ldots, i_{p-1},j_{1}, \ldots, j_{10-p}} f \cdot dt \wedge dx_{i_{1}} \wedge \ldots \wedge dx_{i_{p-1}} = -\varepsilon^{i_{1}, \ldots, i_{p-1},j_{1}, \ldots, j_{10-p}} f \cdot dt \wedge dx_{i_{1}} \wedge \ldots \wedge dx_{i_{p-1}} = -*_{9}f \cdot dt \wedge dx_{i_{1}} \wedge \ldots \wedge dx_{i_{p-1}}$, the minus sign being due to the fact that $dt$ is negative definite (for a review of Hodge-$*$ with minkowskian signature see appendix \ref{HodgeMinkowski}). Then the solutions of $dG_{p} = 0$ and $d*G_{p} = 0$ becomes:
	\[\begin{array}{lllll}
	dG_{p} = 0: & & \frac{\partial G^{s}_{p}}{\partial t} - d_{9}G^{t}_{p-1} = 0 & & d_{9}G^{s}_{p} = 0\\
	d*G_{p} = 0: & & -*_{9}\frac{\partial G^{t}_{p-1}}{\partial t} - (-1)^{p} d_{9}*_{9}G^{s}_{p} = 0 & & d_{9}*_{9}G^{t}_{p-1} = 0
\end{array}\]
which correspond to Maxwell equations for $p = 2$ in dimension 3 if we identify $G^{s}_{p} = *_{3}\varphi(B)$ and $G^{t}_{p-1} = -\varphi(E)$ for $\varphi: T(\mathbb{R}^{3}) \rightarrow T^{*}(\mathbb{R}^{3})$ the isomorphism given by the metric.

There is a difference with respect to the classical electromagnetism theory, due to the fact that the space-time $\mathbb{R}^{1,3} \times X$ can have non-trivial cycles in itself, even before putting the charge source, contrary to $\mathbb{R}^{1,3}$ which is contractible. Thus, the equations $dG_{8-p} = 0$ and $dG_{p+2} = 0$ do not imply that $G_{8-p}$ and $G_{p+2}$ are exact. We briefly analyze this difference. Since $\mathbb{R}^{1,3}$ is contractible, the natural immersion $i: X \rightarrow \mathbb{R}^{1,3} \times X$, defined by $i(x) = (0, x)$, induces an isomorphism in cohomology $i^{*}: H^{*}_{dR}(\mathbb{R}^{1,3} \times X) \overset{\simeq}\longrightarrow H^{*}_{dR}(X)$ sending a class $[\omega]$ in the class $[\omega_{0}]$ for $\omega_{0} := \omega\vert_{\{0\} \times X}$. Thus, for any closed $p$-form $\omega$, we have $\omega = \omega_{0} + d\rho$ with\footnote{We should write $\omega = \pi^{*}\omega_{0} + d\rho$ for $\pi: \mathbb{R}^{1,3} \times X \rightarrow X$ the projection, but for simplicity we identify a form on $X$ (as $\omega_{0}$) with the corresponding form on $\mathbb{R}^{1,3} \times X$ which does not depend on $\mathbb{R}^{3}$ (as $\pi^{*}\omega_{0}$).} $\omega_{0} \in \Lambda^{p}T^{*}X$ \emph{closed} and $\rho \in \Lambda^{p-1}T^{*}(\mathbb{R}^{1,3} \times X)$. Now, since $X$ is compact, we can apply Hodge decomposition theorem \cite{GH} to $\omega_{0}$ so that, being it closed, we obtain $\omega_{0} = h_{0} + d\rho_{0}$ with $h_{0}$ \emph{harmonic} in $X$. We can suppose $d\rho_{0}$ already included in $d\rho$, so we finally get:
\begin{equation}\label{Hodge}
	\omega = h_{0} + d\rho
\end{equation}
with $h_{0} \in \Lambda^{p}T^{*}X$ \emph{harmonic} and $\rho \in \Lambda^{p-1}T^{*}(\mathbb{R}^{1,3} \times X)$. The form $h_{0}$ is uniquely determined by the cohomology class of $\omega$, thus, if we fix such a class, we remain with the freedom of $\rho$. In particular, we have that:
	\[G_{p} = (h_{0})_{p} + d\rho_{p} \qquad *G_{p} = *(h_{0})_{p} + *d\rho_{p}
\]
where $\rho_{p}$ is the analogue of the potential $A$. We remark that $*(h_{0})_{p}$ is exact being $*h_{0} = (-1)^{p}dt \wedge dx_{1} \wedge dx_{2} \wedge dx_{3} \wedge *_{6}h_{0} = d\bigl( (-1)^{p} x_{1} \wedge dx_{2} \wedge dx_{3} \wedge *_{6}h_{0}\bigr)$, where in the last equality we used the fact that $*_{6}h_{0}$ is closed being $h_{0}$ harmonic in $X$. Thus, the non-triviality of the space-time topology is encoded in $(h_{0})_{p}$ for the electric charge and in the possible non-triviality of $*d\rho_{p}$ for the magnetic one. We interpret this physically by the fact the a non-trivial cycle can be thought of as a trivial one removing a charge, so that the charge is encoded in the background. For example, in classical electromagnetism if we consider the background $\mathbb{R}^{1,3} \setminus w$ for $w$ the world-line of a charge, in that background Maxwell equations in empty space are satisfied, but the topology is non-trivial and $*F$ is not exact. The situation is analogous.

\paragraph{}Up to now have assumed the existence of a suitable foliation of space-time in order to reproduce a situation analogue to the one of classical electromagnetism, but we can show that we can solve in general Maxwell equations. We search a static solution $G_{p} = dt \wedge G^{t}_{p-1}$, so that, thanks to \eqref{RRHodge}, we have $*G_{p} = -*_{9}G^{t}_{p-1}$. We use smooth forms instead of currents for simplicity, then it will be immediate to reduce to Maxwell equations formulated with $\delta$-functions. Let us consider a form $G^{t}_{p-1} \in \Lambda^{p-1}(\mathbb{R}^{1,3} \times X)$, decomposed as in \eqref{Hodge}: we now want to study the compactly-supported cohomology class of $d*_{9}G^{t}_{p-1}$. Given a function $e: \mathbb{R} \rightarrow \mathbb{R}$ such that $\int_{-\infty}^{+\infty} e = 1$, for any manifold $A$ there is an isomorphism:
\begin{equation}\label{IsoCptCohom}
\begin{split}
	e_{*}: \;&H_{dR,\cpt}^{n-1}(A) \overset{\simeq}\longrightarrow H_{dR,\cpt}^{n}(\mathbb{R} \times A)\\
	&[\,\eta\,] \longrightarrow [\,e(x)\,dx \wedge \eta\,]
\end{split}
\end{equation}
whose inverse is the pull-back $\pi^{*}$ of the projection $\pi: \mathbb{R} \times A \rightarrow A$ \cite{BT}.\footnote{For currents the isomorphism \eqref{IsoCptCohom} can be described by $[\delta(Y)] \rightarrow [\delta(\{0\} \times Y)]$.} Thus, fixing three functions $e_{1}, e_{2}, e_{3}$ with integral $1$ we obtain an isomorphism $H^{n-3}_{dR}(X) \overset{\simeq}\longrightarrow H_{dR,\cpt}^{n}(\mathbb{R}^{3} \times X)$ given by $[\,\eta\,] \longrightarrow [\,e_{1}(x)\,dx_{1} \wedge e_{2}(x)\,dx_{2} \wedge e_{3}(x)\,dx_{3} \wedge \eta\,]$. If we want to fix the cohomology class $[\,d*G^{t}_{p-1}\,]_{\cpt}$, we can choose $\alpha$ harmonic on $X$ corresponding (uniquely) to the fixed class under the latter isomorphism, and require:
	\[d*G^{t}_{p-1} = e_{1}(x)\,dx_{1} \wedge e_{2}(x)\,dx_{2} \wedge e_{3}(x)\,dx_{3} \wedge \alpha + d\xi_{cpt}
\]
for any compactly-supported form $\xi_{cpt}$. In order to show how to solve this equation, we remark that:
\begin{itemize}
	\item considering \eqref{Hodge}, we have that $d*_{9}h_{0} = 0$, since $*_{9}h_{0} = (-1)^{p}dx_{1} \wedge dx_{2} \wedge dx_{3} \wedge *_{6}h_{0}$ where $*_{6}$ is the Hodge-dual on $X$; hence, being $h_{0}$ harmonic in $X$, $d*_{9}h_{0} = 0$, so that we have to consider the cohomology class $[\,d*_{9}d\rho\,]$;
	\item for $\alpha$ closed, $e(x)\,dx \wedge \alpha = d\bigl(\int_{0}^{x}e \cdot \alpha\bigr)$ as one can see from the Leibniz rule or directly from the definition of exterior differential.
\end{itemize}
Thus we obtain:
	\[\begin{array}{l}
	d*_{9}d\rho = e_{1}(x)\,dx_{1} \wedge e_{2}(x)\,dx_{2} \wedge e_{3}(x)\,dx_{3} \wedge \alpha + d\xi_{cpt}\\
	d*_{9}d\rho = d\bigl(\int_{0}^{x_{1}}e_{1} \cdot e_{2}(x)\,dx_{2} \wedge e_{3}(x)\,dx_{3} \wedge \alpha\bigr) + d\xi_{cpt}\\
	*_{9}d\rho = \int_{0}^{x_{1}}e_{1} \cdot e_{2}(x)\,dx_{2} \wedge e_{3}(x)\,dx_{3} \wedge \alpha + \xi_{cpt} + \eta_{closed}\\
	d\rho = \int_{0}^{x_{1}}e_{1} \cdot *_{9}\bigl(e_{2}(x)\,dx_{2} \wedge e_{3}(x)\,dx_{3} \wedge \alpha\bigr) + *_{9}\xi_{cpt} + *_{9}\eta_{closed}.
\end{array}\]
Let us show that the first term of the r.h.s. is actually exact. Since $*_{9}\bigl(e_{2}(x)\,dx_{2} \wedge e_{3}(x)\,dx_{3} \wedge \alpha\bigr) = (-1)^{p-1}dx_{1} \wedge *_{6}\alpha$ we obtain:
	\[\begin{split}
	\textstyle \int_{0}^{x_{1}}e_{1} \cdot *_{9}\bigl(e_{2}(x)\,dx_{2} \wedge e_{3}(x)\,dx_{3} \wedge \alpha\bigr) &= \textstyle (-1)^{p} (\int_{0}^{x_{1}}e_{1})dx_{1} \wedge *_{6}\alpha\\
	&= \textstyle d\bigl( (-1)^{p} \int_{0}^{x_{1}} \int_{0}^{y_{1}} e_{1} \cdot *_{6}\alpha \bigr)
\end{split}\]
where the last equality is due to the fact that $*_{6}\alpha$ is closed since $\alpha$ has been chosen harmonic on $X$. Hence we obtain:
\begin{equation}\label{Rho}
	\textstyle \rho = (-1)^{p} \int_{0}^{x_{1}} \int_{0}^{y_{1}} e_{1} \cdot *_{6}\alpha + \psi + \lambda_{closed}
\end{equation}
where $d\psi = *_{9}\xi_{cpt} + *_{9}\eta_{closed}$. The form $\psi$, in particular for what concerns $\eta_{closed}$, encodes the freedom of Maxwell equations in empty space.

Asking $[d*_{9}G^{t}_{p-1}]_{cpt} = [e_{1}(x)\,dx_{1} \wedge e_{2}(x)\,dx_{2} \wedge e_{3}(x)\,dx_{3} \wedge \alpha]_{cpt}$, we found no obstructions on $\alpha$: this could seem strange, since the r.h.s. must represent a class which is exact in the ordinary cohomology (not compactly supported), being $d*_{9}\omega$ exact. In particular, $[d*_{9}\omega]_{\cpt}$ lies in the kernel of the natural map $\iota: H^{9-p}_{\cpt}(\mathbb{R}^{3} \times X) \rightarrow H^{9-p}(\mathbb{R}^{3} \times X)$. Actually there is no contradiction, since, for manifolds of the form $\mathbb{R} \times A$, the map $\iota$ is the zero map, i.e.\ every \emph{closed} compactly-supported form on $\mathbb{R} \times A$ is exact, although not necessarily compactly-supported exact. In fact, considering the isomorphism \eqref{IsoCptCohom}, we see that every class in $H^{p}_{\cpt}(\mathbb{R} \times A)$ is represented by $e(x)\,dx \wedge \eta$ for $\eta$ closed, and, as we have already shown, $e(x)\,dx \wedge \eta = d\bigl(\int_{0}^{x}e \cdot \eta\bigr)$. We can also see that $\iota = 0$ considering the following maps:
	\[H^{*-1}_{\cpt}(A) \overset{e_{*}}\longrightarrow H^{*}_{\cpt}(\mathbb{R} \times A) \overset{\iota}\longrightarrow H^{*}(\mathbb{R} \times A) \overset{i^{*}}\longrightarrow H^{*}(A).
\]
The composition is the zero map, since, for a fixed form $\eta$, being $\iota$ the identity on the representative, the composition is $i^{*}(e(x)\,dx \wedge \eta) = (e(x)\,dx \wedge \eta)\vert_{\{0\} \times A}$, but the restriction of $e(x)\,dx$ gives $0$. Since $e_{*}$ and $i^{*}$ are isomorphisms, the only possibility is that $\iota = 0$.

This shows that, fixing the class of $Y_{p}$, we can always solve \eqref{BianchiId}. To obtain exactly \eqref{MaxwellDBrane} we use modify $\xi_{\cpt}$ with a current whose differential is the difference between $\delta(Y_{p})$ and the form $d*G^{t}_{p-1}$ obtained with the previous procedure.

\subsection{Summary}

Summarizing, for a Dp-brane with world volume $WY_{p}$ we have equations:
	\[dG_{8-p} = q \cdot \delta(WY_{p}) \qquad dG_{p+2} = 0
\]
from which we obtain the cohomological relation:
	\[[\, d_{M_{t}}(G_{8-p}\vert_{M_{t}}) \,]_{\cpt} = \PD_{M_{t}}(q \cdot Y_{p,t})
\]
and we compute the charge as:
	\[q = \frac{1}{l} \int_{L} G_{8-p}
\]
for $L$ a linking manifold of $WY_{p}$ in $S$. The solutions of this system are given by one particular solution, which under suitable hypotheses is similar to the static one for classical electromagnetism, and a generic solution of the equations in empty space. The particular solution can be obtained by an exact electric field strength and a magnetic one which is non-trivial only on the cycle obtained removing the charge, while the solutions in empty space can add topologically non-trivial terms in any cycle. We interpret these terms as charges hidden in the hole of the cycles which are not considered in our space-time region.

The questions we would like to improve from this picture are about the cohomogical equations, in particular:
\begin{itemize}
	\item we must assume that the brane volume is \emph{compact} at any instant of time, thus, e.g.\ for $S = \mathbb{R}^{1,3} \times X$, the brane must be a particle in the non-compact directions $\mathbb{R}^{1,3}$; in the other cases we cannot describe the D-brane charge as a homology cycle conserved in time, so that, e.g.\ \emph{we miss the torsion part} in describing D-branes and Ramond-Ramond fields (currents are real, so they do not contain torsion);
	\item the equations are not Lorentz-invariant, since the whole world-volume is non-compact and we cannot have a global formulation.
\end{itemize}
The second question arises also in classical electromagnetism theory, since the world-line of a particle is not compact, while the first is specific of D-brane theory. We now introduce a suitable homology and cohomology theory in order to solve these problems.

\section{Borel-Moore homology and D-branes}

The right tool to solve these problem is Borel-Moore homology with its modified versions, described in section \ref{BMHomology}. In fact, it allows us to deal with non-compact homology cycles, so that we can consider branes with are not necessarily particles in the non-compact space-time directions and also write the charge equations considering the whole world-volume.

\subsection{Generic D-branes}

We recall that the space-time manifold is $S = \mathbb{R} \times M$ with $M = \mathbb{R}^{3} \times X$ for $X$ a 6-dimensional compact manifold. We consider for the moment D-branes which are lines or planes in the non-compact space direction. If $Y_{p,t}$ is the volume at time $t$, we call $V = \pi_{\mathbb{R}^{3}}(Y_{p,t})$ (one simple case is $Y_{p,t} = V \times Y'_{p-k,t}$ with $Y'_{p-k,t} \subset X$, but this is not necessary). In this case, to define their charge we use modified Borel-Moore homology, considering that their volume is compact in the directions $V^{\bot} \times X$. We thus write the charge equations as:
\begin{equation}\label{ChargeBM}
	\left\{ \begin{array}{l}
		[\, d_{M}G^{s}_{8-p} \,] = \PD_{BM(M, V^{\bot}, \pi)}(q \cdot Y_{p,t}) \\
		d*_{9}G^{s}_{8-p} = 0.
	\end{array} \right.
\end{equation}
We have chosen the couple $(M, V^{\bot})$ but it is equivalent to the couple $(M, V^{\bot} \times X)$, since $X$ is compact. We can solve equations \eqref{ChargeBM} in a way analogue to the particle case. We obtain up to isomorphism $H_{\cpt}^{p}(V^{\bot} \times X) \simeq H^{p-3+k}(X)$, the isomorphism being given by $3-k$ applications of \eqref{IsoCptCohom}. Thus, instead of solving
	\[dG^{s}_{8-p} = e_{1}(x)\,dx_{1} \wedge e_{2}(x)\,dx_{2} \wedge e_{3}(x)\,dx_{3} \wedge \alpha + d\xi_{cpt}
\]
as in the ordinary case, we have to solve one of the two equations:
	\[dG^{s}_{8-p} = e_{1}(x)\,dx_{1} \wedge \alpha + d\xi_{cpt} \qquad dG^{s}_{8-p} = e_{1}(x)\,dx_{1} \wedge e_{2}(x)\,dx_{2} \wedge \alpha + d\xi_{cpt}
\]
depending whether $k = 2$ or $k = 1$. Then, the same procedure considered before applies.

\paragraph{}As ordinary homology is homotopy-invariant, similarly the modified Borel-Moore homology of a split-manifold $A \times B$ is invariant under homotopies involving only $B$, i.e.\ under homotopies of the form $F_{t}(a,b) = (a, F'_{t}(b))$. In particular, the modified Borel-Moore homology of $V \times V^{\bot} \times X$, non-compact only on $V$, is isomorphic to the one of $V \times X$ since $V^{\bot}$ retracts to a point. Now the only non-compact directions are the one on which cycles are allowed to be non-compact, thus we reduce to standard Borel-Moore homology. The situation is reversed for cohomology, since the ordinary one has \emph{non}-compact support in general, and that's the one which is homotopy-invariant. Thus, the modified Borel-Moore cohomology of $V \times V^{\bot} \times X$, which is the cohomology with compact support on $V^{\bot}$, is isomorphic to the one of $V^{\bot} \times X$. Since we remained only with compact directions, we reduce to the usual compactly-supported cohomology.

\paragraph{Remarks:}
\begin{itemize}
	\item One might think that we can always use standard Borel-Moore homology, since, having no hypotheses on the compactness of the cycles, it includes any kind of D-brane. This is not correct. In fact, let us consider a particle brane with worldvolume $\mathbb{R} \times Y_{p}$ in a fixed reference frame. Then, if we consider it as a Borel-Moore cycle, it is the boundary of $\mathbb{R} \times H^{3} \times Y_{p}$ for $H^{3} = \{(x,y,z): x,y,z \geq 0\}$, thus it has no charge. In general, if we want a cycle to be non-trivial, we must assume the necessary compactness hypothesis.
	\item We considered only lines or planes and not generic curves or surfaces. In the latter case, if we do not assume that they go at infinity along a fixed plane of the same dimension, we must consider only the direction at infinity of the brane itself, thus we should consider $H_{p}((S \cup (WY_{p})^{+}, \{\infty\}), \mathbb{Z})$ or in general $H_{p}((S \cup \overline{WY_{p}}, (S \cup \overline{WY_{p}}) \setminus (S \cup WY_{p})), \mathbb{Z})$. In terms of cycles in $S$ we must ask that their closure on $\overline{S}$ intersects $\partial S$ only on $\overline{WY_{p}}$. This is less natural but it works without any hypotheses.
\end{itemize}

\subsection{D-brane charge and world-volume}

Using modified Borel-Moore homology we can describe D-brane charges directly from the world-volume, without restricting to a fixed instant of time. In particular, let us consider a brane which is a particle in the non-compact space-time directions. In a fixed reference frame in which it is fixed in the origin we can rewrite the charge equation $(dG_{8-p})\vert_{M_{t}} = \PD_{M_{t}}(q \cdot Y_{p,t})$ as $[dG_{8-p}] = \PD_{BM(S, \mathbb{R}^{3}, \pi)}(q \cdot WY_{p})$. However, with Borel-Moore homology, we can write a Lorentz-invariant expression holding for every brane, independently on the behavior $\mathbb{R}^{1,3}$. We call $V = \pi_{\mathbb{R}^{1,3}}(WY_{p})$ and we get:
\begin{equation}\label{ChargeWWGeneric}
	[dG_{8-p}] = \PD_{BM(S, V^{\bot}, \pi)}(q \cdot WY_{p}).
\end{equation}
We do not have problems in considering always Borel-Moore homology in time-direction since any stable world-volume is non-trivial in that direction. If we would want to define an instanton charge, for a world-volume $\{t\} \times Y_{p}$, we could use analogue equations with ordinary $\PD$ (i.e.\ from ordinary homology to compactly supported cohomology in $\mathbb{R}^{1,3} \times X$); of course this is not a charge conserved in time, it is a trajectory charge but computed at a fixed instant since the trajectory itself is at a fixed instant.

Equation \eqref{ChargeWWGeneric} applies also to classical electromagnetism theory, since we can write $[d*F] = \PD_{BM(\mathbb{R}^{1,3}, w^{\bot}, \pi)}(q \cdot w)$.

\subsection{Space-filling D-branes}

Up to now we have not considered the most common D-branes, i.e.\ the space-filling ones. That's because, in this setting, their total charge would be zero. In fact, we should consider Borel-Moore homology which is non-compact in all $\mathbb{R}^{3}$, but in this case the Poincar\'e dual gives an ordinary cohomology class, thus, if is equal to $dG_{8-p}$, it is necessarily the trivial class, so the charge equations have no solution. The physical reason is that the fluxes has no directions at infinity where they can go, so that there is no charge, as it happens for an electron on a compact space. For an electron, we should put an anti-electron in another point of the space, so that fluxes go from one to the other. For D-branes we have two main possibilities:
\begin{itemize}
	\item There is an anti-brane, so that the total charge is $0$. In this case, we could imagine to compute each of the two opposite charges: do to this, if $Y$ is the brane and $\overline{Y}$ the anti-brane, we should solve the equations for Ramond-Ramond field in $(\mathbb{R}^{1,3} \times X) \setminus W\overline{Y}_{p}$ for $Y_{p}$ and in $(\mathbb{R}^{1,3} \times X) \setminus WY_{p}$ for the anti-brane. The result should give Ramond-Ramond fields extendable on all $(\mathbb{R}^{1,3} \times X)$ and closed, since the Poincar\'e dual is the zero class.
	\item There is an orientifold $O$ absorbing fluxes, so that we compute Ramond-Ramond fields in $(\mathbb{R}^{1,3} \times X) \setminus O$. 
\end{itemize}
In any case, we must consider a manifold which is not of the form $\mathbb{R}^{1,3} \times X$ with $X$ compact. However we can use without problems equation \eqref{ChargeWWGeneric}. In fact, an orientifold or an anti-brane is of the form $O = \mathbb{R}^{1,3} \times O'$, so we consider $(\mathbb{R}^{1,3} \times X \setminus O')$ and we reduce to the previous case with the only difference that the internal manifold is non-compact. Here we must consider the cohomology $BM(S, \mathbb{R}^{3-k} \times (X \setminus O'), \pi)$ and not $BM(S, \mathbb{R}^{3-k}, \pi)$ since the brane must be far from the orientifold in type II superstring theory.

\section{Wess-Zumino action}\label{WessZumino}

\subsection{Definition of the action}

If we consider a small charge $q$ moving in an electromagnetic field, the action of the particle minimally couples to the electromagnetic field via the potential, i.e.\ we add the term $q \int_{\gamma} A$. Such an integral is actually the holonomy of the line bundle over the curve $\gamma$, and in a generic background the field strength $F$ can be topologically non-trivial, so $A$ is locally defined and has gauge transformations. The problem is that, as explained in \cite{BFS}, holonomy is a well-defined function on closed curves, while it is a section of a line bundle over the space of open curves. However, at classical level, when we minimize the action we do it for curves connecting two fixed points (they can be at infinity, in case the bundle extends to the closure $\overline{S}$). In this case, if we fix a trivialization of the bundle near the two points we define holonomy as a number; actually, on a connected component of curves linking $x_{1}$ and $x_{2}$ and homotopic one to the other, we can choose a trivialization along all the curves and this is equivalent to fixing a potential $A$ . In this case, if we change potential by a gauge transformation $A \rightarrow A + \Phi$, then $S'(\gamma) = S(\gamma) + \int_{\gamma} \Phi$, but the summand $\int_{\gamma} \Phi$ is independent on $\gamma$ since $\int_{\gamma} \Phi - \int_{\gamma'} \Phi = \int_{\gamma - \gamma'}\Phi = 0$ being $\Phi$ closed and $\gamma - \gamma'$ contractible. We can have different constants $\int_{\gamma} \Phi$ on each connected component of the space of open curves between $x_{1}$ and $x_{2}$, but this has no influence on the minima or in general on stationary points. At quantum level, since Wilson loop is an observable, on our background we have a fixed holonomy for the connection, then we must also consider the case in which $\gamma - \gamma'$ is a non-trivial cycle: in this case the difference is the Wilson loop of a \emph{geometrically trivial} connection over $\gamma - \gamma'$, which is quantized for bundles, i.e.\ for $F$ quantized, thus the holonomy is zero at the exponential, i.e.\ for the partition function. For the D-brane the same considerations hold, the minimal coupling being the Wess-Zumino action:
	\[S_{WZ} = q\int_{WY_{p}} C_{p+1}.
\]
We must assume that $G_{p+2}$ is closed and quantized, i.e.\ that it represents an integral cohomology class. We see it as the curvature of a $p$-gerbe on $S$, and we assume that this gerbe is endowed with a connection $C_{p+1}$, so that $dC_{p+1} = G_{p+2}$. Actually the local forms $C_{p+1}$ are just the top forms representing the gerbe connection: a complete connection is given by a set of local forms from the degree $p+1$ to degree one, to end with transition functions $g_{\alpha_{0}\cdots \alpha_{p+1}}$, as explained above and summarized in appendix \ref{AppGerbes}. In particular, for $\Omega^{p}_{\mathbb{R}}$ the sheaf of smooth $p$-forms on $S$ and $\underline{S}^{1}$ the sheaf of $S^{1}$-valued smooth functions on $S$, the background data is a gerbe with connection:
	\[\mathcal{G}_{p} \in \check{H}^{p+1}(S, \underline{S}^{1} \rightarrow \Omega^{1}_{\mathbb{R}} \rightarrow \cdots \rightarrow \Omega^{p+1}_{\mathbb{R}})
\]
whose curvature is $G_{p+2}$ and whose holonomy on the corresponding world-volumes are the Wess-Zumino actions. For a brief review about the holonomy of gerbes we refer to \cite{BFS} chap.\ 3 and references therein: that discussion can be immediately generalized to $p$-gerbes, considering triangulations of dimension $p+1$ instead of $2$. In particular, in the definition of the holonomy we must consider all the intermediate forms defining the connection, a $k$-form being integrated on the $k$-faces of the triangulation of $WY_{p}$. The top forms $C_{p+1}$ are only a small piece of information, so that the notation $\int_{WY_{p}} C_{p+1}$ is actually approximate.

\subsubsection{Dirac quantization condition}

We now see in more detail Dirac quantization condition for $Y_{p}$ a brane with small charge moving in a background field $G_{p+2}$. Since $C_{p+1}$ is a local potential for $G_{p+2}$, the theory must be invariant under large gauge transformations $C_{p+1} \rightarrow C_{p+1} + \Phi_{p+1}$: we will see that this invariance requires the quantization of $G_{p+2}$, so that $G_{p+2}$ can be seen as the field strength of a gerbe and $C_{p+1}$ as a connection on such a gerbe (which enters dynamically in the path-integral, so it is not fixed a priori). Hence we need that:
\begin{itemize}
	\item $dG_{p+2} = 0$ in a neighborhood of $WY_{p}$, so that $G_{p+2}$ represents a cohomology class;
	\item $G_{p+2}$ represents an \emph{integral} class (from Dirac quantization condition).
\end{itemize}
For the first point, Maxwell equations for a D$p$-brane $Y_{p}$ as a charge source give:
	\[dG_{8-p} = \delta(q\cdot WY_{p}) \qquad dG_{p+2} = 0.
\]
In particular we deduce that $dG_{8-p}\vert_{S \setminus WY_{p}} = 0$. The need for $G_{8-p}$ to represent an \emph{integral} class outside $WY_{p}$ can be physically seen by the argument of Dirac quantization condition. We consider a charged $(6-p)$-brane moving in $S$, with charge small compared to the one of $Y_{p}$, so that its coupling is $q_{6-p} \cdot \int_{WY_{6-p}} C_{7-p}$. We want the theory to be invariant under large gauge transformations $C_{7-p} \rightarrow C_{7-p} + \Phi_{7-p}$ with $d\Phi_{7-p} = 0$: the action has not such an invariance, but, since we work in a quantum theory, we only need invariance of the path-integral $\int\exp(iS)$. To obtain this, let us consider the set of world-volumes $\Psi = \{WY_{6-p}\}$ with common boundary conditions\footnote{The boundary conditions can be given at two fixed instants of time or at infinity. In the first case, in what follows $\Delta W$ is a cycle in the space-time, otherwise it closes at infinity.}, and let $S$ and $S'$ be the actions with respect to $C_{7-p}$ and $C_{7-p} + \Phi_{7-p}$. Then given an element $WY_{6-p} \in \Psi$:
	\[S'(WY_{6-p}) - S(WY_{6-p}) = q \int_{WY_{6-p}} \Phi_{7-p}.
\]
The path-integral contains the integration:
	\[\int \mathcal{D}(WY_{6-p}) \, e^{iS(WY_{6-p})}
\]
where we denote by $WY_{6-p}$ the embedding $X: WY_{6-p} \rightarrow S$. Under the gauge transformation $C_{7-p} \rightarrow C_{7-p} + \Phi_{7-p}$ we get:
	\[\int \mathcal{D}(WY_{6-p}) \, e^{iS'(WY_{6-p})} = \int \mathcal{D}(WY_{6-p}) \, e^{iS(WY_{6-p})} \cdot e^{iq \int_{WY_{6-p}} \Phi_{7-p}}.
\]
For the theory to be invariant, we require that the new factor $\exp(iq \int_{WY_{6-p}} \Phi_{7-p})$ is an overall constant, i.e.\ it does not depend on $WY_{6-p}$. This means that, for $WY_{6-p}^{1}, WY_{6-p}^{2} \in \Psi$:
	\[\begin{split}
	&q\int_{WY_{6-p}^{2}} \Phi_{7-p} - q\int_{WY_{6-p}^{1}} \Phi_{7-p} \in 2\pi\mathbb{Z}\\
	&q\int_{WY_{6-p}^{2} - WY_{6-p}^{1}} \Phi_{7-p} \in 2\pi\mathbb{Z}.
\end{split}\]
Let us call $\Delta W = WY_{6-p}^{2} - WY_{6-p}^{1}$. Since the boundary conditions of the paths are fixed, one has $\partial(\Delta W) = 0$. Since, viewing $\Phi_{7-p}$ as a functional on homology, one has $q\int_{\Delta W} \Phi_{7-p} = \langle \Phi_{7-p}, q \cdot \Delta W \rangle$, then to achieve $\langle \Phi_{7-p}, q \cdot \Delta W \rangle = 2k\pi$ we must ensure that we are integrating an integral class on an integral cycle, i.e.\ we must quantize both the transition function $\Phi_{7-p}$ and the charge $q$. For the quantization of the charge, we consider $G_{8-p}$ as a current \cite{GH} which is equal to a closed form in $S \setminus WY_{p}$. We still denote that form with $G_{8-p}$. Then, by definition of $q$ as integral over the linking surface (which, of course, does not intersect $Y$), quantization of $q$ follows directly from the quantization of $G_{8-p}$. Similarly, for the brane $WY_{6-p}$ (as a charge source, not as small moving charge) we must quantize $G_{p+2}$.

\paragraph{}We now want to see that the quantization of the transition functions is equivalent to the quantization of the field strength $G_{8-p}$. Let us suppose that $\Delta W$ is contained in the intersection of two contractible local charts $U_{1} \cap U_{2}$. We consider $W_{1} \subset U_{1}$ and $W_{2} \subset U_{2}$ such that $\partial W_{1} = -\partial W_{2} = \Delta W$, and put $W = W_{1} + W_{2}$. We choose the representatives $C_{7-p}$ on $U_{1}$ and $C'_{7-p}$ on $U_{2}$, and, on $U_{1} \cap U_{2}$, we put $\Phi_{7-p} = C'_{7-p} - C_{7-p}$. Then:
	\[\begin{split}
	2k\pi &= q\int_{\Delta W} C_{7-p} - q\int_{\Delta W} \bigl(C_{7-p} + \Phi_{7-p}\bigr) = q\int_{W_{1}} dC_{7-p} + q\int_{W_{2}} dC_{7-p}\\
	&= q\int_{W} G_{8-p}.
\end{split}\]
Hence we must quantize the Ramond-Ramond field strength $G_{8-p}$. We can now see $C_{7-p}$ as connection of a gerbe $\mathcal{G}$ on $(\mathbb{R} \times M) \setminus W_{Z}$ with field strength $G_{8-p}$ \cite{Hitchin}.

\paragraph{Remark:} at the classical level, the dynamics is invariant under gauge transformation of the form $C_{7-p} \rightarrow C_{7-p} + d\Lambda_{6-p}$, i.e.\ topologically trivial transition functions for the gerbe connection $C_{7-p}$. In fact, let us consider the set of world-volumes $\Psi = \{WY_{6-p}\}$ with common boundary conditions, and let $S$ and $S'$ be the actions with respect to $C_{7-p}$ and $C_{7-p} + d\Lambda_{6-p}$. Then given an element $WY_{p} \in \Psi$:
	\[S'(WY_{6-p}) - S(WY_{6-p}) = q \int_{WY_{6-p}} d\Lambda_{6-p} = q \int_{\partial WY_{6-p}} \Lambda_{6-p}
\]
and, since the boundary conditions are fixed, this means that $S'(WY_{6-p}) = S(WY_{6-p}) + \textit{constant}$.

\subsection{Holonomy and boundary conditions}

As for line bundles, the holonomy of a $p$-gerbe is well-defined only on \emph{closed} $(p+1)$-manifolds, i.e.\ on manifolds without boundary, while $WY_{p}$, being the classical trajectory of the D-brane, is in general defined for all times so that it has a boundary at the limit time-coordinates $-\infty$ and $+\infty$, or equivalently it has a boundary contained in the boundary of $S$. As for line bundles, if we fix boundary conditions we have no problems for the partition function. In general, the path-integral gives a section of a bundle, so the there are no problems. If we want to define the holonomy as a number, we must give boundary conditions at infinity, but, being the forms $C_{p+1}$ defined only locally and only up to gauge transformations, what does it exactly mean to give boundary conditions for them at infinity?

We refer to \cite{BFS} chap.\ 4 for a discussion about $\rm\check{C}$ech hypercohomology and trivializations of gerbes. We can generalize the discussion there to $p$-gerbes. In particular, we consider the compactification $\overline{S}$ of space-time making such that $\overline{S}$ is a manifold with boundary and $S$ its interior, so that the infinity of $S$ becomes the boundary $\partial \overline{S}$. For example, for $\mathbb{R}^{1,3} \times X$ such a compactification is $D^{4} \times X$ for $D^{4}$ the $4$-disc, so that the boundary is $S^{4} \times X$. Generalizing what explained in \cite{BFS} to $p$-gerbes, if we have a $p$-gerbe $\mathcal{G}_{p}$ on $S$ we can define its holonomy with respect to $(p+1)$-submanifolds with boundary, but it is not a function, it is a section of a line bundle over the space of maps from open $(p+1)$-manifolds to $S$. In particular, if we fix a $(p+1)$-manifold $\Sigma$ and we endow the space $\Maps(\Sigma, S)$ with a suitable topology, the holonomy of $\mathcal{G}_{p}$ is a section of a line bundle over $\Maps(\Sigma, S)$. However, if we fix a subspace $T \subset S$ such that $\mathcal{G}_{p}\vert_{T}$ is trivial, and we consider only maps such that $\varphi(\partial \Sigma) \subset T$, then the line bundle becomes trivial. This is not enough to have a well-defined holonomy, since we do not have a preferred trivialization. However, a trivialization of $\mathcal{G}_{p}\vert_{T}$ determines canonically a trivialization of the line bundle, making the holonomy a well-defined function. In this case, we consider $\partial S$ as the subset $T$ on which the gerbe must be trivial, since the boundary of the compactified world-volume $WY_{p}$ lies in the boundary of $\overline{S}$. Thus, the background data must be not only a gerbe with connection $\mathcal{G}_{p}$ on $\overline{S}$ which is trivial on $\partial S$, but also a fixed trivialization of it. The $\rm\check{C}$ech double-complex to consider is then:

{ \scriptsize
\[\xymatrix{
	\check{C}^{0}(\overline{S}, \Omega^{p+1}_{\mathbb{R}}) \oplus \check{C}^{0}(\partial S, \Omega^{p}_{\mathbb{R}}) \ar[r]^{\check{\delta}^{0}} & \check{C}^{1}(\overline{S}, \Omega^{p+1}_{\mathbb{R}}) \oplus \check{C}^{1}(\partial S, \Omega^{p}_{\mathbb{R}}) \ar[r]^{\check{\delta}^{1}} & \check{C}^{2}(\overline{S}, \Omega^{p+1}_{\mathbb{R}}) \oplus \check{C}^{2}(\partial S, \Omega^{p}_{\mathbb{R}}) \ar[r]^{\phantom{XXXXXXXI}\check{\delta}^{2}} & \cdots \\
	\qquad\vdots\qquad \ar[r]^{\check{\delta}^{0}} \ar[u]^{d} & \qquad\vdots\qquad \ar[r]^{\check{\delta}^{1}} \ar[u]^{d} & \qquad\vdots\qquad \ar[r]^{\phantom{XXX}\check{\delta}^{2}} \ar[u]^{d} & \cdots \\
	\check{C}^{0}(\overline{S}, \Omega^{1}_{\mathbb{R}}) \oplus \check{C}^{0}(\partial S, \underline{S}^{1}) \ar[r]^{\check{\delta}^{0}} \ar[u]^{d} & \check{C}^{1}(\overline{S}, \Omega^{1}_{\mathbb{R}}) \oplus \check{C}^{1}(\partial S, \underline{S}^{1}) \ar[r]^{\check{\delta}^{1}} \ar[u]^{d} & \check{C}^{2}(\overline{S}, \Omega^{1}_{\mathbb{R}}) \oplus \check{C}^{2}(\partial S, \underline{S}^{1}) \ar[r]^{\phantom{XXXXXXX}\check{\delta}^{2}} \ar[u]^{d} & \cdots \\
	\check{C}^{0}(\overline{S}, \underline{S}^{1}) \ar[r]^{\check{\delta}^{0}} \ar[u]^{\tilde{d}} & \check{C}^{1}(\overline{S}, \underline{S}^{1}) \ar[r]^{\check{\delta}^{1}} \ar[u]^{\tilde{d}} & \check{C}^{2}(\overline{S}, \underline{S}^{1}) \ar[r]^{\phantom{XXX}\check{\delta}^{2}} \ar[u]^{\tilde{d}} & \cdots
	}
\] }
and we denote by $\check{H}^{\bullet}(\overline{S}, \underline{S}^{1} \rightarrow \Omega^{1}_{\mathbb{R}} \rightarrow \cdots \rightarrow \Omega^{p+1}_{\mathbb{R}}, \partial S)$ the hypercohomology of this complex. Thus, if we want to give boundary conditions to field in order to make Wess-Zumino action a well-defined number for a fixed trajectory extending in time from $-\infty$ to $+\infty$, we must give as background data a gerbe with trivialization:
	\[\mathcal{G}_{p} \in \check{H}^{p+1}(\overline{S}, \underline{S}^{1} \rightarrow \Omega^{1}_{\mathbb{R}} \rightarrow \cdots \rightarrow \Omega^{p+1}_{\mathbb{R}}, \partial S).
\]

\chapter{A-field and B-field}

\section{The Freed-Witten anomaly}

Our next aim is to classify the allowed B-field and A-field configurations in type II superstring backgrounds with a fixed set of D-branes. It is well known that to this end the appropriate mathematical framework is represented by gerbes \cite{Hitchin, Brylinski}. As line bundles on a space $X$ are characterized, up to isomorphism, by the first Chern class in $H^{2}(X, \mathbb{Z})$, gerbes are classified by the first Chern class in $H^{3}(X, \mathbb{Z})$. Analogously, as a connection on a line bundle is given by local 1-forms up to gauge transformations, a connection on a gerbe is defined by local 2-forms and 1-forms up to gauge transformations. Definitions and details used in the sequel are given in appendices \ref{AppHyperC} and \ref{AppGerbes}. This chapter is a reproduction of \cite{BFS}, except chap.\ 3.

Let us consider string theory on a smooth space-time $X$ and let us consider a single smooth D-brane with world-volume $Y \subset X$. At first sight, one would expect the background to contain the following data:
\begin{itemize}
	\item on $X$ a gerbe with a connection given by the B-field, with Chern class $\zeta \in H^{3}(X, \mathbb{Z})$ and curvature $H \in \Omega^{3}(X, \mathbb{Z})$, so that $H$ is a de-Rham representative of $\zeta$, i.e.\ $\zeta \otimes_{\mathbb{Z}} \mathbb{R} \simeq [H]_{dR}$;
	\item on $Y$ a line bundle with a connection given by the A-field.
\end{itemize}
However, as pointed out in \cite{FW}, while the assignment of the gerbe on $X$ is always given in the background, the presence of the line bundle is actually consistent only in some specific cases, the most common being the one in which the gerbe restricted to $Y$ is geometrically trivial and $w_{2}(Y) = 0$, i.e.\ $Y$ is spin ($w_{2}(Y)$ is the second Stiefel-Withney class of the tangent bundle of $Y$ \cite{LM}). In general, there is a different object on the brane. To understand what, we start from the world-sheet path-integral.

\paragraph{}In the superstring world-sheet action there are the following terms:
\begin{equation}\label{Action}
	S \supset \biggl(\, \int d\psi \, \psi \,D_{\phi}\, \psi \,\biggr) + 2\pi \cdot \biggl(\, \int_{\Sigma} \phi^{*}B + \int_{\partial \Sigma} \phi^{*}A \,\biggr)
\end{equation}
where $\phi: \Sigma \rightarrow X$ is the embedding of the string world-sheet in the target space. The exponential of the first term is the Pfaffian of the Dirac operator coupled to $TY$ via $\phi$, thus we write:
	\[e^{iS} \supset \pfaff \, D_{\phi} \cdot \exp\biggl(\,2\pi i \cdot \int_{\Sigma} \phi^{*}B\,\biggr) \cdot \exp\biggl(\, 2\pi i \cdot \int_{\partial \Sigma} \phi^{*}A \,\biggr) \,.
\]
The Pfaffian may be problematic. In fact, evaluated in a point $\phi \in \Maps(\Sigma, X)$, it must satisfy $(\pfaff \, D_{\phi})^{2} = \det D_{\phi}$, so we have a sign ambiguity and we need a natural definition of the Pfaffian, up to an overall constant which is immaterial for the path-integral. The problem is that the Pfaffian is not a function, but it is naturally a section of a line bundle over $\Maps(\partial \Sigma, Y)$, called \emph{pfaffian line bundle}, with natural metric and \emph{flat} connection \cite{Freed}. If this bundle is geometrically trivial, we can choose a flat unitary section $1$ up to an overall phase, so that we determine the Pfaffian as $\pfaff\,D_{\phi} \,/\, 1$; otherwise the latter is not well defined as a number. The first Chern class of the Pfaffian line bundle depends on $W_{3}(Y)$ (where $W_{3}(Y)$ is the integral lift of the third Stiefel-Whitney class of the tangent bundle of $Y$, i.e.\ the obstruction to the existence of $U(1)$-charged spinors on $Y$ \cite{LM, Hitchin}), while the holonomy depends on $w_{2}(Y)$. Thus, if the brane is spin the pfaffian is a well-defined function, otherwise the best we can do is to choose \emph{local} parallel sections so that we have a local definition of $\pfaff\,D_{\phi}$.

It turns out that the terms $\exp(\,2\pi i \cdot \int_{\Sigma} \phi^{*}B\,) \cdot \exp(\, 2\pi i \cdot \int_{\partial \Sigma} \phi^{*}A \,)$ can compensate exactly the possible ambiguity of the Pfaffian, giving rise to a well-defined path-integral, if and only if:
\begin{equation}\label{FWAnomaly}
W_{3}(Y) + \zeta\vert_{Y} = 0 \,.
\end{equation}
The class $W_{3}(Y) + \zeta\vert_{Y} \in H^{3}(Y, \mathbb{Z})$ is called \emph{Freed-Witten anomaly} \cite{FW}. In particular, $\zeta\vert_{Y}$ must be a torsion class since $W_{3}(Y)$ is, so that $[\,H\vert_{Y}\,]_{dR} = 0$.

Taking this picture into account, and recalling the geometrical meaning of the terms $\exp(\,2\pi i \cdot \int_{\Sigma} \phi^{*}B\,) \cdot \exp(\, 2\pi i \cdot \int_{\partial \Sigma} \phi^{*}A \,)$ described above, the classifying group of the $B$-field and $A$-field configurations will naturally arise.

\section{Classification by hypercohomology}

We are now ready to describe the classification group for $B$-field and $A$-field configurations in superstring theory with a single D-brane. Our background is specified in particular by a space-time gerbe $\mathcal{G}$ belonging to the following hypercohomology group\footnote{We refer to appendices \ref{AppHyperC} and \ref{AppGerbes} for notations.}:
\begin{equation}\label{GerbeB}
	\mathcal{G} = [\{g_{\alpha\beta\gamma}, -\Lambda_{\alpha\beta}, B_{\alpha}\}] \in \check{H}^{2}(X, \, \underline{S}^{1} \overset{\tilde{d}}\longrightarrow \Omega^{1}_{\mathbb{R}} \overset{d}\longrightarrow \Omega^{2}_{\mathbb{R}}\,)
\end{equation}
where $\tilde{d} = (2\pi i)^{-1}\, d \circ \log\,$, $g_{\alpha\beta\gamma}$ are functions from triple intersections to $S^{1}$, $\Lambda_{\alpha\beta}$ are 1-forms on double intersections and $B_{\alpha}$ are 2-forms on the opens sets of the cover. In \eqref{GerbeB}, we denote by $\underline{S}^{1}$ the sheaf of smooth $S^{1}$-valued functions on $X$ and by $\Omega^{p}_{\mathbb{R}}$ the sheaf of real $p$-forms. On a single brane $Y \subset X$ we consider the restriction of the space-time gerbe, for which we use the same notation $\mathcal{G}\,\vert_{Y} = [\,\{g_{\alpha\beta\gamma}, -\Lambda_{\alpha\beta}, B_{\alpha}\}\,] \in \check{H}^{2}(Y, \, \underline{S}^{1} \rightarrow \Omega^{1}_{\mathbb{R}} \rightarrow \Omega^{2}_{\mathbb{R}}\,)$. To give a meaning to the holonomy for open surfaces with boundary on $Y$, we must fix a specific representative of the class $\mathcal{G}\,\vert_{Y}$, i.e.\ a specific hypercocycle; this operation is analogous to fixing a set of local sections on a line bundle up to pull-back by isomorphism (v.\ section \ref{CocyclesCohomology}). To compensate for the possible non-definiteness of $\pfaff\,D_{\phi}$, this hypercocycle must take the form $\{\eta_{\alpha\beta\gamma}, 0, B+F\}$, with $\eta_{\alpha\beta\gamma}$ representing the class $w_{2} \in H^{2}(Y, S^{1})$, denoting by $S^{1}$ the \emph{constant} sheaf. Here $B+F$ is a 2-form globally defined on $Y$, which we now explain in detail. The choice of the specific cocycle $\eta_{\alpha\beta\gamma}$ in the class $w_{2}$ turns out to be immaterial, as we will show later.

\paragraph{}In order to obtain the hypercocycle $\{\eta_{\alpha\beta\gamma}, 0, B+F\}$ from any gauge representative $\{g_{\alpha\beta\gamma},$ $-\Lambda_{\alpha\beta}, B_{\alpha}\}$ of the gerbe $\mathcal{G}\,\vert_{Y}$, the brane must provide a reparametrization of $\mathcal{G}\,\vert_{Y}$, which, by an active point of view, is a hypercoboundary, i.e.\ a geometrically trivial gerbe. That is, given $\{g_{\alpha\beta\gamma}, -\Lambda_{\alpha\beta}, B_{\alpha}\}$, the brane must provide a coordinate change $\{g_{\alpha\beta\gamma}^{-1} \cdot \eta_{\alpha\beta\gamma}, \Lambda_{\alpha\beta}, dA_{\alpha}\}$, so that:
\begin{equation}
	\{g_{\alpha\beta\gamma}, -\Lambda_{\alpha\beta}, B_{\alpha}\} \cdot \{g_{\alpha\beta\gamma}^{-1} \cdot \eta_{\alpha\beta\gamma}, \Lambda_{\alpha\beta}, dA_{\alpha}\} = \{\eta_{\alpha\beta\gamma}, 0, B+F\}
\end{equation}
for a globally defined $B + F = B_{\alpha} + dA_{\alpha}$. In order for this correction to be geometrically trivial, it must be that:
\begin{equation}\label{CoboundaryA}
	\{g_{\alpha\beta\gamma}^{-1} \cdot \eta_{\alpha\beta\gamma}, \Lambda_{\alpha\beta}, dA_{\alpha}\} = \check{\delta}^{1}\{h_{\alpha\beta}, A_{\alpha}\}
\end{equation}
i.e.\ $\{g_{\alpha\beta\gamma}^{-1} \cdot \eta_{\alpha\beta\gamma}, \Lambda_{\alpha\beta}, dA_{\alpha}\} = \{\check{\delta}^{1}h_{\alpha\beta}, -\tilde{d} h_{\alpha\beta} + A_{\beta} - A_{\alpha}, dA_{\alpha}\}$. For this to hold one must have:
\begin{itemize}
	\item $\{g_{\alpha\beta\gamma}^{-1} \cdot \eta_{\alpha\beta\gamma}\} = \{\check{\delta}^{1}h_{\alpha\beta}\}$: this is precisely the statement of Freed-Witten anomaly, since, considering the Bockstein homomorphism $\beta$ in degree 2 of the sequence $0 \rightarrow \mathbb{Z} \rightarrow \underline{\mathbb{R}} \rightarrow \underline{S}^{1} \rightarrow 0$, this is equivalent to $\beta(\,[\,g_{\alpha\beta\gamma}\,]\,) = \beta(\,[\,\eta_{\alpha\beta\gamma}\,]\,)$, i.e.\ $\zeta\,\vert_{Y} = W_{3}(Y)$; only under this condition is $g_{\alpha\beta\gamma}^{-1} \cdot \eta_{\alpha\beta\gamma}$ trivial in the $\underline{S}^{1}$-cohomology;
	\item $A_{\beta} - A_{\alpha} = \tilde{d} h_{\alpha\beta} + \Lambda_{\alpha\beta}$: these must be the transition relations for $A_{\alpha}$ (coherently with \cite{Kapustin}); this is always possible since $\check{\delta}^{1}\{\tilde{d} h_{\alpha\beta}\} = \{\,\tilde{d}(\,\eta_{\alpha\beta\gamma} - g_{\alpha\beta\gamma}\,)\,\} = \{-\tilde{d} g_{\alpha\beta\gamma} \}= -\check{\delta}^{1}\{\Lambda_{\alpha\beta}\}$ and $\Omega^{1}_{\mathbb{R}}$ is acyclic.
\end{itemize}
From the transition relations of $A_{\alpha}$ we obtain $dA_{\beta} - dA_{\alpha} = d\Lambda_{\alpha\beta}$, thus $B+F$ is globally defined. Of course $B_{\alpha}$ and $A_{\alpha}$ themselves depends on the gauge choices, while $B+F$ is gauge-invariant.\footnote{We remark that, for $W_{3}(Y) = 0$, from the exact sequence $0 \rightarrow \mathbb{Z} \rightarrow \underline{\mathbb{R}} \rightarrow \underline{S}^{1} \rightarrow 0$ it follows that $w_{2}(Y)$, having image $0$ under the degree-2 Bockstein homomorphism, by exactness can be lifted to a real form $G$ on $Y$. Therefore, the gerbe $[\,\{\eta_{\alpha\beta\gamma}, 0, B+F\}\,]$ can be also represented by $[\,\{1,0,B+F+G\}\,]$: however, this is not the cocycle we need, since we need transition function realizing the class $w_{2}(Y)$. These two cocycles are equivalent on closed surfaces, since they represent the same gerbe, but not on open ones.}

\paragraph{}Let us now discuss the role of the representative $\eta_{\alpha\beta\gamma}$ of the class $w_{2}(Y) \in \check{H}^{2}(X, S^{1})$. The choice of a different representative corresponds to changing by \emph{constant} local functions the chosen sections of the bundle over loop space, which define the holonomy for open surfaces. This kind of ambiguity is also present for the Pfaffian, since it also defines a section of a flat bundle with the same holonomy. If $w_{2}(Y) \neq 0$, we have no possibility to eliminate this non-definiteness. We can only choose the sections for the Pfaffian and for the gerbe, in such a way that on the tensor product we have a \emph{global} flat section, up to an immaterial overall constant. Instead, if $w_{2} = 0$, both the pfaffian and the gerbe are geometrically trivial, thus we have a preferred choice, given by a global flat section for both. In this case, we fix the canonical representative $\eta_{\alpha\beta\gamma} = 1$. We will see in the following the consequences of this fact for the gauge theory of the D-brane.

\paragraph{}How can we jointly characterize B-field and A-field taking into account the gauge transformations contained in the previous description? This unifying role is played by a certain hypercohomology group, which is actually a \emph{relative} Deligne cohomology group. We now introduce such a group explicitly constructing the double complex needed, then we show that it can be intrinsically described as a relative cohomology group. Since this construction is not very familiar in the literature, we would like for pedagogical reason to start with the analogous group for line bundles.

\subsection{Line bundles}

Let us consider an embedding of manifolds $i: Y \rightarrow X$: we want to describe the group of line bundles on $X$ which are trivial on $Y$, with a fixed trivialization. We recall that $\underline{S}^{1}$ is the sheaf of smooth functions on $X$: it turns out that the sheaf of smooth functions on $Y$ is its pull-back $i^{*}\underline{S}^{1}$. We thus obtain a cochain map $(i^{*})^{p}: \check{C}^{p}(X, \underline{S}^{1}) \longrightarrow \check{C}^{p}(Y, \underline{S}^{1})$, which can be described as follows: we choose a good cover $\mathfrak{U}$ of $X$ restricting to a good cover $\mathfrak{U} \,\vert_{Y}$ of $Y$, such that every $p$-intersection $U_{i_{0} \cdots i_{p}} \vert_{Y}$ comes from a unique $p$-intersection $U_{i_{0} \cdots i_{p}}$ on $X$. Given a $p$-cochain $\oplus_{i_{0} < \cdots < i_{p}} \, f_{i_{0} \cdots i_{p}}$, we restrict $f_{i_{0} \cdots i_{p}}$ to $U_{i_{0} \cdots i_{p}} \vert_{Y}$ whenever the latter is non-empty. In this way we obtain a double complex:
	\[\xymatrix{
	\check{C}^{0}(Y, \underline{S}^{1}) \ar[r]^{\check{\delta}^{0}} & \check{C}^{1}(Y, \underline{S}^{1}) \ar[r]^{\check{\delta}^{1}} & \check{C}^{2}(Y, \underline{S}^{1}) \ar[r]^{\check{\delta}^{2}} & \cdots\\
	\check{C}^{0}(X, \underline{S}^{1})  \ar[u]^{(i^{*})^{0}} \ar[r]^{\check{\delta}^{0}} & \check{C}^{1}(X, \underline{S}^{1}) \ar[u]^{(i^{*})^{1}} \ar[r]^{\check{\delta}^{1}} & \check{C}^{2}(X, \underline{S}^{1}) \ar[u]^{(i^{*})^{2}} \ar[r]^{\check{\delta}^{2}} & \cdots
	}
\]
We denote by $\check{H}^{\bullet}(X, \underline{S}^{1}, Y)$ the hypercohomology of this double complex. We claim that $\check{H}^{1}(X, \underline{S}^{1}, Y)$ is the group we are looking for. In fact, the latter can be defined in the following way: we choose a line bundle $L$ on $X$ with a fixed set of local sections $\{s_{\alpha}\}$, so that the transition functions are $\{g_{\alpha\beta}\}$ for $g_{\alpha\beta} = s_{\alpha} / s_{\beta}$. We consider $\{s_{\alpha}\,\vert_{Y}\}$ and we express the trivialization by means of local functions $\{f_{\alpha}\}$ on $Y$ such that $f_{\alpha} \cdot s_{\alpha}\,\vert_{Y}$ gives a global section of $L\vert_{Y}$. We have that $\check{C}^{1}(X, \underline{S}^{1}, Y) = \check{C}^{1}(X, \underline{S}^{1}) \oplus \check{C}^{0}(Y, \underline{S}^{1})$, so that we can consider the hypercochain $\{g_{\alpha\beta}, f_{\alpha}\}$. We now claim that this is a hypercocycle: to see this, we describe the cohomology group $\check{H}^{1}(X, \underline{S}^{1}, Y)$.
\begin{itemize}
	\item \emph{Cocycles:} since $\check{\delta}^{1}\{g_{\alpha\beta}, f_{\alpha}\} = \{\check{\delta}^{1}g_{\alpha\beta}, ((i^{*})^{1}g_{\alpha\beta})^{-1} \cdot f_{\beta} f_{\alpha}^{-1}\}$, cocycles are characterized by two conditions: $\check{\delta}^{1}g_{\alpha\beta} = 0$, i.e.\ $g_{\alpha\beta}$ is a line bundle $L$ on $X$, and $(i^{*})^{1}g_{\alpha\beta} = f_{\beta} f_{\alpha}^{-1}$, i.e.\ $f_{\alpha}$ trivializes $L\vert_{Y}$.
	\item \emph{Coboundaries:} $\check{\delta}^{0}\{g_{\alpha}\} = \{\check{\delta}^{0}g_{\alpha}, (i^{*})^{0}g_{\alpha}\}$ thus coboundaries represents line bundles which are trivial on $X$, with a trivialization on $X$ restricting to to chosen one on $Y$.
\end{itemize}
To explain the structure of the coboundaries, let us remark that if we choose different sections $\{s'_{\alpha} = \varphi_{\alpha} \cdot s_{\alpha}\}$, the same trivialization is expressed by $f'_{\alpha} = \varphi_{\alpha}\vert_{Y}^{-1} \cdot f_{\alpha}$. Thus the coordinate change is given by $\{\varphi_{\alpha}^{-1}\varphi_{\beta}, \varphi_{\alpha}\vert_{Y}\}$, which can be seen, by an active point of view, as a $X \times \mathbb{C}$ with the trivialization $Y \times \{1\}$ on $Y$, i.e.\ a trivial bundle with a fixed global section on $X$ restricting to the chosen trivialization on $Y$. Hence, $\check{H}^{1}(X, \underline{S}^{1}, Y)$ is the group we are looking for.

\subsubsection{Line bundles with connection}

Let us now define the analogous group for bundles with connection. The relevant complex is the following:

{\footnotesize
\[\xymatrix{
	\check{C}^{0}(X, \Omega^{1}_{\mathbb{R}}) \oplus \check{C}^{0}(Y, \underline{S}^{1}) \ar[r]^{\check{\delta}^{0} \oplus \check{\delta}^{0}} & \check{C}^{1}(X, \Omega^{1}_{\mathbb{R}}) \oplus \check{C}^{1}(Y, \underline{S}^{1}) \ar[r]^{\check{\delta}^{1} \oplus \check{\delta}^{1}} & \check{C}^{2}(X, \Omega^{1}_{\mathbb{R}}) \oplus \check{C}^{2}(Y, \underline{S}^{1}) \ar[r]^{\phantom{XXXXXXX} \check{\delta}^{2} \oplus \check{\delta}^{2}} & \cdots\\
	\,\phantom{XX} \check{C}^{0}(X, \underline{S}^{1}) \,\phantom{XX} \ar[u]^{\tilde{d} \,\oplus\, (i^{*})^{2}} \ar[r]^{\check{\delta}^{0}} & \,\phantom{XX} \check{C}^{1}(X, \underline{S}^{1}) \,\phantom{XX} \ar[r]^{\check{\delta}^{1}} \ar[u]^{\tilde{d} \,\oplus\, (i^{*})^{1}} & \,\phantom{XX} \check{C}^{2}(X, \underline{S}^{1}) \,\phantom{XX} \ar[u]^{\tilde{d} \,\oplus\, (i^{*})^{2}} \ar[r]^{\phantom{XXXXXX} \check{\delta}^{2}} & \cdots
	}
\] }
We denote by $\check{H}^{\bullet}(X, \underline{S}^{1} \rightarrow \Omega^{1}_{\mathbb{R}}, Y)$ the hypercohomology of this double complex. We claim that the group we are looking for is $\check{H}^{1}(X, \underline{S}^{1} \rightarrow \Omega^{1}_{\mathbb{R}}, Y)$. The cochains are given by $\check{C}^{1}(X, \underline{S}^{1} \rightarrow \Omega^{1}_{\mathbb{R}}, Y) = \check{C}^{1}(X, \underline{S}^{1}) \,\oplus\, \check{C}^{0}(Y, \Omega^{1}_{\mathbb{R}}) \,\oplus\, \check{C}^{0}(Y, \underline{S}^{1})$, so that we consider $\{g_{\alpha\beta}, -A_{\alpha}, f_{\alpha}\}$.
\begin{itemize}
	\item \emph{Cocycles:} since $\check{\delta}^{1}\{g_{\alpha\beta}, -A_{\alpha}, f_{\alpha}\} = \{\check{\delta}^{1}g_{\alpha\beta}, -\tilde{d} g_{\alpha\beta} - A_{\beta} + A_{\alpha}, ((i^{*})^{1}g_{\alpha\beta})^{-1} \cdot f_{\beta} f_{\alpha}^{-1}\}$, cocycles are characterized by three conditions: $\check{\delta}^{1}g_{\alpha\beta} = 0$, i.e.\ $g_{\alpha\beta}$ is a line bundle $L$ on $X$, $A_{\alpha} - A_{\beta} = \tilde{d}g_{\alpha\beta}$, i.e.\ $A_{\alpha}$ is a connection on $L$, and $(i^{*})^{1}g_{\alpha\beta} = f_{\beta} f_{\alpha}^{-1}$, i.e.\ $f_{\alpha}$ trivializes $L\vert_{Y}$.
	\item \emph{Coboundaries:} since $\check{\delta}^{0}\{g_{\alpha}\} = \{\check{\delta}^{0}g_{\alpha}, \tilde{d}g_{\alpha\beta}, (i^{*})^{0}g_{\alpha}\}$, coboundaries represents line bundles which are geometrically trivial on $X$ (v.\ appendix), with a trivialization on $X$ restricting to the chosen one on $Y$.
\end{itemize}

\subsection{Gerbes}

Let us now define the analogous group for gerbes with connection. The relevant complex is the following\footnote{The maps denoted by matrices are supposed to multiply from the right the row vector in the domain.}:

{ \footnotesize
\[\xymatrix{
	\check{C}^{0}(X, \Omega^{2}_{\mathbb{R}}) \oplus \check{C}^{0}(Y, \Omega^{1}_{\mathbb{R}}) \ar[r]^{\check{\delta}^{0} \oplus \check{\delta}^{0}} & \check{C}^{1}(X, \Omega^{2}_{\mathbb{R}}) \oplus \check{C}^{1}(Y, \Omega^{1}_{\mathbb{R}}) \ar[r]^{\check{\delta}^{1} \oplus \check{\delta}^{1}} & \check{C}^{2}(X, \Omega^{2}_{\mathbb{R}}) \oplus \check{C}^{2}(Y, \Omega^{1}_{\mathbb{R}}) \ar[r]^{\phantom{XXXXXXX} \check{\delta}^{2} \oplus \check{\delta}^{2}} & \cdots\\ \\
	\check{C}^{0}(X, \Omega^{1}_{\mathbb{R}}) \oplus \check{C}^{0}(Y, \underline{S}^{1}) \ar[uu]_{\footnotesize{\begin{bmatrix} d & (i^{*})^{0} \\ 0 & -\tilde{d} \end{bmatrix}}} \ar[r]^{\check{\delta}^{0} \oplus \check{\delta}^{0}} & \check{C}^{1}(X, \Omega^{1}_{\mathbb{R}}) \oplus \check{C}^{1}(Y, \underline{S}^{1}) \ar[r]^{\check{\delta}^{1} \oplus \check{\delta}^{1}} \ar[uu]_{\footnotesize{\begin{bmatrix} d & (i^{*})^{1} \\ 0 & -\tilde{d} \end{bmatrix}}} & \check{C}^{2}(X, \Omega^{1}_{\mathbb{R}}) \oplus \check{C}^{2}(Y, \underline{S}^{1}) \ar[uu]_{\footnotesize{\begin{bmatrix} d & (i^{*})^{2} \\ 0 & -\tilde{d} \end{bmatrix}}} \ar[r]^{\phantom{XXXXXXX} \check{\delta}^{2} \oplus \check{\delta}^{2}} & \cdots\\ \\
	\check{C}^{0}(X, \underline{S}^{1}) \ar[uu]_{\tilde{d} \,\oplus\, (i^{*})^{0}} \ar[r]^{\check{\delta}^{0}} & \check{C}^{1}(X, \underline{S}^{1}) \ar[uu]_{\tilde{d} \,\oplus\, (i^{*})^{1}} \ar[r]^{\check{\delta}^{1}} & \check{C}^{2}(X, \underline{S}^{1}) \ar[uu]_{\tilde{d} \,\oplus\, (i^{*})^{2}} \ar[r]^{\phantom{XXXXX} \check{\delta}^{2}} & \cdots
	}
\] }
We denote by $\check{H}^{\bullet}(X, \underline{S}^{1} \rightarrow \Omega^{1}_{\mathbb{R}} \rightarrow \Omega^{2}_{\mathbb{R}}, Y)$ the hypercohomology of this double complex. We claim that the group we are looking for is $\check{H}^{2}(X, \underline{S}^{1} \rightarrow \Omega^{1}_{\mathbb{R}} \rightarrow \Omega^{2}_{\mathbb{R}}, Y)$. The cochains are given by $\check{C}^{2}(X, \underline{S}^{1} \rightarrow \Omega^{1}_{\mathbb{R}} \rightarrow \Omega^{2}_{\mathbb{R}}, Y) = \check{C}^{2}(X, \underline{S}^{1}) \,\oplus\, \check{C}^{1}(X, \Omega^{1}_{\mathbb{R}}) \,\oplus\, \check{C}^{1}(Y, \underline{S}^{1}) \,\oplus\, \check{C}^{0}(X, \Omega^{2}_{\mathbb{R}}) \,\oplus\, \check{C}^{0}(Y, \Omega^{1}_{\mathbb{R}})$, so that we consider $\{g_{\alpha\beta\gamma}, -\Lambda_{\alpha\beta}, h_{\alpha\beta}, B_{\alpha}, -A_{\alpha}\}$.
\begin{itemize}
	\item \emph{Cocycles:} since $\check{\delta}^{2}\{g_{\alpha\beta\gamma}, -\Lambda_{\alpha\beta}, h_{\alpha\beta}, B_{\alpha}, -A_{\alpha}\} = \{\check{\delta}^{2}g_{\alpha\beta\gamma}, \tilde{d}g_{\alpha\beta\gamma} + \check{\delta}^{1}(-\Lambda_{\alpha\beta}),$ $(i^{*})^{2}g_{\alpha\beta\gamma} \cdot \check{\delta}^{2}h_{\alpha\beta},$ $-d(-\Lambda_{\alpha\beta}) + B_{\beta} - B_{\alpha}, -(i^{*})^{1}(-\Lambda_{\alpha\beta}) + \tilde{d}h_{\alpha\beta} + A_{\alpha} - A_{\beta}\}$, cocycles are characterized exactly by the condition we need in order for $\{g_{\alpha\beta\gamma}, -\Lambda_{\alpha\beta}, B_{\alpha}\}$ to be a gerbe with connection and $\{h_{\alpha\beta}, A_{\alpha}\}$ to trivialize it on $Y$;
	\item \emph{Coboundaries:} since $\check{\delta}^{1}\{g_{\alpha\beta}, \Lambda_{\alpha}, h_{\alpha}\} = \{\check{\delta}^{1}g_{\alpha\beta}, -\tilde{d}g_{\alpha\beta} + \Lambda_{\beta} - \Lambda_{\alpha}, ((i^{*})^{1}g_{\alpha\beta})^{-1} \cdot h_{\beta} h_{\alpha}^{-1}, d\Lambda_{\alpha},$ $(i^{*})^{0} \Lambda_{\alpha} - \tilde{d}h_{\alpha}\}$, coboundaries represent gerbes which are geometrically trivial on $X$ (v.\ appendix), with a trivialization on $X$ restricting to the chosen one on $Y$.
\end{itemize}

\paragraph{}There is a last step to obtain the classifying set of $B$-field and $A$-field configurations: in general we do not ask for a trivialization of the gerbe on $Y$, but for a cocycle whose transition functions represent the class $w_{2}(Y) \in H^{2}(Y, S^{1})$. The transition functions of a coboundary in the previous picture represent the zero class, so they are consistent only for $w_{2}(Y) = 0$. Hence, we cannot consider the hypercohomology group, but one of its cosets in the group of \emph{cochains} up to coboundaries. In fact, the condition we need is not cocycle condition, but:
\begin{equation}\label{Coset}
	\check{\delta}^{2}\{g_{\alpha\beta\gamma}, -\Lambda_{\alpha\beta}, h_{\alpha\beta}, B_{\alpha}, -A_{\alpha}\} = \{0, 0, \eta_{\alpha\beta\gamma}, 0, 0\}
\end{equation}
thus we need the coset made by cochains satisfying \eqref{Coset} up to coboundaries. Actually, we need anyone of these cosets for $[\,\{\,\eta_{\alpha\beta\gamma}\,\}\,] = w_{2}(Y) \in \check{H}^{2}(Y, S^{1})$. We denote their union by:
\begin{equation}
	\check{H}^{2}_{w_{2}(Y)}(X, \underline{S}^{1} \rightarrow \Omega^{1}_{\mathbb{R}} \rightarrow \Omega^{2}_{\mathbb{R}}, Y)
\end{equation}
and this is the set of configurations we are looking for.

\subsection{Intrinsic description}

Given a map of complexes $\varphi^{\bullet}: (K^{\bullet}, d_{K}^{\bullet}) \rightarrow (L^{\bullet}, d_{L}^{\bullet})$, the cone of $\varphi$ is the complex:
\begin{equation}\label{Cone}
	C(\varphi)^{i} := K^{i} \oplus L^{i-1} \qquad d_{C(\varphi)}^{i} := \begin{pmatrix} d_{K}^{i} & 0 \\ \varphi^{i} & d_{L}^{i-1} \end{pmatrix}.
\end{equation}
If we consider the cohomology in degree $i$, we see that it is made by classes $[(k^{i}, l^{i-1})]$ where $k^{i}$ represents a cohomology class of $K$ whose image via $\varphi^{i}$ is trivial, and $l^{i-1}$ is a trivialization of $-\varphi^{i}(k^{i})$.
Here we consider the complexes of sheaves:
	\[S_{X,2}^{\bullet} := \underline{U}(1)_{X} \rightarrow \Omega^{1}_{X,\mathbb{R}} \rightarrow \Omega^{2}_{X,\mathbb{R}} \qquad\qquad S_{Y,1}^{\bullet} := \underline{U}(1)_{Y} \rightarrow \Omega^{1}_{Y,\mathbb{R}}
\]
both extended by $0$ on left and right. For $i: Y \rightarrow X$ the embedding of the world-volume in the space-time, we can push forward the complex on $Y$ to a complex of sheaves $i_{*}\underline{U}(1)_{Y} \rightarrow i_{*}\Omega^{1}_{Y,\mathbb{R}}$ on $X$, recalling that, for $\mathcal{F}$ a sheaf on $Y$, the sheaf $i_{*}\mathcal{F}$ on $X$ is defined as $(i_{*}\mathcal{F})(U) := \mathcal{F}(i^{-1}U)$ for any $U \subset X$ open. There is a natural map of complexes:
	\[\varphi_{X,Y,2}^{\bullet}: S_{X,2}^{\bullet} \rightarrow i_{*}S_{Y,1}^{\bullet}
\]
defined defined in the following way: in degree $0$ and $1$ it pulls back via $f$ from $U$ to $f^{-1}U$ the function or differential form in the domain, in degree $2$ it is the zero-map. We can now construct the cone of $\varphi_{X,Y,2}$, which is a complex of sheaves on $X$. The \emph{relative Deligne cohomology groups} of $S_{X,2}^{\bullet}$ with respect to $S_{Y,1}^{\bullet}$ are by definition the hypercohomology groups of the cone of $\varphi_{X,Y,2}$. The group that we called $\check{H}^{2}(X, \underline{U}(1) \rightarrow \Omega^{1}_{\mathbb{R}} \rightarrow \Omega^{2}_{\mathbb{R}}, Y)$ is actually the relative hypercohomology group:
	\[\check{H}^{2}(X, S_{X,2}^{\bullet}, i_{*}S_{Y,1}^{\bullet}).
\]
An element of this group is a couple made by a gerbe on $X$, which is trivial when restricted on $Y$, and an explicit trivialization of that gerbe on $Y$. The gerbe on $X$ is the $B$-field, the trivialization on $Y$ the $A$-field.

\section{Gauge theory on a single D-brane}\label{GaugeTheories}

We are now ready to discuss the possible geometric structures of the gauge theory on the D-brane, arising from the previous picture. The main distinction turns out to be whether or not the B-field is flat when restricted to the D-brane.

\subsection{Generic $B$-field}

We consider the coordinate change given by the D-brane:
\begin{equation}\label{CoordChange}
\begin{split}
	\{g_{\alpha\beta\gamma}, &-\Lambda_{\alpha\beta}, B_{\alpha}\} \cdot \{g_{\alpha\beta\gamma}^{-1} \cdot \eta_{\alpha\beta\gamma}, \Lambda_{\alpha\beta}, dA_{\alpha}\} = \{\eta_{\alpha\beta\gamma}, 0, B+F\}\\
	&\{g_{\alpha\beta\gamma}^{-1} \cdot \eta_{\alpha\beta\gamma}, \Lambda_{\alpha\beta}, dA_{\alpha}\} = \{\check{\delta}^{1}h_{\alpha\beta}, -\tilde{d}h_{\alpha\beta} + A_{\beta} - A_{\alpha}, dA_{\alpha}\}.
\end{split}
\end{equation}
Since, by Freed-Witten anomaly, $[\,\{\,g_{\alpha\beta\gamma}\,\}\,] = [\,\{\,\eta_{\alpha\beta\gamma}\,\}\,] \in \check{H}^{2}(Y, \underline{S}^{1})$ (not the constant sheaf $S^{1}$, the sheaf of functions $\underline{S}^{1}$), we can always choose a gauge $\{\eta_{\alpha\beta\gamma}, 0, B\}$, but we can also consider any gauge $\{\eta_{\alpha\beta\gamma}, 0, B'\}$ with $B'-B$ a closed form representing an integral de-Rham class: for a bundle, this corresponds to the free choice of a global automorphism.\footnote{For gerbes, we directly see this from the fact that $(1, 0, \Phi)$ is a hypercoboundary for $\Phi$ integral. Indeed, we have:
	\[\Phi\vert_{U_{\alpha}} = d\varphi_{\alpha} \qquad \varphi_{\beta} - \varphi_{\alpha} = d\rho_{\alpha\beta} \qquad \rho_{\alpha\beta} + \rho_{\beta\gamma} + \rho_{\gamma\alpha} = c_{\alpha\beta\gamma} \in \mathbb{Z}
\]
thus $\varphi_{\beta} - \varphi_{\alpha} = \tilde{d}h_{\alpha\beta}$ for $h_{\alpha\beta} = \exp(2\pi i \cdot \rho_{\alpha\beta})$ and $\check{\delta}^{1}h_{\alpha\beta} = 1$. Hence, $(1, 0, \Phi) = \check{\delta}^{1}(h_{\alpha\beta}, \varphi_{\alpha})$.} Given a certain gauge of the form $\{\eta_{\alpha\beta\gamma}, 0, B\}$, the brane gives a correction $\{1, 0, F\}$ to arrive at the fixed gauge $\{\eta_{\alpha\beta\gamma}, 0, B+F\}$. In fact, \eqref{CoordChange} becomes:
\begin{equation}
\begin{split}
	\{\eta_{\alpha\beta\gamma}, &0, B\} \cdot \{1, 0, dA_{\alpha}\} = \{\eta_{\alpha\beta\gamma}, 0, B+F\}\\
	&\{1, 0, dA_{\alpha}\} = \{\check{\delta}^{1}h_{\alpha\beta}, -\tilde{d}h_{\alpha\beta} + A_{\beta} - A_{\alpha}, dA_{\alpha}\}.
\end{split}
\end{equation}
We thus get $\check{\delta}^{1}h_{\alpha\beta} = 1$ and $-\tilde{d}h_{\alpha\beta} + A_{\beta} - A_{\alpha} = 0$, so $h_{\alpha\beta}$ give a gauge bundle on the brane with connection $-A_{\alpha}$ and Chern class $[\,-F\,]$. However, since $B$ and $F$ are arbitrary, such a bundle is defined up to large gauge transformations $B \rightarrow B + \Phi$ and $F \rightarrow F - \Phi$ for $\Phi$ integral.\footnote{In particular, we can always choose the gauge $F = 0$, obtaining a flat line bundle.}

\paragraph{}Moreover, we have the freedom to choose a different representative $\eta_{\alpha\beta\gamma} \cdot \check{\delta}^{1}\lambda_{\alpha\beta}$ of $w_{2}(Y) \in \check{H}^{2}(Y, S^{1})$. This is equivalent to consider:
\begin{equation}
\begin{split}
	\{\eta_{\alpha\beta\gamma}, &0, B\} \cdot \{\check{\delta}\lambda_{\alpha\beta}, 0, dA_{\alpha}\} = \{\eta_{\alpha\beta\gamma} \cdot \check{\delta}\lambda_{\alpha\beta}, 0, B+F\}\\
	&\{\check{\delta}\lambda_{\alpha\beta}, 0, dA_{\alpha}\} = \{\check{\delta}h_{\alpha\beta}, -\tilde{d}h_{\alpha\beta} + A_{\beta} - A_{\alpha}, dA_{\alpha}\}.
\end{split}
\end{equation}
We thus obtain that $\check{\delta}h_{\alpha\beta} = \check{\delta}\lambda_{\alpha\beta}$, i.e.\ $\check{\delta}(h_{\alpha\beta} / \lambda_{\alpha\beta}) = 1$. So, instead of $\{h_{\alpha\beta}\}$, we consider the bundle $[\,h_{\alpha\beta} / \lambda_{\alpha\beta}\,]$ instead of $[\,h_{\alpha\beta}\,]$. Since the functions $\lambda_{\alpha\beta}$ are constant, the real image of the Chern class is the same. In fact, if we write $h_{\alpha\beta} = \exp(2\pi i \cdot \tilde{h}_{\alpha\beta})$ and $\lambda_{\alpha\beta} = \exp(2\pi i \cdot \tilde{\lambda}_{\alpha\beta})$, we have that $\tilde{h}_{\alpha\beta} + \tilde{h}_{\beta\gamma} + \tilde{h}_{\gamma\alpha} = \tilde{h}_{\alpha\beta\gamma} \in \mathbb{Z}$ defining the first Chern class, and similarly $\tilde{\lambda}_{\alpha\beta} + \tilde{\lambda}_{\beta\gamma} + \tilde{\lambda}_{\gamma\alpha} = \tilde{\lambda}_{\alpha\beta\gamma} \in \mathbb{Z}$. However, since $\tilde{\lambda}_{\alpha\beta}$ are constant, $\tilde{\lambda}_{\alpha\beta\gamma}$ is a coboundary in the sheaf $\mathbb{R}$ and the real image of the Chern class of $\lambda_{\alpha\beta}$ is $0$.

This means that we fix a line bundle \emph{up to the torsion part}. Thus, the holonomy of $-A_{\alpha}$ is defined also up to the torsion part: this ambiguity is compensated for by the one of the pfaffian, due to the need of obtaining a global section of the tensor product. If $w_{2} = 0$, we can choose the preferred representative $\eta_{\alpha\beta\gamma} = 1$, thus we completely fix a line bundle up to large gauge transformation.

\subsection{Flat $B$-field}

If $B$ is flat, its holonomy is a class $\Hol(B\vert_{Y}) \in H^{2}(Y, S^{1})$ (constant sheaf $S^{1}$). We distinguish three cases:
\begin{itemize}
	\item $\Hol(B\vert_{Y}) = w_{2}(Y) = 0$: as before, we can choose the gauge $\eta_{\alpha\beta\gamma} = 1$, but, via an operation analogous to choosing \emph{parallel} local sections for line bundles, we can obtain $\{1, 0, 0\}$ instead of a generic $\{1, 0, B\}$. The choice $B = 0$ is canonical (it fixes also large gauge transformations). Thus we get $\{1, 0, 0\} \,\cdot\, \{1, 0, dA_{\alpha}\} = \{1, 0, F\}$ with $\{1, 0, dA_{\alpha}\} = \{\check{\delta}^{1}h_{\alpha\beta}, -\tilde{d}h_{\alpha\beta} + A_{\beta} - A_{\alpha}, dA_{\alpha}\}$. Hence we have $\check{\delta}^{1}h_{\alpha\beta} = 1$ and $A_{\beta} - A_{\alpha} = \tilde{d}_{\alpha\beta}$. In this case, we obtain a line bundle $L$ with connection $-A_{\alpha}$ and Chern class $c_{1}(L)$ such that $c_{1}(L) \otimes_{\mathbb{Z}} \mathbb{R} = [\,-F\,]_{dR}$, i.e.\ a gauge theory in the usual sense, canonically fixed. However, we will see in the following that, also in this case, there is a residual freedom in the choice of the bundle.
	\item $\Hol(B\vert_{Y}) = w_{2}(Y)$: as before, we choose $\{\eta_{\alpha\beta\gamma}, 0, 0\}$ instead of a generic $\{\eta_{\alpha\beta\gamma}, 0, B\}$. The choice $B = 0$ is canonical (it fixes also large gauge transformations). Thus we get $\{\eta_{\alpha\beta\gamma}, 0, 0\} \cdot \{1, 0, dA_{\alpha}\} = \{\eta_{\alpha\beta\gamma}, 0, F\}$ with $\{1, 0, dA_{\alpha}\} = \{\check{\delta}^{1}h_{\alpha\beta}, -\tilde{d}h_{\alpha\beta} + A_{\beta} - A_{\alpha}, dA_{\alpha}\}$, or, as discussed before, $\{\eta_{\alpha\beta\gamma}, 0, 0\} \cdot \{\check{\delta}^{1}\lambda_{\alpha\beta}, 0,$ $dA_{\alpha}\} = \{\eta_{\alpha\beta\gamma} \cdot \check{\delta}^{1}\lambda_{\alpha\beta}, 0, F\}$ with $\{\check{\delta}^{1}\lambda_{\alpha\beta}, 0, dA_{\alpha}\} = \{\check{\delta}^{1}h_{\alpha\beta}, -\tilde{d}h_{\alpha\beta} + A_{\beta} - A_{\alpha}, dA_{\alpha}\}$. In this case, we obtain a canonical line bundle with connection $-A_{\alpha}$ up to the torsion part, with real image of the Chern class $[\,-F\,]$.
	\item \emph{$\Hol(B\vert_{Y})$ generic:} in this case, we can use the same picture as for non-flat $B$-fields, obtaining a non-canonical gauge bundle, or we can use flatness to obtain a canonical gauge theory of different nature. In the latter case, we fix a cocycle $\{g_{\alpha\beta\gamma}\}$ such that $[\,\{g_{\alpha\beta\gamma}\}\,] = \Hol(B\vert_{Y}) \in H^{2}(Y, S^{1})$. We thus get a preferred gauge $\{g_{\alpha\beta\gamma}, 0, 0\}$, so that \eqref{CoboundaryA} becomes $\{g_{\alpha\beta\gamma}^{-1} \cdot \eta_{\alpha\beta\gamma}, 0, dA_{\alpha}\} = \{\check{\delta}^{1}h_{\alpha\beta}, -\tilde{d}h_{\alpha\beta} + A_{\beta} - A_{\alpha}, dA_{\alpha}\}$. We obtain $\check{\delta}^{1}h_{\alpha\beta} = g_{\alpha\beta\gamma}^{-1} \cdot \eta_{\alpha\beta\gamma}$ and $A_{\beta} - A_{\alpha} = \tilde{d}h_{\alpha\beta}$. Since $g_{\alpha\beta\gamma}^{-1} \cdot \eta_{\alpha\beta\gamma}$ are constant, we obtain a ``bundle with not integral Chern class'', as explained in the next section.
\end{itemize}

\paragraph{Remark:} We have said above that only for $\Hol(B\vert_{Y}) = 0$ and $w_{2}(Y) = 0$ we are able to recover the torsion of the gauge bundle. Actually, we can still recover the torsion part even if $w_{2}(Y) = 0$ and $B$ is flat. In fact, also in this case we can choose $\eta_{\alpha\beta\gamma} = 1$ fixing the transition function $h_{\alpha\beta}$ of the bundle. Let us consider a fractional bundle $L$ such that $\check{\delta}\{h_{\alpha\beta}\} = \{g_{\alpha\beta\gamma}^{-1}\}$ for $[\,g_{\alpha\beta\gamma}\,] = \Hol(B\vert_{Y}) \in H^{2}(Y, S^{1})$. Then, evaluating the holonomy of $B$ over the generators of $H_{2}(Y, \mathbb{Z})$, we can find a discrete subgroup $\Gamma \leq \mathbb{R}$ such that $c_{1}(L) \in H^{2}(Y, \Gamma)$, so that $c_{1}(L)$ has a torsion part. This is more interesting if we know the fractionality of the brane (see below): for example, if we have a $\frac{1}{n}\,$-fractional gauge theory (e.g.\ fractional branes from $\mathbb{Z}_{n}$-orbifolds), we have $c_{1}(L) \in H^{2}(Y, \frac{1}{n}\mathbb{Z}) \simeq H^{2}(Y, \mathbb{Z})$.\\$\square$

\paragraph{}A comment is in order when $\Hol(B\vert_{Y}) = w_{2}(Y) = 0$: also in this case, the bundle is not completely fixed, but there is a residual gauge freedom. In fact, such a configuration is described by $[\,\{g_{\alpha\beta\gamma}, -\Lambda_{\alpha\beta}, h_{\alpha\beta}, B_{\alpha}, -A_{\alpha}\}\,] \in \check{H}^{2}(X, \underline{S}^{1} \rightarrow \Omega^{1}_{\mathbb{R}} \rightarrow \Omega^{2}_{\mathbb{R}}, Y)$ such that $[\,\{g_{\alpha\beta\gamma}, -\Lambda_{\alpha\beta}, B_{\alpha}\}\,]$ is geometrically trivial on $Y$. As we said, we can choose on $Y$ the preferred gauge $\{1, 0, h_{\alpha\beta}, 0, -A_{\alpha}\}$ so that the cocycle condition gives exactly $\{1, 0, \check{\delta}^{2}h_{\alpha\beta}, 0, \tilde{d}h_{\alpha\beta} + A_{\alpha} - A_{\beta}\} = 0$, i.e.\ $-A_{\alpha}$ is a connection on the bundle $[\,h_{\alpha\beta}\,]$. There is still a question: how are the possible representatives $\{1, 0, h_{\alpha\beta}, 0, -A_{\alpha}\}$ of the same class? Can they all be obtained via a reparametrization of the bundle $[\,h_{\alpha\beta}, A_{\alpha}\,] \in \check{H}^{1}(Y, \underline{S}^{1} \rightarrow \Omega^{1}_{\mathbb{R}})$? The possible reparametrization are given by:
	\[\begin{split}
	\{1, 0, h_{\alpha\beta}, 0, -A_{\alpha}\} \,\cdot\, \{\check{\delta}^{1}g_{\alpha\beta}, -\tilde{d}g_{\alpha\beta} + \Lambda_{\beta} &- \Lambda_{\alpha}, ((i^{*})^{1}g_{\alpha\beta})^{-1} \cdot h_{\beta} h_{\alpha}^{-1}, d\Lambda_{\alpha}, \\
	&(i^{*})^{0} \Lambda_{\alpha} - \tilde{d}h_{\alpha}\} = \{1, 0, h'_{\alpha\beta}, 0, -A'_{\alpha}\}
\end{split}\]
thus we get the conditions:
\begin{equation}\label{ResidualGauge}
	\check{\delta}^{1}g_{\alpha\beta} = 1 \qquad -\tilde{d}g_{\alpha\beta} + \Lambda_{\beta} - \Lambda_{\alpha} = 0 \qquad d\Lambda_{\alpha} = 0.
\end{equation}
If we choose $g_{\alpha\beta} = 1$ and $\Lambda_{\alpha} = 0$ we simply get $h'_{\alpha\beta} = h_{\alpha\beta} \cdot h_{\beta} h_{\alpha}^{-1}$ and $A'_{\alpha} = A_{\alpha} + \tilde{d}h_{\alpha}$, i.e.\ a reparametrization of $[\,h_{\alpha\beta}, A_{\alpha}\,] \in \check{H}^{1}(Y, \underline{S}^{1} \rightarrow \Omega^{1}_{\mathbb{R}})$, and that is what we expected. But what happens in general? Equations \eqref{ResidualGauge} represent any line bundle $g_{\alpha\beta}$ on the whole space-time $X$ with \emph{flat} connection $-\Lambda_{\alpha}$, thus they represent a residual gauge freedom in the choice of the line bundle over $Y$: \emph{any flat bundle on $Y$ which is the restriction of a \emph{flat} line bundle over $X$ is immaterial for the gauge theory on the D-brane}. Can we give a physical interpretation of this fact?

Let us consider a line bundle $L$ over $Y$ with connection $-A_{\alpha}$: it determines the holonomy as a function from the loop space of $Y$ to $S^{1}$. Actually, we are not interested in a generic loop: we always work with $\partial \Sigma$, with $\Sigma$ in general not contained in $Y$: thus, such loops are in general not homologically trivial on $Y$, but they are so on $X$. Let us suppose that $L$ extends to $\tilde{L}$ over $X$: in this case, we can equally consider the holonomy over $\partial \Sigma$ with respect to $\tilde{L}$. If $\tilde{L}$ is flat, such a holonomy becomes an $S^{1}$-cohomology class evaluated over a contractible loop, thus it is $0$. Hence, a bundle extending to a flat one over $X$ gives no contribution to the holonomy over the possible boundaries of the world-sheets. Therefore, also in the case $\Hol(B\vert_{Y}) = w_{2}(Y) = 0$, \emph{we do not have a canonically fixed bundle with connection} on the brane: we rather have an equivalence class of bundles defined up to flat ones extending to flat space-time bundles. For another important comment on this point, see the conclusions.

\section{Real Chern classes}

In the previous section we showed that for $B$ flat we obtain a gauge theory on a generalized bundle: while bundles are represented by cocycles $\{g_{\alpha\beta}\}$ in $\rm\check{C}$ech cohomology, such generalized bundles are represented by cochains whose coboundary $\check{\delta}^{1}\{g_{\alpha\beta}\}$ is made by constant functions (not necessarily $1$), realizing a class in $\check{H}^{2}(X, S^{1})$. We now see that even in these cases we can define connections and first Chern class, but the latter turns out to be any closed form, not necessarily integral.

\paragraph{}Let us consider the definition of Chern class of a trivial bundle: we have a bundle $[\,\{g_{\alpha\beta}\}\,] \in \check{H}^{1}(\mathfrak{U}, \underline{S}^{1})$, so that $g_{\alpha\beta} \cdot g_{\beta\gamma} \cdot g_{\gamma\alpha} = 1$; if $g_{\alpha\beta} = e^{2\pi i \cdot \rho_{\alpha\beta}}$, we have $\rho_{\alpha\beta} + \rho_{\beta\gamma} + \rho_{\gamma\alpha} = \rho_{\alpha\beta\gamma} \in \mathbb{Z}$, so that we obtain a class $[\,\{\rho_{\alpha\beta\gamma}\}\,] \in \check{H}^{2}(\mathfrak{U}, \mathbb{Z})$ which is the first Chern class.

Let us call $\Gamma_{n}$ the subgroup of $S^{1}$ given by the $n$-th root of unity. If we call $\frac{1}{n}\mathbb{Z}$ the subgroup of $\mathbb{R}$ made by the fractions $\frac{k}{n}$ for $k \in \mathbb{Z}$, then $\Gamma_{n} = e^{2\pi i \cdot \frac{1}{n}\mathbb{Z}}$. Let us suppose we have a cochain $\{g_{\alpha\beta}\} \in \check{C}^{1}(\mathfrak{U}, \underline{S}^{1})$ such that $g_{\alpha\beta} \cdot g_{\beta\gamma} \cdot g_{\gamma\alpha} = g_{\alpha\beta\gamma} \in \Gamma_{n}$. Then, for $g_{\alpha\beta} = e^{2\pi i \cdot \rho_{\alpha\beta}}$, we have that $\rho_{\alpha\beta} + \rho_{\beta\gamma} + \rho_{\gamma\alpha} = \rho_{\alpha\beta\gamma} \in \frac{1}{n}\mathbb{Z}$, so that we obtain a rational class $c_{1} = [\,\{\rho_{\alpha\beta\gamma}\}\,] \in \check{H}^{2}(\mathfrak{U}, \mathbb{Q})$ such that $n \cdot c_{1}$ is an integral class. Can we give a geometric interpretation of these classes?

\paragraph{}A 2-cochain can be thought of as a trivialization of a trivialized gerbe, in the same way as a 1-cochain (i.e.\ a set of local functions) is a trivialization of a trivialized line bundle; thus a line bundle is a trivialization of a gerbe represented by the coboundary $1$, in the same way as a global function is a global section of $X \times \mathbb{C}$. We describe first the easier case of local functions trivializing a line bundle, i.e.\ we lower by 1 the degree in cohomology. 

\subsection{Trivializations of line bundles}

\subsubsection{Definition}

As line bundles, which are classes in $\check{H}^{1}(\mathfrak{U}, \underline{S}^{1})$, are trivializations of gerbes represented by the coboundary $1$, likewise a section of a line bundle, represented by transition functions equal to 1, is a class in $\check{H}^{0}(\mathfrak{U}, \underline{S}^{1})$, i.e.\ a function $f: X \rightarrow S^{1}$. A cochain $\{f_{\alpha}\} \in \check{C}^{0}(\mathfrak{U}, \underline{S}^{1})$ is a section of a trivial bundle represented by transition functions $f_{\alpha}^{-1} \cdot f_{\beta}$.

Given a function $f: X \rightarrow S^{1}$, we can naturally define a Chern class $c_{1}(f) \in H^{1}(\mathfrak{U}, \mathbb{Z})$, which is the image under the Bockstein map of $f = [\,\{f_{\alpha}\}\,] \in \check{H}^{0}(\mathfrak{U}, \underline{S}^{1})$. We directly compute it as for bundles: since $f_{\beta} \cdot f_{\alpha}^{-1} = 1$, for $f_{\alpha} = e^{2\pi i \cdot \rho_{\alpha}}$ we have $\rho_{\beta} - \rho_{\alpha} = \rho_{\alpha\beta} \in \mathbb{Z}$, so that we can define a class $c_{1}(f) = [\,\{\rho_{\alpha\beta}\}\,] \in \check{H}^{1}(\mathfrak{U}, \mathbb{Z})$. The geometric interpretation is very simple: $c_{1}(f)$ is the pull-back under $f$ of the generator of $H^{1}(S^{1}, \mathbb{Z}) \simeq \mathbb{Z}$. As we have done for bundles, let us suppose we have a cochain $[\,\{f_{\alpha}\}\,] \in \check{C}^{0}(\mathfrak{U}, \underline{S}^{1})$ such that $f_{\alpha}^{-1} \cdot f_{\beta} = f_{\alpha\beta} \in \Gamma_{n}$. Then $\rho_{\beta} - \rho_{\alpha} = \rho_{\alpha\beta} \in \frac{1}{n} \mathbb{Z}$. Therefore we obtain a class $c_{1} = [\,\{\rho_{\alpha\beta}\}\,] \in \check{H}^{1}(\mathfrak{U}, \mathbb{Q})$ such that $n \cdot c_{1}$ is an integral class.

\paragraph{}From the exact sequences point of view, the Chern class is the image of the Bockstein map of the sequence:
	\[0 \longrightarrow \mathbb{Z} \longrightarrow \underline{\mathbb{R}} \overset{e^{2\pi i \, \cdot}}\longrightarrow \underline{S}^{1} \longrightarrow 0.
\]
In the fractional case, since $\check{\delta}^{0}f_{\alpha}$ takes values in $\Gamma_{n}$, the cochain $\{f_{\alpha}\}$ is a cocycle in $\underline{S}^{1} / \,\Gamma_{n}$. Thus, we consider the sequence:
	\[0 \longrightarrow \textstyle \frac{1}{n} \displaystyle \mathbb{Z} \longrightarrow \underline{\mathbb{R}} \overset{\pi_{\Gamma_{n}} \,\circ\, e^{2\pi i \, \cdot}}\longrightarrow \underline{S}^{1} / \,\Gamma_{n} \longrightarrow 0
\]
and the image of the Bockstein map is exactly the fractional Chern class. We have constructed in this way rational Chern classes, but this is generalizable to any real Chern class. In fact, it is sufficient that $\rho_{\alpha\beta}$ be constant for every $\alpha, \beta$ to apply the previous construction, using the constant sheaf $S^{1}$ instead of $\Gamma_{n}$. The corresponding sequence, which contains all the previous ones by inclusion, is:
	\[0 \longrightarrow \textstyle \mathbb{R} \longrightarrow \underline{\mathbb{R}} \overset{\pi_{S^{1}} \,\circ\, e^{2\pi i \, \cdot}}\longrightarrow \underline{S}^{1} / \,S^{1} \longrightarrow 0.
\]
In other words, if the cochain is a cocycle up to constant functions, we obtain a real Chern class. If these constant functions belong to $\Gamma_{n}$, we obtain a rational Chern class in $\frac{1}{n}\mathbb{Z}$. We now want to give a geometric interpretation of these classes.

\subsubsection{Geometric interpretation}

If we think of the cochain as a trivialization of $X \times \mathbb{C}$, it follows that different trivializations have different Chern classes, depending on the realization of the trivial bundle as $\rm\check{C}$ech coboundary. This seems quite unnatural from a topological point of view, since the particular trivialization should not play any role. However, if we fix a flat connection, we can distinguish a particular class of trivializations, which are parallel with respect to such a connection.

Let us consider a trivial line bundle with a global section and a flat connection $\nabla$, which we think of as $X \times \mathbb{C}$ with a globally defined form $A$, expressing $\nabla$ with respect to the global section $X \times \{1\}$. We know the following facts:
\begin{itemize}
	\item if we choose parallel sections $\{f_{\alpha}\}$, we obtain a trivialization with a real Chern class $c_{1}(\{f_{\alpha}\}) \in \check{H}^{1}(X, \mathbb{R})$, and the local expression of the connection becomes $\{0\}$;
	\item the globally defined connection $A$, expressed with respect to $1$, is closed by flatness, thus it determines a de-Rham cohomology class $[\,A\,] \in H^{1}_{dR}(X)$.
\end{itemize}
We now prove that these two classes coincide under the standard isomorphism between $\rm\check{C}$ech and de-Rham cohomology. This is the geometric interpretation of real Chern classes: \emph{the real Chern class of a trivialization of $X \times \mathbb{C}$ is the cohomology class of a globally-defined flat connection, expressed with respect to $X \times \{1\}$, for which the trivialization is parallel}.

If the trivial bundle has holonomy $1$ (i.e.\ \emph{geometrically trivial}), we can find a global parallel section: thus there exists a function $f \in \check{H}^{1}(X, \underline{S}^{1})$ trivializing the bundle, and the Chern class of a function is integral. If we express the connection with respect to $1$ we obtain an integral class $[\,A\,] = [\,f^{-1} df\,]$, while if we express it with respect to the global section $f \cdot 1$ we obtain $0$.

\paragraph{}We now prove the statement. Given $\{f_{\alpha}\} \in \check{C}^{0}(\mathfrak{U}, \underline{S}^{1})$ such that $\check{\delta}^{0}\{f_{\alpha}\} \in \check{C}^{1}(\mathfrak{U}, S^{1})$, we consider the connection $\nabla$ on $X \times \mathbb{C}$ which is represented by $0$ with respect to $\{f_{\alpha}^{-1}\}$. If we represent $\nabla$ with respect to $X \times \{1\}$ we obtain $A_{\alpha} = \tilde{d} f_{\alpha}$, and $A_{\alpha} - A_{\beta} = \tilde{d} (f_{\beta} \cdot f_{\alpha}^{-1} ) = 0$. We thus realize the 1-form $A$ as a $\rm\check{C}$ech cocycle: we have that $A_{\alpha} = (2\pi i)^{-1} d\log f_{\alpha}$ and $(2\pi i)^{-1}\log f_{\beta} - (2\pi i)^{-1}\log f_{\alpha} = (2\pi i)^{-1}\log g_{\alpha\beta} = \rho_{\alpha\beta}$ which is constant, so that $[\,A\,]_{H^{1}_{dR}(X)} \simeq [\,\{\rho_{\alpha\beta}\}\,]_{\check{H}^{1}(X, \mathbb{R})}$. By definition $c_{1}(\{f_{\alpha}\}) = [\,\{\rho_{\alpha\beta}\}\,]$, thus $[\,A\,]_{H^{1}_{dR}(X)} \simeq c_{1}(\{f_{\alpha}\})_{\check{H}^{1}(X, \mathbb{R})}$.

\paragraph{}Moreover, if we consider the sequence $0 \rightarrow \mathbb{Z} \rightarrow \mathbb{R} \rightarrow S^{1} \rightarrow 0$, for $p_{S^{1}}: H^{1}(X, \mathbb{R}) \rightarrow H^{1}(X, S^{1})$, we have that $p_{S^{1}} \, c_{1}(\{f_{\alpha}\}) = p_{S^{1}} \, [\,\rho_{\alpha\beta}\,] = [\,f_{\beta}f_{\alpha}^{-1}\,]_{S^{1}}$. Thus, for $\check{\delta}^{0}\{f_{\alpha}\} \in \check{C}^{1}(X, S^{1})$ (hence, obviously, $\check{\delta}^{0}\{f_{\alpha}\} \in \check{Z}^{1}(X, S^{1})$) we have that the first Chern class is one of the possible real lifts of $[\,\check{\delta}^{0}\{f_{\alpha}\}\,]_{S^{1}}$. Therefore, $p_{S^{1}} \, c_{1}(\{f_{\alpha}\})$ is the holonomy of the trivial line bundle on which the connection $A$, previously considered, is defined.

\subsubsection{Hypercohomological description}

The trivialized bundle $X \times \mathbb{C}$ with global connection $A$ corresponds to the hypercocycle $\{1, -A\} \in \check{Z}^{1}(X, \underline{S}^{1} \rightarrow \Omega^{1}_{\mathbb{R}})$. For $A$ flat and $\{f_{\alpha}\}$ parallel sections, we have $[\,\{1, -A\}\,] = [\,\{\check{\delta}^{0}f_{\alpha}, 0\}\,]$, thus the difference is a coboundary:
	\[\{1, -A\} \cdot \{\check{\delta}^{0}f_{\alpha}, \tilde{d}f_{\alpha}\} = \{\check{\delta}^{0}f_{\alpha}, 0\}
\]
thus $\tilde{d}f_{\alpha} = A_{\alpha}$ so that, as proven before, $[\,A\,] \simeq c_{1}(\{f_{\alpha}\})$.

If $f$ is globally defined, we get $\{1, -A\} \cdot \{1, \tilde{d}f\} = \{1, 0\}$ so that $[\,A\,] = [\,\tilde{d}f\,]$ which is integral: this corresponds to the choice of a global parallel section $f \cdot 1$ in $X \times \mathbb{C}$.

\subsection{Trivializations of gerbes}

Let us now consider a trivialization of a gerbe $\{h_{\alpha\beta}\} \in \check{C}^{1}(X, \underline{S}^{1})$ such that $\check{\delta}^{1}\{h_{\alpha\beta}\} \in \check{C}^{2}(X, S^{1})$. We can consider a connection $\{-A_{\alpha}\}$ such that $A_{\beta} - A_{\alpha} = \tilde{d}h_{\alpha\beta}$, as for an ordinary bundle. We have $dA_{\alpha} = dA_{\beta}$ so that $-F = -dA_{\alpha}$ is a global closed form whose de-Rham class $[\,-F\,]$ is exactly the fractional Chern class of $[\,\{h_{\alpha\beta}\}\,] \in \check{\delta}^{-1}(\check{C}^{2}(X, S^{1})) \,/\, \check{B}^{1}(X, \underline{S}^{1})$. We define such a trivialization with connection as an element of the hypercohomology group:
	\[\check{H}^{1}\bigl(\, X, \underline{S}^{1} / S^{1} \overset{\tilde{d}}\longrightarrow \Omega^{1}_{\mathbb{R}} \,\bigr).
\]
We interpret the Chern class of such trivializations as before: we consider the flat gerbe $[\,\{\check{\delta}^{1}h_{\alpha\beta}, 0, 0\}\,]$, and we represent it as $[\,\{1, 0, -F\}\,]$:
	\[\{1, 0, -F\} \cdot \{\check{\delta}^{1}h_{\alpha\beta}, -\tilde{d}h_{\alpha\beta} + A_{\beta} - A_{\alpha}, dA_{\alpha}\} = \{\check{\delta}^{1}h_{\alpha\beta}, 0, 0\}
\]
from which we obtain:
	\[A_{\beta} - A_{\alpha} = \tilde{d}h_{\alpha\beta} \qquad dA_{\alpha} = F\vert_{U_{\alpha}}.
\]
From these data we can now realize $F$ as a $\rm\check{C}$ech class: we have $F\vert_{U_{\alpha}} = dA_{\alpha}$ and $A_{\beta} - A_{\alpha} = \tilde{d}h_{\alpha\beta}$, thus $\check{\delta}^{1}\tilde{d}h_{\alpha\beta} = 0$, thus $(2\pi i)^{-1} \check{\delta}^{1}\log h_{\alpha\beta}$ is constant and expresses $[\,F\,]$ as $\rm\check{C}$ech class. The latter is exactly $c_{1}(\{h_{\alpha\beta}\})$.

\paragraph{}What happens for the holonomy of these connections? In general anyone of them is not well-defined as a function on closed curves, but it is a section of a line bundle that, on curves which are boundary of open surfaces, is canonically trivial and coincides with the one determined by the flat gerbe realized by $(1, 0, F)$ but with respect to the sections $\check{\delta}g$. In fact, the expression of the holonomy of $A$ on $\partial\Sigma$ coincides with the holonomy of $(\check{\delta}g, A_{\beta} - A_{\alpha}, dA_{\alpha})$ on $\Sigma$, but $\check{\delta}(g, 0) = (\check{\delta}g, d\log g_{\alpha\beta}, 0)$ and the sum is $(1, 0, dA_{\alpha})$, thus the gerbe is $(1, 0, F)$ but it is realized on open surfaces with respect to $\check{\delta}g$.

\section{Stack of coincident branes}

Up to now we have discussed the case of a single brane. In the case of a stack of coincident D$p$-branes, we need non-abelian cohomology \cite{Brylinski}. However, here we would like to avoid a technical discussion and just state the main differences with respect to the abelian case. We will arrive to the same conclusions as \cite{Kapustin}, taking into account the presence of the Pfaffian.

\paragraph{}Let us consider again the fundamental equation \eqref{CoordChange}:
	\[\begin{split}
		\{g_{\alpha\beta\gamma}, &-\Lambda_{\alpha\beta}, B_{\alpha}\} \cdot \{g_{\alpha\beta\gamma}^{-1} \cdot \eta_{\alpha\beta\gamma}, \Lambda_{\alpha\beta}, dA_{\alpha}\} = \{\eta_{\alpha\beta\gamma}, 0, B+F\}\\
		&\{g_{\alpha\beta\gamma}^{-1} \cdot \eta_{\alpha\beta\gamma}, \Lambda_{\alpha\beta}, dA_{\alpha}\} = \{\check{\delta}^{1}h_{\alpha\beta}, -\tilde{d}h_{\alpha\beta} + A_{\beta} - A_{\alpha}, dA_{\alpha}\}.
	\end{split}
\]
Since $\check{\delta}^{1}h_{\alpha\beta} = g_{\alpha\beta\gamma}^{-1} \cdot \eta_{\alpha\beta\gamma}$, the class $[\,g^{-1}\eta\,] \in H^{1}(Y,\underline{S}^{1})$ must be trivial: this means that $\zeta\vert_{Y} = W_{3}(Y)$, which is the Freed-Witten anomaly equation. Instead, in the case of a stack of branes, $h_{\alpha\beta} \in U(n)$. Then, if we think of $g_{\alpha\beta\gamma}^{-1} \cdot \eta_{\alpha\beta\gamma}$ as a multiple of the identity $I_{n}$, the relation $\check{\delta}^{1}h_{\alpha\beta} = g_{\alpha\beta\gamma}^{-1} \cdot \eta_{\alpha\beta\gamma}$ is not a trivialization of $[\,g^{-1}\eta\,] \in H^{1}(Y,\underline{S}^{1})$ any more and it does not imply that $\zeta\vert_{Y} = W_{3}(Y)$. We thus rewrite the previous equation as:
\begin{equation}
\begin{split}
		&\{g_{\alpha\beta\gamma}, -\Lambda_{\alpha\beta}, B_{\alpha}\} \cdot \{g_{\alpha\beta\gamma}^{-1} \cdot \eta_{\alpha\beta\gamma}, \Lambda_{\alpha\beta}, d\tilde{A}_{\alpha}\} = \{\eta_{\alpha\beta\gamma}, 0, B+\tilde{F}\}\\
		&\{g_{\alpha\beta\gamma}^{-1} \cdot \eta_{\alpha\beta\gamma}, \Lambda_{\alpha\beta}, d\tilde{A}_{\alpha}\} = \textstyle\frac{1}{n} \Tr\, \{\check{\delta}^{1}h_{\alpha\beta}, -h_{\alpha\beta}^{-1}dh_{\alpha\beta} + h^{-1}_{\alpha\beta}A_{\beta}h_{\alpha\beta} - A_{\alpha}, dA_{\alpha} + A_{\alpha} \wedge A_{\alpha}\}
\end{split}
\end{equation}
where the trace is taken in all the components. We thus obtain $\tilde{A} = \frac{1}{n} \Tr A$ and $\tilde{F} = \frac{1}{n} \Tr F$.

\paragraph{}A rank-$n$ bundle $\{h_{\alpha\beta}\}$ such that $\check{\delta}^{1}\{h_{\alpha\beta}\}$ realizes a class in $H^{2}(X, \underline{S}^{1})$ is called a \emph{twisted bundle} or \emph{non-commutative bundle}. For $\beta$ the Bockstein homomorphism in degree 2 of the sequence $0 \rightarrow \mathbb{Z} \rightarrow \underline{\mathbb{R}} \rightarrow \underline{S}^{1} \rightarrow 0$, we define $\beta' = \beta[\,\check{\delta}^{1}\{h_{\alpha\beta}\}\,] \in H^{3}(X, \mathbb{Z})$. Thus, for the relation $\check{\delta}^{1}h_{\alpha\beta} = g_{\alpha\beta\gamma}^{-1} \cdot \eta_{\alpha\beta\gamma}$ to hold, one must have:
\begin{equation}
	\beta' = W_{3}(Y) - \zeta\vert_{Y}.
\end{equation}
This is the Freed-Witten anomaly equation for stack of branes. \emph{We remark that, while in the abelian case the $A$-field corresponds to a reparametrization of the gerbe, in the non-abelian case it provides another non-trivial gerbe}, which tensor-multiplies the gerbe of the B-field.

\paragraph{}The classification of configurations in this case is analogous to the case of a single brane, allowing for the possibility of a non-commutative bundle when $\beta' \neq 0$. For $\beta' = 0$, we have the same situation as before, with irrational Chern classes for non integral bundles.

\section{Conclusions}

We have classified the allowed configurations of $B$-field and $A$-field in type II superstring backgrounds with a fixed set of D-branes, which are free of Freed-Witten anomaly. For a single D-brane $Y \subset X$, we distinguish the following fundamental cases:
\begin{itemize}
	\item \emph{$B$ geometrically trivial, $w_{2}(Y) = 0$:} we fix the preferred gauge $(1, 0, 0)$, so that we have $(1, 0, F) = \delta(h, -A)$ with $(h, -A)$ a line bundle, up to the residual gauge symmetry;
	\item \emph{$B$ flat:} we fix the preferred gauge $(g,0,0)$ so that we have $(g^{-1}\eta,0,F) = \delta(h, -A)$ with $(h, -A)$ a ``bundle'' with, in general, a non-integral Chern class; the image in $S^{1}$ of such a Chern class is given by $\Hol(B\vert_{Y}) - w_{2}(Y)$; even if this bundle has integral Chern class, i.e.\ if $\Hol(B\vert_{Y}) = w_{2}(Y)$, in general it is defined only up to the torsion part; if $\Hol(B\vert_{Y}) = w_{2}(Y) = 0$ we end up with the previous case so that we recover the torsion part up to the residual gauge;
	\item \emph{$B$ generic:} we fix a gauge $(\eta, 0, B)$ so that we have $(1, 0, F) = \delta(h, -A)$ with $(h, -A)$ a \emph{non-canonical} line bundle, where non-canonicity is related to large gauge transformations $B \rightarrow B + \Phi$ and $F \rightarrow F - \Phi$ for $\Phi$ integral.
\end{itemize}
For a stack of coincident branes the situation is analogous, except for the possibility of non-commutative bundles.

\paragraph{}So far we have considered the case of one brane or stack of coincident branes. One may wonder what happens when we have more than one non-coincident branes or stacks of branes: this case is actually already included in the previous discussion, thinking of $Y$ as the disconnected union of all the world-volumes. In particular, the residual gauge symmetry becomes an ambiguity corresponding to the restriction to each brane of a \emph{unique} flat space-time bundle. In physical terms this can be seen as follows: if we choose two cycles, one for each brane, which are homologous in space-time but not necessarily homologically trivial, since the difference is homologically trivial we can link them by an open string loop stretching from one brane to the other. In this way we determine the holonomy on the difference, i.e.\ the difference of the holonomies on the two loops. We thus remain with a global uncertainty, represented by flat space-time line bundles.

\paragraph{}Let us briefly comment on the case of fractional branes coming from orbifolds. Using the notation of \cite{BDM}, let $\Gamma$ be the internal orbifold group, whose regular representation splits into $M$ irreducible representations of dimensions $d_{I}$ for $I = 0, \ldots, M-1$, and let $C_{I}$ be the corresponding cycles in the ADE-resolution of the orbifold singularity. $B$ is taken flat on the internal space and satisfying the formula $\int_{C_{I}} B = d_{I} \,/\, \abs{\Gamma}$ for $I = 1, \ldots, M-1$, while, on the last cycle, $\int_{C_{0}} B = -\sum_{I \neq 0} d_{I}\int_{C_{I}} B$. Moreover one chooses $F$ on a cycle representing $C_{0}$ (to be subsequently shrunk) such that $\int_{C_{0}}F = 1$, while, on the chosen representatives of the other cycles, one chooses $F = 0$. What does this mean in our language? One fixes a gauge $\{1, 0, B\}$ on the whole internal space ($\eta_{\alpha\beta\gamma} = 1$ because the manifold involved is spin), supposing that the hypercocycle fixed on the representatives of the $C_{I}$'s, for $I \neq 0$, is the restriction of the global one: one thus gets $F = 0$. On the representative of $C_{0}$, instead, we consider a hypercocycle corresponding to the restriction of the global one, modified by an automorphism of the gerbe which generates $F$ so that $\int_{C_{0}}F = 1$. In conclusion we obtain, on $C_{0}$, $\{1, 0, B + F\}$. This is \emph{not} the canonical gauge choice adopted in section \ref{GaugeTheories} that gives rise to a fractional bundle: had we made this choice, we would have obtained a bundle with a fractional Chern class $F$, whose imagine in $S^{1}$ is given by $\Hol(B\vert_{C_{I}}) = d_{I} \,/\, \abs{\Gamma}$.

\chapter{D-branes and K-theory}

\section{Overview}

K-theory provides a good tool for classifying D-brane charges in type II superstring theory \cite{Evslin, OS}. In the case of vanishing $B$-field, there are two main approaches in the literature. The first one consists of applying the Gysin map to the gauge bundle of the D-brane, obtaining a K-theory class in the space-time \cite{MM}. This approach is motivated by the Sen's conjecture, stating that a generic configuration of branes and antibranes with gauge bundle is equivalent, via tachyon condensation, to a stack of coincident space-filling brane-antibrane pairs provided with an appropriate K-theory class \cite{Sen}. The second approach consists of applying the Atiyah-Hirzebruch spectral sequence (AHSS, \cite{AH}) to the Poincar\'e dual of the homology class of the D-brane: such a sequence rules out some cycles affected by global world-sheet anomalies, e.g.\ Freed-Witten anomaly \cite{FW}, and quotients out some cycles which are actually unstable, e.g.\ MMS-instantons \cite{MMS}. We start assuming for simplicity that the space-time and the D-brane world-volumes are compact. As we have seen in the first part, for a given filtration of the space-time $S = S^{10} \supset S^{9} \supset \cdots \supset S^{0}$, the second step of AHSS is the cohomology of $S$, i.e.\ $E^{p,\,0}_{2}(S) = H^{p}(S, \mathbb{Z})$, while the last step of AHSS is given by (up to canonical isomorphism):
	\[E^{p,\,0}_{\infty}(S) \simeq \frac{ \Ker ( K^{p}(S) \rightarrow K^{p}(S^{p-1}) ) } {\Ker ( K^{p}(S) \rightarrow K^{p}(S^{p}) )}.
\]
Hence, given a D-brane world-volume $WY_{p}$ of even codimension $10 - (p+1) = 9 - p$, with gauge bundle $E \rightarrow WY_{p}$ of rank $q$, if the Poincar\'e dual of $WY_{p}$ in $S$ survives until the last step of AHSS, it determines a class $\{\PD_{S}[q \cdot WY_{p}]\} \in E^{9-p,\,0}_{\infty}(S)$ whose representatives belong to $\Ker(K^{9-p}(S) \rightarrow K^{9-p}(S^{8-p}))$.

These two approaches give different information, in particular AHSS does not take into account the gauge bundle. Thanks to what we have seen in the first part about the link between Gysin map and AHSS, we have the tools to relate the two approaches, at least assuming that the $B$-field is vanishing. In fact, let us consider a D$p$-brane world-volume $WY_{p} \subset S$ with gauge bundle $E \rightarrow WY_{p}$ of rank $q$, and let $i: WY_{p} \hookrightarrow S$ be the embedding. We know that $i_{!}(E) \in \Ker (K^{9-p}(S) \rightarrow K^{9-p}(S^{8-p}))$ and that:
	\[\{\PD_{S}[q \cdot WY_{p}]\}_{E^{9-p, \,0}_{\infty}} = [i_{!}(E)].
\]
Thus, we must first use AHSS to detect possible anomalies, then we can use the Gysin map to get the charge of a non-anomalous brane: such a charge belongs to the equivalence class reached by AHSS, so that the Gysin map gives more detailed information.

Moreover, we compare this picture with the case of rational coefficients: since the Chern character provides isomorphisms $K(S) \otimes_{\mathbb{Z}} \mathbb{Q} \simeq H^{\ev}(S, \mathbb{Q})$ and $K^{1}(S) \otimes_{\mathbb{Z}} \mathbb{Q} \simeq H^{\odd}(S, \mathbb{Q})$, and since AHSS with rational coefficients degenerates at the second step, i.e.\ at the level of cohomology, we gain a complete equivalence between the two K-theoretical approaches, both being equivalent to the old cohomological classification. We now describe in more detail what we have summarized up to now. This chapter is a reproduction of \cite[Chap.\ 1,2,6]{FS}.

\section{D-brane charges and K-theory - Part I}\label{DBranesKTheoryCpt}

As we already said, for simplicity we start working assuming the ten-dimensional space-time $S$ to be compact, so that also a D-brane world-volumes are compact. This seems not physically reasonable, but it has more meaning if we suppose to have performed the Wick rotation in space-time, so that we work in a euclidean setting. In this setting we loose the physical interpretation of the D-brane world-volume as a volume moving in time and of the charge $q$ (actually all the homology class $[q \cdot Y_{p,t}]$ for $Y_{p,t}$ the restriction of the world-volume at an instant $t$ in a fixed reference frame) as a charge conserved in time. Thus, rather than considering the homology class of the D-brane volume at every instant of time, we prefer to consider the homology class of the entire world-volume in $S$, using standard homology with compact support.

\subsection{Classification}

For a D$p$-brane with $(p+1)$-dimensional world-volume $WY_{p}$ and charge $q$ we consider the corresponding homology class in $S$:
\begin{equation}\label{HomologicalClassification}
	[q \cdot WY_{p}] \in H_{p+1}(S, \mathbb{Z}) = \frac{Z_{p+1}(S, \mathbb{Z})}{B_{p+1}(S, \mathbb{Z})} = \mathbb{Z}^{b_{p+1}} \oplus_{i} \mathbb{Z}_{p_{i}^{n_{i}}}
\end{equation}
where $Z_{p+1}(S, \mathbb{Z})$ denotes the group of singular $(p+1)$-cycles of $S$, $B_{p+1}(S, \mathbb{Z})$ the subgroup of $(p+1)$-boundaries, $b_{p+1}$ the $(p+1)$-th Betti number of $S$, and $p_{i}$ is a prime number for every $i$. For what will follow, it is convenient to consider the cohomology of $S$ rather than the homology. Hence, denoting by $\PD_{S}$ the Poincar\'e duality map on $S$,\footnote{As we said above, we are assuming for simplicity that the space-time is a compact manifold (without singularities), and we also suppose it is orientable, thus Poincar\'e duality holds.} we define the \emph{charge density}:
\begin{equation}\label{PDS}
	\PD_{S}[q \cdot WY_{p}] \in H^{9-p}(S, \mathbb{Z}) = \frac{Z^{9-p}(S, \mathbb{Z})}{B^{9-p}(S, \mathbb{Z})} = \mathbb{Z}^{b_{p+1}} \oplus_{i} \mathbb{Z}_{p_{i}^{n_{i}}}
\end{equation}
where $Z^{9-p}(S, \mathbb{Z})$ is the group of singular $(9-p)$-cocyles and $B^{9-p}(S, \mathbb{Z})$ the subgroup of $(9-p)$-coboundaries. This classification encounters some problems due to the presence of quantum anomalies. Two remarkable examples are the following:
\begin{itemize}
	\item a brane wrapping a cycle $WY_{p} \subset S$ is Freed-Witten anomalous if its third integral Stiefel-Whitney class $W_{3}(WY_{p})$ is not zero, hence not all the cycles are allowed \cite{FW, Evslin};
	\item given a world-volume $WY_{p}$ with $W_{3}(WY_{p}) \neq 0$, it can be interpreted as an MMS-instanton in the minkowskian setting \cite{MMS, Evslin}; in this case there are cycles intersecting $WY_{p}$ in $\PD_{WY_{p}}(W_{3}(WY_{p}))$ which, although homologically non-trivial in general, are actually unstable.
\end{itemize}
The two points above imply that:
\begin{itemize}
	\item the numerator $Z_{p+1}(S, \mathbb{Z})$ of \eqref{HomologicalClassification} is too large, since it contains anomalous cycles;
	\item the denominator $B_{p+1}(S, \mathbb{Z})$ of \eqref{HomologicalClassification} is too small, since it does not cut all the unstable charges.
\end{itemize}
There are other possible anomalies, although not yet completely understood, some of which are probably related to homology classes not representable by a smooth submanifold \cite{ES2, BHK, Evslin}.

\paragraph{}We start by considering the case of world-volumes of \emph{even} codimension in $S$, i.e.\ we start with IIB superstring theory. To solve the problems mentioned above, one possible tool seems to be the Atiyah-Hirzebruch spectral sequence \cite{AH}. Choosing a finite simplicial decomposition \cite{Hatcher} of the space-time manifold $S$, and considering the filtration $S = S^{10} \supset \cdots \supset S^{0}$ for $S^{i}$ the $i$-th dimensional skeleton, such a spectral sequence starts from the even-dimensional simplicial cochains of $S$ and, after a finite number of steps, it stabilizes to the graded group $E_{\infty}^{\ev, \,0}(S) = \bigoplus_{2k}K_{2k}(S) / K_{2k+1}(S)$. Here $K_{q}(S)$ is the kernel of the natural restriction map from the K-theory group of $S$, which we denote by $K(S)$, to the K-theory group of $S^{q-1}$, which we call $K(S^{q-1})$: i.e.\ $K_{q}(S) = \Ker(K(S) \rightarrow K(S^{q-1}))$. We also use the notation $E_{\infty}^{2k, \,0}(S) = K_{2k}(S) / K_{2k+1}(S)$, so that $E_{\infty}^{\ev, \,0}(S) = \bigoplus_{2k} E_{\infty}^{2k, \,0}(S)$. We can start from a representative of the Poincar\'e dual of the D-brane $\PD_{S}[q \cdot WY_{p}]$, which in our hypotheses is even-dimensional, and, if it survives until the last step, we arrive at a class $\{\PD_{S}[q \cdot WY_{p}]\} \in K_{9-p}(S) / K_{9-p+1}(S)$. The even boundaries $d_{2}, d_{4}, \ldots$ of this sequence are 0, hence the important ones are the odd boundaries. In particular, one can prove that:
\begin{itemize}
	\item $d_{1}$ coincides with the ordinary coboundary operator, hence the second step is the even cohomology of $S$ \cite{Segal, AH};
	\item the cocycles not living in the kernel of $d_{3}$ are Freed-Witten anomalous, while the cocycles contained in its image are unstable because of the presence of MMS-instantons \cite{Evslin, MMS}.
\end{itemize}
As we will say in a while, there are good reasons to use K-theory to classify D-brane charges, hence, although the physical meaning of higher order boundaries is not completely clear, the behavior of $d_{3}$ and the fact that the last step is directly related to K-theory suggest that the class $\{\PD_{S}[q \cdot WY_{p}])\} \in E_{\infty}^{9-p,0}(S)$ is a good candidate to be considered as the charge of the D-brane. Summarizing, we saw two ways to classify D-brane cycles and charges:
\begin{itemize}
	\item the homological classification, i.e.\ $[q \cdot WY_{p}] \in H_{p+1}(S, \mathbb{Z})$;
	\item the classification via the Atiyah-Hirzebruch spectral sequence, i.e.\ $\{\PD_{S}[q \cdot WY_{p}]\} \in E_{\infty}^{9-p, \,0}(S)$.
\end{itemize}

\subsection{K-theory from the Sen conjecture}

\subsubsection{Gauge and gravitational couplings}

Up to now we have only considered the cycle wrapped by the D-brane world-volume. There are other important features: the gauge bundle and the embedding in space-time, which enter in the action via the two following couplings:
\begin{itemize}
	\item the gauge coupling through the Chern character \cite{LM} of the Chan-Paton bundle;
	\item the gravitational coupling through the $\hat{A}$-genus \cite{LM} of the tangent and the normal bundle of the world-volume.
\end{itemize}
The unique non-anomalous form of these couplings, computed by Minasian and Moore in \cite{MM}, is:
\begin{equation}\label{action}
	S = \int_{WY_{p}} \textstyle i^{*}C \wedge \ch(E) \wedge e^{\frac{d}{2}} \wedge \frac{\,\sqrt{\hat{A}(T(WY_{p}))}\,} {\sqrt{\hat{A}(N(WY_{p}))}}
\end{equation}
where $i: WY_{p} \rightarrow S$ is the embedding, $E$ is the Chan-Paton bundle, $T(WY_{p})$ and $N(WY_{p})$ are the tangent bundle and the normal bundle of $WY_{p}$ in $S$, and $d \in H^{2}(WY_{p}, \mathbb{Z})$ is a class whose restriction mod 2 is the second Stiefel-Whitney class of the normal bundle $w_{2}(N(WY_{p}))$. The polyform that multiplies $i^{*}C$ has 0-term equal to $\ch_{0}(E) = \rk(E)$, hence \eqref{action} is an extension of the usual minimal coupling $q\int_{WY_{p}} i^{*}C_{p+1}$ for $q = \rk(E)$: the charge of the D-brane coincides with the rank of the gauge bundle (up to a normalization constant). In the case of anti-branes, we have to allow for negative charges, hence the gauge bundle is actually a \emph{K-theory class}: a generic class $E - F$ can be interpreted as a stack of pairs of a brane $Y$ and an anti-brane $\overline{Y}$ with gauge bundle $E$ and $F$ respectively. For $i_{\#}: H^{*}(WY_{p}, \mathbb{Q}) \rightarrow H^{*}(S, \mathbb{Q})$ the Gysin map in cohomology \cite{Hirzebruch, OS}, we define the \emph{charge density}:
\begin{equation}\label{QWY}
	Q_{WY_{p}} = i_{\#} \Bigl( \ch(E) \wedge e^{\frac{d}{2}} \wedge \textstyle\frac{\,\sqrt{\hat{A}(T(WY_{p})})\,} {\sqrt{\hat{A}(N(WY_{p}))}} \Bigr).
\end{equation}
Since new terms have appeared in the charge, we should discuss also their quantization, which is not immediate since the Chern character and the $\hat{A}$-genus are intrinsically rational cohomology classes. To avoid the discussion of these problems \cite{MW}, in the expression \eqref{action} we suppose $C$ to be globally defined, which implies that the field strength $G = dC$ is trivial in the de-Rham cohomology at any degree.\footnote{Actually the assumption that $C$ is globally defined does not solve the problem, since one should take into account the large gauge transformations $C_{p+1} \rightarrow C_{p+1} + \Phi_{p+1}$ with $\Phi_{p+1}$ integral but not necessarily exact. It turns out that the action \eqref{action} is well-defined under these gauge transformations only under the suitable quantization conditions we have mentioned above. Anyway, for a fixed global $C_{p+1}$ formula \eqref{action} is meaningful, and this is enough for our purposes here.} For a general discussion see \cite{Freed2}.

\paragraph{} We put for notational convenience:
	\[\textstyle G(WY_{p}) = e^{\frac{d}{2}} \wedge \frac{\,\sqrt{\hat{A}(T(WY_{p}))}\,} {\sqrt{\hat{A}(N(WY_{p}))}}.
\]
The action \eqref{action} is equal to:
	\[S = \int_{\PD_{WY_{p}}(\ch(E))} i^{*}C \wedge G(WY_{p}).
\]
Let $\{q_{k} \cdot WY_{k}\}$ be the set of D-branes appearing in the Poincar\'e dual of $\ch(E)$ in $WY_{p}$ (we mean that we choose a representative cycle for each homology class in $\PD_{WY_{p}}(\ch(E))$ and we think of it as a subbrane of $WY_{p}$): the first one is $\PD_{WY_{p}}(\ch_{0}(E)) = q \cdot WY_{p}$, so it gives rise to the action without gauge coupling. The other ones are lower dimensional branes. Let us consider the first one, i.e.\ $WY_{(1)} = \PD_{WY_{p}}(\ch_{1}(E))$. Then the corresponding term in the action is $\int_{WY_{(1)}} i^{*}C \wedge G(WY_{p})$, which can be written as $\int_{WY_{(1)}} i^{*}C \wedge G(WY_{(1)}) + \int_{WY_{(1)}} i^{*}C \wedge (G(WY_{p}) - G(WY_{(1)}))$. Since in the second term the sum $G(WY_{p}) - G(WY_{(1)})$ has 0-term equal to $0$, then $\PD_{WY_{(1)}} (G(WY_{p}) - G(WY_{1}))$ is made only by lower-dimensional subbranes. Let $WY_{(1,1)}$ be the first one: we get $\int_{WY_{(1,1)}} i^{*}C$, which is equal to $\int_{WY_{(1,1)}} i^{*}C \wedge G(WY_{(1,1)}) + \int_{WY_{(1)}} i^{*}C \wedge (1 - G(WY_{(1,1)}))$. The second term gives rise only to lower dimensional subbranes. Proceeding inductively until we arrive at D0-branes, whose $G$-term is $1$, we can write:
	\[\int_{WY_{(1)}} i^{*}C \wedge G(WY_{p}) = \sum_{h=0}^{m} \int_{WY_{(1,h)}} i^{*}C \wedge G(WY_{(1,h)})
\]
where, for $h = 0$, $WY_{(1, 0)} = WY_{(1)}$ holds. Proceeding in the same way for every $WY_{(k)}$, we obtain a set of subbranes $\{q_{k,h} \cdot WY_{(k,h)}\}$, which, using only one index, we still denote by $\{q_{k} \cdot WY_{(k)}\}$. Therefore, calling $i_{k}: WY_{(k)} \rightarrow S$ the embedding, we get:
	\[S = \sum_{k} \int_{WY_{(k)}} i_{k}^{*}C \wedge G(WY_{(k)}).
\]
From this expression we see that \emph{the brane $WY_{p}$ with gauge and gravitational couplings is equivalent to the set of sub-branes $WY_{(k)}$ with trivial gauge bundle}. Moreover we now show that the following equality holds:
\begin{equation}\label{SplittingCharge}
	i_{\#} \bigl( \ch(E) \wedge G(WY_{p}) \bigr) = \sum_{k} (i_{k})_{\#} \,G(WY_{(k)})
\end{equation}
i.e.\ the charge densities of the two configurations are the same. In order to prove this, we recall the formulas:
\begin{equation}\label{FormulasGysin}
\begin{split}
	&i_{\#}(\alpha \wedge i^{*}\beta) = i_{\#}(\alpha) \wedge \beta\\
	&\int_{WY_{p}} \alpha = \int_{S} i_{\#}(\alpha)
\end{split}
\end{equation}
for $\alpha \in H^{*}(WY_{p}, \mathbb{Q})$ and $\beta \in H^{*}(S, \mathbb{Q})$. Thus:
	\[\begin{split}
	&\int_{WY_{p}} i^{*}C \wedge \ch(E) \wedge G(WY_{p}) = \int_{S} i_{\#} \bigl[ i^{*}C \wedge \ch(E) \wedge G(WY_{p}) \bigr]\\
	&\qquad\qquad\qquad = \int_{S} C \wedge i_{\#} \bigl( \ch(E) \wedge G(WY_{p}) \bigr)\\
	&\sum_{k} \int_{WY_{p}} i_{k}^{*}C \wedge G(WY_{(k)}) = \sum_{k} \int_{S} (i_{k})_{\#} \bigl[ i_{k}^{*}C \wedge G(WY_{(k)}) \bigr]\\
	&\qquad\qquad\qquad = \int_{S} C \wedge \sum_{k} (i_{k})_{\#} \bigl( G(WY_{(k)}) \bigr).
\end{split}\]
Since the two terms are equal for every form $C$, we get formula \eqref{SplittingCharge}. We thus get:
\begin{quote}
\textbf{Splitting principle:} \emph{a D-brane $WY_{p}$ with gauge bundle is dynamically equivalent to a set of sub-branes $WY_{(k)}$ with trivial gauge bundle, such that the total charge density of the two configurations is the same.}
\end{quote}
The physical interpretation of this conjecture is the phenomenon of tachyon condensation \cite{Sen,Witten,Evslin}: the quantization of strings extending from a brane to an antibrane leads to a tachyonic mode, which represents an instability and generates a process of annihilation of brane and antibrane world-volumes via an RG-flow \cite{APS}, leaving lower dimensional branes. In particular, given a D-brane $WY_{p}$ with gauge bundle $E \rightarrow WY_{p}$, we can write $E = (E - \rk\, E) + \rk\, E$, so that $E - \rk\, E \in \tilde{K}(WY_{p})$, where $\tilde{K}(WY_{p})$ is the reduced K-theory group of $WY_{p}$ \cite{Atiyah}: thus we think of this configuration as a triple made by a D-brane $WZ_{p}$ with gauge bundle $\rk\, E$, a brane $WY_{p}$ with gauge bundle $E$ and an antibrane $\overline{WZ}_{p}$ with gauge bundle $\rk\, E$. By tachyon condensation only $WZ_{p}$ remains (with trivial bundle, i.e.\ only with its own charge), while $WY_{p}$ and $\overline{WZ}_{p}$ annihilate giving rise to lower dimensional branes with trivial bundle, as stated in the splitting principle. Moreover, if we consider a stack of pairs $(WY_{p}, \overline{WY}_{p})$ with gauge bundles $E$ and $F$ respectively, this is equivalent to consider gauge bundles $E \oplus G$ and $F \oplus G$ respectively, since, viewing the factor $G$ as a stack of pairs $(WZ_{p}, \overline{WZ}_{p})$ with the \emph{same} gauge bundle, it happens that by tachyon condensation $WZ_{p}$ and $\overline{WZ}_{p}$ disappear, leaving no other subbranes. This is the physical interpretation of the stable equivalence relation in K-theory. This principle, as we will see, is an inverse of the Sen conjecture, but we will actually use it to show the Sen conjecture in this setting.

\paragraph{Remark:} the splitting principle holds only at rational level, since it involves Chern characters and $\hat{A}$-genus. At integral level, we do not state such a principle.

\subsubsection{K-theory}

Since we are assuming the $H$-flux to vanish, in order not to be Freed-Witten anomalous the D-brane must be spin$^{c}$. Since the whole space-time is spin, in particular also spin$^{c}$, it follows that the normal bundle of the D-brane is spin$^{c}$ too. Therefore we can consider the K-theory Gysin map $i_{!}: K(WY_{p}) \rightarrow K(S)$ \cite{LM}. We recall the differentiable Riemann-Roch theorem \cite{Hirzebruch, OS}:
\begin{equation}\label{RiemannRoch}
	\ch(i_{!}(E)) \wedge \hat{A}(TS) = i_{\#}\bigl(\ch(E) \wedge e^{\frac{d}{2}} \wedge \hat{A}(T(WY_{p}))\bigr).
\end{equation}
Using \eqref{RiemannRoch} and \eqref{FormulasGysin} we obtain:
\[\begin{split}
	\int_{WY_{p}} i^{*}C \wedge &\,\ch(E) \wedge e^{\frac{d}{2}} \wedge \textstyle \frac{\,\sqrt{\hat{A}(T(WY_{p}))}\,} {\sqrt{\hat{A}(N(WY_{p}))}} \displaystyle = \int_{S} C \wedge \ch\bigl(i_{!}(E)\bigr) \wedge \sqrt{\hat{A}(TS)}.
\end{split}\]
Thus we get:
	\[S = \int_{S} C \wedge \ch\bigl(i_{!}(E)\bigr) \wedge \sqrt{\hat{A}(TS)}
\]
hence:
\begin{equation}\label{QWYGysin}
	Q_{WY_{p}} = \ch(i_{!}E) \wedge \sqrt{\hat{A}(TS)}.
\end{equation}
In this way, \eqref{QWYGysin} is another expression for $Q_{WY_{p}}$ with respect to \eqref{QWY}, but with an important difference: the $\hat{A}$-factor does not depend on $WY_{p}$, hence all $Q_{WY_{p}}$ is a function only of $E$. Thus, we can consider $i_{!}E$ as the K-theory analogue of the charge density, considered as an \emph{integral} K-theory class. The use of Chern characters, instead, obliges to consider rational classes, which cannot contain information about the torsion part.
 
\subsubsection{Sen conjecture}

Let us consider the two expressions found for the rational charge density:
	\[\begin{split}
	&Q^{(1)}_{WY_{p}} = i_{\#} \bigl( \ch(E) \wedge G(WY_{p}) \bigr) \\
	&Q^{(2)}_{WY_{p}} = \ch(i_{!}E) \wedge \sqrt{\hat{A}(TS)}.
\end{split}\]
$Q^{(2)}_{WY_{p}}$ is exactly the charge density of a stack of D$9$-branes and anti-branes (whose world-volume coincides with $S$), whose gauge bundle is the K-theory class $i_{!}E$. Hence, expressing the charge in the form $Q^{(2)}_{WY_{p}}$ for each D-brane in our background is equivalent to think that there exists only one stack of couples brane-antibrane of dimension $9$ encoding all the dynamics. Hence we formulate the conjecture \cite{Sen,Witten}:
\begin{quote}
\textbf{Sen conjecture:} \emph{every configuration of branes and anti-branes with any gauge bundle is dynamically equivalent to a configuration with only a stack of coincident pairs brane-antibrane of dimension $9$ with an appropriate K-theory class on it.}
\end{quote}
In order to see that the dynamics is actually equivalent, we use the splitting principle stated above: since $Q^{(1)}_{WY_{p}} = Q^{(2)}_{WY_{p}}$, the brane $WY_{p}$ with the charge $Q^{(1)}_{WY_{p}}$ and the D$9$-brane with charge $Q^{(2)}_{WY_{p}}$ split into the same set of subbranes (with trivial gauge bundle). We remark that in order to state the Sen conjecture is necessary that the $H$-flux vanishes. Indeed, the space-time is spin$^{c}$ (it is spin since space-time spinors exist, therefore also spin$^{c}$), hence Freed-Witten anomaly cancellation for D$9$-branes requires that $H = 0$. Actually, an appropriate stack of D$9$-branes can be consistent for $H$ a torsion class \cite{Kapustin}, but we do not consider this case here.

\paragraph{}In order to formulate both the splitting principle and the Sen conjecture, we have only considered the action, hence only \emph{rational} classes given by Chern characters and $\hat{A}$-genus. Thus, we can classify the charge density in the two following ways:
\begin{itemize}
	\item as a rational cohomology class $i_{\#} (\ch(E) \wedge G(WY_{p})) \in H^{\ev}(S, \mathbb{Q})$;
	\item as a rational K-theory class $i_{!}E \in K_{\mathbb{Q}}(S) := K(S) \otimes_{\mathbb{Z}} \mathbb{Q}$.
\end{itemize}
These two classification schemes are completely equivalent due to the fact that the map:
	\[\ch(\,\cdot\,) \wedge \sqrt{\hat{A}(TS)}:\; K_{\mathbb{Q}}(S) \longrightarrow H^{\ev}(S, \mathbb{Q})
\]
is an isomorphism. This equivalence is lost at the integral level, since the torsion parts of $K(S)$ and $H^{\ev}(S, \mathbb{Z})$ are in general different. Moreover, since at the integral level the splitting principle does not apply, we cannot prove that the Sen conjecture holds: the classification via Gysin map and cohomology are different, and the use of the Gysin map is just \emph{suggested} by the equivalence at rational level, i.e.\ by the equivalence of the dynamics.

Moreover, for the integral case, we have also seen the classification via the Atiyah-Hirzebruch spectral sequence (AHSS). In the rational case, we can build the corresponding sequence AHSS$_{\mathbb{Q}}$ \cite{AH}, ending at the groups $Q_{\infty}^{\ev, \,0}(S)$, but it stabilizes at the second step, i.e.\ at the level of cohomology. Hence, the class $\{i_{\#}(\ch(E) \wedge G(WY_{p}))\} \in Q^{\ev,\,0}_{\infty}(S)$ is completely equivalent to the cohomology class $i_{\#}(\ch(E) \wedge G(WY_{p})) \in H^{\ev}(S, \mathbb{Q})$.

\subsection{Linking the classifications}

To summarize, we are trying to classify the charges of D-branes in a compact euclidean space-time $S$. In order to achieve this, we can use cohomology or K-theory, with integer or rational coefficients, obtaining the possibilities showed in table \ref{fig:Classifications}.
\begin{table*}[h] \footnotesize
	\centering
		\begin{tabular}{|l|l|l|}
			\hline & & \\ & Integer & Rational \\ & & \\ \hline
			& & \\ Cohomology & $\PD_{S}[q \cdot WY_{p}] \in H^{9-p}(S, \mathbb{Z})$ & $i_{\#}\bigl(\ch(E) \wedge G(WY_{p})\bigr) \in H^{\ev}(S, \mathbb{Q})$ \\ & & \\ \hline
			& & \\ K-theory (Gysin map) & $i_{!}(E) \in K(S)$ & $i_{!}(E) \in K_{\mathbb{Q}}(S)$ \\ & & \\ \hline
			& & \\ K-theory (AHSS) & $\{\PD_{S}[q\cdot WY_{p}]\} \in E^{9-p,\,0}_{\infty}(S)$ & $\bigl\{i_{\#}(\ch(E) \wedge G(WY_{p}))\bigr\} \in Q^{\ev,\,0}_{\infty}(S)$ \\ & & \\ \hline
		\end{tabular}
\caption{Classifications}\label{fig:Classifications}
\end{table*}

In the rational case, as we have seen, there is a complete equivalence of the three approaches, since the three groups we consider, i.e.\ $\bigoplus_{2k}H^{2k}(S, \mathbb{Q})$, $K_{\mathbb{Q}}(S)$ and $\bigoplus_{2k}Q^{2k,\,0}_{\infty}(S)$ are canonically isomorphic. Instead, in the integral case there are no such isomorphisms (in general the three groups are all different), and there is a strong asymmetry due to the fact that in the homological and AHSS classifications \emph{the gauge bundle and the gravitational coupling are not considered at all}, while they are of course taken into account in the Gysin map approach. Up to now we discussed the case of even-codimensional branes: that is because the Gysin map requires an even-dimensional normal bundle in order to take value in $K(S)$. We will discuss also the odd-dimensional case, considering the group $K^{1}(S)$, and the picture will be similar.

Since the integral approaches are not equivalent, we have to investigate the relations among them: it is clear how to link the cohomological class and the AHSS class, since the second step of AHSS is exactly the cohomology. Having seen the link between the Gysin map and the Atiyah-Hirzebruch spectral sequence, we can also link these two approaches. In fact, let us consider a D$p$-brane world-volume $WY_{p} \subset S$ with gauge bundle $E \rightarrow WY_{p}$ of rank $q$, and let $i: WY_{p} \hookrightarrow S$ be the embedding. We have proven that $i_{!}(E) \in \Ker (K^{9-p}(S) \rightarrow K^{9-p}(S^{8-p}))$ and that:
	\[\{\PD_{S}[q \cdot WY_{p}]\}_{E^{9-p, \,0}_{\infty}} = [i_{!}(E)].
\]
Thus, we can use AHSS to detect possible anomalies, then we can use the Gysin map to get the charge of a non-anomalous brane: such a charge belongs to the equivalence class reached by AHSS, so that the Gysin map gives richer information. Some comments are in order. One could ask why the additional information provided by the Gysin map has to be considered: in fact, we have proven that it concerns the choice of a representative of the class, while, discussing AHSS in chapter 2, we have seen that one of its advantages is that it quotients out unstable configurations. It seems that such additional information keeps into account only instabilities. Actually, this is not the case. Let us consider a couple $(WY_{p}, i_{!}(E))$ made by a D-brane world-volume and its charge with respect to the Gysin map approach. The charge does not provide complete information about the world-volume, since $i_{!}E$ is a class in the whole space-time, exactly as the charge $q$ of a particle does not provide information about its trajectory. This is true also for the cohomological and AHSS classifications: two homologous world-volumes are not the same trajectory. If we consider two couples $(WY_{p}, i_{!}(E))$ and $(WY_{p}, i_{!}(F))$, we know that $[i_{!}(E) - i_{!}(F)]_{E^{9-p, \,0}_{\infty}} = 0$, which means that $i_{!}(E) - i_{!}(F)$ lies in the image of some boundaries of AHSS. Let us suppose that it lies in the image of $d_{3}$. This means that there exists an unstable world-volume $WU_{p}$ with a gauge bundle, e.g.\ the trivial one, such that $i_{!}(WU_{p} \times \mathbb{C}) = i_{!}(E) - i_{!}(F)$, but the two terms of the latter equality concern different world-volumes with the same zero charge: in fact, $WU_{p}$ has charge $0$ because it lies in the image of $d_{3}$, while $i_{!}(E-F)$ has charge $0$ since, being $\rk(E - F) = 0$, it is a representative of the class reached starting from $0\cdot WY_{p}$. Anyway, the world-volume $WY_{p}$ is not anomalous in general and the fact that the gauge bundle on it is $E$ or $F$ is a meaningful information. Actually the information contained in $i_{!}(E-F)$ is partially contained in the charges of the sub-branes of $WY_{p}$. Thus, we can apply AHSS to the world-volume of the D-brane, then, if it corresponds to the trivial class we consider it as an unstable one, otherwise we can consider each representative of the class as an additional meaningful information.

\section{D-brane charges and K-theory - Part II}

We now rediscuss the topics of section \ref{DBranesKTheoryCpt} without assuming that the space-time is compact. We consider $S = \mathbb{R}^{1,3} \times X$ for $X$ compact. We will consider later also the case of $X$ not compact. In a fixed reference frame, we call $M = \mathbb{R}^{3} \times X$ the space manifold. We assume the background metric $\eta^{\mu\nu}$ to be the classical Minkowskian one on $\mathbb{R}^{1,3}$.

\subsection{Review of the cohomological description}

To calculate the charge $q$ of D-brane, we think of the latter as a magnetic monopole, i.e.\ as a source for violation of the Bianchi identity of the associated magnetic Ramond-Ramond field strength. For the $p$-brane world-volume $WY_{p}$ let us consider the corresponding \emph{magnetic} field strength $G_{8-p}$ and \emph{electric} field strength $G_{p+2} = *G_{8-p}$. The violated Bianchi identity for the world-volume $WY_{p}$ is:
\begin{equation}\label{BianchiId2}
	dG_{8-p} = \delta(q \cdot WY_{p}).
\end{equation}
In cohomology we get:
	\[[dG_{8-p}] = \PD_{BM(S, (\pi_{\mathbb{R}^{1,3}}(WY_{p}))^{\bot}, \pi)}(q \cdot WY_{p}).
\]
For every fixed instant $t$, we put $M_{t} = \{t\} \times M$ and $Y_{p,t} = WY_{p} \cap M_{t}$. Then:
	\[[d_{M_{t}}(G_{8-p}^{s}\vert_{M_{t}})]_{\cpt} = \PD_{M_{t}}(q\cdot Y_{p,t}).
\]
We now consider a linking $(8-p)$-cycle $L$ of $WY_{p}$ in $S$ with linking number $l$: by definition, there exists a $(9-p)$-chain $B$ such that $\#(B, Y) = l$ and $L = \partial B$. Then \cite{Polchinski}:
\begin{equation}\label{Charge}
q = \frac{1}{l} \int_{B} \delta(q\cdot WY_{p}) = \frac{1}{l} \int_{B} dG_{8-p} = \frac{1}{l} \int_{L} G_{8-p}.
\end{equation}
This is the way we recover $q$ from the background data. The case $q < 0$ corresponds to \emph{anti-branes}. The charge $q$ is conserved in time, actually all the homology class of the D-brane is conserved. In fact, in a fixed reference frame, let us consider two volumes $Y_{p,t_{1}}$ and $Y_{p,t_{2}}$. Then we can consider the piece of the world-volume linking them, which is $(WY_{p})\vert_{[t_{1}, t_{2}] \times M}$. If we consider the canonical identification $M_{t_{1}} \simeq M_{t_{2}} \simeq M$, we can consider both $Y_{t_{1}}$ and $Y_{t_{2}}$ as cycles in $M$. If we consider the projection $\pi: [t_{1}, t_{2}] \times M \rightarrow M$, then $\pi((WY_{p})\vert_{[t_{1}, t_{2}] \times M})$ is a singular chain in $M$ which makes $Y_{t_{1}}$ and $Y_{t_{2}}$ homologous. Thus they have the same Poincar\'e dual and they define the same charge.

\subsection{K-theoretical description}

In \cite{FR2}, within the standard homological description of D-brane charges, we described the D-brane charge density as a class:
	\[[\,q \cdot WY_{p}\,] \in H^{BM\left((\pi_{\mathbb{R}^{1,3}}(WY_{p}))^{\bot}, \pi\right)}_{p+1}(\mathbb{R}^{1,3} \times X)
\]
leading, for a fixed reference frame, to a charge conserved in time:
	\[[\,q \cdot Y_{p,t}\,] \in H^{BM\left((\pi_{\mathbb{R}^{3}}(Y_{p,t}))^{\bot}, \pi\right)}_{p}(\mathbb{R}^{3} \times X).
\]
We can describe analogously the D-brane charge within the K-theoretical picture. In particular we define the D-brane charge density as:
	\[[(WY_{p}, E, i)] \in K^{BM\left((\pi_{\mathbb{R}^{1,3}}(WY_{p}))^{\bot}, \pi\right)}_{p+1}(\mathbb{R}^{1,3} \times X)
\]
and we can show that, for a fixed reference frame, it leads to a charge conserved in time:
	\[[(Y_{p,t}, E_{t}, i_{t})] \in K^{BM\left((\pi_{\mathbb{R}^{3}}(Y_{p,t}))^{\bot}, \pi\right)}_{p}(\mathbb{R}^{3} \times X).
\]
The fact that it is conserved in time depends on the fact that if we fix two volumes $Y_{p,t_{1}}$ and $Y_{p,t_{2}}$ we can consider the piece of the world-volume linking them, which is $(WY_{p})\vert_{[t_{1}, t_{2}] \times M}$. If we consider the canonical identification $M_{t_{1}} \simeq M_{t_{2}} \simeq M$, we can see both $Y_{t_{1}}$ and $Y_{t_{2}}$ as cycles in $M$. If we consider the projection $\pi: [t_{1}, t_{2}] \times M \rightarrow M$, then $(WY_{p}\vert_{[t_{1}, t_{2}] \times M}, E\vert_{[t_{1}, t_{2}] \times M}, \pi \circ i\vert_{[t_{1}, t_{2}] \times M})$ makes $Y_{t_{1}}$ and $Y_{t_{2}}$ K-homologous in $M$.

We remark that K-homology is 2-periodic, thus only the parity of the gradation index is relevant. We also remark that the class conserved in time belongs to standard K-homology if the D-brane is a particle in the non-compact space-time directions $\mathbb{R}^{1,3}$.

\paragraph{}We can now consider the corresponding cohomological version, involving in particular K-theory. Actually the approach via Gysin map turns out to be equivalent to the one via K-homology thanks to what we said in subsection \ref{KHomologyGysinMap}. In particular, we define the charge density simply applying the Poincar\'e dual, which we have seen to correspond to the Gysin map. Thus we get the charged density:
	\[[i_{!}E] \in K_{\cpt\left((\pi_{\mathbb{R}^{1,3}}(WY_{p}))^{\bot}, \pi\right)}^{9-p}(\mathbb{R}^{1,3} \times X)
\]
and the charge conserved in time:
	\[[(i_{t})_{!}(E_{t})] \in K_{\cpt\left((\pi_{\mathbb{R}^{3}}(Y_{p,t}))^{\bot}, \pi\right)}^{10-p}(\mathbb{R}^{3} \times X).
\]
The latter is conserved since it is the Poincar\'e dual of a conserved K-homology class. This does not seem so evident if we consider directly the K-theoretical approach without referring to K-homology. Moreover, we immediately see that charges in IIB theory are classified by $K^{0}(S)$ and in IIA theory by $K^{1}(S)$: the K-homology cycle has naturally the same dimension of the D-brane world-volume (since the gauge bundle lives in $K^{0}(WY_{p})$, so the dimension of the cycle is the one of $WY_{p}$), thus by Poincar\'e duality and considering the periodicity we get the analogous result for K-theory. We remark that the class conserved in time belong to usual K-theory if the D-brane is space-filling, which is actually a particular case by this point of view. For this case, the same considerations of \cite{FR2} apply.

\paragraph{}We now consider AHSS. In this case, since to have finite convergence of the spectral sequence we need to work with a finite CW-complex, we consider the one-point compactification in the direction of the brane in $\mathbb{R}^{1,3}$ and the disc compactification in the orthogonal directions, so that the brane becomes a non-trivial cycle with the suitable compactness hypothesis. Thus, calling $V = (\pi_{\mathbb{R}^{1,3}}(WY_{p}))^{\bot}$ we get the charge density:
	\[\{\PD_{S}[q\cdot WY_{p}]\} \in E^{9-p,\,0}_{\infty}(V^{+} \times \overline{V^{\bot}} \times X)
\]
and, calling $W = (\pi_{\mathbb{R}^{3}}(Y_{p,t}))^{\bot}$, the charge conserved in time:
	\[\{\PD_{S}[q\cdot Y_{p,t}]\} \in E^{9-p,\,0}_{\infty}(W^{+} \times \overline{W^{\bot}} \times X).
\]
The same relation between Gysin map and AHSS holds.

\paragraph{}We can now show the analogous table of \ref{fig:Classifications} without assuming compactness. In particular, we get table \ref{fig:Classifications2} for the world-volume charge and table \ref{fig:Classifications3} for the charge conserved in time. In any case the charge conserved in time is the restriction at a fixed instant of the world-volume charge, and the link between the three charges is analogous to the one we found assuming compactness of space-time. \newpage
\begin{table*}[h] \footnotesize
	\centering
		\begin{tabular}{|l|l|}
			\hline & \\ & Integer \\ & \\ \hline
			& \\ Cohomology & $\PD^{BM\left((\pi_{\mathbb{R}^{1,3}}(WY_{p}))^{\bot}, \pi\right)}_{S}[q \cdot WY_{p}] \in H^{9-p}_{\cpt\left((\pi_{\mathbb{R}^{1,3}}(WY_{p}))^{\bot}, \pi\right)}(S, \mathbb{Z})$ \\ & \\ \hline
			& \\ K-theory (Gysin map) & $i_{!}(E) \in K^{9-p}_{\cpt\left((\pi_{\mathbb{R}^{1,3}}(WY_{p}))^{\bot}, \pi\right)}(S)$\\ & \\ \hline
			& \\ K-theory (AHSS) & $\{\PD[q\cdot WY_{p}]\} \in E^{9-p,\,0}_{\infty}((\pi_{\mathbb{R}^{1,3}}(WY_{p}))^{+} \times \overline{(\pi_{\mathbb{R}^{1,3}}(WY_{p}))^{\bot}} \times X)$ \\ & \\ \hline
		\end{tabular}
\caption{World-volume charge}\label{fig:Classifications2}
\end{table*}
\begin{table*}[h] \footnotesize
	\centering
		\begin{tabular}{|l|l|}
			\hline & \\ & Integer \\ & \\ \hline
			& \\ Cohomology & $\PD^{BM\left((\pi_{\mathbb{R}^{3}}(Y_{p,t}))^{\bot}, \pi\right)}_{S}[q \cdot Y_{p,t}] \in H^{9-p}_{\cpt\left((\pi_{\mathbb{R}^{3}}(Y_{p,t}))^{\bot}, \pi\right)}(M, \mathbb{Z})$ \\ & \\ \hline
			& \\ K-theory (Gysin map) & $(i_{t})_{!}(E_{t}) \in K^{9-p}_{\cpt\left((\pi_{\mathbb{R}^{3}}(Y_{p,t}))^{\bot}, \pi\right)}(M)$\\ & \\ \hline
			& \\ K-theory (AHSS) & $\{\PD[q\cdot WY_{p}]\} \in E^{9-p,\,0}_{\infty}((\pi_{\mathbb{R}^{3}}(Y_{p,t}))^{+} \times \overline{(\pi_{\mathbb{R}^{3}}(Y_{p,t}))^{\bot}} \times X)$ \\ & \\ \hline
		\end{tabular}
\caption{Charge conserved in time}\label{fig:Classifications3}
\end{table*}


\part{Pinors and spinors}

\chapter{General theory}

\section{General definitions}

\subsection{Clifford algebras}

\begin{Def} Given a real vector space with a bilinear form $(V, \langle \cdot, \cdot \rangle)$ its \emph{Clifford algebra} $\Cl(V, \langle \cdot, \cdot \rangle)$ is obtained from the free unit algebra generated by $V$ and $1$ quotienting by the relations:
	\[\{v, w\} = -2\langle v, w \rangle.
\]
\end{Def}
We denote:
\begin{itemize}
\item $\Cl(n) = \Cl(\mathbb{R}^{n}, g)$ with the usual euclidean scalar product $g$;
\item $\Cl(p,q) = \Cl(\mathbb{R}^{p+q}, \eta^{p,q})$ with $\eta^{p,q}$ the standard form of signature $(p, q)$. In particular, $\Cl(1, n-1) = \Cl(\mathbb{R}^{n}, \eta)$ with $\eta$ the usual minkowskian metric.
\end{itemize}
Given a Clifford algebra $\Cl(V)$, we have a natural embedding $V \subset \Cl(V)$, and its image generates $\Cl(V)$. Hence we have the following splitting:
\begin{equation}\label{Splitting}
	\Cl(V) = \Cl^{0}(V) \oplus \Cl^{1}(V)
\end{equation}
where $\Cl^{0}(V)$ is the subalgebra generated by products of an even number of vectors and $\Cl^{1}(V)$ is the subspace (not subalgebra!) generated by products of an odd number of vectors. It is easy to verify that this gives a structure of $\mathbb{Z}_{2}$-graded algebra to $\Cl(V)$. If we define the involution:
	\[\iota: \Cl(V) \longrightarrow \Cl(V)
\]
defined by $\iota(v_{1} \cdots v_{n}) = v_{n} \cdots v_{1}$ and extended by linearity, we have that $\Cl^{0}(V)$ and $\Cl^{1}(V)$ are the eigenspaces of $1$ and $-1$ of $\iota$.

\paragraph{} Given a Clifford algebra $\Cl(V)$, we can consider its complexification:
	\[\CCl(V) = \Cl(V) \otimes_{\mathbb{R}} \mathbb{C}.
\]
The complexified algebra still has a natural splitting:
\begin{equation}\label{CSplitting}
	\CCl(V) = \CCl^{0}(V) \oplus \CCl^{1}(V)
\end{equation}
defined in the same way since the vectors still generate (by products and complex linear combinations) the algebra. Clearly $\CCl^{0}(V)$ and $\CCl^{1}(V)$ are the complexifications of $\Cl^{0}(V)$ and $\Cl^{1}(V)$.

\subsection{Pin and Spin}

\begin{Def} Let $V$ be a real vector space and $\Cl(V)$ its associated Clifford algebra. We define a \emph{unit vector} as a vector of $V \subset \Cl(V)$ with square norm $\pm 1$. Then:
\begin{itemize}
\item $\Pin(V)$ is the subgroup of $(\Cl(V), \cdot)$ whose elements are products of unit vectors and $\pm 1$;
\item $\Spin(V) = \Pin(V) \cap \Cl^{0}(V)$ is the subgroup of $(\Cl(V), \cdot)$ whose elements are $\pm 1$ and \emph{even} products of unit vectors.
\end{itemize}
\end{Def}

We can define an action $\Ad$ of the non-null vectors of $V \subset \Cl(V)$ on $V$ itself and this action naturally extends to $\Pin(V)$ and $\Spin(V)$. In particular, we put:
\begin{equation}\label{Ad}
	\Ad_{v}(w) := - v \cdot w \cdot v^{-1}.
\end{equation}
It is easy to verify that, for non-degenerate bilinear forms, $\Ad_{v}$ is the reflection with respect to the hyperplane $v^{\bot}$, so that in particular $\Ad_{v}(V) = V \, \forall v \in V$. In fact, decomposing $w = \alpha v + \beta v^{\bot}$ and considering that $v^{-1} = -\frac{v}{\norm{v}^{2}}$, we have:
	\[\Ad_{v}(w) := \frac{1}{\norm{v}^{2}}\, v \cdot (\alpha v + \beta v^{\bot}) \cdot v = \frac{1}{\norm{v}^{2}}\, (\alpha v^{3} - \beta v^{2} v^{\bot}) = -\alpha v + \beta v^{\bot}.
\]
Since reflections are generators of orthogonal transformations, we have a natural surjective map:
	\[\pi: \Pin(V) \rightarrow O(V).
\]
$\pi$ is a $2:1$ covering, since the reflection with respect to $v$ or $-v$ is the same operation, i.e.\ $\Ker(\pi) = \pm 1$.
Since a product of an even number of reflections is a rotation and viceversa, $\pi$ restricts to a surjection:
	\[\pi: \Spin(V) \rightarrow SO(V)
\]
which is still a $2:1$ covering with kernel $\pm 1$.

\paragraph{Remark:} the groups $O(V)$ and $SO(V)$ are thought with respect to the fixed metric (or pseudo-metric) $\langle \cdot, \cdot \rangle$.

\subsection{The structure of complex Clifford algebras}

We start from the structure of the complexified Clifford algebras, since it is particularly simple. First of all, it is clear that $\CCl(p, q) \simeq \CCl(p+q)$, since $\CCl(p, q)$ is canonically isomorphic to the free algebra obtained from the complex vector space $\mathbb{C}^{p+q}$ with the scalar product extended by linearity\footnote{Thus the extension is a bilinear form, not an hermitian product.} and with relations $\{v, w\} = -2\langle v, w \rangle$: since the extended product depends up to isomorphism only on the rank $p+q$, we get $\CCl(p, q) \simeq \CCl(p+q)$.

\begin{Lemma}\label{PeriodicityC} There is an isomorphism:
	\[\CCl(n+2) \simeq \CCl(n) \otimes_{\mathbb{C}} \CCl(2)
\]
given in the following way:
\begin{itemize}
	\item we consider on $\mathbb{C}^{n}$ the $\mathbb{C}$-linear extension of the \emph{euclidean} scalar product of $\mathbb{R}^{n}$;
	\item we fix $\{e_{1}, \ldots, e_{n}\}$ an orthonormal basis of $\mathbb{C}^{n}$, $\{e'_{1}, e'_{2}\}$ an orthonormal basis of $\mathbb{C}^{2}$ and $\{f_{1}, \ldots, f_{n}, f_{n+1}, f_{n+2}\}$ an orthonormal basis of $\mathbb{C}^{n+2}$;
\end{itemize}
then an isomorphism is:
	\[\begin{array}{lcl}
	1 & \rightarrow & 1\\
	f_{1} & \rightarrow & e_{1} \otimes ie'_{1}e'_{2}\\
	& \vdots & \\
	f_{n} & \rightarrow & e_{n} \otimes ie'_{1}e'_{2}\\
	f_{n+1} & \rightarrow & 1 \otimes e'_{1}\\
	f_{n+2} & \rightarrow & 1 \otimes e'_{2}.
\end{array}\]
\end{Lemma}
\textbf{Proof:} It is easy to prove that the images of the $f_{j}$'s are linearly independent in $\CCl(n) \otimes_{\mathbb{C}} \CCl(2)$. We now verify the behaviour with respect to the product. We first notice that $(ie'_{1}e'_{2})^{2} = -e'_{1}e'_{2}e'_{1}e'_{2} = (e'_{1})^{2} (e'_{2})^{2} = (-1)(-1) = 1$. Thus:
\begin{itemize}
	\item $\{e_{i} \otimes ie'_{1}e'_{2}, e_{j} \otimes ie'_{1}e'_{2}\} = \{e_{i}, e_{j}\} \otimes (ie'_{1}e'_{2})^{2} = -\delta_{ij} \otimes 1 = -\delta_{ij}$;
	\item $\{1 \otimes e'_{i}, 1 \otimes e'_{j}\} = 1 \otimes \{e'_{i}, e'_{j}\} = -\delta_{ij}$;
	\item $\{e_{i} \otimes ie'_{1}e'_{2}, 1 \otimes e'_{j}\} = e_{i} \otimes i(e'_{1}e'_{2}e'_{j} + e'_{j}e'_{1}e'_{2}) = 0$.
\end{itemize}
Thus the linear mapping in the statement can be extended to an algebra homomorphism and, by dimensional reason, it is an isomorphism. $\square$

\begin{Lemma} $ $
\begin{enumerate}
	\item $\CCl(1) \simeq \mathbb{C} \oplus \mathbb{C}$;
	\item $\CCl(2) \simeq M(2, \mathbb{C})$.
\end{enumerate}
\end{Lemma}

\paragraph{Proof:} For the first, $\CCl(1)$ is generated by $1$ and $e_{0}$ with $e_{0}^{2} = -1$. Let us consider the base $v_{0} = \frac{1}{2}(1 + e_{0})$ and $v_{1} = \frac{1}{2}(1 - e_{0})$. Then it is easy to prove that $(\alpha_{0} v_{0} + \alpha_{1} v_{1}) (\beta_{0} v_{0} + \beta_{1} v_{1}) = (\alpha_{0}\beta_{0})v_{0} + (\alpha_{1}\beta_{1})v_{1}$, so that $\CCl(1) \simeq \mathbb{C} \oplus \mathbb{C}$.

For the second, $\CCl(2)$ is generated by $1, e_{0}, e_{1}$ with $e_{i}^{2} = -1$ and $\{e_{0}, e_{1}\} = 0$. Let us consider the following map $\CCl(2) \rightarrow M(2, \mathbb{C})$:
	\[1 \rightarrow \begin{bmatrix} 1 & 0 \\ 0 & 1 \end{bmatrix} \quad e_{0} \rightarrow \begin{bmatrix} i & 0 \\ 0 & i \end{bmatrix} \quad e_{1} \rightarrow \begin{bmatrix} 0 & 1 \\ -1 & 0 \end{bmatrix}.
\]
It is easy to prove that this map is an algebra isomorphism. $\square$

\paragraph{}From the two previous lemmas, we immediately deduce the following theorem:

\begin{Theorem} $ $
\begin{itemize}
	\item $\CCl(2k) \simeq M(2^{k}, \mathbb{C})$;
	\item $\CCl(2k+1) \simeq M(2^{k}, \mathbb{C}) \oplus M(2^{k}, \mathbb{C})$.
\end{itemize}
\end{Theorem}
$\square$

\subsection{The structure of real Clifford algebras}

\subsection{Spinors}

We discuss the even-dimensional case. Since $\CCl(2k) \simeq M(2^{k}, \mathbb{C})$, it has a unique irreducible representation $\Phi$ up to isomorphism, i.e.\ the fundamental one acting on $\mathbb{C}^{2^{k}}$ by matrix multiplication. For $p+q = 2k$, one has $\Spin(p,q) \subset \Cl(p,q) \subset \CCl(2k)$, thus $\Phi$ restricts to a representation $\rho: \Spin(p, q) \rightarrow GL(2^{k}, \mathbb{C})$.

\begin{Def} The elements of $\mathbb{C}^{2^{k}}$, thought as a $\Spin(p, q)$-module, are called \emph{(p,q)-Dirac spinors}.
\end{Def}

Although the representation of $\CCl(2k)$ on $\mathbb{C}^{2^{k}}$ is irreducible, its restriction of $\Spin(p,q)$ is not. In fact, for an orthonormal basis $\{e_{1}, \ldots, e_{2k}\}$ of $\mathbb{R}^{p+q}$, let us consider the product $e = e_{1}\cdots e_{2k} \in \Spin(p,q) \subset \CCl(2k)$. It anti-commutes with every $e_{i}$, since $e_{1}\cdots e_{2k}e_{i} = (-1)^{2k-1}e_{i}e_{1}\cdots e_{2k}$ (the $-1$ at the exponent is due to the fact that when $e_{i}$ encounters itself there is no exchange to do), thus it commutes with every element of $\Spin(p,q)$, since the latter are linear combinations of even products $e_{i_{1}}\cdots e_{i_{2h}}$: hence, $e$ is a Casimir of the representation $\rho$. Moreover:
	\[\begin{split}
	e^{2} &= e_{1}\cdots e_{2k}e_{1}\cdots e_{2k} = (-1)^{2k-1}\cdot (-1)^{2k-2}\cdots(-1)^{1} e_{1}^{2} \cdots e_{2k}^{2} \\
	&= (-1)^{\frac{(2k-1)\cdot 2k}{2}} \cdot (-1)^{p} = (-1)^{k+p}
\end{split}\]
thus the only possible eigenvalues of $\rho(e^{2})$ are $\pm 1$ for $k+p$ even and $\pm i$ for $k+p$ odd.
\begin{Def} The \emph{chirality element} of $\Cl(p,q)$, with $p+q = 2k$, is:
	\[e_{c} := i^{k+p}e_{1}\cdots e_{2k}.
\]
\end{Def}
In this way the eigenvalues of $e_{c}$ are always $\pm 1$. Therefore, the representation $\rho$ splits into two representations $\mathbb{C}^{2^{k}} = S^{+} \oplus S^{-}$, where $S^{\pm}$ are the $\pm 1$-eigenspaces of $e_{c}$. It turns out that $S^{+}$ and $S^{-}$ are irreducible.

\begin{Def} The elements of $S^{+}$ and $S^{-}$, thought as $\Spin(p,q)$-modules, are called \emph{Weyl spinors} or \emph{chiral spinors}.
\end{Def}

Since $\Cl(p,q) \subset \CCl(2k)$, we can restrict the fundamental representation $\Phi$ to a complex representation $\varphi: \Cl(p,q) \rightarrow M(2^{k}, \mathbb{C})$. Identifying $\mathbb{C}^{2^{k}}$ with $\mathbb{R}^{2^{k+1}}$, we can also think $\varphi$ as a real representation $\varphi_{\mathbb{R}}: \Cl(p,q) \rightarrow M(2^{k+1}, \mathbb{R})$. It can happen that $\varphi_{\mathbb{R}}$ is reducible in $\mathbb{R}^{2^{k+1}} = S_{\mathbb{R}} \oplus i S_{\mathbb{R}}$. If, for example, we can represent $\Cl(p,q)$ by real matrices, we can choose $S_{\mathbb{R}} = \mathbb{R}^{2^{k}}$. Since $\Spin(p,q) \subset \Cl(p, q)$, when we have such a decomposition we can restrict $\varphi_{\mathbb{R}}$ to $\rho_{\mathbb{R}}: \Spin(p, q) \rightarrow \End_{\mathbb{R}}(S_{\mathbb{R}})$.

\begin{Def} The elements of $S_{\mathbb{R}}$, thought as a $\Spin(p,q)$-module, are called \emph{Majorana spinors}.
\end{Def}

When $S_{\mathbb{R}}$ exists, it may be compatible with the chirality decomposition or not, i.e.\ it can happen that $S_{\mathbb{R}} = S^{+}_{\mathbb{R}} \oplus S^{-}_{\mathbb{R}}$ for $S^{\pm}_{\mathbb{R}} = S_{\mathbb{R}} \cap S^{\pm}$.

\begin{Def} The elements of $S^{\pm}_{\mathbb{R}}$, thought as a $\Spin(p,q)$-module, are called \emph{Majorana-Weyl spinors}.
\end{Def}

\subsection{Lie algebra of the spin group}

Since $\Spin(p,q) \subset \CCl(p,q)$, the latter being a vector space, we have that the Lie algebra of $\Spin(p,q)$, i.e.\ $\mathfrak{so}(p,q)$, can be naturally embedded in the tangent space $T_{1}\CCl(p,q) \simeq \CCl(p,q)$. We claim that, for an orthonormal basis $e_{1}, \ldots, e_{n}$ of $(\mathbb{R}^{p+q}, \eta^{p,q})$, the Lie algebra is generated as a vector space by the double products $e_{i}e_{j}$ for $i < j$. In fact, if $e_{i}$ and $e_{j}$ have both square norm $1$ or $-1$ we can consider, for any $t \in \mathbb{R}$, the unit vector $u(t) = -\cos(t)e_{1} + \sin(t)e_{2}$ and the curve in $\Spin(p,q)$ given by $\varphi(t) = e_{1}\cdot u(t) = \cos(t) + \sin(t)e_{1}e_{2}$. Then, $\dot{\varphi}(0) = e_{1}e_{2}$. Instead, if $e_{i}$ and $e_{j}$ have square norm with opposite sign, we consider $u(t) = -\cosh(t)e_{1} + \sinh(t)e_{2}$ and we argue in the same way. Thus, the vector space $\langle e_{i}e_{j} \rangle_{i<j}$ is contained in the Lie algebra: for dimensional reason, it coincides with it.

\paragraph{}We now claim that an explicit isomorphism from $\mathfrak{so}(p,q)$ thought as the Lie algebra of $SO(p,q) \subset GL(p+q, \mathbb{R})$, and $\mathfrak{so}(p,q)$ thought as the Lie algebra of $\Spin(p,q) \subset \CCl(p,q)$, is given by:
	\[M^{(ij)} \rightarrow \frac{1}{2}e_{i}e_{j}
\]
for $i < j$. In fact, we just need to check the commutator rules of the r.h.s.:
	\[\textstyle \bigl[\frac{1}{2}e_{\alpha}e_{\beta}, \frac{1}{2}e_{\mu}e_{\nu}\bigr] = \frac{1}{4}(e_{\alpha}e_{\beta} e_{\mu}e_{\nu} - e_{\mu}e_{\nu} e_{\alpha}e_{\beta})
\]
so that we have the following two possibilities:
\begin{itemize}
	\item if $\{\alpha,\beta\} \cap \{\mu,\nu\} = \emptyset$, then $e_{\mu}e_{\nu} e_{\alpha}e_{\beta} = e_{\alpha}e_{\beta} e_{\mu}e_{\nu}$ since we have to perform two exchanges for $e_{\nu}$ and two for $e_{\mu}$, getting $(-1)^{4} = 1$; in this case the commutator is $0$;
	\item if, for example, $\alpha = \mu$, we have $e_{\alpha}e_{\beta} e_{\mu}e_{\nu} = -\eta^{\alpha\alpha} e_{\beta}e_{\nu}$ and $e_{\mu}e_{\nu} e_{\alpha}e_{\beta} = -\eta^{\alpha\alpha} e_{\nu}e_{\beta} = \eta^{\alpha\alpha} e_{\beta}e_{\nu}$, so that the anticommutator gives $-\frac{1}{2}\eta^{\alpha\alpha} e_{\beta}e_{\nu}$, coherently with the fact that $[M^{(\alpha\beta)}, M^{(\alpha\nu)}] = -\eta^{\alpha\alpha}M^{\beta\nu}$. Similarly for the other cases.
\end{itemize}

\paragraph{Remark:} Usually $\frac{1}{2}e_{i}e_{j}$ is written in the form $\frac{1}{4}[e_{i},e_{j}]$.

\section{Two dimensional spinors}

\subsection{Clifford algebras of dimension 1 and 2}

\begin{Lemma}\label{Clifford12} $ $
\begin{itemize}
\item $\Cl(1) = \mathbb{C}$, $\Cl(1,0) = \mathbb{R} \oplus \mathbb{R}$;
\item $\Cl(2) = \mathbb{H}$, $\Cl(1,1) = M(2, \mathbb{R})$.
\end{itemize}
\end{Lemma}

\paragraph{Proof:} $\Cl(1) = \langle 1, e_{1} \rangle$ with $e_{1}^{2} = -1$, hence identifying $e_{1}$ with $i$ we have the isomorphism with $\mathbb{C}$. Moreover, $\Cl(1,0) = \langle 1, e_{1} \rangle$ with $e_{1}^{2} = 1$. If we put $e_{\pm} := \frac{1}{2} (1 \pm e_{1})$, we have that $\Cl(1, 0) = \langle e_{-}, e_{+} \rangle$ with $e_{\pm}^{2} = e_{\pm}$ and $e_{-}e_{+} = 0$. Hence the isomorphism with $\mathbb{R} \oplus \mathbb{R}$ is obtained by the identifications $e_{-} \sim (1,0)$ and $e_{+} \sim (0,1)$.

We have that $\Cl(2) = \langle 1, e_{1}, e_{2}, e_{1}e_{2} \rangle$ with $e_{1}^{2} = e_{2}^{2} = -1$ and $(e_{1}e_{2})^{2} = -1$. It is easy to verify that the identifications $e_{1} \sim i$, $e_{2} \sim j$, $e_{3} \sim k$ give the isomorphism with $\mathbb{H}$. Finally, $\Cl(2) = \langle 1, e_{1}, e_{2}, e_{1}e_{2} \rangle$ with $e_{1}^{2} = 1$, $e_{2}^{2} = -1$ and $(e_{1}e_{2})^{2} = 1$. The isomorphism with $M(2, \mathbb{R})$ can be given for example by the identifications:
	\[1 \sim \begin{bmatrix} 1 & 0 \\ 0 & 1 \end{bmatrix} \quad
	e_{1} \sim \begin{bmatrix} -1 & 0 \\ 0 & 1 \end{bmatrix} \quad
	e_{2} \sim \begin{bmatrix} 0 & 1 \\ -1 & 0 \end{bmatrix} \quad 
	e_{1}e_{2} \sim \begin{bmatrix} 0 & -1 \\ -1 & 0 \end{bmatrix}.
\]
$\square$

\paragraph{} $\Cl(1,1)$ is obviously a real subalgebra of $M(2, \mathbb{C})$. The same is true for $\Cl(2)$, since $\mathbb{H}$ can be described by the matrices of the form:
	\[\begin{bmatrix} \alpha & \beta \\ -\overline{\beta} & \overline{\alpha} \end{bmatrix}, \qquad \alpha, \beta \in \mathbb{C}.
\]
So we have for example the identifications:
	\[1 \sim \begin{bmatrix} 1 & 0 \\ 0 & 1 \end{bmatrix} \quad
	e_{1} \sim \begin{bmatrix} i & 0 \\ 0 & -i \end{bmatrix} \quad
	e_{2} \sim \begin{bmatrix} 0 & 1 \\ -1 & 0 \end{bmatrix} \quad 
	e_{1}e_{2} \sim \begin{bmatrix} 0 & i \\ i & 0 \end{bmatrix}
\]
and $\Cl(2)$ is given by the \emph{real} combinations of these matrices. Hence, we have the following lemma:

\begin{Lemma} $ $
\begin{itemize}
\item $\CCl(1) = \CCl(1,0) = \mathbb{C} \oplus \mathbb{C}$;
\item $\CCl(2) = \CCl(1,1)= M(2,\mathbb{C})$.
\end{itemize}
$\square$
\end{Lemma}

\paragraph{Remark:} The lemma is a particular case of a general fact: the complexification $\CCl(p,q)$ depends only on $p+q$.

\subsection{Pin and Spin in the euclidean case}

The generic unit vector in $\Cl(2)$ is given by $u_{\theta} = \cos(\theta)e_{1} + \sin(\theta)e_{2}$, while the generic vector is $v = \lambda e_{1} + \mu e_{2}$. The product of two unit vector is given by:
\begin{equation}\label{SpinElements}
	\begin{split}
	u_{\psi} \cdot u_{\theta} &= \bigl(\cos(\psi)e_{1} + \sin(\psi)e_{2}\bigr) \cdot \bigl(\cos(\theta)e_{1} + \sin(\theta)e_{2}\bigr)\\
	&= \cos(\pi + \psi - \theta) + \sin(\pi + \psi - \theta)e_{1}e_{2}\\
	&= \cos(\rho) + \sin(\rho)e_{1}e_{2}.
\end{split}
\end{equation}
We have proven that the generic unit vector $u_{\theta}$ acts on $\mathbb{R}^{2}$ as reflection along the hyperplane $u_{\theta}^{\bot}$. One can easily verify (just looking at the action on $e_{1}$ and $e_{2}$) that the representative matrix of such a reflection is $\begin{bmatrix} -\cos(2\theta) & -\sin(2\theta) \\ -\sin(2\theta) & \cos(2\theta) \end{bmatrix}$. Multiplying the representative matrices one can verify that the composition of two reflections $u_{\psi} \cdot u_{\theta}$ is a rotation by $2(\psi - \theta)$, i.e.\ a rotation by $2\rho$.\footnote{To verify the representative matrix we can also check that the vector $(\cos(\theta), \sin(\theta))$ is sent to its opposite and that its orthogonal $(-\sin(\theta), \cos(\theta))$ is sent into itself. Moreover, we can see that $u_{\theta}$ and $-u_{\theta}$ generate the same reflection, since $-u_{\theta} = u_{\pi + \theta}$, but $2(\pi + \theta) \equiv 2\theta$.}

\paragraph{} Now we verify these relation within the Clifford algebra $\Cl(2)$. We choose representative matrices so that the elements of the spin group \eqref{SpinElements} are rotations:
	\[1 \sim \begin{bmatrix} 1 & 0 \\ 0 & 1 \end{bmatrix} \quad
	e_{1} \sim \begin{bmatrix} i & 0 \\ 0 & -i \end{bmatrix} \quad
	e_{2} \sim \begin{bmatrix} 0 & i \\ i & 0 \end{bmatrix} \quad 
	e_{1}e_{2} \sim \begin{bmatrix} 0 & -1 \\ 1 & 0 \end{bmatrix}.
\]
Hence, we have:
\begin{equation}\label{SpinMatrices}
\begin{split}
	&u_{\theta} = \begin{bmatrix} i\cos(\theta) & i\sin(\theta) \\ i\sin(\theta) & -i\cos(\theta) \end{bmatrix}\\
	&s_{\rho} = u_{\psi} \cdot u_{\theta} = \begin{bmatrix} \cos(\rho) & -\sin(\rho) \\ \sin(\rho) & \cos(\rho) \end{bmatrix}
\end{split}
\end{equation}
where $\rho = \pi + \psi - \theta$. In particular, we see that $\Spin(2) \simeq SO(2)$, they are both isomorphic to $U(1)$ via $s_{\rho} \rightarrow e^{i \rho}$. The generic vector $v = \lambda e_{1} + \mu e_{2}$ is given by:
	\[v = \begin{bmatrix} i\lambda & i\mu \\ i\mu & -i\lambda \end{bmatrix}.
\]
We can now explicitly calculate the action of $\Pin(2)$ and $\Spin(2)$ on $\mathbb{R}^{2}$, starting from $\Spin(2)$:
	\[\begin{split}
	s_{\rho}(v) &= \begin{bmatrix} \cos(\rho) & -\sin(\rho) \\ \sin(\rho) & \cos(\rho) \end{bmatrix} \begin{bmatrix} i\lambda & i\mu \\ i\mu & -i\lambda \end{bmatrix} \begin{bmatrix} \cos(\rho) & \sin(\rho) \\ -\sin(\rho) & \cos(\rho) \end{bmatrix}\\
	&= i \begin{bmatrix} \lambda\cos(2\rho) - \mu\sin(2\rho) & \lambda\sin(2\rho) + \mu\cos(2\rho) \\ \lambda\sin(2\rho) + \mu\cos(2\rho) & \mu\sin(2\rho) - \lambda\cos(2\rho) \end{bmatrix}
\end{split}\]
corresponding to the vector:
	\[\begin{bmatrix} \cos(2\rho) & -\sin(2\rho) \\ \sin(2\rho) & \cos(2\rho) \end{bmatrix} \begin{bmatrix} \lambda \\ \mu \end{bmatrix}.
\]
Similarly:
	\[\begin{split}
	u_{\theta}(v) &= -\begin{bmatrix} i\cos(\theta) & i\sin(\theta) \\ i\sin(\theta) & -i\cos(\theta) \end{bmatrix} \begin{bmatrix} i\lambda & i\mu \\ i\mu & -i\lambda \end{bmatrix} \begin{bmatrix} -i\cos(\theta) & -i\sin(\theta) \\ -i\sin(\theta) & i\cos(\theta) \end{bmatrix}\\
	&= i \begin{bmatrix} -\lambda\cos(2\theta) - \mu\sin(2\theta) & -\lambda\sin(2\theta) + \mu\cos(2\theta) \\ -\lambda\sin(2\theta) + \mu\cos(2\theta) & \lambda\cos(2\theta) + \mu\sin(2\theta) \end{bmatrix}
\end{split}\]
and the result corresponds, as expected, to the vector: 
	\[\begin{bmatrix} -\cos(2\theta) & -\sin(2\theta) \\ -\sin(2\theta) & \cos(2\theta) \end{bmatrix} \begin{bmatrix} \lambda \\ \mu \end{bmatrix}.
\]

\paragraph{} $SO(2)$ is connected, while $O(2)$ has two connected component, diffeomorphic one to the other\footnote{Thus, the group manifold of $O(2)$ is $S^{1} \times S^{1}$.}: one is $SO(2)$, which is a subgroup, the other is $\tilde{O}(2)$ and is given by the transformations connected to $-1$. Similarly, $\Spin(2)$ is connected, since it is a 2:1 covering of the connected group $SO(2)$ with kernel $\pm 1$, and $1$ and $-1$ are in the same connected component: in fact, consider the path $\varphi(t) = \cos(\pi t) + \sin(\pi t) e_{1}e_{2}$. Instead, $\Pin(2)$ as two connected component $\Spin(2)$ and $\widetilde{\Pin}(2)$, which are connected coverings of $SO(2)$ and $\tilde{O}(2)$. As $\Spin(2) \simeq SO(2)$, similarly $\Pin(2) \simeq O(2)$.

\subsection{Two dimensional euclidean spinors}

We now construct the \emph{complex spinor representation}. Since $\CCl(2) = M(2, \mathbb{C})$, then $\mathbb{C}^{2}$ is naturally an irreducible $\CCl(2)$-module, which we denote by $S$. By the embedding $\Spin(2) \subset \Cl(2) \subset \CCl(2)$, we can consider $S$ as a representation of $\Spin(2)$. By \eqref{SpinMatrices}, we can see that the generic element $s_{\rho}$ acts on $S$ simply as a rotation by $\rho$, extended by $\mathbb{C}$-linearity.

\paragraph{}$S$ is irreducible as $\CCl(2)$-module, but it become reducible with respect to $\Spin(2)$. In fact, we have two joint eigenspaces for $\Spin(2)$-elements:
\begin{equation}
\begin{split}
	&\begin{bmatrix} \cos(\rho) & -\sin(\rho) \\ \sin(\rho) & \cos(\rho) \end{bmatrix} \begin{bmatrix} \;1\; \\ i \end{bmatrix} = e^{-i\rho} \begin{bmatrix} \;1\; \\ i \end{bmatrix}\\
	&\begin{bmatrix} \cos(\rho) & -\sin(\rho) \\ \sin(\rho) & \cos(\rho) \end{bmatrix} \begin{bmatrix} 1 \\ -i \end{bmatrix} = \;e^{i\rho}\; \begin{bmatrix} 1 \\ -i \end{bmatrix}
\end{split}
\end{equation}
Hence we have a splitting $S = S^{-} \oplus S^{+}$, with $S^{\pm} = \mathbb{C}$ as vector space, on which an element of $\Spin(2)$ acts as a real rotation by $-\rho$ and $\rho$ respectively.

\begin{Def} The elements of $S$ are called (2-dimensional) \emph{spinors} or \emph{Dirac spinors}, the elements of $S^{-}$ and $S^{+}$ are called \emph{chiral spinors} or \emph{Weyl spinors}.
\end{Def}
We have seen two actions of the spin group:
\begin{itemize}
	\item $s_{\rho}$ acts via $\Ad$ on $\mathbb{R}^{2} \subset \Cl(2)$ as a rotation by $2\rho$;
	\item $s_{\rho}$ acts on $S^{\pm}$ by a rotation of $\pm\rho$.
\end{itemize}
The first action is actually an action of $SO(2)$, since $s_{\rho} \equiv s_{\rho+\pi}$, in particular $-1 = s_{\pi} \equiv 1$. The second one, instead, is an action of $\Spin(2)$ not passing to the quotient. As $\Spin(2)$ is a 2:1 covering of $SO(2)$, similarly we can see $S^{\pm}$ as a 2:1 covering of $\mathbb{R}^{2}$, but we have to remove the origin. In particular, we construct a morphism of (real) representations sending a vector $r \cdot e^{i\theta} \in S^{\pm} \simeq \mathbb{C}$ to the vector $r \cdot e^{\pm 2 i\theta} \in \mathbb{R}^{2} \simeq \mathbb{C}$. The explicit morphisms, which are 2:1 covering outside $0$, are given by:
\begin{equation}
\begin{split}
	\varphi^{\pm}: \; & \, S^{\pm} \longrightarrow \mathbb{\mathbb{R}}^{2}\\
	&\!\begin{bmatrix} 1 \\ \mp i \end{bmatrix} \rightarrow \begin{bmatrix} 1 \\ 0 \end{bmatrix} \qquad e^{\pm i \frac{\pi}{4}} \begin{bmatrix} 1 \\ \mp i \end{bmatrix} \rightarrow \begin{bmatrix} 0 \\ 1 \end{bmatrix}.
\end{split}
\end{equation}.

We can thus imagine a Weyl spinor as a vector corresponding to its double-rotated, so that a vector and its opposite have the same image. A Dirac spinor is simply a couple of Weyl spinors rotating in opposite directions.

\paragraph{} Moreover, the representative matrix of $s_{\rho}$ is real. Hence we have a decomposition of $S = \mathbb{C}^{2}$ in two \emph{real} representations $\mathbb{C}^{2} = \mathbb{R}^{2} \oplus i\mathbb{R}^{2}$, which we denote by $S = S_{\mathbb{R}} \oplus i S_{\mathbb{R}}$.

\begin{Def} The elements of $S_{\mathbb{R}}$ and $iS_{\mathbb{R}}$ are called (2-dimensional) \emph{Majorana spinors}.
\end{Def}

Looking at the action of $\Spin(2)$ on $S_{\mathbb{R}}$ and $iS_{\mathbb{R}}$, it is a usual rotation. This is due to the fact that $\Spin(2) \simeq SO(2)$ since they are both isomorphic to $U(1)$. However, it is not correct to reduce this action to rotations via a covering of $\mathbb{R}^{2}$, since there is not a definite chirality.

\paragraph{Remark:} the fact that the components of a Majorana spinor are both real or both imaginary, is a consequence of the fact that we have represented $e_{1}e_{2}$ by a real matrix. Otherwise, we would however have the decomposition of $S$ in two real representations of $\Spin(2)$, each formed by real linear combinations of two fixed complex vectors.

\paragraph{} We obtained two splittings of $S$, given by $S = S^{-} \oplus S^{+}$ and $S = S_{\mathbb{R}} \oplus i S_{\mathbb{R}}$. These two decomposition are not compatible, in the sense that the factors of the Majorana splitting are not themselves splitted by the Weyl decomposition, i.e.\ \emph{Majorana-Weyl spinors does not exist in the two-dimensional euclidean setting}.

\subsection{Pin and Spin in the minkowskian case}

Let us now analyze the structure of $SO(1,1)$ and $O(1,1)$. If we fix the vector $e_{1} = (1, 0) \in \mathbb{R}^{2}$, a euclidean rotation by $\theta$ moves it to the unit vector $e_{\theta}$ such that $\theta$ is the length of the arc in $S^{1}$ between $e_{1}$ and $e_{\theta}$. In the minkowskian setting, let us consider the right component of the hyperbola $H^{1} \equiv x^{2} - y^{2} = 1$: it is the hyperbola with asymptotes $y = x$ and $y = -x$, passing through $(1,0)$. A minkowskian rotation by $\theta$ moves $e_{1}$ to the vector $e_{\theta}$ such that $\theta$ is the length of the arc in $H^{1}$ between $e_{1}$ and $e_{\theta}$. The matrix of such a rotation is:
\begin{equation}\label{HyperbolicRotation}
	\begin{bmatrix} \cosh(\theta) & \sinh(\theta) \\ \sinh(\theta) & \cosh(\theta) \end{bmatrix}.
\end{equation}
The behavior of $e_{2} = (0,1)$ is analogous for the upper component of the hyperbola $x^{2} - y^{2} = -1$. Any vector $(v_{1}, v_{2})$ moves on the hyperbola $x^{2} - y^{2} = \pm a^{2}$ passing through it. Actually $SO(1,1)$ is not connected: the transformations \eqref{HyperbolicRotation} are the component $SO^{0}(1,1)$ connected to the identity.

The other component is $SO^{1}(1,1)$, connected to the inversion $\iota: (v_{1}, v_{2}) \rightarrow (-v_{1}, -v_{2})$. It is made by transformations obtained composing $\iota$ with anyone of \eqref{HyperbolicRotation}. In particular, $\iota$ flips the two components of the hyperbola passing through a point. The matrix of such transformations are:
\begin{equation}
	\begin{bmatrix} -\cosh(\theta) & \sinh(\theta) \\ \sinh(\theta) & -\cosh(\theta) \end{bmatrix}.
\end{equation}

$O(1,1)$ has 4 connected components: two are $SO^{0}(1,1)$ and $SO^{1}(1,1)$, the other two are $\tilde{O}^{0}(1,1)$, connected to the time inversion\footnote{Time is the negative-definite direction.} $\tau: (v_{1}, v_{2}) \rightarrow (v_{1}, -v_{2})$, and $\tilde{O}^{1}(1,1)$, connected to the space inversion\footnote{We could exchange the definitions of $\tilde{O}^{0}(1,1)$ and $\tilde{O}^{1}(1,1)$, since they are not canonical.} $\kappa: (v_{1}, v_{2}) \rightarrow (-v_{1}, v_{2})$. Matrices for such transformations are:
	\[\tilde{O}^{0}(1,1) \sim \begin{bmatrix} \cosh(\theta) & -\sinh(\theta) \\ \sinh(\theta) & -\cosh(\theta) \end{bmatrix} \qquad \tilde{O}^{1}(1,1) \sim \begin{bmatrix} -\cosh(\theta) & \sinh(\theta) \\ -\sinh(\theta) & \cosh(\theta) \end{bmatrix}.
\]
We then define in the obvious way $O^{0}(1,1) = SO^{0}(1,1) \cup \tilde{O}^{0}(1,1)$ and $O^{1}(1,1) = SO^{1}(1,1) \cup \tilde{O}^{1}(1,1)$. Of course, $SO^{0}(1,1)$ and $O^{0}(1,1)$ are subgroups, while $SO^{1}(1,1)$ and $O^{1}(1,1)$ are cosets.

\paragraph{}The generic unit vector in $\Cl(1,1)$ is given by:
\begin{equation}
\begin{array}{ll}
	u_{\theta}^{\pm} = \phantom{+}\sinh(\theta)e_{1} \pm \cosh(\theta)e_{2}, & \norm{u_{\theta}}^{2} = 1\\
	v_{\theta}^{\pm} = \pm\cosh(\theta)e_{1} + \sinh(\theta)e_{2}, & \norm{v_{\theta}}^{2} = -1
\end{array}
\end{equation}
The product of two unit vector can be of the form:
\begin{equation}\label{Spin11Elements}
\begin{array}{lll}
	s_{\rho}^{\pm} & = & \pm\cosh(\rho) + \sinh(\rho) e_{1}e_{2}\\
	t_{\rho}^{\pm} & = & \phantom{\pm}\sinh(\rho) \pm \cosh(\rho) e_{1}e_{2}.
\end{array}
\end{equation}
In particular, the following relations hold:
\[\begin{array}{lll}
	u_{\psi}^{\pm} \cdot u_{\theta}^{\pm} = s_{\pm (\psi - \theta)}^{-} & u_{\psi}^{\pm} \cdot u_{\theta}^{\mp} = s_{\pm (\psi + \theta)}^{+} & v_{\psi}^{\pm} \cdot v_{\theta}^{\pm} \,= s_{\pm (\psi + \theta)}^{-} \\
	v_{\psi}^{\pm} \cdot v_{\theta}^{\mp} \,= s_{\pm (\psi - \theta)}^{+} & u_{\psi}^{\pm} \cdot v_{\theta}^{\pm} = t_{\pm (\psi - \theta)}^{-} & u_{\psi}^{\pm} \cdot v_{\theta}^{\mp} = t_{\pm (\psi + \theta)}^{+}.\\
	s_{\eta}^{\pm} \cdot s_{\rho}^{\pm} = s_{\pm (\eta + \rho)}^{+} & s_{\eta}^{\pm} \cdot s_{\rho}^{\mp} = s_{\pm (\psi + \theta)}^{+} & t_{\eta}^{\pm} \cdot t_{\rho}^{\pm} \,= s_{\pm (\psi + \theta)}^{-} \\
	t_{\eta}^{\pm} \cdot t_{\rho}^{\mp} \,= s_{\pm (\psi - \theta)}^{+} & s_{\eta}^{\pm} \cdot t_{\rho}^{\pm} = t_{\pm (\psi - \theta)}^{-} & s_{\eta}^{\pm} \cdot t_{\rho}^{\mp} = t_{\pm (\psi + \theta)}^{+}.
\end{array}\]
Of course $\Spin(1,1)$ is not connected, since there is at least the topological subdivision $\Spin(1,1) = \Spin^{0}(1,1) \cup \Spin^{1}(1,1)$, but it turns out that \emph{neither $Spin^{0}(1,1)$ is connected}, since \emph{$Spin(1,1)$ is the trivial 2:1 covering $SO(1,1) \times SO(1,1)$}. In particular, $\Spin(1,1)$ has the 4 connected components corresponding to \eqref{Spin11Elements}.

\paragraph{} We have proven that the generic unit vector $u_{\theta}$ acts on $\mathbb{R}^{2}$ as reflection along the hyperplane $u_{\theta}^{\bot}$.\footnote{A non-degenerate bilinear form, as the minkowskian one, defines an orthogonal space even if it is not positive-definite.} We now compute the representative matrix of such a reflection $\tau(u_{\theta}^{\bot})$. For $u_{\theta} = (\pm\cosh\theta, \sinh\theta)$, one has $u_{\theta}^{\bot} = (\sinh\theta, \pm\cosh\theta)$: in particular, $u_{\theta}$ lives on $x^{2} - y^{2} = 1$, while $u_{\theta}^{\bot}$ is its reflection with respect to $y = x$ and lives on $x^{2} - y^{2} = -1$. Moreover, $(1,0) = \pm\cosh\theta \cdot (\pm\cosh\theta, \sinh\theta) - \sinh\theta \cdot (\sinh\theta, \pm\cosh\theta)$, thus it is sent by $\tau(u_{\theta}^{\bot})$ to $\mp\cosh\theta \cdot (\pm\cosh\theta, \sinh\theta) - \sinh\theta \cdot (\sinh\theta, \pm\cosh\theta) = (-\cosh 2\theta, \mp\sinh 2\theta)$. Similarly, $(0,1) = -\sinh\theta \cdot (\pm\cosh\theta, \sinh\theta) \pm \cosh\theta \cdot (\sinh\theta,$ $\pm\cosh\theta)$, thus it is sent by $\tau(u_{\theta}^{\bot})$ to $\sinh\theta \cdot (\pm\cosh\theta, \sinh\theta) \pm \cosh\theta \cdot (\sinh\theta, \pm\cosh\theta)$ $= (\pm\sinh 2\theta, \cosh 2\theta)$. At the end, the representative matrix is:
	\[\begin{bmatrix} -\cosh(\pm 2\theta) & \sinh(\pm 2\theta) \\ -\sinh(\pm 2\theta) & \cosh(\pm 2\theta) \end{bmatrix}.
\]
Since $v_{\theta} = u_{\theta}^{\bot}$, for $v_{\theta}$ we obtain the same result reflecting with respect to the other component of the decomposition of $(1,0)$ and $(0,1)$, thus we get:
	\[\begin{bmatrix} \cosh(\pm 2\theta) & -\sinh(\pm 2\theta) \\ \sinh(\pm 2\theta) & -\cosh(\pm 2\theta) \end{bmatrix}.
\]
Multiplying the representative matrices one can verify that the composition of two reflections $u_{\psi} \cdot u_{\theta}$ is an element of $SO(1,1)$ corresponding to $2(\psi - \theta)$, which we call $2\rho$.\footnote{In particular, one can see that $u_{\theta}$ and $-u_{\theta}$ generate the same reflection.}

\paragraph{} Now we verify these relations within the Clifford algebra $\Cl(1,1)$. We choose representative matrices so that the elements of the spin group \eqref{Spin11Elements} are rotations:
	\[1 \sim \begin{bmatrix} 1 & 0 \\ 0 & 1 \end{bmatrix} \quad
	e_{1} \sim \begin{bmatrix} 1 & 0 \\ 0 & -1 \end{bmatrix} \quad
	e_{2} \sim \begin{bmatrix} 0 & 1 \\ -1 & 0 \end{bmatrix} \quad 
	e_{1}e_{2} \sim \begin{bmatrix} 0 & 1 \\ 1 & 0 \end{bmatrix}.
\]
Hence, we have, for example:
\begin{equation}\label{SpinMatricesMink}
\begin{split}
	&u_{\theta}^{+} = \begin{bmatrix} \sinh(\theta) & \cosh(\theta) \\ -\cosh(\theta) & -\sinh(\theta) \end{bmatrix}\\
	&s^{-}_{\rho} = u^{+}_{\psi} \cdot u^{+}_{\theta} = \begin{bmatrix} -\cosh(\rho) & \sinh(\rho) \\ \sinh(\rho) & -\cosh(\rho) \end{bmatrix}
\end{split}
\end{equation}
where $\rho = \pm(\psi - \theta)$. The generic vector $v = \lambda e_{1} + \mu e_{2}$ is given by:
	\[v = \begin{bmatrix} \lambda & \mu \\ -\mu & -\lambda \end{bmatrix}.
\]
We can now explicitly calculate the action of $\Pin(2)$ and $\Spin(2)$ on $\mathbb{R}^{2}$, starting from the $\Spin$ group action. We have:
	\[\begin{split}
	s_{\rho}^{-}(v) &= \begin{bmatrix} -\cosh(\rho) & \sinh(\rho) \\ \sinh(\rho) & -\cosh(\rho) \end{bmatrix} \begin{bmatrix} \lambda & \mu \\ -\mu & -\lambda \end{bmatrix} \begin{bmatrix} -\cosh(\rho) & -\sinh(\rho) \\ -\sinh(\rho) & -\cosh(\rho) \end{bmatrix}\\
	&= \begin{bmatrix} \lambda\cosh(2\rho) + \mu\sinh(2\rho) & \lambda\sinh(2\rho) + \mu\cosh(2\rho) \\ -\lambda\sinh(2\rho) - \mu\cosh(2\rho) & -\lambda\cosh(2\rho) - \mu\sinh(2\rho) \end{bmatrix}
\end{split}\]
corresponding to the vector:
	\[\begin{bmatrix} \cosh(2\rho) & \sinh(2\rho) \\ \sinh(2\rho) & \cosh(2\rho) \end{bmatrix} \begin{bmatrix} \lambda \\ \mu \end{bmatrix}.
\]
If we do the same computations for the other families of \eqref{Spin11Elements}, we find the following scheme:
\begin{equation}\label{Spin11Scheme}
	\begin{array}{lll}
	$ $\\
	\textit{Spin(1,1) element} & & \textit{Action on vectors} \\ \\
	s_{\rho}^{+} = \begin{bmatrix} \cosh(\rho) & \sinh(\rho) \\ \sinh(\rho) & \cosh(\rho) \end{bmatrix} & \Longrightarrow & \begin{bmatrix} \cosh(-2\rho) & \sinh(-2\rho) \\ \sinh(-2\rho) & \cosh(-2\rho) \end{bmatrix} \\ \\
	s_{\rho}^{-} = \begin{bmatrix} -\cosh(\rho) & \sinh(\rho) \\ \sinh(\rho) & -\cosh(\rho) \end{bmatrix} & \Longrightarrow & \begin{bmatrix} \cosh(2\rho) & \sinh(2\rho) \\ \sinh(2\rho) & \cosh(2\rho) \end{bmatrix} \\ \\
	t_{\rho}^{+} = \begin{bmatrix} \sinh(\rho) & \cosh(\rho) \\ \cosh(\rho) & \sinh(\rho) \end{bmatrix} & \Longrightarrow & \begin{bmatrix} -\cosh(2\rho) & \sinh(2\rho) \\ \sinh(2\rho) & -\cosh(2\rho) \end{bmatrix} \\ \\
	t_{\rho}^{-} = \begin{bmatrix} \sinh(\rho) & -\cosh(\rho) \\ -\cosh(\rho) & \sinh(\rho) \end{bmatrix} & \Longrightarrow & \begin{bmatrix} -\cosh(-2\rho) & \sinh(-2\rho) \\ \sinh(-2\rho) & -\cosh(-2\rho) \end{bmatrix} \\
	$ $
\end{array}
\end{equation}
For the $\Pin$ group action, we also consider the single unit vectors, obtaining the following scheme:
\begin{equation}\label{Pin11Scheme}
	\begin{array}{lll}
	$ $\\
	\textit{Pin(1,1) element} & & \textit{Action on vectors} \\ \\
	u_{\theta}^{+} = \begin{bmatrix} \cosh(\theta) & \sinh(\theta) \\ -\sinh(\theta) & -\cosh(\theta) \end{bmatrix} & \Longrightarrow & \begin{bmatrix} -\cosh(-2\theta) & \sinh(-2\theta) \\ -\sinh(-2\theta) & \cosh(-2\theta) \end{bmatrix} \\ \\
	u_{\theta}^{-} = \begin{bmatrix} -\cosh(\theta) & \sinh(\theta) \\ -\sinh(\theta) & \cosh(\theta) \end{bmatrix} & \Longrightarrow & \begin{bmatrix} -\cosh(2\theta) & \sinh(2\theta) \\ -\sinh(2\theta) & \cosh(2\theta) \end{bmatrix} \\ \\
	v_{\theta}^{+} = \begin{bmatrix} \sinh(\theta) & \cosh(\theta) \\ -\cosh(\theta) & -\sinh(\theta) \end{bmatrix} & \Longrightarrow & \begin{bmatrix} \cosh(-2\theta) & -\sinh(-2\theta) \\ \sinh(-2\theta) & -\cosh(-2\theta) \end{bmatrix} \\ \\
	v_{\theta}^{-} = \begin{bmatrix} \sinh(\theta) & -\cosh(\theta) \\ \cosh(\theta) & -\sinh(\theta) \end{bmatrix} & \Longrightarrow & \begin{bmatrix} \cosh(2\theta) & -\sinh(2\theta) \\ \sinh(2\theta) & -\cosh(2\theta) \end{bmatrix} \\
	$ $
\end{array}
\end{equation}

\subsection{Two dimensional minkowskian spinors}

We now construct the \emph{complex spinor representation}. Since $\CCl(2) = M(2, \mathbb{C})$, then $\mathbb{C}^{2}$ is naturally an irreducible $\CCl(2)$-module, which we denote by $S$. By the embedding $\Spin(1,1) \subset \Cl(1,1) \subset \CCl(2)$, we can consider $S$ as a representation of $\Spin(1,1)$. By the scheme \eqref{Spin11Scheme}, we can see that the generic element $s_{\rho}$ acts on $S$ simply as an element of $O(1,1)$.

\paragraph{}$S$ is irreducible as $\CCl(2)$-module, but it become reducible with respect to $\Spin(1,1)$. In fact, we have two joint eigenspaces for $\Spin(1,1)$-elements, generated by $(1,1)$ and $(1,-1)$ respectively, with the following scheme:

	\[\begin{array}{|c||c|c|} \hline
	& \,(1,1)\, & (1, -1)\\ \hline
	s_{\rho}^{+} & e^{\rho} & e^{-\rho}\\ \hline
	s_{\rho}^{-} & -e^{-\rho} & -e^{\rho}\\ \hline
	t_{\rho}^{+} & e^{\rho} & -e^{-\rho}\\ \hline
	t_{\rho}^{-} & -e^{-\rho} & e^{\rho}\\ \hline
	\end{array}
\]

\paragraph{}Hence we have a splitting $S = S^{-} \oplus S^{+}$, with $S^{\pm} = \mathbb{C}$ as vector space, on which an element of $\Spin(1,1)$ acts as a real rotation by $-\rho$ and $\rho$ respectively.

\begin{Def} The elements of $S$ are called (2-dimensional) \emph{spinors} or \emph{Dirac spinors}, the elements of $S^{-}$ and $S^{+}$ are called \emph{chiral spinors} or \emph{Weyl spinors}.
\end{Def}
We have seen two actions of the spin group:
\begin{itemize}
	\item $s_{\rho}$ acts via $\Ad$ on $\mathbb{R}^{2} \subset \Cl(2)$ as a rotation by $2\rho$;
	\item $s_{\rho}$ acts on $S^{\pm}$ by a rotation of $\pm\rho$.
\end{itemize}
The first action is actually an action of $SO(2)$, since $s_{\rho} \equiv s_{\rho+\pi}$, in particular $-1 = s_{\pi} \equiv 1$. The second one, instead, is an action of $\Spin(2)$ not passing to the quotient. As $\Spin(2)$ is a 2:1 covering of $SO(2)$, similarly we can see $S^{\pm}$ as a 2:1 covering of $\mathbb{R}^{2}$, but we have to remove the origin. In particular, we construct a morphism of (real) representations sending a vector $r \cdot e^{i\theta} \in S^{\pm} \simeq \mathbb{C}$ to the vector $r \cdot e^{\pm 2 i\theta} \in \mathbb{R}^{2} \simeq \mathbb{C}$. The explicit morphisms, which are 2:1 covering outside $0$, are given by:
\begin{equation}
\begin{split}
	\varphi^{\pm}: \; & \, S^{\pm} \longrightarrow \mathbb{\mathbb{R}}^{2}\\
	&\!\begin{bmatrix} 1 \\ \mp i \end{bmatrix} \rightarrow \begin{bmatrix} 1 \\ 0 \end{bmatrix} \qquad e^{\pm i \frac{\pi}{4}} \begin{bmatrix} 1 \\ \mp i \end{bmatrix} \rightarrow \begin{bmatrix} 0 \\ 1 \end{bmatrix}.
\end{split}
\end{equation}.

We can thus imagine a Weyl spinor as a vector corresponding to its double-rotated, so that a vector and its opposite have the same image. A Dirac spinor is simply a couple of Weyl spinors rotating in opposite directions.

\paragraph{} Moreover, the representative matrices of $s^{\pm}_{\rho}$ and $t^{\pm}_{\rho}$ are real. Hence we have a decomposition of $S = \mathbb{C}^{2}$ in two \emph{real} representations $\mathbb{C}^{2} = \mathbb{R}^{2} \oplus i\mathbb{R}^{2}$, which we denote by $S = S_{\mathbb{R}} \oplus i S_{\mathbb{R}}$.

\begin{Def} The elements of $S_{\mathbb{R}}$ and $iS_{\mathbb{R}}$ are called (2-dimensional) \emph{Majorana spinors}.
\end{Def}

Looking at the action of $\Spin(2)$ on $S_{\mathbb{R}}$ and $iS_{\mathbb{R}}$, it is a usual rotation. This is due to the fact that $\Spin(2) \simeq SO(2)$ since they are both isomorphic to $U(1)$. However, it is not correct to reducible this action to rotations via a covering of $\mathbb{R}^{2}$, since there is not a definite chirality.

\paragraph{Remark:} the fact that the components of a Majorana spinor are both real or both imaginary, is a consequence of the fact that we have represented $e_{1}e_{2}$ by a real matrix. Otherwise, we would however have the decomposition of $S$ in two real representations of $\Spin(2)$, each formed by real linear combinations of two fixed complex vectors.

\paragraph{} We obtained two splittings of $S$, given by $S = S^{-} \oplus S^{+}$ and $S = S_{\mathbb{R}} \oplus i S_{\mathbb{R}}$. These two decomposition are not compatible, in the sense that the factors of the Majorana splitting are not themselves split by the Weyl decomposition, i.e.\ \emph{chiral real spinors does not exist in the two-dimensional euclidean setting}.

\section{Four dimensional minkowskian spinors}

We have an analogue of lemma \ref{PeriodicityC} for the real case:

\begin{Lemma}\label{Periodicity11} There is an isomorphism:
	\[\Cl(p+1,q+1) \simeq \Cl(p,q) \otimes_{\mathbb{R}} \Cl(1,1)
\]
given by, for $\{\varepsilon_{0}, \ldots, \varepsilon_{q-1}, e_{0}, \ldots, e_{p-1}\}$ an orthonormal basis of $(\mathbb{R}^{n}, \eta^{p,q})$, $\{\varepsilon'_{0}, e'_{0}\}$ an orthonormal basis of $(\mathbb{R}^{2}, \eta)$ and $\{\varphi_{0}, \ldots, \varphi_{q-1}, \varphi_{q}, f_{0}, \ldots, f_{p-1}, f_{p}\}$ an orthonormal basis of $(\mathbb{R}^{n+2},$ $\eta^{p+1,q+1})$:
	\[\begin{array}{lcl}
	1 & \rightarrow & 1\\
	\varphi_{0} & \rightarrow & \varepsilon_{0} \otimes \varepsilon'_{0}e'_{0}\\
	& \vdots & \\
	\varphi_{q-1} & \rightarrow & \varepsilon_{q-1} \otimes \varepsilon'_{0}e'_{0}\\
	\varphi_{q} & \rightarrow & 1 \otimes \varepsilon'_{0}\\
	f_{0} & \rightarrow & e_{0} \otimes \varepsilon'_{0}e'_{0}\\
	& \vdots & \\
	f_{p-1} & \rightarrow & e_{p-1} \otimes \varepsilon'_{0}e'_{0}\\
	f_{p} & \rightarrow & 1 \otimes e'_{0}
\end{array}\]
\end{Lemma}

\paragraph{Proof:} It is easy to prove that the images of the $f_{j}$'s are independent in $\Cl(p,q) \otimes_{\mathbb{R}} \Cl(1,1)$. We now verify the behavior with respect to the product. We first notice that $(\varepsilon'_{0}e'_{0})^{2} = \varepsilon'_{0}e'_{0} \varepsilon'_{0}e'_{0} = -(\varepsilon'_{0})^{2} (e'_{0})^{2} = -(1)(-1) = 1$. Thus:
\begin{itemize}
	\item $\{\varepsilon_{i} \otimes \varepsilon'_{0}e'_{0}, \varepsilon_{j} \otimes \varepsilon'_{0}e'_{0}\} = \{\varepsilon_{i}, \varepsilon_{j}\} \otimes (\varepsilon'_{0}e'_{0})^{2} = \delta_{ij} \otimes 1 = \delta_{ij}$;
	\item $\{e_{i} \otimes \varepsilon'_{0}e'_{0}, e_{j} \otimes \varepsilon'_{0}e'_{0}\} = \{e_{i}, e_{j}\} \otimes (\varepsilon'_{0}e'_{0})^{2} = -\delta_{ij} \otimes 1 = -\delta_{ij}$;
	\item $\{\varepsilon_{i} \otimes \varepsilon'_{0}e'_{0}, e_{j} \otimes \varepsilon'_{0}e'_{0}\} = \{\varepsilon_{i}, e_{j}\} \otimes (\varepsilon'_{0}e'_{0})^{2} = 0 \otimes 1 = 0$;
	\item $\{e_{i} \otimes \varepsilon'_{0}e'_{0}, 1 \otimes e'_{0}\} = e_{i} \otimes (\varepsilon'_{0}e'_{0}e'_{0} + e'_{0}\varepsilon'_{0}e'_{0}) = 0$;
	\item $\ldots$
\end{itemize}
Thus the linear mapping in the statement can be extended to an algebra homomorphism and, by dimensional reason, it is an isomorphism.\\
$\square$

\paragraph{}In particular, by the previous lemma we have that $\Cl(1,3) \simeq \Cl(1, 1) \otimes_{\mathbb{R}} \Cl(0, 2)$, i.e.\ $\Cl(1,3) \simeq M(2, \mathbb{H})$. We thus construct the isomorphism at matrix level. We consider the identifications for $\Cl(0,2) \simeq \mathbb{H}$ and $\Cl(1,1) \simeq M(2, \mathbb{R})$:
	\[\begin{array}{llll}
	1 \sim \begin{bmatrix} 1 & 0 \\ 0 & 1 \end{bmatrix} &
	e_{0} \sim \begin{bmatrix} i & 0 \\ 0 & -i \end{bmatrix} &
	e_{1} \sim \begin{bmatrix} 0 & i \\ i & 0 \end{bmatrix} &
	e_{0}e_{1} \sim \begin{bmatrix} 0 & -1 \\ 1 & 0 \end{bmatrix}\\ \\
	1 \sim \begin{bmatrix} 1 & 0 \\ 0 & 1 \end{bmatrix} &
	\varepsilon'_{0} \sim \begin{bmatrix} 1 & 0 \\ 0 & -1 \end{bmatrix} &
	e'_{0} \sim \begin{bmatrix} 0 & 1 \\ -1 & 0 \end{bmatrix} &
	\varepsilon'_{0}e'_{0} \sim \begin{bmatrix} 0 & 1 \\ 1 & 0 \end{bmatrix}.
\end{array}\]
Thus we get for $\Cl(1,3)$:
	\[\begin{array}{ll}
	\Gamma_{0} \simeq 1 \otimes \varepsilon'_{0} \sim \begin{bmatrix} 1 & 0 & 0 & 0 \\ 0 & 1 & 0 & 0 \\
	0 & 0 & -1 & 0 \\ 0 & 0 & 0 & -1 \end{bmatrix} & 
	\Gamma_{1} \simeq 1 \otimes e'_{0} \sim \begin{bmatrix} 0 & 0 & 1 & 0 \\ 0 & 0 & 0 & 1 \\
	-1 & 0 & 0 & 0 \\ 0 & -1 & 0 & 0 \end{bmatrix}\\ \\
	\Gamma_{2} \simeq e_{0} \otimes \varepsilon'_{0}e'_{0} \sim \begin{bmatrix} 0 & 0 & i & 0 \\ 0 & 0 & 0 & -i \\
	i & 0 & 0 & 0 \\ 0 & -i & 0 & 0 \end{bmatrix} & 
	\Gamma_{3} \simeq e_{1} \otimes \varepsilon'_{0}e'_{0} \sim \begin{bmatrix} 0 & 0 & 0 & i \\ 0 & 0 & i & 0 \\
	0 & i & 0 & 0 \\ i & 0 & 0 & 0 \end{bmatrix}.
\end{array}\]
The chirality matrix is given by:
	\[\Gamma = \Gamma_{0} \cdots \Gamma_{3} = \begin{bmatrix} 0 & 0 & 0 & -1 \\ 0 & 0 & 1 & 0 \\
	0 & -1 & 0 & 0 \\ 1 & 0 & 0 & 0 \end{bmatrix}.
\]


\chapter{Pin and Spin structures}

We now discuss some topics about spinors and non-orientable manifolds. This chapter contains \cite{BFS2}, and it almost coincides with its extended version on arXiv.

\section{Pinors vs Spinors}

\subsection{Preliminaries on pinors}

We recall that the group $\SO(n)$ has a unique 2-covering $\Spin(n)$, while the group $\OO(n)$ has two inequivalent 2-coverings $\Pin^{\pm}(n)$, obtained from the Clifford algebras with positive and negative signature respectively, as explained in \cite{KT} (for Clifford algebras we use the convention $vw + wv = 2\langle v, w \rangle$, without the minus sign). Let $p^{\pm}: \Pin^{\pm}(n) \rightarrow \OO(n)$ be such 2-coverings with kernel $\{\pm 1\}$, both restricting to $\rho: \Spin(n) \rightarrow \SO(n)$. If we fix a the canonical basis $\{e_{1}, \ldots, e_{n}\}$ of $\mathbb{R}^{n}$ and we denote by $j_{1}$ the reflection with respect to the hyperplane $e_{1}^{\bot}$, we have that $\OO(n) = \langle \SO(n), j_{1} \rangle$, and $(p^{\pm})^{-1}(\{1,j_{1}\}) = \{\pm 1, \pm e_{1}\}$: the latter is isomorphic to $\mathbb{Z}_{4}$ if $e_{1}^{2} = -1$ and to $\mathbb{Z}_{2} \oplus \mathbb{Z}_{2}$ if $e_{1}^{2} = 1$, that's why in general we get non-isomorphic coverings. For details the reader can see \cite{KT}.
\begin{Def} Let $\pi: E \rightarrow M$ be a vector bundle of rank $n$ with a metric and let $\pi_{\OO}: \PP_{\OO}E \rightarrow M$ be the principal $\OO(n)$-bundle of orthonormal frames. A \emph{pin$^{\pm}$ structure} on $E$ is a principal $\Pin^{\pm}(n)$-bundle $\pi_{\Pin^{\pm}}: \PP_{\Pin^{\pm}}E \rightarrow M$ with a 2-covering $\xi: \PP_{\Pin^{\pm}}E \rightarrow \PP_{\OO}E$ such that the following diagram commutes:
	\[\xymatrix{
	\PP_{\Pin^{\pm}}E \times \Pin^{\pm}(n) \ar[r]^{\qquad\cdot} \ar[dd]_{\xi \times p^{\pm}} & \PP_{\Pin^{\pm}}E \ar[dd]^{\xi} \ar[dr]^{\pi_{\Pin^{\pm}}} \\
	& & M. \\
	\PP_{\OO}E \times \OO(n) \ar[r]^{\quad\;\cdot} & \PP_{\OO}E \ar[ur]_{\pi_{\OO}}
}\]
\end{Def}
\begin{Def} Two pin$^{\pm}$ structures $(\pi_{\Pin^{\pm}}, \xi^{\pm})$ and $(\pi'_{\Pin^{\pm}}, \xi'^{\pm})$ are \emph{equivalent} if there exists a principal $\Pin^{\pm}$-bundles isomorphism $\varphi: \PP_{\Pin^{\pm}}E \rightarrow \PP'_{\Pin^{\pm}}E$ such that $\xi' \circ \varphi = \xi$.
\end{Def}
Similarly to the case of spin structures, there is a simply transitive action of $H^{1}(M, \mathbb{Z}_{2})$ on pin$^{\pm}$ structures on a bundle $E \rightarrow M$. Given a good cover $\mathfrak{U} = \{U_{\alpha}\}_{\alpha \in I}$ of $M$, the bundle $E$ is represented by $\OO(n)$-valued transition functions $\{g_{\alpha\beta}\}$. A pin$^{\pm}$ structure is represented by $\Pin^{\pm}(n)$-valued transition functions\footnote{It is not true that the pin structure depends only on $[\{s_{\alpha\beta}\}\,] \in H^{1}(M, \underline{\Pin}^{\pm}(n))$ for $\underline{\Pin}^{\pm}(n)$ the sheaf of $\Pin^{\pm}$-valued smooth functions, because such a cohomology class determines the equivalence class of the principal bundle without considering the projection to the tangent bundle. In particular, two spin lifts can be isomorphic as principal pin$^{\pm}$-bundles, but in such a way that there are no isomorphisms commuting with the projections to $P_{O}M$: in this case they determine the same class in $H^{1}(M, \underline{\Pin}^{\pm}(n))$ but they are not equivalent as pin structures.} $\{s_{\alpha\beta}\}$ such that $p^{\pm}(s_{\alpha\beta}) = g_{\alpha\beta}$; all other pin$^{\pm}$ structures are represented by $\{s_{\alpha\beta} \cdot \varepsilon_{\alpha\beta}\}$ for $\{\varepsilon_{\alpha\beta}\}$ a $\mathbb{Z}_{2}$-cocycle and depend up to equivalence only by $[\{\varepsilon_{\alpha\beta}\}] \in \check{H}^{1}(M, \mathbb{Z}_{2})$. In particular, this implies that if there exist both $\Pin^{+}$ and $\Pin^{-}$ structures, their number is the same. Given a real vector bundle $\pi: E \rightarrow M$ the following conditions hold:
\begin{itemize}
	\item $E$ admits a $\Pin^{+}$-structure if and only if $w_{2}(E) = 0$;
	\item $E$ admits a $\Pin^{-}$-structure if and only if $w_{2}(E) + w_{1}(E) \cup w_{1}(E) = 0$.
\end{itemize}
For the proof the reader is referred to \cite{Karoubi}. As for spin structures, a pin structure on a manifold is by definition a pin structure on its tangent bundle.

\paragraph{}Let $M$ be a manifold of dimension $2n$. Given a pin$^{\pm}$ structure $\xi: P_{\Pin^{\pm}}M \rightarrow P_{\OO}M$ and an isometry $\varphi: M \rightarrow M$, we define the pin$^{\pm}$ structure $\varphi^{*}\xi$ via the following diagram:
	\[\xymatrix{
	& P_{\Pin^{\pm}}M \ar[d]^{\xi} \ar[dl]_{\varphi^{*}\xi} \\
	P_{\OO}M \ar[d]_{p} \ar[r]^{d\varphi} \ar[r]^{d\varphi} & P_{\OO}M \ar[d]^{p}  \ar@/^/[l] \\
	M \ar[r]^{\varphi} & M
}\]
i.e.\ $\varphi^{*}\xi = d\varphi^{-1} \circ \xi$. We remark that total space of the principal bundle $P_{\Pin^{\pm}}M$ is the same for both $\xi$ and $\varphi^{*}\xi$: what changes is the way it covers $P_{\OO}M$. Thus, also the vector bundle of pinors $S := P_{\Pin^{\pm}}M \times_{\rho} \mathbb{C}^{2^{n}}$, being $\rho$ the standard action of $\Pin^{\pm}(2n)$ on $\mathbb{C}^{2^{n}}$ as Clifford module, has the same total space in both cases. The only difference is the projection to $M$, which can be seen ignoring the second line of the previous diagram: we simply have $p_{\varphi^{*}\xi} = \varphi^{-1} \circ p_{\xi}$, i.e.\ if we call $q_{\xi}: S_{\xi} \rightarrow M$ and $q_{\varphi^{*}\xi}: S_{\varphi^{*}\xi} \rightarrow M$ the two bundle of pinors of $\xi$ and $\varphi^{*}\xi$, it follows that $S_{\xi} = S_{\varphi^{*}\xi}$ as total spaces and $(S_{\varphi^{*}\xi})_{x} = (S_{\xi})_{\varphi(x)}$, i.e.\ $S_{\varphi^{*}\xi} = \varphi^{*}S_{\xi}$. This is the well-known geometrical property that pinors, as well as spinors, are scalars under isometries (or in general under diffeomorphisms \cite{DP}). The differential $d\varphi$ does not have any local effect on pinors (contrary to vectors), it just determines the way they must be globally thought of as pinors, i.e.\ the way the corresponding principal bundle covers $P_{\OO}M$.

\paragraph{}We recall that two pin$^{\pm}$ structures $\xi: P_{\Pin^{\pm}}M \rightarrow P_{\OO}M$ and $\xi': P'_{\Pin^{\pm}}M \rightarrow P_{\OO}M$ are equivalent if there exists a principal bundle isomorphism $\rho: P_{\Pin^{\pm}}M \rightarrow P'_{\Pin^{\pm}}M$ such that $\xi = \xi' \circ \rho$. We say that a pin$^{\pm}$ structure $\xi$ is invariant under an isometry $\varphi$ if $\xi \simeq \varphi^{*}\xi$, i.e.\ if there exists a (non-canonical) lift $\widetilde{d\varphi}$ completing the following diagram:
\begin{equation}\label{EquivalencePhiStar}
	\xymatrix{
	P_{\Pin^{\pm}}M \ar[d]_{\xi} \ar@{.>}[r]^{\widetilde{d\varphi}} & P_{\Pin^{\pm}}M \ar[d]^{\xi} \ar[dl]_{\varphi^{*}\xi} \\
	P_{\OO}M \ar[d]_{p} \ar[r]^{d\varphi} \ar[r]^{d\varphi} & P_{\OO}M \ar[d]^{p} \\
	M \ar[r]^{\varphi} & M.
}
\end{equation}
If such a $\widetilde{d\varphi}$ exists, there are only two possibilities, linked by an exchange of the two sheets: in fact, $\widetilde{d\varphi}$ is a lifting of the map $\varphi^{*}\xi$ to a $2:1$ covering of the codomain (v.\ \cite{Hatcher} prop.\ 1.34 pag.\ 62). Calling $\gamma$ the sheet exchange, the two possible liftings are $\widetilde{d\varphi}$ and $\widetilde{d\varphi} \circ \gamma$. Then $\widetilde{d\varphi} \circ \gamma = \gamma \circ \widetilde{d\varphi}$ since, if $\widetilde{d\varphi}(p_{x}) = q_{\varphi(x)}$, then the only possibility is that $\widetilde{d\varphi}(\gamma(p_{x})) = \gamma(q_{\varphi(x)})$ in order to cover $d\varphi$. 

We now consider invariance under isometries of pinors, i.e.\ of sections of the associated vector bundle. Since we just have $S_{\varphi^{*}\xi} = \varphi^{*}S_{\xi}$ so that the total spaces are the same, also the sections of the two bundles, as subset of their total spaces, are the same. In particular, a section $s \in \Gamma(S_{\xi}^{+})$ becomes naturally a section of $\varphi^{*}S_{\xi}$ via the natural map:
	\[\begin{split}
	\eta_{\varphi}:\; & \Gamma(S_{\xi}) \longrightarrow \Gamma(S_{\varphi^{*}\xi})\\
	& s \longrightarrow \eta_{\varphi}(s): \; \eta_{\varphi}(s)_{x} := s_{\varphi^{-1}(x)}.
\end{split}\]
This is the scalar behavior of a pinor field. If $\xi \simeq \varphi^{*}\xi$, from diagram \eqref{EquivalencePhiStar} we have the isomorphism $\widetilde{d\varphi}: P_{\Spin}M \rightarrow P_{\Spin}M$, unique up to sheet exchange, so we have a vector bundle map $\widetilde{d\varphi}: S_{\varphi^{*}\xi} \rightarrow S_{\xi}$: the ambiguity of $\widetilde{d\varphi}$ corresponds via $\rho$ to a sign ambiguity on the action on $S_{\xi}$. Thus we have a map:
	\[S_{\xi} \overset{\id}\longrightarrow S_{\varphi^{*}\xi} \overset{\pm\widetilde{d\varphi}}\longrightarrow S_{\xi}
\]
whose behavior with respect to the base point is:
	\[(S_{\xi})_{x} \overset{\id}\longrightarrow (S_{\xi})_{x} = (S_{\varphi^{*}\xi})_{\varphi^{-1}(x)} \overset{\pm\widetilde{d\varphi}}\longrightarrow (S_{\xi})_{x}
\]
(note that $\widetilde{d\varphi}$ commutes with projections since it is a bundle map, while the identity is not) inducing the natural map of sections:
	\[\xymatrix{
	\Gamma(S_{\xi}) \ar[r]^{\eta_{\varphi}} & \Gamma(S_{\varphi^{*}\xi}) \ar[r]^{\pm\widetilde{d\varphi}} & \Gamma(S_{\xi})
}\]
so that the invariance condition reads:
	\[s = \pm \widetilde{d\varphi} \circ \eta_{\varphi}(s)
\]
i.e.\ $s_{x} = \pm\widetilde{d\varphi}(s_{\varphi^{-1}(x)})$. We remain with a sign ambiguity, contrary to the case of vectors, in which case we can completely define invariance of a section by requiring that $d\varphi(v) = v$. As explained in \cite{DP}, for pinors (as well as for spinors) we have just a \emph{projective} action of $\varphi$. So also the notion of invariance is affected by this.

\subsection{Double covering of a non-orientable manifold}

As is well-known, every non-orientable manifold $X$ has an orientable double-cover $\tilde{X}$ with an orientation-reversing involution $\tau$ such that $X \simeq \tilde{X} \,/\, \tau$. It can be constructed as follows: we choose an atlas $\{(U_{\alpha}, \psi_{\alpha})\}_{\alpha \in I}$ of $X$ with corresponding transition functions $g_{\alpha\beta}$, and we consider the $\mathbb{Z}_{2}$-bundle with charts $U_{\alpha} \times \mathbb{Z}_{2}$ and transition functions $\varepsilon_{\alpha\beta}$ equal to the sign of the Jacobian $J(g_{\alpha\beta})$. The involution $\tau$ is the exchange of the two sheets. If we consider the projection $\pi: \tilde{X} \rightarrow X$, then $\tilde{X}$ as a manifold has an atlas given by couples of charts $\{(\pi^{-1}U_{\alpha}, \psi_{\alpha} \circ \pi)\}_{\alpha \in I}$ so that the transition functions are still $g_{\alpha\beta}$ for both the components of $\pi^{-1}(U_{\alpha\beta})$. We now want to study the behavior of $\pi_{*}: H_{1}(\tilde{X}, \mathbb{Z}_{2}) \rightarrow H_{1}(X, \mathbb{Z}_{2})$ and consequently of $\pi^{*}: H^{1}(X, \mathbb{Z}_{2}) \rightarrow H^{1}(\tilde{X}, \mathbb{Z}_{2})$. We recall the following canonical isomorphisms \cite{Hatcher}:
\begin{equation}\label{CanonicalIso}
\begin{split}
	&H_{1}(X, \mathbb{Z}) \simeq \Ab \, \pi_{1}(X)\\
	&H_{1}(X, \mathbb{Z}_{2}) \simeq H_{1}(X, \mathbb{Z}) \otimes_{\mathbb{Z}} \mathbb{Z}_{2}\\
	&H^{1}(X, \mathbb{Z}_{2}) \simeq \Hom(H_{1}(X, \mathbb{Z}), \mathbb{Z}_{2}) \simeq \Hom(H_{1}(X, \mathbb{Z}_{2}), \mathbb{Z}_{2}).
\end{split}
\end{equation}

\paragraph{}Let us show that, for $\pi_{*}: H_{1}(\tilde{X}, \mathbb{Z}_{2}) \rightarrow H_{1}(X, \mathbb{Z}_{2})$, the image $\IIm \, \pi_{*}$ has always index two in $H_{1}(X, \mathbb{Z}_{2})$, or equivalently that $\Ker \, \pi^{*} \simeq \mathbb{Z}_{2}$ in $H^{1}(X, \mathbb{Z}_{2})$. This can be seen in two ways. One is the following simple algebraic lemma:
\begin{Lemma} Let $G$ be a group and $H$ a subgroup such that $[G:H] = 2$. Then the natural map $\Ab\,H \rightarrow \Ab\,G$ has image of index $2$.
\end{Lemma}
\textbf{Proof:} For $g \in G$, $g$ and $g^{-1}$ lie in the same $H$-coset. Thus, considering a commutator $g_{1}g_{2}g_{1}^{-1}g_{2}^{-1}$ we have that an even number of factors can lie in the coset $G \setminus H$, thus the product lives in $H$; hence $G' \leq H$. Let us prove that the image of the natural map $\psi: H / H' \rightarrow G / G'$ has index $2$. If $g' = gh$ for $g, g' \in G$ and $h \in H$, then $[g']_{G/G'} = [g]_{G/G'}[h]_{G/G'} = [g]_{G/G'}\psi([h]_{H/H'})$. Thus the number of cosets of $\IIm \psi$ in $G/G'$ is less than $2$. Moreover, if $g \in G\setminus H$ it cannot happen that $[g]_{G/G'} = \psi([h]_{H/H'})$, because otherwise $g \in h \cdot G' \subset h \cdot H = H$. $\square$

\paragraph{}The previous lemma for $G = \pi_{1}(X)$ and $H = \pi_{1}(\tilde{X})$ implies that $[H_{1}(X, \mathbb{Z}) : \pi_{*}H_{1}(\tilde{X}, \mathbb{Z})] = 2$. The other way to prove the latter result is by means of the following exact sequence in cohomology, which can be found in \cite{MS}:
	\[\xymatrix{
	\cdots \ar[r] & H^{i-1}(X, \mathbb{Z}_{2}) \ar[r]^{\cup w_{1}(X)} & H^{i}(X, \mathbb{Z}_{2}) \ar[r]^{\pi^{*}} & H^{i}(\tilde{X}, \mathbb{Z}_{2}) \ar[r] & H^{i}(X, \mathbb{Z}_{2}) \ar[r] & \cdots.
}\]
For $i = 1$, since $H^{0}(X, \mathbb{Z}_{2}) = \mathbb{Z}_{2}$ we have that $\IIm(\cup w_{1}(X)) = w_{1}(X)$, thus by exactness $\Ker \, \pi^{*} = \{0, w_{1}(X)\} \simeq \mathbb{Z}_{2}$. This is what we expected: since the double covering is orientable, the pull-back $\pi^{*}$ must kill $w_{1}(X)$.

\paragraph{}Since $\mathbb{Z}_{2}$ is a field, $H^{1}(\tilde{X}, \mathbb{Z}_{2})$ is a vector space, thus every subspace can be complemented. Hence we have:
\begin{equation}\label{SplitH1Z2}
	H^{1}(\tilde{X}, \mathbb{Z}_{2}) = \mathbb{Z}_{2}^{k} \oplus \IIm \, \pi^{*}
\end{equation}
where $\pi^{*}: H^{1}(X, \mathbb{Z}_{2}) \rightarrow \IIm \, \pi^{*}$ is a surjection with kernel $\mathbb{Z}_{2}$. Since $H^{1}(X, \mathbb{Z}_{2}) = H^{1}(X, \mathbb{Z}) \otimes_{\mathbb{Z}} \mathbb{Z}_{2}$, its dimension is equal to the Betti number $b_{1}(X)$ plus the number of $\mathbb{Z}_{2^{k}}$-components of $H^{1}(X, \mathbb{Z})$: we call this number $b_{1}^{(2)}(X)$. The dimension of $\IIm \, \pi^{*}$ is thus $b_{1}^{(2)}(X)-1$ so, in \eqref{SplitH1Z2}, we have $k = b_{1}^{(2)}(\tilde{X})-b_{1}^{(2)}(X)+1$. Thus we get the following general picture:
	\[\xymatrix{
	\mathbb{Z}_{2}^{\oplus ( b_{1}^{(2)}(X)-1 )} \oplus (\mathbb{Z}_{2}^{\oplus ( b_{1}^{(2)}(\tilde{X})-b_{1}^{(2)}(X)+1 )} \simeq \Coker\, \pi^{*}) \; \,\\
	\mathbb{Z}_{2}^{\oplus (b_{1}^{(2)}(X)-1)} \oplus (\mathbb{Z}_{2} = \Ker\, \pi^{*}) \ar@<18ex>[u]^{\pi^{*}}_{\simeq} \;.\qquad\qquad\qquad\qquad
}\]

\paragraph{}In the sequel we will also need another general result: we compare the tangent bundle of $\tilde{X}$ and the tangent bundle of $X$. We recall that if $f: X \rightarrow Y$ is a continuous map and $p: E \rightarrow Y$ a fiber bundle, the pull-back $\pi_{2}: f^{*}E \rightarrow X$ is defined as the fiber product $E \times_{Y} X$ via $\pi$ and $f$, thus its elements are of the form $(e, x)$ with $p(e) = f(x)$. The projection is $\pi_{2}(e,x) = x$.
\begin{Lemma}\label{TangentLemma} For $\pi: \tilde{X} \rightarrow X$ the projection and $p: TX \rightarrow X$, $\tilde{p}: T\tilde{X} \rightarrow \tilde{X}$ the tangent bundles, there is the canonical bundle isomorphism:
	\[\begin{split}
	\varphi: \; &T\tilde{X} \overset{\simeq}\longrightarrow \pi^{*}TX\\
	&\varphi(v) = (d\pi(v), \tilde{p}(v)).
\end{split}\]
Similarly for the orthogonal frame bundles with respect to a metric $g$ on $X$ and its pull-back $\pi^{*}g$ on $\tilde{X}$ there is the canonical isomorphism:
	\[\begin{split}
	\varphi_{O}: \; &P_{O}\tilde{X} \overset{\simeq}\longrightarrow \pi^{*}P_{O}X\\
	&\varphi_{O}(x) = (d\pi(x), \tilde{p}(x)).
\end{split}\]
\end{Lemma}
\textbf{Proof:} It is easy to verify that $\varphi$ is a well-defined bundle map. It also follows from the definition of pull-back:
\begin{equation}\label{PiCartesian}
	\xymatrix{
	T\tilde{X} \ar[d]_{\tilde{p}} \ar@{.>}[r]_{\varphi} \ar@/^1pc/[rr]^{d\pi} & \pi^{*}TX \ar[r]_{\pi_{1}} \ar[d]_{\pi_{2}} & TX \ar[d]^{p}\\
	\tilde{X} \ar@/_1pc/[rr]_{\pi} \ar[r]^{\id} & \tilde{X} \ar[r]^{\pi} & X
}
\end{equation}
By construction the map $\varphi$ lifts the identity of $\tilde{X}$ and it must satisfy $\pi_{1}\varphi(v) = d\pi(v)$, thus $\varphi(v) = (d\pi(v), \tilde{p}(v))$. We have to verify that it is an isomorphism on each fiber, i.e.\ that $d\pi_{x}$ is an isomorphism for each $x$: this is true since $\pi$ is a local diffeomorphism. The same considerations apply for frame bundles. $\square$

\subsection{Pinors on the double covering}\label{PinorsDouble}

We now want to compare pinors on a non-orientable manifold $X$ and pinors on its double covering $\tilde{X}$ which are $\tau$-invariant. We start with the following simple lemma:
\begin{Lemma} If $X$ admits a $\Pin^{+}$-structure \emph{or} a $\Pin^{-}$-structure then $\tilde{X}$ is spin.
\end{Lemma}
\textbf{Proof:} by lemma \ref{TangentLemma} we have that $w_{2}(\tilde{X}) = \pi^{*}w_{2}(X)$. Since $w_{1}(X) \in \Ker\,\pi^{*}$, we obtain $\pi^{*}w_{2}(X) = \pi^{*}(w_{2}(X) + w_{1}(X) \cup w_{1}(X))$, thus if there is a pin$^{\pm}$ structure we get $w_{2}(\tilde{X}) = 0$. $\square$

\paragraph{}Let us suppose that a pin$^{\pm}$ structure on $\tilde{X}$ is $\tau$-invariant. Thus we have two possible liftings of $d\tau$:
	\[\xymatrix{
	P_{\Pin^{\pm}}\tilde{X} \ar[d]_{\tilde{\xi}} \ar[rr]^{\widetilde{d\tau}, \;\widetilde{d\tau} \circ \gamma} & & P_{\Pin^{\pm}}\tilde{X} \ar[d]^{\tilde{\xi}} \ar[dll]_{\tau^{*}\tilde{\xi}} \\
	P_{O}\tilde{X} \ar[d]_{\tilde{p}} \ar[rr]^{d\tau} & & P_{O}\tilde{X} \ar[d]^{\tilde{p}} \\
	\tilde{X} \ar[rr]^{\tau} & & \,\tilde{X}.
}\]
Since $\widetilde{d\tau} \circ \gamma = \gamma \circ \widetilde{d\tau}$, it follows that $(\widetilde{d\tau} \circ \gamma)^{2} = \widetilde{d\tau}^{2}$, and the latter can be only $\id$ or $\gamma$, since it is an auto-equivalence of $\tilde{\xi}$ which covers $d\tau^{2} = \id$.

\paragraph{} We would like to show that the pull-back of a pin$^{\pm}$ structure on $X$ is a pin$^{\pm}$ structure on $\tilde{X}$ which is $\tau$-invariant and such that $\widetilde{d\tau}^{2} = \id$. Then, $\tilde{X}$ being orientable, we will be able to reduce the structure group to $\Spin$. Let us consider the following diagram:
\begin{equation}\label{PinPullback}
	\xymatrix{
	& \pi^{*}P_{\Pin^{\pm}}X \ar[r]^{\pi_{1}} \ar[d]|{(\xi,\id)} \ar[dl]_{\tilde{\xi}} & P_{\Pin^{\pm}}X \ar[d]^{\xi} \\
	P_{O}\tilde{X} \ar[d]_{\tilde{p}} \ar@{.>}[r]_{\varphi} \ar@/^2pc/[rr]_{\qquad\qquad\qquad\qquad\; d\pi} & \pi^{*}P_{O}X \ar[r]_{\pi_{1}} \ar[d]_{\pi_{2}} \ar@/_/@{.>}[l] & P_{O}X \ar[d]^{p}\\
	\tilde{X} \ar@/_2pc/[rr]^{\pi} \ar[r]^{\id} & \tilde{X} \ar[r]^{\pi} & X
}
\end{equation}
where $\tilde{\xi}$ defines the pull-back on $\tilde{X}$ of the spin structure $\xi$ of $X$, for $\varphi$ defined in lemma \ref{TangentLemma}. Thus we consider as total space of the bundle exactly $\pi^{*}P_{\Pin^{\pm}}X$. We now see that $\tilde{\xi}$ is $\tau$-invariant. We recall that $\tau^{*}\tilde{\xi}$ is defined by:
	\[\xymatrix{
	& \pi^{*}\Pin^{\pm}X \ar[d]_{\tilde{\xi}} \ar[dl]_{\tau^{*}\tilde{\xi}} \\
	P_{O}\tilde{X} \ar[d]_{\tilde{p}} \ar[r]^{d\tau} & P_{O}\tilde{X} \ar[d]^{\tilde{p}} \\
	\tilde{X} \ar[r]^{\tau} & \tilde{X}
}\]
and we claim that $\tau^{*}\tilde{\xi} \simeq \tilde{\xi}$ via the two possible equivalences:
	\[\xymatrix{
	\pi^{*}P_{\Pin^{\pm}}X \ar[d]_{\tilde{\xi}} \ar[rr]^{(1,\tau), (\gamma,\tau)} & & \pi^{*}P_{\Pin^{\pm}}X \ar[d]^{\tilde{\xi}} \ar[dll]_{\tau^{*}\xi} \\
	P_{O}\tilde{X} \ar[d]_{\tilde{p}} \ar[rr]^{d\tau} & & P_{O}\tilde{X} \ar[d]^{\tilde{p}} \\
	\tilde{X} \ar[rr]^{\tau} & & \,\tilde{X}
}\]
where $\gamma$ is the exchange of sheets of $P_{\Pin^{\pm}}X$ with respect to $P_{O}X$, while $\tau$ is the exchange of sheets of $\tilde{X}$ with respect to $X$. In fact, by diagram \eqref{PinPullback} we have $\tilde{\xi}(p', \tilde{x}) = \varphi^{-1}\circ (\xi,\id)(p', \tilde{x}) = \varphi^{-1}(p, \tilde{x}) = d\pi^{-1}_{\tilde{x}}(p)$ where $\pi_{\tilde{x}}$ is $\pi$ restricted to a neighborhood of $\tilde{x}$ on which it is a diffeomorphism. Therefore, for $\varepsilon = 1, \gamma$:
	\[\begin{split}
	&d\tau \circ \tilde{\xi}(p', \tilde{x}) = d\tau(d\pi^{-1}_{\tilde{x}}(p)) = d(\tau \circ \pi^{-1}_{\tilde{x}})(p) = d(\pi^{-1}_{\tau(\tilde{x})})(p)\\
	&\tilde{\xi} \circ (\varepsilon, \tau)(p', \tilde{x}) = \tilde{\xi}(\varepsilon(p'), \tau(\tilde{x})) = d(\pi^{-1}_{\tau(\tilde{x})})(p)
\end{split}\]
so that the diagram commutes. In particular, we see that the two possible isomorphisms $\widetilde{d\tau} = (1, \tau), (\gamma, \tau)$ have the property that $\widetilde{d\tau}^{2} = 1$. We have thus constructed a function:
	\[\Phi: \; \{\textnormal{pin$^{\pm}$ structures on } X \} \longrightarrow \{\textnormal{pin$^{\pm}$ structures on } \tilde{X} \textnormal{ $\tau$-invariant with } \widetilde{d\tau}^{2} = 1\}.
\]

\paragraph{}We now show that $\Phi$ is surjective, i.e.\ that a $\tau$-invariant pin$^{\pm}$ structure $\tilde{\xi}$ on $\tilde{X}$ satisfying $\widetilde{d\tau}^{2} = 1$ is the pull-back of a pin$^{\pm}$ structure on $X$. The latter is:
	\[\xi: P_{\Pin^{\pm}}\tilde{X} \, / \, \widetilde{d\tau} \longrightarrow P_{O}\tilde{X} \, / \, d\tau \simeq P_{O}X.
\]
In more detail:
	\[\xymatrix{
		P_{\Pin^{\pm}}\tilde{X} \,/\, \widetilde{d\tau} \ar[d]_{[\tilde{\xi}]} \ar[dr]^{\xi} \\
		P_{O}\tilde{X} \,/\, d\tau \ar[d]_{[\pi_{2}]} \ar[r]_{\quad \simeq}^{\quad \nu} & P_{O}X \ar[d]^{p} \\
		\tilde{X} \,/\, \tau \ar[r]^{[\pi]}_{\simeq} & \; X
	}
\]
where $\nu([p]) = d\pi(p)$. From $\widetilde{d\tau}^{2} = 1$ we get that the quotient is a 2-covering of $P_{O}X$, otherwise we would obtain a 1-covering, i.e.\ a bundle isomorphism, since $\gamma = \widetilde{d\tau}^{2}$ would identify also the two points of the same fiber. To see that $\tilde{\xi} \simeq \pi^{*}\xi$, we use the equivalence:
\begin{equation}\label{Mu}
\xymatrix{
	P_{\Pin^{\pm}}\tilde{X} \ar[dr]_{\tilde{\xi}} \ar[rr]^{\mu \quad}_{\simeq \quad} & & \pi^{*}(P_{\Pin^{\pm}}\tilde{X} \,/\, \widetilde{d\tau}) \ar[dl]^{\varphi^{-1} \circ (\xi, \id)} \\
	& P_{O}\tilde{X}
}
\end{equation}
for $\mu(\tilde{p}_{\tilde{x}}) = ([\tilde{p}_{\tilde{x}}], \tilde{x})$. The inverse of $\mu$ is given by $\mu^{-1}([\tilde{p}_{\tilde{x}}], \tilde{x}) = \tilde{p}_{\tilde{x}}$ or equivalently $\mu^{-1}([\tilde{p}_{\tilde{x}}], \tau(\tilde{x})) = \widetilde{d\tau}(\tilde{p}_{\tilde{x}})$. The diagram is commutative: $\varphi^{-1}\circ (\xi, \id) \circ \mu(\tilde{p}_{\tilde{x}}) = \varphi^{-1}\circ (\xi, \id)([\tilde{p}_{\tilde{x}}], \tilde{x}) = \varphi^{-1}((\nu\circ[\xi])([\tilde{p}_{\tilde{x}}]), \tilde{x}) = \varphi^{-1}(d\pi(\tilde{\xi}(\tilde{p}_{\tilde{x}})), \tilde{x}) = \tilde{\xi}(\tilde{p}_{\tilde{x}})$.

\paragraph{}It is easy to show that $\Phi$ commutes via $\pi^{*}$ with the actions of $H^{1}(X, \mathbb{Z}_{2})$ and $H^{1}(\tilde{X}, \mathbb{Z}_{2})$. In fact, for $\xi: P_{\Pin^{\pm}}X \rightarrow P_{O}X$ a pin$^{\pm}$ structure, up to isomorphism we can view $\Phi(\xi)$ as $\pi^{*}\xi: \pi^{*}P_{\Pin^{\pm}}X \rightarrow \pi^{*}P_{O}X$. We fix a $\rm\check{C}$ech class $[\omega] \in \check{H}^{1}(\mathfrak{U}, \mathbb{Z}_{2})$ for a good cover $\mathfrak{U} = \{U_{\alpha}\}_{\alpha \in I}$ of $X$. If the transition function of $P_{\Pin^{\pm}}X$ are $s_{\alpha\beta}$ and we fix a representative $\omega$, then the new transition functions are $s_{\alpha\beta} \cdot \omega_{\alpha\beta}$. On the two components of $\pi^{-1}U_{\alpha\beta}$, the transition functions were both $s_{\alpha\beta}$ and they become both $s_{\alpha\beta} \cdot \omega_{\alpha\beta}$, i.e.\ $\omega$ acts on the transition functions of $\pi^{*}\xi$ exactly as $\pi^{*}\omega$. Since $[\pi^{*}\omega] = \pi^{*}[\omega]$, we get the claim. Thus we have a diagram:
	\[\begin{array}{ccc}
	\{\Pin^{\pm}\textnormal{-structures on } X \} & \overset{\Phi}\longrightarrow & \{\Pin^{\pm}\textnormal{-structures on } \tilde{X} \textnormal{ $\tau$-invariant with } \widetilde{d\tau}^{2} = 1\} \\
	\circlearrowleft & & \circlearrowleft \\
	\check{H}^{1}(X, \mathbb{Z}_{2}) & \overset{\pi^{*}}\longrightarrow & \check{H}^{1}(\tilde{X}, \mathbb{Z}_{2}).
\end{array}\]
This implies in particular that $\Phi^{-1}(\tilde{\xi})$ is made by two inequivalent pin$^{\pm}$ structures, obtainable from each other via the action of $w_{1}(X) \in \Ker\,\pi^{*}$. We will now show that the two inequivalent counterimages can be recovered as $P_{\Pin^{\pm}}\tilde{X} \, / \, \widetilde{d\tau}$ and $P_{\Pin^{\pm}}\tilde{X} \, / \, (\widetilde{d\tau} \circ \gamma)$, by proving that these two quotients are inequivalent. In fact, let us suppose that there exists an equivalence:
	\[\xymatrix{
	P_{\Pin^{\pm}}\tilde{X} \, / \, \widetilde{d\tau} \ar[rr]^{\rho} \ar[dr]_{\xi} & & P_{\Pin^{\pm}}\tilde{X} \, / \, (\widetilde{d\tau} \circ \gamma) \ar[dl]^{\xi'} \\
	& P_{O}
}\]
then it lifts to an equivalence of the pull-backs:
	\[\xymatrix{
	& P_{O}\tilde{X} \\
	\pi^{*}(P_{\Pin^{\pm}}\tilde{X} \, / \, \widetilde{d\tau}) \ar[rr]^{\tilde{\rho}} \ar[d]_{\pi_{1}} \ar[ur]^{\tilde{\xi}} & & \pi^{*}(P_{\Pin^{\pm}}\tilde{X} \, / \, (\widetilde{d\tau} \circ \gamma)) \ar[d]^{\pi_{1}} \ar[ul]_{\tilde{\xi}'} \\
	P_{\Pin^{\pm}}\tilde{X} \, / \, \widetilde{d\tau} \ar[rr]^{\rho} \ar[dr]_{\xi} & & P_{\Pin^{\pm}}\tilde{X} \, / \, (\widetilde{d\tau} \circ \gamma) \ar[dl]^{\xi'} \\
	& P_{O}X
}\]
but, being both the pull-backs equivalent to $P_{\Pin^{\pm}}\tilde{X}$ via \eqref{Mu}, the only two possibilities for $\tilde{\rho}$ are the following:
	\[\xymatrix{
	P_{\Pin^{\pm}}\tilde{X} \ar[r]^{\id, \gamma} \ar[d]_{\mu} & P_{\Pin^{\pm}}\tilde{X} \ar[d]^{\mu'} \\
	\pi^{*}(P_{\Pin^{\pm}}\tilde{X} \, / \, \widetilde{d\tau}) \ar[r]^{\tilde{\rho} \quad} \ar@/_/[u] & \pi^{*}(P_{\Pin^{\pm}}\tilde{X} \, / \, (\widetilde{d\tau} \circ \gamma)). \ar@/^/[u]
}\]
Let us show that none of the two can be a lift of $\rho$. In fact, if it were so, they would be of the form:
\begin{equation}\label{RhoLift}
\tilde{\rho}([p], \tilde{x}) = (\rho[p], \tilde{x})
\end{equation}
while:
	\[\xymatrix{
	([p_{\tilde{x}}], \tilde{x}) \ar[r]^{\quad \mu^{-1}} & p_{\tilde{x}} \ar[r]^{\id} & p_{\tilde{x}} \ar[r]^{\mu' \quad} & ([p_{\tilde{x}}], \tilde{x}) \\
	([p_{\tilde{x}}], \tau(\tilde{x})) \ar[r]^{\quad \mu^{-1}} & \widetilde{d\tau}(p_{\tilde{x}}) \ar[r]^{\id} & \widetilde{d\tau}(p_{\tilde{x}}) \ar[r]^{\mu' \quad} & ([\widetilde{d\tau}(p_{\tilde{x}})], \tau(\tilde{x}))
}\]
and in the codomain $[p_{\tilde{x}}] \neq [\widetilde{d\tau}(p_{\tilde{x}})]$ since the class is taken with respect to $\widetilde{d\tau} \circ \gamma$, thus \eqref{RhoLift} is inconsistent. The same would happen choosing $\gamma$ instead of the identity. Thus $\tilde{\rho}$ lifts only the autoequivalences of each of the two quotients, not an equivalence between them.

\paragraph{}Now that we have seen the relationship between pin$^{\pm}$ structures on $X$ and the corresponding ones on $\tilde{X}$, we analyze such a relationship at the level of pinors (i.e.\ sections of the associated vector bundles). Let us start from $X$ and pull-back a pin$^{\pm}$ structure as in the following diagram:
	\[\xymatrix{
	\pi^{*}P_{\Pin^{\pm}}X \ar[d]_{\tilde{\xi}} \ar[r]^{\;\; \pi_{1}} & P_{\Pin^{\pm}}X \ar[d]^{\xi} \\
	P_{O}\tilde{X} \ar[r]^{d\pi} & P_{O}X.
}\]
For the associated bundles of pinors, we have that $(\pi^{*}P_{\Pin^{\pm}}X) \times_{\rho} \mathbb{C}^{2^{n}} \simeq \pi^{*}(P_{\Pin^{\pm}}X \times_{\rho} \mathbb{C}^{2^{n}})$ canonically. Thus, given on $X$ a pinor $s \in \Gamma(P_{\Pin^{\pm}}X \times_{\rho} \mathbb{C}^{2^{n}})$, we can naturally consider on $\tilde{X}$ its pull-back $\pi^{*}s \in \Gamma((\pi^{*}P_{\Pin^{\pm}}X) \times_{\rho} \mathbb{C}^{2^{n}})$. The natural equivalence between $\tilde{\xi}$ and $\tau^{*}\tilde{\xi}$ is given by $\widetilde{d\tau}(p, \tilde{x}) = (p, \tau(\tilde{x}))$, and, if we extend it to the associated vector bundles, we have that a section $s' \in \Gamma((\pi^{*}P_{\Pin^{\pm}}X) \times_{\rho} \mathbb{C}^{2^{n}})$ is the pull-back of a section on $X$ if and only if $\widetilde{d\tau}(s') = s'$.

Viceversa, let us start from $\tilde{X}$. We fix a pin$^{\pm}$ structure $\tilde{\xi}$ such that $\tilde{\xi} \simeq \tau^{*}\tilde{\xi}$ with $\widetilde{d\tau}^{2} = 1$. Then there are two natural vector space isomorphisms:
\begin{itemize}
	\item $\widetilde{d\tau}$-invariant sections of the associated bundle correspond to sections of the pin$^{\pm}$ structure $P_{\Pin^{\pm}}\tilde{X} \,/\, \widetilde{d\tau}$ on $X$;
	\item $(\widetilde{d\tau} \circ \gamma)$-invariant sections of the associated bundle correspond to sections of the pin$^{\pm}$ structure $P_{\Pin^{\pm}}\tilde{X} \,/\, (\widetilde{d\tau} \circ \gamma)$ on $X$.
\end{itemize}
In particular, $s = \widetilde{d\tau} \circ \eta_{\tau}(s)$ means that $s_{x} = \widetilde{d\tau}(s_{\tau(x)})$, while $s = \widetilde{d\tau} \circ \gamma \circ \eta_{\tau}(s)$ means that $s_{x} = -\widetilde{d\tau}(s_{\tau(x)})$, since the action of $\gamma$ corresponds to the multiplication by $-1 \in \Pin^{\pm}(n)$. We remark that if we want to describe invariance of pinors under general isometries, we must take into account the sign ambiguity discussed in the first section. In the present case, since we distinguish $\widetilde{d\tau}$ and $\widetilde{d\tau} \circ \gamma$ on the basis of the associated quotient on $X$, we fix this ambiguity.

\subsection{Spinors and orientation-reversing isometries}\label{SpinorsOrientRev}

The discussion of the first paragraph about pin$^{\pm}$ structures and isometries holds also for spin structures and orientation-\emph{preserving} isometries, but it must be modified in order to deal with orientation-\emph{reversing} isometries. In fact, in the previous definition of invariance, if $\varphi$ does not preserve the orientation, the differential $d\varphi$ does not have the same $\SO$-bundle for domain and codomain.\footnote{The two bundles could be isomorphic but not canonically, so we cannot think of $d\varphi$ as an automorphism anyway.} Thus in this case it does not make sense to speak about invariant spin structures. Let us consider what happens on the associated vector bundle of spinors: we have seen that the vector bundles of spinors $q: S \rightarrow M$ and $q': S'\rightarrow M$ corresponding to $\xi$ and $\varphi^{*}\xi$ satisfy $S'_{x} = S_{\varphi^{-1}(x)}$. If we split the bundles into chiral sub-bundles $S = S^{+} \oplus S^{-}$ and $S' = {S'}^{+} \oplus {S'}^{-}$, we have that for $\varphi$ orientation-reversing ${S'}^{+}_{x} = S^{-}_{\varphi^{-1}(x)}$ and viceversa, i.e.\ the chiralities are reversed. The reason for this is that the chirality element of the Clifford algebra of $TM$, which is the product of elements of an oriented orthonormal basis, becomes pointwise its own opposite if we change orientation (it is enough to multiply by $-1$ one of the vectors). So for $\xi$ and $\varphi^{*}\xi$ the chirality elements are opposite. Thus, to define invariance, we must consider the case of different spin structures for the two chiralities, i.e.\ we must deal with \emph{ordered couples of spin structures}.
\begin{Def} The \emph{bundle of spinors} associated to an ordered couple of spin structures $(\xi, \xi')$ is the vector bundle $S_{\xi,\xi'} := S^{+}_{\xi} \oplus S^{-}_{\xi'}$, where $S^{+}_{\xi}$ is the bundle of positive-chiral spinors with structure $\xi$ and $S^{-}_{\xi'}$ is the bundle of negative-chiral spinors with structure $\xi'$.
\end{Def}
\begin{Def} An ordered couple of spin structures $(\xi, \xi')$ is \emph{oriented} if both spin structures lift the bundle of frames relative to the same orientation.
\end{Def}
We can now consider the pull-back of couples of spin structures.
\begin{Def} For $\varphi: M \rightarrow M$ an isometry and $(\xi, \xi')$ an ordered couple of spin structures, we define:
\begin{itemize}
	\item for $\varphi$ orientation-preserving, $\varphi^{*}(\xi, \xi') := (\varphi^{*}\xi, \varphi^{*}\xi')$;
	\item for $\varphi$ orientation-reversing, $\varphi^{*}(\xi, \xi') := (\varphi^{*}\xi', \varphi^{*}\xi)$.
\end{itemize}
We say that $(\xi, \xi')$ is \emph{invariant under $\varphi$} if $\varphi^{*}(\xi, \xi') \simeq (\xi, \xi')$, where $\simeq$ means that they are componentwise equivalent.
\end{Def}
Let us now see that for $\varphi$ orientation-preserving, if $\xi$ is invariant then the couple $(\xi, \xi)$ is invariant. If $\varphi$ is orientation-reversing, when the two components are equal the couple is never invariant; in order for a couple to be invariant in the latter case it must satisfy $\xi' \simeq \varphi^{*}\xi$ and $\xi \simeq \varphi^{*}\xi'$. Denoting by $u$ and $u'$ the two possible orientations, diagram \eqref{EquivalencePhiStar} (page \pageref{EquivalencePhiStar}) becomes now:
	\[\xymatrix{
	P_{\Spin}M \ar[d]_{\xi} \ar@{.>}[r]^{\widetilde{d\varphi}} & P'_{\Spin}M \ar[d]^{\xi'} \ar[dl]_{\varphi^{*}\xi} \\
	P_{\SO_{u}}M \ar[d]_{p} \ar[r]^{d\varphi} \ar[r]^{d\varphi} & P_{\SO_{u'}}M \ar[d]^{p'} \\
	M \ar[r]^{\varphi} & M.
}\]
and it gives the equivalence $\varphi^{*}\xi \simeq \xi'$. We have also the analogous diagram reversing $\xi$ and $\xi'$. Then we have a canonical representative $(\xi, \varphi^{*}\xi)$, satisfying $(\varphi^{*})^{2}\xi \simeq \xi$: then, the previous diagram becomes:
	\[\xymatrix{
	P_{\Spin}M \ar[d]_{\xi} \ar@{.>}[r]^{\id, \gamma} & P_{\Spin}M \ar[d]^{\varphi^{*}\xi} \ar[dl]_{\varphi^{*}\xi} \\
	P_{\SO_{u}}M \ar[d]_{p} \ar[r]^{d\varphi} \ar[r]^{d\varphi} & P_{\SO_{u'}}M \ar[d]^{p'} \\
	M \ar[r]^{\varphi} & M
}\]
where $\gamma$ is the sheet exchange. However, we can canonically choose the identity as equivalence. The ambiguity is left in the choice of $\widetilde{d(\varphi^{2})}$ for the equivalence $(\varphi^{*})^{2}\xi \simeq \xi$. However, when $\varphi$ is an involution, every couple of the form $(\xi, \varphi^{*}\xi)$ is $\varphi$-invariant, and no ambiguity is left in the choice of equivalences.

\paragraph{}We now consider the invariance of spinors for $\varphi$-invariant couples of spin structures. In the orientation-preserving case, since the two elements of a couple are independent, the same considerations of the first paragraph about pinors apply separately for both elements of the couple. For the orientation-reversing case, we have the same ambiguity for couples $(\xi, \xi')$ with $\xi' \simeq \varphi^{*}\xi$ and $\xi \simeq \varphi^{*}\xi'$. If, as discussed before, $\varphi$ is an involution and we choose the canonical representative $(\xi, \varphi^{*}\xi)$, we do not have any ambiguity left, and we have a good notion of $\varphi$-invariant spinor. We call the involution $\tau$ instead of $\varphi$. In this case, the invariance acts as follows: let us consider a section $s \in \Gamma(S_{\xi, \tau^{*}\xi})$, i.e.\ a section $(s^{+}, s^{-}) \in \Gamma(S^{+}_{\xi} \oplus S^{-}_{\tau^{*}\xi})$. Then:
	\[s^{+}_{x} \in (S^{+}_{\xi})_{x} \qquad s^{-}_{x} \in (S^{-}_{\tau^{*}\xi})_{x} = (S^{+}_{\xi})_{\tau(x)}
\]
and we have the identity $(S^{+}_{\xi})_{x} \longrightarrow (S^{+}_{\xi})_{x} = (S^{-}_{\tau^{*}\xi})_{\tau(x)}$ inducing on sections the map $\eta_{\tau}: \Gamma(S^{+}_{\xi}) \rightarrow \Gamma(S^{-}_{\tau^{*}\xi})$. Thus we can ask $s^{-} = \eta_{\tau}(s^{+})$, i.e.\ $s^{-}_{x} = \eta_{\tau}(s^{+})_{x} = s^{+}_{\tau(x)}$. An invariant couple is thus of the form $(s^{+}, \eta_{\tau}(s^{+}))$, which corresponds to $(s^{+}_{x}, s^{+}_{\tau(x)})$.

We remark that given a generic couple $(s^{+}, s^{-})$ we can write it as:
	\[\bigl(\; \textstyle\frac{1}{2}(s^{+} + \eta(s^{-})) + \frac{1}{2}(s^{+} - \eta(s^{-})), \; \frac{1}{2}(s^{-} + \eta(s^{+})) + \frac{1}{2}(s^{-} - \eta(s^{+})) \;\bigr)
\]
so that we can project it to an invariant couple by considering:
	\[\bigl(\; \textstyle\frac{1}{2}(s^{+} + \eta(s^{-})), \; \frac{1}{2}(s^{-} + \eta(s^{+})) \;\bigr).
\]
This is coherent with closed unoriented superstring theory, where the quantum states surviving the projection are of the form $(\psi_{n} + \tilde{\psi}_{n})\vert 0 \rangle$.

\subsection{Spinors on the double covering}

For any spin structure $\xi$ on $\tilde{X}$ the couple $(\xi, \tau^{*}\xi)$ is $\tau$-invariant. Thus, the requirement of invariance does not entail any relationship with pin$^{\pm}$ structures on $X$. Let us suppose that $\xi$ is a $\tau$-invariant pin$^{\pm}$ structure: then, if we restrict it to spin structures $\xi_{u}$ and $\xi_{u'}$ corresponding to the two orientations $u$ and $u'$, we get isomorphisms $\xi_{u} \simeq \tau^{*}\xi_{u'}$ and $\xi_{u'} \simeq \tau^{*}\xi_{u}$ because the diagram:
	\[\xymatrix{
	P_{\Pin^{\pm}}\tilde{X} \ar[r]^{\widetilde{d\tau}} \ar[d]^{\xi} & P_{\Pin^{\pm}}\tilde{X} \ar[d]^{\xi} \ar[dl]_{\tau^{*}\xi} \\
	P_{O}\tilde{X} \ar[r]^{d\tau} & P_{O}\tilde{X}
}\]
restricts to:
	\[\xymatrix{
	P_{\Spin_{u}}\tilde{X} \ar[r]^{\widetilde{d\tau}} \ar[d]^{\xi_{u}} & P_{\Spin_{u'}}\tilde{X} \ar[d]^{\xi_{u'}} \ar[dl]_{\tau^{*}\xi_{u'}} & & P_{\Spin_{u'}}\tilde{X} \ar[r]^{\widetilde{d\tau}} \ar[d]^{\xi_{u'}} & P_{\Spin_{u}}\tilde{X} \ar[d]^{\xi_{u}} \ar[dl]_{\tau^{*}\xi_{u}} \\
	P_{SO_{u}}\tilde{X} \ar[r]^{d\tau} & P_{SO_{u'}}\tilde{X} & & P_{SO_{u'}}\tilde{X} \ar[r]^{d\tau} & P_{SO_{u}}\tilde{X}.
}\]
In particular, we get an isomorphism of couples $\widetilde{d\tau}: (\xi_{u}, \tau^{*}\xi_{u}) \overset{\simeq}\longrightarrow (\tau^{*}\xi_{u'}, \xi_{u'})$.

\paragraph{}We can thus find a correspondence between pin$^{\pm}$ structures on $X$ and suitable couples of spin structures on $\tilde{X}$. Given $\xi_{u}$, we extend it to a pin$^{\pm}$ structure $\xi$ via $P_{\Pin^{\pm}}\tilde{X} = P_{\Spin}\tilde{X} \times \Pin^{\pm}(n) \,/ \sim_{\Spin}$, where $(p, t) \sim_{\Spin} (ps, s^{-1}t)$ for $s \in \Spin(n)$. Then, reducing $\xi$ with respect to the other orientation we get $\xi_{u'}$, and in order to implement this correspondence we require that there is an isomorphism of couples:
\begin{equation}\label{DTauTilde}
	\widetilde{d\tau}: (\xi_{u}, \tau^{*}\xi_{u}) \overset{\simeq}\longrightarrow (\tau^{*}\xi_{u'}, \xi_{u'})
\end{equation}
satisfying $\widetilde{d\tau}^{2} = 1$. Then we can easily see that $\widetilde{d\tau}$ gives an equivalence of pin$^{\pm}$ structures between $\xi$ and $\tau^{*}\xi$, so that $\xi$ is the pull-back of a pin$^{\pm}$ structure on $X$. Summarizing, there are two ways to relate $\Spin$ structures relative to the two different orientations. The first one is the map $\xi_{u} \rightarrow \xi_{u'}$ obtained via the corresponding pin$^{\pm}$ structure as we have just explained, the second one is via $\tau^{*}$. We require that, for fixed $\xi_{u}$, these two maps give the same value up to equivalence, and we also require that the equivalence squares to $1$.

\paragraph{Remark:} This requirement depends upon whether we pass through $\Pin^{+}$ or $\Pin^{-}$ extensions. One may wonder whether the two ways lead to the same result. Actually it is true that, starting from $\alpha_{u}$, we obtain the same structure $\alpha_{u'}$ in both cases (we obtain the same result we would get reversing the orientation as described in \cite{KT}, without referring to pinors), but, as we will explicitly see for surfaces, the fact that $\widetilde{d\tau}^{2} = 1$ depends on the kind of $\Pin$-structure we consider.

\paragraph{}We now analyze the behavior of spinors as sections of the associated bundle. Fixing a pin$^{\pm}$ structure $\xi$ such that $\xi \simeq \tau^{*}\xi$ with $\widetilde{d\tau}^{2} = 1$, we can consider the couple of sections $(s,s)$ relative to the couple of pin$^{\pm}$ structures $(\xi, \tau^{*}\xi \simeq \xi)$. The section $s$ is supposed to satisfy $s = \widetilde{d\tau} \circ \eta_{\tau}(s)$, where the sign of $\widetilde{d\tau}$ is fixed by requiring that $P_{\Pin^{\pm},\xi}\tilde{X} \,/\, \widetilde{d\tau}$ is isomorphic to the pin$^{\pm}$ structure on $X$ we started from. Then we restrict to the \emph{oriented} couple of spin structures $(\xi_{u}, \tau^{*}\xi_{u'})$, so that the bundle of spinors is split in chiralities. Then we restrict $s$ to $S^{+}_{\xi_{u}}$ and $S^{-}_{\tau^{*}\xi_{u'}}$ obtaining $(s^{+}, s^{-})$. Then $s^{-} = \widetilde{d\tau} \circ \eta_{\tau}(s^{+})$, i.e.\footnote{We remark that $\eta_{\tau}$ reverses the chiralities with respect to $\xi$ and $\tau^{*}\xi$, but here $s^{+}$ is positive-chiral with respect to $\xi$, so $\eta_{\tau}(s^{+}_{x}) = s^{+}_{\tau(x)}$, as it is obvious from the fact that $\eta_{\tau}$ is the identity on the total space.} $s^{-}_{x} = \widetilde{d\tau}(s^{+}_{\tau(x)})$, so that we have a natural bijection between invariant pinors and couple of invariant spinors. If we consider $\widetilde{d\tau} \circ \gamma$, we get $s^{-} = \widetilde{d\tau} \circ \gamma \circ \eta_{\tau}(s^{+}) = -\widetilde{d\tau} \circ \eta_{\tau}(s^{+})$. Hence, as $\widetilde{d\tau}$-invariant couples are of the form $(s^{+}, \widetilde{d\tau} \circ \eta_{\tau}(s^{+}))$, similarly $(\widetilde{d\tau} \circ \gamma)$-invariant couples are of the form $(s^{+}, -\widetilde{d\tau} \circ \eta_{\tau}(s^{+}))$. In particular, for a given couple $(s^{+}, t^{-})$, we have the two projectors $(\frac{1}{2}(s^{+} + \widetilde{d\tau} \circ \eta_{\tau}(t^{-}), \frac{1}{2}(t^{-} + \widetilde{d\tau} \circ \eta_{\tau}(s^{+})))$ and $(\frac{1}{2}(s^{+} - \widetilde{d\tau} \circ \eta_{\tau}(t^{-}), \frac{1}{2}(t^{-} - \widetilde{d\tau} \circ \eta_{\tau}(s^{+})))$.

Had we chosen the other orientation $u'$, we would have considered the couple $(\tau^{*}\xi_{u}, \xi_{u'})$, but since the sections of $\xi$ and $\tau^{*}\xi$ are the same as subset of the total space, we had the same result. This is a consequence of the fact that for the couples $(\xi_{u}, \tau^{*}\xi_{u})$ and $(\xi_{u'}, \tau^{*}\xi_{u'})$ we have a canonical notion of $\tau$-invariant spinor.

\paragraph{}Summarizing:
\begin{enumerate}
	\item an equivalence of couples $(\xi_{u}, \tau^{*}\xi_{u}) \simeq (\tau^{*}\xi_{u'}, \xi_{u'})$ via $\widetilde{d\tau}$;
	\item two $\tau$-invariant couples $(\xi_{u}, \tau^{*}\xi_{u})$ and $(\xi_{u'}, \tau^{*}\xi_{u'})$ with a good notion of invariant spinor;
	\item two oriented couples $(\xi_{u}, \tau^{*}\xi_{u'})$ and $(\xi_{u'}, \tau^{*}\xi_{u})$.
\end{enumerate}
What we have done is fixing an orientation $u$ of $\tilde{X}$ and consider the oriented couple $(\xi_{u}, \tau^{*}\xi_{u'})$ of point 3: its components are the first members of the two couples in point 1, so that such components are equivalent via $\widetilde{d\tau}$ and we have a good notion of $\widetilde{d\tau}$-invariant couple of sections. \emph{These are the spinors we consider.} Had we fixed the other orientation $u'$, we should consider the ordered couple $(\xi_{u'}, \tau^{*}\xi_{u})$, but thanks to point 2 we have a canonical bijection between the sections of this couple and the ones of the previous, thus the choice of the orientation is immaterial.

\section{Surfaces}

All non-orientable surfaces can be obtained via connected sum of tori from the real projective plane or the Klein bottle. We use the following notations:
\begin{itemize}
	\item $\Sigma_{g}$ is the connected sum of $g$ tori;
	\item $N_{g,1}$ is the connected sum of $\Sigma_{g}$ and the real projective plane $\mathbb{RP}^{2}$;
	\item $N_{g,2}$ is the connected sum of $\Sigma_{g}$ and the Klein bottle $K^{2}$.
\end{itemize}

\subsection{One cross-cap}

We first consider the case of surfaces with one cross-cap, starting from the real projective plane $\mathbb{RP}^{2}$. Its orientable double covering is the 2-sphere $S^{2}$ with involution $\tau$ given by the reflection with respect to the center. In this case we have the following situation:
	\[H_{1}(S^{2}, \mathbb{Z}) = 0 \qquad H_{1}(\mathbb{RP}^{2}, \mathbb{Z}) = \mathbb{Z}_{2}
\]
so that the projection $\pi: S^{2} \rightarrow \mathbb{RP}^{2}$ induces a trivial immersion in homology, which is not surjective since there is a cycle in $\mathbb{RP}^{2}$ which does not lift to a cycle in $S^{2}$. Passing to $\mathbb{Z}_{2}$ coefficients we thus get:
	\[\begin{array}{lll}
	H_{1}(S^{2}, \mathbb{Z}_{2}) = 0 & & H_{1}(\mathbb{RP}^{2}, \mathbb{Z}_{2}) = \mathbb{Z}_{2}\\
	H^{1}(S^{2}, \mathbb{Z}_{2}) = 0 & & H^{1}(\mathbb{RP}^{2}, \mathbb{Z}_{2}) = \mathbb{Z}_{2}	
\end{array}\]
so that the induced maps are the zero maps:
	\[\pi_{*}: 0 \longrightarrow \mathbb{Z}_{2} \qquad \pi^{*}: \mathbb{Z}_{2} \longrightarrow 0
\]
and, in particular, the only non-trivial cohomology class, which corresponds to the element of $\Hom(H_{1}(X, \mathbb{Z}), \mathbb{Z}_{2})$ assigning $1$ to the class that does not lift, lies in the kernel of $\pi^{*}$.

All other surfaces $N_{g,1}$ with one number of cross-cap can be obtained adding $g$ tori to $\mathbb{RP}^{2}$ via connected sum. The double of $N_{g,1}$ is $\Sigma_{2g}$, namely the connected sum of $2g$-tori. In this case the situation is the following:
	\[H_{1}(\Sigma_{2g}, \mathbb{Z}) = \mathbb{Z}^{\oplus 4g} \qquad H_{1}(N_{g,1}, \mathbb{Z}) = \mathbb{Z}^{\oplus 2g} \oplus \mathbb{Z}_{2}.
\]
If we fix a canonical basis $\{a_{1}, b_{1}, \ldots, a_{2g}, b_{2g}\}$ of $H_{1}(\Sigma_{2g}, \mathbb{Z})$, the involution $\tau$ acts in such a way that $\tau_{*}(a_{i}) = a_{i+g}$ and $\tau_{*}(b_{i}) = b_{i+g}$ for $i = 1, \ldots, g$. There is a trivial cycle $c$ of which $\tau$ exchanges antipodal points: half of it is the lift of a representative of the $\mathbb{Z}_{2}$-generator of $N_{g,1}$ via $\pi: \Sigma_{2g} \rightarrow N_{g,1}$. Passing to $\mathbb{Z}_{2}$-coefficients we have:
	\[H_{1}(\Sigma_{2g}, \mathbb{Z}) = \mathbb{Z}_{2}^{\oplus 4g} \qquad H_{1}(N_{g,1}, \mathbb{Z}) = \mathbb{Z}_{2}^{\oplus 2g+1}.
\]
The push-forward $\pi_{*}$ sends $a_{i}$ and $a_{i+g}$ to the same class, similarly for $b_{i}$ and $b_{i+g}$. Instead of considering the $\mathbb{Z}_{2}$-reduction of the canonical basis, we consider the basis $\{a_{1}, b_{1}, \ldots,$ $a_{g}, b_{g}, a_{g+1} + a_{1}, b_{g+1} + b_{1}, \ldots, a_{2g} + a_{g}, b_{2g} + b_{g}\}$. It is still a basis since it can be obtained from the canonical one via the invertible $(4g \times 4g)$-matrix:
	\[\begin{bmatrix} I_{2g} & 0_{2g} \\ I_{2g} & I_{2g}
\end{bmatrix}.
\]
In this way we split $\mathbb{Z}_{2}^{\oplus 4g} = \mathbb{Z}_{2}^{\oplus 2g} \oplus \Ker \, \pi_{*}$, so that $\pi_{*}$ is injective on the first summand. Moreover, its image has index $2$ in $H_{1}(N_{g,1}, \mathbb{Z})$, since the only generator not lying in the image is the $\mathbb{Z}_{2}$-reduction of the $2$-torsion integral one. Thus we have the following picture for homology:
	\[\xymatrix{
	\mathbb{Z}_{2}^{\oplus 2g} \oplus (\mathbb{Z}_{2}^{\oplus 2g} = \Ker\, \pi_{*}) \ar@<-10ex>[d]^{\pi_{*}}_{\simeq} \; \,\\
	\mathbb{Z}_{2}^{\oplus 2g} \oplus (\mathbb{Z}_{2} \simeq \Coker\, \pi_{*})\;
}\]
which becomes in cohomology:
	\[\xymatrix{
	\mathbb{Z}_{2}^{\oplus 2g} \oplus (\mathbb{Z}_{2}^{\oplus 2g} \simeq \Coker\, \pi^{*}) \; \,\\
	\mathbb{Z}_{2}^{\oplus 2g} \oplus (\mathbb{Z}_{2} = \Ker\, \pi^{*}) . \quad\;\; \ar@<10ex>[u]^{\pi^{*}}_{\simeq}
}\]

\subsection{Two cross-caps}

We now consider the case of surfaces with two cross-caps, starting from the Klein bottle $K^{2} = N_{0,2}$. Its orientable double cover is the torus $T^{2} = \Sigma_{1}$ with involution $\tau$ defined in the following way: if we represent the torus as $\mathbb{C} / (2\pi\mathbb{Z} + 2\pi i\mathbb{Z})$, then we define
	\[\tau(z) = \overline{z} + \pi
\]
or equivalently in real coordinates $\tau(x, y) = (x + \pi, -y)$. We represent the torus as the square $[0, 2\pi] \times [0, 2\pi]$ with $(0, y) \sim (2\pi, y)$ and $(x, 0) \sim (x, 2\pi)$, and the Klein bottle as $[0, 2\pi] \times [0, 2\pi]$ with $(0, y) \sim (2\pi, y)$ and $(x, 0) \sim (2\pi - x, 2\pi)$. We call $a$ the loop $[0, 2\pi] \times \{0\}$ and $b$ the loop $\{0\} \times [0, 2\pi]$. We have that:
\begin{equation}\label{Pi1TK}
\begin{split}
	&\pi_{1}(T^{2}) = \langle \tilde{a}, \tilde{b} \,\vert\, \tilde{a}\tilde{b}\tilde{a}^{-1}\tilde{b}^{-1} = 1 \rangle\\
	&\pi_{1}(K^{2}) = \langle a, b \,\vert\, abab^{-1} = 1 \rangle
\end{split}
\end{equation}
The involution $\tau$ is the antipodal map of the $\tilde{a}$-generator, thus $(\tau_{*})_{\pi_{1}}[\tilde{a}] = [\tilde{a}]$, while it reflects the $\tilde{b}$-generator with respect to $y = \frac{1}{2}$ and apply the antipodal map, thus $(\tau_{*})_{\pi_{1}}[\tilde{b}] = [\tilde{b}]^{-1}$. The injective map induced by the projection is:
\begin{equation}\label{PiStarPi1}
\begin{split}
	(\pi_{*})_{\pi_{1}}: \; &\pi_{1}(T^{2}) \hookrightarrow \pi_{1}(K^{2})\\
	&(\pi_{*})_{\pi_{1}}(\tilde{a}) = b^{2}; \quad (\pi_{*})_{\pi_{1}}(\tilde{b}) = a.
\end{split}
\end{equation}
In homology, the abelianizations of \eqref{Pi1TK} are:
	\[H_{1}(T^{2}, \mathbb{Z}) = \mathbb{Z} \oplus \mathbb{Z} = \langle\!\langle \tilde{a}, \tilde{b} \rangle\!\rangle \qquad H_{1}(K^{2}, \mathbb{Z}) = \mathbb{Z} \oplus \mathbb{Z}_{2} = \langle\!\langle b, a \,\vert\, a^{2} = 1 \rangle\!\rangle
\]
(where $\langle\!\langle \,\cdot\, \rangle\!\rangle$ denotes the \emph{abelian} group with specified generators and relations) thus the map \eqref{PiStarPi1} becomes:
\begin{equation}\label{PiStar}
\begin{split}
	\pi_{*}: \; &\mathbb{Z} \oplus \mathbb{Z} \longrightarrow \mathbb{Z} \oplus \mathbb{Z}_{2}\\
	&\pi_{*}(1,0) = (2,0); \quad \pi_{*}(0,1) = (0,1).
\end{split}
\end{equation}
Contrary to \eqref{PiStarPi1}, the map \eqref{PiStar} is not injective any more, but its image has still index $2$ in the codomain, even if in this case the generator in $H_{1}(K^{2}, \mathbb{Z})$ not lifting to $H_{1}(T^{2}, \mathbb{Z})$ is the non-torsion one (i.e.\ the lifting of its representatives are now half of a non-trivial cycle of the covering, while for one-cross cap they were half of a trivial cycle). Passing to $\mathbb{Z}_{2}$ coefficients we get:
	\[\begin{array}{lll}
	H_{1}(T^{2}, \mathbb{Z}_{2}) = \mathbb{Z}_{2} \oplus \mathbb{Z}_{2} & & H_{1}(K^{2}, \mathbb{Z}_{2}) = \mathbb{Z}_{2} \oplus \mathbb{Z}_{2}\\
	H^{1}(T^{2}, \mathbb{Z}_{2}) = \mathbb{Z}_{2} \oplus \mathbb{Z}_{2} & & H^{1}(K^{2}, \mathbb{Z}_{2}) = \mathbb{Z}_{2}	\oplus \mathbb{Z}_{2}
\end{array}\]
so that the induced map in homology is:
	\[\begin{split}
	\pi_{*}:\;& H_{1}(T^{2}, \mathbb{Z}_{2}) \longrightarrow H_{1}(K^{2}, \mathbb{Z}_{2})\\
	&\pi_{*}(1,0) = 0 \qquad \pi_{*}(0,1) = (0,1)
\end{split}\]
For cohomology, we identify $(1,0) \in H^{1}(K^{2}, \mathbb{Z}_{2})$ with the functional $\varphi: H_{1}(K^{2}, \mathbb{Z}_{2}) \rightarrow \mathbb{Z}_{2}$ such that $\varphi(1,0) = 1$ and $\varphi(0,1) = 0$, and similarly for $(0,1) \in H^{1}(K^{2}, \mathbb{Z}_{2})$ with the functional $\psi$. Then $\pi^{*}(\varphi) = \varphi \circ \pi_{*}$ so that $\pi^{*}(\varphi)(1,0) = 0$ and $\pi^{*}(\varphi)(0,1) = 0$, while $\pi^{*}(\psi)(1,0) = 0$ and $\pi^{*}(\psi)(0,1) = 1$. Hence:
	\[\begin{split}
	\pi^{*}:\;& H^{1}(K^{2}, \mathbb{Z}_{2}) \longrightarrow H^{1}(T^{2}, \mathbb{Z}_{2})\\
	&\pi^{*}(1,0) = 0 \qquad \pi^{*}(0,1) = (0,1).
\end{split}\]
and, in particular, the cohomology class which corresponds to the element of $\Hom(H_{1}$ $(X, \mathbb{Z}), \mathbb{Z}_{2})$ assigning $1$ to the class that does not lift, lies in the kernel of $\pi^{*}$.

All other surfaces $N_{g,2}$ with two cross-caps can be obtained adding $g$ tori to $K^{2}$ via connected sum. The double of $N_{g,2}$ is $\Sigma_{2g+1}$, namely the connected sum of $2g$-tori. In this case the situation is the following:
	\[H_{1}(\Sigma_{2g+1}, \mathbb{Z}) = \mathbb{Z}^{\oplus 4g} \oplus \mathbb{Z} \oplus \mathbb{Z} \qquad H_{1}(N_{g,2}, \mathbb{Z}) = \mathbb{Z}^{\oplus 2g} \oplus \mathbb{Z} \oplus \mathbb{Z}_{2}.
\]
If we fix a canonical basis $\{a_{1}, b_{1}, \ldots, a_{2g+1}, b_{2g+1}\}$ of $H_{1}(\Sigma_{2g+1}, \mathbb{Z})$, the involution $\tau$ acts in such a way that:
\begin{itemize}
	\item $\tau_{*}(a_{i}) = a_{i+g}$ and $\tau_{*}(b_{i}) = b_{i+g}$ for $i = 1, \ldots, g$;
	\item $\tau_{*}(a_{2g+1}) = a_{2g+1}$ and $\tau_{*}(b_{2g+1}) = -b_{2g+1}$
\end{itemize}
and $\tau$ acts on two representatives $\tilde{a}$ of $a_{2g+1}$ and $\tilde{b}$ of $b_{2g+1}$ as in the case of the Klein bottle. Thus, half of $\tilde{a}$ is the lift of a representative of the last $\mathbb{Z}$-generator of $H^{1}(T^{2}, \mathbb{Z})$ via $\pi: \Sigma_{2g} \rightarrow N_{g,1}$, while $\tilde{b}$ is the lift of the $\mathbb{Z}_{2}$-generator. Passing to $\mathbb{Z}_{2}$-coefficients we have:
	\[H_{1}(\Sigma_{2g+1}, \mathbb{Z}_{2}) = \mathbb{Z}_{2}^{\oplus 4g} \oplus \mathbb{Z}_{2} \oplus \mathbb{Z}_{2} \qquad H_{1}(N_{g,2}, \mathbb{Z}_{2}) = \mathbb{Z}_{2}^{\oplus 2g} \oplus \mathbb{Z}_{2} \oplus \mathbb{Z}_{2}.
\]
The push-forward $\pi_{*}$ sends $a_{i}$ and $a_{i+g}$ to the same class, similarly for $b_{i}$ and $b_{i+g}$. As before, instead of considering the $\mathbb{Z}_{2}$-reduction of the canonical basis, we consider the basis $\{a_{1}, b_{1}, \ldots, a_{g}, b_{g},$ $a_{g+1} + a_{1}, b_{g+1} + b_{1}, \ldots, a_{2g} + a_{g}, b_{2g} + b_{g}, a_{2g+1}, b_{2g+1}\}$. In this way we split $\mathbb{Z}_{2}^{\oplus 4g}  \oplus \mathbb{Z}_{2} \oplus \mathbb{Z}_{2} = \mathbb{Z}_{2}^{\oplus 2g}  \oplus \mathbb{Z}_{2} \oplus \Ker \, \pi_{*}$, so that $\pi_{*}$ is injective on the first summand. Moreover, its image has index $2$ in $H_{1}(N_{g,2}, \mathbb{Z})$, since the only generator not lying in the image is the $\mathbb{Z}_{2}$-reduction of the $\mathbb{Z}$-factor of $K^{2}$. Thus we have the following picture:
	\[\xymatrix{
	\mathbb{Z}_{2}^{\oplus 2g} \oplus \mathbb{Z}_{2} \oplus (\mathbb{Z}_{2}^{\oplus 2g} \oplus \mathbb{Z}_{2} = \Ker\, \pi_{*}) \ar@<-12.5ex>[d]^{\pi_{*}}_{\simeq} \; \,\\
	\mathbb{Z}_{2}^{\oplus 2g} \oplus \mathbb{Z}_{2} \oplus (\mathbb{Z}_{2} \simeq \Coker\, \pi^{*}) \qquad\;\;
}\]
which becomes in cohomology:
	\[\xymatrix{
	\mathbb{Z}_{2}^{\oplus 2g} \oplus \mathbb{Z}_{2} \oplus (\mathbb{Z}_{2}^{\oplus 2g} \oplus \mathbb{Z}_{2} \simeq \Coker\, \pi^{*}) \; \,\\
	\mathbb{Z}_{2}^{\oplus 2g} \oplus \mathbb{Z}_{2} \oplus (\mathbb{Z}_{2} = \Ker\, \pi^{*}) \ar@<13ex>[u]^{\pi^{*}}_{\simeq} \;.\qquad\qquad\;
}\]

\subsection{Invariant structures on the sphere}

We think of the sphere $S^{2}$ as the Riemann sphere $\mathbb{CP}^{1}$, with two charts $U_{0} = \mathbb{CP}^{1} \setminus \{N\}$ and $U_{1} = \mathbb{CP}^{1} \setminus \{S\}$ and transition function $g_{01}(z) = -\frac{1}{z}$. The antipodal involution $\tau$ is specified each of the two charts\footnote{Note that $\tau$ commutes with $g_{01}$, that's why the expression is the same in both charts.} by $\tau(z) = -\frac{1}{\overline{z}}$. We compute its Jacobian to find the action $d\tau$ on the tangent bundle. In real coordinates:
	\[\tau(x,y) = \Bigl( \frac{-x}{x^{2} + y^{2}}, \frac{-y}{x^{2} + y^{2}} \Bigr)
\]
so that the Jacobian becomes:
	\[J\tau(x, y) = \frac{1}{x^{2} + y^{2}} \begin{bmatrix} x^{2} - y^{2} & 2xy \\ 2xy & y^{2} - x^{2}
\end{bmatrix} \]
which, on the equator $\abs{z} = 1$ becomes the orthogonal matrix:
\begin{equation}\label{JacTau}
	J\tau(\cos \theta, \sin \theta) = \begin{bmatrix} \cos 2\theta & \sin 2\theta \\ \sin 2\theta & -\cos 2\theta
\end{bmatrix}.
\end{equation}
We now consider the sphere as the union of the two halves glued on the equator, so that we restrict both the charts $U_{0}$ and $U_{1}$ to the disc $\abs{z} \leq 1$, and we glue them via $g_{01}$. Now we consider the trivial spin structure for each of the two discs and we glue via a lift of $dg_{01}$ on $\abs{z} = 1$. On the equator of both charts the transformation \eqref{JacTau} is a reflection with respect to the real line generated by $(\cos \theta, \sin \theta)$, i.e.\ by $(-\sin \theta, \cos \theta)^{\bot}$. Thus, if we consider the point $(\cos \theta, \sin \theta) \in \mathbb{C} \simeq U_{0}$, we get $\tau(\cos \theta, \sin \theta) = -(\cos \theta, \sin \theta)$ and $d\tau_{(\cos \theta, \sin \theta)}$ acts on the tangent bundle as a rotation of $\pi$ along the equator composed with a reflection of the orthogonal direction. Hence its possible lifts to a $\Pin^{\pm}$-principal bundle are:
	\[\widetilde{d\tau}(\theta, p) = (\pi + \theta, \pm (-\sin \theta e_{1} + \cos \theta e_{2}) \cdot p).
\]
Then $\widetilde{d\tau}^{2}$ is given by $(-\sin (\theta + \pi) e_{1} + \cos (\theta + \pi) e_{2})(-\sin \theta e_{1} + \cos \theta e_{2}) = (\sin \theta e_{1} - \cos \theta e_{2})(-\sin \theta e_{1} + \cos \theta e_{2}) = -\sin^{2}\theta e_{1}^{2} - \cos^{2}\theta e_{2}^{2}$. Thus we see that $\widetilde{d\tau}^{2} = 1$ if and only if $e_{1}^{2} = e_{2}^{2} = -1$, namely if the structure is $\Pin^{-}$: this shows that $\mathbb{RP}^{2} \simeq S^{2} \,/\, \tau$ has two pin$^{-}$ structures, lifting to the one of the sphere, but no pin$^{+}$ structures (compare with \cite{KT}).

\subsection{Invariant structures on the torus}

The torus has trivial tangent bundle $T^{2} \times \mathbb{R}^{2} \simeq S^{1} \times S^{1} \times \mathbb{R}^{2}$. The four inequivalent $\Spin$ or pin$^{\pm}$ structures can be all obtained from the trivial principal bundle $S^{1} \times S^{1} \times \Spin(2)$ or $S^{1} \times S^{1} \times \Pin^{\pm}(2)$ in the following way:
	\[\xymatrix{
	(\theta, \varphi, p') \ar[d]^{\tilde{\xi}_{0}} & & (\theta, \varphi, p') \ar[d]^{\tilde{\xi}_{1}} & & (\theta, \varphi, p') \ar[d]^{\tilde{\xi}_{2}} & & (\theta, \varphi, p') \ar[d]^{\tilde{\xi}_{3}}\\
	(\theta, \varphi, p)                           & & (\theta, \varphi, R_{\theta} \cdot p)          & & (\theta, \varphi, R_{\varphi} \cdot p)         & & (\theta, \varphi, R_{\varphi}R_{\theta} \cdot p)
}\]
where $R_{x}$ is the rotation by the angle $x$. To see that, e.g.\ the first two are not equivalent, we notice that we would need a map:
	\[\xymatrix{
	(\theta, \varphi, p') \ar[rr]^{\rho} \ar[dr]_{\tilde{\xi}_{0}} & & (\theta, \varphi, \tilde{R}_{-\theta}p') \ar[dl]^{\tilde{\xi}_{1}} \\
	& (\theta, \varphi, p)
}\]
for $\tilde{R}_{-\theta}$ a lift of $R_{-\theta}$ to $\Spin$ or $\Pin^{\pm}$. But in this way $\rho$ is not well defined, since for $\theta$ and $\theta + 2\pi$ we get two lifts differing by $-1$.

We now see that all these pin$^{\pm}$ structures are $\tau$-invariant, where $\tau$ is the involution giving the Klein bottle, namely $\tau(\theta, \varphi) = (\theta + \pi, -\varphi)$. On the tangent frame bundle we have the action $d\tau(\theta, \varphi, p) = (\theta + \pi, -\varphi, j_{2}p)$ where $j_{2}$ is the reflection along $e_{2}^{\bot}$, i.e.\ $(x, y) \rightarrow (x, -y)$. The equivalence between $\tilde{\xi}_{0}$ and $\tau^{*}\tilde{\xi}_{0}$ is given by the following diagram:
	\[\xymatrix{
	(\theta, \varphi, p') \ar[r]^{\widetilde{d\tau} \qquad} \ar[d]_{\tilde{\xi}_{0}} & (\theta + \pi, -\varphi, e_{2} \cdot p') \ar[d]^{\tilde{\xi}_{0}} \ar[dl]_{\tau^{*}\tilde{\xi}_{0}} \\
	(\theta, \varphi, p) \ar[r]^{d\tau \quad\;\;} & (\theta + \pi, -\varphi, j_{2}p)
}\]
or equivalently by $\widetilde{d\tau} \circ \gamma$ which can be obtained by choosing $-e_{2}$. Here we see that for the $\Pin^{+}$-structure, since $e_{2}^{2} = 1$, we get $\widetilde{d\tau}^{2} = 1$, while for the $\Pin^{-}$-structure we get $\widetilde{d\tau}^{2} = -1$. Thus, only the $\Pin^{+}$-structure is the pull-back of a $\Pin^{+}$-structure of $K^{2}$. For $\tilde{\xi}_{1}$:
	\[\xymatrix{
	(\theta, \varphi, p') \ar[r]^{\widetilde{d\tau} \qquad\qquad} \ar[d]_{\tilde{\xi}_{1}} & (\theta + \pi, -\varphi, \tilde{R}_{-\theta-\pi} e_{2} \tilde{R}_{\theta} p') \ar[d]^{\tilde{\xi}_{1}} \ar[dl]_{\tau^{*}\tilde{\xi}_{1}} \\
	(\theta, \varphi, R_{\theta} p) \ar[r]^{d\tau \qquad} & (\theta + \pi, -\varphi, j_{2} R_{\theta} p)
}\]
and $\widetilde{d\tau}$ is well-defined since with the shift $\theta \rightarrow \theta + 2\pi$ we get a minus sign in both liftings of the rotations. Then $\widetilde{d\tau}^{2} = \tilde{R}_{-(\theta+\pi)-\pi} e_{2} \tilde{R}_{\theta+\pi} \tilde{R}_{-\theta-\pi} e_{2} \tilde{R}_{\theta} = \tilde{R}_{-2\pi}e_{2}^{2} = -e_{2}^{2}$, thus we get opposite results with respect to $\tilde{\xi}_{0}$. For $\tilde{\xi}_{2}$:
	\[\xymatrix{
	(\theta, \varphi, p') \ar[r]^{\widetilde{d\tau} \qquad\qquad\quad} \ar[d]_{\tilde{\xi}_{2}} & (\theta + \pi, -\varphi, \tilde{R}_{\varphi} e_{2} \tilde{R}_{\varphi} p') \ar[d]^{\tilde{\xi}_{2}} \ar[dl]_{\tau^{*}\tilde{\xi}_{2}} \\
	(\theta, \varphi, R_{\varphi} p) \ar[r]^{d\tau \qquad} & (\theta + \pi, -\varphi, j_{2} R_{\varphi} p)
}\]
and $\widetilde{d\tau}$ is well-defined since with the shift $\theta \rightarrow \theta + 2\pi$ we get a minus sign in both liftings of the rotations. Then $\widetilde{d\tau}^{2} = \tilde{R}_{-\varphi} e_{2} \tilde{R}_{-\varphi} \tilde{R}_{\varphi} e_{2} \tilde{R}_{\varphi} = e_{2}^{2}$, thus we get the same results of $\tilde{\xi}_{0}$. It is clear that $\tilde{\xi}_{3}$ behaves as $\tilde{\xi}_{1}$.

\paragraph{}If we consider pinors and spinors as sections of the associated vector bundles, for $\tilde{\xi}_{0}$ the condition on pinors is $s_{(\theta, \varphi)} = e_{2} \cdot s_{(\theta + \pi, -\varphi)}$, while for spinors we consider the couple $(s^{+}, s^{-})$ with structure $((\tilde{\xi}_{0})_{u}, \tau_{*}(\tilde{\xi}_{0})_{u'}))$ where $s^{+}$ is free and $s^{-}_{(\theta, \varphi)} = e_{2} \cdot s^{+}_{(\theta + \pi, -\varphi)}$. Similar conditions for the other spin structures.

\section{Manifolds with boundary}

We now want to give the analogous description in the case of unorientable manifolds \emph{with boundary}. We start with a brief recall of the well-known case of spinors on orientable manifolds with boundary, in order to extend it to pinors and discuss the non-orientable case.

\subsection{Orientable manifolds with boundary}

Let $X$ be an orientable manifold of dimension $2n$ with boundary $\partial X$, and let us consider its double $X^{d}$ obtained considering two disjoint copies of $X$ and identifying the corresponding boundary points. We mark one of the two copies considering an embedding $i: X \rightarrow X^{d}$. In this way, an orientation of $X^{d}$ induces an orientation of $X$ and the opposite one on the other copy. We have a natural orientation-reversing involution $\tau$ identifying corresponding points of the two copies, which is \emph{not} a double covering since the boundary points are fixed.

\paragraph{Remark:} We have a natural projection $\pi: X^{d} \longrightarrow X \simeq X^{d} \,/\, \tau$, but it is in general not smooth, since on a local curve orthogonal to a boundary point the behavior of $\tau$ and $\pi$ is of the form $\tau(x) = -x$ and $\pi(x) = \abs{x}$. This is why in the open case it is more natural to deal with the immersion $i: X \rightarrow X^{d}$ which has no analogue in the closed non-orientable case.

\paragraph{}We consider on $X$ couples of spin structure $(\xi, \xi')$ with an isomorphism $\theta: \xi \vert_{\partial X} \rightarrow \xi' \vert_{\partial X}$, where the restriction is obtained in the following way: we consider the immersion $P_{O}(\partial X) \subset P_{O}X$ sending a basis $\{e_{1}, \ldots e_{2n-1}\}$ of $T_{x}(\partial X)$ to the basis $\{e_{1}, \ldots e_{2n-1}, u\}$ of $T_{x}X$ where $u$ is the outward orthogonal unit vector. We consider two triples $(\xi, \xi', \theta)$ and $(\eta, \eta', \varphi)$ equivalent if there exist equivalences $\rho_{1}: \xi \rightarrow \eta$ and $\rho_{2}: \xi' \rightarrow \eta'$ such that $\rho_{2}\vert_{\partial X}^{-1} \circ \varphi \circ \rho_{1}\vert_{\partial X} = \theta$. On each connected component $Y \subset \partial X$, there are two possibilities for $\theta\vert_{Y}$ linked by $\gamma$. An overall change from $\theta$ to $\theta \circ \gamma$ is irrelevant since $(\xi, \xi', \theta) \simeq (\xi, \xi', \theta \circ \gamma)$ via $\rho_{1} = \id$ and $\rho_{2} = \gamma$; instead, the separate restrictions determined by $\theta$ are meaningful, thus we must fix all of them except one. Let us show that equivalence classes of such triples $(\xi, \xi', \theta)$ correspond bijectively to equivalence classes of spin structures $\tilde{\xi}$ on $X^{d}$ associated only to positive chirality.\\
\textbf{From $X$ to $X^{d}$:} given $(\xi, \xi', \theta)$ we define $\tilde{\xi}\vert_{X} := \xi$ and $\tilde{\xi}\vert_{X^{d}\setminus \Int(X)} := \xi'$, and we glue them on $\partial X$ via the isomorphism $\theta$. We call such a spin structure $\xi \cup_{\theta} \xi'$. We can always restrict ourselves to the case $\theta = \id$ by considering $\bigl( (\xi \cup_{\theta} \xi')\vert_{X}, \tau^{*}((\xi \cup_{\theta} \xi')\vert_{X^{d} \setminus \Int(X)}), \id \bigr)$. \\
\textbf{From $X^{d}$ to $X$:} given $\tilde{\xi}$ we define $\xi := \tilde{\xi}\vert_{X}$ and $\xi' := \tau^{*}(\tilde{\xi}\vert_{X^{d}\setminus \Int(X)})$. The isomorphism $\theta$ is the identity on $\tilde{\xi}\vert_{\partial X}$.

\paragraph{}A few comments are in order. When we define $\xi' := \tau^{*}(\tilde{\xi}\vert_{X^{d}\setminus \Int(X)})$, the set $X^{d}\setminus \Int(X)$ is a copy of $X$ oriented in the opposite way, but $\tau$ recovers the original orientation, thus it reverses the chirality. In particular, considering spinors as sections of the associated vector bundles, we have a section $\tilde{s}^{+}$ on $X^{d}$ and a couple $(s^{+}, s^{-})$ on $X$, with the condition $\tilde{s}^{+}_{x} = s^{+}_{x}$ for $x \in X \setminus \partial X$ and $\tilde{s}^{+}_{x} = s^{-}_{\tau(x)}$ for $x \in X^{d} \setminus \Int(X)$, while on the boundary we consider $s^{+}_{x}$ on one copy, $s^{-}_{x}$ on the other and we glue them on $\partial X$ via $\theta: S^{+}_{x} \rightarrow S^{-}_{x}$ extended to the vector bundles. Of course we could also choose negative chirality on the double, by exchanging the roles of $\xi$ and $\xi'$.

When we double a manifold, we can create new cycles. Let us think of the case of a cylinder whose double is a torus. Two structures $\tilde{\xi}_{1}$ and $\tilde{\xi}_{2}$ on the double can differ by the holonomy of the spin connection along one of those cycles: the corresponding couples $(\xi_{1}, \xi_{1}', \theta_{1})$ and $(\xi_{2}, \xi_{2}', \theta_{2})$ will verify $\xi_{1} \simeq \xi_{2}$ and $\xi_{1}' \simeq \xi_{2}'$, but the difference can be read in $\theta_{1}$ and $\theta_{2}$: if we fix the same representative bundles for $(\xi_{1}, \xi_{2})$ and $(\xi_{1}', \xi_{2}')$, then $\theta_{1}$ and $\theta_{2}$ will differ by a $-1$ in one of the two boundary components intersecting the involved half-cycle (which of the two boundary components depends on the overall sign).

\paragraph{}In this section we have not referred so far to pinors, since an orientation was fixed both for $X$ and $X^{d}$ and there was no reason to relate it to the other orientation. For later use we need however to  consider a pin$^{\pm}$ structure on $X$, forgetting the orientation. In this case we do not consider a couple of pin$^{\pm}$ structures since we have no chirality, but we consider couples $(\xi, \theta)$ where $\theta: \xi\vert_{\partial X} \rightarrow \xi\vert_{\partial X}$ is an automorphism. Then we can glue two copies of $\xi$ on $X^{d}$ to $\xi \cup_{\theta} \xi$. For every connected component $Y \subset \partial X$, since $\theta\vert_{Y}$ lifts the identity of the tangent bundle of $Y$, it must be the identity or $\gamma$. Viceversa, if we have a pin$^{\pm}$ structure $\tilde{\xi}$ on $X^{d}$, then $\tilde{\xi}$ is equivalent to $\tilde{\xi}\vert_{X} \cup_{\id} \tau_{*}(\tilde{\xi}\vert_{X^{d} \setminus \Int(X)})$. If there exists an isomorphism $\widetilde{d\tau}: \tilde{\xi}\vert_{X} \overset{\simeq}\longrightarrow \tau_{*}(\tilde{\xi}\vert_{X^{d} \setminus \Int(X)})$, we restrict the latter to $\widetilde{d\tau}\vert_{\partial X}: \tilde{\xi}\vert_{\partial X} \overset{\simeq}\longrightarrow \tilde{\xi}\vert_{\partial X}$ and if we apply $\widetilde{d\tau}^{-1}$ to the second component, we obtain $\tilde{\xi}\vert_{X} \cup_{\widetilde{d\tau}\vert_{\partial X}^{-1}} \tilde{\xi}\vert_{X}$ (we can freely choose $\widetilde{d\tau}$ or $\widetilde{d\tau} \circ \gamma$ since they differ by an overall sign, thus we can suppose $\widetilde{d\tau}^{-1} = \widetilde{d\tau}$). On sections, we just ask $s_{x} = \widetilde{d\tau}(s_{\tau(x)})$. Thus we have an equivalence of categories:
	\[\left\{ \begin{array}{ll}
	(\xi, \theta) \,: & \xi \textnormal{ pin$^{\pm}$ structure on $X$}\\
	& \theta: \xi\vert_{\partial X} \overset{\simeq}\longrightarrow \xi\vert_{\partial X}
\end{array} \right\} \longleftrightarrow \left\{ \begin{array}{ll}
	\tilde{\xi} \,: & \tilde{\xi} \textnormal{ pin$^{\pm}$ structure on $X^{d}$ s.t.}\\
	& \exists \, \widetilde{d\tau}: \tilde{\xi}\vert_{X} \overset{\simeq}\longrightarrow \tau^{*}(\tilde{\xi}\vert_{X^{d} \setminus \Int(X)})
\end{array} \right\}.
\]

\paragraph{}Here are some remarks about this picture. The condition of the existence of $\widetilde{d\tau}: \tilde{\xi}\vert_{X} \simeq \tau^{*}\tilde{\xi}\vert_{X^{d} \setminus \Int(X)}$ has no analogue for spin structures since for that case we considered only the positive chirality on $X^{d}$. Should we consider also the negative chirality, the couple of spin structures should satisfy this condition. The same for sections. In particular, with spinors we have on $X$ the freedom of choosing positive and negative chiralities, with the condition that they must be isomorphic at the boundary. With pinors, which are extendable to the non-orientable case, there are no distinction between chiralites: this is the analogue of considering the same chirality for positive and negative spinors. In this case, for spin structures we should consider triples $(\xi, \xi, \theta)$, corresponding to the couples $(\xi, \theta)$ for pinors. If we start from the double, in the spin case we can freely choose $\tilde{\xi}$ for positive-chiral spinors, then on $X$ we have $\tilde{\xi}\vert_{X}$ for positive and $\tau_{*}(\tilde{\xi}\vert_{X^{d} \setminus \Int(X)})$ for negative ones: they are not in general isomorphic, but they coincide on the boundary since $\tau\vert_{\partial X}$ is the identity. For pinors (or spinors considering both chiralities) we recover the fact not to have only positive chirality by asking that there is an isomorphism $\widetilde{d\tau}$ between $\tau^{*}(\tilde{\xi}\vert_{X^{d} \setminus \Int(X)})$ and $\tilde{\xi}\vert_{X}$, but we do not ask that such an isomorphism restricts to the identity on the boundary. Thus, if we glue the two pieces with the identity we have an isomorphism which is not in general a restriction of a global one as $\widetilde{d\tau}$: applying $\widetilde{d\tau}^{-1}$ to both members we obtain $\widetilde{d\tau}^{-1}$ itself as gluing isomorphism, but now both the members are equal and the identity between them still does not restrict to $\widetilde{d\tau}^{-1}$. Thus, we have a generic isomorphism at the boundary, not necessarily the restriction of a global one.

\subsection{Unorientable manifolds with boundary}

Let $X$ be an unorientable manifold with boundary. Then we can consider the diagram:
\begin{equation}\label{DiagramCoverings}
	\xymatrix{
	& (\tilde{X}, \tau_{1}) \ar@{->>}[dl]_{\pi_{1}} & \\
	X & & (\tilde{X}^{d}, \tau_{3}, \tau_{4}) \;. \ar@{->>}[ul]_{\pi_{3}} \ar@{->>}[dl]^{\pi_{4}} \\
	& (X^{d}, \tau_{2}) \ar@{->>}[ul]^{\pi_{2}} &
}
\end{equation}
We remark that we have immersions $X \subset X^{d}$ and $\tilde{X} \subset \tilde{X}^{d}$, while $\pi_{1}$ and $\pi_{4}$ are double coverings. In particular $\tau_{2}$ and $\tau_{3}$ have fixed points while $\tau_{1}$ and $\tau_{4}$ do not. By the definition of $\tilde{X}$ and $X^{d}$ with the relevant involutions we easily get the following properties:
\begin{itemize}
	\item $\pi_{1} \circ \pi_{3} = \pi_{2} \circ \pi_{4}$;
	\item $\pi_{4} \vert_{\tilde{X}} = \pi_{1}$ and $\tau_{4} \vert_{\tilde{X}} = \tau_{1}$;
	\item $\tau_{3} \circ \tau_{4} = \tau_{4} \circ \tau_{3}$.
\end{itemize}
As for the open oriented case, we must fix a couple $(\xi, \theta)$ with $\xi$ a pin$^{\pm}$ structure on $X$ and $\theta: \xi\vert_{\partial X} \rightarrow \xi\vert_{\partial X}$ an automorphism. To establish a correspondence with pin$^{\pm}$ structures on $\tilde{X}^{d}$, we can follow the upper or the lower paths of diagram \eqref{DiagramCoverings}. If we follow the lower path, we consider $\xi^{d(\theta)} := \xi \cup_{\theta} \xi$ on $X^{d}$, then we pull it back to $\pi_{4}^{*}(\xi^{d(\theta)})$. Otherwise, following the upper path, we first pull back $\xi$ to $\pi_{1}^{*}\xi$ as in the closed case, so that $\xi\vert_{\partial X}$ pulls-back to $(\pi_{1}^{*}\xi)\vert_{\partial \tilde{X}}$ and the morphism $\theta$ pulls back to a morphism $\pi_{1}^{*}\theta: (\pi_{1}^{*}\xi)\vert_{\partial \tilde{X}} \rightarrow (\pi_{1}^{*}\xi)\vert_{\partial \tilde{X}}$. Then we double $\pi_{1}^{*}\xi$ on $\tilde{X}^{d}$ putting it on both copies of $\tilde{X}$ and using $\pi_{1}^{*}\theta$ as the isomorphism on $\partial \tilde{X}$, i.e.\ we consider $\pi_{1}^{*}\xi \cup_{\pi_{1}^{*}\theta} \pi_{1}^{*}\xi$, which we call $(\pi_{1}^{*}\xi)^{d(\pi_{1}^{*}\theta)}$. The two results are the same, in fact $(\pi_{4}^{*}(\xi^{d(\theta)}))\vert_{\tilde{X}} = (\pi_{4}\vert_{\tilde{X}})^{*}(\xi^{d(\theta)}\vert_{X}) = \pi_{1}^{*}(\xi) = (\pi_{1}^{*}\xi)^{d(\pi_{1}^{*}\theta)}\vert_{\tilde{X}}$, and the same for the other half of $\tilde{X}^{d}$ and for the isomorphism $\theta$. Considering sections of the associated vector bundles of pinors, since under pull-back of pin$^{\pm}$ structures we pull-back also sections and under doubling we ask invariance of the sections, we obtain sections $s \in \Gamma(P_{\Pin^{\pm}, (\pi_{1}^{*}\xi)^{d(\pi_{1}^{*}\theta)}}(\tilde{X}^{d}) \times_{\rho} \mathbb{C}^{2^{n}})$, such that $s_{x} = s_{\tau_{3}(x)} = s_{\tau_{4}(x)} = s_{\tau_{3}\tau_{4}(x)}$. Here we do not have $\widetilde{d\tau_{3}}$ and $\widetilde{d\tau_{4}}$ since we are working with explicit pull-backs.

Viceversa, if we are given a pin$^{\pm}$ structure $\xi'$ on $\tilde{X}^{d}$, such that there exists $\widetilde{d\tau_{3}}: \xi'\vert_{\tilde{X}} \simeq (\tau_{3})_{*}(\xi'\vert_{\tilde{X}^{d} \setminus \Int(\tilde{X})})$ restricting to the boundary, and $\widetilde{d\tau_{4}}: \xi' \overset{\simeq}\longrightarrow (\tau_{4})_{*}\xi'$ with $\widetilde{d\tau_{4}}^{2} = 1$, then we can find a pin$^{\pm}$ structure on $X$ such that $\xi' \simeq (\pi_{1}^{*}\xi)^{d(\theta)}$. We can find it using the two paths of the diagram. If we follow the upper path, we consider the couple $(\xi'\vert_{\tilde{X}}, \id)$ where $\id: \xi'\vert_{\partial \tilde{X}} \rightarrow \xi'\vert_{\partial \tilde{X}}$ is the restriction of $\widetilde{d\tau_{3}}$. Then, since $\tau_{4} \vert_{\tilde{X}} = \tau_{1}$, if follows that $\xi'\vert_{\tilde{X}}$ is $\tau_{1}$-invariant with $\widetilde{d\tau_{1}}^{2} = 1$, thus we can consider $\xi = (\xi'\vert_{\tilde{X}}) \,/\, \widetilde{d\tau_{1}}$ as in the closed case. For sections on $\tilde{X}^{d}$, we must ask $s_{x} = \widetilde{d\tau_{4}}(s_{\tau_{4}(x)}) = \widetilde{d\tau_{3}}(s_{\tau_{3}(x)})$. If we follow the lower path of the diagram, we first quotient by $\widetilde{d\tau_{4}}$ and then we use the projection of $\widetilde{d\tau_{3}}$ to $X^{d}$ by $\pi_{4}$.

\paragraph{}Given the invariant structure $\xi'$, we can consider the couple of spin structures $(\xi'_{u}, (\tau_{4})_{*}\xi'_{u'})$ and the relative spinors $(s^{+}, s^{-})$: in this way the conditions become $s^{-}_{x} = \widetilde{d\tau_{4}}(s^{+}_{\tau_{4}(x)})$ and $s^{-}_{x} = \widetilde{d\tau_{3}}(s^{+}_{\tau_{3}(x)})$. Thus, $s^{-}$ is completely determined by $s^{+}$, and we have one condition $s^{+}_{x} = \widetilde{d\tau_{4}}\circ \widetilde{d\tau_{3}} (s^{+}_{\tau_{3}\tau_{4}(x)})$ (necessarily $\widetilde{d\tau_{3}}^{2} = 1$ and $\widetilde{d\tau_{3}} \circ \widetilde{d\tau_{4}} = \widetilde{d\tau_{4}} \circ \widetilde{d\tau_{3}}$).

If we follow the upper path starting from $\xi$ on $X$, instead of $\pi_{1}^{*}\xi$ we can directly consider a pin$^{\pm}$ structure $\xi$ with an isomorphism $\widetilde{d\tau_{1}}: \xi \rightarrow (\tau_{1})_{*}\xi$, such that $\widetilde{d\tau_{1}}^{2} = 1$, and the corresponding couple $(\xi_{u}, (\tau_{1})_{*}\xi_{u'})$. If we double such a couple to a spin structure on $\tilde{X}^{d}$ via $\theta = \widetilde{d\tau_{1}}\vert_{\partial X}$, we obtain exactly the structure of positive-chiral spinors obtained from $\xi'$ which is the double of $\xi$.

\paragraph{}We can also consider the orientation-\emph{preserving} involution $\tau_{34} = \tau_{3} \circ \tau_{4}$ on $\tilde{X}^{d}$. We now show that $X' = \tilde{X}^{d} \,/\, \tau_{34}$ is an oriented and closed manifold with an orientation-reversing involution $\tau'$ such that $X' \,/\, \tau' \simeq X$. In fact, $\tau_{34}$ has no fixed points: $\tau_{34}(x) = x$ is equivalent to $\tau_{3}(x) = \tau_{4}(x)$, but if $x \notin \partial \tilde{X} \subset \tilde{X}^{d}$, then $\tau_{3}$ maps it to a point of the other copy of $\tilde{X}$, while $\tau_{4}$ exchanges the sheets of the covering of the same copy of $X$; instead, if $x \in \partial \tilde{X} \subset \tilde{X}^{d}$, then $\tau_{3}(x) = x$ while $\tau_{4}$ has no fixed points. Therefore $\tau_{3}(x) = \tau_{4}(x)$ is impossible. Hence $X' = \tilde{X}^{d} \,/\, \tau_{34}$ is a smooth closed orientable manifolds double-covered by $\tilde{X}^{d}$. Then $\tau_{3}$ and $\tau_{4}$ projects at the quotient to the same involution $\tau'$. We can thus complete the diagram:
	\[\xymatrix{
	& (\tilde{X}, \tau_{1}) \ar@{->>}[dl]_{\pi_{1}} & \\
	X & (X', \tau') \ar@{->>}[l]_{\pi'} & (\tilde{X}^{d}, \tau_{3}, \tau_{4}) \;. \ar@{->>}[ul]_{\pi_{3}} \ar@{->>}[dl]^{\pi_{4}} \ar@{->>}[l]_{\pi_{34}} \\
	& (X^{d}, \tau_{2}). \ar@{->>}[ul]^{\pi_{2}} &
}\]
The previous picture is analogous to considering a pin$^{\pm}$ structure $\xi'$ on $X'$ which is $\tau'$-invariant with $\widetilde{d\tau'}^{2} = 1$. Via $\pi_{34}$ we pull-back it to a structure on $\tilde{X}^{d}$ satisfying the previous requirements.

\subsection{Moebius strip}

We now study as an example pinors on the Moebius strip. In this case diagram \eqref{DiagramCoverings} becomes (calling $\Cyl$ the finite cylinder or annulus and $M^{2}$ the Moebius strip):
	\[\xymatrix{
	& (\Cyl, \tau_{1}) \ar@{->>}[dl]_{\pi_{1}} & \\
	M^{2} & (T^{2}, \tau') \ar@{->>}[l]_{\pi'} & (T^{2}, \tau_{3}, \tau_{4}) \;. \ar@{->>}[ul]_{\pi_{3}} \ar@{->>}[dl]^{\pi_{4}} \ar@{->>}[l]_{\pi_{34}} \\
	& (K^{2}, \tau_{2}). \ar@{->>}[ul]^{\pi_{2}} &
}\]
with the involutions we now describe. We represent all the four surfaces involved as the square $[0, 2\pi] \times [0, 2\pi]$ with suitable identifications on the edges. In particular, for $M^{2}$ we identify $(0, y) \sim (2\pi, 2\pi - y)$, for $\Cyl$ $(0, y) \sim (2\pi, y)$, for $T^{2}$ $(0, y) \sim (2\pi, y)$ and $(x, 0) \sim (x, 2\pi)$, and for $K^{2}$ $(0, y) \sim (2\pi, y)$ and $(x, 0) \sim (2\pi - x, 2\pi)$. When two edges are identified with the same direction (\emph{and only in this case}), we think of the orthogonal coordinate as a $2\pi$-periodical coordinate $\mathbb{R} \,/\, 2\pi \mathbb{Z}$. With these conventions a possible choice of involutions is:
	\[\begin{array}{lll}
	\tau_{1}(x,y) = (x + \pi, 2\pi - y) & & \tau_{2}(x,y) = (y - x, y) \\
	\tau_{3}(x,y) = (-x, y) & & \tau_{4}(x,y) = (\pi - x, y + \pi).
	\end{array}
\]
We now analyze pin$^{\pm}$ structures on $T^{2}$. We can prove as before that they are all $\tau_{4}$-invariant. In the $(\theta, \varphi)$-coordinates $\tau_{4}$ becomes $\tau_{4}(\theta, \varphi) = (-\theta + \pi, \varphi + \pi)$. Then:
	\[\xymatrix{
	(\theta, \varphi, p') \ar[r]^{\widetilde{d\tau_{4}} \qquad\quad} \ar[d]_{\tilde{\xi}_{0}} & (-\theta + \pi, \varphi + \pi, e_{1} \cdot p') \ar[d]^{\tilde{\xi}_{0}} \ar[dl]_{(\tau_{4})^{*}\tilde{\xi}_{0}} \\
	(\theta, \varphi, p) \ar[r]^{d\tau_{4} \qquad\;} & (-\theta + \pi, \varphi + \pi, j_{1}p)
}\]
	\[\xymatrix{
	(\theta, \varphi, p') \ar[r]^{\widetilde{d\tau_{4}\;} \qquad\qquad\quad} \ar[d]_{\tilde{\xi}_{1}} & (-\theta + \pi, \varphi + \pi, \tilde{R}_{\theta-\pi} \cdot e_{1} \cdot \tilde{R}_{\theta} \cdot p') \ar[d]^{\tilde{\xi}_{1}} \ar[dl]_{(\tau_{4})^{*}\tilde{\xi}_{1}} \\
	(\theta, \varphi, R_{\theta} \cdot p) \ar[r]^{d\tau_{4} \qquad\quad} & (-\theta + \pi, \varphi + \pi, j_{1} \cdot R_{\theta} \cdot p)
}\]
	\[\xymatrix{
	(\theta, \varphi, p') \ar[r]^{\widetilde{d\tau_{4}} \qquad\qquad\qquad} \ar[d]_{\tilde{\xi}_{2}} & (-\theta + \pi, \varphi + \pi, \tilde{R}_{-\varphi-\pi} \cdot e_{1} \cdot \tilde{R}_{\varphi} \cdot p') \ar[d]^{\tilde{\xi}_{2}} \ar[dl]_{(\tau_{4})^{*}\tilde{\xi}_{2}} \\
	(\theta, \varphi, R_{\varphi} \cdot p) \ar[r]^{d\tau_{4} \qquad\quad} & (-\theta + \pi, \varphi + \pi, j_{1} \cdot R_{\varphi} \cdot p)
}\]
so that $\widetilde{d\tau_{4}}^{2}$ becomes respectively:
\begin{itemize}
	\item $e_{1}^{2}$ for $\xi_{0}$;
	\item $\tilde{R}_{-\theta} \cdot e_{1} \cdot \tilde{R}_{-\theta+\pi} \cdot \tilde{R}_{\theta-\pi} \cdot e_{1} \cdot \tilde{R}_{\theta} = e_{1}^{2}$ for $\xi_{1}$;
	\item $\tilde{R}_{-\varphi-2\pi} \cdot e_{1} \cdot \tilde{R}_{\varphi+\pi} \cdot \tilde{R}_{\theta-\pi} \cdot e_{1} \cdot \tilde{R}_{\theta} = -e_{1}^{2}$ for $\xi_{2}$.
\end{itemize}
The structures $\xi_{1}$ and $\xi_{2}$ have opposite behavior with respect to the involution previously considered, since in this case the variable changing sign is $x$ and not $y$.

We now analyze the situation for $\tau_{3}$. First of all we can show that all the four structures satisfy $\xi_{i}\vert_{\Cyl} \simeq (\tau_{3})_{*}(\xi_{i}\vert_{T^{2} \setminus \Int(\Cyl)})$ exactly in the same way as for $\tau_{4}$, and in this case we do not have to require that the isomorphism squares to $1$. Actually, we will now prove that $\xi_{0}$ and $\xi_{1}$ (and similarly $\xi_{2}$ and $\xi_{3}$) restrict to equivalent structures on $\Cyl$, but they differ by the isomorphism $\theta$ at the boundary. In fact, the equivalence:
\begin{equation}\label{IsoCyl}
\xymatrix{
	(\theta, \varphi, p') \ar[rr]^{\rho} \ar[dr]_{\xi_{0}} & & (\theta, \varphi, \tilde{R}_{-\theta}p') \ar[dl]^{\xi_{1}} \\
	& (\theta, \varphi, p)
}
\end{equation}
is not well defined on $T^{2}$ since $\tilde{R}_{\theta} = -\tilde{R}_{\theta+2\pi}$, but if we restrict $\theta$ to the interval $[0, \pi]$, corresponding to the cylinder, there is no ambiguity left. This reasoning does not work between $\xi_{0}$ and $\xi_{2}$ since the interval of $\varphi$ is not halved. For $\xi_{0}$ we have the diagram:
	\[\xymatrix{
	(\theta, \varphi, p') \ar[r]^{(\widetilde{d\tau_{3}})_{0} \quad} \ar[d]_{\tilde{\xi}_{0}} & (-\theta, \varphi, e_{1} \cdot p') \ar[d]^{\tilde{\xi}_{0}} \ar[dl]_{(\tau_{3})^{*}\tilde{\xi}_{0}} \\
	(\theta, \varphi, p) \ar[r]^{d\tau_{3} \quad} & (-\theta, \varphi, j_{1}p)
}\]
while for $\xi_{1}$:
	\[\xymatrix{
	(\theta, \varphi, p') \ar[r]^{(\widetilde{d\tau_{3}})_{1} \qquad} \ar[d]_{\tilde{\xi}_{1}} & (-\theta, \varphi, \tilde{R}_{\theta} e_{1} \tilde{R}_{\theta} p') \ar[d]^{\tilde{\xi}_{1}} \ar[dl]_{(\tau_{3})^{*}\tilde{\xi}_{1}} \\
	(\theta, \varphi, R_{\theta} p) \ar[r]^{\quad d\tau_{3} \qquad} & (-\theta, \varphi, j_{1} R_{\theta} p)
}\]
and we can show that the two couples $(\xi_{0}, (\widetilde{d\tau_{3}})_{0}\vert_{\partial X})$ and $(\xi_{1}, (\widetilde{d\tau_{3}})_{1}\vert_{\partial X})$ are not equivalent. In fact, they are equivalent to the triples $(\xi_{0}, (\tau_{3})_{*}\xi_{0}, \id)$ and $(\xi_{1}, (\tau_{3})_{*}\xi_{1}, \id)$ via the equivalences $\rho$ of diagram \eqref{IsoCyl} and $(\tau_{3})_{*}\rho$ of the following diagram:
\begin{equation}\label{IsoCylTauStar}
\xymatrix{
	(-\theta, \varphi, e_{1}p') \ar[rr]^{(\tau_{3})_{*}\rho} \ar[dr]_{(\tau_{3})_{*}\xi_{0}} & & (-\theta, \varphi, \tilde{R}_{\theta}e_{1}p') \ar[dl]^{(\tau_{3})_{*}\xi_{1}} \\
	& (\theta, \varphi, p).
}
\end{equation}
Comparing \eqref{IsoCyl} and \eqref{IsoCylTauStar} we can see that the diagram:
	\[\xymatrix{
	(\tau_{3})_{*}\xi_{0}\vert_{\partial X} \ar[rr]^{(\tau_{3})_{*}\rho\vert_{\partial X}} & & (\tau_{3})_{*}\xi_{1}\vert_{\partial X} \\
	\xi_{0}\vert_{\partial X} \ar[rr]^{\rho\vert_{\partial X}} \ar[u]^{\id} & & \xi_{1}\vert_{\partial X} \ar[u]^{\id}
}\]
does \emph{not} commute or anti-commute. In fact, for $\theta = 0$ we get $\rho(0, \varphi, q') = (\tau_{3})_{*}\rho(0, \varphi,$ $q')$ while for $\theta = \pi$ we get $\rho(\pi, \varphi, q') = -(\tau_{3})_{*}\rho(\pi, \varphi, q')$ since $\tilde{R}_{-\pi} = -\tilde{R}_{\pi}$. The diagram would not commute either by choosing $\rho \circ \gamma$ or $(\tau_{3})_{*}\rho \circ \gamma$ or both.

Some comments about the behavior of $\xi_{0}, \xi_{1}, (\tau_{3})_{*}\xi_{0}, (\tau_{3})_{*}\xi_{0}$ at the boundary are needed in order to avoid possible confusion. If we embed $P_{\SO}(\partial \Cyl) \subset P_{\SO}(T^{2})$ via the outward orthogonal normal unit vector, it follows that $\{(e_{1}, e_{2}\}$ the only orthonormal oriented basis\footnote{Since the boundary has dimension $1$ there is only one oriented orthonormal basis.} at a boundary point $(0, \varphi)$ with $e_{1}$ outward, while for $(\pi, \varphi)$ the only embedded basis is $\{-e_{1}, -e_{2}\}$. Thus, since the principal bundle is $P_{\SO}(T^{2})$ is the bundle of isomorphisms from the trivial bundle $T^{2} \times \mathbb{R}^{2}$ to the tangent bundle $T(T^{2})$, which is also trivial, it follows that the embedded basis for $\theta = 0$ corresponds to $(0, \varphi, \id) \in S^{1} \times S^{1} \times \SO(2)$, while the embedded basis for $\theta = \pi$ corresponds to $(\pi, \varphi, -\id) \in S^{1} \times S^{1} \times \SO(2)$. Thus, is we consider the $\Spin$-bundles $P_{\Spin}(\partial \Cyl) \subset P_{\Spin}(T^{2})$ we have that the lifts of the embedded basis are:
	\[\begin{array}{|l|l|l|} \hline
	& \theta = 0 & \theta = \pi\\ \hline
	\textnormal{$\tilde{\xi}_{0}$-lift:} & (0, \varphi, \pm 1) & (\pi, \varphi, \pm e_{1}e_{2})\\
	\textnormal{$\tilde{\xi}_{1}$-lift:} & (0, \varphi, \pm 1) & (\pi, \varphi, \pm 1)\\
	\textnormal{$(\tau_{3})^{*}\tilde{\xi}_{0}$-lift:} & (0, \varphi, \pm e_{1}) & (\pi, \varphi, \pm (e_{1})^{2}e_{2})\\
	\textnormal{$(\tau_{3})^{*}\tilde{\xi}_{1}$-lift:} & (0, \varphi, \pm e_{1}) & (\pi, \varphi, \pm e_{1})\\ \hline
\end{array}\]
It may seem strange that at the boundary, whose tangent space is generated only by $e_{2}$, also the outward vector $e_{1}$ is involved, but that's due to the fact that on $\pi$ there is a $-1$ to lift due to the orientation and for all the structures different from $\tilde{\xi}_{0}$ there is a twist in the projection of the third factor $\Spin(2) \rightarrow \SO(2)$ which makes $e_{1}$ enter in the lifting. The isomorphisms of spin structures we dealt with until now are then at the boundary:
	\[\begin{array}{lll}
	\rho(0, \varphi, \pm 1) = (0, \varphi, \pm 1) & & \rho(\pi, \varphi, \pm e_{1}e_{2}) = (\pi, \varphi, \mp 1)\\
	(\widetilde{d\tau_{3}})_{0}(0, \varphi, \pm 1) = (0, \varphi, \pm e_{1}) & & (\widetilde{d\tau_{3}})_{0}(\pi, \varphi, \pm e_{1}e_{2}) = (\pi, \varphi, \pm (e_{1})^{2}e_{2})\\
	(\widetilde{d\tau_{3}})_{1}(0, \varphi, \pm 1) = (0, \varphi, \pm e_{1}) & & (\widetilde{d\tau_{3}})_{1}(\pi, \varphi, \pm 1) = (\pi, \varphi, \pm e_{1}).
\end{array}\]
We know that $\tilde{\xi}_{0} \cup_{\id} (\tau_{3})^{*}\tilde{\xi}_{0} \simeq \tilde{\xi}_{0} \cup_{(\widetilde{d\tau_{3}})_{0}} \tilde{\xi}_{0}$ and the same for $\tilde{\xi}_{1}$. Some representatives of the two equivalence classes are then:
	\[\begin{array}{ll}
	\textnormal{Class inducing $\tilde{\xi}_{0}$ on $T^{2}$:} & \tilde{\xi}_{0} \cup_{(\widetilde{d\tau_{3}})_{0}} \tilde{\xi}_{0} \; \simeq \; \tilde{\xi}_{0} \cup_{\id} (\tau_{3})^{*}\tilde{\xi}_{0} \; \simeq \; \tilde{\xi}_{1} \cup_{\id} \tilde{\xi}_{1} \\
	\textnormal{Class inducing $\tilde{\xi}_{1}$ on $T^{2}$:} & \tilde{\xi}_{1} \cup_{(\widetilde{d\tau_{3}})_{1}} \tilde{\xi}_{1} \; \simeq \; \tilde{\xi}_{1} \cup_{\id} (\tau_{3})^{*}\tilde{\xi}_{1} \; \simeq \; \tilde{\xi}_{0} \cup_{\id} \tilde{\xi}_{0}.
\end{array}\]
It is easy to find the invariance conditions for pinors and spinors as sections of the associated vector bundles. Moreover, all this picture is equivalent to considering $(T^{2}, \tau')$; we leave the details to the reader.


\appendix

\chapter{Appendices of Part I}

\section{Direct sum and direct product}\label{DirectSumProd}

We consider abelian groups, but all the discussion applies equally to the case of rings, vector spaces, or in general objects of a fixed abelian category. Given a family of abelian groups $\{G_{\alpha}\}_{\alpha \in I}$, we define the \emph{direct sum}:
	\[\bigoplus_{\alpha \in I} G_{\alpha}
\]
as the group whose elements are families made by one element for each group $G_{\alpha}$, such that only finitely many of them are non-zero; the sum is defined componentwise. Thus, an element of $G$ is a collection $\{g_{\alpha}\}_{\alpha \in I}$ for $g_{\alpha} \in G_{\alpha} \,\forall \alpha \in I$ and such that there exists a \emph{finite} set $J \subset I$ such that $g_{\alpha} = 0 \, \forall \alpha \in I \setminus J$. Instead, we define the \emph{direct product}:
	\[\prod_{\alpha \in I} G_{\alpha}
\]
as the group whose elements are families made by one element for each group $G_{\alpha}$, without any restriction. The direct sum is naturally a subgroup of the direct product; when the family is finite they coincide (in particular, the direct sum and the direct product of two groups coincide).

For $G^{*} := \Hom(G, \mathbb{Z})$ the following relations hold:
	\[\Bigl(\bigoplus_{\alpha \in I} G_{\alpha}\Bigr)^{*} = \prod_{\alpha \in I} G_{\alpha}^{*} \qquad\qquad \Bigl(\prod_{\alpha \in I} G_{\alpha}\Bigr)^{*} \supset \bigoplus_{\alpha \in I} G_{\alpha}^{*}.
\]
In fact, in order to give a homomorphism $\varphi$ from $\bigoplus_{\alpha \in I} G_{\alpha}$ to $\mathbb{Z}$ it is enough to specify its restriction on each single group $G_{\alpha}$, since such groups generates their direct sum; thus, the homomorphism $\varphi$ is specified by a collection $\{\varphi_{\alpha}\}_{\alpha \in I}$ for $\varphi_{\alpha} \in G_{\alpha}^{*} \,\forall \alpha \in I$. We do not have to impose a finiteness condition, since, even if there are infinitely many non-zero homomorphisms in the family, when we apply them to an element of the direct sum they can assume a non-zero value only on the non-zero elements, which are a finite set. That's why every element of $\prod_{\alpha \in I} G_{\alpha}^{*}$ gives a well-defined homomorphism from $\bigoplus_{\alpha \in I} G_{\alpha}$ to $\mathbb{Z}$. Instead, for the direct product, given a family $\{\varphi_{\alpha}\}_{\alpha \in I}$ for $\varphi_{\alpha} \in G_{\alpha}^{*} \,\forall \alpha \in I$, it gives a well-defined homomorphism from $\prod_{\alpha \in I} G_{\alpha}$ to $\mathbb{Z}$ if and only $\varphi_{\alpha} \neq 0$ only for finitely many elements. In fact, let us define $J \subset I$ as the set such that $\varphi_{\alpha} \neq 0$ if and only if $\alpha \in J$ and let us suppose that $J$ is infinite. Then, for each $\alpha \in J$, there exists $g_{\alpha} \in G_{\alpha}$ such $\varphi_{\alpha}(g_{\alpha}) = n_{\alpha} > 0$. If we choose any element $g_{\alpha}$ for $\alpha \in I \setminus J$, we obtain that $\{\varphi_{\alpha}\}(\{g_{\alpha}\})$ is an infinite sum, thus it is not well-defined. That's why only the elements of $\bigoplus_{\alpha \in I} G_{\alpha}^{*}$ give a well-defined homomorphism from $\prod_{\alpha \in I} G_{\alpha}$ to $\mathbb{Z}$. In this case we have just an inclusion, since it is not true that the single groups $G_{\alpha}$ generate the direct product: actually, the subgroup of the direct product generated by the single groups is exactly the direct sum, since in a group we allow only finite sums.

\section{Compactifications}\label{AppCompactifications}

We briefly discuss the notion of compactification of a topological space. Given a topological space $X$, a \emph{compactification} of $X$ is a couple $(\overline{X}, f)$ where $\overline{X}$ is a \emph{compact} topological space and $f: X \rightarrow \overline{X}$ is a continuous map with two properties:
\begin{itemize}
	\item $f$ is a homeomorphism between its domain and its image;
	\item the image of $f$ is \emph{dense} in $\overline{X}$, i.e.\ its closure if the whole $\overline{X}$.
\end{itemize}
Thus, a compactification is an embedding of a topological space in a compact one such that the complement is small with respect to the space itself. A topological space has in general much different compactifications. For example, if $X = \mathbb{R}^{n}$ we can compactify it in at least three natural ways: we can add one point at infinity considering $\overline{X} = S^{n}$ and $f$ the embedding of $\mathbb{R}^{n}$ in $S^{n}$ whose image does not contain the north pole; we can add one point for each half-line going from the origin to infinity, obtaining $\overline{X} = D^{n}$ and $f$ the embedding of $\mathbb{R}^{n}$ as the interior of $D^{n}$; we can add one point for each line passing through the origin, obtaining the real projective plane $\overline{X} = \mathbb{RP}^{n}$ and $f$ the natural embedding of $\mathbb{R}^{n}$ in $\mathbb{RP}^{n}$ excluding the ``projective hyperplane at infinity''.

While uniqueness definitely fails, existence is instead guaranteed. In fact, for every space $X$ there exists its \emph{one-point compactification} $X^{+}$, also called \emph{Alexandrov compactification}, which consists in adding one point at infinity, as in the first example considered for $\mathbb{R}^{n}$. It is defined as follows: we consider as as set the disjoint union of $X$ with a point, i.e.\ $\overline{X} := X \sqcup \{\infty\}$, as we give $\overline{X}$ the topology such that $A \subset \overline{X}$ is open if and only if one of the two following conditions is satisfied:
\begin{itemize}
	\item $A \subset X$ and $A$ is open in $X$;
	\item $A = Y \cup \{\infty\}$ for $Y \subset X$ and $X \setminus Y$ compact.
\end{itemize}
If $X$ is locally compact, as any finite-dimensional manifold, then $\overline{X}$ is a Hausdorff space. We consider as $f$ the immersion. This is a well-defined compactification, but, for example, if the space-time is $\mathbb{R}^{1,3}$ we prefer to consider one point at infinity for each direction, so it seems more natural to consider $D^{4}$ instead of $S^{4}$ as a compactification.

If we analyze the three examples considered for $\mathbb{R}^{n}$, we see that $D^{n}$ is in a certain sense bigger than $S^{n}$ and $\mathbb{RP}^{n}$: more precisely, both $S^{n}$ and $\mathbb{RP}^{n}$ can be obtained as a quotient of $D^{n}$ by an equivalence relation on the infinity part. For example, $S^{n} \simeq D^{n} \,/\, \partial D^{n}$ and $\partial D^{n}$ is exactly the infinity with respect to the embedding considered; similarly $\mathbb{RP}^{n} \simeq D^{n} \,/ \sim$ where $\sim$ identifies antipodal points of $\partial D^{n}$. Thus one might suspect that $D^{n}$ is maximal in the family of compactifications of $\mathbb{R}^{n}$, in the sense that it adds the biggest set of points in order for $X$ to be dense in $\overline{X}$, and that this maximal compactification is the one we can consider to definite infinity for a generic space-time. Actually the situation is different: in particular, such a maximal compactification, called \emph{Stone-$\check{C}$ech compactification}, exists for any space, but it is more complicated than one might think; in particular, the maximal compactification of $\mathbb{R}^{n}$ is much bigger than $D^{n}$, so it is not physically reasonable at all. We just sketch the reason, for details the reader can see \cite{Dugundji}. To define a maximal compactification of a space $X$ we ask that any continuous function $\varphi: X \rightarrow Y$ can be extended to $\overline{\varphi}: \overline{X} \rightarrow Y$ such that $\varphi = f \circ \overline{\varphi}$, so that the maximally compactified space contains one point for each possible direction at infinity of any continuous function. That's why, if we consider for example $X = (0, 1]$, its maximal compactification cannot be $[0,1]$ as we naively expect: the function $\varphi: (0,1] \rightarrow \mathbb{R}$ given by $\varphi(t) = \sin\frac{1}{t}$ cannot be extended to $[0,1]$, thus on $\overline{X}$ we need one point also for such irregular functions. Instead, if we compactify $\mathbb{R}^{n}$ to $D^{n}$, we add a point at infinity only for lines or functions going to infinity as lines, which are surely not the most generic continuous functions.

We are thus forced to avoid maximal compactification of a generic space, so we consider the specific case of manifolds and search a compactification which is still a manifold (with boundary). If we compactify $\mathbb{R}^{n}$ to $D^{n}$ we obtain a manifold with boundary such that $\mathbb{R}^{n} = D^{n} \setminus \partial D^{n}$. We search for a generic situation like this. For a manifold $X$ without boundary we define the \emph{collar compactification} of $X$ as a manifold with boundary $\overline{X}$ with a diffeomorphism $\varphi: X \rightarrow \overline{X} \setminus \partial \overline{X}$. Under this assumption $X$ and $\overline{X}$ are homotopic: in fact, by the collar neighborhood theorem \cite{Hirsh} there is a collar neighborhood of $\partial \overline{X}$ in $\overline{X}$, which is by definition a neighborhood $U$ diffeomorphic to $\partial \overline{X} \times [0,1)$. Now we can retract both $X$ and $\overline{X}$ to the same compact submanifold obtained retracting $U$ to the image of $\partial \overline{X} \times [\frac{1}{2},1)$ under the diffeomorphism, thus $X$ and $\overline{X}$ are homotopic. Not every open manifold admits a collar compactification. In fact, if it exists, we have shown that $X$ is homotopic to $\overline{X}$, thus, being $\overline{X}$ compact, it is homotopic to a finite CW-complex. There are manifold not homotopic to a finite CW-complex: one counterexample is given by a surface with infinite genus. In this case $H_{1}(X, \mathbb{Z}) = \mathbb{Z}^{\aleph_{0}}$, where $\aleph_{0}$ is the cardinality of countable infinite sets as $\mathbb{N}$ or $\mathbb{Z}$, in particular $H_{1}(X, \mathbb{Z})$ is not finitely generated, hence $X$ cannot be homotopic to a finite CW-complex.

Although existence is not guaranteed, so that we must assume it as an hypothesis for our background manifolds, we can actually prove uniqueness, so that, if the hypothesis holds, infinity is intrinsically determined by the space-time itself, it is not an additional data. In fact, let us consider a manifold $X$ and a collar compactification $\overline{X}$. We call $U$ a collar neighborhood of $\partial \overline{X}$ in $\overline{X}$ and $\varphi: \partial \overline{X} \times [0,1) \rightarrow U$ the diffeomorphism realizing the definition. We call $\tilde{X} := \varphi(\partial \overline{X} \times (\frac{1}{2},1))$. Then $\tilde{X}$ is a sumbanifold with boundary and $X = \tilde{X} \cup \varphi\bigl(\partial \overline{X} \times (0,\frac{1}{2}]\bigr)$ and $\overline{X} = \tilde{X} \cup \varphi\bigl(\partial \overline{X} \times [0,\frac{1}{2}]\bigr)$. Let us consider another collar compactification $\overline{X'}$. Then its boundary $\partial \overline{X'}$ is made by the limit points of the segments $\varphi(\{p\} \times (0,\frac{1}{2}])$ near $0$, so that $\overline{X'} = \tilde{X} \cup \varphi\bigl(\partial \overline{X} \times (0,\frac{1}{2}]\bigr) \cup \partial \overline{X'}$. Thus it is enough to prove that:
\begin{equation}\label{CollarCpt}
	\textstyle \bigl(\partial \overline{X} \times [0,\frac{1}{2}]\bigr) \simeq \bigl(\partial \overline{X} \times (0,\frac{1}{2}]\bigr) \cup \partial \overline{X'}.
\end{equation}
The r.h.s., which we call $Y$, is a cylinder with lower base $\partial \overline{X}$ and upper base $\partial \overline{X'}$ which is a disjoint union of manifolds with boundary (one component for each end of $X$). We consider the function $f: Y \rightarrow [0,\frac{1}{2}]$ given by $f(\varphi(p, t)) = t$ for $p \in \partial \overline{X}$ and $f(q) = \frac{1}{2}$ for $q \in \partial \overline{X'}$. It is smooth since it extends naturally the function $f(\varphi(p, t)) = t$ to the compactification (one can see in local charts near the upper boundary that it is smooth), thus it is a Morse function for $Y$ \cite{Milnor}. It has no critical points since along the vertical direction of the cylinder it has derivative $1$: then, as proven in \cite{Milnor}, $Y$ must be the trivial cobordism, i.e.\ it must be diffeomorphic to $\textstyle \bigl(\partial \overline{X} \times [0,\frac{1}{2}]\bigr)$ which is the l.h.s. of \eqref{CollarCpt}. \footnote{We remark that in order to find the collar compactification of a manifold it is not always correct to embed it as subset of $\mathbb{R}^{n}$ for some $n$ and consider its closure in $D^{n}$. In fact, for example, if we consider the unit ball with one hole $B_{1}^{n} \setminus \{0\}$, it is naturally embedded in $\mathbb{R}^{n}$ by the immersion, but its closure becomes $D_{1}^{n}$ and $D_{1}^{n} \setminus \partial D_{1}^{n} = B_{1}^{n}$ which is not diffeomorphic to $B_{1}^{n} \setminus \{0\}$: that's because the origin becomes an interior point of the closure. Instead, if we embed $B_{1}^{n} \setminus \{0\}$ in $\mathbb{R}^{n}$ as $B_{1}^{n} \setminus D_{\frac{1}{2}}^{n}$, then we obtain the right closure.} That's why the infinity manifold, if we suppose it exists, is intrinsically determined by the space-time manifold itself.

\chapter{Appendices of Part II}

\section{$\rm\bf\check{C}$ech Hypercohomology}\label{AppHyperC}

We refer to \cite{Brylinski} for a comprehensive treatment of hypercohomology. Given a sheaf $\mathcal{F}$ on a topological space $X$ with a good cover $\mathfrak{U} = \{U_{i}\}_{i \in I}$, we construct the complex of $\rm\check{C}$ech cochains:
	\[\check{C}^{0}(\mathfrak{U}, \mathcal{F}) \overset{\check{\delta}^{0}}\longrightarrow \check{C}^{1}(\mathfrak{U}, \mathcal{F}) \overset{\check{\delta}^{1}}\longrightarrow \check{C}^{2}(\mathfrak{U}, \mathcal{F}) \overset{\check{\delta}^{2}}\longrightarrow \cdots
\]
whose cohomology is by definition $\rm\check{C}$ech cohomology of $\mathcal{F}$. We recall, in particular, that $\check{\delta}^{p}: \check{C}^{p}(\mathfrak{U}, \mathcal{F}) \rightarrow \check{C}^{p+1}(\mathfrak{U}, \mathcal{F})$ is defined by $(\check{\delta}^{p}g)_{\alpha_{0} \cdots \alpha_{p+1}} = \sum_{i=0}^{p+1} (-1)^{i}g_{\alpha_{0} \cdots \check{\alpha}_{i} \cdots \alpha_{p+1}}$. If, instead of a single sheaf, we have a complex of sheaves:
	\[\cdots \overset{d^{i-2}}\longrightarrow \mathcal{F}^{i-1} \overset{d^{i-1}}\longrightarrow \mathcal{F}^{i} \overset{d^{i}}\longrightarrow \mathcal{F}^{i+1} \overset{d^{i+1}}\longrightarrow \cdots
\]
we can still associate to it a cohomology, called \emph{hypercohomology} of the complex. To define it, we consider the double complex made by the $\rm\check{C}$ech complexes of each sheaf:
\[\xymatrix{
	\vdots & \vdots & \vdots & \\
	\check{C}^{0}(\mathfrak{U}, \mathcal{F}^{q+1}) \ar[r]^{\check{\delta}^{0}} \ar[u]^{d^{q+1}} & \check{C}^{1}(\mathfrak{U}, \mathcal{F}^{q+1}) \ar[r]^{\check{\delta}^{1}} \ar[u]^{d^{q+1}} & \check{C}^{2}(\mathfrak{U}, \mathcal{F}^{q+1}) \ar[r]^{\phantom{XXX}\check{\delta}^{2}} \ar[u]^{d^{q+1}} & \cdots \\
	\check{C}^{0}(\mathfrak{U}, \mathcal{F}^{q}) \ar[r]^{\check{\delta}^{0}} \ar[u]^{d^{q}} & \check{C}^{1}(\mathfrak{U}, \mathcal{F}^{q}) \ar[r]^{\check{\delta}^{1}} \ar[u]^{d^{q}} & \check{C}^{2}(\mathfrak{U}, \mathcal{F}^{q})  \ar[r]^{\phantom{XXX}\check{\delta}^{2}} \ar[u]^{d^{q}} & \cdots \\
	\check{C}^{0}(\mathfrak{U}, \mathcal{F}^{q-1}) \ar[r]^{\check{\delta}^{0}} \ar[u]^{d^{q-1}} & \check{C}^{1}(\mathfrak{U}, \mathcal{F}^{q-1}) \ar[r]^{\check{\delta}^{1}} \ar[u]^{d^{q-1}} & \check{C}^{2}(\mathfrak{U}, \mathcal{F}^{q-1}) \ar[r]^{\phantom{XXX}\check{\delta}^{2}} \ar[u]^{d^{q-1}} & \cdots\\
	\vdots \ar[u]^{d^{q-2}} & \vdots \ar[u]^{d^{q-2}} & \vdots \ar[u]^{d^{q-2}} & 
	}
\]
We now consider the associated total complex\footnote{We use notation of \cite{Brylinski}, in which the two boundaries of the double complex commute, so that the boundary of the total complex has a factor $(-1)^{p}$. In the most common notation the two boundaries anticommute.}:
	\[T^{n} = \bigoplus_{p+q = n} \check{C}^{p}(\mathfrak{U}, \mathcal{F}^{q}) \qquad\quad d^{n} = \bigoplus_{p+q = n} \bigl(\,\check{\delta}^{p} + (-1)^{p}\,d^{q}\,\bigr)
\]
By definition, the \emph{$\rm\check{C}$ech hypercohomology} of the complex of sheaves is the cohomology of the total complex $H^{\bullet}(T^{n}, d^{n})$. It is denoted by:
	\[\check{H}^{\bullet}\bigl( \, \mathfrak{U}, \, \cdots \overset{d^{i-1}}\longrightarrow \mathcal{F}^{i} \overset{d^{i}}\longrightarrow \mathcal{F}^{i+1} \overset{d^{i+1}}\longrightarrow \cdots \, \bigr).
\]

Using hypercohomology we can describe the group of line bundles with connection, up to isomorphism and pull-back of the connection, on a space $X$. We recall that a bundle with connection is specified by a couple $(\{h_{\alpha\beta}\}, \{A_{\alpha}\})$ where $\check{\delta}\{h_{\alpha\beta}\} = 1$ and $A_{\alpha} - A_{\beta} = (2\pi i)^{-1} d \log h_{\alpha\beta}$. The bundle is trivial if there exists a 0-cochain $\{f_{\alpha}\}$ such that $\check{\delta}^{0}\{f_{\alpha}\} = \{h_{\alpha\beta}\}$. Let us consider the complex of sheaves on $X$:
	\[\underline{S}^{1} \overset{\tilde{d}}\longrightarrow \Omega^{1}_{\mathbb{R}}
\]
where $\underline{S}^{1}$ is the sheaf of smooth $S^{1}$-valued functions, $\Omega^{1}_{\mathbb{R}}$ the sheaf of $1$-forms and $\tilde{d} = (2\pi i)^{-1} \, d\circ \log$. (The complex is trivially extended on left and right by $0$.) The associated $\rm\check{C}$ech double complex is given by:
\[\xymatrix{
	\check{C}^{0}(\mathfrak{U}, \Omega^{1}_{\mathbb{R}}) \ar[r]^{\check{\delta}^{0}} & \check{C}^{1}(\mathfrak{U}, \Omega^{1}_{\mathbb{R}}) \ar[r]^{\check{\delta}^{1}} & \check{C}^{2}(\mathfrak{U}, \Omega^{1}_{\mathbb{R}})  \ar[r]^{\phantom{XXX}\check{\delta}^{2}} & \cdots \\
	\check{C}^{0}(\mathfrak{U}, \underline{S}^{1}) \ar[r]^{\check{\delta}^{0}} \ar[u]^{\tilde{d}} & \check{C}^{1}(\mathfrak{U}, \underline{S}^{1}) \ar[r]^{\check{\delta}^{1}} \ar[u]^{\tilde{d}} & \check{C}^{2}(\mathfrak{U}, \underline{S}^{1}) \ar[r]^{\phantom{XXX}\check{\delta}^{2}} \ar[u]^{\tilde{d}} & \cdots
	}
\]
Thus we have that $\check{C}^{1}(\mathfrak{U}, \underline{S}^{1} \rightarrow \Omega^{1}_{\mathbb{R}}) = \check{C}^{1}(\mathfrak{U}, \underline{S}^{1}) \oplus \check{C}^{0}(\mathfrak{U}, \Omega^{1}_{\mathbb{R}})$. Given a line bundle $L \rightarrow X$ we fix a set of local sections, with respect to $\mathfrak{U}$, determining transition functions $\{g_{\alpha\beta}\}$ and local representation of the connection $\{A_{\alpha}\}$. We claim that $(g_{\alpha\beta}, -A_{\alpha}) \in \check{C}^{1}(\mathfrak{U}, \underline{S}^{1} \rightarrow \Omega^{1}_{\mathbb{R}})$ is a cocycle. In fact, by definition, $\check{\delta}^{1}(g_{\alpha\beta}, -A_{\alpha}) = (\check{\delta}^{1}g_{\alpha\beta}, -\tilde{d}g_{\alpha\beta} + \check{\delta}^{0}(-A_{\alpha}))$, thus cocycle condition gives $\check{\delta}^{1}g_{\alpha\beta} = 0$, i.e.\ $g_{\alpha\beta}$ must be transition functions of a line bundle, and $A_{\alpha} - A_{\beta} = (2\pi i)^{-1}d\log g_{\alpha\beta}$, the latter being exactly the gauge transformation of a connection. Moreover, coboundaries are of the form $\check{\delta}^{0}(g_{\alpha}) = (\check{\delta}^{0}g_{\alpha}, \tilde{d}g_{\alpha})$ and it is easy to prove that these are exactly the possible local representations of the trivial connection $\partial_{X}$ on the trivial bundle $X \times \mathbb{C}$, i.e.\ the unit element of the group of line bundles with connection. Thus, such a group is isomorphic to:
	\[\check{H}^{1}(\mathfrak{U}, \underline{S}^{1} \overset{\tilde{d}}\longrightarrow \Omega^{1}_{\mathbb{R}}).
\]

\section{Gerbes}\label{AppGerbes}

We refer to \cite{Hitchin} for a clear introduction to gerbes. A gerbe with connection is defined by a triple $(\{g_{\alpha\beta\gamma}\}, \{\Lambda_{\alpha\beta}\}, \{B_{\alpha}\})$ where $\check{\delta}\{g_{\alpha\beta\gamma}\} = 1$, $\check{\delta}^{1}\{\Lambda_{\alpha\beta}\} = \{(2\pi i)^{-1}d \log g_{\alpha\beta\gamma}\}$ and $B_{\alpha} - B_{\beta} = d\Lambda_{\alpha\beta}$. The gerbe is trivial if there exists a 1-cochain $\{f_{\alpha\beta}\}$ such that $\check{\delta}\{f_{\alpha\beta}\} = \{g_{\alpha\beta\gamma}\}$. We use the approach of \cite{Brylinski}. As the group of isomorphism classes of line bundles on $X$ is isomorphic to $\check{H}^{1}(X, \underline{S}^{1})$, the group of gerbes on $X$ up to isomorphism can be identified with $\check{H}^{2}(X, \underline{S}^{1})$. Here we consider this as the definition of gerbe.

We consider the complex of sheaves:
	\[\underline{S}^{1} \overset{\tilde{d}}\longrightarrow \Omega^{1}_{\mathbb{R}} \overset{d}\longrightarrow \Omega^{2}_{\mathbb{R}}
\]
where $\underline{S}^{1}$ is the sheaf of smooth $S^{1}$-valued functions, $\Omega^{p}_{\mathbb{R}}$ the sheaf of $p$-forms and $\tilde{d} = (2\pi i)^{-1} \, d\circ \log$. (The complex is trivially extended on left and right by $0$.) In analogy with the case of line bundles, we define the equivalence classes of gerbes with connection as the elements of the group:
	\[\check{H}^{2}(X, \underline{S}^{1} \rightarrow \Omega^{1}_{\mathbb{R}} \rightarrow \Omega^{2}_{\mathbb{R}}).
\]
The $\rm\check{C}$ech double complex is given by:
\[\xymatrix{
	\check{C}^{0}(\mathfrak{U}, \Omega^{2}_{\mathbb{R}}) \ar[r]^{\check{\delta}^{0}} & \check{C}^{1}(\mathfrak{U}, \Omega^{2}_{\mathbb{R}}) \ar[r]^{\check{\delta}^{1}} & \check{C}^{2}(\mathfrak{U}, \Omega^{2}_{\mathbb{R}})  \ar[r]^{\phantom{XXX}\check{\delta}^{2}} & \cdots \\
	\check{C}^{0}(\mathfrak{U}, \Omega^{1}_{\mathbb{R}}) \ar[r]^{\check{\delta}^{0}} \ar[u]^{d} & \check{C}^{1}(\mathfrak{U}, \Omega^{1}_{\mathbb{R}}) \ar[r]^{\check{\delta}^{1}} \ar[u]^{d} & \check{C}^{2}(\mathfrak{U}, \Omega^{1}_{\mathbb{R}}) \ar[r]^{\phantom{XXX}\check{\delta}^{2}} \ar[u]^{d} & \cdots \\
	\check{C}^{0}(\mathfrak{U}, \underline{S}^{1}) \ar[r]^{\check{\delta}^{0}} \ar[u]^{\tilde{d}} & \check{C}^{1}(\mathfrak{U}, \underline{S}^{1}) \ar[r]^{\check{\delta}^{1}} \ar[u]^{\tilde{d}} & \check{C}^{2}(\mathfrak{U}, \underline{S}^{1}) \ar[r]^{\phantom{XXX}\check{\delta}^{2}} \ar[u]^{\tilde{d}} & \cdots
	}
\]
Thus we have that $\check{C}^{2}(\mathfrak{U}, \underline{S}^{1} \rightarrow \Omega^{1}_{\mathbb{R}} \rightarrow \Omega^{2}_{\mathbb{R}}) = \check{C}^{2}(\mathfrak{U}, \underline{S}^{1}) \oplus \check{C}^{1}(\mathfrak{U}, \Omega^{1}_{\mathbb{R}}) \oplus \check{C}^{0}(\mathfrak{U}, \Omega^{2}_{\mathbb{R}})$. By definition, $\check{\delta}^{1}(g_{\alpha\beta\gamma}, -\Lambda_{\alpha\beta}, B_{\alpha}) = (\check{\delta}^{2}g_{\alpha\beta\gamma}, \tilde{d}g_{\alpha\beta\gamma} + \check{\delta}^{1} (-\Lambda_{\alpha\beta}), -d(-\Lambda_{\alpha\beta}) + \check{\delta}^{0}B_{\alpha})$. Thus cocycle condition gives $\check{\delta}^{2}g_{\alpha\beta\gamma} = 0$, i.e.\ $g_{\alpha\beta\gamma}$ must be transition functions of a gerbe, and:
	\[\begin{split}
	&B_{\alpha} - B_{\beta} = d\Lambda_{\alpha\beta}\\
	&\Lambda_{\alpha\beta} + \Lambda_{\beta\gamma} + \Lambda_{\gamma\alpha} = (2\pi i)^{-1} d \log g_{\alpha\beta\gamma}.
\end{split}\]
Coboundaries are of the form $\check{\delta}^{1}(h_{\alpha\beta}, -A_{\alpha}) = (\check{\delta}^{1}h_{\alpha\beta}, -\tilde{d}h_{\alpha\beta} + \check{\delta}^{0}(-A_{\alpha}), d(-A_{\alpha}))$, thus gerbes of this form are geometrically trivial.

\chapter{Appendices of Part III}

\section{p-Gerbes}\label{AppPGerbes}

We refer to appendix B of \cite{BFS} and references therein for an introduction to gerbes. Here we just generalize the discussion to $p$-gerbes. In particular, we recall that a gerbe with connection is given by an element of the $\rm\check{C}$ech hypercohomology group:
	\[\check{H}^{2}(X, \underline{S}^{1} \rightarrow \Omega^{1}_{\mathbb{R}} \rightarrow \Omega^{2}_{\mathbb{R}}).
\]
We thus define a $p$-gerbe with connection as an element of the $\rm\check{C}$ech hypercohomology group:
	\[\check{H}^{p+1}(X, \underline{S}^{1} \rightarrow \Omega^{1}_{\mathbb{R}} \rightarrow \cdots \rightarrow \Omega^{p+1}_{\mathbb{R}}).
\]
The $\rm\check{C}$ech double complex with respect to a good cover $\mathfrak{U}$ is given by:
\[\xymatrix{
	\check{C}^{0}(\mathfrak{U}, \Omega^{p+1}_{\mathbb{R}}) \ar[r]^{\check{\delta}^{0}} & \check{C}^{1}(\mathfrak{U}, \Omega^{p+1}_{\mathbb{R}}) \ar[r]^{\check{\delta}^{1}} & \check{C}^{2}(\mathfrak{U}, \Omega^{p+1}_{\mathbb{R}})  \ar[r]^{\phantom{XXX}\check{\delta}^{2}} & \cdots \\
	\qquad\vdots\qquad \ar[r]^{\check{\delta}^{0}} \ar[u]^{d} & \qquad\vdots\qquad \ar[r]^{\check{\delta}^{1}} \ar[u]^{d} & \qquad\vdots\qquad \ar[r]^{\phantom{XXX}\check{\delta}^{2}} \ar[u]^{d} & \cdots \\
	\check{C}^{0}(\mathfrak{U}, \Omega^{1}_{\mathbb{R}}) \ar[r]^{\check{\delta}^{0}} \ar[u]^{d} & \check{C}^{1}(\mathfrak{U}, \Omega^{1}_{\mathbb{R}}) \ar[r]^{\check{\delta}^{1}} \ar[u]^{d} & \check{C}^{2}(\mathfrak{U}, \Omega^{1}_{\mathbb{R}}) \ar[r]^{\phantom{XXX}\check{\delta}^{2}} \ar[u]^{d} & \cdots \\
	\check{C}^{0}(\mathfrak{U}, \underline{S}^{1}) \ar[r]^{\check{\delta}^{0}} \ar[u]^{\tilde{d}} & \check{C}^{1}(\mathfrak{U}, \underline{S}^{1}) \ar[r]^{\check{\delta}^{1}} \ar[u]^{\tilde{d}} & \check{C}^{2}(\mathfrak{U}, \underline{S}^{1}) \ar[r]^{\phantom{XXX}\check{\delta}^{2}} \ar[u]^{\tilde{d}} & \cdots
	}
\]
so that $\check{C}^{p+1}(\mathfrak{U}, \underline{S}^{1} \rightarrow \Omega^{1}_{\mathbb{R}} \rightarrow \cdots \rightarrow \Omega^{p+1}_{\mathbb{R}}) = \check{C}^{p+1}(\mathfrak{U}, \underline{S}^{1}) \oplus \check{C}^{p}(\mathfrak{U}, \Omega^{1}_{\mathbb{R}}) \oplus \cdots \oplus \check{C}^{0}(\mathfrak{U}, \Omega^{p+1}_{\mathbb{R}})$. Thus, a representative hypercocycle of a gerbe with connection is a sequence $(g_{\alpha_{0} \cdots \alpha_{p+1}},$ $(C_{1})_{\alpha_{0} \cdots \alpha_{p}}, \ldots, (C_{p+1})_{\alpha_{0}})$, while summing an hypercoboundary represents a gauge transformation. It is easy to verify that for hypercocycles the local forms $dC_{p+1}$ glue to a global one $G_{p+2}$ which is the curvature of the gerbe. Thus, the data of the superstring background must be an equivalence class like this one, not only $C_{p+1}$.

Given a $p$-gerbe with connection $[(g_{\alpha_{0} \cdots \alpha_{p+1}}, (C_{1})_{\alpha_{0} \cdots \alpha_{p}}, \ldots, (C_{p+1})_{\alpha_{0}})]$, we can forget the connection and consider just the $p$-gerbe $\mathcal{G} = [g_{\alpha_{0} \cdots \alpha_{p+1}}] \in \check{H}^{p+1}(X, \underline{S}^{1})$. Then we can define the \emph{first Chern class} $c_{1}(\mathcal{G}) \in \check{H}^{p+2}(X, \mathbb{Z})$: we write the transition functions as $g_{\alpha_{0} \cdots \alpha_{p+1}} = e^{2\pi i \rho_{\alpha_{0} \cdots \alpha_{p+1}}}$ so that $\check{\delta}\{\rho_{\alpha_{0} \cdots \alpha_{p+1}}\} = \{c_{\alpha_{0} \cdots \alpha_{p+1} \alpha_{p+1}}\}$ with $c_{\alpha_{0} \cdots \alpha_{p+1} \alpha_{p+1}} \in \mathbb{Z}$. We then consider $c_{1}(\mathcal{G}) := [\{c_{\alpha_{0} \cdots \alpha_{p+1} \alpha_{p+1}}\}] \in \check{H}^{p+2}(X, \mathbb{Z})$. One can see that the de-Rham cohomology class of the curvature $G_{p+2}$ corresponds to $c_{1}(\mathcal{G}) \otimes_{\mathbb{Z}} \mathbb{R}$ under the canonical isomorphism between de-Rham cohomology and $\rm\check{C}$ech cohomology of the constant sheaf $\mathbb{R}$.

\section{Hodge-$*$ with minkowskian signature}\label{HodgeMinkowski}

Let $V$ be an oriented vector space of dimension $n$ with a fixed \emph{euclidean} metric. We recall that Hodge-$*$ operation is defined on the exterior algebra $\Lambda^{\bullet}V^{*}$ by:
\begin{equation}\label{HodgeStar}
	\alpha \wedge *\beta = \langle \alpha, \beta \rangle \cdot \vol
\end{equation}
where $\vol$ is the unit oriented volume form, given by $\vol = e_{1}^{*} \wedge \ldots \wedge e_{n}^{*}$ for $\{e_{1}, \ldots, e_{n}\}$ an oriented orthonormal basis. In particular, for $\alpha = e_{i_{1}}^{*} \wedge \ldots \wedge e_{i_{p}}^{*}$, equation \eqref{HodgeStar} with $\beta = \alpha$ gives $*\alpha = \varepsilon^{i_{1} \cdots i_{p} j_{1} \cdots j_{n-p}} e_{j_{1}}^{*} \wedge \ldots \wedge e_{j_{n-p}}^{*}$.

If the metric is minkowskian, definition of Hodge-$*$ via \eqref{HodgeStar} still holds. Moreover, the volume form is the same, i.e.\ $\vol = e_{0}^{*} \wedge \ldots \wedge e_{n-1}^{*}$ for $\{e_{0}, \ldots, e_{n-1}\}$ an oriented orthonormal basis, although it has square-norm $-1$ (to correct this we should multiply it by $i$, but we are on a real vector space). We use the convention $\norm{e_{0}}^{2} = -1$. In this case, there is sometimes, but not always, a sign change with respect to the euclidean case. For example, $*(e_{1}^{*} \wedge \ldots \wedge e_{n-1}^{*}) = (-1)^{n-1}e_{0}^{*}$, exactly as in the euclidean case, since \eqref{HodgeStar} becomes, for $\alpha = \beta = (e_{1}^{*} \wedge \ldots \wedge e_{n-1}^{*})$:
	\[(e_{1}^{*} \wedge \ldots \wedge e_{n-1}^{*}) \wedge (-1)^{n-1}e_{0}^{*} = \langle (e_{1}^{*} \wedge \ldots \wedge e_{n-1}^{*}), (e_{1}^{*} \wedge \ldots \wedge e_{n-1}^{*}) \rangle \cdot \vol
\]
which is true since both the l.h.s. and the r.h.s. are equal to the volume form. Instead, $*(e_{0}^{*}) = -e_{1}^{*} \wedge \ldots \wedge e_{n-1}^{*}$, while in the euclidean case there is no minus sign. In fact, \eqref{HodgeStar} becomes, for $\alpha = \beta = e_{0}^{*}$:
	\[e_{0}^{*} \wedge (-e_{1}^{*} \wedge \ldots \wedge e_{n-1}^{*}) = \langle e_{0}^{*}, e_{0}^{*} \rangle \vol
\]
and this is true because $\langle e_{0}^{*}, e_{0}^{*} \rangle = -1$ in the minkoskian case. Thus, given a summand $x \cdot e_{i_{1}}^{*} \wedge \ldots \wedge e_{i_{p}}^{*}$, its Hodge duals in the euclidean and minkowskian cases are equal if $0$ is not one of the indices $i_{1}, \ldots, i_{p}$, and they are opposite otherwise. In particular, since applying $*^{2}$ to any summand of this form we get the index $0$ one of the two times, it follows that $*^{2}$ in the minkowskian and euclidean cases are always opposite (we recall that in the eulidean case $*^{2}\vert_{\Lambda^{p}V^{*}} = (-1)^{p(n-p)}$, as it is easy to verify).


\end{document}